\documentclass[10pt,a4paper,english,amssymb,pra,nobalancelastpage,nofootinbib,noshowpacs,notitlepage]{revtex4-1}
\usepackage[latin9]{inputenc}
\usepackage{color}
\usepackage{babel}
\usepackage{amsmath}
\usepackage{amssymb}
\usepackage{graphicx}
\usepackage[unicode=true,pdfusetitle,
 bookmarks=true,bookmarksnumbered=false,bookmarksopen=false,
 breaklinks=false,pdfborder={0 0 0},pdfborderstyle={},backref=false,colorlinks=true]
 {hyperref}
\hypersetup{
 linkcolor=blue, urlcolor=blue, citecolor=blue, pdfstartview={FitH}, hyperfootnotes=blue}
\usepackage{breakurl}

\makeatletter

\usepackage{amsfonts}
\providecommand{\U}[1]{\protect\rule{.1in}{.1in}}
\ifx\pdfoutput\relax\let\pdfoutput=\undefined\fi
\newcount\msipdfoutput
\ifx\pdfoutput\undefined\else
\ifcase\pdfoutput\else
\ifx\paperwidth\undefined\else
\ifdim\paperheight=0pt\relax\else\pdfpageheight\paperheight\fi
\ifdim\paperwidth=0pt\relax\else\pdfpagewidth\paperwidth\fi
\fi\fi\fi
\makeatother

\begin{document}

\title{Introduction to quantum optics}

\author{Carlos Navarrete-Benlloch}
\email{derekkorg@gmail.com}

\homepage{www.carlosnb.com}

\affiliation{Wilczek Quantum Center - School of Physics and Astronomy - Shanghai Jiao Tong University, China
\\
Max Planck Institute for the Science of Light - Erlangen, Germany}

\date{Spring Semester 2022 {[}last updated on March 8, 2022{]}}

\begin{abstract}
These are the lecture notes for a course that I am teaching at Zhiyuan College of Shanghai Jiao Tong University (available at www.youtube.com/derekkorg), though the first draft was created for a previous course I taught at the University of Erlangen-Nuremberg in Germany. It has been designed for students who have only had basic training on quantum mechanics, and hence, the course is suited for people at all levels (say, from the end of the bachelor all the way into the PhD). The notes are a work in progress, meaning that some proofs and many figures are still missing. However, I've tried my best to write everything in such a way that a reader can follow naturally all arguments and derivations even with these missing bits. Also a few chapters are left to add, including one on mathematical methods to analyze the dynamics of open systems, and another introducing the plethora of current experimental platforms where the tools and ideas developed in these notes are being currently implemented.

Let me start with a few words about the topic of the lectures. Quantum
optics treats the interaction between light and matter. We may think
of light as the optical part of the electromagnetic spectrum, and
matter as atoms. However, modern quantum optics covers a wild variety
of systems, so that a more timely definition could be ``quantum electrodynamics
at low energies''. Such scenario includes, for example, superconducting
circuits, confined electrons, excitons in semiconductors, defects
in solid state, or the center-of-mass motion of micro-, meso-, and
macroscopic systems. Moreover, quantum optics is at the heart of the
exponentially-growing field of quantum information processing and
communication, both at the conceptual level and at the level of technological
implementations. The ideas and experiments developed in quantum optics
have also allowed us to take a fresh look at many-body problems of
relevance for condensed matter and even high-energy physics. In addition,
quantum optics holds the promise of testing foundational problems
in quantum mechanics as well as physics beyond the standard model
in table-sized experiments. One of the distinct features of quantum
optics is that it deals with systems that are not isolated, that is,
they leak out energy and information to their surrounding environment.
While this is actually the most common situation in real physical
systems, it is not the one students usually encounter in their standard
quantum mechanics courses. A big part of this course is devoted to
fill this gap: it goes through many of the tools and methods that
have been developed to describe open quantum optical systems. Apart
from their practical use, these methods also have deep physical interpretations
which will make students understand quantum mechanics much better.
Quantum optics and open systems are therefore topics that no future
researcher in quantum physics should miss.

I cannot emphasize enough how important it is to read as much as possible
about these topics in order to mature into the best version of a quantum
physicist that one can be. Therefore, I conclude with a list of references
that I have found especially useful at different points of my career
\cite{QO9,QO1,QO10,QO11,QO12,QO13,QO14,QO2,QO3,QO4,QO5,QO6,QO7,QO8,SchleichBook,ShapiroNotes,Weedbrook12,BraunsteinVanLoockReview,CerfBook07,ParisBook,KokBook10}.

Finally, I would like to acknowledge the many students who have thoroughly gone through these lecture notes over the last years and help me smoothen them, as well as several colleagues that have made suggestions to improve them, or that have spread them among their students.

\end{abstract}
\maketitle
\tableofcontents{}

\bigskip{}
\newpage{}

\section{Brief review of quantum mechanics}

In this short introductory chapter we will quickly review the basic
elements of quantum mechanics required to follow the rest of the notes.
This summary is just intended as a reminder for students who have
already studied a course on quantum mechanics, as well as courses
in linear algebra and classical mechanics. Nevertheless, in order
to make the course self-contained and make sure that all students
are familiar with all the concepts and notation used throughout the
course, in Appendix \ref{QuantumMechanics} these are reviewed in
full detail. It is advised that all students go slowly through that
appendix, making sure they understand each and every concept that
it introduces.

\subsection{Quantum mechanics with pure states\label{QuantumMehcanicsPureStates}}

\textbf{Setting the stage: states and observables. }Within the mathematical
formalism of quantum mechanics, physical degrees of freedom are associated
with a complex Hilbert space, that is, a vector space endorsed with
an inner product (structure known as a \emph{Euclidean space}), composed
of \emph{normalizable} vectors. We will denote Hilbert spaces with
a calligraphic symbol, say $\mathcal{H}$, for example.

We will use the `bra-ket' notation introduced by Dirac, where vectors
are denoted by $|\psi\rangle\in\mathcal{H}$ (ket), and the inner
product of two vectors $|\psi\rangle$ and $|\phi\rangle$ is denoted
by $\langle\psi|\phi\rangle\in\mathbb{C}$ (bra-ket). Note that `$\psi$'
and `$\phi$' act just as labels for the vectors (their \emph{names},
if you will).

The inner-product structure of the Hilbert space $\mathcal{H}$ allows
us to define its dual, denoted by $\mathcal{H}^{+}$, whose elements
are denoted by $\langle\psi|\in\mathcal{H}^{+}$ (bra). Whenever a
bra meets a ket, a complex number is produced: $\langle\psi|\;|\phi\rangle\rightarrow\langle\psi|\phi\rangle$.
Hence, ket-bra structures such as $|\psi\rangle\langle\phi|$ are
operators, that is, objects that map kets onto kets: $|\psi\rangle\langle\phi|\:|\varphi\rangle\rightarrow\langle\phi|\varphi\rangle\:|\psi\rangle\in\mathcal{H}$
(and similarly for bras).

In quantum mechanics, the state of an isolated system is completely
specified by a normalized vector in its Hilbert space, say $|\psi\rangle\in\mathcal{H}$
with $\langle\psi|\psi\rangle=1$. Observables, on the other hand,
are represented by self-adjoint operators, say $\hat{A}=\hat{A}^{\dagger}$
(we will always denote operators by a `hat'). The spectral theorem
tells us that the eigenvalues of such an operator are always real,
while their eigenvectors form a basis of the Hilbert basis, that is,
\begin{equation}
\hat{A}=\sum_{j=1}^{d}a_{j}|a_{j}\rangle\langle a_{j}|,
\end{equation}
with $a_{j}\in\mathbb{R}$ and\begin{subequations}
\begin{align}
\sum_{j=1}^{d}|a_{j}\rangle\langle a_{j}| & =\hat{I},\;\text{(resolution of the identity)}\label{Resolution_Identity}\\
\langle a_{j}|a_{l}\rangle & =\delta_{jl},\;\text{(orthonormality)}
\end{align}
\end{subequations}where $\hat{I}$ is the identity operator and we
have denoted by $d$ the dimension of the Hilbert space (which can
be infinite, as we will discuss shortly). Applying (\ref{Resolution_Identity})
to any vector $|\psi\rangle$, we obtain its expansion in the basis
$\{|a_{j}\rangle\}_{j=1,...,d}$,
\begin{equation}
|\psi\rangle=\sum_{j=1}^{d}\psi_{j}|a_{j}\rangle,\qquad\text{with }\psi_{j}=\langle a_{j}|\psi\rangle.
\end{equation}
The column vector with $\psi_{j}$ as components, that is, $\vec{\psi}=(\psi_{1},...,\psi_{d})^{T}$
is known as the representation of the ket $|\psi\rangle$ in the basis
$\{|a_{j}\rangle\}_{j=1,...,d}$ (the upper index `$T$' denotes `transpose').
The representation of a bra $\langle\psi|$ is then given by the row
vector $\vec{\psi}^{\dagger}=(\psi_{1}^{*},...,\psi_{d}^{*})$. On
the other hand, the representation of an operator $\hat{B}$ is a
$d\times d$ matrix $B$ with elements $B_{jl}=\langle a_{j}|\hat{B}|a_{l}\rangle$.
Abstract expressions such as $\langle\phi|\hat{B}|\psi\rangle$ can
then be evaluated as $\vec{\phi}^{\dagger}B\vec{\psi}$.

The case of a Hilbert space with infinite dimension, $d\rightarrow\infty$,
deserves special attention. Since one can build infinite sequences
that do not converge, there exist elements of the vector space underlaying
the Hilbert space that are not normalizable. While these vectors are
not elements of the Hilbert space itself, they can still be eigenvectors
of operators acting on the Hilbert space. Specifically, one can find
self-adjoint operators with a continuous spectrum, say,
\begin{equation}
\hat{X}=\int_{\mathbb{R}}dx\;x|x\rangle\langle x|,
\end{equation}
where the eigenvectors are not normalizable, and therefore are not
elements of the Hilbert space, $|x\rangle\notin\mathcal{H}$. Remarkably,
one can still use these eigenvectors to build a resolution of the
identity, provided that they are chosen to satisfy the so-called `Dirac-delta
orthonormalization condition'
\begin{equation}
\langle x|y\rangle=\delta(x-y)\quad\Rightarrow\quad\hat{I}=\int_{\mathbb{R}}dx\;|x\rangle\langle x|.
\end{equation}
Using this property, we can now expand any vector $|\psi\rangle$
(belonging to the Hilbert space or not), as
\begin{equation}
|\psi\rangle=\int_{\mathbb{R}}dx\psi(x)|x\rangle,\qquad\text{with }\psi(x)=\langle x|\psi\rangle,
\end{equation}
where now $\psi(x)$ is a continuous representation of the ket $|\psi\rangle$
in the \emph{generalized }or\emph{ continuous basis} $\{|x\rangle\}_{x\in\mathbb{R}}$.

Note that while here we have discussed only operators with a purely
continuous spectrum, in general, one can find observables with mixed
spectrum, that is, part discrete and part continuous. The former are
always normalizable while the latter are Dirac-delta normalizable,
such that the collection of all eigenvectors forms a generalized basis
that can be used to to form a resolution of the identity. Indeed,
this is the case for most Hamiltonian operators of real systems. For
example, atoms present a discrete set of low-energy states where the
electrons are bound to the nucleus, but also a continuum of high-energy
states where the atom is ionized and the electron is set free. 

An important question is how to build observables in quantum mechanics,
that is, how to define the self-adjoint operator associated to a physically
measurable magnitude. In some cases, such as the spin of the electron,
the observable isn't even present in classical physics, and it needs
to be built based either on empirical observations or reasonable symmetry
principles (for example, the electron's spin was postulated by Pauli
to fit experiments, but it was finally shown to emerge naturally in
quantum field theory by demanding the framework to be consistent with
special relativity). However, for observables with a classical counterpart,
there is a well-defined quantum mechanical prescription that generally
works. First, note that in classical physics one can always describe
the system through a set of generalized positions $\mathbf{q}$ and
momenta $\mathbf{p}$, which define \emph{phase space}. A classical
state of the system is then associated with a point in phase space.
On the other hand, observables are real functions on phase space,
say, $A(\mathbf{q},\mathbf{p})$. We can then build the corresponding
quantum observable by symmetrizing $A$ with respect to the possible
orderings of positions and momenta\footnote{For example, given $A=pq^{2}$, we write it in symmetric form as $A=\frac{1}{3}(q^{2}p+qpq+pq^{2})\equiv A^{(s)}$,
such that the expression contains all the possible orderings of $pq^{2}$
with equal weights.}, what we denote by $A^{(s)}$, and then replacing them by operators
$\hat{\mathbf{q}}$ and $\hat{\mathbf{p}}$, whose components satisfy
\emph{canonical commutation relations}
\begin{equation}
[\hat{q}_{m},\hat{q}_{n}]=0=[\hat{p}_{m},\hat{p}_{n}],\qquad[\hat{q}_{m},\hat{p}_{n}]=\mathrm{i}\hbar\delta_{mn}.\label{CCR}
\end{equation}
The symmetrization ensures that the resulting operator $\hat{A}\equiv A^{(s)}(\hat{\mathbf{q}},\hat{\mathbf{p}})$
is self-adjoint, and other subtle properties that make this prescription
reasonable \cite{Groenewold46}. Note that it is the first time that
we introduce Planck's constant, which plays a fundamental role in
quantum mechanics. In this case, it prevents positions and momenta
from commuting, if only by a little amount ($\hbar$ is a very small
action `by classical standards'), just enough to bring about a change
of paradigm in physics!

\textbf{Observing quantum systems. }One of the most striking features
of quantum mechanics concerns measurements of observables. When measuring
an observable $\hat{A}$, quantum mechanics says that the outcome
can be any of its eigenvalues $\{a_{j}\}_{j=1,...,d}$, which will
appear with probability 
\begin{equation}
\{p_{j}=|\langle a_{j}|\psi\rangle|^{2}\}_{j=1,...,d},\label{pj}
\end{equation}
where $|\psi\rangle$ is the state of the system. Hence, it is in
general impossible to predict the outcome of a single measurement,
but just the frequency with which the outcomes will appear when the
experiment is repeated many times (e.g., either the process of state-preparation
and measurement are repeated many times on a single system, or many
copies of the same system in the same state are independently measured
once).

Moreover, suppose that after the single measurement we obtain eigenvalue
$a_{j}$ as the outcome. Then, quantum mechanics says that right after
the measurement, the state of the system \emph{collapses} to $|a_{j}\rangle$.
This is arguably the most radical difference with respect to classical
physics: while observables can be measured without perturbing the
state of classical systems, in quantum mechanics the measurement process
irreversibly changes the state of the system. This collapse of the
system's state is arguably the quantum-mechanical concept that has
ignited the most fierce fights among physicists, and it keeps doing
so \cite{WhitakerBook}.

On the practical side, there are two important statistical objects
that roughly characterize what's to be expected from a state in a
measurement process. First, the expectation value of the observable,
defined as
\begin{equation}
\langle\hat{A}\rangle\equiv\langle\psi|\hat{A}|\psi\rangle=\sum_{j=1}^{d}a_{j}p_{j},
\end{equation}
which coincides with the statistical \emph{mean} or \emph{average}
of the probability distribution, that is, the average of the outcomes
expected after many repetitions of the measurement process. On the
other hand, we also need a figure of the spread of the outcomes around
the mean. Defining the \emph{variance}
\begin{equation}
V(A)\equiv\langle\psi|\hat{A}^{2}|\psi\rangle-\langle\psi|\hat{A}|\psi\rangle^{2},
\end{equation}
such a spread can be quantified by the \emph{standard deviation }or
\emph{uncertainty} $\Delta A=\sqrt{V(A)}$. It is common to define
the fluctuation operator $\delta\hat{A}=\hat{A}-\langle\hat{A}\rangle$,
whose square gives us direct access to the variance as $V(A)=\langle(\delta\hat{A})^{2}\rangle$.
A shocking result is that of the \emph{uncertainty relations}: given
two observables $\hat{A}$ and $\hat{B}$, it is possible to prove
that
\begin{equation}
\Delta A\Delta B\geq\frac{1}{2}|\langle[\hat{A},\hat{B}]\rangle|.
\end{equation}
Hence, while in classical physics nothing prevents us from preparing
states where we have perfect knowledge of all the observables of the
system, the situation is radically different in quantum mechanics:
a state with negligible variance in one observable, necessarily comes
with a large variance in other observables that do not commute with
the first one.

\textbf{Unsupervised evolution.} In the previous lines we discussed
how the system behaves when it is observed. But what about the evolution
of the system when it is not subject to any measurements? Quantum
mechanics says that the evolution of a quantum system that is left
\emph{unsupervised} is completely characterized by the \emph{evolution
operator}, defined from the Hamiltonian $\hat{H}$ of the system as
\begin{equation}
\hat{U}(t)=e^{\hat{H}t/\mathrm{i}\hbar},\label{TimeEvoOp}
\end{equation}
where for simplicity we consider a time-independent Hamiltonian (in
Section \ref{Sec:ChangingPictures} we generalize it to time-dependent
Hamiltonians). Recall that the Hamiltonian is easy to construct from
the classical one when that's available, but has to be formulated
empirically or with reasonable theoretical arguments when the system
involves degrees of freedom with no classical analog. In any case,
the Hamiltonian is a self-adjoint operator, and therefore the evolution
operator is unitary, $\hat{U}^{\dagger}(t)=\hat{U}^{-1}(t)$.

We can formulate the evolution induced by this operator in two apparently
different, but equivalent ways:
\begin{itemize}
\item \emph{Schrödinger picture}. In this formulation states evolve, while
operators remain fixed in time. In particular, denoting by $|\psi(t)\rangle$
the state of the system at any time, quantum mechanics says
\begin{equation}
|\psi(t)\rangle=\hat{U}(t)|\psi(0)\rangle\quad\Leftrightarrow\quad\mathrm{i}\hbar\partial_{t}|\psi\rangle=\hat{H}|\psi\rangle,\label{SchrodingerEq}
\end{equation}
where the latter is known as the \emph{Schrödinger equation}.
\item \emph{Heisenberg picture}. An alternative formulation consists on
the following one, where states remain fixed, and it is operators
that evolve in time. Specifically, denoting by $\hat{A}(t)$ the time-evolved
of an operator that is $\hat{A}(0)$ originally, quantum mechanics
says that
\begin{equation}
\hat{A}(t)=\hat{U}^{\dagger}(t)\hat{A}(0)\hat{U}(t)\quad\Leftrightarrow\quad\mathrm{i}\hbar\partial_{t}\hat{A}=[\hat{A},\hat{H}],\label{HeisenbergEq}
\end{equation}
where the latter is known as \emph{Heisenberg equation.}
\end{itemize}
The equivalence between this two pictures is obvious from the following
argument. In real experiments, states and operators are not directly
observable. Instead, experimental observations are only sensible to
expectation values of operators\textemdash note that even probabilities
such as (\ref{SchrodingerEq}) can be written as expectation values
$p_{j}=\langle\hat{P}_{j}\rangle$ of projectors $\hat{P}_{j}=|a_{j}\rangle\langle a_{j}|$\textemdash ,
which are written in both pictures as
\begin{equation}
\langle\psi(t)|\hat{A}|\psi(t)\rangle=\langle\psi(0)|\hat{U}^{\dagger}(t)\hat{A}(0)\hat{U}(t)|\psi(0)\rangle=\langle\psi|\hat{A}(t)|\psi\rangle.
\end{equation}
Note that, for consistency, the original Heisenberg operator $\hat{A}(0)$
coincides with the one that remains fixed in the Schrödinger picture,
simply denoted by $\hat{A}$ in the latter picture. Similarly, the
original Schrödinger state $|\psi(0)\rangle$ coincides with the one
that remains fixed in the Heisenberg picture, denoted in the latter
picture by $|\psi\rangle$. Having these two alternative descriptions
of quantum mechanics is very handy, as we will see throughout the
notes.

\textbf{Composite systems.} Consider a system formed by two degrees
of freedom or \emph{subsystems}\footnote{Here we speak about distinguishable subsystems, which is all we will
care about in this course. The indistinguishable case is a bit more
subtle. In particular, quantum mechanics says that, in such case,
the Hilbert space of the total system is not all $\mathcal{A}\otimes\mathcal{B}$,
but only the subspace composed by vectors which are either symmetric
or antisymmetric under the exchange of the labels of the subsystems.
In the first case we say we are dealing with bosons, and in the second
with fermions. For example, the state $|e_{j}\rangle\otimes|e_{l\neq j}\rangle$,
which is a basis state for distinguishable systems as we mention below,
is not a basis state for indistinguishable subsystems. Instead, states
$(|e_{j}\rangle\otimes|e_{l}\rangle\pm|e_{l}\rangle\otimes|e_{j}\rangle)/\sqrt{2}$
are correct basis states for bosons and fermions, respectively.} (everything is trivially generalized to more subsystems). Can we
build the Hilbert space $\mathcal{H}$ of the total system, just from
knowledge of the individual Hilbert spaces $\mathcal{A}$ and $\mathcal{B}$
of its subsystems? Indeed we can, and quantum mechanics tells us that
the total Hilbert space will be the tensor product of those spaces,
that is, $\mathcal{H}=\mathcal{A}\otimes\mathcal{B}$.

The definition and properties of the tensor product map are reviewed
in detail in Section \ref{CompositeHilbertSpaces} of Appendix \ref{QuantumMechanics}.
Suffice here to say that: 
\begin{itemize}
\item It puts in (unique) correspondence pairs of vectors $(|\psi_{\mathcal{A}}\rangle\in\mathcal{H}_{\mathcal{A}},|\psi_{\mathcal{B}}\rangle\in\mathcal{H}_{\mathcal{B}})$,
with a vector $|\psi\rangle\in\mathcal{H}$ in the total Hilbert space.
We will denote such vector by $|\psi\rangle=|\psi_{\mathcal{A}}\rangle\otimes|\psi_{\mathcal{B}}\rangle$,
or simply by the more economic way $|\psi\rangle=|\psi_{\mathcal{A}},\psi_{\mathcal{B}}\rangle$
when there is no room for confusion.
\item Of particular relevance is the correspondence between bases in the
different Hilbert spaces. Given the bases $\{|e_{j}^{\mathcal{A}}\rangle\}_{j=1,...,d_{\mathcal{A}}}$
and $\{|e_{j}^{\mathcal{B}}\rangle\}_{j=1,...,d_{\mathcal{B}}}$ of
$\mathcal{A}$ and $\mathcal{B}$, respectively, the set $\{|e_{j}^{\mathcal{A}},e_{l}^{\mathcal{B}}\rangle=|e_{j}^{\mathcal{A}}\rangle\otimes|e_{l}^{\mathcal{B}}\rangle\}_{l=1,...,d_{\mathcal{B}}}^{j=1,...,d_{\mathcal{A}}}$
forms a basis of $\mathcal{H}$, which has then dimension $d=d_{\mathcal{A}}\times d_{\mathcal{B}}$.
\item Given two kets $|\psi\rangle=|\psi_{\mathcal{A}}\rangle\otimes|\psi_{\mathcal{B}}\rangle$
and $|\phi\rangle=|\phi_{\mathcal{A}}\rangle\otimes|\phi_{\mathcal{B}}\rangle$,
their inner product is defined as $\langle\phi|\psi\rangle=\langle\phi_{\mathcal{A}}|\psi_{\mathcal{A}}\rangle\langle\phi_{\mathcal{B}}|\psi_{\mathcal{B}}\rangle$.
\item Given two operators $\hat{L}_{\mathcal{A}}$ and $\hat{L}_{\mathcal{B}}$,
we define the action of their tensor product $\hat{L}=\hat{L}_{\mathcal{A}}\otimes\hat{L}_{\mathcal{B}}$
onto a vector $|\psi\rangle=|\psi_{\mathcal{A}}\rangle\otimes|\psi_{\mathcal{B}}\rangle$
by $\hat{L}|\psi\rangle=(\hat{L}_{\mathcal{A}}|\psi_{\mathcal{A}}\rangle)\otimes(\hat{L}_{\mathcal{B}}|\psi_{\mathcal{B}}\rangle)$.
We will omit the tensor product between the operators when the notation
makes it clear onto which Hilbert space they act. Hence, for example,
we will simply write $\hat{L}_{\mathcal{A}}\hat{L}_{\mathcal{B}}|\psi\rangle=(\hat{L}_{\mathcal{A}}|\psi_{\mathcal{A}}\rangle)\otimes(\hat{L}_{\mathcal{B}}|\psi_{\mathcal{B}}\rangle)$
or $\hat{L}_{\mathcal{A}}|\psi\rangle=(\hat{L}_{\mathcal{A}}|\psi_{\mathcal{A}}\rangle)\otimes|\psi_{\mathcal{B}}\rangle$.
\end{itemize}
A remarkable concept appears in composite systems: \emph{entanglement}.
It refers to correlations between the subsystems that exploit the
concept of quantum superposition, and cannot therefore appear in classical
physics. As a concrete example, consider the state $|\psi\rangle=(|E_{1}^{\mathcal{A}}\rangle\otimes|E_{1}^{\mathcal{B}}\rangle+|E_{2}^{\mathcal{A}}\rangle\otimes|E_{2}^{\mathcal{B}}\rangle)/\sqrt{2}$,
where $\left\{ |E_{j}^{\mathcal{S}}\rangle\right\} _{j=1,2}^{\mathcal{S}=\mathcal{A},\mathcal{B}}$
are energy eigenstates of the corresponding subsystem (we assume that
the subsystems do not interact after preparation of the state). Upon
a measurement of the energy of subsystem $\mathcal{A}$, we cannot
predict the outcome with certainty, which can be either $E_{1}^{\mathcal{A}}$
or $E_{2}^{\mathcal{A}}$ with 50\% probability. However, we know
that once one outcome appears, e.g., $E_{1}^{\mathcal{A}}$, then
the full system collapses to the state $|E_{1}^{\mathcal{A}}\rangle\otimes|E_{1}^{\mathcal{B}}\rangle$,
and therefore, we are now able to predict with certainty the outcome
of an energy measurement on $\mathcal{B}$, without ever having interrogated
$\mathcal{B}$ itself. Hence, $\mathcal{A}$ and $\mathcal{B}$ are
correlated, since a measurement on one affects the state of the other.
However, since the correlations are encoded in a quantum superposition,
these are a purely quantum mechanical type of correlations that we
cannot generate classically. This type of quantum correlations that
make use of the superposition principle are known as entanglement,
and are the heart of the field of quantum information that has emerged
in recent decades, and holds great promise for technological applications
in computation, communication, and sensing. Remarkably, this type
of states have allowed us to show experimentally that nature is incompatible
with determinism, unless we are willing to give up causality, and
hence the probabilistic character of quantum mechanics is inherent
to nature, not a product of our ignorance. You can learn a bit more
about entanglement in Appendix \ref{Sec:Entanglement}, and hopefully
through some examples that we will study in the course.

\subsection{Quantum mechanics with mixed states\label{vonNeumannEntropy}}

\textbf{Need for mixed states: noisy systems.} Consider a system for
which our state-preparation device is not perfect (as it often occurs
in real experiments), and we cannot be sure that we prepared a given
state $|\psi\rangle$, but only know that we prepared some state among
a set $\{|\psi_{m}\rangle\}_{m=1,...,M}$ with associated probability
distribution $\{w_{m}\}_{m=1,...,M}$, where $w_{m}\geq0$ and $\sum_{m=1}^{M}w_{m}=1$.
We will call \emph{ensemble} to the set $\{w_{m},|\psi_{m}\rangle\}_{m=1,...,M}$.
Note that the states of the ensemble do not need to be orthogonal,
and actually $M$ needs not be the dimension of the Hilbert space.

What will be the mean or expectation value of an operator $\hat{A}$
in this setup? The expectation value in a given state of the ensemble
is $\langle\psi_{m}|\hat{A}|\psi_{m}\rangle$, so we simply need to
average these over their probability distribution, that is
\begin{equation}
\langle\hat{A}\rangle=\sum_{m=1}^{M}w_{m}\langle\psi_{m}|\hat{A}|\psi_{m}\rangle.\label{AmeanMixed}
\end{equation}
By defining an operator
\begin{equation}
\hat{\rho}=\sum_{m=1}^{M}w_{m}|\psi_{m}\rangle\langle\psi_{m}|,\label{rhoMixed}
\end{equation}
we can rewrite (\ref{AmeanMixed}) in the compact form
\begin{equation}
\langle\hat{A}\rangle=\text{tr}\{\hat{\rho}\hat{A}\},\label{meanArho}
\end{equation}
where `tr' is the trace operation, defined in terms of a basis $\{|e_{j}\rangle\}_{j=1...d}$
as $\text{tr}\{\hat{D}\}=\sum_{j=1}^{d}\langle e_{j}|\hat{D}|e_{j}\rangle$
for any operator $\hat{D}$. It is very interesting to have been able
to write expectation values in terms of a trace, since this is independent
of the basis chosen to compute it, as is easily proved by using its
cyclic property\footnote{In finite dimension, the cyclic property of the trace always holds.
However, in infinite dimension, the cyclic property (as presented
in the main text) only holds when the operator $\hat{D}_{1}\hat{D}_{2}...\hat{D}_{N}$
has finite trace (indeed, one can go crazy trying to compute the trace
of $[\hat{q},\hat{p}]=\mathrm{i}\hbar\hat{I}$; is it infinity or
zero?). In practical terms, a finite trace is ensured when at least
one of the operators is a mixed state $\hat{\rho}$.} $\text{tr}\{\hat{D}_{1}\hat{D}_{2}...\hat{D}_{N}\}=\text{tr}\{\hat{D}_{2}...\hat{D}_{N}\hat{D}_{1}\}$,
and remembering that two bases are necessarily connected by a unitary
transformation (see Appendix \ref{Sec:Entanglement} for more details).
Hence, this shows that the operator $\hat{\rho}$ is an important
object on its own right, that contains all the information about the
state of the \emph{noisy} system. We then call it the \emph{mixed
state} of the system. Whenever $\hat{\rho}$ can be written as a rank\footnote{The rank of an operator is defined as its number of eigenvalues different
than zero.}-1 projector $\hat{\rho}=|\psi\rangle\langle\psi|$, we say that the
system is in \emph{pure state} $|\psi\rangle$, and we can apply the
formalism introduced in the previous section.

As a specific, instructive case of this type, consider the following
situation.\textbf{ }We are given a system in some known state $|\psi\rangle$,
and we perform a measurement of some observable $\hat{A}=\sum_{j=1}^{d}a_{j}|a_{j}\rangle\langle a_{j}|$
as the initial step of some protocol that might involve some more
operations later. Suppose, however, that the display of our measurement
apparatus is broken, and doesn't show the outcome, what is known as
a \emph{non-selective measurement}. In other words, we know that a
measurement took place, but we don't know the resulting outcome. Can
we still describe the quantum statistics of the protocol? The answer
is yes: even though we don't know the outcome, we know that after
the measurement the state would have collapsed into state $|a_{j}\rangle$
with probability $p_{j}=|\langle a_{j}|\psi\rangle|^{2}$. This corresponds
to the ensemble $\{p_{j},|a_{j}\rangle\}_{j=1,...,d}$ or the mixed
state $\hat{\rho}=\sum_{j=1}^{d}p_{j}|a_{j}\rangle\langle a_{j}|$.
The only difference with the general case exposed above is that in
this case the states of the ensemble form an orthonormal set, and
hence the mixed state is diagonal in this basis.

\textbf{Need for mixed states: state of a subsystem.} Consider now
a composite system such as the one introduced at the end of the previous
section, and let it be in an arbitrary mixed state $\hat{\rho}$ acting
on the full space $\mathcal{H}=\mathcal{A}\otimes\mathcal{B}$. Suppose
that we hand in subsystem $\mathcal{A}$ to Alice, who is then allowed
to perform experiments onto it, perhaps not even knowing that it is
correlated with subsystem $\mathcal{B}$. The question now is: can
Alice describe her experiments by using operators and states defined
solely on $\mathcal{A}$? Below we prove that, indeed, all Alice needs
to know in order to reproduce the statistics of her experiments is
that her subsystem starts in the \emph{reduced} \emph{state}
\begin{equation}
\hat{\rho}_{\mathcal{A}}=\sum_{l=1}^{d_{\mathcal{B}}}\langle e_{l}^{\mathcal{B}}|\hat{\rho}|e_{l}^{\mathcal{B}}\rangle\equiv\text{tr}_{\mathcal{B}}\{\hat{\rho}\},\label{rhoA}
\end{equation}
without having to make any reference to subsystem $\mathcal{B}$.
The notation $\text{tr}_{\mathcal{B}}$ means that we perform the
trace only within the basis of subspace $\mathcal{B}$, operation
known as \emph{partial trace}, which leaves us with an operator acting
on subspace $\mathcal{B}$. In more detail, if we expand the state
of the full system as
\begin{equation}
\hat{\rho}=\sum_{mm'=1}^{d_{\mathcal{A}}}\sum_{nn'=1}^{d_{\mathcal{B}}}\rho_{mm';nn'}|e_{m}^{\mathcal{A}}\rangle\langle e_{m'}^{\mathcal{A}}|\otimes|e_{n}^{\mathcal{B}}\rangle\langle e_{n'}^{\mathcal{B}}|,
\end{equation}
 then, the reduced state (\ref{rhoA}) can be written as
\begin{equation}
\hat{\rho}_{\mathcal{A}}=\sum_{lnn'=1}^{d_{\mathcal{B}}}\left(\sum_{mm'=1}^{d_{\mathcal{A}}}\rho_{mm';nn'}|e_{m}^{\mathcal{A}}\rangle\langle e_{m'}^{\mathcal{A}}|\right)\underbrace{\langle e_{l}^{\mathcal{B}}|e_{n}^{\mathcal{B}}\rangle}_{\delta_{ln}}\underbrace{\langle e_{n'}^{\mathcal{B}}|e_{l}^{\mathcal{B}}\rangle}_{\delta_{ln'}}=\sum_{mm'=1}^{d_{\mathcal{A}}}\left(\sum_{l=1}^{d_{\mathcal{B}}}\rho_{mm';ll}\right)|e_{m}^{\mathcal{A}}\rangle\langle e_{m'}^{\mathcal{A}}|,
\end{equation}
which is indeed an operator acting on $\mathcal{A}$ alone.

In order to prove that (\ref{rhoA}) is the correct `local' state
for Alice, let us compute the expectation value of an arbitrary operator
$\hat{A}\otimes\hat{I}$, acting as the identity on subspace $\mathcal{B}$.
According to (\ref{meanArho}), we can write the expectation value
as
\begin{align}
\langle\hat{A}\otimes\hat{I}\rangle & =\text{tr}\{\hat{\rho}(\hat{A}\otimes\hat{I})\}=\sum_{j=1}^{d_{\mathcal{A}}}\sum_{l=1}^{d_{\mathcal{B}}}\left(\langle e_{j}^{\mathcal{A}}|\otimes\langle e_{l}^{\mathcal{B}}|\right)\hat{\rho}(\hat{A}\otimes\hat{I})\left(|e_{j}^{\mathcal{A}}\rangle\otimes|e_{l}^{\mathcal{B}}\rangle\right),\\
 & =\sum_{j=1}^{d_{\mathcal{A}}}\sum_{l=1}^{d_{\mathcal{B}}}\left(\langle e_{j}^{\mathcal{A}}|\otimes\langle e_{l}^{\mathcal{B}}|\right)\hat{\rho}\left(\hat{A}|e_{j}^{\mathcal{A}}\rangle\otimes|e_{l}^{\mathcal{B}}\rangle\right)=\sum_{j=1}^{d_{\mathcal{A}}}\langle e_{j}^{\mathcal{A}}|\underbrace{\left(\sum_{l=1}^{d_{\mathcal{B}}}\langle e_{l}^{\mathcal{B}}|\hat{\rho}|e_{l}^{\mathcal{B}}\rangle\right)}_{\hat{\rho}_{\mathcal{A}}}\hat{A}|e_{j}^{\mathcal{A}}\rangle=\text{tr}\{\hat{\rho}_{\mathcal{A}}\hat{A}\}.\nonumber 
\end{align}
Hence, Alice is able to refer any of her observations, which depend
only on observables $\hat{A}$ acting on $\mathcal{A}$, to expectation
values $\langle\hat{A}\rangle=\text{tr}\{\hat{\rho}_{\mathcal{A}}\hat{A}\}$
in which $\hat{\rho}_{\mathcal{A}}$ is the state of her subsystem.

Consider now the special case in which the full system is in some
pure state $|\psi\rangle$, that is, $\hat{\rho}=|\psi\rangle\langle\psi|$.
Referred to some basis $\{|e_{j}^{\mathcal{B}}\rangle\}_{j=1,...,d_{\mathcal{B}}}$
of subsystem $\mathcal{B}$, the state of the full system can always
be written in the form
\begin{equation}
|\psi\rangle=\sum_{m=1}^{d_{\mathcal{B}}}\sqrt{w_{m}}|\psi_{m}\rangle\otimes|e_{m}^{\mathcal{B}}\rangle,\label{psiMixed}
\end{equation}
where $\{w_{m}\}_{m=1,...,d_{\mathcal{B}}}$ is a probability distribution,
and $\{|\psi_{m}\rangle\}_{m=1,...,d_{\mathcal{B}}}$ are some states
in $\mathcal{A}$ (as above, they do not need to be orthogonal). The
reduced state (\ref{rhoA}) then reads 
\begin{equation}
\hat{\rho}_{\mathcal{A}}=\sum_{lmm=1}^{d_{\mathcal{B}}}\sqrt{w_{m}w_{m'}}|\psi_{m}\rangle\langle\psi_{m'}|\underbrace{\langle e_{l}^{\mathcal{B}}|e_{m}^{\mathcal{B}}\rangle}_{\delta_{lm}}\underbrace{\langle e_{m'}^{\mathcal{B}}|e_{l}^{\mathcal{B}}\rangle}_{\delta_{lm'}}=\sum_{m=1}^{d_{\mathcal{B}}}w_{m}|\psi_{m}\rangle\langle\psi_{m}|,\label{rhoAmixed}
\end{equation}
which is of the same form as the mixed state we defined in (\ref{rhoMixed})
for noisy systems. Hence, even though the state $|\psi\rangle$ of
the whole system is pure, in general we need to describe the reduced
state of the subsystems by a mixed state.

This situation allows us to discuss an important property of mixed
states: their ensemble decomposition is not unique. In particular,
note that we could have chosen any other basis of subspace $\mathcal{B}$,
for example, a basis $\{|d_{k}^{\mathcal{B}}\rangle\}_{k=1,...,d_{\mathcal{B}}}$,
related to the previous basis by a unitary matrix $U$ with elements
$U_{jl}$, that is, $\{|e_{m}^{\mathcal{B}}\rangle=\sum_{k=1}^{d_{\mathcal{B}}}U_{mk}|d_{k}^{\mathcal{B}}\rangle\}_{j=1,...,d_{\mathcal{B}}}$.
In terms of this new basis, state (\ref{psiMixed}) is written as
\begin{equation}
|\psi\rangle=\sum_{m=1}^{d_{\mathcal{B}}}\sqrt{w_{m}}|\psi_{m}\rangle\otimes\left(\sum_{k=1}^{d_{\mathcal{B}}}U_{mk}|d_{k}^{\mathcal{B}}\rangle\right)=\sum_{k=1}^{d_{\mathcal{B}}}\sqrt{v_{k}}|\phi_{k}\rangle\otimes|d_{k}^{\mathcal{B}}\rangle,
\end{equation}
where we have defined
\begin{equation}
\sqrt{v_{k}}|\phi_{k}\rangle=\sum_{m=1}^{d_{\mathcal{B}}}U_{mk}\sqrt{w_{m}}|\psi_{m}\rangle,
\end{equation}
where $|\phi_{k}\rangle\in\mathcal{A}$ are states, and the new probability
distribution $\{v_{k}\}_{k=1,...,d_{\mathcal{B}}}$ is found by taking
the norm square of the last expression:
\begin{equation}
v_{k}\underbrace{\langle\phi_{k}|\phi_{k}\rangle}_{1}=\sum_{mm'=1}^{d_{\mathcal{B}}}U_{mk}^{*}U_{m'k}\underbrace{\sqrt{w_{m}w_{m'}}\langle\psi_{m}|\psi_{m'}\rangle}_{\text{define it as }S_{mm'}}=(USU^{\dagger})_{kk},
\end{equation}
and is easily shown to be normalized (note that we use the cyclic
property of the trace in the second step)
\begin{equation}
\sum_{k=1}^{d_{\mathcal{B}}}v_{k}=\text{tr}\{USU^{\dagger}\}=\text{tr}\{S\}=\sum_{k=1}^{d_{\mathcal{B}}}w_{k}\underbrace{\langle\psi_{k}|\psi_{k}\rangle}_{1}=1.
\end{equation}
Hence, taking the partial trace with respect to the new basis, the
reduced state reads now
\begin{equation}
\hat{\rho}_{\mathcal{A}}=\sum_{k=1}^{d_{\mathcal{B}}}v_{k}|\phi_{k}\rangle\langle\phi_{k}|,
\end{equation}
which is formed from an ensemble $\{v_{k},|\phi_{k}\rangle\}_{k=1,...,d_{\mathcal{B}}}$
different than that of (\ref{rhoAmixed}).

\textbf{Informational interpretation and quantification of mixedness.
}There is a common thread to both the case of a noisy system and a
subsystem of a composite system: in both cases, the mixture appears
because we are not monitoring some degree of freedom that our system
shares correlations with (the noise in the first case, subsystem $\mathcal{B}$
in the second case). In other words, the \emph{mixture} simply \emph{reflects}
our \emph{ignorance} about some information of the system that has
\textquoteleft leaked\textquoteright{} into another system that we
cannot interrogate. This suggests an interpretation of mixedness as
lack of information, making the concept more intuitive.

Given this informational interpretation of mixed states, it is fairly
natural to ask which is the state that is \emph{maximally mixed},
that is, that has leaked out the maximum amount of information. Let
us consider here a Hilbert space $\mathcal{H}$ with finite dimension
$d$ (we will consider the infinite dimensional case in the next chapter,
when discussing states of the harmonic oscillator). It is easy to
argue that the maximally-mixed state in this case corresponds to
\begin{equation}
\hat{\rho}_{\mathrm{MM}}=\frac{\hat{I}}{d}.\label{RhoMM}
\end{equation}
Indeed, when the state of the system is $\hat{\rho}_{\mathrm{MM}}$
and an arbitrary observable $\hat{A}$ is measured, all its eigenvalues
are equally likely to appear as an outcome of the measurement, that
is, 
\begin{equation}
p_{j}=\langle a_{j}|\hat{\rho}_{\text{MM}}|a_{j}\rangle=\frac{1}{d}\hspace{3mm}\forall j.\hspace{5mm}\text{(flat distribution)}
\end{equation}
Hence, this state maximizes the overall uncertainty of the system's
observables, as expected for a maximally-mixed state. In other words:
with the system in such state, the outcome of any measurement is completely
random!

The mixedness of a state $\hat{\rho}$ can be quantified by the \textit{von
Neumann entropy}, which is defined as 
\begin{equation}
S[\hat{\rho}]=-\mathrm{tr}\{\hat{\rho}\log\hat{\rho}\}.
\end{equation}
Given the diagonal representation of the state 
\begin{equation}
\hat{\rho}=\sum_{j=1}^{d}\lambda_{j}|r_{j}\rangle\hspace{-0.4mm}\langle r_{j}|,
\end{equation}
where its eigenvectors $\{|r_{j}\rangle\}_{j=1,2,...,d}$ form an
orthonormal basis of the Hilbert space $\mathcal{H}$, the von Neumann
entropy reads then (proven from the spectral theorem)
\begin{equation}
S[\hat{\rho}]=-\sum_{j=1}^{d}\lambda_{j}\log\lambda_{j}.
\end{equation}
This is just the Shannon entropy of the distribution $\{\lambda_{j}\}_{j=1,2,...,d}$,
which is one of the most fundamental quantities in classical information
theory. You can easily check that the entropy is 0 for a pure state,
while it has a maximum $\log d$ for the \textit{\emph{maximally-mixed
state}}.

It is interesting to note (and easy to prove mathematically \cite{NielsenChuangBook})
that the entropy does not change by unitary evolution, cannot decrease
by non-selective measurements, and cannot increase by selective measurements
(indeed, when no degeneracies are present, it collapses to zero, as
the state becomes pure). These properties are indeed expected from
a purely informational point of view: while evolving unsupervised,
the system does not exchange any information with any other system;
when we perform a selective measurement indeed we gain information
about the system and its post-measurement state; and when the measurement
is non-selective, not only we don't gain any information, but it can
scramble the information that was contained in the state.

\textbf{Reformulation of quantum mechanics for mixed states.} The
previous discussions have served to illustrate how incredibly useful
mixed states are when dealing with quantum systems. For this reason,
it is important to take a step back, and reformulate the basic laws
of quantum mechanics that we saw in the previous section, but with
a formalism adapted to mixed states.

Let us start by enunciating that: the state of a general quantum system
is completely specified by a \emph{density operator} $\hat{\rho}$
acting on its Hilbert space, that is, a self-adjoint operator ($\hat{\rho}=\hat{\rho}^{\dagger}$),
with non-negative eigenvalues ($\hat{\rho}\geq0$), and unit trace
($\text{tr}\{\hat{\rho}\}=1$). Because projectors are positive, self-adjoint
operators, density operators always allow for an ensemble decomposition
of the type (\ref{rhoMixed}), which is not unique as explained above.
The condition for two ensemble decompositions $\{w_{m},|\psi_{m}\rangle\}_{m=1,...,M}$
and $\{v_{k},|\phi_{k}\rangle\}_{k=1,...,K}$ to represent the same
density operator, is that they are connected by a unitary matrix $U$
(with elements $U_{km}$) through
\begin{equation}
\sqrt{v_{k}}|\phi_{k}\rangle=\sum_{m=1}^{M}U_{km}\sqrt{w_{m}}|\psi_{m}\rangle,
\end{equation}
where if $N\neq M$ we can add $|N-M|$ arbitrary states with zero
probability to the ensemble with smaller number of elements, so that
$U$ is a square matrix.

The combination of the three properties that define density operators
ensure that the diagonal entries of the representation of $\hat{\rho}$
form a probability distribution in any basis. In fact, when measuring
an observable with eigen-expansion $\hat{A}=\sum_{j=1}^{d}a_{j}|a_{j}\rangle\langle a_{j}|$,
the probability of obtaining outcome $a_{j}$ is given by
\begin{equation}
p_{j}=\langle a_{j}|\hat{\rho}|a_{j}\rangle=\text{tr}\{\hat{\rho}|a_{j}\rangle\langle a_{j}|\},
\end{equation}
that is, by the corresponding diagonal element of the state's representation
in the observable's eigenbasis. The expectation value of any operator
$\hat{A}$ is then given by
\begin{equation}
\langle\hat{A}\rangle=\sum_{j=1}^{d}a_{j}p_{j}=\text{tr}\{\hat{\rho}\hat{A}\}.
\end{equation}
Note that we have written all observable quantities in terms of traces
of operators multiplied by the density operator. This is very useful
because traces do not depend on the basis, as mentioned above. The
property $\text{tr}\{\hat{\rho}|\psi\rangle\langle\phi|\}=\langle\phi|\hat{\rho}|\psi\rangle$
is sometimes useful in this context.

Finally, in the Schrödinger picture, states of unsupervised systems
evolve according to
\begin{equation}
\hat{\rho}(t)=\hat{U}(t)\hat{\rho}(0)\hat{U}^{\dagger}(t)\quad\Leftrightarrow\quad\mathrm{i}\hbar\partial_{t}\hat{\rho}=[\hat{H},\hat{\rho}],\label{VonNeumannPicture}
\end{equation}
where $\hat{U}(t)$ is the time-evolution operator defined in (\ref{TimeEvoOp}),
and the latter is known as the \emph{von Neumann equation}. Note that
(\ref{VonNeumannPicture}) is trivial to prove by writing the initial
state in some initial ensemble, and letting each state of the ensemble
evolve according the Schrödinger-picture expression (\ref{SchrodingerEq}).

These are the laws of quantum mechanics as we will used them throughout
the lectures, where we will see that mixed states appear very naturally
in many interesting situations.

\subsection{Change of picture and time-dependent Hamiltonians\label{Sec:ChangingPictures}}

As we have seen, unsupervised evolution in quantum mechanics admits
two alternative, but equivalent descriptions: one in which states
evolve (Schrödinger picture) and another where operators evolve (Heisenberg
picture). Having these two alternatives is actually quite useful,
both from the conceptual and practical points of view. Here, however,
we will discuss how these are just two limits of an infinite number
of equivalent or \emph{intermediate pictures}, in which both states
and operators evolve.

In order to see this, consider a completely general unitary operator
$\hat{U}_{\text{c}}(t)$, which might even be time dependent. The
`c' subindex stands for `change' (of picture). As we have stressed
several times already, in quantum mechanics, all connections to observable
phenomena are made through expectation values. On the other hand,
starting from the Schrödinger picture, we can rewrite the expectation
value of a generic operator $\hat{A}$ at any time as 
\begin{equation}
\langle\hat{A}\rangle(t)=\text{tr}\{\hat{\rho}(t)\hat{A}\}=\text{tr}\Big\{\hat{U}_{\text{c}}^{\dagger}(t)\hat{\rho}(t)\hat{U}_{\text{c}}(t)\hat{U}_{\text{c}}^{\dagger}(t)\hat{A}\hat{U}_{\text{c}}(t)\Big\},
\end{equation}
where we have used $\hat{U}_{c}(t)\hat{U}_{c}^{\dagger}(t)=\hat{I}$
and the cyclic property of the trace. This is a very suggestive expression
that allows us to define new pictures for quantum dynamics as follows.
Let's define the state and operator in the new picture as\begin{subequations}
\begin{align}
\hat{\rho}_{\mathrm{I}}(t) & =\hat{U}_{\text{c}}^{\dagger}(t)\hat{\rho}(t)\hat{U}_{\text{c}}(t),\label{rhoI}\\
\hat{A}_{\text{I}}(t) & =\hat{U}_{\text{c}}^{\dagger}(t)\hat{A}\hat{U}_{\text{c}}(t),\label{AI}
\end{align}
\end{subequations}where the `I' subindex stands for `intermediate'
(picture). The interpretation of these operators is very interesting.
Recall that $\hat{\rho}(t)=\hat{U}(t)\hat{\rho}(0)\hat{U}^{\dagger}(t)$,
where $\hat{U}(t)$ is the time-evolution operator of the system.
We see that the state (\ref{rhoI}) in the new picture can be written
as $\hat{\rho}_{\mathrm{I}}(t)=\hat{U}_{\text{I}}(t)\hat{\rho}(0)\hat{U}_{\text{I}}^{\dagger}(t)$,
in terms of a unitary $\hat{U}_{\text{I}}(t)=\hat{U}_{\text{c}}^{\dagger}(t)\hat{U}(t)$,
which can be interpreted as the time-evolution operator of the state
in the new picture (with the subtlety that we might have $\hat{U}_{\text{c}}(0)\neq\hat{I}$).
Moreover, the form of $\hat{U}_{\text{I}}(t)$ allows us to interpret
the new picture in a very intuitive way: we are simply removing some
part $\hat{U}_{\text{c}}(t)$ from the total evolution $\hat{U}(t)$
of the system. And since we are removing it from the state, we add
it to the operators, as shown in (\ref{AI}), in order for expectation
values to remain unaffected, that is,
\begin{equation}
\langle\hat{A}\rangle(t)=\text{tr}\{\hat{\rho}_{\text{I}}(t)\hat{A}_{\text{I}}(t)\}.
\end{equation}
As you may imagine, it is very interesting to have a tool like this,
that allows us to remove parts of the evolution of the system that
we are not interested in, for example because they induce trivial
fast dynamics that simply overshadow more subtle slowly-varying phenomena.
Note also that taking $\hat{U}_{\text{c}}=\hat{I}$ we remain in the
Schrödinger picture, while taking $\hat{U}_{c}(t)=\hat{U}(t)$ we
move all the way to the Heisenberg picture.

It's interesting to write down the evolution equation of the state
in the new intermediate picture. Taking the time derivative of (\ref{rhoI}),
and denoting the Hamiltonian of the system by $\hat{H}(t)$, where
we even allow for some explicit time dependence, so that $\mathrm{i}\hbar\partial_{t}\hat{\rho}=[\hat{H}(t),\hat{\rho}]$,
we easily obtain
\begin{equation}
\mathrm{i}\hbar\partial_{t}\hat{\rho}_{\text{I}}=[\hat{H}_{\text{I}}(t),\hat{\rho}_{\text{I}}],\quad\text{with }\hat{H}_{\text{I}}(t)=\hat{U}_{\text{c}}^{\dagger}(t)\hat{H}(t)\hat{U}_{\text{c}}(t)-\mathrm{i}\hbar\hat{U}_{\text{c}}^{\dagger}(t)\partial_{t}\hat{U}_{\text{c}}(t),\label{VonNeumannIntermediate}
\end{equation}
where we have used $\partial_{t}(\hat{U}_{\text{c}}^{\dagger}\hat{U}_{\text{c}})=0$,
so that $(\partial_{t}\hat{U}_{\text{c}}^{\dagger})\hat{U}_{\text{c}}=-\hat{U}_{\text{c}}^{\dagger}\partial_{t}\hat{U}_{\text{c}}$.
Hence, states still evolve according to the von Neumann equation,
but with a modified Hamiltonian $\hat{H}_{\text{I}}(t)$. Note that
this Hamiltonian is in general time dependent, even if the original
one $\hat{H}$ is not. The case in which we choose $\hat{U}_{\text{c}}(t)=\exp(\hat{H}_{\text{c}}t/\mathrm{i}\hbar)$,
with $\hat{H}_{\text{c}}$ an arbitrary Hermitian time-independent
operator with energy units, is especially interesting, since in that
case the Hamiltonian in the new picture takes the form
\begin{equation}
\hat{H}_{\text{I}}(t)=\hat{U}_{\text{c}}^{\dagger}(t)\hat{H}(t)\hat{U}_{\text{c}}(t)-\hat{H}_{\text{c}}.\label{HI}
\end{equation}
Moreover, in such case, operators evolve according to a Heisenberg
equation, $\mathrm{i}\hbar\partial_{t}\hat{A}_{\text{I}}=[\hat{A}_{\text{I}},\hat{H}_{\text{c}}]$.

As mentioned, it is clear that Hamiltonians in intermediate pictures
are time dependent, even if the original Hamiltonian is not. It is
then interesting to consider which form the time-evolution operator
takes in such a case. In order to find a compact expression, we will
combine the so-called Dyson expansion with the time-ordering symbol.
The Dyson expansion is just a formal solution of the von Neumann equation
(\ref{VonNeumannIntermediate}), or its equivalent Schrödinger equation
$\mathrm{i}\hbar\partial_{t}|\psi\rangle_{\text{I}}=\hat{H}_{\text{I}}(t)|\psi\rangle_{\text{I}}$.
In particular, we are looking for the unitary operator that connects
the states at times $0$ and $t$, that is, $|\psi(t)\rangle_{\text{I}}=\hat{U}_{\text{I}}(t)|\psi(0)\rangle_{\text{I}}$.
Inserting this expression in the Schrödinger equation, and noting
that it should be satisfied for all initial states $|\psi(0)\rangle_{\text{I}}$,
we then see that the time-evolution operator itself satisfies the
Schrödinger equation
\begin{equation}
\mathrm{i}\hbar\partial_{t}\hat{U}_{\text{I}}(t)=\hat{H}_{\text{I}}(t)\hat{U}_{\text{I}}(t),\qquad\text{with }\hat{U}_{\text{I}}(0)=\hat{I}.\label{SchrodingerU}
\end{equation}
The formal integration of this equation reads
\begin{equation}
\hat{U}_{\text{I}}(t)=\hat{I}+\frac{1}{\mathrm{i}\hbar}\int_{0}^{t}dt_{1}\hat{H}_{\text{I}}(t_{1})\hat{U}_{\text{I}}(t_{1}).
\end{equation}
Iterating this solution, by plugging it in $\hat{U}_{\text{I}}(t_{1})$
on the right-hand-side, we obtain
\begin{equation}
\hat{U}_{\text{I}}(t)=\hat{I}+\frac{1}{\mathrm{i}\hbar}\int_{0}^{t}dt_{1}\hat{H}_{\text{I}}(t_{1})+\frac{1}{(\mathrm{i}\hbar)^{2}}\int_{0}^{t}dt_{1}\int_{0}^{t_{1}}dt_{2}\hat{H}_{\text{I}}(t_{1})\hat{H}_{\text{I}}(t_{2})\hat{U}_{\text{I}}(t_{2}).
\end{equation}
Iterating this process an infinite number of times, we find the formal
expression of the time-evolution operator in the form of the following
\emph{Dyson series} expansion
\begin{equation}
\hat{U}_{\text{I}}(t)=\hat{I}+\sum_{n=1}^{\infty}\frac{1}{(\mathrm{i}\hbar)^{n}}\int_{0}^{t}dt_{1}\int_{0}^{t_{1}}dt_{2}...\int_{0}^{t_{n-1}}dt_{n}\hat{H}_{\text{I}}(t_{1})\hat{H}_{\text{I}}(t_{2})...\hat{H}_{\text{I}}(t_{n}).
\end{equation}
This is not a very practical expression, except for perturbation theory,
where the series can be truncated to a desired order. However, it
can be written in a very compact and useful form by making use of
the time-ordering symbol $\mathcal{T}$, which is defined as the operation
that reorders products of time-dependent operators in chronological
order (from right to left). For example, for a product of two operators,
we have
\begin{equation}
\mathcal{T}\left\{ \hat{A}(t)\hat{B}(t')\right\} =\left\{ \begin{array}{cc}
\hat{A}(t)\hat{B}(t') & \text{for }t>t'\\
\hat{B}(t')\hat{A}(t) & \text{for }t<t'
\end{array}\right..
\end{equation}
The Dyson series can then be written in the simpler form
\begin{equation}
\hat{U}_{\text{I}}(t)=\mathcal{T}\left\{ e^{\int_{0}^{t}dt'\hat{H}_{\text{I}}(t')/\mathrm{i}\hbar}\right\} ,\label{TimeEvoOp-TimeOrdered}
\end{equation}
which is specially useful if we know the explicit time dependence
of the Hamiltonian, so the integral can be carried out, after which
the time-ordering symbol is not required anymore. We have an example
of this far along the course, see Eq. (\ref{Uc-OpenSystems}). Note
also that the time-ordering symbol can also be removed if the Hamiltonian
commutes with itself at all times, $[\hat{H}_{\text{I}}(t),\hat{H}_{\text{I}}(t')]=0$
$\forall(t,t')$.

Let us finally remark that, in the literature, what we have called
`intermediate' pictures are sometimes called `interaction' pictures.
However, we will reserve such name for the specific situation of starting
from a Hamiltonian $\hat{H}=\hat{H}_{0}(t)+\hat{H}_{\text{int}}(t)$,
in which $\hat{H}_{0}(t)$ contains the free evolution of two subsystems
interacting through $\hat{H}_{\text{int}}(t)$, and moving to an intermediate
picture defined by $\hat{U}_{\text{c}}(t)=\mathcal{T}\left\{ e^{\int_{0}^{t}dt'\hat{H}_{0}(t')/\mathrm{i}\hbar}\right\} $,
that is, a picture where free evolution is `discounted' in the sense
described above, and the Hamiltonian in the new picture simply reads
as $\hat{H}_{\text{I}}(t)=\hat{U}_{\text{c}}^{\dagger}(t)\hat{H}_{\text{int}}(t)\hat{U}_{\text{c}}(t)$.

\newpage

\section{Quantization of the electromagnetic field as a collection of harmonic
oscillators.\label{Sec:QuantizationEMfield}}

The main goal of this chapter is the quantization of the electromagnetic
field. We will follow a heuristic, but physically intuitive approach
in which, starting from Maxwell equations, the electromagnetic field
is put in correspondence with a mechanical model consisting of collection
of harmonic oscillators. Given the connection between the electromagnetic
field and the harmonic oscillator, we also study in detail here the
latter. We first explain how the one-dimensional harmonic oscillator
is described in a classical context by a trajectory in phase space.
The first step in the quantum description will be finding the Hilbert
space by which it is described. We will then introduce a convenient
way of representing the quantum state of the oscillator in phase space,
the so-called Wigner function, and then put forward a few important
states: coherent, squeezed, and thermal states.

\subsection{Light as an electromagnetic wave}

Nowadays it feels quite natural to say that \textit{light} is an \textit{electromagnetic
wave}. Arriving to this conclusion, however, was not trivial at all.
The history of such a discovery starts in the first half of the XIX
century with Faraday, who showed that the polarization of light can
change when subject to a magnetic field; this was the first hint suggesting
that there could be a connection between light and electromagnetism,
and he was the first to propose that light could be an electromagnetic
disturbance of some kind, able to propagate without the need of a
reference medium. However, this qualitative idea did not find a rigorous
mathematical formulation until the second half of the century, when
Maxwell developed a consistent theory of electromagnetism, and showed
how the theory was able to predict the existence of electromagnetic
waves propagating at a speed which was in agreement with the speed
measured for light at the time \cite{Maxwell1865}. A couple of decades
after his proposal, the existence of electromagnetic waves was experimentally
demonstrated by Hertz \cite{Hertz1893}, and the theory of light as
an electromagnetic wave found its way towards being accepted.

Our starting point are Maxwell's equations formulated as partial differential
equations for the \textit{electric} and \textit{magnetic}\footnote{\textit{As we won't deal with materials sensitive to the magnetic
field, we will use the term ``magnetic field\textquotedblright{}
for the $\mathbf{B}$-field, which is usually denoted by ``magnetic
induction field\textquotedblright{} when it needs to be distinguished
from the $\mathbf{H}$-field (which we won't be using in this notes).}} vector \textit{fields} $\mathbf{E}\left(\mathbf{r},t\right)$ and
$\mathbf{B}\left(\mathbf{r},t\right)$, respectively, where $\mathbf{r}$
($t$) is the position (time) where (when) the fields are observed.
This formulation is due to Heaviside \cite{Heaviside893Book}, as
Maxwell originally proposed his theory in terms of \textit{quaternions}.
The theory consists of four equations. The first two are called the
\textit{homogeneous Maxwell equations} and read
\begin{equation}
\boldsymbol{\nabla}\cdot\mathbf{B}=0\text{ \ \ \ and \ \ \ }\boldsymbol{\nabla}\times\mathbf{E}=-\partial_{t}\mathbf{B},\label{MaxwellHomoEqs}
\end{equation}
where $\boldsymbol{\nabla}=\left(\partial_{x},\partial_{y},\partial_{z}\right)$.
The other two are called the \textit{inhomogeneous Maxwell equations}
and are written as
\begin{equation}
\boldsymbol{\nabla}\cdot\mathbf{E}=\rho/\varepsilon_{0}\text{ \ \ \ and \ \ \ }\boldsymbol{\nabla}\times\mathbf{B}=\mu_{0}\varepsilon_{0}\partial_{t}\mathbf{E}+\mu_{0}\mathbf{j},\label{MaxwellInhomoEqs}
\end{equation}
where any electric or magnetic source is introduced in the theory
by a \textit{charge density} function $\rho\left(\mathbf{r},t\right)$
and a \textit{current distribution} vector $\mathbf{j}\left(\mathbf{r},t\right)$,
respectively; the parameters $\varepsilon_{0}=8.8\times10^{-12}$
F/m and $\mu_{0}=1.3\times10^{-6}$ H/m are the so-called \textit{electric
permittivity} and the \textit{magnetic permeability} of vacuum, respectively.

We will show the process of quantization of the electromagnetic field
in the absence of sources ($\rho=0$ and $\mathbf{j}=\mathbf{0}$).
Under these circumstances, the inhomogeneous equations are simplified
to
\begin{equation}
\boldsymbol{\nabla}\cdot\mathbf{E}=0\text{ \ \ \ and \ \ \ }c^{2}\boldsymbol{\nabla}\times\mathbf{B}=\partial_{t}\mathbf{E}\text{,}
\end{equation}
where $c=1/\sqrt{\varepsilon_{0}\mu_{0}}\simeq3\times10^{8}\mathrm{m/s}$.

The homogeneous equations (\ref{MaxwellHomoEqs}) allow us to derive
the fields from a \textit{scalar potential} $\phi\left(\mathbf{r},t\right)$
and a \textit{vector potential} $\mathbf{A}\left(\mathbf{r},t\right)$
as
\begin{equation}
\mathbf{B}=\boldsymbol{\nabla}\times\mathbf{A}\text{ \ \ and \ \ }\mathbf{E}=-\boldsymbol{\nabla}\phi-\partial_{t}\mathbf{A}\text{,}\label{AtoEB}
\end{equation}
hence reducing to four the degrees of freedom of the electromagnetic
field. These potentials, however, are not unique: we can always use
an arbitrary function $\Lambda\left(\mathbf{r},t\right)$ to change
them as
\begin{equation}
\mathbf{A\rightarrow A}+\boldsymbol{\nabla}\Lambda\text{ \ \ and \ \ }\phi\rightarrow\phi-\partial_{t}\Lambda\text{,}
\end{equation}
what is known as the \textit{gauge invariance} of Maxwell's equations.

Introducing (\ref{AtoEB}) into the inhomogeneous equations we get
the equations satisfied by the potentials\begin{subequations} 
\begin{align}
\left(c^{2}\boldsymbol{\nabla}^{2}-\partial_{t}^{2}\right)\mathbf{A} & =\partial_{t}\boldsymbol{\nabla}\phi+c^{2}\boldsymbol{\nabla}\left(\boldsymbol{\nabla}\cdot\mathbf{A}\right),\\
\boldsymbol{\nabla}^{2}\phi+\partial_{t}\boldsymbol{\nabla}\cdot\mathbf{A} & =0.
\end{align}
\end{subequations}The problem is highly simplified if we exploit
the gauge invariance and choose $\boldsymbol{\nabla}\cdot\mathbf{A}=0$.
In such case, the second equation becomes the Laplace equation $\boldsymbol{\nabla}^{2}\phi=0$,
whose solution can always be chosen\footnote{In particular, note that the most general solution of this equation
is $\phi(\mathbf{r},t)=a(t)+\mathbf{b}(t)\cdot\mathbf{r}$. However,
given this solution, we can always transform to a different gauge
defined by $\Lambda(\mathbf{r},t)=-\alpha(t)-\boldsymbol{\beta}(t)\cdot\mathbf{r}$,
with the choices $\partial_{t}\alpha=a$ and $\partial_{t}\boldsymbol{\beta}=\mathbf{b}$,
such that the new potentials read $\phi'=\phi-\partial_{t}\Lambda=0$
and $\mathbf{A}'=\mathbf{A}-\boldsymbol{\nabla}\Lambda=\mathbf{A}-\boldsymbol{\beta}$.
Thus, the new scalar potential vanishes, while the vector potential
is still divergence-free, $\boldsymbol{\nabla}\cdot\mathbf{A}'=0$,
and satisfies the homogeneous wave equation $\left(c^{2}\boldsymbol{\nabla}^{2}-\partial_{t}^{2}\right)\mathbf{A}'=\mathbf{0}$
as we wanted. } as $\phi=0$. Hence the only equations left are 
\begin{equation}
\left(c^{2}\boldsymbol{\nabla}^{2}-\partial_{t}^{2}\right)\mathbf{A}=\mathbf{0}\label{WaveEq}
\end{equation}
which are wave equations with speed $c$ for the components of the
vector potential. Note that the condition $\boldsymbol{\nabla}\cdot\mathbf{A}=0$
(known as the \textit{Coulomb condition}) relates the three components
of $\mathbf{A}$, and hence, only two degrees of freedom of the initial
six (the electric and magnetic vector fields) remain.

It is finally important to note that the wave equation (\ref{WaveEq})
has a unique solution inside a given spatio-temporal region only if
both the vector potential and its derivative along the direction normal
to the boundary of the region are specified at any point of the boundary
\cite{Jackson62book}; these are known as \textit{Dirichlet} and \textit{Neumann}
conditions, respectively. In general, however, physical problems do
not impose so many constraints, and hence there coexist several solutions
of the wave equation, which we will call \textit{spatiotemporal} \textit{modes}.
Each of these modes can in principle be excited independently, so
that the total electromagnetic field takes the form of a superposition
of all of them. 

\subsection{Quasi-1D approximation and quantization inside a cavity}

\begin{figure}
\includegraphics[width=0.8\textwidth]{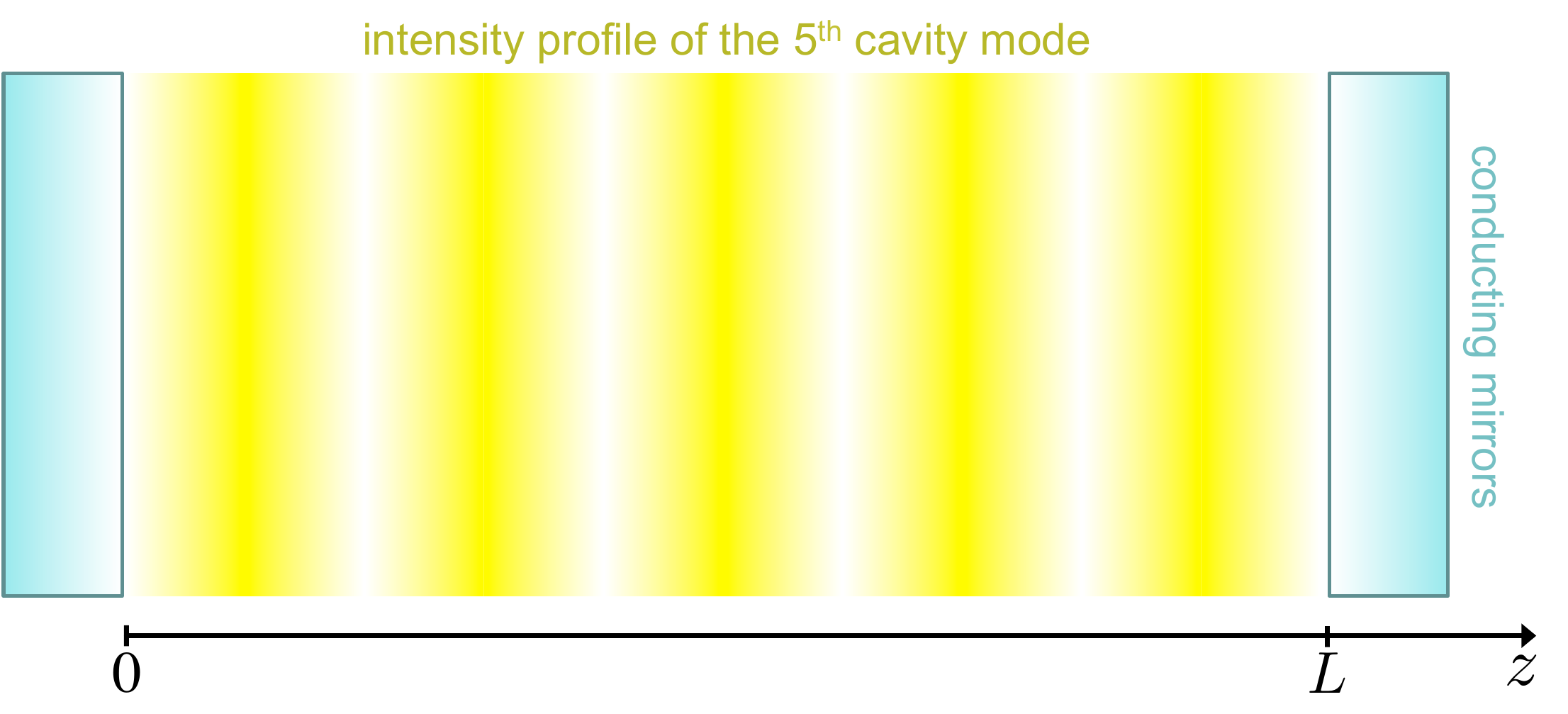}\caption{Simplified cavity used in the notes. It is formed by two perfectly
conducting plane mirrors facing each other. We neglect variations
in polarization or transverse ($x$,$y$) profile, while the vanishing
of the electric field at the mirror's surface provides the cavity
mode structure. We show the intensity pattern corresponding to the
5$^{\text{th}}$ mode, proportional to $\sin^{2}5\pi z/L$. \label{fCavity}}
\end{figure}

In order to simplify the derivations and get to the core of the physical
problem without many spurious technicalities, we will consider a simplified
one-dimensional model for the light field\footnote{Quantization in three dimensions and in a realistic cavity with spherical
dielectric mirrors can be found in \cite{NavarretePhDthesis}.}. In particular, we assume that the field propagates along the $z$
direction, with neither its polarization nor its transverse profile
along the $x$ and $y$ directions changing upon propagation. Moreover,
we take the transverse profile as homogeneous, that is, independent
of $x$ and $y$, and a linear polarization along the $x$ axis. Under
such conditions, the vector potential can be expanded in terms of
linearly polarized plane waves $\mathbf{e}_{x}e^{\mathrm{i}kz}$ with
$k\in\mathbb{R}$ and $\mathbf{e}_{x}$ the unit vector along the
$x$ direction. Hence, we can generally write
\begin{equation}
\mathbf{A}(z,t)=\mathbf{e}_{x}\sum_{k}\mathcal{N}_{k}q_{k}(t)e^{\mathrm{i}kz},
\end{equation}
where from now on we omit the dependence on $x$ and $y$ in the argument
of the fields, $q_{k}$ are the expansion coefficients, for which
we allow a general normalization factor $\mathcal{N}_{k}$ for future
convenience, and the sum runs over all the allowed wave vectors, which
are all the real numbers in free space, but get restricted whenever
the boundary conditions are not open. Specifically, let us consider
a simple cavity consisting on two perfectly conducting plane mirrors
facing each other, see Fig. \ref{fCavity}. The components of the
electric field parallel to the conducting mirrors must vanish \cite{Jackson62book,Griffiths99book}.
Taking $z=0$ at the mirror on the left, and $z=L$ at the mirror
on the right, and focusing on a pair of $\pm k$ components, the $z=0$
boundary conditions imply
\begin{align}
\mathbf{E}(0,t) & =-\partial_{t}\mathbf{A}(0,t)=0\Rightarrow\mathcal{N}_{k}\dot{q}_{k}(t)=-\mathcal{N}_{-k}\dot{q}_{-k}(t).
\end{align}
Choosing $\mathcal{N}_{k}=\mathcal{N}_{-k}$ and integrating the previous
equation, we obtain $q_{k}(t)=b-q_{-k}(t),$ where furthermore $q_{k}(t)\in\mathbb{R}$
and $b=0$, because otherwise $\mathbf{A}(z,t)$ is not real in all
spacetime. On the other hand, the boundary conditions at $z=L$ imply
\[
\mathbf{E}(L,t)=-\partial_{t}\mathbf{A}(L,t)=0\Rightarrow\dot{q}_{k}(t)\sin(kL)=0\Rightarrow kL=n\pi,\text{with }n\in\mathbb{N}.
\]
Putting everything together, we obtain the relation $q_{k}(t)=-q_{-k}(t)\in\mathbb{R}$
between the expansion coefficients, and $k=\pm\pi n/L\equiv\pm k_{n}$
with $n\in\mathbb{N}$ as the only allowed wave vectors. The vector
potential, and the electric and magnetic fields are then written as\begin{subequations}\label{FieldExpansions}
\begin{align}
\mathbf{A}(z,t) & =\mathbf{e}_{x}\sum_{n=1}^{\infty}\mathcal{N}_{n}q_{n}(t)\sin\left(k_{n}z\right),\label{Acavity}\\
\mathbf{E}(z,t) & =-\partial_{t}\mathbf{A}(z,t)=-\mathbf{e}_{x}\sum_{n=1}^{\infty}\mathcal{N}_{n}\dot{q}_{n}(t)\sin\left(k_{n}z\right),\\
\mathbf{B}(z,t) & =\boldsymbol{\nabla}\times\mathbf{A}(z,t)=\mathbf{e}_{y}\sum_{n=1}^{\infty}\mathcal{N}_{n}k_{n}q_{n}(t)\cos\left(k_{n}z\right),
\end{align}
\end{subequations}so that the electric field vanishes at $z=0$ and
$z=L$ as required. Note that we have changed the notation slightly
from $q_{k}$ to $q_{n}$, since the wave vectors are specified by
the integer $n$. Also, note that we have introduced an irrelevant
factor $2\mathrm{i}$ into the normalization constants, denoted by
$\mathcal{N}_{n}$, since these arbitrary factors will conveniently
chosen later.

It is interesting to note that the choices $q_{k}(t)=-q_{-k}(t)$
and $k=\pi n/L$ can be understood from a different point of view,
more general than the boundary conditions imposed by the conducting
mirrors. The choice for the amplitudes $q_{\pm k}$ simply reflects
the fact that inside a cavity there is no reason why we should distinguish
left moving and right moving waves, as they cannot be excited separately
(if we excite one, the other will appear by reflection in the mirror,
picking up a $\pi$ phase). On the other hand, the quantization of
the wave number appears from the natural requirement that a cavity
mode has to reproduce itself after a cavity roundtrip, so that the
phase accumulated after a roundtrip, $2kL$, must be an integer multiple
of $2\pi$.

At this point it is important to highlight the following orthogonality
relations between the cavity mode functions
\begin{equation}
\int_{0}^{L}dz\sin(k_{n}z)\sin(k_{l}z)=\int_{0}^{L}dz\cos(k_{n}z)\cos(k_{l}z)=\frac{L}{2}\delta_{nl}.\label{eq:OrthogonalityModes}
\end{equation}
Introducing (\ref{Acavity}) in the wave equation and acting with
$\int_{0}^{L}dz\sin(k_{n}z)$ from the left, we get
\begin{equation}
\int_{0}^{L}dz\sin(k_{n}z)\left(c^{2}\partial_{z}^{2}-\partial_{t}^{2}\right)\mathbf{A}=-\mathbf{e}_{x}\sum_{l=1}^{\infty}\mathcal{N}_{l}\int_{0}^{L}dz\sin(k_{n}z)\sin\left(k_{l}z\right)\left(k_{l}^{2}c^{2}q_{l}(t)+\ddot{q}_{l}(t)\right)=\mathbf{0},
\end{equation}
which using the previous orthogonality relations leads us the following
evolution equations for the amplitudes $q_{n}(t)$:
\begin{equation}
\ddot{q}_{n}+\omega_{n}^{2}q_{n}=0,
\end{equation}
with $\omega_{n}=ck_{n}$. This is precisely the evolution equation
for a harmonic oscillator of frequency $\omega_{n}$. Hence, this
suggests that the modes of the electromagnetic field behave as harmonic
oscillators, so that field quantization can be carried out simply
by quantizing each of these oscillators. The way we will proceed to
prove this is by showing that the electromagnetic energy indeed coincides
with the Hamiltonian of these oscillators. In particular, note that
the previous equations are associated with a Lagrangian\footnote{see Section \ref{ClassicalMechanics} for a reminder of classical
mechanics}
\begin{equation}
L_{\text{em}}=\sum_{n=1}^{\infty}\frac{m_{n}}{2}\left(\dot{q}_{n}^{2}-\omega_{n}^{2}q_{n}\right),
\end{equation}
where $m_{n}$ are the masses of the oscillators (arbitrary at this
point). The corresponding canonical momenta are $p_{n}=\partial L_{\text{em}}/\partial\dot{q}_{n}=m_{n}\dot{q}_{n}$,
leading to a Hamiltonian
\begin{equation}
H_{\text{em}}=\sum_{n=1}^{\infty}\left(\frac{p_{n}^{2}}{2m_{n}}+\frac{m_{n}\omega_{n}^{2}}{2}q_{n}\right).\label{Hem}
\end{equation}
Next we prove that this is indeed compatible with the electromagnetic
energy contained in the cavity, which at time $t$ reads \cite{Griffiths99book,Jackson62book}
\begin{equation}
E_{\text{em}}(t)=\frac{1}{2}\int_{\text{cavity}}d^{3}\mathbf{r}\left[\varepsilon_{0}\mathbf{E}^{2}(z,t)+\frac{1}{\mu_{0}}\mathbf{B}^{2}(z,t)\right].\label{Eem}
\end{equation}
Introducing the expansions (\ref{FieldExpansions}) in this expression,
and using the orthogonality relations (\ref{eq:OrthogonalityModes}),
we obtain
\begin{eqnarray}
E_{\text{em}}(t) & = & \frac{1}{2}\int_{\text{cavity}}dxdy\sum_{n,m=1}^{\infty}\mathcal{N}_{n}\mathcal{N}_{m}\left[\varepsilon_{0}\dot{q}_{n}(t)\dot{q}_{m}(t)\int_{0}^{L}dz\sin\left(k_{n}z\right)\sin\left(k_{m}z\right)+\frac{k_{n}k_{m}}{\mu_{0}}q_{n}(t)q_{m}(t)\int_{0}^{L}dz\cos\left(k_{n}z\right)\cos\left(k_{m}z\right)\right]\nonumber \\
 & = & \frac{\varepsilon_{0}LS}{4}\sum_{n=1}^{\infty}\mathcal{N}_{n}^{2}\left[\dot{q}_{n}^{2}(t)+c^{2}k_{n}^{2}q_{n}^{2}(t)\right].
\end{eqnarray}
where we have assumed that the transverse integral contributes with
a finite area $\int_{\text{cavity}}dxdy=S$ (as would happen in a
physical scenario where the electromagnetic field is confined also
laterally). Using the definition of the canonical momenta presented
above and choosing the normalization constants as $\mathcal{N}_{n}=\sqrt{2m_{n}/\varepsilon_{0}LS}$,
we get
\begin{equation}
E_{\text{em}}(t)=\sum_{n=1}^{\infty}\left[\frac{p_{n}^{2}(t)}{2m_{n}}+\frac{m_{n}\omega_{n}^{2}}{2}q_{n}^{2}(t)\right],
\end{equation}
which is precisely the Hamiltonian derived in (\ref{Hem}).

We have been able to make a formal correspondence between the electromagnetic
field contained inside a cavity and a mechanical model consisting
of a collection of independent Harmonic oscillators. We can then quantize
the electromagnetic field by following the quantization prescription
that we learned in Section \ref{QuantumMehcanicsPureStates}. In particular,
we just need to replace positions and momenta with self-adjoint operators
$\hat{q}_{n}$ and $\hat{p}_{n}$ satisfying canonical commutation
relations $[\hat{q}_{n},\hat{p}_{m}]=\mathrm{i}\hbar\delta_{nm}$
and $[\hat{q}_{n},\hat{q}_{m}]=0=[\hat{p}_{n},\hat{p}_{m}]$. Expressions
(\ref{FieldExpansions}) turn then into the quantized fields\begin{subequations}\label{FieldExpansionsQuantum}
\begin{align}
\hat{\mathbf{A}}(z) & =\mathbf{e}_{x}\sum_{n=1}^{\infty}\mathcal{N}_{n}\hat{q}_{n}\sin\left(k_{n}z\right),\label{AcavityQuantum}\\
\hat{\mathbf{E}}(z) & =-\mathbf{e}_{x}\sum_{n=1}^{\infty}\frac{\mathcal{N}_{n}}{m_{n}}\hat{p}_{n}\sin\left(k_{n}z\right),\\
\hat{\mathbf{B}}(z) & =\mathbf{e}_{y}\sum_{n=1}^{\infty}\mathcal{N}_{n}k_{n}\hat{q}_{n}\cos\left(k_{n}z\right),
\end{align}
\end{subequations}where removing the time dependence is equivalent
to the implicit assumption that we are working in the Schrödinger
picture, where operators do not evolve in time. Of course, the fields
recover their time dependence in the Heisenberg picture. In this lectures
we will alternate between these pictures, depending on the particular
situation. Note further that, throughout the process, the masses $m_{n}$
of the electromagnetic oscillators have remain arbitrary and we are
free to choose them as desired. While this seems puzzling at first,
we will show at the end of the chapter that they pose no observational
consequences whatsoever.

Let us finally remark that, while we have presented the highly idealized
and simplified situation of a perfect optical cavity within the quasi
1-D approximation, quantization in more realistic settings proceeds
in a similar way, see for example \cite{NavarretePhDthesis} for a
still pretty ideal cavity, but with spherical mirrors. In particular,
it's always a matter of finding the normal modes of Maxwell's equations
and mapping them to a collection of harmonic oscillators, which can
be technically challenging, but is conceptually the same we have done.
Moreover, even in the context of general quantum field theory one
proceeds exactly in the same fashion, first finding or proposing the
Lagrangian of the theory, and then imposing canonical commutation
relations between the fields and their canonical momenta. In such
case, there are a few important subtleties though. For example, when
dealing with fields with half-integer spin (fermions), canonical \emph{anticommutation}
relations must be used in order to obtain a bounded Hamiltonian (spin-statistics
theorem). Also, at high energies, one needs to use a formalism that
explicitly shows Lorentz invariance, in order to make sure that the
theory is free of incompatibilities with special relativity. In our
case, while Maxwell equations are indeed Lorentz invariant, the Coulomb-gauge
condition $\boldsymbol{\nabla}\cdot\mathbf{A}=0$ is not, which is
fine for our low-energy (optical) purposes, but can lead to inconsistencies
in higher-energy contexts. If you are interested on the quantization
of the electromagnetic field in a relativistic setting and/or in the
presence of matter,\textit{ }\cite{QO2} is a good place to start.

\subsection{Classical analysis of the harmonic oscillator\label{ClassHO-1}}

We thus see that there exists a direct relation between the electromagnetic
field and the one-dimensional harmonic oscillator. Let us then discuss
along the next sections the physics of the latter. We start here by
its classical description (remember that Section \ref{ClassicalMechanics}
offers a review of the required classical mechanics).

Consider the basic mechanical model of a \textit{one-dimensional harmonic
oscillator}: A particle of mass $m$ is at rest at some equilibrium
position which we take as $x=0$; when displaced from this position
by some amount $a$, a restoring force $F=-kx$ starts acting on the
particle, trying to bring it back to $x=0$. Newton's equation of
motion for the particle is therefore $m\ddot{x}=-kx$, which together
with the initial conditions $x(0)=a$ and $\dot{x}(0)=v$ gives the
solution $x(t)=a\cos\omega t+(v/\omega)\sin\omega t$, being $\omega=\sqrt{k/m}$
the so-called \textit{angular frequency}. Therefore the particle will
be bouncing back and forth between positions $-\sqrt{a^{2}+v^{2}/\omega^{2}}$
and $\sqrt{a^{2}+v^{2}/\omega^{2}}$ with time period $2\pi/\omega$
(hence the name `harmonic oscillator').

\begin{figure}
\includegraphics[width=0.35\textwidth]{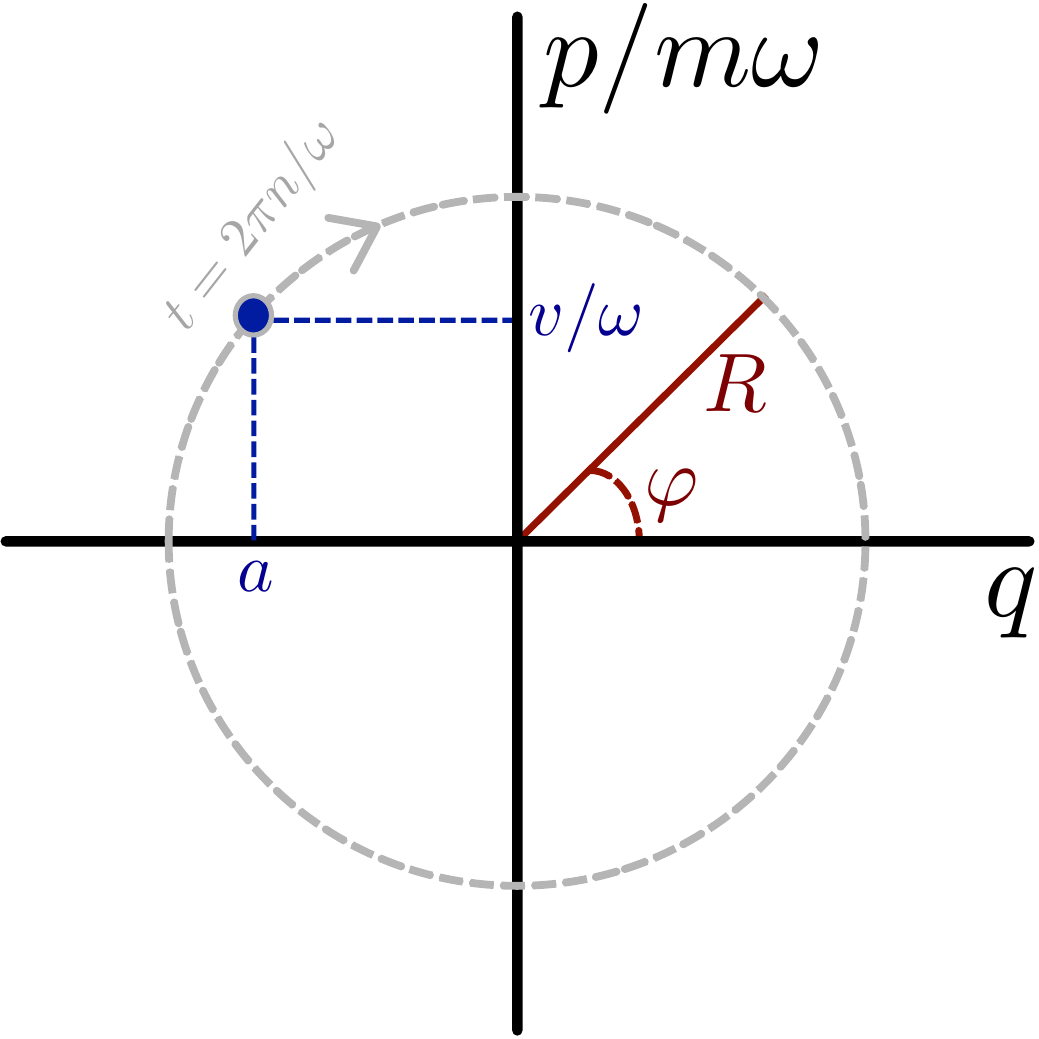}\caption{Phase space trajectory of the classical harmonic oscillator. It starts
at a point $(a,v/\omega)$ and describes a circular trajectory of
radius $R$ coming back to this initial point at times $t_{n}$. Defining
the amplitude and phase of the oscillator, the motion is described
by a fixed amplitude and a phase $\varphi$ which decreases linearly
with time, as explained in the main text. \label{Fig-OsciClass}}
\end{figure}

Let us study now the problem from a Hamiltonian point of view. For
this one-dimensional problem with no constraints, we can take the
position of the particle and its momentum as the generalized coordinate
and momentum, that is, $q=x$ and $p=m\dot{x}$. The restoring force
derives from a potential $V(x)=kx^{2}/2$, and hence the Hamiltonian
takes the form
\begin{equation}
H_{\text{o}}=\frac{p^{2}}{2m}+\frac{m\omega^{2}}{2}q^{2}\text{.}\label{ClassHOham-1}
\end{equation}
The canonical equations read 
\begin{equation}
\dot{q}=\frac{p}{m}\hspace{8mm}\text{and}\hspace{8mm}\dot{p}=-m\omega^{2}q,\label{CanonicalEquationsHO}
\end{equation}
which together with some initial conditions $q(0)=a$ and $p(0)=mv$
give the trajectory 
\begin{equation}
\left(q,\frac{p}{m\omega}\right)=\left(a\cos\omega t+\frac{v}{\omega}\sin\omega t,\frac{v}{\omega}\cos\omega t-a\sin\omega t\right),
\end{equation}
where we normalize the momentum to $m\omega$ for convenience. Starting
at the phase space point $(a,v/\omega)$ the system evolves periodically
drawing a circle of radius $R=\sqrt{a^{2}+v^{2}/\omega^{2}}$ as shown
in Fig. \ref{Fig-OsciClass}, returning to its initial point at times
$t_{n}=2\pi n/\omega$, with $n\in\mathbb{N}$. This circular trajectory
could have been derived without even solving the equations of motion,
as the conservation of the Hamiltonian $H_{\text{o}}(t)=H_{\text{o}}(0)$
leads directly to $q^{2}+p^{2}/m^{2}\omega^{2}=R^{2}$, which is exactly
the circumference of Fig. \ref{Fig-OsciClass}. This is a simple manifestation
of the power of the Hamiltonian formalism. The trajectory is most
easily represented by defining the so-called \emph{normal} variable
$\nu(t)=q(t)+\mathrm{i}p(t)/m\omega=R\exp[\mathrm{i}\varphi(t)]$,
which in this case has a constant amplitude $|\nu|$ and a phase that
decreases linearly with time, $\varphi(t)=\varphi(0)-\omega t$.

\subsection{The quantum harmonic oscillator: number states, energy quantization,
and quadrature eigenstates}

The harmonic oscillator is the prototype of a system described quantum
mechanically by an infinite-dimensional Hilbert space. In order to
see this, we just find the eigenstates of its Hamiltonian, which is
given by the operator 
\begin{equation}
\hat{H}_{\text{o}}=\frac{\hat{p}^{2}}{2m}+\frac{m\omega^{2}}{2}\hat{q}^{2},
\end{equation}
by virtue of the discussion after Principle III in Section \ref{QuantumMechanics},
with the \textit{position} $\hat{q}$ and \textit{momentum} $\hat{p}$
satisfying the commutation relation 
\begin{equation}
[\hat{q},\hat{p}]=\mathrm{i}\hbar.
\end{equation}
We will always work with dimensionless versions of them, the so-called
\textsf{X} and \textsf{P} \textit{quadratures} (although we may keep
using the names `position' and `momentum' most of the time) 
\begin{equation}
\hat{X}=\hat{q}/q_{\text{zpf}}\hspace{8mm}\text{and}\hspace{8mm}\hat{P}=\hat{p}/p_{\text{zpf}},\label{QuadNorm}
\end{equation}
with $q_{\text{zpf}}=\sqrt{\hbar/2\omega m}$ and $p_{\text{zpf}}=\sqrt{\hbar\omega m/2}$,
which satisfy the commutation relation 
\begin{equation}
[\hat{X},\hat{P}]=2\mathrm{i},\label{XPcommutator}
\end{equation}
and therefore the uncertainty relation 
\begin{equation}
\Delta X\Delta P\geq1.\label{XPuncertainty}
\end{equation}
We will see later that $q_{\text{zpf}}$ and $p_{\text{zpf}}$ provide
the uncertainty of position and momentum when the oscillator is in
its ground state, the so-called \emph{zero-point fluctuations}. In
terms of these quadratures, the Hamiltonian reads 
\begin{equation}
\hat{H}_{\text{o}}=\frac{\hbar\omega}{4}\left(\hat{X}^{2}+\hat{P}^{2}\right).
\end{equation}
In order to find the eigensystem of this operator, we decompose the
quadratures as 
\begin{equation}
\hat{X}=\hat{a}^{\dagger}+\hat{a}\hspace{8mm}\text{and}\hspace{8mm}\hat{P}=\mathrm{i}(\hat{a}^{\dagger}-\hat{a}),
\end{equation}
where the operators $\hat{a}$ and $\hat{a}^{\dagger}$, known as
the \textit{annihilation} and \textit{creation} \textit{operators},
satisfy the commutation relation 
\begin{equation}
[\hat{a},\hat{a}^{\dagger}]=1.
\end{equation}
In terms of these operators, the Hamiltonian is rewritten as 
\begin{equation}
\hat{H}_{\text{o}}=\hbar\omega(\hat{a}^{\dagger}\hat{a}+1/2),\label{HosciAnnihilationCreation}
\end{equation}
and hence the problem has been reduced to finding the eigensystem
of the so-called \textit{number operator} $\hat{N}=\hat{a}^{\dagger}\hat{a}$.

Let us denote by $n$ a generic real number contained in the spectrum
of $\hat{N}$, whose corresponding eigenvector we denote by $|n\rangle$,
so that, $\hat{N}|n\rangle=n|n\rangle$. The eigensystem of $\hat{N}$
is readily found from the following four properties:
\begin{itemize}
\item $\hat{N}$ is a positive semidefinite operator, as for any vector
$|\psi\rangle$ it is satisfied $\langle\psi|\hat{N}|\psi\rangle=|\hat{a}|\psi\rangle|^{2}\geq0$.
When applied to its eigenvectors, $|\psi\rangle=|n\rangle$, this
property forbids the existence of negative eigenvalues, that is, 
\begin{equation}
n\geq0.
\end{equation}
\item Applying the commutation relation\footnote{This is straightforward to find by using the property $[\hat{A}\hat{B},\hat{C}]=\hat{A}[\hat{B},\hat{C}]+[\hat{A},\hat{C}]\hat{B}$,
valid for any three operators $\hat{A}$, $\hat{B}$, and $\hat{C}$.} $[\hat{N},\hat{a}]=-\hat{a}$ to $\vert n\rangle$, it is straightforward
to show that the vector $\hat{a}|n\rangle$ is also an eigenvector
of $\hat{N}$ with eigenvalue $n-1$. Similarly, from the commutation
relation $[\hat{N},\hat{a}^{\dagger}]=\hat{a}^{\dagger}$ it is found
that the vector $\hat{a}^{\dagger}|n\rangle$ is an eigenvector of
$\hat{N}$ with eigenvalue $n+1$.\\
Hence, we have $\hat{a}|n\rangle=k_{1}|n-1\rangle$ and $\hat{a}^{\dagger}|n\rangle=k_{2}|n+1\rangle$,
with some constants $k_{1}$ and $k_{2}$ that can be found as follows.
We just calculate the absolute value squared of these expressions,
obtaining\begin{subequations}
\begin{align}
|k_{1}|^{2}\langle n-1|n-1\rangle & =\langle n|\hat{a}^{\dagger}\hat{a}|n\rangle=n\langle n|n\rangle,\\
|k_{2}|^{2}\langle n+1|n+1\rangle & =\underset{\hat{a}^{\dagger}\hat{a}+1}{\langle n|\underbrace{\hat{a}\hat{a}^{\dagger}}|n\rangle}=(n+1)\langle n|n\rangle.
\end{align}
\end{subequations}Assuming that the eigenvectors can be normalized
(which is another of the properties that we introduce next), and taking
the constants positive for definiteness, we then obtain $k_{1}=\sqrt{n}$
and $k_{2}=\sqrt{n+1}$, and finally
\begin{equation}
\hat{a}|n\rangle=\sqrt{n}|n-1\rangle\hspace{8mm}\text{and}\hspace{8mm}\hat{a}^{\dagger}|n\rangle=\sqrt{n+1}|n+1\rangle.\label{LadderHO}
\end{equation}
\item The identities $\langle n|\hat{N}|m\rangle=n\langle n|m\rangle=m\langle n|m\rangle$,
allow us to write $(n-m)\langle n|m\rangle=0$, hence showing that
eigenvectors corresponding to different eigenvalues are arthogonal,
that is,
\begin{equation}
\langle n|m\rangle=0\hspace{1em}\text{when}\ n\neq m.
\end{equation}
\item We will show later that the eigenstate with $n=0$ is normalizable,
and hence, so are all the other eigenvectors with $n\in\mathbb{N}$
by virtue of (\ref{LadderHO}).
\end{itemize}
These four properties are compatible only with a spectrum formed by
non-negative integers $n=0,1,2,3,...$; otherwise (\ref{LadderHO})
would allow us to find negative eigenvalues, which are not allowed
by the first property. Note that (\ref{LadderHO}) ensures that $\hat{a}|0\rangle=0$.
Thus, the set of eigenvectors $\left\{ |n\rangle\right\} _{n=0,1,...}$
is an infinite, countable set of orthonormal eigenvectors, that is,
$\langle n|m\rangle=\delta_{nm}$. Finally, according to the principles
of quantum mechanics only the vectors normalized to one are physically
accessible states, and hence we conclude that the vector space spanned
by the eigenvectors of $\hat{N}$ is an infinite-dimensional Hilbert
space (it is isomorphic to $l^{2}(\infty)$, the prototype of infinite-dimensional
Hilbert space, see Section \ref{InfiniteHilbert}). The orthonormal
basis $\left\{ |n\rangle\right\} _{n=0,1,...}$ is known as \emph{Fock
basis}.

Let us now explain some physical consequences. The vectors $\left\{ |n\rangle\right\} _{n=0,1,...}$
are eigenvectors of the energy (the Hamiltonian) with eigenvalues
$\{E_{n}=\hbar\omega(n+1/2)\}_{n=0,1,...}$, and hence quantum theory
predicts that the energy of the oscillator is quantized: Only a discrete
set of energies separated by $\hbar\omega$ can be measured in an
experiment. The number of \textit{quanta} or \textit{excitations}
is given by $n$, and that's why $\hat{n}$ is called the `number'
operator, as it `counts' the number of excitations. Similarly, the
creation and annihilation operators receive their names because they
add and subtract excitations. As these vectors have a well defined
number of excitations, $\Delta N=0$, we will call them \textit{number
states}. Consequently, $|0\rangle$ will be called the \textit{vacuum
state} of the oscillator, as it has no quanta.

On the other hand, while in classical mechanics the harmonic oscillator
can have zero energy (what happens when it is resting in its equilibrium
position), quantum mechanics predicts that the minimum energy that
the oscillator can have is $E_{0}=\hbar\omega/2>0$. One way to understand
where this \textit{zero\textendash point} energy comes from is by
minimizing the expectation value of the Hamiltonian, which can be
written as 
\begin{equation}
\langle\hat{H}_{\text{o}}\rangle=\frac{\hbar\omega}{4}\left(\Delta X^{2}+\Delta P^{2}+\langle\hat{X}\rangle^{2}+\langle\hat{P}\rangle^{2}\right),\label{MeanEnergyQuadratures-1}
\end{equation}
subject to the constraint $\Delta X\Delta P\geq1$ imposed by the
uncertainty principle. It is easy to argue that the minimum value
of $\langle\hat{H}_{\text{o}}\rangle$ is obtained for the state satisfying
$\Delta X=\Delta P=1$ and $\langle\hat{X}\rangle=\langle\hat{P}\rangle=0$,
which corresponds, not surprisingly, to the vacuum state $|0\rangle$.
Hence, the energy present in the ground state of the oscillator comes
from the fact that the uncertainty principle does not allow its position
and momentum to be exactly zero, they have some fluctuations even
in the vacuum state, and this \textit{vacuum }\textit{\emph{or }}\textit{zero-point
fluctuations} contribute to the energy of the oscillator. Indeed,
since the quadrature fluctuations are equal to 1 with our normalization
of position and momentum, see (\ref{QuadNorm}), this means that $q_{\text{zpf}}=\sqrt{\hbar/2\omega m}$
and $p_{\text{zpf}}=\sqrt{\hbar\omega m/2}$ are the real position
and momentum zero-point fluctuations for the particular oscillator
we work with. 

In contrast to the number operator, which has a discrete spectrum,
the quadrature operators possess a pure continuous spectrum. Let us
focus on the $\hat{X}$ operator, whose eigenvectors we denote by
$\{|x\rangle\}_{x\in\mathbb{R}}$ with corresponding eigenvalues $\{x\}_{x\in\mathbb{R}}$,
that is, 
\begin{equation}
\hat{X}|x\rangle=x|x\rangle.
\end{equation}
In order to prove that $\hat{X}$ has a pure continuous spectrum,
just note that, from the relation 
\begin{equation}
e^{\frac{\mathrm{i}}{2}y\hat{P}}\hat{X}e^{-\frac{\mathrm{i}}{2}y\hat{P}}=\hat{X}+y,
\end{equation}
which is easily found via the Baker-Campbell-Haussdorf lemma\footnote{This lemma reads 
\begin{equation}
e^{\hat{B}}\hat{A}e^{-\hat{B}}=\sum_{n=0}^{\infty}\frac{1}{n!}\underset{n}{\underbrace{[\hat{B},[\hat{B},...[\hat{B},}}\hat{A}\underset{n}{\underbrace{]...]]}},\label{BCHlemma-1}
\end{equation}
and is valid for two general operators $\hat{A}$ and $\hat{B}$.}, it follows that if $|x\rangle$ is an eigenvector of $\hat{X}$
with $x$ eigenvalue, then the vector $\exp(-\mathrm{i}y\hat{P}/2)|x\rangle$
is also an eigenvector of $\hat{X}$ with eigenvalue $x+y$. Now,
as this holds for any real $y$, we conclude that the spectrum of
$\hat{X}$ is the whole real line. Moreover, as a self-adjoint operator,
one can use its eigenvectors as a continuous basis of the Hilbert
space of the oscillator by using the Dirac normalization $\langle x|y\rangle=\delta(x-y)$.
The same results can be obtained for the $\hat{P}$ operator, whose
eigenvectors we denote\footnote{Truth is that it doesn't look very smart to differentiate eigenstates
of different operators ($\hat{N},$ $\hat{X}$, $\hat{P}$,...) just
by the label ($n$, $x$, $p$,...), but it will always be clear which
state we are referring to from the context.} by $\{|p\rangle\}_{p\in\mathbb{R}}$ with corresponding eigenvalues
$\{p\}_{p\in\mathbb{R}}$, that is, 
\begin{equation}
\hat{P}|p\rangle=p|p\rangle.
\end{equation}
Note that this results rely only on the canonical commutation relations,
and hence are completely general, valid for any system, not only for
the harmonic oscillator. Note also that not being vectors contained
in the Hilbert space of the oscillator (they cannot be properly normalized),
the position and momentum eigenvectors cannot correspond to physical
states. Nevertheless, we will see that they can be understood as an
unphysical limit of some physical states (the squeezed states).

Let us now prove that the vacuum state can be normalized. We will
proceed by constructing explicitly its representation in the position
eigenbasis (\emph{wave function}). Let us first prove that there exists
a Fourier transform relation between the position and momentum bases,
that is, 
\begin{equation}
|p\rangle=\int_{-\infty}^{+\infty}\frac{dx}{\sqrt{4\pi}}\exp\left(\frac{\mathrm{i}}{2}px\right)|x\rangle\hspace{5mm}\Longleftrightarrow\hspace{5mm}|x\rangle=\int_{-\infty}^{+\infty}\frac{dp}{\sqrt{4\pi}}\exp\left(-\frac{\mathrm{i}}{2}px\right)|p\rangle.
\end{equation}
To this aim we now prove that 
\begin{equation}
\langle x|p\rangle=\frac{1}{\sqrt{4\pi}}\exp(\mathrm{i}xp/2).\label{xy-scalar}
\end{equation}
First note that the commutator $[\hat{X},\hat{P}]=2\mathrm{i}$ implies
that 
\begin{equation}
\langle x|\hat{P}|x^{\prime}\rangle=\frac{2\mathrm{i}\delta\left(x-x^{\prime}\right)}{x-x^{\prime}},
\end{equation}
and hence 
\begin{eqnarray}
\langle x|\hat{P}|\psi\rangle & = & \int_{\mathbb{R}}dx'\langle x|\hat{P}|x'\rangle\langle x'|\psi\rangle=\int_{\mathbb{R}}dx'\frac{2\mathrm{i}\delta(x-x')}{x-x'}\langle x'|\psi\rangle\\
 & = & \int_{\mathbb{R}}dx^{\prime}\frac{2\mathrm{i}\delta\left(x-x^{\prime}\right)}{x-x^{\prime}}\left[\langle x|\psi\rangle+(x^{\prime}-x)\frac{d\langle x|\psi\rangle}{dx}+\sum_{n=2}^{\infty}\frac{(x^{\prime}-x)^{n}}{n!}\frac{d^{n}\langle x|\psi\rangle}{dx^{n}}\right].\nonumber 
\end{eqnarray}
The order zero of the Taylor expansion is zero because the kernel
is antisymmetric around $x$, while the terms of order two or above
give zero as well after integrating them. This means that 
\begin{equation}
\langle x|\hat{P}|\psi\rangle=-2\mathrm{i}\frac{d\langle x|\psi\rangle}{dx},\label{MomentumDerivative}
\end{equation}
which applied to $|\psi\rangle=|p\rangle$ yields the differential
equation 
\begin{equation}
p\langle x|p\rangle=-2\mathrm{i}\frac{d\langle x|p\rangle}{dx},
\end{equation}
with solution $\langle x|p\rangle=C\exp(\mathrm{i}xp/2)$. We find
the normalization constant $C$ by demanding the states to satisfy
the Dirac-delta normalization\footnote{When working with quadratures, the following form of the Dirac delta
is useful:
\begin{equation}
\delta(x)=\int_{\mathbb{R}}\frac{dz}{2\pi}e^{\mathrm{i}zx}=\int_{\mathbb{R}}\frac{dp}{4\pi}e^{\mathrm{i}px/2}.
\end{equation}
}:
\[
\langle x|x'\rangle=\int_{\mathbb{R}}dp\langle x|p\rangle\langle p|x'\rangle=|C|^{2}\int_{\mathbb{R}}dpe^{\mathrm{i}(x-x')p/2}=4\pi|C|^{2}\delta(x-x')\Rightarrow C=1/\sqrt{4\pi},
\]
which leads to (\ref{xy-scalar}).

As an example of the use of these continuous representations, we now
find the position representation of the number states, which we write
as 
\begin{equation}
|n\rangle=\int_{\mathbb{R}}dx\psi_{n}(x)|x\rangle.
\end{equation}
As a first step we find the projection of vacuum onto a position eigenstate,
the so-called \textit{ground state wave function} $\psi_{0}(x)=\langle x|0\rangle$.
We do it from 
\begin{equation}
0=\langle x|\hat{a}|0\rangle=\frac{1}{2}\langle x|(\hat{X}+\mathrm{i}\hat{P})|0\rangle=\frac{1}{2}\left(x+2\frac{d}{dx}\right)\psi_{0}(x),
\end{equation}
where we have used (\ref{MomentumDerivative}), which is a simple
differential equation for $\psi_{0}(x)$ having 
\begin{equation}
\psi_{0}(x)=\frac{1}{(2\pi)^{1/4}}e^{-x^{2}/4},
\end{equation}
as its solution. The factor $(2\pi)^{-1/4}$ is found by imposing
the normalization 
\begin{equation}
\langle0|0\rangle=\langle0|\underbrace{\int_{\mathbb{R}}dx|x\rangle\langle x|}_{\hat{I}}|0\rangle=\int_{\mathbb{R}}dx|\psi_{0}(x)|^{2}=1,\label{NormVacuum}
\end{equation}
which proves that the vacuum state is normalizable.

Let us also write explicit wave functions for an arbitrary number
state $|n\rangle$ (the $n^{\mathrm{th}}$ \textit{excited wave function}).
This is found from the ground state wave function as 
\begin{equation}
\psi_{n}(x)=\langle x|n\rangle=\frac{1}{\sqrt{n!}}\langle x|\hat{a}^{\dagger n}|0\rangle=\frac{1}{\sqrt{n!}2^{n}}\langle x|\left(\hat{X}-\mathrm{i}\hat{P}\right)^{n}|0\rangle=\frac{1}{\sqrt{n!}2^{n}}\left(x-2\frac{d}{dx}\right)^{n}\psi_{0}(x),
\end{equation}
which, reminding the Rodrigues formula for the Hermite polynomials
\begin{equation}
H_{n}\hspace{-1mm}\left(\frac{x}{\sqrt{2}}\right)=2^{-n/2}e^{x^{2}/4}\left(x-2\frac{d}{dx}\right)^{n}e^{-x^{2}/4},
\end{equation}
leads to the simple expression 
\begin{equation}
\psi_{n}(x)=\frac{1}{\sqrt{2^{n+1/2}\pi^{1/2}n!}}H_{n}\hspace{-1mm}\left(\frac{x}{\sqrt{2}}\right)e^{-x^{2}/4}.\label{FockWF}
\end{equation}

\subsection{Visualizing quantum states in phase space: The Wigner function}

As the position and momentum do not have common eigenstates, and moreover,
their eigenvectors cannot correspond to physical states of the oscillator,
one concludes that these observables cannot take definite values in
quantum mechanics. Given the state $\hat{\rho}$, the best one can
offer are\textit{ probability density functions} that will dictate
the statistics of a measurement of these observables, that is, $\langle x|\hat{\rho}|x\rangle$
and $\langle p|\hat{\rho}|p\rangle$. In other words, quantum mechanically,
well defined trajectories in phase space do not exist: the position
and momentum of the oscillator are always affected by some (\textit{quantum})
\textit{noise}.

The following question arises naturally: is it then possible to describe
quantum mechanics as a probability distribution in phase space which
simply blurs classical trajectories? As we are about to see, the answer
is only partially positive, as quantum noise is much more subtle than
common classical noise.

Let us denote such probability distribution by $W_{\rho}(\mathbf{r})$,
where we combine all phase-space variables into the vector $\mathbf{r}=(x,p)^{T}$
and the subindex indicates the quantum state $\hat{\rho}$ it corresponds
to. A logical way of building this distribution is as that having
the position and momentum probability density functions as its marginals,
that is,
\begin{equation}
\langle x|\hat{\rho}|x\rangle=\int_{\mathbb{R}}dpW_{\rho}(\mathbf{r})\hspace{7mm}\text{and}\hspace{7mm}\langle p|\hat{\rho}|p\rangle=\int_{\mathbb{R}}dxW_{\rho}(\mathbf{r}).\label{Marginals}
\end{equation}
It's possible to show that these conditions uniquely define $W_{\rho}(\mathbf{r})$,
which receives the name of \emph{Wigner function}.

Before presenting it, let us combine the quadrature operators in the
vector $\hat{\mathbf{R}}=(\hat{X},\hat{P})^{T}$, so that the canonical
commutation relations can be combined into the single expression
\[
[\hat{R}_{m},\hat{R}_{n}]=2\mathrm{i}\Omega_{mn},\quad\text{where }\Omega=\left(\begin{array}{cc}
0 & 1\\
-1 & 0
\end{array}\right),
\]
is the so-called \emph{symplectic form}, which satisfies $\Omega^{T}=-\Omega=\Omega^{-1}$.
It is also convenient to define the so-called \emph{displacement operator}
\begin{equation}
\hat{D}(\mathbf{r})=\exp\left[\frac{\mathrm{i}}{2}\hat{\mathbf{R}}^{T}\Omega\mathbf{r}\right]=\exp\left[\frac{\mathrm{i}}{2}(p\hat{X}-x\hat{P})\right],\label{DisplacementOp}
\end{equation}
 of which we will learn a lot more later, and its expectation value\emph{
}
\begin{equation}
\chi_{\rho}(\mathbf{s})=\text{tr}\{\hat{\rho}\hat{D}(\mathbf{s})\}=\langle\hat{D}(\mathbf{s})\rangle.\label{CharacteristicFunction}
\end{equation}
which is known as \emph{quantum characteristic} \emph{function}. Consider
then the distribution defined as the Fourier transform of the characteristic
function
\begin{equation}
W_{\rho}(\mathbf{r})=\int_{\mathbb{R}^{2}}\frac{d^{2}\mathbf{s}}{(4\pi)^{2}}e^{-\frac{\mathrm{i}}{2}\mathbf{r}^{T}\Omega\mathbf{s}}\chi_{\rho}(\mathbf{s}),\label{GenWigner}
\end{equation}
which is dubbed the \emph{Wigner function} of state $\hat{\rho}$.
In the next section we prove that this distribution satisfies the
following properties:
\begin{enumerate}
\item It has the right marginals, as defined by (\ref{Marginals}).
\item It is real at all points in phase space, that is, $W_{\rho}(\mathbf{r})\in\mathbb{R}$
for all $\mathbf{r}$ and $\hat{\rho}$.
\item It is normalized, that is,
\begin{equation}
\int_{\mathbb{R}^{2}}d^{2}\mathbf{r}W_{\rho}(\mathbf{r})=1.
\end{equation}
\item Averages in phase space correspond to quantum expectation values of
symmetrically-ordered operators, that is,
\begin{equation}
\langle(\hat{X}^{m}\hat{P}^{n})^{(s)}\rangle=\int_{\mathbb{R}^{2}}d^{2}\mathbf{r}W_{\rho}(\mathbf{r})x^{m}p^{n},\label{SymExpectWigner}
\end{equation}
where we remind that $(\hat{X}^{m}\hat{P}^{n})^{(s)}$ refers to the
symmetrized version of the corresponding product with respect to position
and momentum, e.g., $(\hat{X}^{2}\hat{P})^{(s)}=(\hat{X}^{2}\hat{P}+\hat{P}\hat{X}^{2}+\hat{X}\hat{P}\hat{X})/3$.
Taking into account that a reasonable prescription for finding the
quantum operator associated to a classical observable $A(x,p)$ consists
precisely of symmetrizing it with respect to $x$ and $p$, and then
change the position and momentum by the corresponding self-adjoint
operators (what guarantees the self-adjointness of the remaining operator,
as explained in Secs. \ref{QuantumMehcanicsPureStates} and \ref{Axiom4}),
this result seems to reinforce the interpretation of $W_{\rho}(\mathbf{r})$
as a probability density function dictating how quantum fluctuations
are distributed in phase space.
\item The Wigner function admits the following alternative expression in
terms of the transition amplitude $\langle x_{2}|\hat{\rho}|x_{1}\rangle$
between positions $x_{1}$ and $x_{2}$:
\begin{equation}
W_{\rho}(\mathbf{r})=\int_{\mathbb{R}}\frac{dy}{4\pi}e^{-\frac{\mathrm{i}}{2}py}\left\langle x+\frac{y}{2}\right|\hat{\rho}\left|x-\frac{y}{2}\right\rangle .\label{WignerMidPoint}
\end{equation}
This is indeed the original form first introduced by Wigner \cite{SchleichBook},
and it is sometimes useful for calculations.
\item The trace product of two states can be written as the overlap between
their corresponding Wigner functions, that is,
\begin{equation}
\text{tr}\{\hat{\rho}_{1}\hat{\rho}_{2}\}=4\pi\int_{\mathbb{R}^{2}}d^{2}\mathbf{r}W_{\rho_{1}}(\mathbf{r})W_{\rho_{2}}(\mathbf{r}).\label{TraceProductWigner}
\end{equation}
\item When particularizing this expression to two orthogonal pure states,
that is, $\hat{\rho}_{j}=|\psi_{j}\rangle\langle\psi_{j}|$ with $\langle\psi_{1}|\psi_{2}\rangle=0$,
we obtain\footnote{At this point, it is interesting to remind that for any vector $|\psi\rangle$
and operator $\hat{A}$, the following identity holds $\text{tr}\{|\psi\rangle\langle\psi|\hat{A}\}=\langle\psi|\hat{A}|\psi\rangle$.
This is easy to prove just by writing the trace in a basis of the
Hilbert space, and using that the sum of the corresponding projectors
is a resolution of the identity.}
\begin{equation}
\int_{\mathbb{R}^{2}}d^{2}\mathbf{r}W_{|\psi_{1}\rangle}(\mathbf{r})W_{|\psi_{2}\rangle}(\mathbf{r})=\frac{1}{4\pi}|\langle\psi_{1}|\psi_{2}\rangle|^{2}=0.
\end{equation}
This identity is only possible if the Wigner function is negative
at some points of phase space, as otherwise the product of two Wigner
functions would always add positively to the integral. Hence, in general,
the Wigner function is not a true probability density function in
the classical-statistical sense!
\item When particularizing the trace-product expression to $\hat{\rho}_{1}=\hat{\rho}_{2}\equiv\hat{\rho}$,
we get
\begin{equation}
\int_{\mathbb{R}^{2}}d^{2}\mathbf{r}W_{\rho}^{2}(\mathbf{r})=\frac{1}{4\pi}\text{tr}\{\hat{\rho}^{2}\}\leq\frac{1}{4\pi},
\end{equation}
where the last inequality follows from $\text{tr}\{\hat{\rho}^{2}\}\leq1$,
a condition satisfied by all density operators, since they have positive
eigenvalues smaller than or equal to one. This expression shows that
the Wigner function cannot have divergences, and hence, it can always
be plotted in phase space to visualize quantum states.
\item It uniquely determines the quantum state $\hat{\rho}$ of the system
as
\begin{equation}
\hat{\rho}=\int_{\mathbb{R}^{2}}\frac{d^{2}\mathbf{s}}{4\pi}\hat{D}^{\dagger}(\mathbf{s})\chi_{\rho}(\mathbf{s}),\label{WtoRho}
\end{equation}
and hence the Wigner function or its characteristic function contain
the same information as the quantum state, providing an alternative
representation of the latter. At the end of the next section we explain
in detail how this expression is facilitated by the fact that the
set of displacement operators $\{\hat{D}(\mathbf{r})\}_{\mathbf{r}\in\mathbb{R}^{2}}$
forms a basis in the space of operators normalizable with respect
the trace norm, which is a Hilbert space.
\end{enumerate}
These properties show that the Wigner function has almost all the
right properties expected from a probability density function in phase
space. In fact, it only differs from it in the fact that it can be
negative. This is indeed crucial, as it means that quantum mechanics
cannot be simulated with classical means, and therefore, it really
goes beyond anything we could predict with classical physics. Despite
not being a true probability density function in general, the Wigner
function is still a very useful and rigorous way of visualizing quantum
fluctuations in phase space. Moreover, as we will see, most of the
easily experimentally accessible states of the harmonic oscillator
have a positive Wigner function, and therefore, for these states we
can apply all the intuition behind standard probability theory to
how quantum fluctuations are distributed in phase space. In addition,
whenever negativities appear in an experiment, these offer a smoking
gun that something genuinely quantum is happening, something that
cannot be simulated by adding classical noise in the system.

\begin{figure}
\includegraphics[width=0.6\textwidth]{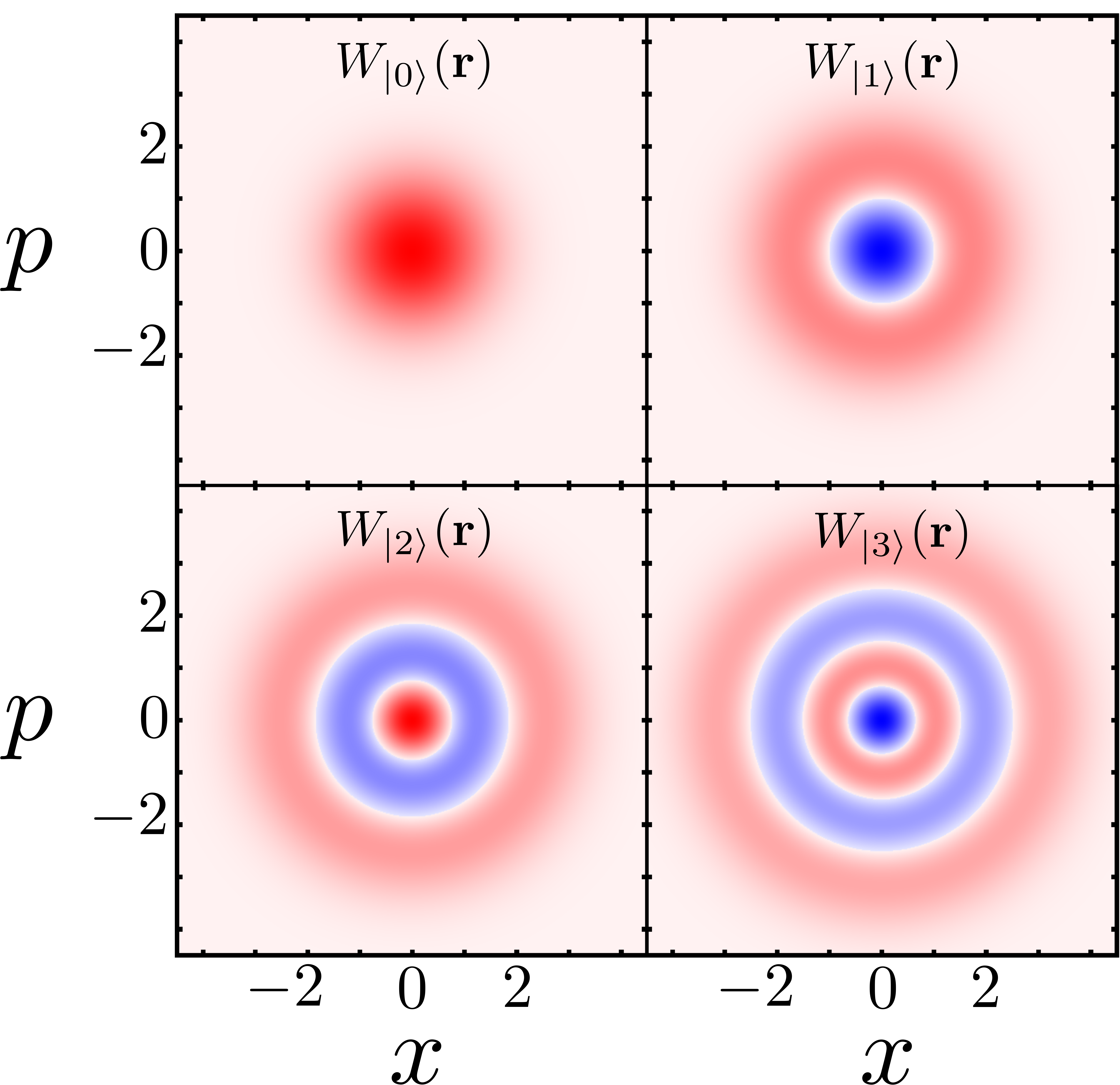}\caption{Density plot of the Wigner functions corresponding to the first 4
Fock states. Red and blue regions correspond to positive and negative
values of the function, respectively. In both cases regions with higher
contrast correspond to larger absolute value.\label{Fig-WignersFock}}
\end{figure}

To conclude this section, let us evaluate the Wigner function of a
number state $\hat{\rho}=|n\rangle\hspace{-0.4mm}\langle n|$ as an
example. It is not difficult to do so by using expression (\ref{WignerMidPoint})
for the Wigner function, together with the wave function (\ref{FockWF})
that we found for that state, obtaining
\begin{align}
W_{|n\rangle}(\mathbf{r})= & \int_{\mathbb{R}}\frac{dy}{4\pi}e^{-\frac{\mathrm{i}}{2}py}\langle x+y/2|n\rangle\langle n|x-y/2\rangle=\int_{\mathbb{R}}\frac{dy}{4\pi}e^{-\frac{\mathrm{i}}{2}py}\psi_{n}(x+y/2)\psi_{n}^{*}(x-y/2)\\
= & \frac{1}{2^{n+1/2}\pi^{1/2}n!}\int_{\mathbb{R}}\frac{dy}{4\pi}e^{-\frac{\mathrm{i}}{2}py-\frac{(x+y/2)^{2}+(x-y/2)^{2}}{4}}H_{n}\hspace{-1mm}\left(\frac{x+y/2}{\sqrt{2}}\right)H_{n}\hspace{-1mm}\left(\frac{x-y/2}{\sqrt{2}}\right)\nonumber \\
= & \frac{1}{2^{n+1/2}\pi^{1/2}n!}e^{-\frac{x^{2}}{2}}\int_{\mathbb{R}}\frac{dy}{4\pi}e^{-\frac{\mathrm{i}}{2}py-\frac{y^{2}}{8}}H_{n}\hspace{-1mm}\left(\frac{x+y/2}{\sqrt{2}}\right)H_{n}\hspace{-1mm}\left(\frac{x-y/2}{\sqrt{2}}\right)\nonumber \\
= & \frac{1}{2^{n+1/2}\pi^{1/2}n!}e^{-\frac{x^{2}+p^{2}}{2}}\int_{\mathbb{R}}\frac{dy}{4\pi}e^{-\frac{1}{8}(y+2\mathrm{i}p)^{2}}H_{n}\hspace{-1mm}\left(\frac{x+y/2}{\sqrt{2}}\right)H_{n}\hspace{-1mm}\left(\frac{x-y/2}{\sqrt{2}}\right),\nonumber 
\end{align}
where in the last step we have just completed the square in the exponential.
Let us now make the variable change $y=2(z-\mathrm{i}p)$, which is
a shift of the integration variable along the imaginary line, so that
the integral now is still performed parallel to the real line in the
complex-$y$ plane. The previous expression is then turned into
\begin{align}
W_{|n\rangle}(\mathbf{r}) & =\frac{1}{2^{n+1/2}\pi^{1/2}n!}e^{-\frac{x^{2}+p^{2}}{2}}\int_{\mathrm{i}p-\infty}^{\mathrm{i}p+\infty}\frac{dz}{2\pi}e^{-z^{2}/2}H_{n}\hspace{-1mm}\left(\frac{x+z-\mathrm{i}p}{\sqrt{2}}\right)H_{n}\hspace{-1mm}\left(\frac{x-z+\mathrm{i}p}{\sqrt{2}}\right)\\
 & =\frac{(-1)^{n}}{2^{n+1/2}\pi^{1/2}n!}e^{-\frac{x^{2}+p^{2}}{2}}\int_{\mathrm{i}p-\infty}^{\mathrm{i}p+\infty}\frac{dz}{2\pi}e^{-z^{2}/2}H_{n}\hspace{-1mm}\left(\frac{z+x-\mathrm{i}p}{\sqrt{2}}\right)H_{n}\hspace{-1mm}\left(\frac{z-x-\mathrm{i}p}{\sqrt{2}}\right),\nonumber 
\end{align}
where in the second equality we have used $H_{n}(-x)=(-1)^{n}H_{n}(x)$.
Finally, using the following relation between the Hermite and Laguerre
polynomials
\begin{equation}
\int_{\mathrm{i}p-\infty}^{\mathrm{i}p+\infty}dze^{-z^{2}/2}H_{n}\hspace{-1mm}\left(\frac{z+\xi_{1}}{\sqrt{2}}\right)H_{n}\hspace{-1mm}\left(\frac{z-\xi_{2}}{\sqrt{2}}\right)=2^{n}\sqrt{2\pi}n!L_{n}(\xi_{1}\xi_{2}),
\end{equation}
where the Laguerre polynomial of order $n$ is defined by the Rodrigues
formula 
\begin{equation}
L_{n}(z)=\frac{\exp(z)}{n!}\frac{d^{n}}{dz^{n}}\left[z^{n}\exp(-z)\right],
\end{equation}
we arrive to the simple expression
\begin{equation}
W_{|n\rangle}(\mathbf{r})=\frac{(-1)^{n}}{2\pi}L_{n}(x^{2}+p^{2})e^{-\frac{x^{2}+p^{2}}{2}}.\label{NumberWigner}
\end{equation}
For any $n>0$, this function has negative regions. For example, for
odd $n$ it is always negative at the origin of phase space, since
$L_{n}(0)=1\hspace{2mm}\forall n$. The Wigner functions of the first
4 Fock states are plotted in Fig. \ref{Fig-WignersFock}.

\subsection{Proof of the Wigner function properties and discussion about the
Hilbert space of operators}

In this section we first prove the properties of the Wigner function
introduced above, and then provide some interesting facts about the
structure of the space of operators. Let us start with property 1.
In particular, we next prove that integrating the Wigner function
over momenta leads to the probability density function associated
to position measurements. We start from
\begin{equation}
\int_{\mathbb{R}}dpW_{\rho}(\mathbf{r})=\int_{\mathbb{R}^{2}}\frac{d^{2}\mathbf{r}'}{(4\pi)^{2}}\underset{4\pi\delta(x')}{\underbrace{\left[\int_{\mathbb{R}}dpe^{\frac{\mathrm{i}}{2}x'p}\right]}}e^{-\frac{\mathrm{i}}{2}p'x}\chi(\mathbf{r}')=\int_{\mathbb{R}}\frac{dp'}{4\pi}e^{-\frac{\mathrm{i}}{2}p'x}\underbrace{\chi(0,p')}_{\text{tr}\{\hat{\rho}\hat{D}(0,p')\}},
\end{equation}
and then write the trace in the position eigenbasis obtaining
\begin{equation}
\int_{\mathbb{R}}dpW_{\rho}(\mathbf{r})=\int_{\mathbb{R}}dy\int_{\mathbb{R}}\frac{dp'}{4\pi}e^{-\frac{\mathrm{i}}{2}p'x}\langle y|\hat{\rho}e^{\frac{\mathrm{i}}{2}p'\hat{X}}|y\rangle=\int_{\mathbb{R}}dy\underset{4\pi\delta(x-y)}{\underbrace{\left[\int_{\mathbb{R}}\frac{dp'}{4\pi}e^{-\frac{\mathrm{i}}{2}p'(x-y)}\right]}}\langle y|\hat{\rho}|y\rangle=\langle x|\hat{\rho}|x\rangle,
\end{equation}
just as we wanted to prove. Similarly, you can prove that the integration
of the Wigner function over position leads to the probability density
function for momentum measurements.

Property 3, that the Wigner function is normalized, is proven immediately
from the previous expression, since
\begin{equation}
\int_{\mathbb{R}^{2}}d^{2}\mathbf{r}W_{\rho}(\mathbf{r})=\int_{\mathbb{R}}dx\underset{\langle x|\hat{\rho}|x\rangle}{\underbrace{\int_{\mathbb{R}}dpW_{\rho}(\mathbf{r})}}=\text{tr}\left\{ \hat{\rho}\int_{\mathbb{R}}dx|x\rangle\langle x|\right\} =\text{tr}\{\hat{\rho}\}=1.
\end{equation}

On the other hand, the Wigner function is real (property 2) by construction,
since using $\hat{D}^{\dagger}(\mathbf{r})=\hat{D}(-\mathbf{r})$,
we get
\begin{align}
W_{\rho}^{*}(\mathbf{r}) & =\int_{\mathbb{R}^{2}}\frac{d^{2}\mathbf{r}'}{(4\pi)^{2}}e^{\frac{\mathrm{i}}{2}\mathbf{r}^{T}\Omega\mathbf{r}'}\langle\hat{D}(\mathbf{r}')\rangle^{*}=\int_{\mathbb{R}^{2}}\frac{d^{2}\mathbf{r}'}{(4\pi)^{2}}e^{\frac{\mathrm{i}}{2}\mathbf{r}^{T}\Omega\mathbf{r}'}\langle\hat{D}(-\mathbf{r}')\rangle=\int_{\mathbb{R}^{2}}\frac{d^{2}\mathbf{r}''}{(4\pi)^{2}}e^{-\frac{\mathrm{i}}{2}\mathbf{r}^{T}\Omega\mathbf{r}''}\langle\hat{D}(\mathbf{r}'')\rangle=W_{\rho}(\mathbf{r})
\end{align}
where in the second to last step we made the integration variable
change $\mathbf{r}'=-\mathbf{r}''$.

In order to prove property 4 regarding expectation values of operators
in symmetric order, we need some preliminary results. First, note
that the exponentiation of a linear combination of position and momentum
follows Newton's binomial expansion, but with products of operators
in symmetric order, that is,
\begin{equation}
(a\hat{X}+b\hat{P})^{n}=\sum_{k=0}^{n}\left(\begin{array}{c}
n\\
k
\end{array}\right)a^{n-k}b^{k}\left(\hat{X}^{n-k}\hat{P}^{k}\right)^{(s)}.
\end{equation}
You can easily convince yourself by trying out some values of $n$.
This leads to the following expansion of the displacement operator
\begin{equation}
\hat{D}(\mathbf{r})=\sum_{n=0}^{\infty}\frac{1}{n!}\left(\frac{\mathrm{i}}{2}\right)^{n}\left(p\hat{X}-x\hat{P}\right)^{n}=\sum_{n=0}^{\infty}\left(\frac{\mathrm{i}}{2}\right)^{n}\sum_{k=0}^{n}\frac{p^{n-k}(-x)^{k}}{k!(n-k)!}\left(\hat{X}^{n-k}\hat{P}^{k}\right)^{(s)}.\label{DisplacementBinomialExpansion}
\end{equation}
Then, we will also need to keep in mind the action of derivatives
of the Dirac delta function,
\begin{equation}
\int_{-\infty}^{+\infty}dx\left[\frac{d^{n}\delta(x)}{dx^{n}}\right]f(x)=(-1)^{n}\int_{-\infty}^{+\infty}dx\delta(x)\frac{d^{n}f(x)}{dx^{n}},
\end{equation}
which is easily proven integrating by parts and using the fact that
the delta function and its derivatives to all orders vanish at infinity,
that is, $[d^{n}\delta(x)/dx^{n}]|_{x=\pm\infty}=0\hspace{1em}\forall n$.
Using these properties, we can then prove the property we seek as
\begin{align}
\int_{\mathbb{R}^{2}}d^{2}\mathbf{r}W_{\rho}(\mathbf{r})x^{m}p^{r} & =\int_{\mathbb{R}^{2}}d^{2}\mathbf{r}\int_{\mathbb{R}^{2}}\frac{d^{2}\mathbf{r}'}{(4\pi)^{2}}\left[\left(-\frac{2}{i}\right)^{m}\left(\frac{2}{i}\right)^{r}\partial_{p'}^{m}\partial_{x'}^{r}e^{\frac{\mathrm{i}}{2}(x'p-p'x)}\right]\langle\hat{D}(\mathbf{r}')\rangle\\
 & =\left(-\frac{2}{i}\right)^{m}\left(\frac{2}{i}\right)^{r}\int_{\mathbb{R}^{2}}d^{2}\mathbf{r}'\left\langle \hat{D}(\mathbf{r}')\right\rangle \partial_{p'}^{m}\partial_{x'}^{r}\underset{\delta(x')\delta(p')}{\underbrace{\int_{\mathbb{R}^{2}}\frac{d^{2}\mathbf{r}}{(4\pi)^{2}}e^{\frac{\mathrm{i}}{2}(x'p-p'x)}}}\nonumber \\
 & =\left(-\frac{2}{i}\right)^{m}\left(\frac{2}{i}\right)^{r}(-1)^{r+m}\int_{\mathbb{R}^{2}}d^{2}\mathbf{r}'\delta(x')\delta(p')\partial_{p'}^{m}\partial_{x'}^{r}\langle\hat{D}(\mathbf{r}')\rangle,\nonumber 
\end{align}
where in the last step we have integrated by parts in order to bring
the derivatives to the characteristic function. The derivatives acting
on the expansion (\ref{DisplacementBinomialExpansion}) for the displacement
operator, together with the action of the delta functions, force $n=m+r$
and $k=r$ in that expression, leading to the desired expression
\begin{equation}
\int_{\mathbb{R}^{2}}d^{2}\mathbf{r}pW_{\rho}(\mathbf{r})x^{m}p^{r}=\left(\hat{X}^{m}\hat{P}^{r}\right)^{(s)}.
\end{equation}

Next we prove that the definition of the Wigner function in terms
of a quantum characteristic function (\ref{GenWigner}) coincides
with Wigner's original formulation (\ref{WignerMidPoint}). For this,
we will rely on some additional properties of the displacement and
translation operators. In the case of the displacement operator, we
use the weaker (sometimes called \emph{disentangling}) form of the
Baker-Campbell-Haussdorf lemma\footnote{This form says that given two operators $\hat{A}$ and $\hat{B}$
that commute with their commutator, we can ``disentangle'' the exponential
of their sum as
\begin{equation}
e^{\hat{A}+\hat{B}}=e^{-[\hat{A},\hat{B}]/2}e^{\hat{A}}e^{\hat{B}}.\label{DisentanglingBCH}
\end{equation}
} (\ref{DisentanglingBCH}) to write it as
\begin{equation}
\hat{D}(\mathbf{r})=e^{\frac{\mathrm{i}}{4}xp}e^{-\frac{\mathrm{i}}{2}x\hat{P}}e^{\frac{\mathrm{i}}{2}p\hat{X}},\label{DisplacementDisentangled}
\end{equation}
proven by setting $\hat{A}=-\mathrm{i}x\hat{P}/2$ and $\hat{B}=\mathrm{i}p\hat{X}/2$
in (\ref{DisentanglingBCH}). In the case of the translation operator,
we need its action on the position eigenstates
\begin{equation}
e^{-\frac{\mathrm{i}}{2}y\hat{P}}|x\rangle=|x+y\rangle,\label{TranslationPosition}
\end{equation}
which is easily proven by inserting a representation of the identity
in the momentum basis and using (\ref{xy-scalar}),
\begin{equation}
e^{-\frac{\mathrm{i}}{2}y\hat{P}}|x\rangle=\int_{\mathbb{R}}dp\langle p|x\rangle e^{-\frac{\mathrm{i}}{2}y\hat{P}}|p\rangle=\int_{\mathbb{R}}dp\underset{\langle p|x+y\rangle}{\underbrace{\frac{1}{\sqrt{4\pi}}e^{-\frac{\mathrm{i}}{2}(x+y)p}}}|p\rangle=|x+y\rangle.
\end{equation}
With this properties at hand, we can now turn our definition of the
Wigner function (\ref{GenWigner}) into Wigner's original definition
(\ref{WignerMidPoint}). For this, we simply write the trace in the
position eigenbasis, obtaining
\begin{align}
W_{\rho}(\mathbf{r}) & =\int_{\mathbb{R}^{2}}\frac{d^{2}\mathbf{r}'}{(4\pi)^{2}}e^{\frac{\mathrm{i}}{2}(x'p-p'x)}\int_{\mathbb{R}}dy\langle y|\hat{\rho}e^{\frac{\mathrm{i}}{4}x'p'}e^{-\frac{\mathrm{i}}{2}x'\hat{P}}e^{\frac{\mathrm{i}}{2}p'\hat{X}}|y\rangle=\int_{\mathbb{R}}\frac{dx'}{4\pi}e^{\frac{\mathrm{i}}{2}x'p}\int_{\mathbb{R}}dy\underset{\delta(y-x+x'/2)}{\underbrace{\int_{\mathbb{R}}\frac{dp'}{4\pi}e^{\frac{\mathrm{i}}{2}p'(y-x+x'/2)}}}\langle y|\hat{\rho}|y+x'\rangle\nonumber \\
 & =\int_{\mathbb{R}}\frac{dx'}{4\pi}e^{\frac{\mathrm{i}}{2}x'p}\langle x-x'/2|\hat{\rho}|x+x'/2\rangle,
\end{align}
which making the integration variable change $x'=-y$ leads to the
desired expression (\ref{WignerMidPoint}).

The next property, specifically the trace product rule of expression
(\ref{TraceProductWigner}), is easily proven by using Wigner's original
formulation (\ref{WignerMidPoint}). In particular, we write
\begin{align}
\int_{\mathbb{R}^{2}}d^{2}\mathbf{r}W_{\rho_{1}}(\mathbf{r})W_{\rho_{2}}(\mathbf{r}) & =\int_{\mathbb{R}}dx\int_{\mathbb{R}}\frac{dy_{1}}{4\pi}\int_{\mathbb{R}}\frac{dy_{2}}{4\pi}\underset{4\pi\delta(y_{1}+y_{2})}{\underbrace{\int_{\mathbb{R}}dpe^{-\frac{\mathrm{i}}{2}(y_{1}+y_{2})p}}}\langle x+y_{1}/2|\hat{\rho}_{1}|x-y_{1}/2\rangle\langle x+y_{2}/2|\hat{\rho}_{2}|x-y_{2}/2\rangle\nonumber \\
 & =\frac{1}{4\pi}\int_{\mathbb{R}}dx\int_{\mathbb{R}}dy\langle x+y/2|\hat{\rho}_{1}|x-y/2\rangle\langle x-y/2|\hat{\rho}_{2}|x+y/2\rangle,
\end{align}
so that making the variable changes $\{x_{+}=x+y/2,x_{-}=x-y/2\}$,
we prove the desired result
\begin{equation}
\int_{\mathbb{R}^{2}}d^{2}\mathbf{r}W_{\rho_{1}}(\mathbf{r})W_{\rho_{2}}(\mathbf{r})=\frac{1}{4\pi}\int_{\mathbb{R}^{2}}dx_{+}dx_{-}\langle x_{+}|\hat{\rho}_{1}|x_{-}\rangle\langle x_{-}|\hat{\rho}_{2}|x_{+}\rangle=\frac{1}{4\pi}\int_{\mathbb{R}}dx_{+}\langle x_{+}|\hat{\rho}_{1}\hat{\rho}_{2}|x_{+}\rangle=\frac{1}{4\pi}\text{tr}\{\hat{\rho}_{1}\hat{\rho}_{2}\}.
\end{equation}

In order to prove the last property, which allows us to recover the
quantum state $\hat{\rho}$ from the characteristic function through
expression (\ref{WtoRho}), we simply need two more useful properties
of the displacement operator. The first one is
\begin{equation}
\text{tr}\{\hat{D}(\mathbf{r})\}=4\pi\delta^{(2)}(\mathbf{r}),\label{DisplacementTrace}
\end{equation}
easily proven by using the form (\ref{DisplacementDisentangled})
of the displacement operator, performing the trace in the position
eigenbasis, using the translation property (\ref{TranslationPosition}),
and the fact that position eigenstates are Dirac-delta orthonormal:
\begin{equation}
\text{tr}\{\hat{D}(\mathbf{r})\}=e^{\frac{\mathrm{i}}{4}xp}\text{tr}\left\{ e^{-\frac{\mathrm{i}}{2}x\hat{P}}e^{\frac{\mathrm{i}}{2}p\hat{X}}\right\} =e^{\frac{\mathrm{i}}{4}xp}\int_{\mathbb{R}}dy\langle y|e^{-\frac{\mathrm{i}}{2}x\hat{P}}e^{\frac{\mathrm{i}}{2}p\hat{X}}|y\rangle=e^{\frac{\mathrm{i}}{4}xp}\underbrace{\int_{\mathbb{R}}dye^{\frac{\mathrm{i}}{2}py}}_{4\pi\delta(p)}\underbrace{\langle y|\underbrace{e^{-\frac{\mathrm{i}}{2}x\hat{P}}|y\rangle}_{|y+x\rangle}}_{\delta(x)}=4\pi\delta^{(2)}(\mathbf{r}).
\end{equation}
The second one refers to the composition of two displacements
\begin{equation}
\hat{D}(\mathbf{r})\hat{D}(\mathbf{s})=e^{-\frac{\mathrm{i}}{4}\mathbf{r}^{T}\Omega\mathbf{s}}\hat{D}(\mathbf{r}+\mathbf{s}),\label{DisplacementConcatenation}
\end{equation}
which is easily proven by applying the disentangling Baker-Campbell-Haussdorf
lemma (\ref{DisentanglingBCH}) with $\hat{A}=\frac{\mathrm{i}}{2}\hat{\mathbf{R}}^{T}\Omega\mathbf{r}$
and $\hat{B}=\frac{\mathrm{i}}{2}\hat{\mathbf{R}}^{T}\Omega\mathbf{r}$,
noting that
\begin{equation}
[\hat{A},\hat{B}]=\sum_{mnjl=1}^{2}\left(\frac{\mathrm{i}}{2}\right)^{2}\underbrace{[\hat{R}_{m},\hat{R}_{j}]}_{2\mathrm{i}\Omega_{mj}}\Omega_{mn}r_{n}\Omega_{jl}s_{l}=-\frac{\mathrm{i}}{2}\sum_{nl=1}^{2}r_{n}\underbrace{\left(\sum_{mj=1}^{2}\Omega_{mn}\Omega_{mj}\Omega_{jl}\right)}_{(\Omega^{T}\Omega\Omega)_{ml}=\Omega_{ml}}s_{l}=-\frac{\mathrm{i}}{2}\mathbf{r}^{T}\Omega\mathbf{s}.
\end{equation}

Using these properties, we then see that expression (\ref{WtoRho})
is correct by applying the displacement operator onto it, and taking
the trace 
\begin{equation}
\text{tr}\{\hat{D}(\mathbf{s})\hat{\rho}\}=\int_{\mathbb{R}}d^{2}\mathbf{z}\underbrace{\text{tr}\{\hat{D}(\mathbf{s})\underbrace{\hat{D}^{\dagger}(\mathbf{z})}_{\hat{D}(-\mathbf{z})}\}}_{e^{\frac{\mathrm{i}}{4}\mathbf{s}^{T}\Omega\mathbf{z}}\text{tr}\{\hat{D}(\mathbf{s}-\mathbf{z})\}}\chi_{\rho}(\mathbf{z})=\int_{\mathbb{R}}d^{2}\mathbf{z}e^{\frac{\mathrm{i}}{4}\mathbf{s}^{T}\Omega\mathbf{z}}\delta^{(2)}(\mathbf{s}-\mathbf{z})\chi_{\rho}(\mathbf{z})=\chi_{\rho}(\mathbf{s}),
\end{equation}
where we have used $\mathbf{s}^{T}\Omega\mathbf{s}=0$.

As mentioned in the previous section, this last property is connected
to the fact that the set of all displacement operators $\{\hat{D}(\mathbf{r})\}_{\mathbf{r}\in\mathbb{R}^{2}}$
forms a basis in the space of operators. More concretely, note that
the set of operators, together with the usual operations defined on
them (addition, multiplication by a complex number,...), satisfies
all the properties of a complex vector space\footnote{See Section \ref{Section:HilbertSpaces} for a reminder of linear
algebra}. We can elevate it to the category of Hilbert space by introducing
an inner product known in this context as `trace product', which takes
two operators $\hat{A}$ and $\hat{B}$ to build the complex number
$\text{tr}\{\hat{A}^{\dagger}\hat{B}\}$. To distinguish this space
from the original Hilbert space onto which the operators upon, we
will call it \emph{operator space} in the following. It is clear that
if the Hilbert space dimension is $d$, the operator space dimension
is $d^{2}$, since given a basis $\left\{ \vert e_{j}\rangle\right\} _{j=1,2,..,d}$
of the Hilbert space, the set of $d^{2}$ operators $\left\{ \hat{\Lambda}_{jl}=\vert e_{j}\rangle\langle e_{l}\vert\right\} _{j,l=1,2,..,d}$
is a basis of the operator space, since it can be used span any operator
as $\hat{A}=\sum_{j,l=1}^{d}A_{jl}\vert e_{j}\rangle\langle e_{l}\vert$,
with $A_{jl}=\text{tr}\{\hat{\Lambda}_{jl}^{\dagger}\hat{A}\}=\langle e_{j}\vert\hat{A}\vert e_{l}\rangle$.
You can also certify that the elements of such basis are orthonormal
with respect the trace product, that is, $\text{tr}\{\hat{\Lambda}_{jl}^{\dagger}\hat{\Lambda}_{mn}\}=\delta_{jm}\delta_{ln}$.

The special properties of infinite-dimensional Hilbert spaces are
of course inherited by their corresponding operator spaces. In particular,
in infinite dimension there exist operators that cannot be normalized
with respect the trace product. In the specific case that occupies
this chapter, these include, for example, the quadratures, the annihilation
operator, the number operator, and the displacement operator, since
$\text{tr}\{\hat{N}^{\dagger}\hat{N}\}=\sum_{n=0}^{\infty}n^{2}$
is not finite for the number operator, and similarly for the rest.
The operators which are normalizable in this sense are said to belong
to the \emph{trace class}. Now, even though displacement operators
are not in the trace class, properties (\ref{DisplacementTrace})
and (\ref{DisplacementConcatenation}) show that they are `Dirac-delta
orthonormalized', since, $\text{tr}\{\hat{D}^{\dagger}(\mathbf{r})\hat{D}(\mathbf{s})\}=4\pi\delta^{(2)}(\mathbf{r}-\mathbf{s})$.
Hence, they form a continuous or generalized basis in the operator
space, so any operator $\hat{A}$ belonging to the trace class can
be represented as $\hat{A}=\int_{\mathbb{R}^{2}}\frac{d^{2}\mathbf{r}}{4\pi}C(\mathbf{r})\hat{D}(\mathbf{r}),$
with expansion coefficients $C(\mathbf{r})=\text{tr}\{\hat{D}^{\dagger}(\mathbf{r})\hat{A}\}$.
Applied to the density operator representing the state of the system,
these coefficients define the characteristic function that we introduced
in the previous section (with a change of sign in the argument, $C(-\mathbf{s})=\chi_{\rho}(\mathbf{s})$,
chosen for convenience).

\subsection{Gaussian states}

Let us now introduce an important class of states, the so-called \emph{Gaussian
states}. As we will argue, these are the type of states that appear
most naturally in experiments, and indeed experimentalists need to
work very hard to go away of the Gaussian-state manifold.

\subsubsection{Definition and interpretation}

Gaussian states are defined as those states for whose Wigner function
is a Gaussian distribution, which can always be parametrized as\footnote{When dealing with Gaussian states, the following integral is quite
useful: 
\begin{equation}
\int_{\mathbb{R}^{N}}d^{N}\mathbf{r}\exp\left(-\frac{1}{2}\mathbf{r}^{T}A\mathbf{r}+\mathbf{x}^{T}\mathbf{r}\right)=\sqrt{\frac{(2\pi)^{N}}{\text{det}A}}\exp\left(\frac{1}{2}\mathbf{x}^{T}A^{-1}\mathbf{x}\right)\label{GaussianIntegral}
\end{equation}
 where $\mathbf{x}\in\mathbb{R}^{N}$ and A is a non-singular $N\times N$
matrix. For example, check with it that the Gaussian Wigner function
in (\ref{GaussianWigner}) is indeed normalized.}
\begin{equation}
W_{\rho}(\mathbf{r})=\frac{1}{2\pi\sqrt{\text{det}V}}\exp\left[-\frac{1}{2}(\mathbf{r}-\mathbf{d})^{T}V^{-1}(\mathbf{r}-\mathbf{d})\right],\label{GaussianWigner}
\end{equation}
in terms of a so-called \emph{mean vector} $\mathbf{d}$, and the
\emph{covariance matrix} $V$. The quantum statistics of such states
are completely defined by first and second order moments such as $\langle\hat{R}_{j}\rangle$
and $\langle\hat{R}_{j}\hat{R}_{l}\rangle$, respectively. In particular,
it is not difficult to prove (and we do it below, in Section \ref{ProofOfMoments})
that\begin{subequations}\label{GaussianMoments}
\begin{align}
\langle\hat{R}_{j}\rangle=\int_{\mathbb{R}^{2}}d^{2}\mathbf{r}W_{\rho}(\mathbf{r})r_{j}=d_{j} & ,\\
\frac{1}{2}\langle\delta\hat{R}_{j}\delta\hat{R}_{l}+\delta\hat{R}_{l}\delta\hat{R}_{j}\rangle=\int_{\mathbb{R}^{2}}d^{2}\mathbf{r}W_{\rho}(\mathbf{r})\delta r_{j}\delta r_{l} & =V_{jl}.
\end{align}
\end{subequations}where we have defined the fluctuation vector $\delta\mathbf{r}=\mathbf{r}-\mathbf{d}$,
which measures how far away the phase-space coordinate is from the
mean of the distribution, and we remind that we already defined in
Section \ref{QuantumMehcanicsPureStates} the fluctuation of any operator
$\hat{A}$ as $\delta\hat{A}=\hat{A}-\langle\hat{A}\rangle$, which
allows us to find its variance as $V(A)=\langle\delta\hat{A}^{2}\rangle$.

Any higher-order moment can be generated from these ones by using
the Gaussian-moment theorem \cite{QO5}
\begin{equation}
\int_{\mathbb{R}^{2}}d^{2}\mathbf{r}W_{\rho}(\mathbf{r})\delta r_{j_{1}}\delta r_{j_{2}}...\delta r_{j_{N}}=\left\{ \begin{array}{cc}
0 & \text{if }N\text{ is odd}\\
\sum_{\text{all }(N-1)!!\text{ pairings}}^{\{i_{1},i_{2},...,i_{N}\}\in}V_{i_{1}i_{2}}V_{i_{3}i_{4}}...V_{i_{N-1}i_{N}} & \text{if }N\text{ is even}
\end{array}\right.,\label{GaussianFactorization}
\end{equation}
where the sum is restricted to the $(N-1)!!$ different ways in which
we can pair the $\{j_{1},j_{2},...,j_{N}\}$ indices. An example with
$N=4$ is
\begin{equation}
\int_{\mathbb{R}^{2}}d^{2}\mathbf{r}W_{\rho}(\mathbf{r})\underbrace{\delta x\delta p\delta x\delta p}_{j_{1}j_{2}j_{3}j_{4}=1,2,1,2}=V_{12}V_{12}+V_{11}V_{22}+V_{12}V_{21}=V_{11}V_{22}+2V_{12}^{2},
\end{equation}
where in the last step we used the symmetry of the covariance matrix.
It's important to remark that this theorem implies that for any set
of variables $\{L_{1}(\mathbf{r}),L_{2}(\mathbf{r}),...,L_{M}(\mathbf{r})\}$
that are linear combinations of the phase-space fluctuations $\delta\mathbf{r}$,
the same type of expression applies:
\begin{equation}
\overline{L_{1}L_{2}...L_{M}}=\left\{ \begin{array}{cc}
0 & \text{if }M\text{ is odd}\\
\sum_{\text{all }(M-1)!!\text{ pairings}}^{\{m_{1},m_{2},...,m_{M}\}\in}\overline{L_{m_{1}}L_{m_{2}}}\;\overline{L_{m_{3}}L_{m_{4}}}...\overline{L_{m_{M-1}}L_{m_{M}}} & \text{if }M\text{ is even}
\end{array}\right.,
\end{equation}
where for any function $F(\mathbf{r})$ we have defined the phase-space
average
\begin{equation}
\overline{F}=\int_{\mathbb{R}^{2}}d^{2}\mathbf{r}W_{\rho}(\mathbf{r})F(\mathbf{r}),
\end{equation}
for brevity. Most importantly, the Gaussian-state theorem also applies
to quantum expectation values when the system is in a Gaussian state.
In particular, consider any set of operators $\{\hat{L}_{1},\hat{L}_{2},...,\hat{L}_{M}\}$,
all linear combination of quadrature fluctuations $\delta\hat{\mathbf{R}}$.
Then, we have
\begin{equation}
\langle\hat{L}_{1}\hat{L}_{2}...\hat{L}_{M}\rangle=\left\{ \begin{array}{cc}
0 & \text{for odd }M\\
\sum_{\text{all }(M-1)!!\text{ pairings}}^{\{m_{1},m_{2},...,m_{M}\}\in}\langle\hat{L}_{m_{1}}\hat{L}_{m_{2}}\rangle...\langle\hat{L}_{m_{M-1}}\hat{L}_{m_{M}}\rangle & \text{for even }M
\end{array}\right..\label{GaussianMomentTheoremQuantum}
\end{equation}
This allows evaluating any higher-order quantum moment of a Gaussian
state in a simple fashion. For example,
\begin{equation}
\langle\hat{X}^{3}\rangle=\langle(\delta\hat{X}+\langle\hat{X}\rangle)^{3}\rangle=\underbrace{\langle\delta\hat{X}^{3}\rangle}_{0}+3\langle\hat{X}\rangle\langle\delta\hat{X}^{2}\rangle+3\langle\hat{X}\rangle^{2}\underbrace{\langle\delta\hat{X}\rangle}_{0}+\langle\hat{X}\rangle^{3}=\left[\langle\hat{X}\rangle^{2}+3V(X)\right]\langle\hat{X}\rangle=\left[d_{1}^{2}+3V_{11}\right]d_{1}.
\end{equation}

\begin{figure}
\includegraphics[width=0.5\textwidth]{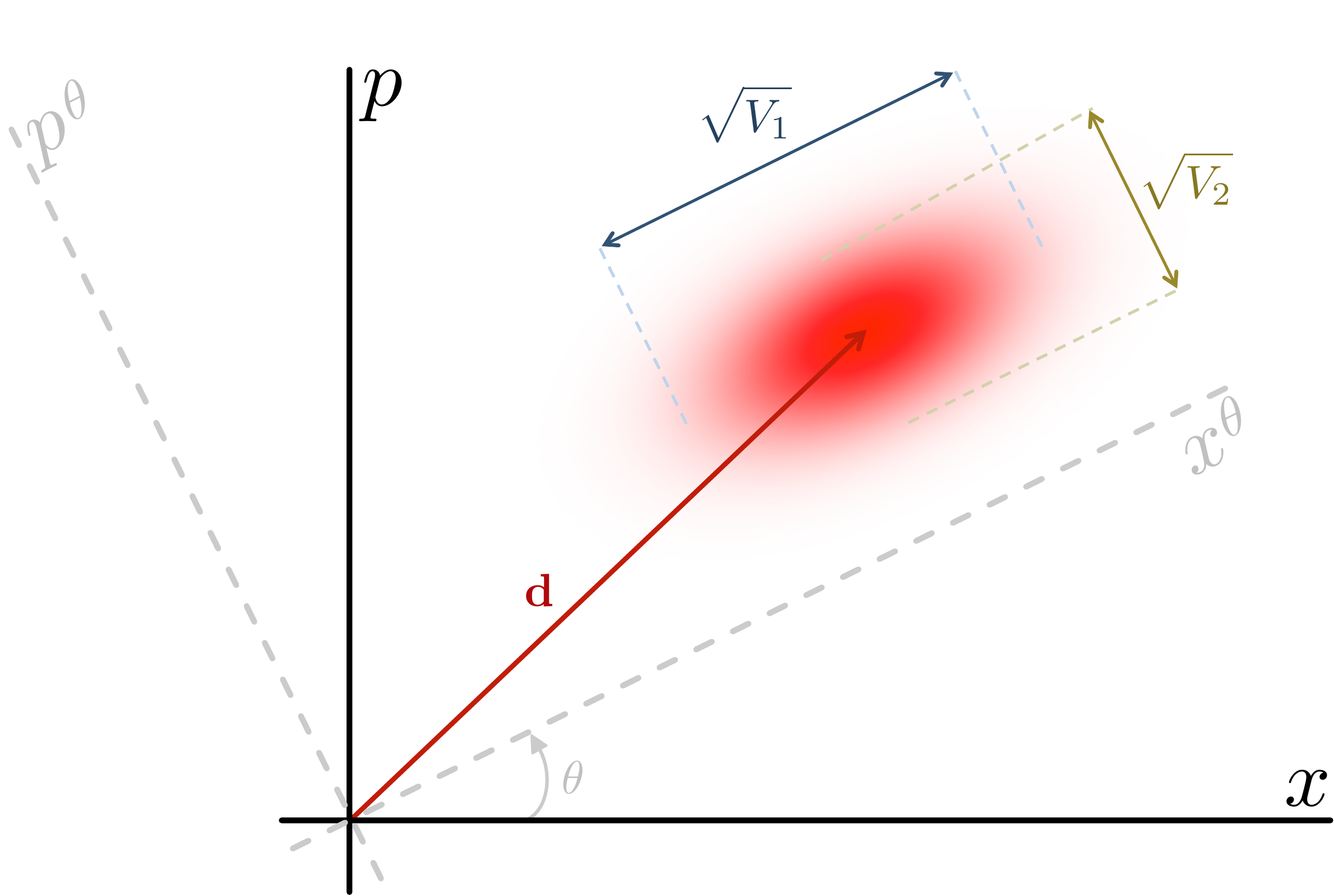}\caption{Wigner function of a generic Gaussian state (\ref{GaussianWigner}).
It consists of a single lob with the shape of an ellipse, centered
at the position given by the mean vector $\mathbf{d}$. The covariance
matrix $V$ can be diagonalized through a rotation of angle $\theta$.
Its eigenvalues provide the uncertainty of the quadratures along the
principal axes of the ellipse. \label{fig-GeneralGaussian}}
\end{figure}

Let us now discuss about the interpretation of Gaussian states. As
we mentioned above, the mean vector and covariance matrix are related
to quantum expectation values by\begin{subequations}
\begin{align}
\mathbf{d} & =\langle\hat{\mathbf{R}}\rangle=\left(\begin{array}{c}
\langle\hat{X}\rangle\\
\langle\hat{P}\rangle
\end{array}\right),\\
V & =\frac{1}{2}\left[\langle\delta\hat{\mathbf{R}}\delta\hat{\mathbf{R}}^{T}\rangle+\left(\langle\delta\hat{\mathbf{R}}\delta\hat{\mathbf{R}}^{T}\rangle\right)^{T}\right]=\left(\begin{array}{cc}
\langle\delta\hat{X}^{2}\rangle & \langle(\delta\hat{X}\delta\hat{P})^{(s)}\rangle\\
\langle(\delta\hat{X}\delta\hat{P})^{(s)}\rangle & \langle\delta\hat{P}^{2}\rangle
\end{array}\right).
\end{align}
\end{subequations}Now, since Gaussian Wigner functions are positive
everywhere, they can be interpreted as probability density functions
for quadrature measurements. The mean vector encodes the expectation
values describing the measurements, while the covariance matrix encodes
the distribution of the measurements around the mean. It is then interesting
to plot a typical Gaussian Wigner function, which is what can be seen
in Fig. \ref{fig-GeneralGaussian}. It has the shape of an ellipse
centered at the mean vector. The widths along the principal axes of
the ellipse inform us about the quadrature variances along those directions.
Being symmetric and real, the covariance matrix can always be diagonalized
via a proper phase-space rotation, say
\begin{equation}
V=R^{T}(\theta)V^{\theta}R(\theta),\quad\text{with }R(\theta)=\left(\begin{array}{cc}
\cos\theta & \sin\theta\\
-\sin\theta & \cos\theta
\end{array}\right),\text{ and }V^{\theta}=\left(\begin{array}{cc}
V_{1} & 0\\
0 & V_{2}
\end{array}\right).\label{Vdiagonalization}
\end{equation}
The eigenvalues then inform us about the variance of the quadratures
along the two principal axes of the ellipse, the angle of the rotation
providing the orientation of the ellipse. In particular, defining
the rotated coordinate system $\mathbf{r}^{\theta}=R(\theta)\mathbf{r}\equiv(x^{\theta},p^{\theta})$,
the Gaussian Wigner function (\ref{GaussianWigner}) takes the separable
form
\begin{equation}
W_{\rho}(\mathbf{r}^{\theta})=e^{-(x^{\theta}-d_{1}^{\theta})^{2}/2V_{1}}e^{-(p^{\theta}-d_{2}^{\theta})^{2}/2V_{2}},
\end{equation}
where we have defined the rotated mean vector $\mathbf{d}^{\theta}=R(\theta)\mathbf{d}$.
This expression indeed shows that $\sqrt{V_{j}}$ corresponds to the
standard deviation of the distribution along the corresponding direction
$r_{j}^{\theta}$. On the other hand, defining the rotated quadratures
$\hat{\mathbf{R}}^{\theta}=R(\theta)\hat{\mathbf{R}}\equiv(\hat{X}^{\theta},\hat{P}^{\theta})$,
such that $\mathbf{d}^{\theta}=\langle\hat{\mathbf{R}}^{\theta}\rangle$,
the eigenvalues of the covariance matrix correspond to the variances
of the quadratures: $V(X^{\theta})=\langle(\delta\hat{X}^{\theta})^{2}\rangle=V_{1}$
and $V(P^{\theta})=\langle(\delta\hat{P}^{\theta})^{2}\rangle=V_{2}$.

It is also interesting to note that, in a classical world, Gaussian
states are physical for any choice of the mean vector and covariance
matrix as long as they are real, and the covariance matrix is symmetric
and positive semidefinite (the variances along the principal axes
cannot be negative). However, quantum mechanics imposes the extra
constrain $\det\{V\}\geq1$, what comes from the uncertainty principle
between quadratures (note that the determinant is invariant under
rotations, so this proof is completely general): 
\begin{align}
\det\{V\} & =V(X^{\theta})V(P^{\theta})\geq\underset{2\mathrm{i}}{\frac{1}{4}|\underbrace{\langle[\delta\hat{X}^{\theta},\delta\hat{P}^{\theta}]\rangle}|^{2}=1}.
\end{align}
A Gaussian state corresponding to any real, symmetric, positive definite
covariance matrix satisfying this condition is physically achievable,
as we will see during the course.

Let us finally remark that, according to (\ref{NumberWigner}), number
states are not Gaussian, except for the vacuum state $|0\rangle$,
whose mean vector and covariance matrix are given by 
\[
\mathbf{d}=\left(\begin{array}{c}
0\\
0
\end{array}\right),\qquad V=\left(\begin{array}{cc}
1 & 0\\
0 & 1
\end{array}\right).
\]

\subsubsection{Generation and characterization}

So far we have introduced Gaussian states, but we still haven't provided
a criterion to characterize whether a given state $\hat{\rho}$ is
indeed Gaussian. Of course, we can always compute its Wigner function
and check if it has the Gaussian form of (\ref{GaussianWigner}).
But this is not a very practical method, and it would be nice to have
simpler criteria. In the next section, after we have introduced the
so-called coherent states, we will provide a simple criterion based
on them. Here, however, we will discuss a weaker criterion, which
is not of general practical applicability, but it's very useful in
many cases. Moreover, it is a criterion with interesting experimental
implications.

The criterion says that the state $\hat{\rho}$ is Gaussian, if and
only if it can be generated from another Gaussian state $\hat{\rho}_{0}$
by evolving it with a Hamiltonian that is, at most, quadratic in quadrature
operators. More precisely,
\begin{equation}
\hat{\rho}\text{ is Gaussian}\;\Leftrightarrow\;\hat{\rho}=\hat{U}\hat{\rho}_{0}\hat{U}^{\dagger},\quad\text{with }\hat{U}=e^{\mathrm{i}\left(\sum_{j}b_{j}\hat{R}_{j}+\sum_{jl}h_{jl}\hat{R}_{j}\hat{R}_{l}\right)}\text{ and }\hat{\rho}_{0}\text{ Gaussian.}\label{GaussianConnection}
\end{equation}
Note that the unitary $\hat{U}$ can be seen as the time-evolution
operator associated to a quadratic Hamiltonian $\hat{H}=-\hbar(\sum_{j}b_{j}\hat{R}_{j}+\sum_{jl}h_{jl}\hat{R}_{j}\hat{R}_{l})/T$
acting during a time $T$. Note that self-adjointness of the Hamiltonian
implies that $b_{j}\in\mathbb{R}$ and $h_{jl}=h_{lj}^{*}$. We sometimes
refer to $\hat{U}$ and $\hat{H}$ as \emph{Gaussian unitaries} and
\emph{Gaussian Hamiltonians}, respectively.

We will prove (\ref{GaussianConnection}) shortly, but first, let
us discuss something concerning its meaning for experiments. As we
will see during the course, the most natural states of the electromagnetic
field are Gaussian. For example, at zero temperature, a free harmonic
oscillator freezes into its ground state, which we have shown to be
Gaussian, and even at any other temperature, we will show at the end
of this chapter that the corresponding thermal state is Gaussian as
well. On the other hand, in optics it is not easy to induce Hamiltonians
with terms beyond quadratic in quadrature operators, which requires
materials that have a strong nonlinear reaction to the applied fields.
Hence, (\ref{GaussianConnection}) tells us that we have to work very
hard in the lab in order to generate non-Gaussian states of light,
coming up with clever strategies beyond what's naturally available
in optical labs.

Let us now proceed with the proof of (\ref{GaussianConnection}).
Let's start by denoting the mean vector and covariance matrix of the
Gaussian state $\hat{\rho}_{0}$ by $\mathbf{d}_{0}$ and $V_{0}$,
respectively, such that the corresponding Wigner function reads
\begin{equation}
W_{\rho_{0}}(\mathbf{r})=\frac{1}{2\pi\sqrt{\text{det}V_{0}}}e^{-\frac{1}{2}(\mathbf{r}-\mathbf{d}_{0})^{T}V_{0}^{-1}(\mathbf{r}-\mathbf{d}_{0})}.
\end{equation}
On the other hand, if the state $\hat{\rho}=\hat{U}\hat{\rho}_{0}\hat{U}^{\dagger}$
is Gaussian, then its Wigner function can also be expressed in the
Gaussian form (\ref{GaussianWigner}). Hence, both $W_{\rho_{0}}(\mathbf{r})$
and $W_{U\rho_{0}U^{\dagger}}(\mathbf{r})$ are represented by the
(normalized) exponential of some quadratic form. But we know that
two different quadratic forms can always be related by a linear transformation.
Hence, it is clear that\begin{subequations}
\begin{align}
\sqrt{\text{det}V_{0}}W_{\rho_{0}}(\mathbf{r}) & =\sqrt{\text{det}V}W_{U\rho_{0}U^{\dagger}}(S\mathbf{r}+\mathbf{a}),\\
 & \Downarrow\nonumber \\
e^{-\frac{1}{2}(\mathbf{r}-\mathbf{d}_{0})^{T}V_{0}^{-1}(\mathbf{r}-\mathbf{d}_{0})} & =e^{-\frac{1}{2}(S\mathbf{r}+\mathbf{a}-\mathbf{d})^{T}V^{-1}(S\mathbf{r}+\mathbf{a}-\mathbf{d})}=e^{-\frac{1}{2}\left[\mathbf{r}+S^{-1}(\mathbf{a}-\mathbf{d})\right]^{T}S^{T}V^{-1}S\left[\mathbf{r}+S^{-1}(\mathbf{a}-\mathbf{d})\right]},
\end{align}
\end{subequations}for some $\mathbf{a}\in\mathbb{R}^{2}$ and some
$2\times2$ matrix $S$, whose properties will naturally unveil shortly.
Comparing the leftmost and rightmost sides of the last equation, we
then conclude that
\begin{equation}
\mathbf{d}=S\mathbf{d}_{0}+\mathbf{a},\qquad V=SV_{0}S^{T},\label{GaussianUnitaryTransformation}
\end{equation}
and hence, the effect of the unitary transformation is simplified
in phase space to this simple transformation of the mean vector and
covariance matrix. Using (\ref{GaussianUnitaryTransformation}), we
show below that, seen in the Heisenberg picture, the unitary induces
a linear transformation on the quadrature operators, given by
\begin{equation}
\hat{U}^{\dagger}\hat{\mathbf{R}}\hat{U}=S\hat{\mathbf{R}}+\mathbf{a}.\label{GaussianUnitaryTransformation_Heisenberg}
\end{equation}
On the other hand, given that the commutator of the quadratures is
proportional to the identity, according to the Baker-Campbell-Haussdorf
lemma (\ref{BCHlemma-1}), the only way that the unitary $\hat{U}=\exp(\mathrm{i}\hat{B})$
can induce a linear transformation (\ref{GaussianUnitaryTransformation_Heisenberg})
is if the operator $\hat{B}$ is at most quadratic in quadrature operators.
Moreover, while the vector $\mathbf{a}$ is arbitrary, the matrix
$S$ is easily shown to be constrained by the relation
\begin{equation}
S\Omega S^{T}=\Omega,\label{SymplecticCondition}
\end{equation}
easily proven as
\begin{equation}
\hat{U}^{\dagger}\underbrace{[\hat{R}_{m},\hat{R}_{n}]}_{2\mathrm{i}\Omega_{mn}}\hat{U}=\left[\sum_{j=1}^{2}S_{mj}\hat{R}_{j}+a_{m},\sum_{l=1}^{2}S_{nl}\hat{R}_{l}+a_{n}\right]=\sum_{jl=1}^{2}S_{mj}S_{nl}\underbrace{[\hat{R}_{j},\hat{R}_{l}]}_{2\mathrm{i}\Omega_{jl}},
\end{equation}
which is equal to (\ref{SymplecticCondition}) when written in matrix
form. The set of all matrices satisfying condition (\ref{SymplecticCondition})
forms a group structure with many interesting properties that are
heavily exploited both in classical and quantum mechanics. We call
it the \emph{symplectic group}, and then say that $S$ is a \emph{symplectic
matrix}. Note that the symplectic condition implies that $\det^{2}S=1$,
and therefore, $\det V=\det V_{0}$ from (\ref{GaussianUnitaryTransformation}).

It is also very useful and interesting to note that Gaussian unitaries
induce a very simple transformation of the Wigner function even for
non-Gaussian states. In particular, the following expression is true
for any state $\hat{\rho}$ as long as $\hat{U}$ is a Gaussian unitary:
\begin{equation}
W_{U\rho U^{\dagger}}(\mathbf{r})=W_{\rho}\left(S^{-1}(\mathbf{r-\mathbf{a})}\right).\label{GaussianUnitaryGeneralTransform}
\end{equation}
This is easily proven as follows
\begin{align}
W_{U\rho U^{\dagger}}(\mathbf{r}) & =\int_{\mathbb{R}^{2}}\frac{d^{2}\mathbf{s}}{(4\pi)^{2}}e^{-\frac{\mathrm{i}}{2}\mathbf{r}^{T}\Omega\mathbf{s}}\text{tr}\{\hat{U}\hat{\rho}\hat{U}^{\dagger}\hat{D}(\mathbf{s})\}=\int_{\mathbb{R}^{2}}\frac{d^{2}\mathbf{s}}{(4\pi)^{2}}e^{-\frac{\mathrm{i}}{2}\mathbf{r}^{T}\Omega\mathbf{s}}\text{tr}\{\hat{\rho}\hat{U}^{\dagger}\hat{D}(\mathbf{s})\hat{U}\}\\
 & =\int_{\mathbb{R}^{2}}\frac{d^{2}\mathbf{s}}{(4\pi)^{2}}e^{-\frac{\mathrm{i}}{2}\mathbf{r}^{T}\Omega\mathbf{s}}\text{tr}\{\hat{\rho}e^{\frac{\mathrm{i}}{2}(\hat{U}^{\dagger}\hat{\mathbf{R}}\hat{U})^{T}\Omega\mathbf{s}}\}\overset{(\ref{GaussianUnitaryTransformation_Heisenberg})}{=}=\int_{\mathbb{R}^{2}}\frac{d^{2}\mathbf{s}}{(4\pi)^{2}}e^{-\frac{\mathrm{i}}{2}\mathbf{r}^{T}\Omega\mathbf{s}}\text{tr}\{\hat{\rho}e^{\frac{\mathrm{i}}{2}(S\hat{\mathbf{R}}+\mathbf{a})^{T}\Omega\mathbf{s}}\}\nonumber \\
 & =\int_{\mathbb{R}^{2}}\frac{d^{2}\mathbf{s}}{(4\pi)^{2}}e^{-\frac{\mathrm{i}}{2}(\mathbf{r}-\mathbf{a})^{T}\Omega\mathbf{s}}\text{tr}\{\hat{\rho}e^{\frac{\mathrm{i}}{2}\hat{\mathbf{R}}^{T}S^{T}\Omega\mathbf{s}}\}\overset{(\ref{SymplecticCondition})}{=}\int_{\mathbb{R}^{2}}\frac{d^{2}\mathbf{s}}{(4\pi)^{2}}e^{-\frac{\mathrm{i}}{2}(\mathbf{r}-\mathbf{a})^{T}\Omega\mathbf{s}}\text{tr}\{\hat{\rho}e^{\frac{\mathrm{i}}{2}\hat{\mathbf{R}}^{T}\Omega S^{-1}\mathbf{s}}\}\nonumber \\
 & =\int_{\mathbb{R}^{2}}\frac{d^{2}\mathbf{s}}{(4\pi)^{2}}e^{-\frac{\mathrm{i}}{2}(\mathbf{r}-\mathbf{a})^{T}\Omega\mathbf{s}}\chi_{\rho}(S^{-1}\mathbf{s})\overset{\mathbf{s}=S\mathbf{z}}{=}\int_{\mathbb{R}^{2}}\frac{d^{2}\mathbf{z}}{(4\pi)^{2}}e^{-\frac{\mathrm{i}}{2}(\mathbf{r}-\mathbf{a})^{T}\Omega S\mathbf{z}}\chi_{\rho}(\mathbf{z})\overset{(\ref{SymplecticCondition})}{=}\int_{\mathbb{R}^{2}}\frac{d^{2}\mathbf{z}}{(4\pi)^{2}}e^{-\frac{\mathrm{i}}{2}(\mathbf{r}-\mathbf{a})^{T}S^{-1T}\Omega\mathbf{z}}\chi_{\rho}(\mathbf{z}),\nonumber 
\end{align}
which from (\ref{GenWigner}) we identify as the expression (\ref{GaussianUnitaryGeneralTransform})
we wanted to prove.

To conclude, we still need to prove (\ref{GaussianUnitaryTransformation_Heisenberg})
from (\ref{GaussianUnitaryTransformation}). In order to do this,
consider first the expression
\begin{equation}
\text{tr}\{\hat{\rho}_{0}\hat{U}^{\dagger}\hat{\mathbf{R}}\hat{U}\}\overset{\text{(cyclic)}}{=}\text{tr}\{\underbrace{\hat{U}\hat{\rho}_{0}\hat{U}^{\dagger}}_{\hat{\rho}}\hat{\mathbf{R}}\}=\mathbf{d}\overset{(\ref{GaussianUnitaryTransformation})}{=}S\mathbf{d}_{0}+\mathbf{a}=S\text{tr}\{\hat{\rho}_{0}\hat{\mathbf{R}}\}+\mathbf{a}=\text{tr}\{\hat{\rho}_{0}(S\hat{\mathbf{R}}+\mathbf{a})\}.
\end{equation}
Comparing the left-most and right-most expressions, this relation
seems to suggest that (\ref{GaussianUnitaryTransformation_Heisenberg})
is right. However, such a strong conclusion cannot be reached from
just this relation, since many different Gaussian states have the
same mean. But since Gaussian states are completely characterized
by first and second order moments, we simply need to prove a similar
expression for the covariance matrix, which we do next. In particular,
let us first note that, using the commutation relations of the quadrature
operators, we can rewrite the elements of the covariance matrix as
$V_{mn}=\langle\delta\hat{R}_{m}\delta\hat{R}_{n}\rangle-\mathrm{i}\Omega_{mn}$,
and therefore 
\begin{align}
\text{tr} & \{\hat{\rho}_{0}(\hat{U}^{\dagger}\hat{\mathbf{R}}\hat{U}-\mathbf{d})(\hat{U}^{\dagger}\hat{\mathbf{R}}\hat{U}-\mathbf{d})^{T}\}-\mathrm{i}\Omega\overset{\text{(cyclic)}}{=}\text{tr}\{\overbrace{\hat{U}\hat{\rho}_{0}\hat{U}^{\dagger}}^{\hat{\rho}}(\hat{\mathbf{R}}-\mathbf{d})(\hat{\mathbf{R}}-\mathbf{d})^{T}\}-\mathrm{i}\Omega=V\\
\overset{(\ref{GaussianUnitaryTransformation})}{=} & SV_{0}S^{T}=\text{tr}\{\hat{\rho}_{0}S(\hat{\mathbf{R}}-\mathbf{d}_{0})(\hat{\mathbf{R}}-\mathbf{d}_{0})^{T}S^{T}\}-\mathrm{i}S\Omega S^{T}\overset{(\ref{SymplecticCondition})}{=}\text{tr}\{\hat{\rho}_{0}(S\hat{\mathbf{R}}-\underbrace{S\mathbf{d}_{0}}_{\mathbf{d}-\mathbf{a}})(S\hat{\mathbf{R}}-\underbrace{S\mathbf{d}_{0}}_{\mathbf{d}-\mathbf{a}})^{T}\}-\mathrm{i}\Omega.\nonumber 
\end{align}
Comparing again the left-most and out-most expressions, together with
the previous calculation with the mean vector, we are now in conditions
to conclude that (\ref{GaussianUnitaryTransformation_Heisenberg})
is correct.

\subsubsection{Proof of (\ref{GaussianMoments})\label{ProofOfMoments}}

In this final section, let us prove that Eqs. (\ref{GaussianMoments}),
which connect first and second order moments of the quadrature operators
to the mean vector and covariance matrix appearing in the Gaussian
Wigner function. The first equality in these equations, connecting
quantum expectation values with phase-space integrals, is just a consequence
of the general relation (\ref{SymExpectWigner}) that we proved for
arbitrary Wigner functions, not only Gaussian ones. Then, all that
is left to prove is the second equality, specifically\begin{subequations}\label{GaussianMomentsProof}
\begin{align}
\int_{\mathbb{R}^{2}}d^{2}\mathbf{r}W_{\rho}(\mathbf{r})r_{j} & =d_{j},\\
\int_{\mathbb{R}^{2}}d^{2}\mathbf{r}W_{\rho}(\mathbf{r})\delta r_{j}\delta r_{l} & =V_{jl}.
\end{align}
\end{subequations}In order to do this, it is useful to determine
first the characteristic function $\chi_{\rho}(\mathbf{s})$ associated
to the Gaussian Wigner function (\ref{GaussianWigner}). Note that
the we can easily invert the Fourier transform in (\ref{GenWigner})
to write the characteristic function in terms of the Wigner function
as\footnote{Note that the following representation of the two-dimensional Dirac
delta is easily proven
\[
\delta^{(2)}(\mathbf{z})=\int_{\mathbb{R}^{2}}\frac{d^{2}\mathbf{r}}{(4\pi)^{2}}e^{\frac{\mathrm{i}}{2}\mathbf{r}^{T}\Omega\mathbf{z}},
\]
with $\mathbf{r},\mathbf{z}\in\mathbb{R}^{2}$.}
\begin{equation}
\chi_{\rho}(\mathbf{s})=\int_{\mathbb{R}^{2}}d^{2}\mathbf{r}e^{\frac{\mathrm{i}}{2}\mathbf{r}^{T}\Omega\mathbf{s}}W_{\rho}(\mathbf{r}).
\end{equation}
 Then, the characteristic function of a general Gaussian Wigner function
is found as
\begin{equation}
\chi_{\rho}(\mathbf{s})=\frac{1}{2\pi\sqrt{\text{det}V}}\int_{\mathbb{R}^{2}}d^{2}\mathbf{r}e^{-\frac{1}{2}(\mathbf{r}-\mathbf{d})^{T}V^{-1}(\mathbf{r}-\mathbf{d})+\frac{\mathrm{i}}{2}\mathbf{r}^{T}\Omega\mathbf{s}}=\frac{e^{\frac{\mathrm{i}}{2}\mathbf{d}^{T}\Omega\mathbf{s}}}{2\pi\sqrt{\text{det}V}}\int_{\mathbb{R}^{2}}d^{2}\mathbf{z}e^{-\frac{1}{2}\mathbf{z}^{T}V^{-1}\mathbf{z}+\frac{\mathrm{i}}{2}\mathbf{z}^{T}\Omega\mathbf{s}}=e^{-\frac{1}{8}\mathbf{s}^{T}\Omega^{T}V\Omega\mathbf{s}+\frac{\mathrm{i}}{2}\mathbf{d}^{T}\Omega\mathbf{s}},
\end{equation}
where we made the variable change $\mathbf{r}=\mathbf{z}+\mathbf{d}$,
and used the Gaussian integral formula (\ref{GaussianIntegral}) in
the last step. Hence, we see that the characteristic function is also
a Gaussian function. It is convenient for the upcoming derivations
to write it in terms of a new variable $\mathbf{t}=\Omega\mathbf{s}$,
so that
\begin{equation}
\chi_{\rho}(\mathbf{t})=\int_{\mathbb{R}^{2}}d^{2}\mathbf{r}e^{\frac{\mathrm{i}}{2}\mathbf{r}^{T}\mathbf{t}}W_{\rho}(\mathbf{r})=e^{-\frac{1}{8}\mathbf{t}^{T}V\mathbf{t}+\frac{\mathrm{i}}{2}\mathbf{d}^{T}\mathbf{t}}.\label{Characteristic_t}
\end{equation}
We can prove what we seek for by taking derivatives of this expression
with respect to $\mathbf{t}$, and then particularizing to $\mathbf{t}=0$.
In particular, from the first equality we see\begin{subequations}
\begin{align}
\left.\partial_{t_{j}}\chi_{\rho}(\mathbf{t})\right|_{\mathbf{t}=0} & =\left.\frac{\mathrm{i}}{2}\int_{\mathbb{R}^{2}}d^{2}\mathbf{r}\,r_{j}e^{\frac{\mathrm{i}}{2}\mathbf{r}^{T}\mathbf{t}}W(\mathbf{r})\right|_{\mathbf{t}=0}=\frac{\mathrm{i}}{2}\int_{\mathbb{R}^{2}}d^{2}\mathbf{r}\,r_{j}W(\mathbf{r}),\\
\left.\partial_{t_{l}}\partial_{t_{j}}\chi_{\rho}(\mathbf{t})\right|_{\mathbf{t}=0} & =\left.\left(\frac{\mathrm{i}}{2}\right)^{2}\int_{\mathbb{R}^{2}}d^{2}\mathbf{r}\,r_{l}r_{j}e^{\frac{\mathrm{i}}{2}\mathbf{r}^{T}\mathbf{t}}W(\mathbf{r})\right|_{\mathbf{t}=0}=\left(\frac{\mathrm{i}}{2}\right)^{2}\int_{\mathbb{R}^{2}}d^{2}\mathbf{r}\,r_{l}r_{j}W(\mathbf{r}).
\end{align}
\end{subequations}On the other hand, taking derivatives of the right-hand-side
of (\ref{Characteristic_t}) instead, we get\begin{subequations}
\begin{align}
\left.\partial_{t_{j}}\chi_{\rho}(\mathbf{t})\right|_{\mathbf{t}=0} & =\left.\left(-\frac{1}{4}\sum_{n=1}^{2}t_{n}V_{jn}+\frac{\mathrm{i}}{2}d_{j}\right)e^{-\frac{1}{8}\mathbf{t}^{T}V\mathbf{t}+\frac{\mathrm{i}}{2}\mathbf{d}^{T}\mathbf{t}}\right|_{\mathbf{t}=0}=\frac{\mathrm{i}}{2}d_{j},\\
\left.\partial_{t_{l}}\partial_{t_{j}}\chi_{\rho}(\mathbf{t})\right|_{\mathbf{t}=0} & =\left.\left[-\frac{1}{4}V_{jl}+\left(-\frac{1}{4}\sum_{n=1}^{2}t_{n}V_{jn}+\frac{\mathrm{i}}{2}d_{j}\right)\left(-\frac{1}{4}\sum_{n=1}^{2}t_{m}V_{lm}+\frac{\mathrm{i}}{2}d_{l}\right)\right]e^{-\frac{1}{8}\mathbf{t}^{T}V\mathbf{t}+\frac{\mathrm{i}}{2}\mathbf{d}^{T}\mathbf{t}}\right|_{\mathbf{t}=0}=-\frac{1}{4}(V_{jl}+d_{j}d_{l}).
\end{align}
\end{subequations}Comparing these equations and the previous ones,
we then prove expression (\ref{GaussianMoments}).

\subsection{Coherent states}

Let us now discuss a very important class of states of the harmonic
oscillator, the so-called \emph{coherent states}.\emph{ }These are
the most simple type of Gaussian states, and yet, they were introduced
by Roy J. Glauber \cite{Glauber63a,Glauber63b,Glauber63c} in order
to explain quantum-mechanically the outstanding coherence properties
of the laser, in a series of works which marked the birth of modern
theoretical quantum optics for many people (including the Nobel committee,
who awarded him with the Nobel prize in 2005 \cite{GlauberNobelLecture}).
Moreover, as we discuss below, these states will allow us to reconcile
the quantum and classical descriptions of the harmonic oscillator.

\subsubsection{Definition and phase-space description}

In the previous section we saw that Gaussian states are mapped into
other Gaussian states through the evolution induced by Hamiltonians
which are at most quadratic in the quadrature operators. As a first
example of a relevant Gaussian state, we then study here the transformation
of the simplest Gaussian state, vacuum, by the simplest type of Hamiltonian,
one that is linear in the quadratures. In fact, the time-evolution
induced by such Hamiltonian is described by a unitary operator which
we have already introduced: the \emph{displacement operator} of Eq.
(\ref{DisplacementOp}). Let us here write it in the alternative (so-called
\emph{complex}) form
\begin{equation}
\hat{D}\left(\alpha\right)=\exp\left(\alpha\hat{a}^{\dagger}-\alpha^{\ast}\hat{a}\right),\label{DisplacementComplex}
\end{equation}
which is exactly the same as (\ref{DisplacementOp}), but written
in terms of creation and annihilation operators, and a complex parameter
$\alpha=(x+\mathrm{i}p)/2$. We define then \emph{coherent states}
as `displaced' vacuum states
\begin{equation}
|\alpha\rangle=\hat{D}\left(\alpha\right)|0\rangle.
\end{equation}
Note that when a displacement is applied to the annihilation and creation
operators, we get\footnote{This is trivially proved by using again the Baker\textendash Campbell\textendash Haussdorf
lemma (\ref{BCHlemma-1}).}
\begin{equation}
\hat{D}^{\dagger}\left(\alpha\right)\hat{a}\hat{D}\left(\alpha\right)=\hat{a}+\alpha\text{ \ \ \ and \ \ }\hat{D}^{\dagger}\left(\alpha\right)\hat{a}^{\dagger}\hat{D}\left(\alpha\right)=\hat{a}^{\dagger}+\alpha^{\ast}\text{,}\label{Displacement_Boson}
\end{equation}
which shows where the name `displacement' comes from. Applying the
first of these equations to the vacuum state, and noting that the
displacement operator is unitary, we obtain $\hat{a}\hat{D}\left(\alpha\right)|0\rangle=\alpha\hat{D}\left(\alpha\right)|0\rangle$,
and hence coherent states are (right) eigenvectors of the annihilation
operator, which has therefore complex eigenvalues, that is, $\hat{a}|\alpha\rangle=\alpha|\alpha\rangle$
with $\alpha\in\mathbb{C}$ (remember that $\hat{a}$ is not Hermitian).
Note that this immediately implies that coherent states are left eigenstates
of the creation operator, that is, $\langle\alpha|\hat{a}^{\dagger}=\alpha^{*}\langle\alpha|$.

\begin{figure}
\includegraphics[width=1\textwidth]{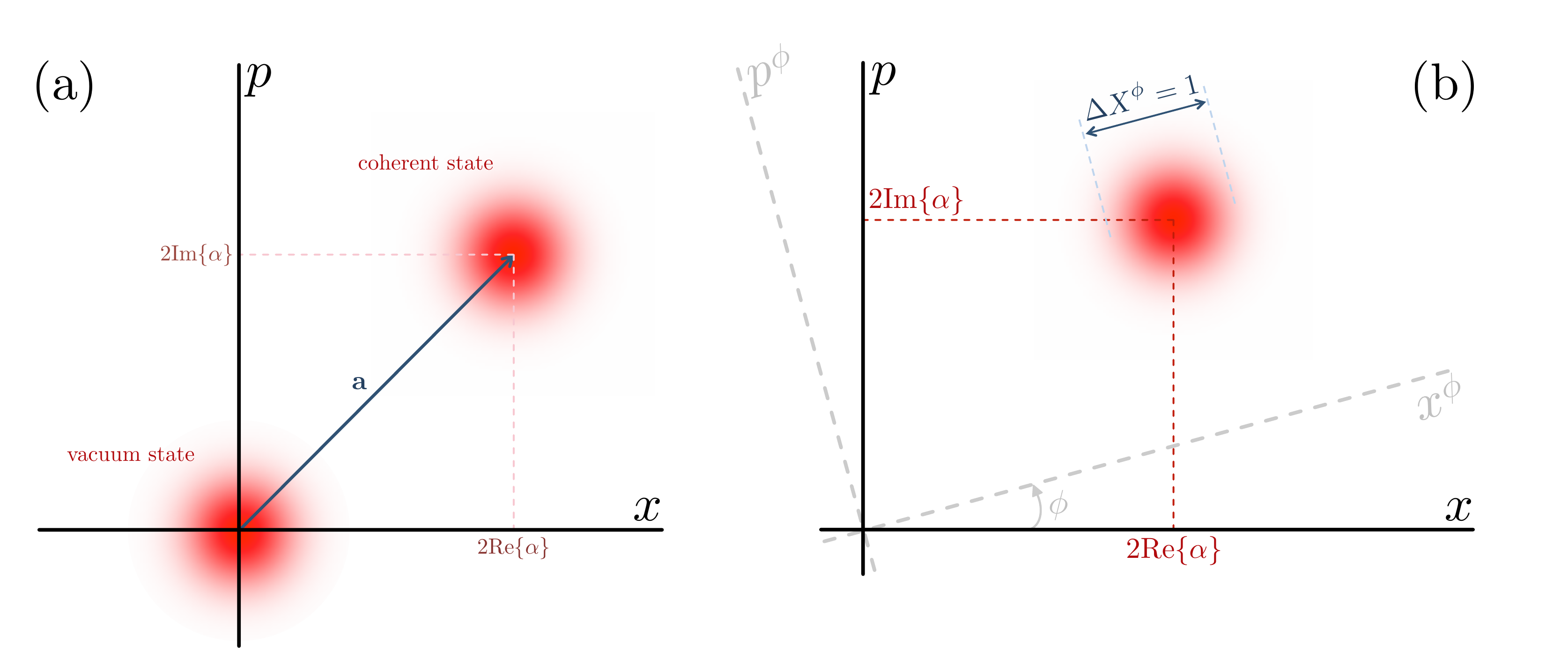}\caption{(a) A coherent state of arbitrary amplitude can be generated by applying
a displacement to the vacuum state. (b) Wigner function of a coherent
state. The function is rotationally symmetric around its center, reflecting
the fact that quadratures are affected by vacuum fluctuations in all
directions.\label{fOsci2}}
\end{figure}

In terms of the quadratures, the displacement transformation reads
\begin{equation}
\hat{D}^{\dagger}\left(\alpha\right)\hat{X}\hat{D}\left(\alpha\right)=\hat{X}+2\text{Re}\{\alpha\}\text{ \ \ \ and \ \ }\hat{D}^{\dagger}\left(\alpha\right)\hat{P}\hat{D}\left(\alpha\right)=\hat{P}+2\text{Im}\{\alpha\}\text{.}
\end{equation}
This transformation changes the expectation value of the quadratures
but not their variance. More specifically, in the notation of the
previous section, see Eq. (\ref{GaussianUnitaryTransformation_Heisenberg}),
the linear transformation has $\mathbf{a}(\alpha)=2(\text{Re}\{\alpha\},\text{Im}\{\alpha\})^{T}$
and $S=\tiny{\left(\begin{array}{cc}
1 & 0\\
0 & 1
\end{array}\right)}$, so that for a Gaussian state, it shifts (`displaces') its mean vector,
but has no effect on the covariance matrix. Hence, the Wigner function
of a coherent state has the same form as that of vacuum, but centered
in a different point of phase space,
\[
W_{|\alpha\rangle}(\mathbf{r})=W_{|0\rangle}[\mathbf{r}-\mathbf{a}(\alpha)],\quad\text{with }\mathbf{a}(\alpha)=2\left(\begin{array}{c}
\text{Re}[\alpha]\\
\text{Im}[\alpha]
\end{array}\right).
\]
We sketch this in Fig. \ref{fOsci2}a. In other words, $W_{|\alpha\rangle}(\mathbf{r})$
is a Gaussian function of the type (\ref{GaussianWigner}) with mean
vector $\mathbf{d}_{|\alpha\rangle}=2(\text{Re}\{\alpha\},\text{Im}\{\alpha\})^{T}$
and the covariance matrix of vacuum $V_{|\alpha\rangle}=\tiny{\left(\begin{array}{cc}
1 & 0\\
0 & 1
\end{array}\right)}$.

Let us now move to the interpretation of this specific type of Gaussian
Wigner function. As we already did in the previous section, let us
first define a \textit{general quadrature} defined along an arbitrary
direction $\phi$ of phase space as
\begin{equation}
\hat{X}^{\phi}=\hat{X}\cos\phi+\hat{P}\sin\phi=e^{-\mathrm{i}\phi}\hat{a}+e^{\mathrm{i}\phi}\hat{a}^{\dagger}\text{.}\label{GenQuad}
\end{equation}
We will denote the quadrature defined along its orthogonal direction
by $\hat{P}^{\phi}=\hat{X}^{\phi+\pi/2}$. These orthogonal quadratures
define a new phase-space coordinate system $(x^{\phi},p^{\phi})$
rotated by an angle $\phi$ respect to the original $(x,p)$ system
(see Fig. \ref{fOsci2}b). They also satisfy the commutation relation
$[\hat{X}^{\phi},\hat{P}^{\phi}]=2\mathrm{i}$, and hence must satisfy
the uncertainty relation $\Delta X^{\phi}\Delta P^{\phi}\geq1$. The
statistics of a measurement of one of such general quadratures $\hat{X}^{\phi}$
will be described by a Gaussian distribution with mean $\langle\hat{X}^{\phi}\rangle=2|\alpha|\cos(\phi-\varphi)$,
where $\varphi$ is the phase of $\alpha$, and uncertainty $\Delta X^{\phi}=1$,
irrespective of $\phi$ (note that $V$ is invariant under rotations
for a coherent state). Hence, the statistics of a measurement of the
quadratures will be spread around their mean equally in any direction
of phase space, as can be appreciated from the fact that the Wigner
function is invariant under rotations around its mean (it has a circular
shape).

\begin{figure}
\includegraphics[width=0.6\textwidth]{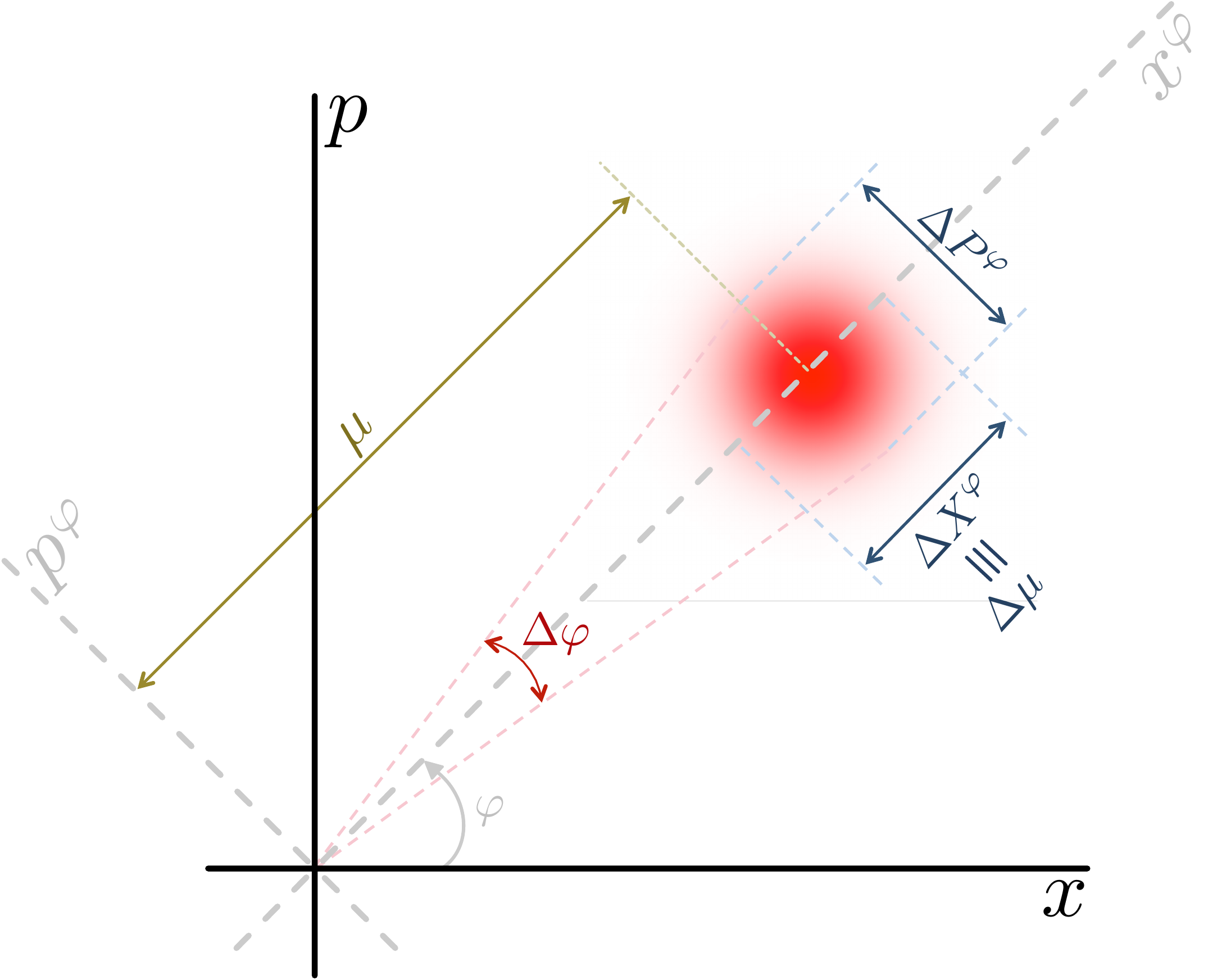}\caption{Quantum amplitude\textendash phase properties of a coherent state.
$\Delta\mu$ and $\Delta\varphi$ represent the uncertainties in the
amplitude and phase, $\mu$ and $\varphi$, of the oscillator, which
are clearly related to the uncertainties of the amplitude and phase
quadratures, $\hat{X}^{\varphi}$ and $\hat{P}^{\varphi}$, respectively.\label{fOsci3}}
\end{figure}

The phase-space representation of coherent states allows us to understand
its amplitude and phase properties (Fig. \ref{fOsci3}). This is not
trivial for general states, since these are two observables which
still lack a satisfactory description in terms of self-adjoint operators
\cite{QO10,QO5}; but for coherent states of sufficiently large amplitude,
we now show that it is possible to find a proper description. For
general states, the expectation values of the quadratures $\langle\hat{\mathbf{R}}\rangle$
define a phase $\varphi=$ $\arctan\langle\hat{P}\rangle/\langle\hat{X}\rangle$
and an amplitude $\mu=\sqrt{\langle\hat{P}\rangle^{2}+\langle\hat{X}\rangle^{2}}$,
which in the case of a coherent state $|\alpha\rangle$, are directly
related to the coherent-state amplitude by $\alpha=\mu e^{\mathrm{i}\varphi}/\sqrt{2}$.
In the classical limit, neglecting quantum noise, these are exactly
the phase and amplitude that would be measured for the oscillator.
When quantum uncertainties in the quadratures cannot be neglected,
it is reasonable to think that $\varphi$ and $\mu$ will still be
the mean values measured for the phase and amplitude of the oscillator,
but now they will also be affected by some uncertainties, say $\Delta\varphi$
and $\Delta\mu$. It seems obvious from Fig. \ref{fOsci3} that these
amplitude and phase uncertainties are related to the uncertainties
of the quadratures $\hat{X}^{\varphi}$ and $\hat{P}^{\varphi}$ in
the direction of the state's phase $\varphi$. In particular, we have
$\Delta\mu=\Delta X^{\varphi}$ and $\Delta\varphi=\Delta P^{\varphi}/\mu$.
Consequently, we call \textit{amplitude} and \textit{phase }quadratures,
to the quadratures in this direction of phase space. Hence, even if
at the quantum level we still don't know how assign assign self-adjoint
operators to the (absolute) phase and amplitude observables \cite{QO10,QO5},
we can somehow relate their properties to those of the amplitude and
phase quadratures, at least for Gaussian states with a well defined
amplitude, that is, whenever $\mu>\Delta X^{\varphi}$.

\subsubsection{Properties and general Gaussianity criterion}

In the previous subsection we have introduced coherent states as displaced
vacuum states, discussing their Wigner function, as well as showing
that they are eigenstates of the annihilation operators. Here we provide
several interesting properties of these states, and introduce a general
criterion to determine whether a given state $\hat{\rho}$ is gaussian.

It is simple to find an explicit representation of the coherent states
in terms of the basis of number states. For that, we simply write
the displacement operator as\footnote{This is easy to prove by using the disentangling Baker-Campbell-Haussdorf
lemma (\ref{DisentanglingBCH}) with $\hat{A}=\alpha\hat{a}^{\dagger}$
and $\hat{B}=-\alpha^{*}\hat{a}$.}
\begin{equation}
\hat{D}\left(\alpha\right)=\exp\left(-|\alpha|^{2}/2\right)\exp\left(\alpha\hat{a}^{\dagger}\right)\exp\left(-\alpha^{\ast}\hat{a}\right),\label{DisplacementNormalForm}
\end{equation}
and Taylor-expand the exponentials to obtain
\begin{equation}
|\alpha\rangle=e^{-|\alpha|^{2}/2}\sum_{n=0}^{\infty}\frac{\alpha^{n}}{n!}\hat{a}^{\dagger n}|0\rangle=\sum_{n=0}^{\infty}e^{-|\alpha|^{2}/2}\frac{\alpha^{n}}{\sqrt{n!}}|n\rangle,\label{NumToCoh}
\end{equation}
where we have used $\hat{a}^{\dagger n}|0\rangle=\sqrt{n!}|n\rangle$
by virtue of (\ref{LadderHO}). Note that for $\alpha=0$, coherent
states are nothing but the vacuum state. For any other value of $\alpha$
coherent states do not have a well defined number of excitations;
instead, the number of \textit{quanta} is distributed according to
a Poisson probability distribution
\begin{equation}
P_{n}\left(\alpha\right)=|\langle n|\alpha\rangle|^{2}=e^{-|\alpha|^{2}}\frac{|\alpha|^{2n}}{n!},\label{eq:Poisson}
\end{equation}
with mean photon number $\langle\hat{N}\rangle=|\alpha|^{2}$ and
uncertainty $\Delta N=|\alpha|$. The shape of this distribution is
heavily dependent on the value of $|\alpha|^{2}$, as shown in Fig.
\textbf{ToDo}. When $|\alpha|^{2}\gg1$, the distribution looks quite
symmetric and normal, just peaked at the mean $|\alpha|^{2}$ with
a relatively thin width $|\alpha|$. In this limit, one can even make
a Gaussian approximation for the distribution in many situations,
without loosing to much accuracy. However, when $|\alpha|^{2}$ is
not so large, the distribution is highly asymmetric, and it's even
peaked at 0 when $|\alpha|^{2}\leq1$.

It is easy to show that coherent states form a resolution of the identity.
For this, we consider the corresponding sum of projectors and use
(\ref{NumToCoh}), obtaining
\begin{equation}
\int_{\mathbb{C}}d^{2}\alpha|\alpha\rangle\langle\alpha|\underset{\tiny{\alpha=re^{\mathrm{i}\phi}}}{\underbrace{=}}\sum_{m,n=0}^{\infty}\int_{0}^{\infty}drre^{-r^{2}}\frac{r^{m+n}}{\sqrt{m!n!}}\underset{2\pi\delta_{mn}}{\underbrace{\int_{0}^{2\pi}d\phi e^{\mathrm{i}\phi(m-n)}}}|m\rangle\langle n|=\pi\sum_{n=0}^{\infty}\frac{1}{n!}\int_{0}^{\infty}dr2r^{2n+1}e^{-r^{2}}|n\rangle\langle n|.
\end{equation}
The integral is easily carried out\footnote{For example, using the variable change $z=x^{2}$, we can write
\begin{align}
\int_{0}^{\infty}dr2r^{2n+1}e^{-r^{2}} & =\int_{0}^{\infty}dzz^{n}e^{-z}=\lim_{\kappa\rightarrow1}\int_{0}^{\infty}dz(-1)^{n}\partial_{\kappa}^{n}e^{-\kappa z}=(-1)^{n}\lim_{\kappa\rightarrow1}\partial_{\kappa}^{n}\int_{0}^{\infty}dze^{\kappa z}\\
 & =(-1)^{n}\lim_{\kappa\rightarrow1}\partial_{\kappa}^{n}\frac{1}{-\kappa}(\underset{0}{\underbrace{e^{-\kappa\infty}}}-1)=(-1)^{n}\lim_{\kappa\rightarrow1}\partial_{\kappa}^{n}\frac{1}{\kappa}=(-1)^{n}n!\lim_{\kappa\rightarrow1}(-1)^{n}\frac{1}{\kappa^{n+1}}=n!\nonumber 
\end{align}
} as $\int_{0}^{\infty}dr2r^{2n+1}e^{-r^{2}}=n!$, obtaining then the
desired completeness relation
\begin{equation}
\int_{\mathbb{C}}\frac{d^{2}\alpha}{\pi}|\alpha\rangle\langle\alpha|=\sum_{n=0}^{\infty}|n\rangle\langle n|=\hat{I}.
\end{equation}

However, coherent states are not orthogonal, and hence, they do not
form a standard basis of the Hilbert space. Instead, they form an
overcomplete basis, which can still be useful for many calculations,
as we shall see later. Let us here prove that coherent states are
not orthogonal. To this aim, note that (\ref{DisplacementConcatenation})
allows us to write the concatenation of two displacements as
\begin{equation}
\hat{D}(\alpha_{1})\hat{D}(\alpha_{2})=e^{\mathrm{i}\text{Im}\{\alpha_{1}\alpha_{2}^{*}\}}\hat{D}(\alpha_{1}+\alpha_{2}),
\end{equation}
and also that $\hat{D}^{\dagger}(\alpha)=\hat{D}(-\alpha)$. Hence,
the overlap between two coherent states can be written as\footnote{An alternative proof of this expression can be obtained by using the
Fock-state representation (\ref{NumToCoh}) as follows:
\begin{equation}
\langle\alpha|\beta\rangle=\sum_{n,m=0}e^{-(|\alpha|^{2}-|\beta|^{2})/2}\frac{\alpha^{*n}\beta^{m}}{\sqrt{n!m!}}\underset{\delta_{nm}}{\underbrace{\langle n|m\rangle}}=e^{-(|\alpha|^{2}-|\beta|^{2})/2}\underset{e^{\alpha^{*}\beta}}{\underbrace{\sum_{n=0}\frac{(\alpha^{*}\beta)^{n}}{n!}}}=e^{-\mathrm{i}\text{Im}\{\alpha\beta^{*}\}}e^{-|\alpha-\beta|^{2}/2}.
\end{equation}
}
\begin{equation}
\langle\alpha|\beta\rangle=\langle0|\hat{D}(-\alpha)\hat{D}(\beta)|0\rangle=e^{-\mathrm{i}\text{Im}\{\alpha\beta^{*}\}}\langle0|\beta-\alpha\rangle=e^{-\mathrm{i}\text{Im}\{\alpha\beta^{*}\}}e^{-|\alpha-\beta|^{2}/2},
\end{equation}
where we have used the Fock-state representation (\ref{NumToCoh})
of coherent states.

Finally, coherent states allow us to find a criterion for a given
state $\hat{\rho}$ to be Gaussian. To this aim, just note that, using
the completeness of the coherent states and the displacement decomposition\footnote{Again easy to prove by using the (\ref{DisentanglingBCH}) with $\hat{A}=-\alpha^{*}\hat{a}$
and $\hat{B}=\alpha\hat{a}^{\dagger}$.}
\begin{equation}
\hat{D}(\alpha)=\exp\left(|\alpha|^{2}/2\right)\exp\left(-\alpha^{\ast}\hat{a}\right)\exp\left(\alpha\hat{a}^{\dagger}\right),
\end{equation}
the quantum characteristic function can be written as
\begin{align}
\text{tr}\{\hat{\rho}\hat{D}(x,p)\}\underset{\beta=(x+\mathrm{i}p)/2}{\underbrace{=}} & e^{|\beta|^{2}/2}\text{tr}\left\{ e^{\beta\hat{a}^{\dagger}}\hat{\rho}e^{-\beta^{\ast}\hat{a}}\right\} =e^{|\beta|^{2}/2}\text{tr}\left\{ \int_{\mathbb{C}}\frac{d^{2}\nu}{\pi}|\nu\rangle\langle\nu|e^{\beta\hat{a}^{\dagger}}\hat{\rho}e^{-\beta^{\ast}\hat{a}}\right\} \label{CharacteristicGaussianCond}\\
= & \int_{\mathbb{C}}\frac{d^{2}\nu}{\pi}e^{|\beta|^{2}/2}\langle\nu|e^{\beta\hat{a}^{\dagger}}\hat{\rho}e^{-\beta^{\ast}\hat{a}}|\nu\rangle=\int_{\mathbb{C}}\frac{d^{2}\nu}{\pi}e^{\beta\nu^{*}-\beta^{*}\nu+|\beta|^{2}/2}\langle\nu|\hat{\rho}|\nu\rangle.\nonumber 
\end{align}
The Wigner function, being a Fourier transform of the characteristic
function, will be Gaussian whenever the latter can be written as an
exponential of a quadratic form of the phase-space variables. On other
hand, it is clear from (\ref{CharacteristicGaussianCond}) that this
will happen whenever $\langle\nu|\hat{\rho}|\nu\rangle$ also admits
such a form in terms of $\nu$. This gives us an easy way of checking
whether the state is Gaussian or not: we just need to evaluate its
overlap with a coherent state. More precisely:
\[
\hat{\rho}\text{ is Gaussian}\Leftrightarrow\langle\nu|\hat{\rho}|\nu\rangle=e^{f(\nu)},\text{ with }f(\nu)\text{ at most quadratic in }\ensuremath{\nu}\in\mathbb{C},\text{and }|\nu\rangle\text{ a coherent state}.
\]

Once we are sure that the state is Gaussian, it is typically useful
to evaluate the expectation values $\langle\hat{a}\rangle$, $\langle\delta\hat{a}^{2}\rangle=\langle\hat{a}^{2}\rangle-\langle\hat{a}\rangle^{2}$,
and $\langle\delta\hat{a}^{\dagger}\delta\hat{a}\rangle=\langle\hat{a}^{\dagger}\hat{a}\rangle-|\langle\hat{a}\rangle|^{2}$,
from which we can write down the mean vector and covariance matrix
as\begin{subequations}\label{GaussianMomentsComplex}
\begin{align}
\mathbf{d} & =2(\text{Re}\{\langle\hat{a}\rangle\},\text{Im}\{\langle\hat{a}\rangle\})^{T},\\
V & =(1+2\langle\delta\hat{a}^{\dagger}\delta\hat{a}\rangle)\left(\begin{array}{cc}
1 & 0\\
0 & 1
\end{array}\right)+2\left(\begin{array}{cc}
\text{Re}\{\langle\delta\hat{a}^{2}\rangle\} & \text{Im}\{\langle\delta\hat{a}^{2}\rangle\}\\
\text{Im}\{\langle\delta\hat{a}^{2}\rangle\} & -\text{Re}\{\langle\delta\hat{a}^{2}\rangle\}
\end{array}\right).
\end{align}
\end{subequations}

Of course, coherent states themselves satisfy the criterion established
above. In particular, for $\hat{\rho}=|\alpha\rangle\langle\alpha|$
we can write $\langle\nu|\hat{\rho}|\nu\rangle=|\langle\nu|\alpha\rangle|^{2}=\exp(|\nu-\alpha|^{2})$,
which has an exponent quadratic in $\nu$. Moreover, since coherent
states are eigenstates of the annihilation operator, in this case
the required expectation values are incredibly simple to evaluate,
obtaining $\langle\alpha|\hat{a}|\alpha\rangle=\alpha$ and $\langle\alpha|\delta\hat{a}^{2}|\alpha\rangle=\langle\alpha|\delta\hat{a}^{\dagger}\delta\hat{a}|\alpha\rangle=0$,
and leading to the mean vector and covariance matrix discussed in
the previous section.

\subsubsection{Bridge between quantum and classical physics}

We have seen that the classical and quantum formalisms seem dramatically
different: While classically the oscillator can have any positive
value of the energy and has a definite trajectory in phase space,
quantum mechanics allows only for discrete values of the energy and
introduces position and momentum uncertainties which prevent the existence
of well defined trajectories. In this section we show that coherent
states are the easiest way of reconciling these two seemingly contradictory
descriptions.

Quantum mechanics is all about predicting the statistics of experiments.
Hence, the way of connecting it to classical mechanics is by finding
the quantum state that predicts that the statistics obtained in the
experiment will coincide with what is classically expected. In the
following we show that coherent states with $|\alpha|\gg1$ satisfy
the desired conditions.

Let us start by showing how the classical well-defined trajectories
can be recovered. For this, we start from a coherent state $|\psi(0)\rangle=|\alpha\rangle$
at time zero, and let it evolve by the Hamiltonian of the harmonic
oscillator (\ref{HosciAnnihilationCreation}), that is, by the corresponding
time-evolution operator. We obtain
\begin{equation}
|\psi(t)\rangle=e^{-\mathrm{i}\omega t\hat{N}}|\alpha\rangle=\sum_{n=0}^{\infty}e^{-|\alpha|^{2}/2}\frac{\alpha^{n}}{\sqrt{n!}}e^{-\mathrm{i}n\omega t}|n\rangle=\left|e^{-\mathrm{i}\omega t}\alpha\right\rangle ,
\end{equation}
that is, another coherent state, with a time-dependent amplitude $\alpha(t)=e^{-\mathrm{i}\omega t}\alpha$,
which is equal to the expectation value of the annihilation operator,
$\langle\hat{a}(t)\rangle=\alpha(t)$. Reminding that this is precisely
the solution for the harmonic oscillator in the complex representation,
this means that, on average, the oscillator indeed describes the classical
trajectory. Moreover, since the state is coherent at all times, we
know that the uncertainty of any quadrature $\hat{X}^{\phi}$ is independent
of is amplitude, specifically $\Delta X^{\phi}(t)=1$. Hence, the
uncertainty becomes less and less relevant the bigger is the radius
of the trajectory, that is, the \emph{signal-to-noise ratio} satisfies
$|\alpha|/\Delta X^{\phi}\stackrel{|\alpha|\rightarrow\infty}{\longrightarrow}\infty$.
In other words, \emph{for all practical purposes}, in this limit the
oscillator seems to describe the classical trajectory when measured
in the laboratory, where other sources of classical noise and imprecision
will dominate over the tiny quantum noise.

Let us now move on to the energy-quantization issue. The ratio between
the energies of two consecutive number states is $E_{n+1}/E_{n}=\left(n+3/2\right)/(n+1/2)$.
Hence, as the number of excitations increases, the discrete character
of the energy becomes barely perceptive, that is, $E_{n+1}/E_{n}\sim1$
if $n\gg1$. Thus, \emph{for all practical purposes}, we can recover
a seemingly continuous energy spectrum if the state of the oscillator
projects only on high-order (approximately contiguous) number states.
As we saw in (\ref{eq:Poisson}), this is precisely the case of coherent
states with $|\alpha|\gg1$, which lead to a narrowly-peaked Poisson
number distribution centered at $\langle\hat{N}\rangle=|\alpha|^{2}$
with width $\Delta N=|\alpha|$.

\begin{figure}
\includegraphics[width=1\textwidth]{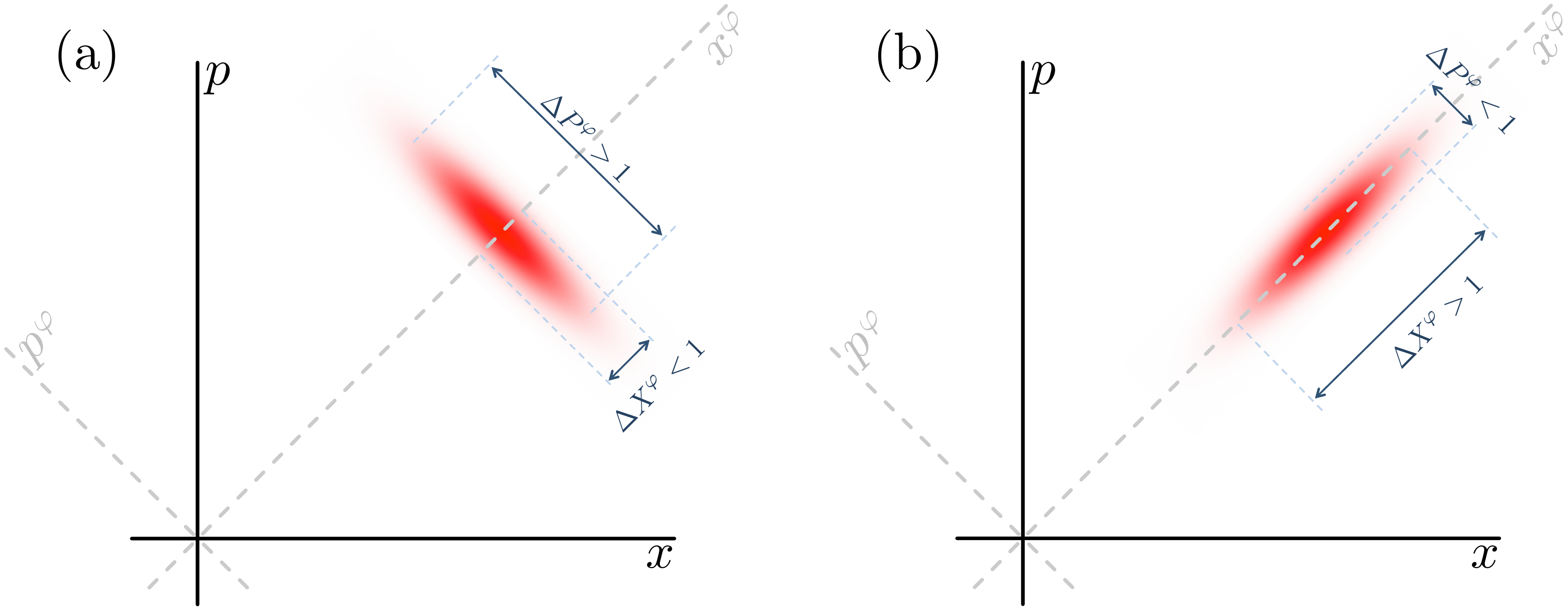}\caption{Wigner functions of two squeezed states. In (a) the uncertainty of
the amplitude quadrature is reduced below the vacuum level at the
expense of increasing the uncertainty of the phase quadrature. In
(b), on the other hand, it is the phase quadrature the one which is
squeezed, while the amplitude quadrature becomes more noisy.\label{fOsci4}}
\end{figure}

\subsection{Squeezed states\label{Sec:SqueezedStates}}

\subsubsection{Definition and relevance}

Two important applications of harmonic oscillators are \textit{sensing}
and \emph{information encoding/trasmission}: The oscillator is put
in contact with the system or signal that we want to study or transmit,
and it gets codified as phase or amplitude modulations in the oscillator
(think, for example, about gravitational waves encoded as amplitude
or phase modulations in the light circulating at the LIGO interferometer).
We have seen in the previous section that when the oscillator is in
a coherent state both its amplitude and phase suffer from uncertainties,
and hence the encoded signal cannot be perfectly retrieved from measurements
on the oscillator. When any other source of technical noise is removed,
that is, when the measurement equipment behaves basically as ideal,
this so-called \textit{quantum shot noise} becomes the main limitation.
Moreover, when the signal generated by the system that we want to
study is tiny (as gravitational waves are), it can even be hidden
below the quantum noise, so that it might not be perceived at all.

Squeezed states are the solution to this problem. Suppose that the
signal is encoded in the amplitude of the oscillator. In a coherent
state the amplitude and phase quadratures are affected by equal uncertainties
$\Delta X^{\varphi}=\Delta P^{\varphi}=1$. However, we can conceive
a state of the oscillator in which the uncertainty of the amplitude
quadrature is reduced, while that of the phase quadrature is increased,
say $\Delta X^{\varphi}\ll1$ and $\Delta P^{\varphi}\gg1$, so that
the product of uncertainties keeps lower bounded by one, $\Delta X^{\varphi}\Delta P^{\varphi}\geq1$.
The Wigner function of one of such states is depicted in Fig. \ref{fOsci4}a.
In this case the amplitude quadrature is well defined, and one can
in principle monitor its modulations with arbitrary accuracy. Of course,
this is accomplished at the expense of not being able to retrieve
any information from the phase quadrature, but that is not a problem
if we only care about the signal encoded in the amplitude quadrature.

We will then define \textit{squeezed states} as those in which some
quadrature, say $\hat{X}^{\phi}$, has an uncertainty below the vacuum
or coherent level, that is, $\Delta X^{\phi}<1$. We then say that
quadrature $\hat{X}^{\phi}$ is \textit{squeezed}\textit{\emph{, while
its conjugate $\hat{P}^{\phi}$}} is \emph{antisqueezed} ($\Delta P^{\phi}>1$).
Together with the \textit{amplitude squeezed state} already introduced
(for which $\phi=\varphi$), we show a \textit{phase squeezed state
}(for which $\phi=\varphi+\pi/2$) in Fig. \ref{fOsci4}b. Let us
now study one specially relevant type of squeezed states.

\subsubsection{Minimum uncertainty squeezed states}

\textit{Minimum uncertainty states} are states which satisfy the lower
bound of the quadrature uncertainty relation, that is, $\Delta X^{\phi}\Delta P^{\phi}=1$
$\forall\phi$. The simplest of these states is the vacuum state $|0\rangle$;
any other number state $|n\neq0\rangle$ is not contained in this
class, as it is easily checked that it satisfies $\Delta X^{\phi}=2n+1$
for any $\phi$. Coherent states, on the other hand, are minimum uncertainty
states, as they are obtained from vacuum by the displacement transformation,
which does not change the quadrature variances.

It is possible to generate squeezed states of this kind by using the
\textit{squeezing operator}
\begin{equation}
\hat{S}\left(z\right)=\exp\left(\frac{z^{\ast}}{2}\hat{a}^{2}-\frac{z}{2}\hat{a}^{\dagger2}\right)\text{,}\label{Ssq}
\end{equation}
where $z\in\mathbb{C}$ is called the \textit{squeezing parameter}.
Note that this can be seen as generalization of the displacement operator
(\ref{DisplacementComplex}), but with $\hat{a}^{2}$ instead of $\hat{a}$
in the exponent. Note that since the exponent is still quadratic in
the quadratures, the squeezing operator is a Gaussian unitary.

Let us analyze how this transformation acts on the annihilation operators.
We will use again the Baker-Campbell-Haussdorf lemma (\ref{BCHlemma-1}),
which defining $\hat{B}=\frac{z}{2}\hat{a}^{\dagger2}-\frac{z^{\ast}}{2}\hat{a}^{2}$,
requires the commutators
\begin{align}
[\hat{B},\hat{a}] & =-z\hat{a}^{\dagger},\\{}
[\hat{B},[\hat{B},\hat{a}]] & =|z|^{2}\hat{a},\nonumber \\{}
[\hat{B},[\hat{B},[\hat{B},\hat{a}]]] & =-z|z|^{2}\hat{a}^{\dagger},\nonumber \\{}
[\hat{B},[\hat{B},[\hat{B},[\hat{B},\hat{a}]]]] & =|z|^{4}\hat{a},\nonumber \\
 & \vdots\nonumber 
\end{align}
so that writing $z$ in the polar form $z=r\exp\left(\mathrm{i}\theta\right)$,
with $r\in[0,\infty]$ and $\theta\in[0,2\pi[$, and splitting the
sum of the lemma into even and odd terms, we obtain
\begin{align}
\hat{S}^{\dagger}\left(z\right)\hat{a}\hat{S}\left(z\right) & =\sum_{n=0}^{\infty}\frac{1}{(2n)!}\underset{2n}{\underbrace{[\hat{B},[\hat{B},...[\hat{B},}}\hat{a}\underset{2n}{\underbrace{]...]]}}+\frac{1}{(2n+1)!}\underset{2n+1}{\underbrace{[\hat{B},[\hat{B},...[\hat{B},}}\hat{a}\underset{2n+1}{\underbrace{]...]]}}\label{SqueezingTransformationBoson}\\
 & =\left[\sum_{n=0}^{\infty}\frac{r^{2n}}{(2n)!}\right]\hat{a}-e^{\mathrm{i}\theta}\left[\sum_{n=0}^{\infty}\frac{r^{2n+1}}{(2n+1)!}\right]\hat{a}^{\dagger}\nonumber \\
 & =\hat{a}\cosh r-e^{\mathrm{i}\theta}\hat{a}^{\dagger}\sinh r.\nonumber 
\end{align}
In terms of quadratures, this expression is easily rewritten as
\begin{equation}
\hat{S}^{\dagger}\left(z\right)\hat{X}^{\theta/2}\hat{S}\left(z\right)=e^{-r}\hat{X}^{\theta/2}\text{ \ \ \ \ and \ \ \ }\hat{S}^{\dagger}\left(z\right)\hat{P}^{\theta/2}\hat{S}\left(z\right)=e^{r}\hat{P}^{\theta/2},\label{SqueezingTransformation}
\end{equation}
which shows that the squeezing transformation corresponds to the contraction
and dilation of two orthogonal directions of phase space. In the notation
of Eq. (\ref{GaussianUnitaryTransformation_Heisenberg}), the linear
transformation has $\mathbf{a}=(0,0)^{T}$ and $S(z)=R^{T}(\theta/2)Q(r)R(\theta/2)$,
where we have defined the matrix $Q(r)=\tiny{\left(\begin{array}{cc}
e^{-r} & 0\\
0 & e^{r}
\end{array}\right)}$, and $R(\theta/2)$ is just a rotation matrix as introduced in Eq.
(\ref{Vdiagonalization}).

\begin{figure}
\includegraphics[width=1\textwidth]{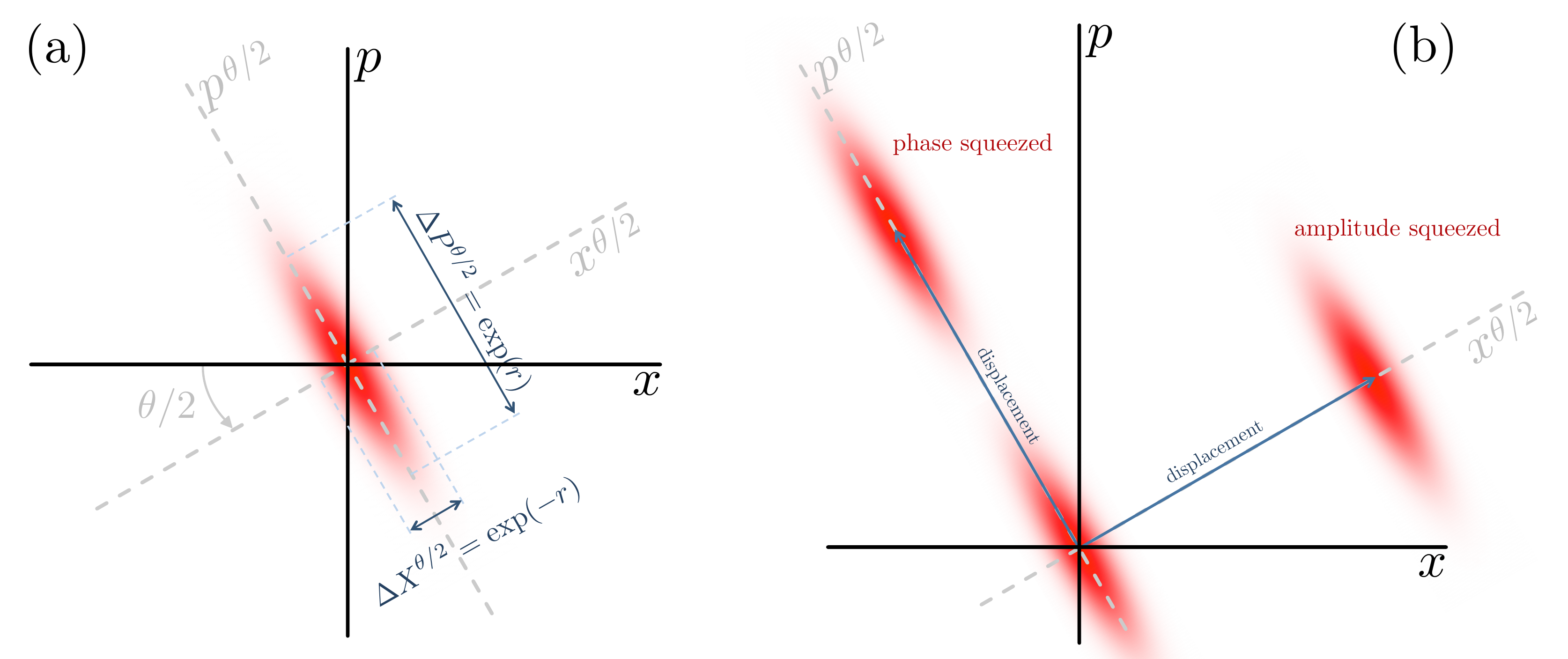}\caption{(a) Wigner function of a squeezed vacuum state. (b) Applying displacements
in the direction of the squeezed or antisqueezed quadratures, one
obtains amplitude or phase squeezed states, respectively.\label{fOsci5}}
\end{figure}

Suppose now that prior to the transformation the state of the system
was vacuum, defined by the statistical properties $\langle\hat{X}^{\phi}\rangle=0$
and $\Delta X^{\phi}=1$ $\forall\phi$ as already seen. After the
transformation (\ref{SqueezingTransformation}), the mean of any quadrature
is still zero, but the uncertainty of quadrature $\hat{X}^{\theta/2}$
has decreased to $\Delta X^{\theta/2}=\exp\left(-r\right)$, while
that of quadrature $\hat{P}^{\theta/2}$ has increased to $\Delta P^{\theta/2}=\exp\left(r\right)$.
Hence, the squeezing operator creates a \textit{\emph{minimum uncertainty
squeezed state}}, since $\Delta X^{\theta/2}\Delta P^{\theta/2}=1$.
This state $|z\rangle=\hat{S}(z)|0\rangle$ is usually called the
\emph{squeezed vacuum state}. It is a Gaussian state centered at the
origin of phase space, $\mathbf{d}_{|z\rangle}=(0,0)^{T}$, and with
a covariance matrix
\begin{equation}
V_{|z\rangle}(z)=S(z)V_{|0\rangle}S^{T}(z)=R^{T}(\theta/2)Q(2r)R(\theta/2),
\end{equation}
where we have used (\ref{GaussianUnitaryTransformation}) and $V_{|0\rangle}=\tiny{\left(\begin{array}{cc}
1 & 0\\
0 & 1
\end{array}\right)}$. We show the corresponding Wigner function in Fig. \ref{fOsci5}a.
The uncertainty circle associated to the vacuum state has turned into
an ellipse, with the principal axis oriented along axes rotated by
$\theta/2$ in phase space, as shown explicitly by the form we used
for the covariance matrix above. An amplitude squeezed state can be
then created by applying a subsequent displacement along the $\theta/2$
axis as shown in Fig. \ref{fOsci5}b. If the displacement is applied
along the $(\theta+\pi)/2$ direction, then a phase squeezed state
is obtained. As displacements do not change the uncertainty properties
of the state, these amplitude or phase squeezed states are still minimum
uncertainty states.

It is instructive to show that the squeezed vacuum state is Gaussian
by using the criterion that we developed above, based on $\langle\nu|\hat{\rho}|\nu\rangle$,
with $|\nu\rangle$ a coherent state. In order to evaluate this, we
use the following expansion for the squeezing operator, which is of
general practical use:\footnote{The proof of this identity is not entirely trivial, and can be checked
in \cite{PuriBook01}. Regarding this expression, the most important
idea to keep is the following. Whenever one has a set of operators
$\{\hat{Q}_{j}\}_{j=1,2,...,J}$ which forms a closed algebra (that
is, the commutator of any two operators of the set is proportional
to another operator of the set), then, it is possible to \emph{disentangle}
any exponential as
\begin{equation}
e^{\sum_{j=1}^{J}c_{j}\hat{Q}_{j}}=\prod_{j=1}^{J}e^{f_{j}(\mathbf{c})\hat{Q}_{j}},
\end{equation}
where $\mathbf{c}=(c_{1},c_{2},...,c_{J})$ collects a set of complex
coefficients, and the disentangling coefficients $f_{j}(\mathbf{c})$
can be found with the procedure explained in \cite{PuriBook01}. }
\begin{equation}
\hat{S}(z)=e^{-\frac{1}{2}e^{\mathrm{i}\theta}\tanh(r)\hat{a}^{\dagger2}}(\cosh r)^{-\hat{a}^{\dagger}\hat{a}-1/2}e^{\frac{1}{2}e^{-\mathrm{i}\theta}\tanh(r)\hat{a}^{2}},\label{SqueezingOpExpansion}
\end{equation}
leading to
\begin{equation}
\langle\nu|\hat{\rho}|\nu\rangle=\left|\langle\nu|\hat{S}(z)|0\rangle\right|^{2}=\frac{1}{\cosh r}\left|e^{-\frac{1}{2}\tanh(r)e^{\mathrm{i}\theta}\nu^{*2}}\langle\nu|0\rangle\right|^{2}=\frac{1}{\cosh r}e^{-\frac{1}{2}\tanh(r)\left(e^{-\mathrm{i}\theta}\nu^{2}+e^{\mathrm{i}\theta}\nu^{*2}\right)-|\nu|^{2}},
\end{equation}
which is indeed the exponential of a quadratic form. On the other
hand, using (\ref{SqueezingTransformationBoson}) it is simple find
\begin{subequations}
\begin{align}
\langle\hat{a}\rangle & =\langle0|\hat{S}^{\dagger}(z)\hat{a}\hat{S}(z)|0\rangle=0,\\
\langle\hat{a}^{2}\rangle & =\langle0|(\hat{a}\cosh r-e^{\mathrm{i}\theta}\hat{a}^{\dagger}\sinh r)^{2}|0\rangle=-e^{\mathrm{i}\theta}\cosh r\sinh r,\\
\langle\hat{a}^{\dagger}\hat{a}\rangle & =\langle0|(\hat{a}^{\dagger}\cosh r-e^{-\mathrm{i}\theta}\hat{a}\sinh r)(\hat{a}\cosh r-e^{\mathrm{i}\theta}\hat{a}^{\dagger}\sinh r)|0\rangle=\sinh^{2}r,
\end{align}
 \end{subequations}which using (\ref{GaussianMomentsComplex}) lead
to the mean vector and covariance matrix that we already introduced
above.

It is finally interesting to write the squeezed vacuum state in the
number state basis. Using (\ref{SqueezingOpExpansion}) and expanding
the exponentials in Taylor series, it is immediate to find 
\begin{equation}
\hat{S}(z)|0\rangle=\sum_{n=0}^{\infty}\frac{(-1)^{n}}{2^{n}n!}\sqrt{\frac{(2n)!}{\cosh r}}e^{\mathrm{i}n\theta}\tanh^{n}\hspace{-0.6mm}r\hspace{0.8mm}|2n\rangle.\label{SqueezedState}
\end{equation}
The most characteristic feature of this state is that it contains
only even Fock states. This comes physically from the fact that the
squeezing operator can generate excitations only by pairs, unlike
the displacement operator. The mean of the number operator has been
provided above, $\langle\hat{N}\rangle=\sinh^{2}r$, while we can
evaluate its variance from the Gaussian moment theorem (\ref{GaussianFactorization}),
which holds as well for annihilation and creation operators (since
they are just linear in the quadratures), so that, in this case
\begin{equation}
\langle\hat{N}^{2}\rangle=\langle\hat{a}^{\dagger}\hat{a}\hat{a}^{\dagger}\hat{a}\rangle=\langle\hat{a}^{\dagger}\hat{a}\rangle\langle\hat{a}^{\dagger}\hat{a}\rangle+\langle\hat{a}^{\dagger2}\rangle\langle\hat{a}^{2}\rangle+\langle\hat{a}^{\dagger}\hat{a}\rangle\underbrace{\langle\hat{a}\hat{a}^{\dagger}\rangle}_{\langle\hat{a}^{\dagger}\hat{a}\rangle+1}=2\langle\hat{N}\rangle^{2}+\langle\hat{N}\rangle+|\langle\hat{a}^{2}\rangle|^{2},
\end{equation}
leading to a variance
\begin{equation}
V(N)=\langle\hat{N}^{2}\rangle-\langle\hat{N}\rangle^{2}=\sinh^{4}r+\sinh^{2}r+\frac{1}{4}\sinh^{2}2r.
\end{equation}
Remarkably, the uncertainty of the number of quanta is always of the
same order of their mean, specifically,
\begin{equation}
\frac{\langle\hat{N}\rangle}{\Delta N}=\frac{\sinh^{2}r}{\sqrt{\sinh^{4}r+\sinh^{2}r+\frac{1}{4}\sinh^{2}2r}}\overset{r\rightarrow\infty}{\longrightarrow}\frac{1}{\sqrt{2}}.
\end{equation}
Hence, unlike for coherent states, the width of the distribution of
quanta is large, no matter the number of quanta.

\subsection{Thermal states\label{Sec:ThermalStates}}

So far we have only talked about pure states. However, in general
the state of the harmonic oscillator can be mixed. As highlighted
in the introduction to quantum mechanics, this will happen whenever
the oscillator gets correlated with another system (either classical
or quantum) to which we don't have access. This means that information
about the oscillator leaks out to this inaccessible system, and we
are then forced to use a mixed state for the oscillator that reflects
our ignorance.

Given a system whose associated Hilbert space has dimension $d$,
we have already seen that its maximally-mixed state is $\hat{\rho}_{\mathrm{MM}}=\hat{I}/d$.
As it is proportional to the identity, this state is invariant under
changes of basis, and hence, the eigenvalues of any observable of
the system are equally likely. This agrees with what one expects intuitively
from a state which has leaked the maximum amount of information to
another system: the joint uncertainty of all observables is maximized.
In particular, the von Neumann entropy, which is a measure of the
mixedness of the state, reaches its maximum $S[\hat{\rho}_{\text{MM}}]=\log d$.

For infinite-dimensional Hilbert spaces ($d\rightarrow\infty$) this
state is not physical since it has infinite energy, e.g., $\mathrm{tr}\{\hat{\rho}\hat{N}\}=\lim_{d\rightarrow\infty}\sum_{n=0}^{d}n/d\rightarrow\infty$
for the harmonic oscillator. Hence, finding the maximally-mixed state
in infinite dimension makes sense only if one adds an energy constraint
such as $\mathrm{tr}\{\hat{\rho}\hat{H}\}=\bar{E}$, where $\bar{E}$
is a positive real. Demanding in addition that the state is stationary,
that is, it does not evolve with time, it is easy to prove, and we
do so below, that the state that maximizes the von Neumann entropy
subject to this constraint is
\begin{equation}
\hat{\rho}_{\mathrm{th}}(\bar{E})=\frac{\exp(-\beta\hat{H})}{\text{tr}\{\exp(-\beta\hat{H})\}},\label{ThStateGen}
\end{equation}
where $\beta$ can be found from
\begin{equation}
\bar{E}=\frac{\text{tr}\{\hat{H}\exp(-\beta\hat{H})\}}{\text{tr}\{\exp(-\beta\hat{H})\}}=-\partial_{\beta}\ln\left[\text{tr}\left\{ e^{-\beta\hat{H}}\right\} \right]=-\partial_{\beta}\ln\left[\sum_{n}e^{-\beta E_{n}}\right],\label{ConstrainEqua}
\end{equation}
where in the last equality we have performed the trace in the basis
of eigenstates of the Hamiltonian (assumed countable for simplicity),
denoting by $E_{n}$ the corresponding eigenenergies. The expression
between square brackets is known in statistical physics as the \emph{partition
function}, $Z=\sum_{n}e^{-\beta E_{n}}$, and it's \emph{the} central
object in that field. it is remarkable that it appears naturally in
this quantum context when maximizing the entropy subject to an energy
constraint. But the connection to statistical physics doesn't end
there. The probability of obtaining eigenvalue $E_{n}$ in an energy
measurement decreases exponentially with the energy, that is,
\begin{equation}
p(E_{n})=\langle E_{n}|\hat{\rho}_{\text{th}}|E_{n}\rangle=e^{-\beta E_{n}}/Z,
\end{equation}
where $|E_{n}\rangle$ are is the corresponding eigenvector of the
Hamiltonian. This is precisely the Boltzman distribution for the energies,
as predicted in statistical physics for a system at thermal equilibrium
with temperature $T=1/k_{B}\beta$\emph{, }where $k_{B}$ is the Boltzmann
constant. Hence, the state $\hat{\rho}_{\text{th}}$ is state is known
as \emph{thermal }state, and can be parametrized either by a mean
energy $\bar{E}$ or, equivalently, by a temperature $T$ (or its
inverse $\beta$), related by the constraint equation (\ref{ConstrainEqua}).
It is interesting to remark that this result is valid for any dimension
$d$, and we recover the unconstrained maximally-mixed state $\hat{I}/d$
for infinite temperature ($\beta\rightarrow0$).

Let us differ the proof of $\hat{\rho}_{\text{th}}(\bar{E})$ as the
energy-constrained maximally-mixed state to the end of the section,
and focus now on what this means for the harmonic oscillator, whose
Hamiltonian is $\hat{H}=\hbar\omega(\hat{N}+1/2)$. First, taking
into account that $\{E_{n}=\hbar\omega(n+1/2)\}_{n=0,1,...}$, the
sum in (\ref{ConstrainEqua}), that is, the partition function, is
easily carried out as\footnote{Note that this is a simple geometric sum $S=\sum_{m=0}^{M}c^{n}=1+c+...+c^{M}$
with $c\in[0,1[$. The sum can be performed with a very simple trick.
We multiply by it $c$, obtaining $cS=c+c^{2}+...+c^{M+1}$. Hence,
we get $S-cS=1-c^{M+1}$, leading to $S=(1-c^{M+1})/(1-c)\underset{M\rightarrow\infty}{\rightarrow}1/(1-c)$. }
\begin{equation}
Z=\text{tr}\left\{ e^{-\beta\hat{H}}\right\} =e^{-\beta\hbar\omega/2}\sum_{n=0}^{\infty}\left(e^{-\beta\hbar\omega}\right)^{n}=\frac{e^{-\beta\hbar\omega/2}}{1-e^{-\beta\hbar\omega}}.\label{PartitionFunctionHO}
\end{equation}
Hence, equation (\ref{ConstrainEqua}) reads
\begin{equation}
\bar{E}=\partial_{\beta}\left[\frac{\beta\hbar\omega}{2}+\ln\left(1-e^{-\beta\hbar\omega}\right)\right]=\frac{\hbar\omega}{2}+\frac{\hbar\omega}{e^{\beta\hbar\omega}-1}.
\end{equation}
Writing the mean energy in terms of the mean number of excitations
$\bar{n}=\text{tr}\{\hat{\rho}\hat{N}\}$ as $\bar{E}=\hbar\omega(\bar{n}+1/2)$,
we can rewrite the previous expression as 
\begin{equation}
\bar{n}=\frac{\bar{E}}{\hbar\omega}-\frac{1}{2}=\frac{1}{e^{\beta\hbar\omega}-1},
\end{equation}
which is the Bose-Einstein distribution for an equilibrium temperature
$T=1/k_{B}\beta$.

In addition, it is straightforward to show that the thermal state
is written in the Fock basis as 
\begin{equation}
\hat{\rho}_{\mathrm{th}}(\bar{n})=\sum_{n=0}^{\infty}\underbrace{\frac{\bar{n}^{n}}{\left(1+\bar{n}\right)^{1+n}}}_{p_{n}(\bar{n})}|n\rangle\hspace{-0.4mm}\langle n|,\label{ThermalStateFock}
\end{equation}
where we parametrize the state in terms of the mean number of quanta
$\bar{n}$. As in finite dimension without energy constraints, this
state is diagonal, but the distribution is not flat, although it looks
flatter the larger is $\bar{n}$, as expected, that is $p_{n}(\bar{n})\approx p_{n+1}(\bar{n})$
for $\bar{n}\gg n$.

\begin{figure}
\includegraphics[width=1\textwidth]{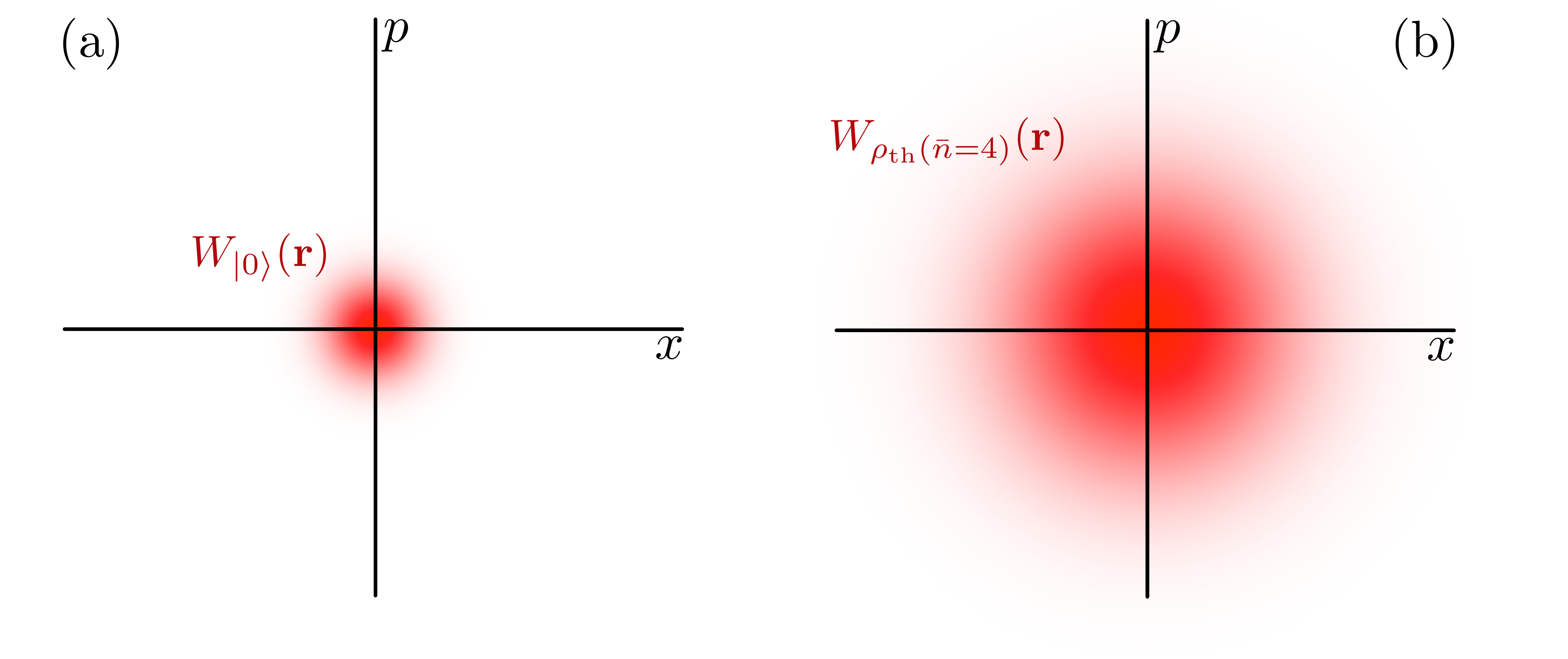}\caption{(b) Wigner function of a thermal state with mean excitation number
$\bar{n}=4$. For comparison, we show in (a) the Wigner function of
the vacuum state. It can be appreciated that both are rotationally
symmetric, but the thermal state has $\sqrt{2\bar{n}+1}$ times more
uncertainty than vacuum.\label{fOsci6}}
\end{figure}

It is easy to see that the thermal state is Gaussian using the criterion
we developed above for the expectation value of the state in a coherent
state, $\langle\nu|\hat{\rho}|\nu\rangle$. In this case, we do so
by introducing a resolution of the identity $\sum_{n=0}^{\infty}|n\rangle\langle n|=1$
into this expression, and using the Fock-state representation of coherent
states (\ref{NumToCoh}). In particular, we get
\begin{equation}
\langle\nu|e^{-\beta\hbar\omega\hat{a}^{\dagger}\hat{a}}|\nu\rangle=\sum_{n=0}^{\infty}e^{-\beta\hbar\omega n}\langle\nu|n\rangle\langle n|\nu\rangle=\sum_{n=0}^{\infty}e^{-\beta\hbar\omega n}|\langle\nu|n\rangle|^{2}=\sum_{n=0}^{\infty}e^{-\beta\hbar\omega n}\frac{|\nu|^{2n}}{n!}e^{-|\nu|^{2}}=e^{(e^{-\beta\hbar\omega}-1)|\nu|^{2}},
\end{equation}
which is the exponential of a quadratic form in $\nu$. As for the
relevant expectation values, we obtain\begin{subequations}
\begin{align}
\langle\hat{a}\rangle & =\text{tr}\{\hat{a}\hat{\rho}_{\mathrm{th}}(\bar{n})\}=\sum_{n=0}^{\infty}\frac{\bar{n}^{n}}{\left(1+\bar{n}\right)^{1+n}}\langle n|\hat{a}|n\rangle=0,\\
\langle\hat{a}^{2}\rangle & =\text{tr}\{\hat{a}^{2}\hat{\rho}_{\mathrm{th}}(\bar{n})\}=\sum_{n=0}^{\infty}\frac{\bar{n}^{n}}{\left(1+\bar{n}\right)^{1+n}}\langle n|\hat{a}^{2}|n\rangle=0,\\
\langle\hat{a}^{\dagger}\hat{a}\rangle & =\bar{n}\text{ (this is by construction, since it is the energy constrain)}.
\end{align}
\end{subequations}Hence, using (\ref{GaussianMomentsComplex}), we
see that the Wigner function of a thermal state is given by a Gaussian
with mean vector $\mathbf{d}_{\text{th}}=(0,0)^{T}$ and a covariance
matrix 
\begin{equation}
V_{\text{th}}=(2\bar{n}+1)\left(\begin{array}{cc}
1 & 0\\
0 & 1
\end{array}\right).
\end{equation}
We show it in Fig. \ref{fOsci6}, where it can be appreciated that
thermal states are similar to a vacuum state (rotationally symmetric),
but with $\sqrt{2\bar{n}+1}$ times more noise. Consequently, the
vacuum state can be seen as a thermal state with zero mean photon
number.

Let us conclude this section by proving that the thermal state is
the one maximizing the von Neumann entropy of any system when subject
to the energy constrain $\mathrm{tr}\{\hat{\rho}\hat{H}\}=\bar{E}$.
We do so by Lagrangian optimization, defining the Lagrange function
\begin{equation}
\mathcal{L}[\hat{\rho}]=-\text{tr}\left\{ \hat{\rho}\ln\hat{\rho}\right\} +\lambda(1-\text{tr}\{\hat{\rho}\})+\beta(\bar{E}-\text{tr}\{\hat{H}\hat{\rho}\}),
\end{equation}
which includes the energy constraint and the normalization constraint
$\text{tr}\{\hat{\rho}\}=1$ of the state through Lagrange multipliers
$\beta$ and $\lambda$, respectively. We use a principle of least
action by setting to zero the variation of the Lagrangian function,
\begin{equation}
\delta\mathcal{L}[\hat{\rho}]=\mathcal{L}[\hat{\rho}+\delta\hat{\rho}]-\mathcal{L}[\hat{\rho}]=-\text{tr}\{\delta\hat{\rho}(\ln\hat{\rho}+1+\lambda+\beta\hat{H})\}=0,
\end{equation}
where we have kept only terms linear in the variation\footnote{Note that we can assume that $[\delta\hat{\rho},\hat{\rho}]=0$, in
other words, we can restrict variations that only perturb the eigenvalues
of the density operator, but not its eigenstates. The reason for this
is that, since we are seeking for a time-independent state, the von-Neumann
equation implies $[\hat{H},\hat{\rho}]=0$, and therefore, we know
from the start that the eigenvectors of $\hat{\rho}$ are just the
energy eigenstates.} $\delta\hat{\rho}$ and used\footnote{This expression is found as follows. Define $\hat{V}=\ln\hat{\rho}$.
Then, taking the variation of this expression, we get $\delta\hat{V}=\ln(\hat{\rho}+\delta\hat{\rho})-\ln\hat{\rho}=\ln(1+\delta\hat{\rho}\hat{\rho}^{-1})$,
where we have used the fact that $\delta\hat{\rho}$ and $\hat{\rho}$
commute. Now, taking the exponential of this expression and expanding
it to first order in $\delta\hat{V}$, we obtain $1+\delta\hat{\rho}\hat{\rho}^{-1}=e^{\delta\hat{V}}\approx1+\delta\hat{V}$,
which provides the expression we were looking for.} $\delta(\ln\hat{\rho})=\delta\hat{\rho}\hat{\rho}^{-1}$. In order
to satisfy this condition for arbitrary variations, it must then be
satisfied 
\begin{equation}
\hat{\rho}=e^{-(\lambda+1)}e^{-\beta\hat{H}}.
\end{equation}
The Lagrange multipliers are found from the constraints. First, the
normalization constraint gives us
\begin{equation}
\text{tr}\{\hat{\rho}\}=e^{-(\lambda+1)}\text{tr}\left\{ e^{-\beta\hat{H}}\right\} =1\Rightarrow e^{-(\lambda+1)}=1/\text{tr}\left\{ e^{-\beta\hat{H}}\right\} ,
\end{equation}
leading to the form (\ref{ThStateGen}) that we wrote for the state.
Then, the energy constraint gives us Eq. (\ref{ConstrainEqua}).

\subsection{Final remarks about the quantized expression for the electromagnetic
field}

Let us conclude this chapter by coming back to the electromagnetic
field. It will be convenient to write the quantum fields in terms
of annihilation and creation operators, rather than position and momenta
as we did in (\ref{FieldExpansionsQuantum}). Then for each of the
oscillators, we define annihilation and creation operators through
the relations
\begin{equation}
\hat{a}_{n}=\sqrt{\frac{m_{n}\omega_{n}}{2\hbar}}\left(\hat{q}_{n}+\frac{\mathrm{i}}{m_{n}\omega_{n}}\hat{p}_{n}\right)\hspace{1em}\Longleftrightarrow\hspace{1em}\begin{cases}
\hat{q}_{n}=\sqrt{\frac{\hbar}{2m_{n}\omega_{n}}}(\hat{a}_{n}+\hat{a}_{n}^{\dagger})\\
\hat{p}_{n}=-\mathrm{i}\sqrt{\frac{\hbar m_{n}\omega_{n}}{2}}(\hat{a}_{n}-\hat{a}_{n}^{\dagger})
\end{cases}
\end{equation}
satisfying canonical commutation relations
\begin{equation}
[\hat{a}_{n},\hat{a}_{l}^{\dagger}]=\delta_{nl},
\end{equation}
leading to following forms of the quantum fields (\ref{FieldExpansionsQuantum}):\begin{subequations}
\begin{align}
\hat{\mathbf{A}}(z,t) & =\mathbf{e}_{x}\sum_{n=1}^{\infty}\sqrt{\frac{\hbar}{\varepsilon_{0}LS\omega_{n}}}[\hat{a}_{n}(t)+\hat{a}_{n}^{\dagger}(t)]\sin\left(k_{n}z\right),\label{Acavity-1}\\
\hat{\mathbf{E}}(z,t) & =\mathrm{i}\mathbf{e}_{x}\sum_{n=1}^{\infty}\sqrt{\frac{\hbar\omega_{n}}{\varepsilon_{0}LS}}[\hat{a}_{n}(t)-\hat{a}_{n}^{\dagger}(t)]\sin\left(k_{n}z\right),\label{Ecavity}\\
\hat{\mathbf{B}}(z,t) & =\mathbf{e}_{y}\sum_{n=1}^{\infty}\sqrt{\frac{\mu_{0}\hbar\omega_{n}}{LS}}[\hat{a}_{n}(t)+\hat{a}_{n}^{\dagger}(t)]\cos\left(k_{n}z\right).
\end{align}
\end{subequations}Note how the arbitrary masses $m_{n}$ have vanished
from the fields when written in terms of annihilation and creation
operators, which are the operators with real physical significance,
unlike the more arbitrary positions and momenta. Let us remark that
in in contrast with (\ref{FieldExpansionsQuantum}), here we have
implicitly adopted the Heisenberg picture, since we allow operators
to evolve in time.

Along the notes, we will use a notation that has become quite standard
in quantum optics: we split the fields in their annihilation and creation
parts. For example, for the vector potential we write $\hat{\mathbf{A}}(z,t)=\hat{\mathbf{A}}^{(+)}(z,t)+\hat{\mathbf{A}}^{(-)}(z,t)$,
with
\begin{equation}
\hat{\mathbf{A}}^{(+)}(z,t)=\mathbf{e}_{x}\sum_{n=1}^{\infty}\sqrt{\frac{\hbar}{\varepsilon_{0}LS\omega_{n}}}\hat{a}_{n}(t)\sin\left(k_{n}z\right)=\left[\hat{\mathbf{A}}^{(-)}(z,t)\right]^{\dagger},
\end{equation}
and we do the same for any other field. This will make some of our
derivations more economic in the next chapters.

In terms of annihilation and creation operators, the Hamiltonian for
the light field reads then
\begin{equation}
\hat{H}_{\text{L}}=\sum_{n=1}^{\infty}\hbar\omega_{n}\hat{a}_{n}^{\dagger}\hat{a}_{n},\label{Hl}
\end{equation}
where we have removed the ground-state energy $\hbar\omega_{n}/2$
of each mode because it will not play any role in future derivations,
since the corresponding Hamiltonian term commutes with any other operator.
Let us finally remark that the excitations associated to this Hamiltonian
are what we call \emph{photons}.

\newpage

\section{Quantum theory of atoms and the two-level approximation\label{Sec:Atoms}}

The electromagnetic field is composed of degrees of freedom characterized
by a uniform energy spectrum, as represented in Fig. \textbf{ToDo}.
In contrast, with full generality, in quantum optics we call \emph{matter}
to any system that interacts with the electromagnetic field, and is
composed by degrees of freedom characterized by a non-uniform energy
spectrum, which interact with the electromagnetic field. In order
to fix ideas, we will focus here on simple atoms, but more complicated
systems such as molecules, confined electrons, or even solid-state
systems can be approached with similar tools as the ones we will learn
here.

\subsection{Atomic energy spectrum\label{Sec:AtomicEnergySpectrum}}

We start by considering the simplest type of matter possible, apart
from fundamental particles: the \emph{hydrogen atom}. It consists
of a proton and an electron interacting with each other in three dimensions.
Let's start first by discussing the degrees of freedom of these system.
We first consider the spin 1/2 of the electron and the proton, denoted
by the three-component vector operators $\hat{\mathbf{S}}$ and $\hat{\mathbf{I}}$,
respectively. These operators commute with each other, and satisfy
the angular momentum algebra $[\hat{S}_{j},\hat{S}_{k}]=\mathrm{i}\hbar\sum_{l=1}^{3}\hat{S}_{l}$,
and similarly for the components of $\hat{\mathbf{I}}$. The other
degrees of freedom correspond to the relative coordinates between
the proton and the electron in 3D, denoted by $\hat{\mathbf{r}}$,
which satisfy the canonical commutation relations $[\hat{r}_{j},\hat{p}_{l}]=\mathrm{i}\hbar\delta_{jl}$
and $[\hat{r}_{j},\hat{r}_{l}]=0=[\hat{p}_{j},\hat{p}_{l}]$, where
$\hat{\mathbf{p}}$ is the proton-electron relative's momentum. The
total Hilbert space associated to the system is then $\mathcal{H}_{\text{3D}}\otimes\mathcal{H}_{S=1/2}\otimes\mathcal{H}_{I=1/2}$,
where the last two Hilbert spaces are 2-dimensional, while $\mathcal{H}_{\text{3D}}$
is composed of three infinite-dimensional spaces (one for each direction
in real space), each the same as the one we used to describe the harmonic
oscillator in the previous chapter. We will ignore the center-of-mass
motion, which essentially behaves as a free particle, unless we expose
the atom to extra external fields or other particles, as we will see
towards the end of the notes.

Within the simplest, so-called \emph{Bohr approximation}, the Hamiltonian
is the sum of a kinetic term and a Coulomb potential term, that is,
\begin{equation}
\hat{H}_{\text{A}}=\frac{\mathbf{\hat{p}}^{2}}{2m}-\frac{e^{2}}{4\pi\epsilon_{0}|\hat{\mathbf{r}}|},\label{HydroH}
\end{equation}
where $m$ is the reduced mass of the system and $e$ the charge of
the electron. Of course, since the mass of the proton is much larger
than that of the electron, to a good approximation it can be assumed
that the proton is sitting motionless at the origin, so that $\hat{\mathbf{r}}$
and $m$ are just the coordinates and mass of the electron. The spectrum
of this Hamiltonian is well studied \cite{CohenTannoudjiBookI,CohenTannoudjiBookII,GreinerQuantumBook1,GreinerQuantumBook2,Basdevant02,Ballentine98},
and next we briefly review it.

The spectrum of this Hamiltonian has both countable and continuous
parts. The continuous part has only positive eigenenergies $\{\hbar|\mathbf{k}|^{2}/2m\}_{\mathbf{k}\in\mathbb{R}^{3}}$,
corresponding essentially to scattering states in electron-proton
collisions, with a wave function that behaves as a plane wave $e^{\mathrm{i}\mathbf{k}\cdot\mathbf{r}}$
at large distances $|\mathbf{k}\cdot\mathbf{r}|\rightarrow\infty$.
We focus then on the negative eigenvalues, which correspond to bound
states that we would characterize as truly atomic states.

The structure of the Hilbert space requires 5 quantum numbers to uniquely
identify a basis element, 3 for each direction in $\mathcal{H}_{\text{3D}}$,
and 1 for each value of the spins along the chosen quantization axis
(see below). In the case of the Hydrogen atom, it is indeed very simple
to find four operators that commute with Hamiltonian (\ref{HydroH}).
First, because it is independent of the spin operators, it commutes
with any of the components of $\hat{\mathbf{S}}$ and $\hat{\mathbf{I}}$.
In addition, because it is invariant under rotations, it commutes
with the \emph{orbital angular momentum} operator $\hat{\mathbf{L}}=\hat{\mathbf{r}}\times\hat{\mathbf{p}}$,
as well as with $\hat{\mathbf{L}}^{2}$, which also commutes with
$\hat{\mathbf{L}}$. Of course, the spin operators commute with the
orbital angular momentum. Hence, choosing the $z$ direction as the
so-called \emph{quantization} \emph{axis} for any angular momentum
operator (that is, we choose to work with eigenstates of $\hat{L}_{3}$,
$\hat{S}_{3}$, and $\hat{I}_{3}$), the eigenstates of the Hamiltonian
can be labeled by an index containing 5 quantum numbers, $\mathbf{n}=(n,l,m_{L},m_{S},m_{I})$,
and satisfy the eigenvalue expressions\begin{subequations}\label{AtomicEigenstates}
\begin{align}
\hat{H}_{\text{A}}|\mathbf{n}\rangle & =E_{n}|\mathbf{n}\rangle,\qquad\text{with }E_{n}=-E_{0}/n^{2},\quad n=1,2,3...,\\
\hat{\mathbf{L}}^{2}|\mathbf{n}\rangle & =l(l+1)\hbar^{2}|\mathbf{n}\rangle,\qquad l=0,1,...,n-1,\\
\hat{L}_{3}|\mathbf{n}\rangle & =m_{L}\hbar|\mathbf{n}\rangle,\qquad m_{L}=-l,-l+1,...,l-1,l,\\
\hat{S}_{3}|\mathbf{n}\rangle & =m_{S}\hbar|\mathbf{n}\rangle,\qquad m_{S}=\pm1/2,\\
\hat{I}_{3}|\mathbf{n}\rangle & =m_{I}\hbar|\mathbf{n}\rangle,\qquad m_{I}=\pm1/2,
\end{align}
\end{subequations}where $E_{0}=me^{4}/2(4\pi\epsilon_{0})^{2}\hbar$
the so-called \emph{Rydberg energy} that we need to give to the ground-state
Hydrogen in order to free its constituents and reach unbounded states.
On the other hand,\emph{ }$n$, $l$, $m_{L}$, and $m_{S,I}$ are
known as \emph{principal}, \emph{orbital}, \emph{magnetic}, and \emph{spin}
quantum numbers. Note that the eigenenergies depend only on the principal
quantum number, and are then $2\times2\times n^{2}$ degenerate, as
represented in Fig. \textbf{ToDo}. This is owed to the so-called \emph{accidental
degeneracy} in $l$ of the Coulomb potential, since general rotationally-symmetric
potentials $V(|\hat{\mathbf{r}}|)$ lead to energy eigenvalues that
depend also on $l$. In contrast, the independence on $m_{L}$ is
characteristic of all $V(|\hat{\mathbf{r}}|)$, and is thus dubbed
\emph{essential degeneracy}.

The situation is much more complicated for heavier atoms \cite{Basdevant02}.
However, in quantum optics experiments we usually work with atoms
or ions that have all their electrons in closed shells, except for
a single one, the so-called \emph{valence electron} (think of alkaline
atoms such us Li, Na, K, or Rb, or alkaline-earth ions such as Be$^{+}$,
Mg$^{+}$, Ca$^{+}$, or Sr$^{+}$). Under such circumstances, one
can adopt a hydrogen-like model for the atom, which differs from the
one introduced above just by the fact that the closed-shell electrons
screen the Coulomb interaction between the valence electron and the
nucleus, turning it into a different potential $V(|\hat{\mathbf{r}}|)=-Z_{\text{eff}}(|\hat{\mathbf{r}}|)e^{2}/4\pi\epsilon_{0}|\hat{\mathbf{r}}|$,
with an effective position-dependent nucleus charge that is fully
screened at large distances, $Z_{\text{eff}}\overset{|\mathbf{r}|\rightarrow\infty}{\longrightarrow}1$,
and not screened at all at very short ones, $Z_{\text{eff}}\overset{|\mathbf{r}|\rightarrow0}{\longrightarrow}Z$,
where $Z$ is the atomic number of the atom. In such case, the accidental
degeneracy in $l$ is broken, as represented in Fig. \textbf{ToDo},
and the eigenenergies must be labeled with two quantum numbers, that
is, $E_{nl}$. Their explicit expressions are not important for our
purposes, and can be checked in \cite{Basdevant02}.

Before proceeding, let us remark a small detail: heavier atoms will
no longer have necessarily total nuclear spin $i=1/2$, where we denote
the eigenvalue of $\hat{\mathbf{I}}^{2}$ by $i(i+1)\hbar^{2}$. Instead,
their nucleons (protons and neutrons) can add up to any of the values
of $i$ allowed by the rules of composition of angular momenta, although
some make the atoms more stable than others, of course (the lowest
ones, intuitively). However, note that quantum optics experiments
typically occur at energy scales in which nuclear processes capable
of changing $i$ are not possible. Therefore, in the following we
don't to include it as a quantum number, since it remains fixed during
experiments. 

Most of the other degeneracies mentioned above are broken once a more
accurate atomic Hamiltonian is considered. In particular, there is
a hierarchy of corrections, which can be mainly collected in two sets,
the so-called \emph{fine} and \emph{hyperfine} corrections \cite{Basdevant02,CohenTannoudjiBookII}.
Let us briefly discuss them now. The fine corrections include relativistic
effects, corresponding mainly to corrections of the kinetic energy
term and to a new term that couples the orbital angular momentum and
the electronic spin as $\hat{\mathbf{S}}\cdot\hat{\mathbf{L}}$, the
so-called \emph{spin-orbit} \emph{coupling}. Let us define $\hat{\mathbf{J}}=\hat{\mathbf{L}}+\hat{\mathbf{S}}$,
sometimes called \emph{electronic angular momentum}, which is readily
shown to commute with $\hat{\mathbf{L}}^{2}$ and $\hat{\mathbf{S}}^{2}$.
Noting that $\hat{\mathbf{S}}\cdot\hat{\mathbf{L}}\propto\hat{\mathbf{J}}^{2}-\hat{\mathbf{S}}^{2}-\hat{\mathbf{L}}^{2}$,
we then see that the eigenenergies of the Hamiltonian will require
now an additional label $j$, the quantum number associated to the
total electronic angular momentum. Hence, the quantum numbers $m_{L}$
and $m_{S}$ are not good ones in the presence of relativistic effects,
but instead we can use $\mathbf{n}=(n,l,j,m_{J},m_{I})$, with\begin{subequations}
\begin{align}
\hat{\mathbf{J}}^{2}|\mathbf{n}\rangle & =j(j+1)\hbar^{2}|\mathbf{n}\rangle,\qquad j=|l\pm1/2|,\\
\hat{J}_{3}|\mathbf{n}\rangle & =m_{J}\hbar|\mathbf{n}\rangle,\qquad m_{J}=-j,-j+1,...,j-1,j,
\end{align}
\end{subequations}added to the previous eigenvalue equations (\ref{AtomicEigenstates})
for the other quantum numbers, and keeping in mind that the eigenenergies
have three indices now $E_{nlj}$, whose explicit expressions can
be checked in \cite{Basdevant02,CohenTannoudjiBookII}. Hence, only
the degeneracy in $m_{J}$ and $m_{I}$ remains. The fine correction
to the spectrum is sketched in Fig. \textbf{ToDo}. 

The hyperfine corrections, called that way because they typically
produce an even smaller correction than the fine ones, include interactions
between the electron and the nucleus that go beyond the Coulomb one.
The most important correction is due to interaction between the electronic
and nuclear magnetic dipoles, that is, an interaction of the $\hat{\mathbf{J}}\cdot\hat{\mathbf{I}}$
type. It is then convenient to define now the \emph{atomic angular
momentum} operator $\hat{\mathbf{F}}=\hat{\mathbf{J}}+\hat{\mathbf{I}}$,
which commutes with $\hat{\mathbf{J}}^{2}$ and $\hat{\mathbf{I}}^{2}$.
Noting again that $\hat{\mathbf{J}}\cdot\hat{\mathbf{I}}\propto\hat{\mathbf{F}}^{2}-\hat{\mathbf{J}}^{2}-\hat{\mathbf{I}}^{2}$,
now the eigenenergies will require an additional label $f$, the quantum
number associated to $\hat{\mathbf{F}}^{2}$. Hence, in the presence
of hyperfine interactions, the quantum numbers $m_{J}$ and $m_{I}$
must be replaced, and instead we can use $\mathbf{n}=(n,l,j,f,m_{F})$,
with\begin{subequations}
\begin{align}
\hat{\mathbf{F}}^{2}|\mathbf{n}\rangle & =f(f+1)\hbar^{2}|\mathbf{n}\rangle,\qquad f=|j-i|,|j-i|+1,...,j+i,\\
\hat{F}_{3}|\mathbf{n}\rangle & =m_{F}\hbar|\mathbf{n}\rangle,\qquad m_{F}=-f,-f+1,...,f-1,f,
\end{align}
\end{subequations}The hyperfine correction to the spectrum is sketched
in Fig. \textbf{ToDo}, where we show the lowest values of $E_{nljf}$.
Note that only the degeneracy in $m_{F}$ remains after all these
corrections, but this can be lifted by applying an external magnetic
field $\boldsymbol{e}B$, whose direction $\boldsymbol{e}$ defines
the quantization axis in experiments, leading to a term proportional
to $B\hat{F}_{3}$ in the Hamiltonian (Zeeman effect). For this reason,
$m_{F}$ is sometimes called \emph{atomic magnetic quantum number}.

From this discussion and the sketches of Fig. \textbf{ToDo}, it is
clear that atoms have a highly non-uniform energy spectrum. Of course,
more complicated matter systems such as molecules or materials have
even more intricate spectra.

A final important property that will become very relevant when studying
the interaction of atoms with light refers to the parity of the atomic
eigenstates. The parity operator $\hat{\Pi}$ is a unitary operator
defined by its action on the coordinate operator $\hat{\Pi}^{\dagger}\hat{\mathbf{r}}\hat{\Pi}=-\hat{\mathbf{r}}$
(which automatically implies $\hat{\Pi}^{\dagger}\hat{\mathbf{p}}\hat{\Pi}=-\hat{\mathbf{p}}$,
as easily proven\footnote{A simple proof is based on the canonical commutation relations. Let
us do it for one component of the position and momentum, say $\hat{q}$
and $\hat{p}$, with $[\hat{q},\hat{p}]=\mathrm{i}\hbar$. Applying
the parity operator from the left and right, we can rewrite this commutator
as -$\hat{q}\hat{\Pi}^{\dagger}\hat{p}\hat{\Pi}+\hat{\Pi}^{\dagger}\hat{p}\hat{\Pi}\hat{q}=\mathrm{i}\hbar$,
or $[\hat{q},\hat{p}]=-[\hat{q},\hat{\Pi}^{\dagger}\hat{p}\hat{\Pi}]$,
which implies $\hat{\Pi}^{\dagger}\hat{p}\hat{\Pi}=-\hat{p}+c$. In
this expression, $c$ is a constant that we can determine by, for
example, evaluating the matrix element $\langle q|\hat{\Pi}^{\dagger}\hat{p}\hat{\Pi}|q'\rangle$,
where $|q\rangle$ and $|q'\rangle$ are eigenstates of the coordinate
$\hat{q}$. Using the previous expression, we obtain $\langle q|\hat{\Pi}^{\dagger}\hat{p}\hat{\Pi}|q'\rangle=c\langle q|q'\rangle-\langle q|\hat{p}|q'\rangle=c\delta(q-q')+\mathrm{i}\hbar\partial_{x}\delta(q-q')$,
where we have used (\ref{MomentumDerivative}), which in terms of
position and momentum (instead of quadratures) reads $\langle q|\hat{p}|\psi\rangle=-\mathrm{i}\partial_{q}\langle q|\psi\rangle$.
On the other hand, using $\hat{\Pi}|q\rangle=|-q\rangle$, we can
also write $\langle q|\hat{\Pi}^{\dagger}\hat{p}\hat{\Pi}|q\rangle=\langle-q|\hat{p}|-q'\rangle=\mathrm{i}\hbar\partial_{(-q)}\langle-q|-q'\rangle=\mathrm{i}\hbar\partial_{q}\delta(q-q')$.
Finally, comparing both expressions we obtain $c=0$.}). Note that this implies that $\hat{\Pi}^{2}=\hat{I}$, which means
that $\hat{\Pi}^{-1}=\hat{\Pi}^{\dagger}=\hat{\Pi}$ and that the
parity operator has eigenvalues $\pm1$. Atomic Hamiltonians are generally
invariant under parity transformations (all the interactions and fundamental
particles are spherically symmetric), and hence, the parity operator
commutes with the Hamiltonian. Therefore, the eigenstates of the Hamiltonian
can be chosen with well-defined parity. This has the consequence that
the expectation value of the coordinate operator $\hat{\mathbf{r}}$
in any of the eigenstates of the Hamiltonian is zero. In order to
prove this, simply consider one of such eigenstates $|a\rangle$,
which will also be an eigenstate of the parity operator, $\hat{\Pi}|a\rangle=\pi_{a}|a\rangle$,
with $\pi_{a}=\pm1$. Then, it follows that $\langle a|\hat{\mathbf{r}}|a\rangle=\langle a|\hat{\Pi}^{2}\hat{\mathbf{r}}\hat{\Pi}^{2}|a\rangle=-\pi_{a}^{2}\langle a|\hat{\mathbf{r}}|a\rangle=-\langle a|\hat{\mathbf{r}}|a\rangle$,
which can hold only if $\langle a|\hat{\mathbf{r}}|a\rangle=\mathbf{0}$.
In general, it is clear from this derivation that the coordinate operator
can only connect states with opposite parity, that is, $\langle a|\hat{\mathbf{r}}|b\rangle\neq\mathbf{0}$
only if $\pi_{a}=-\pi_{b}$.

\subsection{Two-level approximation: Pauli pseudo-spin operators, atomic states,
and Bloch space\label{Sec:TwoLevelApprox}}

The fact that the spectrum of atoms is so complicated seems to suggest
that describing their interaction with light will be extremely difficult.
However, we argue here that in most situations of interest in quantum
optics, all atomic levels but a few can be ignored. The argument proceeds
as follows. Light can provide atoms with the energy to perform transitions
from some state to another with higher energy, say $|g\rangle\rightarrow|e\rangle$,
where the labels refer to \emph{ground} and \emph{excited}. In the
simplest case, this happens because a photon gives its energy $\hbar\omega$
to the electron\footnote{Note that inelastic (Compton) scattering in which the photon cedes
only part of its energy to the electron and changes its frequency,
is irrelevant at optical frequencies. In particular, the change of
wavelength induced by such process is at most on the order of the
electron's de Broglie wavelength, which is on the order of $10^{-3}$nm,
very far away from optical wavelengths, typically on the order of
hundreds of nanometres.}, which can therefore use that energy to effect the transition ($\omega$
is the frequency of the photon). However, energy conservation tells
us that the transition will only be possible when the energy of the
photon matches (at least approximately) the energy difference between
the atomic states, $E_{e}-E_{g}$. Now, since the atomic energy spectrum
is non-uniform, each transition has essentially a unique energy difference,
a unique \emph{spectral fingerprint}. Therefore, if we shine the atom
with monochromatic light, we only expect two specific levels to react,
the ones whose energy difference matches the frequency of light. Any
other level can be simply ignored, as photons cannot generate any
dynamics involving them. Below we indeed give explicit mathematical
support to this intuitive arguments.

In order to be more precise, let us collect all the atomic quantum
numbers into an index $\mathbf{n}$ as we did in the previous section.
From the previous discussion, we see that if we shine light matching
the energy difference between two atomic states labeled as $|g\rangle$
and $|e\rangle$, we can make the approximation
\begin{equation}
\hat{H}_{\text{A}}=\sum_{\mathbf{n}}E_{\mathbf{n}}|\mathbf{n}\rangle\langle\mathbf{n}|\approx E_{g}|g\rangle\langle g|+E_{e}|e\rangle\langle e|,\label{Ha}
\end{equation}
which has a considerably simpler form than that of the original atomic
Hamiltonian. This is known as the \emph{two-level} \emph{approximation}.

Once the Hilbert space of the problem has been reduced to a two-dimensional
one, we have a lot of mathematical machinery developed for spin-1/2
angular momentum that we can use. In particular, let us rewrite the
atomic Hamiltonian as
\begin{equation}
\hat{H}_{\text{A}}=\frac{E_{e}+E_{g}}{2}(|e\rangle\langle e|+|g\rangle\langle g|)+\frac{E_{e}-E_{g}}{2}(|e\rangle\langle e|-|g\rangle\langle g|).
\end{equation}
Since $|e\rangle\langle e|+|g\rangle\langle g|=\hat{I}$ is the identity
of the two-dimensional Hilbert subspace we are working on, the first
term is just a constant shift of the energy that plays no role in
the dynamics of the system. We thus remove it, or, in other words,
we set the energy origin at the center of the transition for convenience.
The second term, on the other hand, has a very suggestive form, as
we discuss next.

If we identify the excited and ground states with the $\pm1/2$ states
of a fictitious spin-1/2 system (a so-called \emph{pseudo-spin}),
it is then natural to define the Pauli operators
\begin{equation}
\hat{\sigma}_{x}=|g\rangle\langle e|+|e\rangle\langle g|,\hspace{1em}\hat{\sigma}_{y}=\mathrm{i}(|g\rangle\langle e|-|e\rangle\langle g|),\hspace{1em}\hat{\sigma}_{z}=|e\rangle\langle e|-|g\rangle\langle g|,
\end{equation}
which satisfy the commutation and anticommutation relations\footnote{Note that $\epsilon_{jkl}$ is the so-called Levi-Civita symbol, defined
by being completely antisymmetric under the exchange of any two indices
and $\epsilon_{xyz}=1$.} $[\hat{\sigma}_{j},\hat{\sigma}_{k}]=2\mathrm{i}\sum_{l=x,y,z}\epsilon_{jkl}\hat{\sigma}_{l}$
and $\hat{\sigma}_{j}\hat{\sigma}_{k}+\hat{\sigma}_{k}\hat{\sigma}_{j}=2\delta_{jk}\hat{I}$,
apart from the useful properties $\text{tr}\{\hat{\sigma}_{j}\}=0$
and $\det\{\hat{\sigma}_{j}\}=-1.$ The property $\hat{\sigma}_{j}\hat{\sigma}_{k}=\delta_{jk}\hat{I}+\mathrm{i}\sum_{l=x,y,z}\epsilon_{jkl}\hat{\sigma}_{l}$
is also useful.

It is also important to remark that the Pauli operators, together
with the identity, form a basis in the space of operators acting on
two-dimensional Hilbert spaces (which itself is a four-dimensional
Hilbert space with respect the inner product defined by the trace
product $\text{tr}\{\hat{A}^{\dagger}\hat{B}\}$ for any two operators
$\hat{A}$ and $\hat{B}$). While this sounds a bit technical, the
proof is actually completely trivial by making use of the representation
of the Pauli operators in the $\{|e\rangle,|g\rangle\}$ basis, the
so-called \emph{Pauli matrices}
\[
\sigma_{x}=\left(\begin{array}{cc}
0 & 1\\
1 & 0
\end{array}\right),\quad\sigma_{y}=\left(\begin{array}{cc}
0 & -\mathrm{i}\\
\mathrm{i} & 0
\end{array}\right),\quad\sigma_{z}=\left(\begin{array}{cc}
1 & 0\\
0 & -1
\end{array}\right).
\]
Together with the representation of the identity, it is obvious the
linear combination
\[
c_{0}I+c_{1}\sigma_{x}+c_{2}\sigma_{y}+c_{3}\sigma_{z}=\left(\begin{array}{cc}
c_{0}+c_{3} & c_{1}-\mathrm{i}c_{2}\\
c_{1}+\mathrm{i}c_{2} & c_{0}-c_{3}
\end{array}\right),
\]
with parameters $c_{j}\in\mathbb{C}$ allows writing any complex $2\times2$
matrix.

Using the Pauli operators, the atomic Hamiltonian is then turned into
\begin{equation}
\hat{H}_{\text{A}}=\frac{\hbar\varepsilon}{2}\hat{\sigma_{z}},\label{HaTLS}
\end{equation}
where we have defined $\varepsilon=(E_{e}-E_{g})/\hbar$, the so-called
\emph{transition frequency}.

It is also convenient to define \emph{raising} and \emph{lowering}
operators that effect transitions in between the atomic levels,
\begin{equation}
\hat{\sigma}=|g\rangle\langle e|=(\hat{\sigma}_{x}-\mathrm{i}\hat{\sigma}_{y})/2,\hspace{1em}\hat{\sigma}^{\dagger}=|e\rangle\langle g|=(\hat{\sigma}_{x}+\mathrm{i}\hat{\sigma}_{y})/2,
\end{equation}
which obey the commutation relations $[\hat{\sigma},\hat{\sigma}^{\dagger}]=-\hat{\sigma}_{z}$,
$[\hat{\sigma},\hat{\sigma}_{z}]=2\hat{\sigma}$, and $[\hat{\sigma}^{\dagger},\hat{\sigma}_{z}]=-2\hat{\sigma}^{\dagger}$. 

Let's move on now to the description of atomic states. As mentioned
above, Together with the identity, the Pauli operators form a basis
for any other operator. Hence, the density operator representing the
atomic state can be expanded as $\hat{\rho}=(b_{0}\hat{I}+\mathbf{b}^{T}\hat{\boldsymbol{\sigma}})/2$,
where we have introduced the Pauli vector $\hat{\boldsymbol{\sigma}}=(\hat{\sigma}_{x},\hat{\sigma}_{y},\hat{\sigma}_{z})^{T}$,
the expansion coefficients $b_{0}$ and $\mathbf{b}=(b_{x,}b_{y},b_{z})^{T}$,
and we have divided by two for future convenience. The properties
that physical density operators must satisfy restrict the values that
these coefficients can take. For example, they must be real, since
$\hat{\rho}^{\dagger}=b_{0}^{*}\hat{I}+\mathbf{b}^{\dagger}\hat{\boldsymbol{\sigma}}=\hat{\rho}$
can hold only if that is the case. Next, the normalization of the
state fixes $b_{0}=1$, since $\text{tr}\{\hat{\rho}\}=b_{0}\text{tr}\{\hat{I}\}/2=b_{0}$
must be equal to 1. Finally, the eigenvalues of the density operator
must be positive, but smaller than or equal to 1. Since the matrix
representation of the operator is only $2\times2$, these are trivial
to find. In particular, the characteristic equation reads
\begin{equation}
\det\{\hat{\rho}-\lambda\hat{I}\}=\det\left\{ \frac{1}{2}\left(\begin{array}{cc}
1-2\lambda+b_{z} & b_{x}-\mathrm{i}b_{y}\\
b_{x}+\mathrm{i}b_{y} & 1-2\lambda-b_{z}
\end{array}\right)\right\} =\frac{1}{4}\left[(1-2\lambda)^{2}-|\mathbf{b}|{}^{2}\right]=0,
\end{equation}
leading to eigenvalues $(1\pm|\mathbf{b}|)/2$, and hence to the condition
$|\mathbf{b}|\leq1$. Therefore, a general atomic state within the
two-level approximation can be written as
\begin{equation}
\hat{\rho}=\frac{1}{2}(\hat{I}+\mathbf{b}^{T}\hat{\boldsymbol{\sigma}}),\hspace{1em}\text{with }|\mathbf{b}|\leq1.\label{AtomicStateGen}
\end{equation}
$\mathbf{b}$ is known as the \emph{Bloch vector}, whose components
can be interpreted as the expectation values of the Pauli operators,
as readily seen from
\begin{equation}
\langle\hat{\sigma}_{j}\rangle=\text{tr}\{\hat{\rho}\hat{\sigma}_{j}\}=\frac{1}{2}\underset{0}{\underbrace{\text{tr}\{\hat{\sigma}_{j}\}}}+\sum_{k=x,y,z}\frac{b_{k}}{2}\underbrace{\text{tr}\{\hat{\sigma}_{k}\hat{\sigma}_{j}\}}_{2\delta_{kj}}=b_{j}.
\end{equation}
Of particular relevance is $b_{z}$, which provides the probability
of being in the excited or ground states as $p_{e}=\langle e|\hat{\rho}|e\rangle=(1+b_{z})/2$
and $p_{g}=\langle g|\hat{\rho}|g\rangle=(1-b_{z})/2$, respectively.
These are usually called the excited and ground state \emph{populations}.

This way of writing density operators naturally leads to a way of
visualizing them in the space generated by the expectation values
of the Pauli operators, which we call \emph{Bloch space}. State (\ref{AtomicStateGen})
corresponds to a point $\mathbf{b}$ in that space. This is represented
in Figure \textbf{ToDo}.

Note that, since the eigenvalues of the density matrix depend solely
on $\mathbf{b}^{2}$, states of equal entropy live all in a sphere
centered at the origin of Bloch space (see Figure \textbf{ToDo}).\textbf{
}In particular, note that the maximally mixed state $\hat{\rho}=\hat{I}/2$
corresponds to $\mathbf{b}=\mathbf{0}$, while pure states have $|\mathbf{b}|=1$,
since only then the density operator has eigenvalues 1 and 0, as corresponds
to a rank-1 projector (that is, a single state contributing to the
mixture). Hence, pure states are specified in the Bloch space by a
point in a sphere of radius one centered at the origin, the so-called
\emph{Bloch sphere}. In particular, we can use the polar $\phi$ and
azimuthal $\theta$ angles of spherical coordinates to pinpoint them,
$\mathbf{b}=(\sin\theta\cos\phi,\sin\theta\sin\phi,\cos\theta)$,
with $\theta\in[0,\pi]$ and $\phi\in[0,2\pi[$. In the Hilbert space,
this means that any pure state can be written as
\begin{equation}
|\psi\rangle=\cos\left(\frac{\theta}{2}\right)|e\rangle+e^{\mathrm{i}\phi}\sin\left(\frac{\theta}{2}\right)|g\rangle,
\end{equation}
as can be readily checked. We thus see that the north pole of the
Bloch sphere ($\theta=0$, $\phi=0$) corresponds to the excited state
$|e\rangle$, while the south pole ($\theta=\pi$, $\phi=0$) corresponds
to the ground state $|g\rangle$. The eigenstates of the other Pauli
operators are located then in the equator $\theta=\pi/2$.

\subsection{Unsupervised evolution of two-level systems}

\subsubsection{General Hamiltonian and Bloch equations}

Using again the fact that the Pauli matrices together with the identity
are a basis for any operator acting on a two-dimensional Hilbert space,
we can write the most general type of Hamiltonian acting on a two-level
system as
\begin{equation}
\hat{H}(t)=\frac{\hbar}{2}\boldsymbol{\alpha}^{T}(t)\hat{\boldsymbol{\sigma}},
\end{equation}
where $\boldsymbol{\alpha}(t)$ is a vector with arbitrary real functions
of time as entries and we have dismissed terms proportional to the
identity, since they don't play any role in the dynamics of the atom.

One convenient way of analyzing the dynamics of the system is based
on the evolution of the Bloch vector in the Bloch space. The evolution
equation of the Bloch vector can be easily found in many ways. Since
we know that its components are the expectation values of the Pauli
operators, possibly the simplest way consists in finding the Heisenberg
equations of these operators, and then take their expectation values.
Let's proceed this way. We get
\begin{equation}
\frac{d\hat{\sigma}_{j}}{dt}=\left[\hat{\sigma}_{j},\frac{\hat{H}(t)}{\mathrm{i}\hbar}\right]=-\frac{\mathrm{i}}{2}\sum_{k=x,y,z}\alpha_{k}(t)[\hat{\sigma}_{j},\hat{\sigma}_{k}]=\sum_{k,l=x,y,z}\epsilon_{jkl}\alpha_{k}(t)\hat{\sigma}_{l},
\end{equation}
and taking the expectation value
\begin{equation}
\frac{db_{j}}{dt}=\sum_{k,l=x,y,z}\epsilon_{jkl}\alpha_{k}(t)b_{l}.
\end{equation}
These are known as the \emph{Bloch equations} of the two-level system,
and can be written in matrix form as
\begin{equation}
\dot{\mathbf{b}}=\mathcal{B}(t)\mathbf{b},\hspace{1em}\text{with }\mathcal{B}(t)=\left[\begin{array}{ccc}
0 & -\alpha_{z}(t) & \alpha_{y}(t)\\
\alpha_{z}(t) & 0 & -\alpha_{x}(t)\\
-\alpha_{y}(t) & \alpha_{x}(t) & 0
\end{array}\right].
\end{equation}
In principle, being a $3\times3$ linear system, it is not difficult
to solve these equations for any choice of $\boldsymbol{\alpha}(t)$,
numerically in the worst case. In the following we will consider a
couple of examples of special interest for quantum optics. But before
that, let us show that $|\mathbf{b}|$ is a conserved quantity, and
therefore, the trajectory $\mathbf{b}(t)$ in Bloch space is restricted
to a sphere with radius equal to $|\mathbf{b}(0)|$. We simply evaluate
\begin{equation}
\frac{d|\mathbf{b}|^{2}}{dt}=\sum_{j=x,y,z}\frac{db_{j}^{2}}{dt}=2\sum_{j=x,y,z}b_{j}\frac{db_{j}}{dt}=2\sum_{j,k,l=x,y,z}\epsilon_{jkl}b_{j}\alpha_{k}b_{l}=0,
\end{equation}
where in the last equality we have used that $\epsilon_{jkl}$ is
antisymmetric in $(jl)$ while $b_{j}\alpha_{k}b_{l}$ is symmetric.
Note that this result is actually quite intuitive once we remember
that unitary transformations cannot change the entropy of the state,
which in the case of two-level systems depends solely on $|\mathbf{b}|$.

We will see that most of the time it is useful to work with the raising
and lowering operators instead of $\hat{\sigma}_{x}$ and $\hat{\sigma}_{y}$.
We can rewrite the Hamiltonian in terms of these as
\begin{equation}
\hat{H}(t)=\frac{\hbar}{2}[\alpha_{z}(t)\hat{\sigma}_{z}+\alpha(t)\hat{\sigma}+\alpha^{*}(t)\hat{\sigma}^{\dagger}],
\end{equation}
with $\alpha(t)=\alpha_{x}(t)+\mathrm{i}\alpha_{y}(t)$. Defining
then $b=(b_{x}-\mathrm{i}b_{y})/2=\langle\hat{\sigma}\rangle$, the
Bloch equations are rewritten as\begin{subequations}\label{ComplexBloch}
\begin{align}
\dot{b} & =-\mathrm{i}\alpha_{z}(t)b+\frac{\mathrm{i}}{2}\alpha^{*}(t)b_{z},\\
\dot{b}_{z} & =\mathrm{i}\alpha(t)b-\mathrm{i}\alpha^{*}(t)b^{*},
\end{align}
\end{subequations}which we will call \emph{complex Bloch equations}.

\begin{figure}
\includegraphics[width=0.4\textwidth]{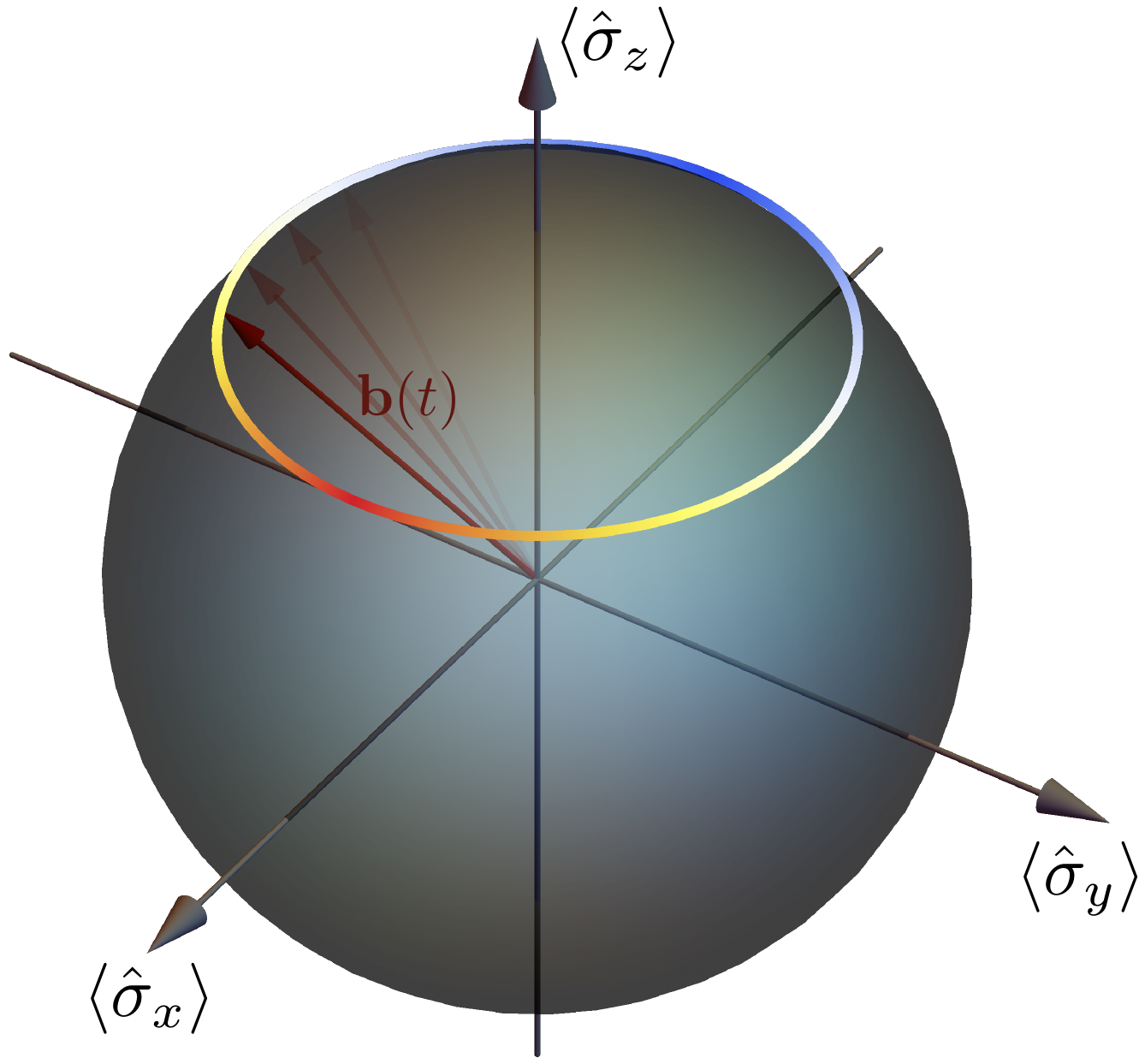}\caption{Evolution of the Bloch vector when the atom is subject to its free
Hamiltonian (\ref{HaTLS}). The trajectory is shown with a color gradient,
where the initial condition is shown in red. The trail of the Bloch
vector indicates the direction of its motion. As explained in the
main text, the evolution induced by any Hamiltonian is constrained
to a sphere of radius $|\mathbf{b}(0)|$, shown in black in the figure.
Hence, we see that when allowed to evolve freely, the atom describes
right-handed circular rotation around the z axis. \label{fBlochSpace-SpinPrecession}}
\end{figure}

\subsubsection{Free evolution}

Let us consider as a first, trivial example the free evolution of
the two-level system. This corresponds to the choice $\boldsymbol{\alpha}=(0,0,\varepsilon)^{T}$,
leading to complex Bloch equations
\begin{equation}
\dot{b}=-\mathrm{i}\varepsilon b,\hspace{1em}\dot{b}_{z}=0\hspace{1em}\Longrightarrow\hspace{1em}b(t)=b(0)e^{-\mathrm{i}\varepsilon t},\;b_{z}(t)=b_{z}(0).\label{FreeAtomSolution}
\end{equation}
Note how simple the solution looks in terms of the complex representation
of the Bloch vector. In terms of the cartesian components of the Bloch
vector, we then get\begin{subequations}
\begin{align}
b_{x}(t) & =b_{x}(0)\cos(\varepsilon t)-b_{y}(0)\sin(\varepsilon t),\\
b_{y}(t) & =b_{y}(0)\cos(\varepsilon t)+b_{x}(0)\sin(\varepsilon t),\\
b_{z}(t) & =b_{z}(0).
\end{align}
\end{subequations}As expected, the populations of the excited and
ground states, $[1\pm b_{z}(t)]/2$, do not change in time ($\hat{\sigma}_{z}$
commutes with $\hat{H}$). The Bloch vector evolves along a circular
trajectory parallel to the $x-y$ plane, undergoing right-handed harmonic
rotation around the $z$ axis, as shown in Figure \ref{fBlochSpace-SpinPrecession}.
This effect is known as (\emph{pseudo}-)\emph{spin precession}.

\subsubsection{Rabi oscillations and the rotating-wave approximation}

Let us now take one step forward, and consider a Hamiltonian
\begin{equation}
\hat{H}(t)=\frac{\hbar\varepsilon}{2}\hat{\sigma}_{z}+\hbar\Omega\cos(\omega t)\hat{\sigma}_{x},\label{SemiclassRabiH}
\end{equation}
described by $\boldsymbol{\alpha}(t)=(2\Omega\cos\omega t,0,\varepsilon)$.
As we will see in the next chapter, this Hamiltonian, known as the
\emph{semiclassical Rabi Hamiltonian}, describes an atom subject to
a (classical) monochromatic light field of frequency $\omega$. Hence,
we expect the atom to react to this field only when $\omega$ is close
to $\varepsilon$, as discussed in the first section of this chapter.
Among other things, this example will actually allow us to specify
what ``close'' means. We will also assume that $\Omega\ll\omega$,
a natural assumption for optical frequencies ($\omega\sim10^{15}\text{Hz}$),
which are quite large, and would require unconceivable large field
intensities, as we shall see in the next chapter.

The complex Bloch equations (\ref{ComplexBloch}) read in this case
as\begin{subequations}\label{ComplexBlochRabi}
\begin{align}
\dot{b} & =-\mathrm{i}\varepsilon b+\mathrm{i}\Omega\cos(\omega t)b_{z},\\
\dot{b}_{z} & =2\mathrm{i}\Omega\cos(\omega t)(b-b^{*}).
\end{align}
\end{subequations} It is convenient to define a \emph{slowly-varying}
variable $\tilde{b}(t)=e^{\mathrm{i}\omega t}b(t)$. Since $\omega$
must be close to $\varepsilon$ for the arguments exposed above, this
transformation removes a large part of the spin precession around
the $z$ axis induced by the free term of the Hamiltonian, so that
we expect the new variable to vary slowly compared to $\varepsilon$.
We will later check that this is indeed the case. In terms of this
new variable, and expanding $2\cos\omega t=e^{\mathrm{i\omega t}}+e^{-\mathrm{i}\omega t}$,
the previous equations read\begin{subequations}\label{ComplexBlochRabiSlowExact}
\begin{align}
\dot{\tilde{b}} & =\mathrm{i}(\omega-\varepsilon)\tilde{b}+\mathrm{i}\frac{\Omega}{2}\left(1+e^{2\mathrm{i}\omega t}\right)b_{z},\\
\dot{b}_{z} & =\mathrm{i}\Omega\left(1+e^{-2\mathrm{i}\omega t}\right)\tilde{b}-\text{c.c.}\hspace{1em}.
\end{align}
\end{subequations}These equations are exact, but not easy to analyze
analytically because of the time-dependent coefficients $e^{2\mathrm{i}\omega t}$.
However, next we argue that this coefficients are negligible. This
is known as the \emph{rotating-wave approximation}, and is of paramount
importance in quantum optics, as we shall see through many examples
over the next chapters. An easy way of understanding the conditions
under which such an approximation works consists in integrating the
equations above over one optical cycle, obtaining\begin{subequations}\label{ComplexBlochRabiSlowAveraged}
\begin{align}
\int_{t-\pi/\omega}^{t+\pi/\omega}d\tau\dot{\tilde{b}}(\tau) & =\mathrm{i}(\omega-\varepsilon)\int_{t-\pi/\omega}^{t+\pi/\omega}d\tau\tilde{b}(\tau)+\mathrm{i}\frac{\Omega}{2}\left(\int_{t-\pi/\omega}^{t+\pi/\omega}d\tau b_{z}(\tau)+\int_{t-\pi/\omega}^{t+\pi/\omega}d\tau b_{z}(\tau)e^{2\mathrm{i}\omega\tau}\right),\\
\int_{t-\pi/\omega}^{t+\pi/\omega}d\tau\dot{b}_{z}(\tau) & =\mathrm{i}\Omega\left(\int_{t-\pi/\omega}^{t+\pi/\omega}d\tau\tilde{b}(\tau)+\int_{t-\pi/\omega}^{t+\pi/\omega}d\tau\tilde{b}(\tau)e^{-2\mathrm{i}\omega\tau}\right)-\text{c.c.}\hspace{1em}.
\end{align}
\end{subequations}The key idea now is that once we perform the rotating-wave
approximation, the evolution of $\tilde{b}(t)$ and $b_{z}(t)$ will
depend only on $|\omega-\varepsilon|$ and $\Omega$, which are assumed
to be much smaller than $\omega$. In other words, these variables
are approximately constant over one optical cycle. Hence, approximating
these variables by their value at the center of the integration domain,
so that, for example,\begin{subequations}
\begin{align}
\int_{t-\pi/\omega}^{t+\pi/\omega}d\tau\tilde{b}(\tau) & \approx\frac{2\pi}{\omega}\tilde{b}(t),\\
\int_{t-\pi/\omega}^{t+\pi/\omega}d\tau\tilde{b}(\tau)e^{2\mathrm{i}\omega\tau} & \approx\tilde{b}(t)\int_{t-\pi/\omega}^{t+\pi/\omega}d\tau e^{2\mathrm{i}\omega\tau}=0,
\end{align}
 \end{subequations}we obtain\begin{subequations}\label{ComplexBlochRabiSlowRWA}
\begin{align}
\dot{\tilde{b}} & =\mathrm{i}\Delta\tilde{b}+\mathrm{i}\frac{\Omega}{2}b_{z},\\
\dot{b}_{z} & =\mathrm{i}\Omega(\tilde{b}-\tilde{b}^{*}),
\end{align}
\end{subequations}where we have defined the \emph{detuning} $\Delta=\omega-\varepsilon$.
These are precisely the equations (\ref{ComplexBlochRabiSlowExact})
but without the time-dependent terms, and will be a good approximation
to these as long as $|\omega-\varepsilon|\ll\omega$ and $\Omega\ll\omega$.
The arguments leading to this rotating-wave approximation can be put
in more rigorous mathematical terms by using time-dependent perturbation
theory, which we shall do later as an exercise for completeness.

\begin{figure}
\includegraphics[width=0.85\textwidth]{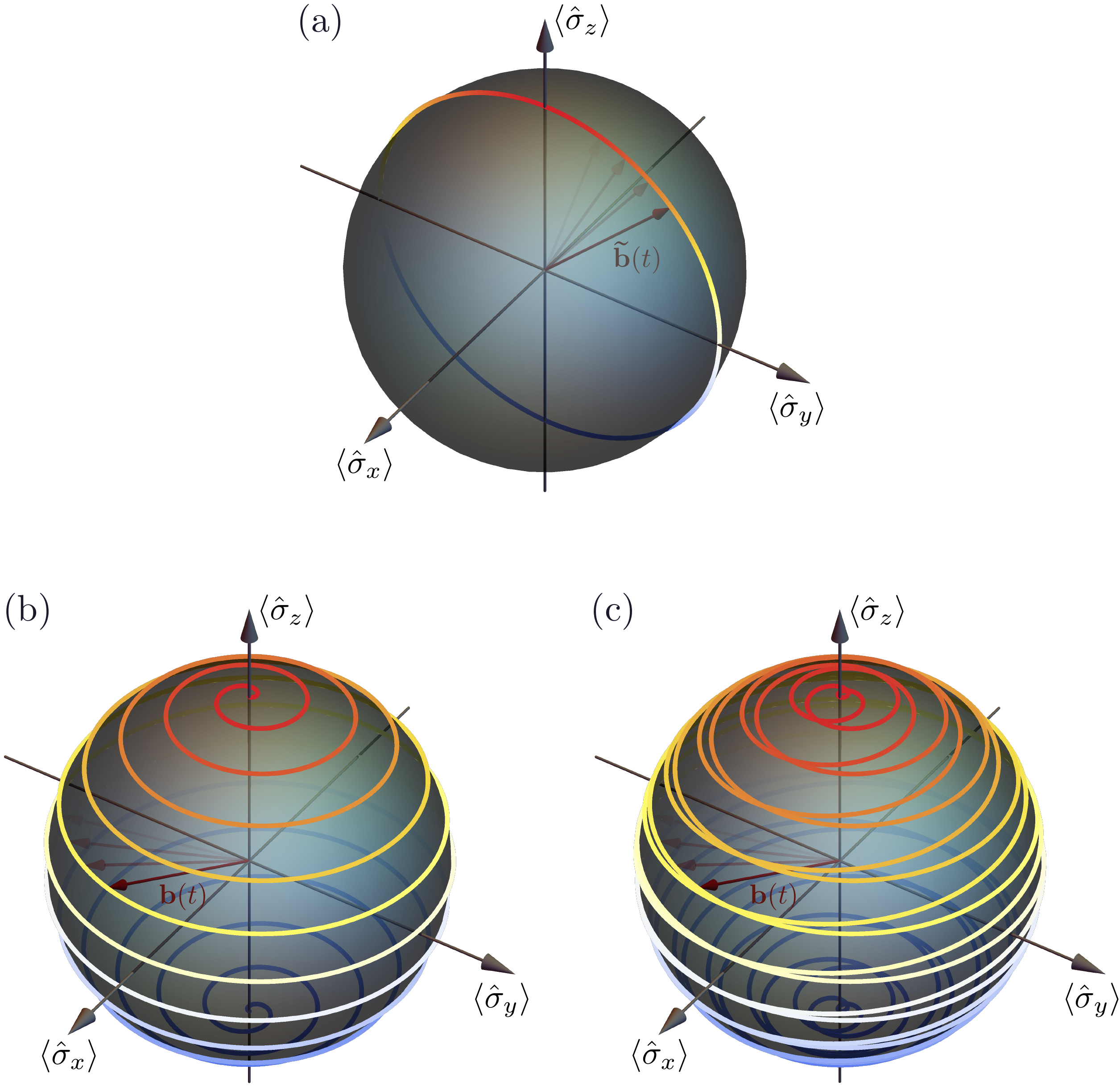}\caption{Evolution of the Bloch vector when the atom is subject to the semiclassical
Rabi Hamiltonian (\ref{SemiclassRabiH}). As in the previous figure,
the trajectory is shown with a color gradient, where the initial condition
is marked in red, here chosen to be the excited state $|e\rangle$
(north pole). In (a) we show the evolution of the slowly-varying Bloch
vector $\tilde{\mathbf{b}}=(\tilde{b}_{x},\tilde{b}_{y},b_{z})^{T}$,
which undergoes right-handed precession around the $x$ axis. In contrast,
the trajectory of the original Bloch vector, which includes the precession
around the $z$ axis, describes a spiral trajectory as shown in (b)
and (c), for half a period and a full period, respectively (note that
on the second half of the evolution the trajectory is also spiral,
but not the same as the spiral described by the first half, as that
would require the precession along the $z$ axis to change from right-handed
to left-handed after half a period). We have chosen $\varepsilon=25\Omega$
in order to be able to see the spiral trajectory to the naked eye,
but keep in mind that in common quantum optics experiments atomic
transitions can get much larger than that. \label{fBlochSpace-RabiOsci}}
\end{figure}

Now that we have removed explicit time dependences, these equations
can be solved analytically. For the sake of keeping the discussion
going, we will do this at the end of the section, and here we just
discuss the solution, which reads\begin{subequations}\label{GenSolsRabi}
\begin{align}
b_{z}(t) & =\frac{\delta^{2}b_{z}(0)-\delta b_{x}(0)+[b_{z}(0)+\delta b_{x}(0)]\cos\left(\Omega_{R}t\right)+\sqrt{1+\delta^{2}}b_{y}(0)\sin\left(\Omega_{R}t\right)}{1+\delta^{2}},\\
\tilde{b}(t) & =\frac{b_{x}(0)-\delta b_{z}(0)+[\delta b_{z}(0)+2\delta^{2}b(0)-\mathrm{i}b_{y}(0)]\cos\left(\Omega_{R}t\right)+\mathrm{i}\sqrt{1+\delta^{2}}[b_{z}(0)+2\delta b(0)]\sin\left(\Omega_{R}t\right)}{2(1+\delta^{2})},
\end{align}
\end{subequations}where we have introduced the normalized detuning
$\delta=\Delta/\Omega$, the \emph{Rabi frequency }$\Omega_{R}=\sqrt{\Omega^{2}+\Delta^{2}}$,
and used the relation $2b(0)=b_{x}(0)-\mathrm{i}b_{y}(0)$ to shorten
a bit the expressions. Note that, importantly, both $b_{z}(t)$ and
$\tilde{b}(t)$ oscillate at frequency $\Omega_{R}$, and are therefore
slowly varying variables as compared to $\omega$, which proves the
consistency of the rotating-wave approximation performed above\footnote{As a general lesson, never forget how important it is to check that
the solution obtained at the end of a calculation is consistent with
any approximations performed on the way to finding it. }. Note as well that the populations $[1\pm b_{z}(t)]$ oscillate in
time, which is a phenomenon known as \emph{Rabi oscillations}.

We consider now some specific limits of this solution. First, let
us consider the $|\delta|\gg1$ limit (remember that $|\mathbf{b}|\leq1$),
which leads to $b_{z}(t)=b_{z}(0)$ and $\tilde{b}(t)=b(0)e^{\mathrm{i}\Delta t}$,
or $b(t)=b(0)e^{-\mathrm{i}\varepsilon t}$ in terms of the original
variable. This is precisely the solution (\ref{FreeAtomSolution})
that we found for an atom evolving freely. Hence, as expected, the
atom won't feel the light field if $|\Delta|\gg\Omega$. Most importantly,
this also justifies the two-level approximation, that is, neglecting
all transitions which are far from resonance with respect to the light
frequency.

The opposite limit, $\delta=0,$ is also interesting. This corresponds
to a light field resonant with the atomic transition, where, writing
$2\tilde{b}(t)=\tilde{b}_{x}(t)+\mathrm{i}\tilde{b}_{y}(t)$, the
solution for the original variables reads\begin{subequations}
\begin{align}
\tilde{b}_{x}(t) & =b_{x}(0),\\
\tilde{b}_{y}(t) & =b_{y}(0)\cos\left(\Omega t\right)-b_{z}(0)\sin\left(\Omega t\right),\\
b_{z}(t) & =b_{z}(0)\cos\left(\Omega t\right)+b_{y}(0)\sin\left(\Omega t\right),
\end{align}
\end{subequations}which corresponds to a (slowly-varying) Bloch vector
precessing left-handedly around the $x$ axis, as shown in Fig. \ref{fBlochSpace-RabiOsci}a.
Of course, the true Bloch vector $\mathbf{b}(t)$ combines this precession
with a faster optical precession around the $z$ axis at frequency
$\omega$, adding up to a double-spiraled trajectory in Bloch space,
as shown in Figure\textbf{s} \ref{fBlochSpace-RabiOsci}b and c.

\begin{figure}
\includegraphics[width=1\textwidth]{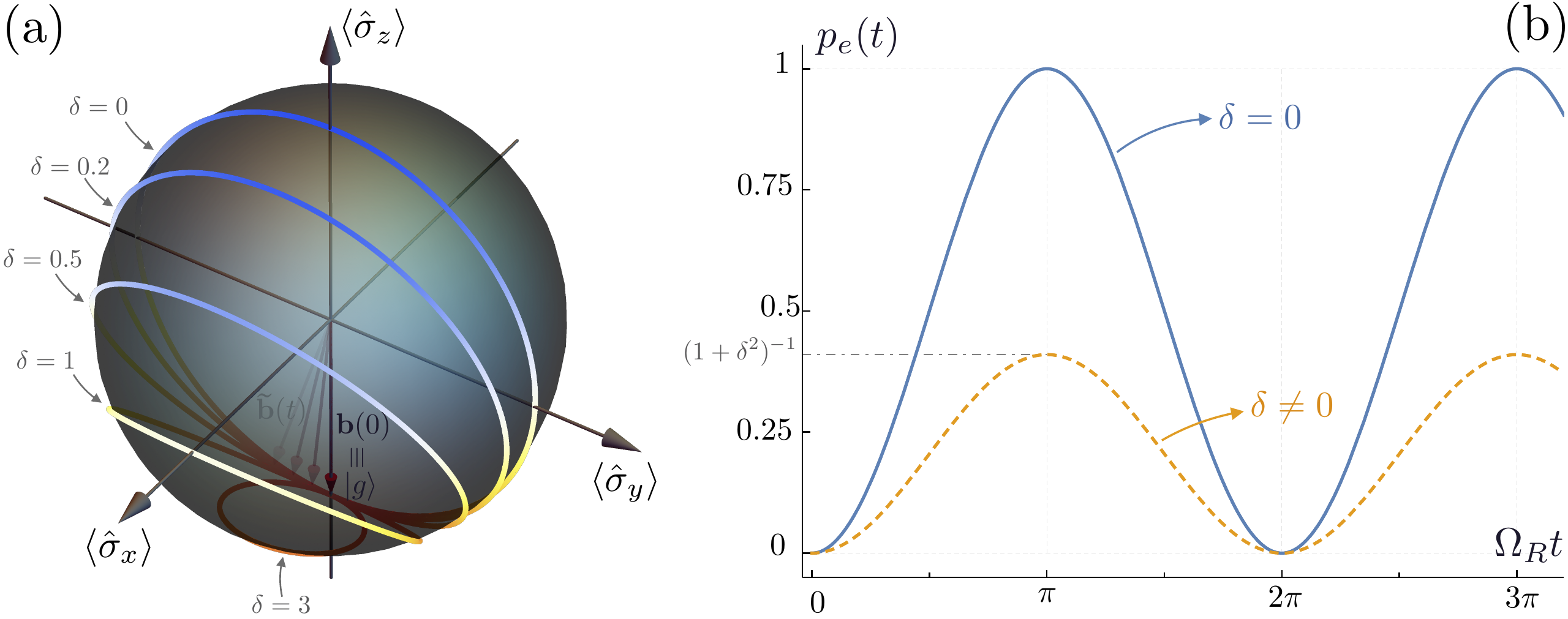}\caption{(a) Evolution of the slowly-varying Bloch vector when the atom is
initialized in the ground state $|g\rangle$ (south pole) and then
subject to the semiclassical Rabi Hamiltonian (\ref{SemiclassRabiH}),
for different values of the normalized detuning $\delta$. Conventions
are as in the previous figures. All the trajectories are circular
and touch the south pole, but their radius is smaller the larger $\delta$
is. We show in (b) the excited state population as a function of time
for $\delta=0$ and a generic $\delta\protect\neq0$. It displays
clear Rabi oscillations that periodically transfer population from
the ground to the excited state, but with an efficiency that decreases
as the detuning increases. \label{fBlochSpace-RabiOsciGround}}
\end{figure}

A final interesting limit corresponds to the situation where the atom
starts in the ground state, that is, $\mathbf{b}(0)=(0,0,-1)^{T}$.
In this case, equations (\ref{GenSolsRabi}) are simplified to
\begin{equation}
\tilde{b}_{x}(t)=\frac{\delta}{1+\delta^{2}}\left[1-\cos\left(\Omega_{R}t\right)\right],\hspace{1em}\tilde{b}_{y}(t)=\frac{1}{\sqrt{1+\delta^{2}}}\sin\left(\Omega_{R}t\right),\hspace{1em}b_{z}(t)=-\frac{\cos\left(\Omega_{R}t\right)+\delta^{2}}{1+\delta^{2}}.
\end{equation}
The corresponding trajectory in the Bloch sphere is shown in Fig.
\ref{fBlochSpace-RabiOsciGround}a for different values of the detuning
$\delta$. Note that trajectory starts at the south pole, and describes
a circle of smaller radius the larger is $\delta$. This is to be
expected, since for $\delta=0$ we should recover the precession around
the $x$ axis that we analyzed in the previous figure, while in the
limit $|\delta|\rightarrow\infty$ we already saw that the free-atom
limit is recovered, so that starting from an eigenstate of the bare
atomic Hamiltonian (\ref{HaTLS}) we should not move at all.

It is interesting to track the population of the excited state in
time for different values of $|\delta|$, which is given by 
\begin{equation}
p_{e}(t)=\frac{1-\cos\left(\Omega_{R}t\right)}{2(1+\delta^{2})},\label{RabiOsciPure}
\end{equation}
and shown in Fig. \ref{fBlochSpace-RabiOsciGround}b. For $\delta=0$
the population is completely transferred from the ground to the excited
state in the middle of the Rabi cycle. The effect of a finite detuning
$\delta\neq0$ is to induce an incomplete population transfer, as
now the maximum population available in the excited state is $(1+\delta^{2})^{-1}$,
which is negligible when $|\delta|\gg1$.

Let us finish this section by explaining in detail how to obtain the
solution (\ref{GenSolsRabi}) of the complex Bloch equations (\ref{ComplexBlochRabiSlowRWA}).
We will do it with a method that is useful for general linear problems,
certainly for many different problems in quantum optics. Let us consider
the system of $N$ coupled linear equations
\begin{equation}
\dot{\mathbf{x}}=\mathcal{B}\mathbf{x}+\mathbf{y}(t),
\end{equation}
where $\mathbf{x}\in\mathbb{C}^{N}$ is a vector of complex variables,
$\mathbf{y}(t)\in\mathbb{C}^{N}$ a vector of complex functions of
time which we assume to be given, and $\mathcal{B}$ is an $N\times N$
complex matrix. In the particular case of (\ref{ComplexBlochRabiSlowRWA}),
we have $N=3$, $\mathbf{x}=(\tilde{b},\tilde{b}^{*},b_{z})^{T}$,
$\mathbf{y}=(0,0,0)^{T}$, and
\begin{equation}
\mathbf{x}=\left(\begin{array}{c}
\tilde{b}\\
\tilde{b}^{*}\\
b_{z}
\end{array}\right),\quad\mathbf{y}=\left(\begin{array}{c}
0\\
0\\
0
\end{array}\right),\quad\text{and }\mathcal{B}=\mathrm{i}\left(\begin{array}{ccc}
\Delta & 0 & \mathrm{\Omega/2}\\
0 & -\Delta & -\Omega/2\\
\Omega & -\Omega & 0
\end{array}\right).
\end{equation}
The whole method relies on our ability to find a matrix $\mathcal{S}$
that diagonalizes $\mathcal{B}$ through a similarity transformation
\begin{equation}
\mathcal{S}^{-1}\mathcal{B}\mathcal{S}=\mathcal{D},
\end{equation}
where $\mathcal{D}$ is a diagonal matrix containing the eigenvalues
$\{\lambda_{j}\}_{j=1,...,N}$ of $\mathcal{B}$. Note that for all
practical purposes, in finite dimension we can assume that all matrices
are diagonalizable\footnote{More precisely, the set of non-diagonalizable matrices of finite dimension
has zero measure. This means that the probability of drawing a non-diagonalizable
matrix at random is zero. It also means that all non-diagonalizable
matrices have diagonalizable matrices infinitely close to them. These
properties are actually obvious once we remember that the condition
for a matrix to be non-diagonalizable is that its eigenvectors must
not be linearly independent. This means that matrix $\mathcal{S}$
satisfying $\mathcal{B}\mathcal{S}=\mathcal{S}\mathcal{D}$ is not
invertible, that is, $\det\mathcal{S}=0$. For finite dimension, there
are a finite number of terms adding up to match this condition, which
therefore requires a lot of fine tuning. In practical terms, this
means that by slightly perturbing some of the elements of matrix $\mathcal{B}$,
we move away from this condition, and obtain a diagonalizable matrix.
Now, since an infinitesimal change of $\mathcal{B}$ should only change
the physical results infinitesimally as well (otherwise it means that
we are in some exotic physical regime that is probably not well described
by our model, such as a phase transition that cannot be perfectly
described by a linear model), we can confidently state that finite-dimensional
linear problems such as the one presented here can always be solved
with this method.}. In the case of our linear system, we have
\begin{equation}
\mathcal{D}=\left(\begin{array}{ccc}
0 & 0 & 0\\
0 & -\mathrm{i}\Omega_{R} & 0\\
0 & 0 & \mathrm{i}\Omega_{R}
\end{array}\right),\qquad\mathcal{S}=\frac{1}{2}\left(\begin{array}{ccc}
1 & \delta-\sqrt{1+\delta^{2}} & \delta+\sqrt{1+\delta^{2}}\\
1 & \delta+\sqrt{1+\delta^{2}} & \delta-\sqrt{1+\delta^{2}}\\
-\delta & 1 & 1
\end{array}\right)\label{DSgenRabi}
\end{equation}
which is easily checked or found with the help of some symbolic mathematics
software. Defining then the projected vectors $\mathbf{c}(t)=\mathcal{S}^{-1}\mathbf{x}(t)$
and $\mathbf{d}(t)=\mathcal{S}^{-1}\mathbf{y}(t)$, the coupled linear
system turns into a decoupled one,
\begin{equation}
\dot{\mathbf{c}}=\mathcal{D}\mathbf{c}+\mathbf{d}(t)\quad\Longrightarrow\quad\dot{c}_{j}=\lambda_{j}c_{j}+d_{j}(t),\label{ProjectedLinearSystem}
\end{equation}
with generic solutions
\begin{equation}
c_{j}(t)=c_{j}(0)e^{\lambda_{j}t}+\int_{0}^{t}dt'e^{\lambda_{j}(t-t')}d_{j}(t'),
\end{equation}
as shown below. Finally, we can retrieve the original variables from
these projections as $\mathbf{x}(t)=\mathcal{S}\mathbf{c}(t)$. In
our simple case, in which $\mathbf{y}=\mathbf{0}$, this can be rewritten
as $\mathbf{x}(t)=\mathcal{S}e^{\mathcal{D}t}\mathcal{S}^{-1}\mathbf{x}(0)$,
which using (\ref{DSgenRabi}) directly leads to the solutions (\ref{GenSolsRabi})
after some algebra.

As a final detail, and since the type of linear equations (\ref{ProjectedLinearSystem})
with forcing will appear many times, let us show how to deal with
them in general. Consider the equation $\dot{c}=\lambda c+d(t)$.
In order to solve it, we make the variable change $z(t)=c(t)e^{-\lambda t}$,
and proceed as $\dot{z}=(\dot{c}-\lambda c)e^{-\lambda t}=d(t)e^{-\lambda t}\Rightarrow z(t)=z(t_{0})+\int_{t_{0}}^{t}dt'e^{-\lambda t'}d(t')$.
Undoing the variable change lead us to the final solution
\begin{equation}
c(t)=c(t_{0})e^{\lambda(t-t_{0})}+\int_{t_{0}}^{t}dt'e^{\lambda(t-t')}d(t').\label{GenLinEq}
\end{equation}

\newpage

\section{Light-matter interaction}

\subsection{Interacting Hamiltonians and the dipole approximation\label{DipoleH}}

In the previous chapters we have quantized light and matter (through
the example of atoms) under the assumption that they are isolated,
non-interacting systems. We introduced interesting quantum states
and dynamics though, but without explaining how these can be obtained
in the laboratory. As we will see, state preparation and the generation
of dynamics require using the interaction between light and matter.
In this chapter we introduce the basic theory that describes light-matter
interaction. We do it through two especially relevant examples that
contain already many generic properties of more complex systems: the
interaction of light with a single atom and with a nonlinear dielectric
material.

Let us start by introducing some general ideas behind the description
of light-matter interactions. When the frequency of the electromagnetic
field is above $10^{20}\text{Hz}$, electron-positron pair production
can start occurring, and interactions must be necessarily described
by relativistic quantum electrodynamics, considering an interacting
(local $U(1)$ gauge) quantum field theory of vector and fermionic
fields \cite{PeskinSchroeder,GreinerQED,GreinerQFT}. Fortunately,
optical frequencies sit around $10^{15}\text{Hz}$, safely away from
such high-energy scenarios. Hence, quantization of the light-matter
interaction can be done in a non-relativistic way. The most rigorous
non-relativistic quantization scheme consists in a procedure similar
to what we did in the previous chapters, but where now the charged
particles forming matter enter the Maxwell equations as sources, while
the electromagnetic field is added through minimal coupling to the
multi-particle Schrödinger equation describing matter \cite{QO2}.
It is easy to understand that such a procedure is still incredibly
complex in most situations.

Therefore, it is interesting to consider simpler approximate approaches,
even if they work only under restricted, but well-specified conditions.
We will use one of such approaches, usually referred to as light-matter
interaction within the \emph{dipole approximation}. This approach
works quite well whenever the interactions can be treated as perturbations,
that is, when they are much smaller than the characteristic energy
of the interacting degrees of freedom. It is also incredibly intuitive,
because it builds on how light-matter interaction is treated classically
in most situations. The approach relies on the following ideas. With
a lot of generality, we assume that matter consists of a collection
of massive charged particles (e.g., electrons and nuclei) in some
equilibrium configuration, possibly charge neutral. To the lowest
order, the effect of an external electromagnetic field consists in
pulling apart positive and negative charges, without ripping them
off from the material. In essence, this means that matter is described
as a collection of electric dipoles that we introduce based on reasonable
and general assumptions. The interaction is then brought about by
the coupling between these dipoles and the electric field of light.
Except for the fact that we will quantize the interaction, this is
actually how the classical description works as well, in particular,
through the so-called Lorentz model \cite{BornWolf,Boyd}, which provides
the macroscopic optical properties of matter starting from a microscopic
model for the electric dipoles.

Being more concrete, let us denote by $\hat{H}_{\text{L}}$ and $\hat{H}_{\text{M}}$
the free Hamiltonians of light and matter in the absence of interaction.
With full generality, consider a continuous model of matter characterized
by an electric polarization density $\mathbf{P}(\mathbf{r},t)$ at
every point $\mathbf{r}$ (electric dipole moment per unit volume).
Note that the discrete case of matter formed by $N$ dipoles $\{\mathbf{d}_{n}(t)\}_{n=1,2,...,N}$
localized at positions $\{\mathbf{r}_{n}\}_{n=1,2,...,N}$, can be
recovered by taking $\mathbf{P}(\mathbf{r},t)=\sum_{n=1}^{N}\delta^{(3)}(\mathbf{r}-\mathbf{r}_{n})\mathbf{d}_{n}$.
Within the dipole approximation, the light-matter system is then described
by a Hamiltonian $\hat{H}=\hat{H}_{\text{L}}+\hat{H}_{\text{M}}+\hat{H}_{\text{LM}}$,
with interaction Hamiltonian
\begin{equation}
\hat{H}_{\text{LM}}=-\int_{\mathbb{R}^{3}}d^{3}\mathbf{r}\hat{\mathbf{E}}(\mathbf{r})\cdot\hat{\mathbf{P}}(\mathbf{r}).\label{HLM-general}
\end{equation}
This Hamiltonian comes from the classical expression of the electromagnetic
energy of a collection of dipoles \cite{Jackson62book,Griffiths99book}.
The operator associated to the electric field is the one we built
in Chapter \ref{Sec:QuantizationEMfield}, see (\ref{Ecavity}). As
for the operator associated to the polarization density, it is found
simply by quantizing the degrees of freedom that are used in the description
of matter. We will see this through specific examples next.

As a final note, let us point out that we could try a description
where matter is quantized, but not the electromagnetic field, which
is known as the \emph{semiclassical limit }of quantum optics. In this
case, we would proceed in the same way, just removing the light's
Hamiltonian $\hat{H}_{\text{L}}$ and substituting the operator $\hat{\mathbf{E}}(\mathbf{r})$
by its classical version $\mathbf{E}(\mathbf{r},t)$ governed by Maxwell
equations (which might incorporate the matter dipoles as sources if
we intend to model the backaction of matter onto the classical field
as well). In this limit, the Hamiltonian simply reads
\begin{equation}
\text{\ensuremath{\hat{H}}}_{\text{semiclass}}=\text{\ensuremath{\hat{H}}}_{\text{M}}-\int_{\mathbb{R}^{3}}d^{3}\mathbf{r}\mathbf{E}(\mathbf{r},t)\cdot\hat{\mathbf{P}}(\mathbf{r}).
\end{equation}

\subsection{Interaction between light and a single atom\label{Sec:LigheAtomInteraction}}

\subsubsection{Single-mode, two-level, and rotating-wave approximations: the Jaynes-Cummings
Hamiltonian}

Let us put the previous discussion to practice with the specific example
of a single atom (the simplest example of a matter system) interacting
with the light of an optical cavity. We know the free Hamiltonians
for both the light field and the atom, see (\ref{Hl}) and (\ref{Ha})\textemdash or
(\ref{HaTLS}) within the two-level approximation. Hence, all that
is left is introducing a dipole model for the atom. In the case of
atoms with a single valence electron such as the ones we discussed
in the previous chapter, this is fairly simple: the core (nucleus
plus the closed-shell electrons) and the valence electron gives us,
respectively, a positive and negative charge separated by their relative
coordinate, which we called just $\mathbf{r}$ in the previous chapter.
Here, however, we will denote this relative coordinate by $\mathbf{r}_{\text{A}}$
to avoid confusion with the coordinate vector $\mathbf{r}$ appearing
in the fields $\hat{\mathbf{E}}(\mathbf{r})$ and $\hat{\mathbf{P}}(\mathbf{r})$.
This type of atoms are then the paradigm of an electric dipole, which
is given by $\mathbf{d}=-e\mathbf{r}_{\text{A}}$, where we have taken
into account that electric dipoles are defined from negative to positive
charges \cite{Griffiths99book,Jackson62book}, while the relative
coordinate $\mathbf{r}_{\text{A}}$ points from the atomic core to
the valence electron. The quantized dipole is obtained by quantizing
the relative coordinate, that is, $\mathbf{\hat{d}}=-e\hat{\mathbf{r}}_{\text{A}}$.
Assuming without loss of generality that the center of mass of the
atom is located in some position $\mathbf{r}_{0}=(0,0,z_{0})$ within
the cavity, so that $\hat{\mathbf{P}}(\mathbf{r})=\hat{\mathbf{d}}\delta^{(3)}(\mathbf{r}-\mathbf{r}_{0})$,
the interaction Hamiltonian (\ref{HLM-general}) is then given by
\begin{equation}
\hat{H}_{\text{LM}}=e\hat{\mathbf{E}}(z_{0})\cdot\hat{\mathbf{r}}_{\text{A}},\label{HLM_SingleAtom}
\end{equation}
where we remind that we only specify the $z$ coordinate in the electric
field because we are using a quasi-1D approximation, as explained
in detailed in Chapter \ref{Sec:QuantizationEMfield}. Note that we
have also made the implicit assumption that the electric field does
not vary within the size of the atom, which is very reasonable since
optical variations (wavelengths) occur on the 100nm scale while atomic
sizes are typically three order of magnitude below, on the 100pm scale.

We can now use the expression of the electric field (\ref{Ecavity}),
together with the generic expansion of the atomic relative coordinate
in the atomic basis $\hat{\mathbf{r}}_{\text{A}}=\sum_{\mathbf{n},\mathbf{n}'}\langle\mathbf{n}|\hat{\mathbf{r}}_{\text{A}}|\mathbf{n}'\rangle|\mathbf{n}\rangle\langle\mathbf{n}'|$,
to write
\begin{equation}
\hat{H}_{\text{LM}}=\sum_{m=1}^{\infty}\sum_{\mathbf{n}\mathbf{n'}}\mathrm{i}e\sqrt{\frac{\hbar\omega_{m}}{\varepsilon_{0}LS}}\sin(k_{m}z_{0})\langle\mathbf{n}|\hat{x}_{A}|\mathbf{n}'\rangle(\hat{a}_{m}-\hat{a}_{m}^{\dagger})|\mathbf{n}\rangle\langle\mathbf{n}'|,
\end{equation}
where $\hat{x}_{A}=\mathbf{e}_{x}\cdot\hat{\mathbf{r}}_{\text{A}}$.
Finally, we assume that all the cavity frequencies are far off-resonant
with all the atomic transitions except for one, denoted by $\omega$,
which is close to resonance with a single atomic transition, for which
we use the two-level notation of the previous chapter. Moreover, for
the sake of simplicity we take the matrix element $\langle g|\hat{x}_{A}|e\rangle$
real\footnote{This choice reflects in the appearance of $\hat{\sigma}_{x}$ in the
interaction Hamiltonian. Taking a different phase of the matrix element
would simply introduce a linear combination of $\hat{\sigma}_{x}$
and $\hat{\sigma}_{y}$ instead, that would complicate the analysis
mathematically, but would not introduce any new physics. } (remember that $\langle g|\hat{x}_{A}|g\rangle=0=\langle e|\hat{x}_{A}|e\rangle$
from parity arguments), arriving to the the Hamiltonian
\begin{equation}
\hat{H}=\hbar\omega\hat{a}^{\dagger}\hat{a}+\frac{\hbar\varepsilon}{2}\hat{\sigma}_{z}+\mathrm{i}\hbar g(\hat{a}-\hat{a}^{\dagger})\hat{\sigma}_{x},
\end{equation}
where we have defined the coupling frequency $g=e\sqrt{\frac{\omega}{\hbar\varepsilon_{0}LS}}\langle g|\hat{x}_{A}|e\rangle\sin(\omega z_{0}/c)$,
assumed to be much smaller than $\varepsilon$ as happens in most
quantum optics experiments and for consistency with the dipole approximation.
This is known as the \emph{quantum Rabi Hamiltonian}. Despite being
one of the simplest Hamiltonians in quantum physics, since it describes
a single optical mode interacting with a single atomic transition,
it wasn't found to be analytically solvable until recently \cite{BraakRabi,QRM-review},
with techniques that definitely go beyond the scope of these notes.

Fortunately, the condition $g\ll\varepsilon$ allows us to make one
further rotating-wave approximation similar to the one introduced
in the previous chapter. In order to see this, we just move to the
interaction picture, defined by the transformation operator $\hat{U}_{\text{c}}(t)=e^{\hat{H}_{\text{c}}t/\mathrm{i}\hbar}$,
with $\hat{H}_{\text{c}}=\hbar\omega\hat{a}^{\dagger}\hat{a}+\hbar\varepsilon\hat{\sigma}_{z}/2$
(see Section \ref{rhoI}, where it is summarized how to move in between
pictures). Using the Baker-Campbell-Haussdorf lemma (\ref{BCHlemma-1}),
we obtain the transformation properties
\begin{equation}
\hat{U}_{\text{c}}^{\dagger}(t)\hat{a}\hat{U}_{\text{c}}(t)=e^{-\mathrm{i}\omega t}\hat{a}\hspace{1em}\text{and}\hspace{1em}\hat{U}_{\text{c}}^{\dagger}(t)\hat{\sigma}\hat{U}_{\text{c}}(t)=e^{-\mathrm{i}\varepsilon t}\hat{\sigma},
\end{equation}
leading to the interaction-picture Hamiltonian
\begin{equation}
\hat{H}_{\text{I}}(t)=\hat{U}_{\text{c}}^{\dagger}(t)\hat{H}\hat{U}_{\text{c}}(t)-\hat{H}_{\text{c}}=\underbrace{\mathrm{i}\mathrm{\hbar}g\left(e^{-\mathrm{i}(\omega-\varepsilon)t}\hat{a}\hat{\sigma}^{\dagger}-e^{\mathrm{i}(\omega-\varepsilon)t}\hat{a}^{\dagger}\hat{\sigma}\right)}_{\hat{H}_{\text{R}}}+\underbrace{\mathrm{i}\mathrm{\hbar}g\left(e^{-\mathrm{i}(\omega+\varepsilon)t}\hat{a}\hat{\sigma}-e^{\mathrm{i}(\omega+\varepsilon)t}\hat{a}^{\dagger}\hat{\sigma}^{\dagger}\right)}_{\hat{H}_{\text{NR}}},
\end{equation}
where we have used the expansion $\hat{\sigma}_{x}=\hat{\sigma}+\hat{\sigma}^{\dagger}$.
The $\hat{H}_{\text{R}}$ term oscillates slowly, since the atom and
the cavity mode are assumed to be nearly resonant. On the other hand,
the $\hat{H}_{\text{NR}}$ term oscillates very fast compared with
the coupling $g$. Hence, according to what we saw in the previous
chapter, we can neglect the latter within the rotating-wave approximation.
Note that since $\hat{H}_{\text{LM}}$ is just a perturbation onto
$\hat{H}_{\text{L}}+\hat{H}_{\text{M}}$, we have an alternative way
of looking at the rotating-wave approximation based on energy-conservation
arguments. In particular, note that $\hat{H}_{\text{L}}+\hat{H}_{\text{M}}$
sets the energy scales of the problem and available transitions, which
are only slightly perturbed by the interaction of order $g$. The
terms in $\hat{H}_{\text{R}}$ induce processes in which the atom
is excited by absorbing a photon (or de-excited by emitting one),
so they preserve the energy as long as the frequency mismatch $\omega-\varepsilon$
is at most on the order of the perturbation $g$. On the contrary,
the terms in $\hat{H}_{\text{NR}}$ induce processes where the atomic
excitation (de-excitation) is accompanied by the creation (annihilation)
of a photon, and will be highly suppressed by energy conservation
requirements as long as $g$ is small compared with $\omega+\varepsilon$,
the energy required to ignite the process.

Coming back to the original picture, we can then write the Hamiltonian
within the rotating-wave approximation as
\begin{equation}
\hat{H}_{\text{JC}}=\hbar\omega\hat{a}^{\dagger}\hat{a}+\frac{\hbar\varepsilon}{2}\hat{\sigma}_{z}+\mathrm{i}\hbar g(\hat{a}\hat{\sigma}^{\dagger}-\hat{a}^{\dagger}\hat{\sigma}),\label{JCH}
\end{equation}
which is known as the \emph{Jaynes-Cummings Hamiltonian}. As we are
about to see, this Hamiltonian is easily diagonalizable and predicts
some interesting quantum phenomena.

\subsubsection{Dressed states}

Before diagonalizing this Hamiltonian, let us remark that the Hilbert
space of the system is the tensor product of the one for a harmonic
oscillator with Fock basis $\{|n\rangle\}_{n=0,1,...}$ and a two-level
system with basis $\{|g\rangle,|e\rangle\}$. We then find a basis
of the composite Hilbert space as $\{|n,g\rangle,|n,e\rangle\}_{n=0,1,...}$,
where we use the notation $|n\rangle\otimes|a\rangle=|n,a\rangle$.
The reason why the Jaynes-Cummings Hamiltonian is so easy to diagonalized
is that it connects the basis states by pairs $|n,g\rangle\leftrightharpoons|n-1,e\rangle$,
and hence, the Hamiltonian is a direct sum of two-dimensional problems.

More rigorously, let us define the operator $\hat{E}=\hat{a}^{\dagger}\hat{a}+\hat{\sigma}^{\dagger}\hat{\sigma}$.
This operator ``counts'' the total number of excitations. In particular,
we have $\hat{E}|n,g\rangle=n|n,g\rangle$ and $\hat{E}|n,e\rangle=(n+1)|n,e\rangle$.
Hence, except for the 0 eigenvalue, corresponding to the ground state
$|0,g\rangle$, all the rest of eigenvalues $n\in\mathbb{N}$ are
doubly degenerate, with subspaces spanned by the vectors $\{|n,g\rangle,|n-1,e\rangle\}$.
It is straightforward to check that this operator commutes with the
Hamiltonian, $[\hat{E},\hat{H}_{\text{JC}}]=0$. Hence, they have
common eigenstates. This means that $|0,g\rangle$ is necessarily
an eigenstate of the Hamiltonian (indeed, with $E_{0}=-\hbar\varepsilon/2$
eigenvalue, remember that the energy origin is in the center of the
atomic transition). On the other hand, the rest of energy eigenstates
must be superpositions of the eigenstates of $\hat{E}$ living in
the same degenerate subspace (otherwise, $\hat{H}_{\text{JC}}$ and
$\hat{E}$ would not have common eigenstates). In order to find these
states, we simply represent $\hat{H}_{\text{JC}}$ in the basis $\{|0,g\rangle,|1,g\rangle,|0,e\rangle,|2,g\rangle,|1,e\rangle,...\}$,
obtaining a matrix that can be written as a direct sum of low-dimensional
matrices (we say it has `box structure')
\begin{equation}
H_{\text{JC}}=\bigoplus_{n=0}^{\infty}H^{(n)}=\left(\begin{array}{cccc}
H^{(0)}\\
 & H^{(1)}\\
 &  & H^{(2)}\\
 &  &  & \ddots
\end{array}\right),
\end{equation}
where $H^{(0)}=\langle0,g|\hat{H}_{\text{JC}}|0,g\rangle=-\hbar\varepsilon/2$,
while for $n\geq1$
\begin{equation}
H^{(n)}=\left(\begin{array}{cc}
\langle n,g|\hat{H}_{\text{JC}}|n,g\rangle & \langle n,g|\hat{H}_{\text{JC}}|n-1,e\rangle\\
\langle n-1,e|\hat{H}_{\text{JC}}|n,g\rangle & \langle n-1,e|\hat{H}_{\text{JC}}|n-1,e\rangle
\end{array}\right)=\hbar\left(\begin{array}{cc}
n\omega-\varepsilon/2 & \mathrm{i}\sqrt{n}g\\
-\mathrm{i}\sqrt{n}g & (n-1)\omega+\varepsilon/2
\end{array}\right).\label{H(n)}
\end{equation}
These are Hermitian $2\times2$ matrices, and we show how to diagonalize
them easily at the end of the section. Applied to our current case,
we find the eigenenergies 
\begin{equation}
E_{\pm}^{(n)}=(n-1/2)\hbar\omega\pm\hbar\Omega_{n}/2,
\end{equation}
where we have defined the \emph{quantum Rabi frequency} $\Omega_{n}=\sqrt{\Delta^{2}+4ng^{2}}$,
whose name will get meaning soon, and the \emph{detuning }$\Delta=\omega-\varepsilon$.
The corresponding eigenvectors can be written as\begin{subequations}\label{DressedStates}
\begin{align}
|\psi_{+}^{(n)}\rangle & =\cos(\theta_{n})|n,g\rangle-\mathrm{i}\sin(\theta_{n})|n-1,e\rangle,\\
|\psi_{-}^{(n)}\rangle & =\sin(\theta_{n})|n,g\rangle+\mathrm{i}\cos(\theta_{n})|n-1,e\rangle,
\end{align}
\end{subequations}where $\theta_{n}=\frac{1}{2}\text{arg}\{\Delta+\mathrm{i}2\sqrt{n}g\}\in[0,\pi/2]$.
Together with the $|0,g\rangle$ state, these so-called \emph{dressed
states} are the energy eigenstates associated to the (rotating-wave)
interaction between a single optical mode and a single atomic transition,
and form a basis of their combined Hilbert space. The expressions
$\cos(2\theta_{n})=\Delta/\sqrt{\Delta^{2}+4ng^{2}}$ and $\sin(2\theta_{n})=2\sqrt{n}g/\sqrt{\Delta^{2}+4ng^{2}}$
will also be useful for some calculations, and are obvious from the
definition of $\theta_{n}$.

\begin{figure}
\includegraphics[width=1\textwidth]{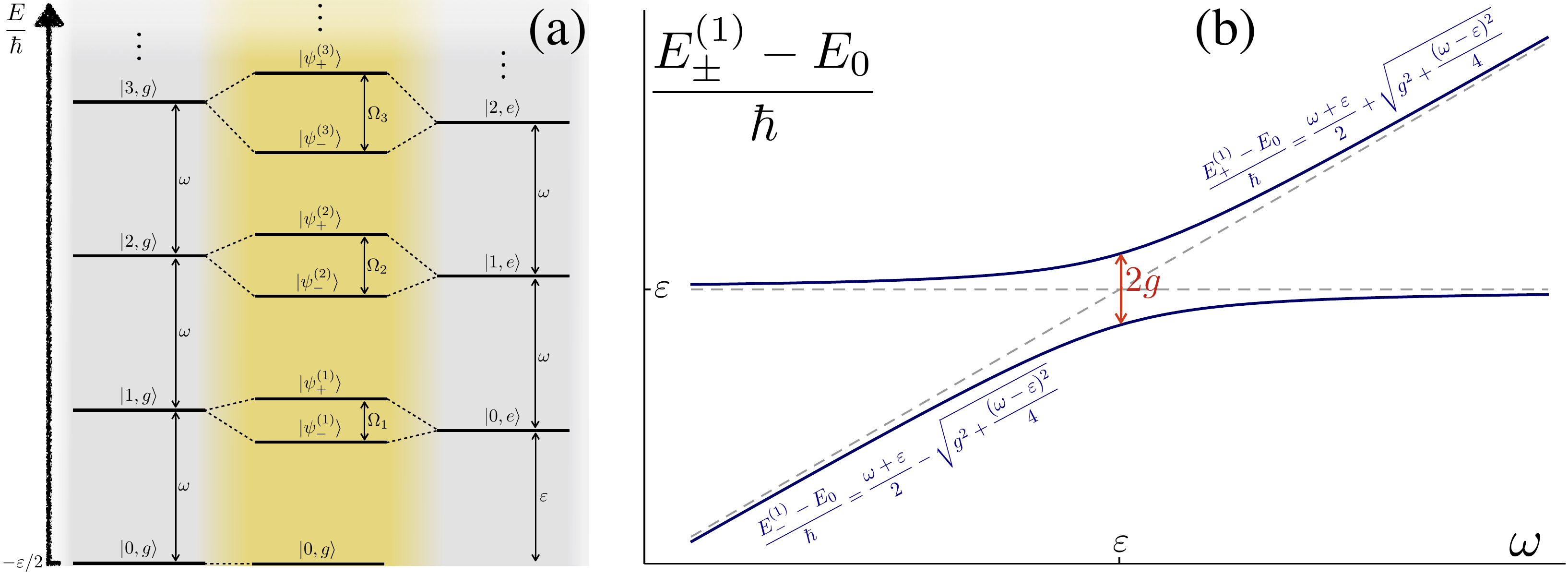}\caption{(a) Representation of the energy spectrum of the Jaynes-Cummings model.
In the absence of coupling, it contains two harmonic ladders shifted
by the atomic transition $\varepsilon$, corresponding to the states
$|n,g\rangle$ and $|n,e\rangle$ (grey levels at both sides in the
figure). Except for the ground state $|0,g\rangle$, the coupling
hybridizes states $|n,g\rangle$ and $|n-1,e\rangle$, generating
the states $|\psi_{\pm}^{(n)}\rangle$, whose energies show a splitting
$\hbar\Omega_{n}$ (yellow levels in the middle). (b) First two transitions
of the Jaynes-Cummings model as a function of the cavity frequency.
For large detunings, the transitions tend to the free ones, $\varepsilon$
and $\omega$, shown as dashed grey lines. The effect of the coupling
is to create an avoided crossing with $2g$ splitting. \label{fJC1}}
\end{figure}

In Fig. \ref{fJC1}a we represent the energy spectrum. In the absence
of coupling ($g=0$) we find two harmonic ladders with uniform level
separation $\hbar\omega$, shifted by $\hbar\varepsilon$ (grey regions
at the right and left sides of the figure). The effect of the coupling
is to hybridize the states with the same number of excitations, $|n,g\rangle$
and $|n-1,e\rangle$, which in the absence of interaction have a splitting
$\hbar|\Delta|$, but after switching on the interaction acquire a
splitting $\hbar\Omega_{n}$ (yellow region in the center of the figure).

It is also interesting to analyze the level splitting as a function
of the detuning. In order to do this, we make the following thought
experiment: for a fixed atomic frequency $\varepsilon$ we vary the
cavity frequency $\omega$ and measure the first two energy transitions
$E_{\pm}^{(1)}-E_{0}=(\omega+\varepsilon)/2\pm\sqrt{g^{2}+(\omega-\varepsilon)^{2}/4}$.
The result is shown in Fig. \ref{fJC1}b. In the absence of coupling,
these transitions are simply given by $\omega$ and $\varepsilon$,
and hence, they cross at $\omega=\varepsilon$. The coupling turns
this crossing into an \emph{avoided crossing} with splitting $2g$.
The observation of such an avoided crossing is quite general for many
types of interactions, and is one of the usual experimental methods
by which couplings are measured in quantum optics.

Let us now, for completeness, show how to find the eigensystem of
a general Hermitian matrix
\begin{equation}
H=\left(\begin{array}{cc}
a & e^{\mathrm{i}\psi}c\\
e^{-\mathrm{i}\psi}c & b
\end{array}\right),
\end{equation}
where all the parameters are real, and we take $c>0$. We denote the
eigenvectors and eigenvalues by $\mathbf{v}_{\pm}$ and $E_{\pm}$,
respectively, so that $H\mathbf{v}_{\pm}=E_{\pm}\mathbf{v}_{\pm}$.
Let us start by noting that the matrix can be decomposed as $H=FSF^{\dagger}$,
where $F$ is diagonal and $S$ a real symmetric matrix, specifically
\begin{equation}
F=\left(\begin{array}{cc}
1 & 0\\
0 & e^{-\mathrm{i}\psi}
\end{array}\right),\quad\text{and}\quad S=\left(\begin{array}{cc}
a & c\\
c & b
\end{array}\right).
\end{equation}
On the other hand, any real symmetric matrix can be diagonalized with
a rotation, and then, $S=R(\theta)DR^{T}(\theta)$, with 
\begin{equation}
R(\theta)=\left(\begin{array}{cc}
\cos\theta & -\sin\theta\\
\sin\theta & \cos\theta
\end{array}\right),\quad\text{and}\quad D=\left(\begin{array}{cc}
E_{+} & 0\\
0 & E_{-}
\end{array}\right),
\end{equation}
where the eigenvalues $E_{\pm}$ are easily found from the characteristic
polynomial of $S$,
\begin{equation}
\det\left\{ \left(\begin{array}{cc}
a-E_{\pm} & c\\
c & b-E_{\pm}
\end{array}\right)\right\} =0\quad\Rightarrow\quad E_{\pm}=\frac{a+b\pm\sqrt{(a-b)^{2}+4c^{2}}}{2},\label{Epm}
\end{equation}
while the angle $\theta$ is found from the condition
\begin{equation}
R^{T}(\theta)SR(\theta)=D\quad\Rightarrow\quad\left(\begin{array}{cc}
a\cos^{2}\theta+b\sin^{2}\theta+c\sin2\theta & \frac{1}{2}(b-a)\sin2\theta+c\cos2\theta\\
\frac{1}{2}(b-a)\sin2\theta+c\cos2\theta & a\sin^{2}\theta+b\cos^{2}\theta-c\sin2\theta
\end{array}\right)=\left(\begin{array}{cc}
E_{+} & 0\\
0 & E_{-}
\end{array}\right),
\end{equation}
whose anti-diagonal terms imply $2c=(a-b)\tan2\theta$, while from
the diagonal ones we get $E_{+}-E_{-}=(a-b)\cos2\theta+2c\sin2\theta=(a-b)/\cos2\theta$,
leading to
\begin{equation}
\cos2\theta=\frac{a-b}{\sqrt{(a-b)^{2}+4c^{2}}},\quad\text{and}\quad\sin2\theta=\frac{2c}{\sqrt{(a-b)^{2}+4c^{2}}}.\label{twotheta}
\end{equation}
These imply that $2\theta\in[0,\pi]$, which allows us to write $2\theta=\arg\{a-b+2\mathrm{i}c\}$.

Combining the diagonalization of $S$ and the initial decomposition
of $H$, we then see that $HU=UD$, where $U=FR(\theta)$ is unitary.
Hence, up to an arbitrary phase, the columns of $U$ are the eigenvectors
of $H$, that is, 
\begin{equation}
U=\left(\begin{array}{cc}
\cos\theta & -\sin\theta\\
e^{-\mathrm{i}\psi}\sin\theta & e^{-\mathrm{i}\psi}\cos\theta
\end{array}\right)=(e^{\mathrm{i}\phi_{+}}\mathbf{v}_{+}\quad e^{\mathrm{i}\phi_{-}}\mathbf{v}_{-}),
\end{equation}
where $\phi_{\pm}$ are arbitrary. For example, by choosing $e^{\mathrm{i}\phi_{\pm}}=\pm1$,
we obtain the eigenvectors
\begin{equation}
\mathbf{v}_{+}=\left(\begin{array}{c}
\cos\theta\\
e^{-\mathrm{i}\psi}\sin\theta
\end{array}\right),\quad\text{and}\quad\mathbf{v}_{-}=\left(\begin{array}{c}
\sin\theta\\
-e^{-\mathrm{i}\psi}\cos\theta
\end{array}\right).\label{vpm}
\end{equation}
Particularizing (\ref{H(n)}), (\ref{twotheta}), and (\ref{vpm})
to $a=n\omega-\varepsilon/2$, $b=(n-1)\omega+\varepsilon/2$, $c=\sqrt{n}g$,
and $\psi=\pi/2$, we obtain the eigensystem of $H^{(n)}$ as we presented
it after (\ref{H(n)}).

\subsubsection{Quantum Rabi oscillations}

As an example of the use of the dressed-state basis, let's consider
the evolution of the system when starting from the atom in the ground
state with the field in an arbitrary pure state, that is,
\begin{equation}
|\psi(0)\rangle=\sum_{n=0}^{\infty}c_{n}|n,g\rangle=c_{0}|0,g\rangle+\sum_{n=1}^{\infty}c_{n}\left[\cos(\theta_{n})|\psi_{+}^{(n)}\rangle+\sin(\theta_{n})|\psi_{-}^{(n)}\rangle\right],
\end{equation}
where in the second equality we have used ($n>0$)\begin{subequations}
\begin{align}
|n,g\rangle & =\cos(\theta_{n})|\psi_{+}^{(n)}\rangle+\sin(\theta_{n})|\psi_{-}^{(n)}\rangle,\\
|n-1,e\rangle & =\mathrm{i}\sin(\theta_{n})|\psi_{+}^{(n)}\rangle-\mathrm{i}\cos(\theta_{n})|\psi_{-}^{(n)}\rangle,
\end{align}
\end{subequations}found by inverting (\ref{DressedStates}). Since
we have written the initial state in terms of eigenstates of the Hamiltonian,
it is straightforward to find the state at any other time,
\begin{equation}
|\psi(t)\rangle=e^{\hat{H}_{\text{JC}}t/\mathrm{i}\hbar}|\psi(0)\rangle=c_{0}e^{\mathrm{i}\varepsilon t/2}|0,g\rangle+\sum_{n=1}^{\infty}c_{n}e^{-\mathrm{i}(n-1/2)\omega t}[\cos(\theta_{n})e^{-\mathrm{i}\Omega_{n}t/2}|\psi_{+}^{(n)}\rangle+\sin(\theta_{n})e^{\mathrm{i}\Omega_{n}t/2}|\psi_{-}^{(n)}\rangle].
\end{equation}

Using this state, we can make predictions for measurements of any
observable at time $t$. For example, consider the probability of
finding the atom in the excited state (irrespective of the optical
state), which can be evaluated as the expectation value of the projector
$\hat{I}\otimes|e\rangle\langle e|$, that is,
\begin{equation}
p_{e}(t)=\langle\psi(t)|\left(\hat{I}\otimes|e\rangle\langle e|\right)|\psi(t)\rangle=\langle\psi(t)|\left(\sum_{n=0}^{\infty}|n,e\rangle\langle n,e|\right)|\psi(t)\rangle=\sum_{n=0}^{\infty}|\langle n,e|\psi(t)\rangle|^{2},
\end{equation}
which is the sum of the probabilities of finding the atom in the excited
state for all the possible photon numbers. Noting that $\langle n,e|\psi_{+}^{(m)}\rangle=-\mathrm{i}\sin(\theta_{m})\delta_{n+1,m}$
and $\langle n,e|\psi_{-}^{(m)}\rangle=\mathrm{i}\cos(\theta_{m})\delta_{n+1,m}$,
so that
\[
\langle n,e|\psi(t)\rangle=c_{n+1}e^{\mathrm{i}(n+1/2)\omega t}\sin(2\theta_{n+1})\sin(\Omega_{n+1}t/2),
\]
we easily obtain
\begin{equation}
p_{e}(t)=\sum_{n=1}^{\infty}|c_{n}|^{2}\sin^{2}(2\theta_{n})\sin^{2}(\Omega_{n}t/2)=\sum_{n=1}^{\infty}|c_{n}|^{2}\frac{1-\cos(\Omega_{n}t)}{2(1+\Delta^{2}/4ng^{2})},\label{PeGen}
\end{equation}
where we have used $\sin^{2}\phi=(1-\cos2\phi)/2$ and the expression
$\sin(2\theta_{n})=2\sqrt{n}g/\Omega_{n}$ that we found in the previous
section.

As a simple example, consider first that the initial state is the
absolute ground state $|0,g\rangle$, so that $c_{n}=\delta_{n0}$.
In such case, we obtain $p_{e}(t)=0$ as expected. On the other hand,
consider an initial Fock state $|N\rangle$ for the field, so that
$c_{n}=\delta_{nN}$, obtaining the excited-state population 
\begin{equation}
p_{e}(t)=\frac{1-\cos(\Omega_{N}t)}{2(1+\Delta^{2}/4Ng^{2})}.
\end{equation}
Starting from the ground state, when interacting with a field with
$N$ photons the atom then oscillates between the ground and excited
states at frequency $\Omega_{N}=\sqrt{\Delta^{2}+4Ng^{2}}$, with
a maximum population $p_{e}^{\text{max}}=(1+\Delta^{2}/4Ng^{2})^{-1}$
which is equal to 1 on resonance ($\Delta=0$) but tends to 0 for
large detuning ($|\Delta|\gg4ng^{2}$). This is exactly the same as
the Rabi oscillations that we studied in the previous chapter. In
fact, the expression above is the same as (\ref{RabiOsciPure}), just
with the identification $\Omega=2\sqrt{N}g$. 

\subsubsection{Coherent light: Rabi oscillations (revisited), collapses, and revivals\label{RabiCollapsesRevivals}}

We find ourselves now in a very peculiar situation. In the previous
chapter, we argue that Rabi oscillations occur when the atom interacts
with classical light. Indeed, let us consider the interaction (\ref{HLM_SingleAtom}),
but with a classical monochromatic field 
\begin{equation}
\mathbf{E}(z_{0},t)=\mathbf{e}_{x}\sqrt{\frac{4\hbar\omega\bar{n}}{\varepsilon_{0}LS}}\cos(\omega t)\sin\left(kz_{0}\right),\label{ClassicalE}
\end{equation}
instead of a quantized one, where we have chosen $\cos(\omega t)$
for convenience, and we parametrize the electric field amplitude by
a real number $\sqrt{\bar{n}}$. In the quantum description, this
is equivalent to assuming that the field is in the coherent state
$|-\mathrm{i}\sqrt{\bar{n}}e^{-\mathrm{i}\omega t}\rangle$ with average
photon number $\bar{n}$, that is, $\langle-\mathrm{i}\sqrt{\bar{n}}e^{-\mathrm{i}\omega t}|\hat{\mathbf{E}}(z_{0})|-\mathrm{i}\sqrt{\bar{n}}e^{-\mathrm{i}\omega t}\rangle$.
Note that we are neglecting any back-action of the atom onto this
field, whose amplitude remains unaffected over time. Proceeding in
the same way as we did, we would obtain the semiclassical Rabi Hamiltonian
(\ref{SemiclassRabiH}) introduced in the previous chapter with the
identification $\Omega=2\sqrt{\bar{n}}g$. Since coherent states are
the states that link quantum and classical physics, we would then
expect equation (\ref{PeGen}) to predict Rabi oscillations when the
field starts in a large-amplitude coherent state. However, it seems
highly unlikely that when introducing in that equation the Poisson
distribution $|c_{n}|^{2}=e^{-\bar{n}}\bar{n}{}^{n}/n!$ corresponding
to the coherent state, we will find the simple Rabi oscillations that
we introduced at the end of the previous chapter in (\ref{RabiOsciPure}).
Moreover, in the previous section we have seen that a fully quantum
mechanical model predicts Rabi oscillations when the field is in a
Fock state, which is actually one of the most quantum states one can
think of. Therefore, it seems that we found a bit of a puzzle that
needs to be understood.

\begin{figure}
\includegraphics[width=1\textwidth]{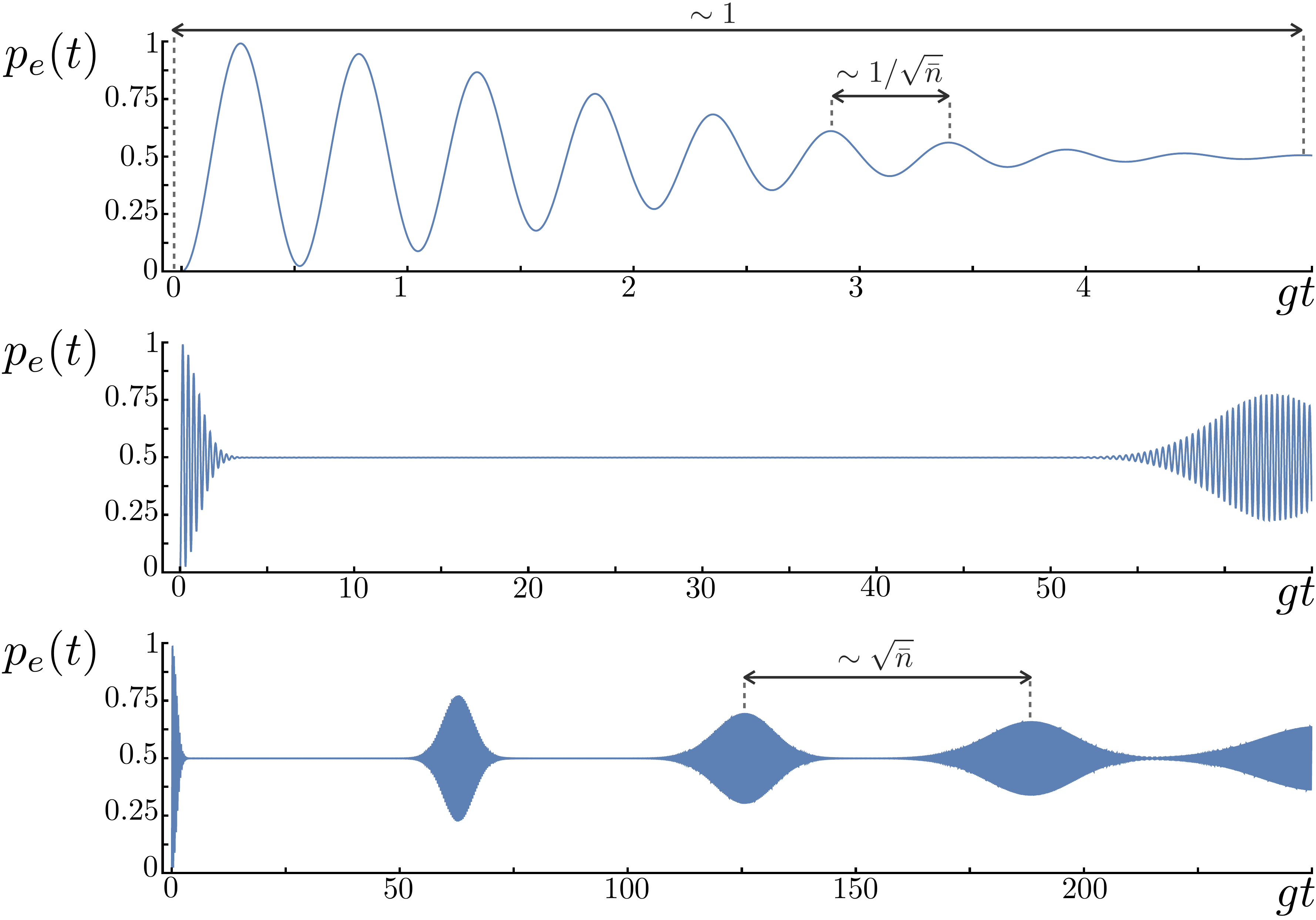}\caption{Excited-state population $p_{e}(t)$ as a function of time for an
initial state $|-\mathrm{i}\sqrt{\bar{n}}\rangle\otimes|g\rangle$
with the field in a coherent state and the atom in the ground state.
We take $\bar{n}=100$ and show the evolution for three different
time domains: $gt\in[0,3]$ (upper figure), $gt\in[0,65]$ (middle
figure), and $gt\in[0,250]$ (lower figure). We also show how the
different relevant time scales scale with the parameters: Rabi oscillations
($gt\sim1/\sqrt{\bar{n}}$), collapses ($gt\sim1$), and revivals
($gt\sim\sqrt{\bar{n}}$). \label{fJC3}}
\end{figure}

In order to do so, let us work in resonance ($\Delta=0$) for simplicity,
so that, with the field in the initial coherent state considered above,
the atomic population reads
\begin{equation}
p_{e}(t)=\sum_{n=0}^{\infty}|c_{n}|^{2}\frac{1-\cos(2\sqrt{n}gt)}{2}=\frac{1}{2}-\frac{1}{2}\sum_{n=0}^{\infty}e^{-\bar{n}}\frac{\bar{n}{}^{n}}{n!}\cos(2\sqrt{n}gt),\label{PeCoherent}
\end{equation}
where we have extended the sum to zero from below for convenience,
since the $n=0$ term does not contribute when $\Delta=0$. In Figure
\ref{fJC3} we plot the population as a function of time with the
help of a computer to perform the sum. We certainly observe oscillations,
but contrary to the ones found in the previous chapter, they are damped
towards $p_{e}=1/2$ and reamplified cyclically. In order to understand
the puzzle exposed above, we need to figure out where these \emph{collapses}
and \emph{revivals} come from.

In order to look for initial clues, let us find some approximate expression
for the sum (\ref{PeCoherent}) that will give us the collapse rate.
We are interested in the large-photon-number limit $\bar{n}\gg1$,
which is when coherent states link quantum and classical physics.
In such case we can take the continuous limit and approximate the
Poisson distribution by a Gaussian of width $\sqrt{\bar{n}}$. Specifically,
we have
\begin{equation}
e^{-\bar{n}}\frac{\bar{n}{}^{n}}{n!}\approx\frac{1}{\sqrt{2\pi\bar{n}}}e^{-(n-\bar{n})^{2}/2\bar{n}}.
\end{equation}
We can then replace the sum by an integral and even extend the lower
integration limit to $-\infty$, since the Gaussian ensures that the
negative values will not contribute effectively. On the other hand,
writing $\cos(2\sqrt{n}gt)=[\exp(2\mathrm{i}\sqrt{n}gt)+\text{c.c.]/2}$
and expanding $\sqrt{n}$ around the mean photon number $\bar{n}$
as $\sqrt{n}\approx\sqrt{\bar{n}}+(n-\bar{n})/2\sqrt{\bar{n}}=\sqrt{\bar{n}}/2+n/2\sqrt{\bar{n}}$,
we get a simple Gaussian integral that can be analytically carried
out\footnote{In general:
\[
\int_{\mathbb{R}}dze^{Bz-z^{2}/2A}=\sqrt{2\pi A}e^{AB^{2}/2},
\]
for any $B>0$ and $A\in\mathbb{C}$.}. In particular, we obtain
\begin{equation}
p_{e}(t)\approx\frac{1}{2}-\frac{1}{4}\left[\int_{-\infty}^{+\infty}dn\frac{1}{\sqrt{2\pi\bar{n}}}e^{-(n-\bar{n})^{2}/2\bar{n}}e^{\mathrm{i}\left(\sqrt{\bar{n}}+n/\sqrt{\bar{n}}\right)gt}+\text{c.c.}\right]=\frac{1}{2}-\frac{1}{2}e^{-g^{2}t^{2}/2}\cos(2g\sqrt{\bar{n}}t).
\end{equation}
This expression contains two interesting features. First, we see that
it provides the expected Rabi frequency and an expression for the
population that matches the one we obtained in the previous chapter
for short times $t\ll\sqrt{2}/g$, see (\ref{RabiOsciPure}). However,
we find that the oscillations are damped at a rate $\gamma_{c}=g/\sqrt{2}$,
which provides the collapse rate. Interestingly, this rate depends
solely on the coupling $g$ and not on the coherent photon number
$\bar{n}$. Now, note that in the classical limit $\bar{n}\gg1$ and
therefore the frequency of the Rabi oscillations $\Omega=2\sqrt{\bar{n}}g$
is much larger than the collapse rate $\gamma_{c}=g/\sqrt{2}$. Hence,
the classical picture studied in the previous chapter is indeed correct
during many Rabi cycles, and quantum effects can only be appreciated
on longer time scales.

Now that we have managed to reconcile our previous semiclassical result
(\ref{RabiOsciPure}) with the fully quantum expression (\ref{PeCoherent}),
let us now discuss the origin of the collapses and revivals. In fact,
they are relatively simple to understand based on expression (\ref{PeCoherent}).
First, note that the Poisson distribution in the sum cuts off essentially
all Fock numbers $n$ outside the interval $\left[\bar{n}-\sqrt{\bar{n}},\bar{n}+\sqrt{\bar{n}}\right]$.
On the other hand, as $\bar{n}$ becomes larger and larger, the quantum
Rabi frequencies $\Omega_{n}=2\sqrt{n}g$ in that interval become
closer and closer. Hence, initially it will look as if all the terms
in the sum oscillate in phase at a similar rate $\Omega_{\bar{n}}=2\sqrt{\bar{n}}g$.
However, as time goes by, the terms will start dephasing more and
more, eventually getting completely out of phase and averaging to
zero. If this intuition is true, then the rate of collapse should
be proportional to the difference between the Rabi frequencies at
the edge of the interval, that is,
\begin{equation}
\gamma_{c}\sim\Omega_{\bar{n}+\sqrt{\bar{n}}}-\Omega_{\bar{n}-\sqrt{\bar{n}}}=2g\left(\sqrt{\bar{n}+\sqrt{\bar{n}}}-\sqrt{\bar{n}-\sqrt{\bar{n}}}\right)=2g\sqrt{\bar{n}}\Biggl(\underset{\approx1+\frac{1}{2\sqrt{\bar{n}}}}{\underbrace{\sqrt{1+\frac{1}{\sqrt{\bar{n}}}}}}-\underset{\approx1-\frac{1}{2\sqrt{\bar{n}}}}{\underbrace{\sqrt{1-\frac{1}{\sqrt{\bar{n}}}}}}\Biggr)\approx2g.
\end{equation}
Indeed, apart from a numerical prefactor, this is in agreement with
the collapse rate $\gamma_{c}=g/\sqrt{2}$ found above.

According to this picture, the revivals can then be interpreted as
the re-phasing of the elements of the sum. We can estimate the times
when the revivals appear by evaluating the times $t_{r}$ for which
neighboring terms re-phase, which in turn can be estimated as
\begin{equation}
\left(\Omega_{\bar{n}+1}-\Omega_{\bar{n}-1}\right)t_{r}=2\pi m,\hspace{1em}\text{with }m\in\mathbb{N}.
\end{equation}
Using the approximation $\sqrt{\bar{n}\pm1}=\sqrt{\bar{n}}\pm1/2\sqrt{\bar{n}}$,
we then obtain
\begin{equation}
t_{r}\approx\frac{\pi\sqrt{\bar{n}}}{g}m,\hspace{1em}\text{with }m\in\mathbb{N}.
\end{equation}
This predicts that the revivals are uniformly spaced in time, which
is exactly what we observe in Fig. \ref{fJC3}. On the other hand,
it predicts that the spacing will increase as the square root of the
coherent photon number, which is easily checked to be true.

\subsection{Light in a nonlinear dielectric\label{Sec:NonlinearDielectric}}

\subsubsection{Dielectric media in the dipole approximation and the refractive index\label{DielectricLinear}}

After studying the interaction of light with a single atom, now we
analyze another paradigmatic example of light-matter interaction in
a completely different regime. In particular, we consider now dielectric
media, which are insulating materials with all their charges bound
(with no free charges they cannot support electrical currents), but
where light can still propagate by polarizing those bound charges
(that is, pulling apart positive and negative charges). Most optical
elements including, for example, beam splitters and mirrors, are made
out of this type of materials. When the coupling between light and
the dielectric is not very strong, we can assume that the electric
polarization density acquired by the medium depends on the applied
electric field as a low-order polynomial,
\begin{equation}
P_{j}(\mathbf{r},t)=\underset{P_{j}^{(1)}(\mathbf{r},t)}{\underbrace{\sum_{k=x,y,z}\varepsilon_{0}\chi_{jk}^{(1)}E_{k}(\mathbf{r},t)}}+\underset{P_{j}^{(2)}(\mathbf{r},t)}{\underbrace{\sum_{k,l=x,y,z}\varepsilon_{0}\chi_{jkl}^{(2)}E_{k}(\mathbf{r},t)E_{l}(\mathbf{r},t)}}+...,\label{NonlinearPolarization}
\end{equation}
so that the medium is treated as a passive system whose information
is all contained in the $\chi_{j_{1}j_{2}...j_{n}}^{(n)}$ coefficients
(assumed larger than zero for all orders for simplicity), called $n^{\text{th}}$-order
susceptibilities. Note that we are further assuming an instantaneous
and homogeneous response of the medium, as the electric polarization
at time $t$ depends only on the electric fields at that time and
the susceptibilities are independent of $\mathbf{r}$.

Let us now discuss how Maxwell equations are modified inside the dielectric
medium. It is a well-known result in electrodynamics \cite{Griffiths99book}
that an electric polarization density $\mathbf{P}(\mathbf{r},t)$
creates a charge density given by $\rho(\mathbf{r},t)=-\boldsymbol{\nabla}\cdot\mathbf{P}(\mathbf{r},t)$.
On the other hand, the continuity equation $\partial_{t}\rho(\mathbf{r},t)=-\boldsymbol{\nabla}\cdot\mathbf{j}(\mathbf{r},t)$
tells us that an effective current $\mathbf{j}(\mathbf{r},t)=\partial_{t}\mathbf{P}(\mathbf{r},t)$
is also induced in the material. These induced charge and current
densities, which are responsible for allowing the electromagnetic
field to propagate in the otherwise isolating medium, must be included
in the inhomogeneous Maxwell equations (\ref{MaxwellInhomoEqs}).
We obtain then the so-called \emph{macroscopic Maxwell equations}
\begin{align}
\boldsymbol{\nabla}\cdot\mathbf{D}(\mathbf{r},t) & =0,\hspace{1em}\hspace{1em}\boldsymbol{\nabla}\times\mathbf{B}(\mathbf{r},t)=\mu_{0}\partial_{t}\mathbf{D}(\mathbf{r},t),\label{MacroscopicMaxwell}\\
\boldsymbol{\nabla}\cdot\mathbf{B}(\mathbf{r},t) & =0,\hspace{1em}\hspace{1em}\boldsymbol{\nabla}\times\mathbf{E}(\mathbf{r},t)=-\partial_{t}\mathbf{B}(\mathbf{r},t),\nonumber 
\end{align}
where we have defined the \emph{displacement field
\begin{equation}
\mathbf{D}(\mathbf{r},t)=\varepsilon_{0}\mathbf{E}(\mathbf{r},t)+\mathbf{P}(\mathbf{r},t).\label{DisplacementField}
\end{equation}
}This expression, together with (\ref{NonlinearPolarization}), is
known as the \emph{constitutive relation} of the dielectric medium.

Let us now neglect higher order terms, and discuss the effect that
the linear term $\mathbf{P}^{(1)}(\mathbf{r},t)$ has on a field propagating
in the dielectric medium. For simplicity, we assume a diagonal susceptibility
that doesn't couple different components of the electric field, that
is
\begin{equation}
\chi_{jk}^{(1)}=\chi_{j}^{(1)}\delta_{jk}.
\end{equation}
Note that this can always be obtained in the experiment by properly
orienting the nonlinear medium, so that its principal axes (the ones
that diagonalize their linear susceptibility matrix, which for crystalline
materials with a cubic lattice simply follow the crystal axes) follow
the directions that we define as $x$, $y$, and $z$. The displacement
field can then be written as
\begin{equation}
D_{j}(\mathbf{r},t)=\underset{\varepsilon_{j}}{\underbrace{\varepsilon_{0}(1+\chi_{j}^{(1)})}}E_{j}(\mathbf{r},t),
\end{equation}
leading to a set of Maxwell equations with the same form as those
in the absence of charges and currents, but with modified electric
permeabilities $\varepsilon_{j}$ along the corresponding coordinate
axis instead of $\varepsilon_{0}$. We discuss now the three most
dramatic effects that this modified electric permeability has: wavelength
and amplitude reduction, as well as the possibility of phase shifts,
see Fig. \textbf{ToDo}. In order to illustrate these, we consider
a monochromatic plane wave with vector potential $\mathbf{A}(z,t)=\mathbf{e}_{j}Ae^{\mathrm{i}(kz-\omega t)}+\mathrm{c.c.}$
and $A\in\mathbb{R}$ for simplicity, propagating from vacuum into
the dielectric, and analyze the changes that it suffers when entering
the medium:
\begin{itemize}
\item Defining the refractive index $n_{j}=\sqrt{1+\chi_{j}^{(2)}}$ along
$\mathbf{e}_{j}$, and defining the Coulomb gauge by the condition
$\sum_{j=x,y,z}\varepsilon_{j}\partial_{j}A_{j}=0$ (which turns into
the usual $\boldsymbol{\nabla}\cdot\mathbf{A}=0$ when $\varepsilon_{x}=\varepsilon_{y}=\varepsilon_{z}$),
the wave equation for the vector potential is now written as
\begin{equation}
\left(\frac{c^{2}}{n_{j}^{2}}\boldsymbol{\nabla}^{2}-\partial_{t}^{2}\right)A_{j}(\mathbf{r},t)=0,
\end{equation}
which shows that the speed of light is reduced in the medium, following
intuition (now the field has to go through the dipoles in order to
advance). Note that this means that when using plane waves to expand
the fields we must be careful to modify accordingly the wave vector\footnote{An important side note is in order. Sometimes there is a bit of confusion
regarding the relation between wavelength and frequency. The frequency
of a generated monochromatic wave, that is, the speed of oscillations
in time, is an intrinsic property of the wave that doesn't change
no matter where it propagates. On the other hand, the wavelength refers
to the periodicity of the wave in space. Clearly, for a fixed temporal
oscillation (with frequency $\nu=\omega/2\pi$), this will depend
on the speed of the wave, say $v$. The wavelength will then be $\lambda=v/\nu$,
and hence, for the monochromatic wave it is larger in vacuum ($v=c$)
than in the dielectric ($v=c/n$).} by a factor $n_{j}$, so that the monochromatic plane wave is modified
as
\begin{equation}
\mathbf{e}_{j}e^{\mathrm{i}(kz-\omega t)}\hspace{1em}\rightarrow\hspace{1em}\mathbf{e}_{j}e^{\mathrm{i}(n_{j}kz-\omega t)},
\end{equation}
where we still define $k=\omega/c$, writing the refractive index
explicitly in the equation.
\item The next effect is more subtle, and has to do with how the amplitude
$A$ is modified when crossing the dielectric interface. In general,
part of the amplitude will be transmitted and part will be reflected,
as specified by the well-known Fresnel relations \cite{BornWolf}.
Therefore, adding a dielectric inside an optical cavity leads in general
to nontrivial interference effects between the various waves being
transmitted and reflected at the dielectric and the mirror surfaces.
For this reason, in experiments one usually uses some anti-reflecting
coating on the surface of the dielectric, so that all the incoming
power enters the dielectric, and interference effects between the
incoming wave and the wave reflected at the dielectric are minimized.
Fortunately, in such case the effect of the dielectric on the amplitude
$A$ can be understood very easily, just by conservation of power
arguments.\\
Let us set the position of the input plane of the dielectric material
at $z=0$ for simplicity, and denote by $z=0^{-}$ and $z=0^{+}$
its left and right sides, where we have vacuum and the dielectric
material, respectively (see Fig. \textbf{ToDo}). The vector potential,
electric and magnetic fields at either side are then\begin{subequations}
\begin{align}
\mathbf{A}(0^{\pm},t) & =\mathbf{e}_{x}A_{\pm}e^{\varphi_{\pm}-\mathrm{i}\omega t}+\mathrm{c.c.},\\
\mathbf{E}(0^{\pm},t) & =-\left.\partial_{t}\mathbf{A}(z,t)\right|_{z=0^{\pm}}=2\mathbf{e}_{x}\omega A_{\pm}\sin(\omega t-\varphi_{\pm}),\\
\mathbf{B}(0^{\pm},t) & =\left.\boldsymbol{\nabla}\times\mathbf{A}(z,t)\right|_{z=0^{\pm}}=2\mathbf{e}_{y}n_{\pm}kA_{\pm}\sin(\omega t-\varphi_{\pm}),
\end{align}
\end{subequations}where $A_{-}=A$, $\varphi_{+}=0$, $n_{-}=1$,
and $n_{+}=n_{x}$, while $A_{+}$ and $\varphi_{+}$ (amplitude and
phase of the transmitted wave) are the only parameters left to determine.
In order to do this, we evaluate the instantaneous power by integrating
the absolute value of the Poynting vector $[\mathbf{E}(z,t)\times\mathbf{B}(z,t)]/\mu_{0}$
in the transverse plane. At the either side of the input plane of
the dielectric we then obtain the power
\begin{equation}
P_{\pm}(t)=\int_{\mathbb{R}^{2}}dxdy\frac{|\mathbf{E}(0^{\pm},t)\times\mathbf{B}(0^{\pm},t)|}{\mu_{0}}=\frac{4S\omega^{2}n_{\pm}}{c\mu_{0}}A_{\pm}^{2}\sin^{2}(\omega t-\varphi_{\pm}).
\end{equation}
Then, imposing that power is conserved upon transmission through the
dielectric interface, so that $P_{-}(t)=P_{+}(t)$, we obtain
\begin{equation}
A_{+}=A/\sqrt{n_{x}},\quad\text{and}\quad\varphi_{-}=0\text{ or }\pi.
\end{equation}
Hence, we have been able to determine the amplitude of the transmitted
wave, which is reduced by a factor $\sqrt{n_{x}}$ as anticipated
above. As for the phase, the conservation of power argument fully
determines it up to a $\pi$ indeterminacy, which will not play any
relevant role in our future derivations.
\end{itemize}
Finally, we need to consider how the quantization of the electromagnetic
field that we introduced in Chapter \ref{Sec:QuantizationEMfield}
gets affected when a dielectric medium is introduced in the optical
cavity.
\begin{itemize}
\item In principle, with a dielectric medium inside the cavity, we would
also need to modify the cavity modes and the quantization procedure.
In order to see this, consider for definiteness the situation depicted
in Figure \textbf{ToDo}. The phase picked up during a round trip is
not $2kL$ but $2k(L-l+n_{j}l)$ for a wave polarized along $\mathbf{e}_{j}$,
and hence, the resonant frequencies obtained when asking this phase
to be an integer multiple of $2\pi$ are modified in the presence
of the dielectric. Moreover, the electric contribution to the electromagnetic
energy contained inside the dielectric is given by the volume integral
of $\mathbf{D}(\mathbf{r},t)\cdot\mathbf{E}(\mathbf{r},t)$ not $\varepsilon_{0}\mathbf{E}^{2}(\mathbf{r},t)$
as in (\ref{Eem}). However, in order to keep things as simple as
possible, we will assume that the dielectric is much smaller than
the cavity length ($l\ll L$), so we can neglect these contributions
to the cavity modes and frequencies, as well as to the electromagnetic
energy.
\end{itemize}
With these considerations in mind, in the following we will write
the cavity electric field inside the dielectric as $\hat{\mathbf{E}}(z,t)=\hat{\mathbf{E}}^{(+)}(z,t)+\hat{\mathbf{E}}^{(-)}(z,t)$,
with
\begin{equation}
\hat{\mathbf{E}}^{(+)}(z,t)=\mathrm{i}\sum_{j=x,y}\mathbf{e}_{j}\sum_{m=1}^{\infty}\sqrt{\frac{\hbar\omega_{m}}{\varepsilon_{0}n_{j}LS}}\hat{a}_{m,j}(t)\sin\left(n_{j}k_{m}z+\varphi_{m,j}\right),\label{EfieldDielectricCavity}
\end{equation}
where for future purposes we allow the electric field to be polarized
either along the $\mathbf{e}_{x}$ or $\mathbf{e}_{y}$ directions,
which see different refractive indices in general. Moreover, we allow
different modes to have different phases $\varphi_{m}$ within the
dielectric, which is what naturally happens in an optical cavity owed
to many other elements that conform it (lenses, phase plates, polarizers,
etc...). In general, different cavity geometries would lead to different
mode profiles inside the crystal, and this is our way to include in
a simple manner such effects. Indeed, we'll see that optimizing the
geometry is crucial in order to obtain the desired effects.

\subsubsection{Basic second-order processes: frequency conversion}

Before proceeding with the quantum description of nonlinear dielectrics,
it is convenient to understand at least conceptually, the classical
phenomena they lead to. We will now discuss this on the basis of a
wave equation for the electric field that we can easily find from
the macroscopic Maxwell equations (working with potentials in nonlinear
dielectrics is now more complicated, and an equation for the electric
field will suffice for our purposes). In order to find such equation,
we take the curl of the homogeneous Maxwell equation $\boldsymbol{\nabla}\times\mathbf{E}(\mathbf{r},t)=-\partial_{t}\mathbf{B}(\mathbf{r},t)$,
and use the identity $\boldsymbol{\nabla}\times\boldsymbol{\nabla}\times\mathbf{E}=\boldsymbol{\nabla}(\boldsymbol{\nabla}\cdot\mathbf{E})-\boldsymbol{\nabla}^{2}\mathbf{E}$,
together with the macroscopic Maxwell equations (\ref{MacroscopicMaxwell})
and the form $D_{j}(\mathbf{r},t)=\varepsilon_{j}E_{j}(\mathbf{r},t)+P_{j}^{(2)}(\mathbf{r},t)$
of the displacement field (we consider effects up to second order
for now). Using a quasi-1D approximation where the fields propagate
along the $z$ axis with transverse polarization and no dependence
on the transverse variables, we finally arrive to
\begin{equation}
\partial_{z}^{2}E_{j}(z,t)-\mu_{0}\varepsilon_{j}\partial_{t}^{2}E_{j}(z,t)=\mu_{0}\partial_{t}^{2}P_{j}^{(2)}(z,t).\label{NonlinearWaveEq}
\end{equation}
This is just a wave equation for the electric field, in which the
nonlinear polarization acts as a source for electromagnetic waves.
As we are about to see, this gives rise to the phenomenon of frequency
conversion, that is, the possibility of generating light at frequencies
different than the one we feed the dielectric with.

We can see this with a simple example. Consider a monochromatic light
wave at some frequency $\omega_{0}$ entering the dielectric material.
Right at the input plane inside the dielectric material, which here
we take as $z=0$ for convenience (see Fig. \textbf{ToDo}), we write
the corresponding electric field as $\mathbf{E}(0,t)=\mathbf{e}_{x}E_{\omega_{0}}\cos(\omega_{0}t)$.
The quadratic polarization density (\ref{NonlinearPolarization})
takes then the form
\begin{equation}
P_{j}^{(2)}(0,t)=\varepsilon_{0}\chi_{jxx}^{(2)}E_{\omega_{0}}^{2}\cos^{2}(\omega_{0}t)=\frac{1}{2}\varepsilon_{0}\chi_{jxx}^{(2)}E_{\omega_{0}}^{2}[1+\cos(2\omega_{0}t)],
\end{equation}
which, apart from an irrelevant constant term, oscillates at frequency
$2\omega_{0}$. When introduced in the wave equation (\ref{NonlinearWaveEq}),
this polarization will then act as a source for waves at frequency
$2\omega_{0}$, that is, the medium will generate the \emph{second
harmonic} of the original light wave (Figure \textbf{ToDo}).

Consider now an input light wave with two frequencies with different
polarization, say $\mathbf{E}(0,t)=\mathbf{e}_{x}E_{\omega_{1}}\cos(\omega_{1}t)+\mathbf{e}_{y}E_{\omega_{2}}\cos(\omega_{2}t)$
with $\omega_{1}>\omega_{2}$ for definiteness. Furthermore, in order
to simplify things, let us assume $\chi_{jxx}^{(2)}=0=\chi_{jyy}^{(2)}$,
so that $\chi_{jxy}^{(2)}=\chi_{jyx}^{(2)}$ are the only susceptibilities
different than zero. In such case, the quadratic polarization density
reads
\begin{equation}
P_{j}^{(2)}(0,t)=2\varepsilon_{0}\chi_{jxy}^{(2)}E_{\omega_{1}}E_{\omega_{2}}\cos(\omega_{1}t)\cos(\omega_{2}t)=\varepsilon_{0}\chi_{jxy}^{(2)}E_{\omega_{1}}E_{\omega_{2}}[\cos(\omega_{+}t)+\cos(\omega_{-}t)],
\end{equation}
where we have defined the sum and difference frequencies $\omega_{\pm}=\omega_{1}\pm\omega_{2}$.
These terms will then act as sources of new waves at those frequencies,
leading to the processes known as \emph{sum-frequency} and \emph{difference-frequency}
generation. Obviously, second-harmonic generation can be understood
as a particular case of sum-frequency generation with $\omega_{1}=\omega_{2}=\omega_{0}$.
In general, these frequency-conversion processes generated by $\mathbf{P}^{(2)}(\mathbf{r},t)$
are known as \emph{three-wave mixing}, because two waves combine to
generate a third one.

For our quantum purposes, the most interesting three-wave mixing process
is that in which we feed both a frequency and its second harmonic,
that is, $\omega_{1}=2\omega_{0}$ and $\omega_{2}=\omega_{0}$. Let
us focus in particular on the frequency difference term, which reads
\begin{equation}
P_{j}^{(2)}(0,t)=\varepsilon_{0}\chi_{jxy}^{(2)}E_{2\omega_{0}}E_{\omega_{0}}\cos(\omega_{0}t),\label{PolarizationDownConversion}
\end{equation}
and acts then a source for waves at frequency $\omega_{0}$. The most
interesting regime in this case is $E_{\omega_{0}}\ll E_{2\omega_{0}}$.
In such situation, the process induced by this term can be understood
as the transfer of energy from the $2\omega_{0}$ wave to the $\omega_{0}$
wave, which will gain intensity as the field propagates in the medium
(Figure \textbf{ToDo}). Such a process is known as \emph{down-conversion}\footnote{Or \emph{parametric down-conversion}, though the term ``parametric''
has an obscure origin, and we will not use it here (especially because
later we will introduce what we will call the ``parametric approximation'',
and we don't want to confuse these terms).}, which can be understood as the dual of second-harmonic generation.
In this context, it is common to called \emph{pump} to the $2\omega_{0}$
wave. Note that down-conversion cannot start without some light at
the down-converted frequency $\omega_{0}$ already present in the
medium, as for $E_{\omega_{0}}=0$ the relevant term (\ref{PolarizationDownConversion})
of the polarization density vanishes. However, this trigger need not
be a monochromatic field that we feed, but could simply be light generated
by random fluctuations of thermal or even quantum origin. In such
situation, we talk then about \emph{spontaneous }down-conversion (Figure
\textbf{ToDo}).

In the remaining of the chapter we focus on this type of process,
which leads to very interesting quantum properties of the down-converted
light, and squeezing in particular.

\subsubsection{Down-conversion Hamiltonian}

We turn now our attention to the quantum mechanical description of
nonlinear optics by means of the light-matter interaction model within
the dipole approximation introduced in Section \ref{DipoleH}. As
explained above, dielectric media admit an effective description in
terms of an electric polarization density that can be expanded in
powers of the electric field, so that matter does not have dynamical
degrees of freedom. The linear term can be included in the free description
of the electric field as explained in Section \ref{DielectricLinear},
see Eq. (\ref{EfieldDielectricCavity}). Hence, for the interaction
terms we only consider the nonlinear terms of the polarization density,
focusing here on the quadratic one. The Hamiltonian is then written
as $\hat{H}=\hat{H}_{\text{L}}+\hat{H}_{\text{LM}}$, with
\begin{equation}
\hat{H}_{\text{LM}}=-\int_{\text{medium}}d^{3}\mathbf{r}\hat{\mathbf{E}}(\mathbf{r})\cdot\hat{\mathbf{P}}^{(2)}(\mathbf{r}),\label{HLMP2}
\end{equation}
where we have implicitly chosen the Schrödinger picture to write the
expression, as operators do not evolve. In order to favor down-conversion,
we assume that the medium is inside a cavity where all three-wave
mixing processes are far off resonant except for the one involving
two cavity modes at frequencies $\omega_{0}$ and $\omega_{2}$ (close
to $2\omega_{0}$), with orthogonal polarizations and annihilation
operators $\hat{a}$ and $\hat{b}$, respectively. Considering only
these relevant modes, the electric field inside the dielectric medium
can be written within our usual quasi-1D approximation as $\hat{\mathbf{E}}(z)=\hat{\mathbf{E}}_{0}(z)+\hat{\mathbf{E}}_{2}(z)$,
with\begin{subequations}\label{E0E2}
\begin{align}
\hat{\mathbf{E}}_{0}(z) & =\mathrm{i}\mathbf{e}_{y}\sqrt{\frac{\hbar\omega_{0}}{\varepsilon_{0}n_{y}LS}}(\hat{a}-\hat{a}^{\dagger})\sin\left(n_{y}k_{0}z+\varphi_{0}\right),\\
\hat{\mathbf{E}}_{2}(z) & =\mathrm{i}\mathbf{e}_{x}\sqrt{\frac{\hbar\omega_{2}}{\varepsilon_{0}n_{x}LS}}(\hat{b}-\hat{b}^{\dagger})\sin\left(n_{x}k_{2}z+\varphi_{2}\right).\label{E2}
\end{align}
\end{subequations}The free Hamiltonian reads in this case as
\begin{equation}
\hat{H}_{\text{L}}=\hbar\omega_{0}\hat{a}^{\dagger}\hat{a}+\hbar\omega_{2}\hat{b}^{\dagger}\hat{b}.
\end{equation}

Let us now find an expression of the interaction Hamiltonian in terms
of annihilation and creation operators. In order to simplify the calculation,
we assume that the only non-vanishing second-order susceptibility
coefficient is $\chi_{xyy}^{(2)}$ (you can convince yourself easily
that a more general case would lead to the same final form of the
Hamiltonian, just with more effort), so that the quadratic polarization
density is written as
\begin{equation}
\hat{\mathbf{P}}^{(2)}(z)=\mathbf{e}_{x}\varepsilon_{0}\chi_{xyy}^{(2)}\hat{\mathbf{E}}_{0}^{2}(z)=-\mathbf{e}_{x}\chi_{xyy}^{(2)}\frac{\hbar\omega_{0}}{n_{y}LS}\sin^{2}\left(n_{y}k_{0}z+\varphi_{0}\right)\left(\hat{a}-\hat{a}^{\dagger}\right)^{2}.
\end{equation}
Introducing this expression into (\ref{HLMP2}) we obtain
\begin{align}
\hat{H}_{\text{LM}} & =-S\int_{z_{0}-l/2}^{z_{0}+l/2}\hat{\mathbf{E}}_{2}(z)\cdot\hat{\mathbf{P}}^{(2)}(z)\label{HLM_chi2}\\
 & =\mathrm{i}\chi_{xyy}^{(2)}\sqrt{\frac{\hbar^{3}\omega_{2}\omega_{0}^{2}}{\varepsilon_{0}n_{x}n_{y}^{2}L^{3}S}}\underbrace{(\hat{b}-\hat{b}^{\dagger})\left(\hat{a}-\hat{a}^{\dagger}\right)^{2}}_{\hat{h}}\underbrace{\int_{z_{0}-l/2}^{z_{0}+l/2}dz\sin^{2}\left(n_{y}k_{0}z+\varphi_{0}\right)\sin\left(n_{x}k_{2}z+\varphi_{2}\right)}_{J},\nonumber 
\end{align}
where $z_{0}$ is the position of the medium's center in the cavity,
see Figure \textbf{ToDo}, and we have assumed that the medium is as
large as the cavity along the transverse direction, so that $\int_{\text{medium}}dxdy=S$.
This Hamiltonian can be greatly simplified by using energy and momentum
conservation. Let's see how these come about.

First we bring our attention to the operator part of the expression,
which we can write as $\hat{h}=\hat{b}\hat{a}^{\dagger2}-\hat{b}(\hat{a}^{\dagger}\hat{a}+\hat{a}\hat{a}^{\dagger})+\hat{b}\hat{a}^{2}-\text{H.c.}$.
Keeping in mind that $\hat{H}_{\text{LM}}$ is supposed to be just
a perturbation onto $\hat{H}_{\text{L}}$, the energies associated
with creation or annihilation of down-converted and pump photons are
$\hbar\omega_{0}$ and $\hbar\omega_{2}\approx2\hbar\omega_{0}$,
respectively. Hence, while the first term in $\hat{h}$ is consistent
with energy conservation (two down-converted photons appear from the
annihilation of a pump photon), the other two are not, since we annihilate
energies $\hbar(\omega_{2}+\omega_{0})$ and $\hbar(\omega_{2}+2\omega_{0})$,
but create only energies $\hbar\omega_{0}$ and 0, respectively. Hence,
within the rotating-wave approximation we can approximate the operator
part of this interaction Hamiltonian by $\hat{h}\approx\hat{b}\hat{a}^{\dagger2}-\hat{b}^{\dagger}\hat{a}^{2}$.

Let us now consider the integral along $z$, which is easily found
to be\footnote{Just write all the sine functions as complex exponentials, perform
the integrals, and simplify.}
\begin{align}
\frac{4}{l}J= & -\sin\left[\varphi_{2}-2\varphi_{0}+(n_{x}k_{2}-2n_{y}k_{0})z_{0}\right]\text{sinc}\left[(n_{x}k_{2}-2n_{y}k_{0})l/2\right]\label{Jintegral}\\
 & -\sin\left[\varphi_{2}+2\varphi_{0}+(n_{x}k_{2}+2n_{y}k_{0})z_{0}\right]\text{sinc}\left[(n_{x}k_{2}+2n_{y}k_{0})l/2\right]+2\sin\left(\varphi_{2}+n_{x}k_{2}z_{0}\right)\text{sinc}\left(n_{x}k_{2}l/2\right)\nonumber 
\end{align}
where $\text{sinc}\,x=\frac{\sin x}{x}$ is a symmetric function that
starts at 1 for $x=0$ and decays as $1/|x|$ with oscillations. Hence,
we see that the interaction will be non-negligible only as long as
at least one among $(2n_{y}k_{0}\pm n_{x}k_{2})l/2$ or $n_{x}k_{2}l/2$
is small. Reminding that typical optical wave vectors are on the order
of $(100\text{nm})^{-1}$ and the refractive index for dielectric
materials is order 1, while dielectric media in nonlinear experiments
is usually never as small as $100\text{nm}$, we see that the terms
in the second line of (\ref{Jintegral}) will always be negligible.
Only the term in the first line can provide a non-negligible contribution,
but it requires the so-called \emph{phase-matching condition}
\begin{equation}
\phi=(2n_{y}k_{0}-n_{x}k_{2})l/2\ll1,
\end{equation}
where $\phi$ is known as the \emph{phase mismatch}. Note that this
condition can be understood as momentum conservation of the photons
involved in the three-wave mixing process: one pump photon with momentum
$n_{x}\hbar k_{2}$ generates two down-converted photons with momenta
$n_{y}\hbar k_{0}$. Note also that this condition is not only achieved
by adjusting the mode frequencies, but also by a proper tuning of
the refractive indices along the different axes, which can be done
experimentally in many ways (e.g., changing the temperature of the
dielectric medium or applying a static electric field on it). Since
this term is rooted in momentum conservation, it will always appear
no matter the cavity geometry or the optical elements we put around
the dielectric. In contrast, the sine term multiplying it is dependent
on the cavity geometry, as evidenced by the fact that it depends on
the phase factor $\varphi_{2}-2\varphi_{0}$. Hence, this term will
take different forms for different configurations, and one has to
be careful to optimize it. For example, in our case, we simply let
the phases $\varphi_{0}$ and $\varphi_{2}$ be zero, we obtain a
negligible interaction. In the following we will assume that the geometry
is optimized (e.g., $\varphi_{2}-2\varphi_{0}=-\pi/2$ in our setting)
and that perfect phase matching has been obtained ($\phi=0$), so
that the integral can be approximated by $J\approx l/4$.

Taking into account all these considerations, we write the final Hamiltonian
as
\begin{equation}
\hat{H}_{\text{DC}}=\hbar\omega_{0}\hat{a}^{\dagger}\hat{a}+\hbar\omega_{2}\hat{b}^{\dagger}\hat{b}+\mathrm{i\hbar}\frac{g_{0}}{2}(\hat{b}\hat{a}^{\dagger2}-\hat{b}^{\dagger}\hat{a}^{2}),\label{HDC}
\end{equation}
where we have defined the \emph{down-conversion rate}
\begin{equation}
g_{0}=\chi_{xyy}^{(2)}\frac{l}{2}\sqrt{\frac{\hbar\omega_{2}\omega_{0}^{2}}{\varepsilon_{0}n_{x}n_{y}^{2}L^{3}S}},
\end{equation}
which is typically several orders of magnitude below optical frequencies.

Despite its seemingly simple form, this Hamiltonian cannot be diagonalized
analytically. This is indeed a common feature of bosonic Hamiltonians
with terms beyond quadratic in the annihilation and creation operators,
except in very specific cases. However, in the following we perform
a simple approximation that will turn the Hamiltonian into a quadratic
one that we can treat analytically.

\subsubsection{Parametric approximation: Bogoliubov physics, squeezing generation,
and bosonic instabilities.}

When we are interested in the situation depicted in Figure \textbf{ToDo},
in which the down-converted field is generated from a pump field of
very large amplitude, we can make one more approximation: the so-called
\emph{parametric approximation}, in which the pump is treated as a
classical source field that doesn't get depleted. Let us perform now
such an approximation, and analyze the Hamiltonian we are left with.

Similarly to the semiclassical approximation for the light-atom interaction
discussed in Section \ref{RabiCollapsesRevivals}, specifically Eq.
(\ref{ClassicalE}), we can perform the parametric approximation by
replacing the operator $\hat{\mathbf{E}}_{\text{2}}(z)$ in (\ref{HLM_chi2})
by the classical field
\begin{equation}
\mathbf{E}_{2}(z,t)=-\mathbf{e}_{x}\sqrt{\frac{4\hbar\omega_{2}\bar{N}_{2}}{\varepsilon_{0}n_{x}LS}}\cos(\omega_{2}t)\sin\left(n_{x}k_{2}z+\varphi_{2}\right),
\end{equation}
where we again parametrize the field amplitude by the square of the
mean photon number $\bar{N}_{2}$, and we choose the $-\cos(\omega_{2}t)$
oscillations for furutre convenience. Equivalently, we are just assuming
that the pump field is in a coherent state $|\mathrm{i}\sqrt{\bar{N}_{2}}e^{-\mathrm{i}\omega_{2}t}\rangle$,
so that $\mathbf{E}_{2}(z,t)=\langle\mathrm{i}\sqrt{\bar{N}_{2}}e^{-\mathrm{i}\omega_{2}t}|\hat{\mathbf{E}}_{\text{2}}(z)|\mathrm{i}\sqrt{\bar{N}_{2}}e^{-\mathrm{i}\omega_{2}t}\rangle$.
The Hamiltonian can be written in this case as
\begin{equation}
\hat{H}_{\text{PDC}}=\hbar\omega_{0}\hat{a}^{\dagger}\hat{a}+\mathrm{\hbar}g_{0}\sqrt{\bar{N}_{2}}\sin(\omega_{2}t)\left(\hat{a}-\hat{a}^{\dagger}\right)^{2}\approx\hbar\omega_{0}\hat{a}^{\dagger}\hat{a}-\frac{\mathrm{\hbar}g}{2}(e^{-\mathrm{i}\omega_{2}t}\hat{a}^{\dagger2}+e^{\mathrm{i}\omega_{2}t}\hat{a}^{2}),\label{HPDC}
\end{equation}
where in the last step we have performed the rotating-wave approximation
and defined a \emph{dressed down-conversion rate}
\begin{equation}
g=\sqrt{\bar{N}_{2}}g_{0},
\end{equation}
which can be many orders of magnitude above the bare rate $g_{0}$,
but still well below $\omega_{0}$ for typical quantum optics experiments.
Note that the final Hamiltonian (\ref{HPDC}) within the parametric
approximation is equivalent to taking the expectation value of the
full Hamiltonian (\ref{HDC}) in the pump's coherent state $|\mathrm{i}\sqrt{\bar{N}_{2}}e^{-\mathrm{i}\omega_{2}t}\rangle$
(and removing the free energy contribution of the pump, which is just
a constant shift).

In contrast to (\ref{HDC}), this Hamiltonian is quadratic in annihilation
and creation operators, and can be treated analytically. In order
to do so, first we move to a picture where it becomes time independent,
which can be accomplished by using the unitary transformation $\hat{U}_{\text{c}}(t)=e^{\hat{H}_{\text{c}}t/\mathrm{i}\hbar}$,
with $\hat{H}_{\text{c}}=\hbar\omega_{2}\hat{a}^{\dagger}\hat{a}/2$
(see Section \ref{Sec:ChangingPictures}). Using the Baker-Campbell-Haussdorf
lemma (\ref{BCHlemma-1}), it's easy to show that the annihilation
operator is transformed as $\hat{U}_{\text{c}}^{\dagger}\hat{a}\hat{U}_{\text{c}}=e^{-\mathrm{i}\omega_{2}t/2}\hat{a}$,
leading to a so-called \emph{rotating picture}, where the Hamiltonian
reads
\begin{equation}
\hat{H}_{\text{I}}=\hat{U}_{\text{c}}^{\dagger}\hat{H}_{\text{PDC}}\hat{U}_{\text{c}}-\hat{H}_{\text{c}}=\hbar\Delta\hat{a}^{\dagger}\hat{a}-\frac{\mathrm{\hbar}g}{2}(\hat{a}^{\dagger2}+\hat{a}^{2}),\label{HPDCti}
\end{equation}
where we define the \emph{detuning} as $\Delta=\omega_{0}-\omega_{2}/2$.
This Hamiltonian is the simplest non-trivial one that allows us to
introduce the so-called \emph{Bogoliubov-deGennes theory}, which appears
in any branch of physics involving bosons or fermions (e.g., condensed
matter and high-energy physics).

We can understand this Hamiltonian analytically by rewriting it in
terms of a new bosonic operator $\hat{c}=\hat{S}^{\dagger}(r)\hat{a}\hat{S}(r)$,
where $\hat{S}(r)=e^{r(\hat{a}^{2}-\hat{a}^{\dagger2})/2}$ is a squeezing
unitary operator, as defined in Section \ref{Sec:SqueezedStates}.
We will choose the parameter $r$ such that the Hamiltonian takes
a simple form that we are familiar with in terms of the new annihilation
and creation operators, which obviously satisfy canonical commutation
relations, $[\hat{c},\hat{c}^{\dagger}]=1$. We will call \emph{Bogoliubov
mode} to the one associated to these bosonic operators. Using (\ref{SqueezingTransformationBoson}),
we obtain the following relations between the original and Bogoliubov
operators
\begin{equation}
\hat{c}=\hat{a}\cosh r-\hat{a}^{\dagger}\sinh r\qquad\Leftrightarrow\qquad\hat{a}=\hat{c}\cosh r+\hat{c}^{\dagger}\sinh r.\label{aTOcToa}
\end{equation}
This is known as a \emph{Bogoliubov transformation}, although for
our single-mode case, it is nothing but the simple squeezing transformation
that we saw in Section \ref{Sec:SqueezedStates}. The real power of
Bogoliubov transformations is evidenced in multi-mode situations,
and in particular, it generalizes to particle non-conserving situations
the \emph{normal mode theory} that is usually studied in the context
of coupled oscillators.

Inserting (\ref{aTOcToa}) into (\ref{HPDCti}), we can rewrite the
Hamiltonian into the form
\begin{equation}
\hat{H}_{\text{I}}=E_{0}(r)+\hbar\Omega(r)\hat{c}^{\dagger}\hat{c}-\frac{\hbar\kappa(r)}{2}(\hat{c}^{2}+\hat{c}^{\dagger2}),\label{HPDCi_c}
\end{equation}
with parameters\begin{subequations}
\begin{align}
\Omega(r) & =\Delta\cosh2r-g\sinh2r,\\
\kappa(r) & =g\cosh2r-\Delta\sinh2r,\\
E_{0}(r) & =\hbar[\Omega(r)-\Delta]/2.
\end{align}
\end{subequations}This calls for the choice $\tanh2r=g/\Delta$ for
the parameter $r$, such that, in terms of the Bogoliubov mode, the
Hamiltonian takes the simple free-harmonic oscillator form\footnote{Note that $\tanh2r=g/\Delta$ implies that (keep in mind that, by
definition, $\cosh x>0$ $\forall x$, so the sign of $\tanh x$ is
always encoded in $\sinh x$)
\begin{equation}
\sinh2r=\text{sign(\ensuremath{\Delta})}\frac{g}{\sqrt{\Delta^{2}-g^{2}}}\qquad\text{and}\qquad\cosh2r=\frac{|\Delta|}{\sqrt{\Delta^{2}-g^{2}}}.
\end{equation}
From these, we can also relate $\sinh^{2}r=[\cosh(2r)-1]/2$ and $\cosh^{2}r=[\cosh(2r)+1]/2$
with $\Delta$ and $g$. } $\hat{H}_{\text{I}}=E_{0}+\hbar\Omega\hat{c}^{\dagger}\hat{c}$,
with $\Omega=\text{sign}(\Delta)\sqrt{\Delta^{2}-g^{2}}$. We know
the spectrum of this Hamiltonian very well from previous chapters:
$\{E_{0}+\hbar\Omega n\}_{n=0,1,...}$. An interesting property of
the spectrum is that it is lower or upper bounded depending on the
sign of $\Delta$, since that's also the sign of the energy step $\hbar\Omega$.
As for the eigenvectors, they are the Fock states associated to the
annihilation operator $\hat{c}$. It is interesting to understand
the meaning of these eigenstates in relation to the original mode
$\hat{a}$, which is the one that describes the down-converted field
$\hat{\mathbf{E}}_{0}(z)$. Let us denote by $|n\rangle_{c}$ the
eigenstates of $\hat{c}^{\dagger}\hat{c}$, and by $|n\rangle_{a}$
the eigenstates of $\hat{a}^{\dagger}\hat{a}$. It is then easy to
prove that $|n\rangle_{c}=\hat{S}(-r)|n\rangle_{a}$, so that the
eigenstates of the Hamiltonian are squeezed Fock states of the down-converted
field. In order to prove this, we first proceed by proving the relation
for the vacuum state:
\begin{equation}
0=\hat{c}|0\rangle_{c}=\hat{S}^{\dagger}(r)\hat{a}\hat{S}(r)|0\rangle_{c}\quad\Rightarrow\quad\hat{a}\hat{S}(r)|0\rangle_{c}=0\quad\Rightarrow\quad|0\rangle_{a}=\hat{S}(r)|0\rangle_{c},
\end{equation}
which proves the relation for $n=0$. On the other hand,
\[
|n\rangle_{c}=\frac{1}{\sqrt{n!}}\hat{c}^{\dagger n}|0\rangle_{c}=\frac{1}{\sqrt{n!}}\left[\hat{S}^{\dagger}(r)\hat{a}^{\dagger n}\hat{S}(r)\right]\hat{S}(-r)|0\rangle_{a}=\hat{S}^{\dagger}(r)\frac{1}{\sqrt{n!}}\hat{a}^{\dagger n}|0\rangle_{a}=\hat{S}(-r)|n\rangle_{a},
\]
as we wanted to prove. Hence, this shows the strong connection that
exists between down-conversion and the generation of squeezing, as
we shall see in more detail later in this section, and along the next
chapters.

The theory above is neat, but we have omitted a huge implicit assumption:
because $|\tanh x|<1$ for any finite $x$, the theory we developed
above works only when $|\Delta|>g$. In the limit $|\Delta|=g$ the
squeezing parameter tends to infinity, $|r|\rightarrow\infty$. In
fact, writing the annihilation operator as $\hat{a}=(\hat{X}+\mathrm{i}\hat{P})/2$
in terms of the position and momentum quadratures ($[\hat{X},\hat{P}]=2\mathrm{i}$),
in this case the Hamiltonian (\ref{HPDCti}) reads
\begin{equation}
\hat{H}_{\text{I}}=\frac{\hbar g}{4}\left[\text{sign}(\Delta)+1\right]\hat{P}^{2}+\frac{\hbar g}{4}\left[\text{sign}(\Delta)-1\right]\hat{X}^{2}=\left\{ \begin{array}{cc}
\hbar g\hat{P}^{2}/2, & \text{for }\Delta>0\\
-\hbar g\hat{X}^{2}/2, & \text{for }\Delta<0
\end{array}\right.,
\end{equation}
showing that the energy eigenstates are position or momentum eigenstates
depending on the sign of the detuning $\Delta$, with the spectrum
becoming continuous (and lower or upper bounded for $\Delta>0$ or
$\Delta<0$, respectively). 

Past this point, that is, for $|\Delta|<g$, there is no choice of
$r$ in Eq. (\ref{HPDCi_c}) that leads to a diagonalizable Hamiltonian.
In contrast, the best we can do to simplify the Hamiltonian is setting
$\Omega=0$ by choosing $\tanh2r=\Delta/g$, implying\footnote{Note that in this case $\tanh2r=\Delta/g$ implies that
\begin{equation}
\sinh2r=\frac{\Delta}{\sqrt{g^{2}-\Delta^{2}}}\qquad\text{and}\qquad\cosh2r=\frac{g}{\sqrt{g^{2}-\Delta^{2}}}.
\end{equation}
} $\kappa=\sqrt{g^{2}-\Delta^{2}}$. In this case, the Hamiltonian
reads $\hat{H}_{\text{I}}=E_{0}-\hbar\kappa(\hat{c}^{2}-\hat{c}^{\dagger2})/2$.
This Hamiltonian is very special, and, in particular, it cannot be
diagonalized, since it is completely unbounded. In order to see this,
just note that writing $\hat{c}=(\hat{x}+\mathrm{i}\hat{p})/2$, in
terms of the position and momentum quadratures of the Bogoliubov mode
($[\hat{x},\hat{p}]=2\mathrm{i}$), we obtain the Hamiltonian $\hat{H}_{\text{I}}=E_{0}+\hbar\kappa(\hat{p}^{2}-\hat{x}^{2})/4$.
This corresponds to a particle moving in an inverted parabolic potential,
which is obviously an unbounded problem, lacking normalizable stationary
eigenstates (not even in the Dirac continuous sense) and real eigenvalues.
In particular, the time evolution induced by this Hamiltonian will
simply exponentially push any initial state towards states with larger
and larger number of excitations, as we are about to see now through
an example.

It is then common to denote the $|\Delta|>g$ regime as the \emph{stable
phase,} and the $|\Delta|<g$ regime as the \emph{unstable phase}.
Hence, the simple down-conversion Hamiltonian (\ref{HPDCti}) leads
to a wide variety of behaviors or phases depending on the ratio $\Delta/g$,
see Figure \textbf{ToDo}: standard harmonic spectrum (uniformly-spaced,
lower-bounded) for $\Delta>g$, standard free-particle spectrum (continuous,
lower-bounded) for $\Delta=g$, instability for $g>\Delta>-g$, continuous
upper-bounded spectrum for $\Delta=-g$, and upper-bounded uniformly-spaced
spectrum for $\Delta<-g$.

In order to get a grip of the meaning and interpretation of the instability,
let us consider now the evolution of the number of excitations $\bar{N}=\langle\hat{a}^{\dagger}\hat{a}\rangle$
when we start from the vacuum state $|0\rangle_{a}$. Note that the
number operator is invariant under the change of picture, that is,
$\hat{U}_{\text{c}}^{\dagger}(t)\hat{a}^{\dagger}\hat{a}\hat{U}_{\text{c}}(t)=\hat{a}^{\dagger}\hat{a}$,
while in the rotating picture the state evolves as $|\psi(t)\rangle_{\text{I}}=\hat{U}_{\text{I}}(t)|0\rangle_{a}$,
being $\hat{U}_{\text{I}}(t)=\exp(\hat{H}_{\text{I}}t/\mathrm{i}\hbar)$
the time-evolution operator in the rotating picture, and where we
have also used $|\psi(0)\rangle_{\text{I}}=|\psi(0)\rangle$. Hence,
we can evaluate the evolution of the number of excitations as
\begin{equation}
\bar{N}(t)=\left._{a}\right\langle 0|\hat{U}_{\text{I}}^{\dagger}(t)\hat{a}^{\dagger}\hat{a}\hat{U}_{\text{I}}(t)|0\rangle_{a}=\left|\hat{U}_{\text{I}}^{\dagger}(t)\hat{a}\hat{U}_{\text{I}}(t)|0\rangle_{a}\right|^{2}.
\end{equation}
In other words, we just need to time-evolve the annihilation operator,
apply it to vacuum, and evaluate the norm of the resulting state.
Let's do this for the stable and unstable phases. In both cases the
trick to evaluate the quantity above is the same: since the Hamiltonian
takes a simple form in terms of $\hat{c}$, we use (\ref{aTOcToa})
to write $\hat{a}$ in terms of $\hat{c}$ and $\hat{c}^{\dagger}$,
time-evolve these, and then use (\ref{aTOcToa}) to bring the expression
back to the original bosonic operators (and we only need to keep the
$\hat{a}^{\dagger}$ term, since $\hat{a}$ annihilates $|0\rangle_{a}$).
In the stable case, $|\Delta|>g$, and taking into account that $\hat{U}_{\text{I}}(t)=\exp(E_{0}t/\mathrm{i}\hbar)\exp(-\mathrm{i}\Omega t\hat{c}^{\dagger}\hat{c})$,
such that $\hat{U}_{\text{I}}^{\dagger}(t)\hat{c}\hat{U}_{\text{I}}(t)=e^{-\mathrm{i}\Omega t}\hat{c}$,
we easily get
\begin{align}
\hat{U}_{\text{I}}^{\dagger}(t)\hat{a}\hat{U}_{\text{I}}(t) & =\left(e^{-\mathrm{i}\Omega t}\hat{c}\cosh r+e^{\mathrm{i}\Omega t}\hat{c}^{\dagger}\sinh r\right)=\left(e^{\mathrm{i}\Omega t}-e^{-\mathrm{i}\Omega t}\right)\hat{a}^{\dagger}\cosh r\sinh r\;\text{+ \ensuremath{\hat{a}}-terms},\\
 & =\mathrm{i}\sinh(2r)\sin(\Omega t)\hat{a}^{\dagger}\;\text{+ \ensuremath{\hat{a}}-terms}
\end{align}
leading to
\begin{equation}
\bar{N}(t)=\left|\mathrm{i}\sinh(2r)\sin(\Omega t)|1\rangle_{a}\right|^{2}=\frac{g^{2}}{\Delta^{2}-g^{2}}\sin^{2}(\Omega t).
\end{equation}
We plot this in Fig. \textbf{ToDo}. Note that the number of excitations
starts at zero, and proceeds with bounded periodic oscillations reaching
maxima at $g^{2}/(\Delta^{2}-g^{2})$.

On the other hand, in the unstable phase $|\Delta|<g$ the time evolution
operator $\hat{U}_{\text{I}}(t)=\exp(E_{0}t/\mathrm{i}\hbar)\exp[\mathrm{i}\kappa t(\hat{c}^{\dagger2}+\hat{c}^{2})/2]$
is equivalent to a squeezing operator, such that $\hat{U}_{\text{I}}^{\dagger}(t)\hat{c}\hat{U}_{\text{I}}(t)=\hat{c}\cosh(\kappa t)+\mathrm{i}\hat{c}^{\dagger}\sinh(\kappa t)$
according to (\ref{SqueezingTransformationBoson}), leading to
\begin{align}
\hat{U}_{\text{I}}^{\dagger}(t)\hat{a}\hat{U}_{\text{I}}(t) & =\hat{U}_{\text{I}}^{\dagger}(t)\hat{c}\hat{U}_{\text{I}}(t)\cosh r+\hat{U}_{\text{I}}^{\dagger}(t)\hat{c}^{\dagger}\hat{U}_{\text{I}}(t)\sinh r\\
 & =\left[\cosh(\kappa t)\cosh r-\mathrm{i}\sinh(\kappa t)\sinh r\right]\hat{c}+\left[\cosh(\kappa t)\sinh r+\mathrm{i}\sinh(\kappa t)\cosh r\right]\hat{c}^{\dagger}\\
 & =\mathrm{i}\cosh(2r)\sinh(\kappa t)\hat{a}^{\dagger}\;\text{+ \ensuremath{\hat{a}}-terms},
\end{align}
so that
\begin{equation}
\bar{N}(t)=\left|\mathrm{i}\cosh(2r)\sinh(\kappa t)|1\rangle_{a}\right|^{2}=\frac{g^{2}}{g^{2}-\Delta^{2}}\sinh^{2}(\kappa t).
\end{equation}
As shown in Fig. \textbf{ToDo}, in this case the number of excitations
is not bounded, and simply increases exponentially with time, following
the hyperbolic sine of the expression. This is indeed generic for
any initial state, since the time evolution operator is a squeezing
transformation with a squeezing parameter $\kappa t$ that increases
with time. Hence, we see that in this case there are no stationary
eigenstates.

At the physical level, the most interesting message that we can get
from these results is that down-conversion can extract energy from
the pump mode to amplify the quantum vacuum fluctuations of the signal
mode, populating it in the form of a squeezed state.

\newpage{}

\section{Quantum optics in open systems\label{Sec:OpenSystems}}

In the previous chapters we have dealt with \emph{closed} quantum
systems, that is, systems that are fully described by a Hamiltonian.
However, most physical systems are \emph{open}, in the sense that
they interact with other systems to which experimentalists do not
have access to. The paradigmatic example is \emph{dissipation}, by
which the system under study leaks energy to a substrate or \emph{environment}
which is too complex to be monitored in experiments. In this section
we will learn how to model quantum systems subject to such type of
situation. In particular, we will learn that there exists a description
of the system's dynamics involving only system operators and states,
with the environmental degrees of freedom effectively integrated out.
We will introduce this description through two paradigmatic examples
of outmost quantum-optical relevance: an optical cavity with a partially-transmitting
mirror and an atom in free space. We start with the former, introducing
a model for the system (cavity), the environment (external electromagnetic
modes), and their interaction (photons tunneling in and out of the
cavity through the mirror). Then we proceed to integrate out the environmental
degrees of freedom either in the Heisenberg or the Schrödinger pictures.
We will see that the system's dynamics will no longer be ruled by
the Heisenberg or von Neumann equations, but by some generalization
of these called, respectively, \emph{quantum Langevin }and \emph{master
equations.}

\subsection{Open optical cavities}

In Chapter \ref{Sec:QuantizationEMfield} we study the quantum description
of the electromagnetic field inside an optical cavity with perfectly
reflecting mirrors. However, in reality optical cavities must have
at least one partially transmitting mirror allowing us to both inject
light inside it and observe or use the light that comes out of it.
In this section we explain how to deal with such an open cavity within
the quantum formalism.

\subsubsection{The open cavity model\label{Sec:OpenCavitymodel}}

For simplicity, we consider only one cavity mode with frequency $\omega_{\mathrm{c}}$,
whose annihilation operator we denote by $\hat{a}$. In addition,
we assume that there are no intracavity processes, just free evolution
(e.g., no atoms or nonlinear dielectric media in the cavity). We will
discuss the generalization to many modes and nontrivial intracavity
dynamics later.

We propose a model in which the cavity mode is coupled through the
partially transmitting mirror to the external modes. These external
modes can be modeled as the modes of a second cavity which shares
the partially transmitting mirror with the main cavity, but has the
second mirror placed at infinity. According to the previous chapter,
the field corresponding to such cavity can be written as (note that
we still take the origin of the $z$ axis in the perfectly reflecting
mirror of the main cavity)
\begin{equation}
\mathbf{\hat{A}}_{\text{ext}}\left(z\right)=\lim_{L_{\text{ext}}\rightarrow\infty}\mathbf{e}_{x}\sum_{m=1}^{\infty}\sqrt{\frac{\hbar}{\varepsilon_{0}L_{\text{ext}}S\omega_{m}}}(\hat{b}_{m}+\hat{b}_{m}^{\dagger})\sin\left[\omega_{m}(z-L)/c\right],\hspace{1cm}\text{with }z\in[L,L+L_{\text{ext}}]
\end{equation}
where the boson operators satisfy the commutation relations $[\hat{b}_{m},\hat{b}_{m^{\prime}}^{\dagger}]=\delta_{mm^{\prime}}$,
and $\omega_{m}=(\pi c/L_{\text{ext}})m$. Now, as the length of this
auxiliary cavity goes to infinity, the set of longitudinal modes becomes
infinitely dense in frequency space (the distance between them, $\pi c/L_{\text{ext}}$,
goes to zero), so that the sum over $m$ can be replaced by an integral
over continuous frequencies\footnote{Note that going as low in frequency as $\omega=0$ or as high as $\omega\rightarrow\infty$
makes no physical sense. However, these fine details are irrelevant
within our level of description, since we will see that only frequencies
around the transitions of the cavity Hamiltonian play a non-negligible
role effectively.}:
\begin{equation}
\lim_{L_{\text{ext}}\rightarrow\infty}\sum_{m=1}^{\infty}=\frac{L_{\text{ext}}}{\pi c}\int_{0}^{\infty}d\omega.
\end{equation}
Accordingly, the Kronecker delta, which must preserve the property
$\lim_{L_{\text{ext}}\rightarrow\infty}\sum_{m=1}^{\infty}\delta_{mm'}=1$,
converges to a Dirac delta as
\begin{equation}
\lim_{L_{\text{ext}}\rightarrow\infty}\delta_{mm^{\prime}}=\frac{\pi c}{L_{\text{ext}}}\delta(\omega-\omega^{\prime}).
\end{equation}
It is then convenient to define a continuous set of annihilation operators
\begin{equation}
\hat{b}(\omega)=\sqrt{\frac{L_{\text{ext}}}{\pi c}}\lim_{L_{\text{ext}}\rightarrow\infty}\hat{b}_{m},
\end{equation}
which satisfy continuous canonical commutation relations 
\begin{equation}
[\hat{b}(\omega),\hat{b}(\omega^{\prime})]=0,\qquad[\hat{b}(\omega),\hat{b}^{\dagger}(\omega^{\prime})]=\delta(\omega-\omega^{\prime}).\label{ExternalCommutators}
\end{equation}
Then, we can finally write the vector potential of the field outside
the cavity as
\begin{equation}
\mathbf{\hat{A}}_{\text{ext}}\left(z\right)=\mathbf{e}_{x}\int_{0}^{\infty}d\omega\sqrt{\frac{\hbar}{\pi c\varepsilon_{0}S\omega}}[\hat{b}(\omega)+\hat{b}^{\dagger}(\omega)]\sin\left[\omega(z-L)/c\right].\label{Aext}
\end{equation}
At this point, it is important to remark that the frequency dependence
of the continuous annihilation operators is by no means a Fourier
of the time-dependent operators. It is merely a label for the external
available electromagnetic modes, just like the discrete index $m$
is for the cavity with finite length. In fact, note that we have implicitly
written the above expressions in the Schrödinger picture, since operators
are time independent.

Now that we have a description of the field outside the cavity, we
can proceed to model the evolution of the whole system through a suitable
Hamiltonian. From Chapter \ref{Sec:QuantizationEMfield}, we know
that the free evolution of the cavity mode and the external modes
is ruled by the Hamiltonian $\hat{H}_{\mathrm{cav}}+\hat{H}_{\mathrm{ext}}$
with
\begin{equation}
\hat{H}_{\mathrm{cav}}=\hbar\omega_{\mathrm{c}}\hat{a}^{\dagger}\hat{a}\text{ \ \ \ \ \ and \ \ \ \ \ }\hat{H}_{\mathrm{ext}}=\int_{0}^{\infty}d\omega\hbar\omega\hat{b}^{\dagger}(\omega)\hat{b}(\omega).\label{HcavHext}
\end{equation}
The coupling between the cavity and the external modes can also be
modeled via a simple Hamiltonian, namely 
\begin{equation}
\hat{H}_{\mathrm{int}}=\mathrm{i}\hbar\int_{0}^{\infty}d\omega g(\omega)[\hat{b}^{\dagger}(\omega)\hat{a}-\hat{b}(\omega)\hat{a}^{\dagger}],
\end{equation}
corresponding to the tunneling of photons between the cavities via
the partially transiting mirror, at a rate determined by the parameter
$g(\omega)$, which is allowed to depend on the frequency of the external
modes. This form for interaction Hamiltonian is supported by many
arguments\footnote{Actually, any form of the type $\hbar\int_{0}^{\infty}d\omega g(\omega)[e^{\mathrm{i}\varphi}\hat{b}^{\dagger}(\omega)\hat{a}+e^{-\mathrm{i}\varphi}\hat{a}^{\dagger}\hat{b}(\omega)]$
is perfectly acceptable. The phase $\varphi$ of the interaction is
actually irrelevant for the final results, and it has been conveniently
chosen here as $\pi/2$ for notational simplicity in upcoming derivations. }. Perhaps the most general one consists in noting that mirrors are
judiciously chosen to be absent of nonlinear optical effects (e.g.,
they are typically built of dielectrics on the linear regime), as
otherwise they would not only transmit/reflect light, but also change
its fundamental properties (e.g., through the frequency conversion
effects that we saw in the previous chapter). We then expect the contribution
of the mirror to the Hamiltonian to be quadratic in annihilation and
creation operators. In addition, since the mirror is a passive element
and we will take the interaction as perturbative, so that the energy
scales are approximately set by $\hat{H}_{0}$, terms such as $\hat{b}^{\dagger}(\omega)\hat{a}^{\dagger}$
are not allowed, which would require an energy $\hbar(\omega_{\text{c}}+\omega)$
coming from some kind of source. A more pragmatic argument is that
we know how mirrors and linear dielectric plates (beam splitters)
act on classical laser beams, in particular splitting them into reflected
and transmitted beams with the same spatiotemporal properties as the
original beam (except for the propagation direction), and whose intensities
add up to the intensity of the original beam. Now, since at the quantum
level these classical beams are described by coherent states, the
only Hamiltonian that provides an evolution consistent with that classical
picture has the properties mentioned above.

We can simplify the the model even further. Note that the coupling
$g(\omega)$ depends essentially on the transmissivity of the mirror
at the corresponding frequency, and we are assuming that $g^{2}(\omega)\ll\omega_{\mathrm{c}}$
for optical frequencies, as the mirror is still assumed to have large
reflectivity (otherwise, we cannot even consider a model with two
separate cavities, but need to quantize the electromagnetic field
of the full space with appropriate boundary conditions as a whole).
On the other hand, from previous chapters we expect energy conservation
to imply that only frequencies close to the cavity frequency will
contribute to the interaction (\textit{resonant interaction}), what
allows us to rewrite $g(\omega)=\sqrt{\gamma/\pi}$ in terms of a
frequency-independent constant $\gamma$, as the transmissivity of
typical mirrors is pretty flat\footnote{A comment on a typical misconception is in order: $g(\omega)$ does
not refer to the transmissivity of the whole cavity, but just to the
coupling induced by the mirror between the field to its right and
the field to its left, irrespective of what it has around it. As we
will see, the transmission coefficient of the whole cavity has a Lorentzian
shape as a function of the external probe's frequency. However, this
will come naturally from the model we are developing, and in particular
from a model with a flat $g(\omega)$. Giving a Lorentzian shape to
$g(\omega)$ is a totally wrong move.} as a function of the frequency at least within the interval $[\omega_{\mathrm{c}}-\gamma,\omega_{\mathrm{c}}+\gamma]$.
In addition, it allows us to extend the integration limits to $[-\infty,+\infty]$
to simplify the upcoming integrals, because any fictitious unphysical
negative frequency mode $\hat{b}(-|\omega|)$ will not contribute
to the physics, as they are very off resonant. These approximations
are one version of the more general \emph{Markov approximation} that
we will comment on later, at the end of the chapter. Hence, this allows
us to finally write the interaction as
\begin{equation}
\hat{H}_{\mathrm{int}}=\mathrm{i}\hbar\sqrt{\frac{\gamma}{\pi}}\int_{-\infty}^{+\infty}d\omega[\hat{b}^{\dagger}(\omega)\hat{a}-\hat{b}(\omega)\hat{a}^{\dagger}].\label{HintExtendedNegativeFreqs}
\end{equation}
In Section \ref{Sec:PhysicalParametersDrivingDamping}, we show that
the explicit relation between $\gamma$ and the cavity parameters
is
\begin{equation}
\gamma=\frac{c\mathcal{T}}{4L},\label{gamma_fromT}
\end{equation}
where $\mathcal{T}$ is the (intensity) transmissivity of the mirror,
and we remind that $L$ is the cavity length.

This is the basic model that we will use to `open the cavity'. In
the following we show how to derive reduced evolution equations for
the intracavity mode alone both at the level of operators (Heisenberg
picture) and states (Schrödinger picture), leading, respectively,
to the so-called \emph{quantum Langevin} and \emph{master equations}.

\begin{figure}
\includegraphics[width=0.8\textwidth]{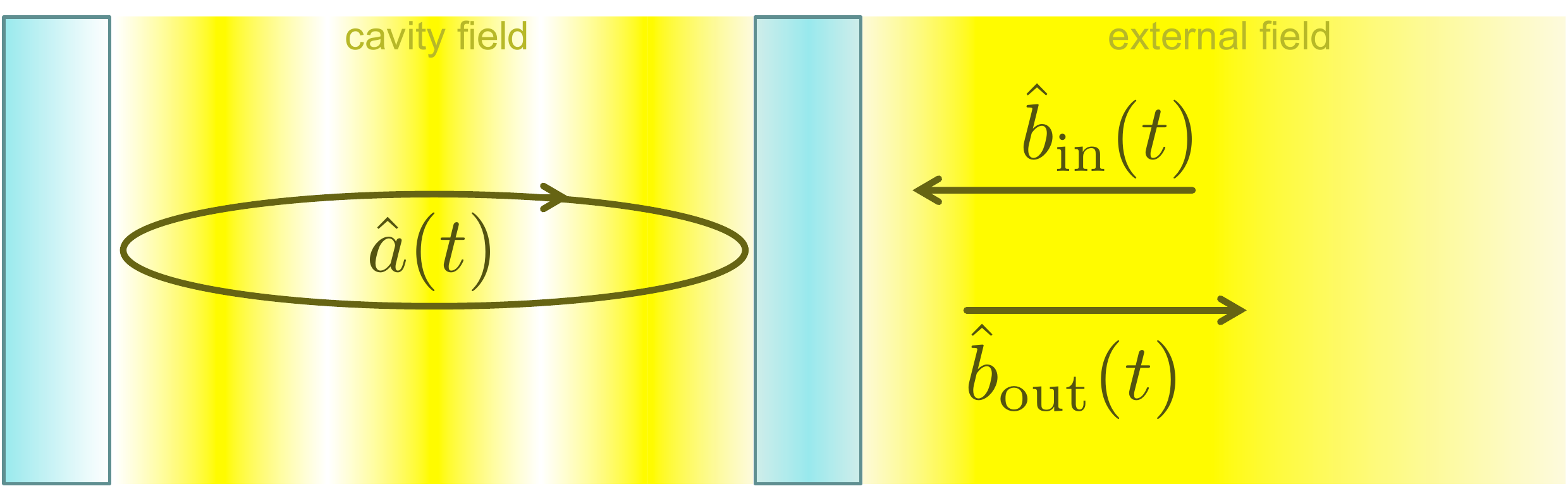}\caption{Schematic view of the open cavity model. Each cavity mode (we show
only the third one) interacts with the external modes through a partially
transmitting mirror. The external modes form a continuum that can
be modeled as a second cavity sharing the partially transmitting mirror
with the main cavity, but with the other mirror placed at infinity
(hence not shown in the figure). As explained in this and the next
chapter, the external modes can be described by an input field driving
the cavity and an output field coming out of the cavity. \label{fOpenCavity}}
\end{figure}

\subsubsection{Heisenberg picture approach: The quantum Langevin equation\label{HeisenbergOpen}}

Given the Hamiltonian written above, $\hat{H}_{0}+\hat{H}_{\text{int }}$,
the Heisenberg equations of motion of the annihilation operators are\begin{subequations}\label{EqsCavityAndExt}
\begin{align}
\partial_{t}\hat{a} & =-\mathrm{i}\omega_{\mathrm{c}}\hat{a}-\sqrt{\frac{\gamma}{\pi}}\int_{-\infty}^{+\infty}d\omega\hat{b}(\omega),\\
\partial_{t}\hat{b}(\omega) & =-\mathrm{i}\omega\hat{b}(\omega)+\sqrt{\frac{\gamma}{\pi}}\hat{a}\text{.}\label{EqsCavityAndExt2}
\end{align}
\end{subequations}We now proceed to reduce these equations to an
evolution equation solely for $\hat{a}$. For this, reminding the
general solution (\ref{GenLinEq}) of a linear differential equation,
we can formally integrate the second equation as 
\begin{equation}
b(\omega;t)=b_{0}(\omega)e^{-\mathrm{i}\omega t}+\sqrt{\frac{\gamma}{\pi}}\int_{0}^{t}dt^{\prime}e^{\mathrm{i}\omega(t^{\prime}-t)}\hat{a}(t^{\prime}),\label{bTOa}
\end{equation}
where $b_{0}(\omega)\equiv b(\omega;0)$ is a shorter notation for
the external annihilation operators at the initial time, which we
take as $t=0$. Introducing this solution into the evolution equation
for $\hat{a}$, we get
\begin{equation}
\partial_{t}\hat{a}=-\mathrm{i}\omega_{\mathrm{c}}\hat{a}-\sqrt{\frac{\gamma}{\pi}}\int_{-\infty}^{+\infty}d\omega b_{0}(\omega)e^{-\mathrm{i}\omega t}-\frac{\gamma}{\pi}\int_{0}^{t}dt^{\prime}\underset{2\pi\delta(t-t')}{\underbrace{\int_{-\infty}^{+\infty}d\omega e^{\mathrm{i}\omega(t^{\prime}-t)}}}\hat{a}(t^{\prime}).
\end{equation}
Next we use the property 
\begin{equation}
\int_{\tau_{-}}^{\tau_{+}}d\tau f(\tau)\delta(\tau-\tau_{\pm})=f(\tau_{\pm})/2,\label{DiracDeltaExtremeInterval}
\end{equation}
 valid for any function $f(\tau)$ that is continuous in all the integration
domain\footnote{A convincing argument in pro of this property comes from considering
the integral $\int_{\tau_{0}}^{\tau_{2}}d\tau f(\tau)\delta(\tau-\tau_{1})=f(\tau_{1})$
for $\tau_{0}<\tau_{1}<\tau_{2}$. Now, we also require the natural
property of integrals $\int_{\tau_{0}}^{\tau_{2}}d\tau f(\tau)\delta(\tau-\tau_{1})=\int_{\tau_{0}}^{\tau_{1}}d\tau f(\tau)\delta(\tau-\tau_{1})+\int_{\tau_{1}}^{\tau_{2}}d\tau f(\tau)\delta(\tau-\tau_{1})$.
On the other hand, if we want this expression to be independent of
the function to be integrated and the integration limits, we must
choose $\int_{\tau_{0}}^{\tau_{1}}d\tau f(\tau)\delta(\tau-\tau_{1})=\int_{\tau_{1}}^{\tau_{2}}d\tau f(\tau)\delta(\tau-\tau_{1})=f(\tau_{1})/2$.
For example, it is enough to consider the particular case $\tau_{0}=-\tau_{2}$
(hence $\tau_{2}>0$), $\tau_{1}=0$, and a symmetric function $f(\tau)=f(-\tau)$;
we then have $\int_{-\tau_{2}}^{0}d\tau f(\tau)\delta(\tau)=\int_{0}^{\tau_{2}}d\tau'f(-\tau')\delta(-\tau')=\int_{0}^{\tau_{2}}d\tau f(\tau)\delta(\tau)$,
in agreement with the choice we made.}, and define the \textit{input operator}
\begin{equation}
\hat{b}_{\mathrm{in}}(t)=-\frac{1}{\sqrt{2\pi}}\int_{-\infty}^{+\infty}d\omega e^{-\mathrm{i}\omega t}\hat{b}_{0}(\omega),\label{bin}
\end{equation}
which is easily shown to satisfy the commutation relations 
\begin{align}
[\hat{b}_{\mathrm{in}}(t),\hat{b}_{\mathrm{in}}(t^{\prime})]=0,\qquad[\hat{b}_{\mathrm{in}}(t),\hat{b}_{\mathrm{in}}^{\dagger}(t^{\prime})] & =\delta(t-t^{\prime}),\label{InCCR}
\end{align}
which correspond to (continuous) canonical commutation relations \emph{in
time}. We then turn the previous equation into 
\begin{equation}
\partial_{t}\hat{a}=-\left(\gamma+\mathrm{i}\omega_{\text{c}}\right)\hat{a}+\sqrt{2\gamma}\hat{b}_{\text{in}}(t),\label{QuantumLangevinBasic}
\end{equation}
which is an evolution equation for the intracavity mode. This equation
is known as the \textit{quantum Langevin equation} for its similarity
with stochastic Langevin equations \cite{QO8} (first order differential
equations forced by noise), the operator $\hat{b}_{\mathrm{in}}(t)$
playing the role of the external noisy force. $\hat{b}_{\mathrm{in}}(t)$
is interpreted as an operator accounting for the input field driving
the cavity at each instant (see Fig. \ref{fOpenCavity}). It is important
to remark that $\hat{b}_{\text{in}}(t)$ is not a dynamical variable,
it is just the Fourier transform of the external annihilation operators
at the origin of time. Hence, it is completely determined by the initial
quantum state of the external modes, as we will see next through some
examples. Additionally, we obtain a damping term $-\gamma\hat{a}$,
much expected since energy is leaking out of the cavity.

In practical terms, the solution for the intracavity operator $\hat{a}(t)$
is found in terms of integrals of $\hat{b}_{\text{in}}(t)$, and since
we know the statistics of this operator (assuming that we know the
initial state of the external modes), this allows us to calculate
the statistics of the intracavity field. We will see this in action
with several examples along the next sections.

It is then interesting to understand the properties of $\hat{b}_{\text{in}}(t)$
for different initial states for the external modes, which we will
denote by $|\psi_{\text{ext}}\rangle$ and $\hat{\rho}_{\text{ext}}$,
for pure and mixed states, respectively:
\begin{itemize}
\item \textbf{Vacuum. }Consider first that they are in the vacuum state
initially, that is $|\psi_{\text{ext}}\rangle=\bigotimes_{\omega}|0\rangle$,
where $\bigotimes_{\omega}$ must be understood as a symbolic notation,
since $\omega$ is a continuous index. We then have\footnote{Note that for Gaussian states, we only need to evaluate first and
second order moments, since higher order moments are completely determined
by these.} $\langle\hat{b}_{0}(\omega)\rangle=\langle\hat{b}_{0}(\omega)\hat{b}_{0}(\omega^{\prime})\rangle=\langle\hat{b}_{0}^{\dagger}(\omega)\hat{b}_{0}(\omega^{\prime})\rangle=0$,
leading to the following statistical properties of the input operator:
\begin{equation}
\langle\hat{b}_{\mathrm{in}}(t)\rangle=\langle\hat{b}_{\mathrm{in}}^{\dagger}(t)\hat{b}_{\mathrm{in}}(t^{\prime})\rangle=0,\text{ \ \ \ \ and \ \ \ \ }\langle\hat{b}_{\mathrm{in}}(t)\hat{b}_{\mathrm{in}}^{\dagger}(t^{\prime})\rangle=\delta(t-t^{\prime}),\label{VacCorr}
\end{equation}
which are actually quite reminiscent of the statistical properties
of complex white Gaussian noises in stochastic Langevin equations
\cite{QO8}.
\item \textbf{Coherent (laser driving). }Let us next consider the injection
of a classical (laser) beam, which means that the external field is
in a coherent state
\[
|\psi_{\text{ext}}\rangle=e^{\int_{-\infty}^{+\infty}d\omega[\alpha(\omega)\hat{b}^{\dagger}(\omega)-\alpha^{*}(\omega)\hat{b}(\omega)]}\bigotimes_{\omega}|0\rangle\equiv\bigotimes_{\omega}|\alpha(\omega)\rangle,
\]
that is, each external mode of frequency $\omega$ is in a coherent
state $|\alpha(\omega)\rangle$. In this case, we obtain
\begin{equation}
\langle\hat{b}_{0}(\omega)\rangle=\alpha(\omega),\hspace{1em}\langle\hat{b}_{0}(\omega)\hat{b}_{0}(\omega^{\prime})\rangle=\alpha(\omega)\alpha(\omega'),\hspace{1em}\text{and}\hspace{1em}\langle\hat{b}_{0}^{\dagger}(\omega)\hat{b}_{0}(\omega^{\prime})\rangle=\alpha^{*}(\omega)\alpha(\omega'),
\end{equation}
so that the input operator satisfies the properties\begin{subequations}
\begin{align}
\langle\hat{b}_{\text{in}}(t)\rangle & =-\frac{1}{\sqrt{2\pi}}\underbrace{\int_{-\infty}^{+\infty}d\omega e^{-\mathrm{i}\omega t}\alpha(\omega)}_{\tilde{\alpha}(t)}=-\frac{\tilde{\alpha}(t)}{\sqrt{2\pi}},\\
\langle\hat{b}_{\text{in}}(t)\hat{b}_{\text{in}}(t')\rangle & =\frac{1}{2\pi}\int_{-\infty}^{+\infty}d\omega e^{-\mathrm{i}\omega t}\int_{-\infty}^{+\infty}d\omega'e^{-\mathrm{i}\omega't)}\underset{\alpha(\omega)\alpha(\omega')}{\underbrace{\langle\hat{b}_{0}(\omega)\hat{b}_{0}(\omega')\rangle}}=\frac{\tilde{\alpha}(t)\tilde{\alpha}(t')}{2\pi}=\langle\hat{b}_{\text{in}}(t)\rangle\langle\hat{b}_{\text{in}}(t')\rangle,\\
\langle\hat{b}_{\text{in}}^{\dagger}(t)\hat{b}_{\text{in}}(t')\rangle & =\frac{1}{2\pi}\int_{-\infty}^{+\infty}d\omega e^{\mathrm{i}\omega t}\int_{-\infty}^{+\infty}d\omega'e^{-\mathrm{i}\omega't'}\underset{\alpha^{*}(\omega)\alpha(\omega')}{\underbrace{\langle\hat{b}_{0}^{\dagger}(\omega)\hat{b}_{0}(\omega')\rangle}}=\frac{\tilde{\alpha}^{*}(t)\tilde{\alpha}(t')}{2\pi}=\langle\hat{b}_{\text{in}}(t)\rangle^{*}\langle\hat{b}_{\text{in}}(t')\rangle.
\end{align}
\end{subequations}In other words, each of the input modes (which
form a continuous set of harmonic oscillators obeying canonical commutation
relations), is in a coherent state $|\alpha_{\text{in}}(t)\rangle$
with an amplitude $\alpha_{\text{in}}(t)=-\tilde{\alpha}(t)/\sqrt{2\pi}$,
which is just proportional to the Fourier transform of the spectral
components $\alpha(\omega)$ of the laser. Hence, the external state
admits an alternative description in the time domain as $|\psi_{\text{ext}}\rangle=\bigotimes_{t}|\alpha_{\text{in}}(t)\rangle$,
with $\hat{b}_{\text{in}}(t)|\alpha_{\text{in}}(t)\rangle=\alpha_{\text{in}}(t)|\alpha_{\text{in}}(t)\rangle$,
and where $\bigotimes_{t}$ is again a symbolic notation, since $t$
is a continuous index.\\
In this case, it is recommendable to define a new input operator
\begin{equation}
\hat{a}_{\mathrm{in}}(t)=\hat{b}_{\mathrm{in}}(t)-\langle\hat{b}_{\mathrm{in}}(t)\rangle,\label{aINdef}
\end{equation}
which using the previous equations is easily shown to satisfy the
vacuum correlators (\ref{VacCorr}). In terms of this operator the
quantum Langevin equation for the intracavity mode reads
\begin{equation}
\partial_{t}\hat{a}=\mathcal{A}(t)-\left(\gamma+\mathrm{i}\omega_{\mathrm{c}}\right)\hat{a}+\sqrt{2\gamma}\hat{a}_{\mathrm{in}}(t),\label{ReducedHeisenberg}
\end{equation}
with $\mathcal{A}=-\sqrt{\gamma/\pi}\tilde{\alpha}(t)$. Hence, the
injection of laser field is equivalent to the addition of the following
time-dependent term in the Hamiltonian
\begin{equation}
\hat{H}_{\mathrm{inj}}(t)=\mathrm{i}\hbar\left(\mathcal{A}(t)a^{\dagger}-\mathcal{A}^{\ast}(t)a\right),\label{Hinj}
\end{equation}
while considering the external modes in vacuum.\\
As a common example, consider a laser containing $L$ frequency components
$\{\omega_{\ell}\}_{\ell=1,2,...,L}$, so that $\alpha(\omega)=\sum_{\ell=1}^{L}\alpha_{\ell}\delta(\omega-\omega_{\ell})$.
This leads to a driving term in the quantum Langevin equation with
\begin{equation}
\mathcal{A}(t)=\sum_{\ell=1}^{L}\mathcal{E}_{\ell}e^{-\mathrm{i}\omega_{\ell}t},\quad\text{with }\mathcal{E}_{\ell}=-\sqrt{\frac{\gamma}{\pi}}\alpha_{\ell}.\label{A_polychromatic}
\end{equation}
The most relevant case is that in which each spectral component is
resolved by the cavity, that is, their spectral separation is larger
than $\gamma$, but still small enough compared with the free spectral
range $\pi c/L$ such that they address the same cavity mode of frequency
$\omega_{\text{c}}$. In such case, we show in Section \ref{Sec:PhysicalParametersDrivingDamping}
the explicit relation of the rates $\mathcal{E}_{\ell}$ and the physical
or experimental parameters is
\begin{equation}
\mathcal{E}_{\ell}=\sqrt{\frac{2\gamma}{\hbar\omega_{\text{c}}}P_{\text{inj},\ell}}e^{\text{i}\phi_{\ell}},\label{El_fromPinj}
\end{equation}
where $P_{\text{inj},\ell}$ and $\phi_{\ell}$ are, respectively,
the power (averaged over an optical cycle) and phase of the $\ell$'th
spectral component.
\item \textbf{Thermal. }Consider now also the case in which the external
field is at thermal equilibrium at temperature $T$, corresponding,
as we saw in Section \ref{Sec:ThermalStates}, to the state 
\begin{equation}
\hat{\rho}_{\text{ext}}=\frac{e^{-\hat{H}_{\text{ext}}/k_{B}T}}{\text{tr}\left\{ e^{-\hat{H}_{\text{ext}}/k_{B}T}\right\} }\equiv\bigotimes_{\omega}\hat{\rho}_{\text{th}}[\bar{n}(\omega)],\label{ExtThermalState}
\end{equation}
where $\hat{\rho}_{\text{th}}(\bar{n})$ is a thermal Gaussian state
for each mode, with an average number of excitations (photons) given
by the Bose-Einstein distribution
\begin{equation}
\bar{n}(\omega)=\frac{1}{\exp(\hbar\omega/k_{B}T)-1}.
\end{equation}
Again, note that the right-hand-side of (\ref{ExtThermalState}) is
a symbolic notation. The first and second moments of the external
field are then
\begin{equation}
\langle\hat{b}_{0}(\omega)\rangle=0=\langle\hat{b}_{0}(\omega)\hat{b}_{0}(\omega')\rangle,\text{ \ \ \ \ and \ \ \ \ }\langle\hat{b}_{0}^{\dagger}(\omega)\hat{b}_{0}(\omega^{\prime})\rangle=\bar{n}(\omega)\delta(\omega-\omega')\approx\bar{n}(\omega_{\text{c}})\delta(\omega-\omega').\label{ThermalStatIni}
\end{equation}
In the last step we have made the approximation $\bar{n}(\omega)\approx\bar{n}(\omega_{\text{c}})\equiv\bar{n}$,
since only a narrow band of frequencies in the interval $[\omega_{\mathrm{c}}-\gamma,\omega_{\mathrm{c}}+\gamma]$
contribute to the problem as argued above, where the Bose-Einstein
distribution can be considered flat to a good approximation. The statistical
properties of the input operators are then enormously simplified,
obtaining
\begin{equation}
\langle\hat{a}_{\mathrm{in}}(t)\rangle=0=\langle\hat{a}_{\mathrm{in}}(t)\hat{a}_{\mathrm{in}}(t')\rangle,\hspace{1em}\langle\hat{a}_{\mathrm{in}}^{\dagger}(t)\hat{a}_{\mathrm{in}}(t^{\prime})\rangle=\bar{n}\delta(t-t'),\hspace{1em}\text{and}\hspace{1em}\langle\hat{a}_{\mathrm{in}}(t)\hat{a}_{\mathrm{in}}^{\dagger}(t^{\prime})\rangle=(\bar{n}+1)\delta(t-t'),\label{ThermalAin}
\end{equation}
where, by using the modified input operator $\hat{a}_{\text{in}}(t)$
of Eq. (\ref{aINdef}), we are implicitly allowing for the injection
of a laser field on top of the thermal excitations.
\end{itemize}
As a final remark, let us extend the model a bit, in particular by
considering the possibility of having nontrivial intracavity processes
given by a Hamiltonian $\hat{H}_{\text{intra}}(t)$. As we saw in
the previous chapter, these might account for the down-conversion
in a nonlinear dielectric or the interaction with a single atom, for
example. Taking a look at the derivation of the quantum Langevin equation,
the only place where the intracavity process could make a difference
is in the assumption that only external modes around the cavity resonance
$\omega_{\text{c}}$ will contribute, since now the energy needed
to annihilate or create an intracavity photon is not $\hbar\omega_{\text{c}}$
but the energy transitions associated to the total intracavity Hamiltonian
$\hat{H}_{\text{cav}}+\hat{H}_{\text{intra}}(t)$, that is, differences
between its eigenvalues. Hence, as long as $\hat{H}_{\text{intra}}(t)$
can be treated as a perturbation with respect to optical energies
$\hbar\omega_{\text{c}}$, we can still simply introduce it in the
quantum Langevin equation without further modifications to a very
good approximation. We will assume that this is the case, so that
the final form of the quantum Langevin equation of a mode in an open
cavity reads
\begin{equation}
\partial_{t}\hat{a}=\mathcal{A}(t)-\left(\gamma+\mathrm{i}\omega_{\mathrm{c}}\right)\hat{a}+\left[\hat{a},\frac{\hat{H}_{\text{intra}}(t)}{\mathrm{i}\hbar}\right]+\sqrt{2\gamma}\hat{a}_{\mathrm{in}}(t),\label{ReducedHeisenbergGeneral}
\end{equation}
where, we further assume an external electromagnetic field at thermal
equilibrium, so the input operator satisfies the statistical properties
(\ref{ThermalAin}).

It is important to note that, in general, the interplay between damping,
driving, and interactions makes the state settle into a well-defined
\emph{asymptotic state} at long times, irrespective of the initial
state. This state is commonly denoted in the literature by \emph{steady}
or \emph{stationary state}, although it's important to keep in mind
that it is time dependent in general.

We will discuss the extension of this description to a multi-mode
cavity coupled to more than one environment in Section \ref{Sec:MultimodeCavity}.
Now, let us consider one example to fix ideas.

\subsubsection{Example in the Heisenberg picture: asymptotic states of an driven
empty cavity\label{CavityExampleSS}}

In order to learn how to use the quantum Langevin equation, let us
consider now the simplest example possible: that of an empty cavity
with no intracavity processes other than free evolution and driven
by a monochromatic laser, that is, $\hat{H}_{\text{intra}}=0$ and
$\mathcal{A}(t)=\mathcal{E}e^{-\mathrm{i}\omega_{\text{L}}t}$. Note
that in this case the evolution equations of the whole ``cavity +
external field'' (\ref{EqsCavityAndExt}) are linear, and hence,
if initially all the fields are in a Gaussian state, the state will
remain Gaussian at all times (a linear combination of Gaussian variables
is Gaussian). This is also clear from the quantum Langevin equation
itself, since a linear equation, with a Gaussian input field, will
lead to Gaussian variables at all times, provided that these are Gaussian
initially. Otherwise, the variables will be Gaussian only asymptotically
(since the asymptotic state is independent of the initial conditions).

The solution of the resulting linear quantum Langevin equation (\ref{ReducedHeisenbergGeneral})
is, according to Eq. (\ref{GenLinEq}),
\begin{equation}
\hat{a}(t)=e^{-(\gamma+\mathrm{i}\omega_{\text{c}})t}\hat{a}(0)+\int_{0}^{t}dt'\left[\mathcal{E}e^{-\text{i}\omega_{\text{L}}t'}+\sqrt{2\gamma}\hat{a}_{\text{in}}(t')\right]e^{-(\gamma+\mathrm{i}\omega_{\text{c}})(t-t')}.
\end{equation}
Making the variable change $t'=t-\tau$ in the integral we obtain
\begin{equation}
\hat{a}(t)=e^{-(\gamma+\mathrm{i}\omega_{\text{c}})t}\hat{a}(0)+\mathcal{E}e^{-\text{i}\omega_{\text{L}}t}\underset{\frac{1-e^{-(\gamma-\mathrm{i}\Delta)t}}{(\gamma-\mathrm{i}\Delta)}}{\underbrace{\int_{0}^{t}d\tau e^{-(\gamma+\mathrm{i}\omega_{\text{c}}-\text{i}\omega_{\text{L}})\tau}}}+\sqrt{2\gamma}\int_{0}^{t}d\tau e^{-(\gamma+\mathrm{i}\omega_{\text{c}})\tau}\hat{a}_{\text{in}}(t-\tau),\label{a(t)_HeisenbergExample}
\end{equation}
where $\Delta=\omega_{\text{L}}-\omega_{\text{c}}$ is the cavity
detuning. For $\gamma t\rightarrow\infty$, two terms vanish, the
one containing the initial condition and the second term of the definite
integral in the middle, obtaining
\begin{equation}
\lim_{\gamma t\rightarrow\infty}\hat{a}(t)=\frac{\mathcal{E}}{\gamma-\mathrm{i}\Delta}e^{-\text{i}\omega_{\text{L}}t}+\sqrt{2\gamma}\lim_{t\rightarrow\infty}\int_{0}^{t}d\tau e^{-(\gamma+\mathrm{i}\omega_{\text{c}})\tau}\hat{a}_{\text{in}}(t-\tau).
\end{equation}
Note that all the information about the initial state of the cavity
mode has disappeared, as mentioned above. In order to understand the
type of asymptotic state that the cavity reaches, we consider now
the first and second order moments of the intracavity mode. In the
case of the first order moment, we get
\begin{equation}
\lim_{\gamma t\rightarrow\infty}\langle\hat{a}(t)\rangle=\frac{\mathcal{E}}{\gamma-\mathrm{i}\Delta}e^{-\text{i}\omega_{\text{L}}t}.
\end{equation}
Defining the fluctuation operator $\delta\hat{a}(t)=\hat{a}(t)-\langle\hat{a}(t)\rangle$,
the second order moments then read\begin{subequations}
\begin{align}
\lim_{\gamma t\rightarrow\infty}\langle\delta\hat{a}^{2}(t)\rangle & =2\gamma\lim_{\gamma t\rightarrow\infty}\int_{0}^{t}d\tau\int_{0}^{t}d\tau'e^{-(\gamma+\mathrm{i}\omega_{\text{c}})(\tau+\tau')}\underset{0}{\underbrace{\langle\hat{a}_{\text{in}}(t-\tau)\hat{a}_{\text{in}}(t-\tau')\rangle}}=0,\\
\lim_{\gamma t\rightarrow\infty}\langle\delta\hat{a}^{\dagger}(t)\delta\hat{a}(t)\rangle & =2\gamma\lim_{\gamma t\rightarrow\infty}\int_{0}^{t}d\tau\int_{0}^{t}d\tau'e^{-\gamma(\tau+\tau')+\mathrm{i}\omega_{\text{c}}(\tau-\tau')}\underset{\bar{n}\delta(\tau-\tau')}{\underbrace{\langle\hat{a}_{\text{in}}^{\dagger}(t-\tau)\hat{a}_{\text{in}}(t-\tau')\rangle}}=2\gamma\bar{n}\lim_{\gamma t\rightarrow\infty}\underset{\frac{1-e^{-2\gamma t}}{2\gamma}}{\underbrace{\int_{0}^{t}d\tau e^{-2\gamma\tau}}}=\bar{n}.
\end{align}
\end{subequations}Using then expression (\ref{GaussianMomentsComplex}),
these moments lead to a Gaussian state with a mean vector with components
given by the real and imaginary parts of $[2\mathcal{E}/(\gamma-\mathrm{i}\Delta)]e^{-\text{i}\omega_{\text{L}}t}$,
and a thermal covariance matrix $V=(2\bar{n}+1)I$, see Figure \textbf{ToDo}.
In other words, we obtain a displaced thermal state, rotating at frequency
$\omega_{\text{L}}$ in phase space.

It is interesting to analyze total photon number, which is given by
\begin{equation}
\lim_{\gamma t\rightarrow\infty}\langle\hat{a}^{\dagger}(t)\hat{a}(t)\rangle=\lim_{\gamma t\rightarrow\infty}\left[\langle\delta\hat{a}^{\dagger}(t)\delta\hat{a}(t)\rangle+\langle\hat{a}(t)\rangle^{*}\langle\hat{a}(t)\rangle\right]=\frac{|\mathcal{E}|^{2}}{\gamma^{2}+\Delta^{2}}+\bar{n},
\end{equation}
which has a classical (or coherent) contribution coming from the laser
field, and a thermal background coming from an equilibration with
the environment at temperature $T$. It is interesting to plot this
quantity as a function of the laser frequency $\omega_{\text{L}}$,
see Figure \textbf{ToDo}. We obtain the characteristic Lorentzian
shape of a cavity, where it can be appreciated that only laser frequencies
close to the cavity resonance (within a bandwidth of order $\gamma$)
can excite the cavity. This proves that only the external frequencies
around the cavity resonance are relevant, and hence the approximations
that we made in the model are consistent, see Section \ref{Sec:OpenCavitymodel}.
It is interesting to note that using the relation of the amplitude
$|\mathcal{E}|$ with the power of the injected laser $P_{\text{inj}}$,
$|\mathcal{E}|^{2}=2\gamma P_{\text{inj}}/\hbar\omega_{\mathrm{c}}$
(we prove this in a later section), we can interpret the maximum coherent
contribution as 
\begin{equation}
\frac{|\mathcal{E}|^{2}}{\gamma^{2}}=\frac{P_{\text{inj}}}{\hbar\omega_{\text{c}}}\times\frac{2}{\gamma}\sim\left(\frac{\text{energy/time}}{\text{energy/photon}}\right)\times(\sim\text{damping time})\equiv\begin{array}{c}
\text{photons accumulated}\\
\text{in a damping cycle}
\end{array}.
\end{equation}
It is also interesting to consider the expression for the coherent
amplitude, which writing the injection parameter as $\mathcal{E}=|\mathcal{E}|e^{\mathrm{i}\phi}$,
can be written as
\begin{equation}
\lim_{t\rightarrow\infty}\langle\hat{a}(t)\rangle=\frac{|\mathcal{E}|}{\sqrt{\gamma^{2}+\Delta^{2}}}e^{\mathrm{i}[\phi+\delta\phi_{\Delta}]}e^{-\text{i}\omega_{\text{L}}t},
\end{equation}
with $\delta\phi_{\Delta}=\text{arg}(\gamma+\mathrm{i}\Delta)\in[-\pi/2,\pi/2]$.
There are two interesting features here. First, note that the field
oscillates at the laser frequency, not the cavity frequency. Second,
note that the intracavity field is in phase with the laser only when
injecting on resonance ($\Delta=0$), otherwise, they have a phase
difference of $\delta\phi_{\Delta}$. This provides a way to measure
the cavity resonance (and hence, its length), simply by analyzing
the relative phase between the injected laser and the output light.
This has many applications in the world of high precision measurements.

\subsubsection{Schrödinger picture approach: The master equation\label{SchrodingerOpen}}

In the previous sections we have treated the open cavity within a
Heisenberg-picture formalism. Now we will do it in the Schrödinger
picture. Hence, we consider now the density operator $\hat{\rho}(t)$
corresponding to the state of the whole system ``cavity mode + external
modes\textquotedblright , which evolves according to the von Neumann
equation
\begin{equation}
\mathrm{i}\hbar\partial_{t}\hat{\rho}=[\hat{H}_{\text{cav}}+\hat{H}_{\text{ext}}+\hat{H}_{\text{int}},\hat{\rho}],
\end{equation}
where the Hamiltonian terms are provided in (\ref{HcavHext}) and
(\ref{HintExtendedNegativeFreqs}). Our goal now is finding an evolution
equation for the reduced state of the cavity mode $\hat{\rho}_{\text{cav}}(t)=\text{tr}_{\text{ext}}\{\hat{\rho}(t)\}$,
by tracing out the external degrees of freedom. As we are about to
see, this goal is relatively easy to achieve if we keep effects only
up to second order on the weak interaction.

The derivation is easier to handle by moving to a more convenient
picture. Specifically, it is a picture composed of a sequence of two
picture changes, both of which appear very naturally from the following
arguments. First, as we did in the Heisenberg picture, we are going
to assume that the external field is initially at thermal equilibrium,
except for the possibility of a coherent laser contribution of amplitude
$\alpha(\omega)$ for the external modes of frequency $\omega$. Specifically,
we assume that
\begin{equation}
\hat{\rho}_{\text{ext}}(0)=\hat{D}(0)\frac{e^{-\hat{H}_{\text{ext}}/k_{B}T}}{\text{tr}_{\text{ext}}\left\{ e^{-\hat{H}_{\text{ext}}/k_{B}T}\right\} }\hat{D}^{\dagger}(0),\qquad\text{with }\hat{D}(0)=e^{\int_{-\infty}^{+\infty}d\omega[\alpha(\omega)\hat{b}^{\dagger}(\omega)-\alpha^{*}(\omega)\hat{b}(\omega)]},\label{RhoExt0}
\end{equation}
where we denote by $\hat{\rho}_{\text{ext}}(t)$ the state of the
external field. As it will be clear shortly, it is then convenient
to move to a picture that discounts the displacement, which will simplify
the statistical properties of the environment (similarly to what we
saw in the Heisenberg picture) and will remove the part of $\hat{H}_{\text{int}}$
that could be non-perturbative for large $\alpha(\omega)$. This is
accomplished by using a time-dependent displacement operator
\begin{equation}
\hat{D}(t)=e^{\int_{-\infty}^{+\infty}d\omega[\beta(\omega,t)\hat{b}^{\dagger}(\omega)-\beta^{*}(\omega,t)\hat{b}(\omega)]},\label{D(t)-FirstChange}
\end{equation}
as the unitary transformation to the new picture, where we show at
the end of the section that $\beta(\omega,t)=\alpha(\omega)e^{-\mathrm{i}\omega t}$
is the most appropriate choice, which corresponds precisely to the
evolution of $\langle\hat{b}(\omega)\rangle$ in the absence of interaction.
In particular, we show at the end of the section that the Hamiltonian
in the new picture is given by
\begin{equation}
\hat{H}_{D}(t)=\underbrace{\hat{H}_{\text{cav}}+\hat{H}_{\text{inj}}(t)+\hat{H}_{\text{ext}}}_{\hat{H}_{0}(t)}+\hat{H}_{\text{int}},\label{HD}
\end{equation}
where $\hat{H}_{\text{inj}}(t)=\mathrm{i}\hbar[\mathcal{A}(t)\hat{a}^{\dagger}-\mathcal{A}^{*}(t)\hat{a}]$,
with $\mathcal{A}(t)=-\sqrt{\gamma/\pi}\int_{-\infty}^{+\infty}d\omega e^{-\mathrm{i}\omega t}\alpha(\omega)$,
is the same Hamiltonian that appeared naturally in the Heisenberg
picture when we discounted the laser contribution from the input operator,
see Eq. (\ref{Hinj}). Next, in order to set the stage for a proper
perturbation theory, the second change of picture will discount the
evolution induced by $\hat{H}_{0}(t)$, which defines the interaction
picture. This is accomplished by the unitary operator satisfying the
equation $\mathrm{i}\hbar\partial_{t}\hat{U}_{\text{c}}=\hat{H}_{0}(t)\hat{U}_{\text{c}}$
with $\hat{U}_{\text{c}}(0)=\hat{I}$. As we saw in Section \ref{Sec:ChangingPictures},
Eq. (\ref{TimeEvoOp-TimeOrdered}), using the Dyson series and the
time-ordering symbol, this operator is written explicitly as
\begin{equation}
\hat{U}_{\text{c}}(t)=\mathcal{T}\left\{ e^{\int_{0}^{t}dt'\hat{H}_{0}(t')/\mathrm{i}\hbar}\right\} =e^{-\mathrm{i}\omega_{\text{c}}t\hat{a}^{\dagger}\hat{a}+B(t)\hat{a}^{\dagger}-B^{*}(t)\hat{a}}e^{-\mathrm{i}\int_{-\infty}^{+\infty}d\omega\,\omega\hat{b}^{\dagger}(\omega)\hat{b}(\omega)},\label{Uc-OpenSystems}
\end{equation}
with $B(t)=\int_{0}^{t}dt'\mathcal{A}(t)$. The interaction-picture
Hamiltonian takes then the form
\begin{align}
\hat{H}_{\text{I}}(t) & =\hat{U}_{\text{c}}^{\dagger}(t)\hat{H}_{D}(t)\hat{U}_{\text{c}}(t)-\mathrm{i}\hbar\hat{U}_{\text{c}}^{\dagger}(t)\partial_{t}\hat{U}_{\text{c}}(t)=\hat{U}_{\text{c}}^{\dagger}(t)\hat{H}_{\mathrm{int}}\hat{U}_{\text{c}}(t)\label{Htilde}\\
 & =\mathrm{i}\hbar\sqrt{\frac{\gamma}{\pi}}\int_{-\infty}^{+\infty}d\omega\left[e^{\mathrm{i}\omega t}\hat{b}^{\dagger}(\omega)\hat{a}_{\text{I}}(t)-e^{-\mathrm{i}\omega t}\hat{b}(\omega)\hat{a}_{\text{I}}^{\dagger}(t)\right],\nonumber 
\end{align}
where we have used $\hat{U}_{\text{c}}^{\dagger}(t)\hat{b}(\omega)\hat{U}_{\text{c}}(t)=e^{-\mathrm{i}\omega t}\hat{b}(\omega)$,
easily found from the Baker-Campbell-Haussdorf lemma (\ref{BCHlemma-1})
as usual, and we have defined the interaction-picture operators $\hat{a}_{\text{I}}(t)=\hat{U}_{\text{c}}^{\dagger}(t)\hat{a}\hat{U}_{\text{c}}(t)$,
whose explicit form in terms of the Schrödinger-picture operators
$\hat{a}$ is easy to find, but we won't need it. In this final interaction
picture, the state $\hat{\rho}_{\text{I}}(t)=\hat{U}_{\text{c}}^{\dagger}(t)\hat{D}^{\dagger}(t)\hat{\rho}(t)\hat{D}(t)\hat{U}_{\text{c}}(t)$
evolves then according to the von Neumann equation
\begin{equation}
\mathrm{i}\hbar\partial_{t}\hat{\rho}_{\text{I}}=[\hat{H}_{\text{I}}(t),\hat{\rho}_{\text{I}}],\qquad\text{with }\hat{\rho}_{\text{I}}(0)=\hat{\rho}_{\text{c}}(0)\bigotimes_{\omega}\hat{\rho}_{\text{th}}(\bar{n}),\label{VonNeumann_InteractionPicture_OpenCavity}
\end{equation}
where we have already made the approximation that the number of thermal
excitations is independent of the frequency and given by the Bose-Einstein
distribution at the cavity frequency, $\bar{n}=[\exp(\hbar\omega_{\text{c}}/k_{B}T)-1]^{-1}$,
just as we did in the Heisenberg picture, see the discussion around
Eq. (\ref{ExtThermalState}).

We are now in conditions to find an equation for the reduced state
of the intracavity field. Since, as already mentioned, we will be
doing so in the style of a perturbation theory up to second order
on the interaction, it is first convenient to manipulate the von Neumann
equation such that terms quadratic in the interaction Hamiltonian
$\hat{H}_{\text{I}}(t)$ appear explicitly. This is easily accomplished
by formally integrating the von Neumann equation (\ref{VonNeumann_InteractionPicture_OpenCavity})
as
\begin{equation}
\hat{\rho}_{\text{I}}(t)=\hat{\rho}_{\text{I}}(0)+\frac{1}{\mathrm{i}\hbar}\int_{0}^{t}dt^{\prime}[\hat{H}_{\text{I}}(t^{\prime}),\hat{\rho}_{\text{I}}(t^{\prime})],
\end{equation}
and reinserting this expression back into the von Neumann equation
(\ref{VonNeumann_InteractionPicture_OpenCavity}), obtaining
\begin{equation}
\partial_{t}\hat{\rho}_{\text{I}}(t)=\frac{1}{\mathrm{i}\hbar}[\hat{H}_{\text{I}}(t),\hat{\rho}_{\text{I}}(0)]-\frac{1}{\hbar^{2}}\int_{0}^{t}dt^{\prime}[\hat{H}_{\text{I}}(t),[\hat{H}_{\text{I}}(t^{\prime}),\hat{\rho}_{\text{I}}(t^{\prime})]].
\end{equation}
Since we are interested in the reduced state of the cavity, we can
now make the partial trace over the external modes, leading to
\begin{equation}
\partial_{t}\hat{\rho}_{\text{cav,I}}(t)=\frac{1}{\mathrm{i}\hbar}\text{tr}_{\text{ext}}\{[\hat{H}_{\text{I}}(t),\hat{\rho}_{\text{I}}(0)]\}-\frac{1}{\hbar^{2}}\int_{0}^{t}dt^{\prime}\text{tr}_{\text{ext}}\{[\hat{H}_{\text{I}}(t),[\hat{H}_{\text{I}}(t^{\prime}),\hat{\rho}_{\text{I}}(t^{\prime})]]\},
\end{equation}
where\footnote{Beware of the loose notation: $\hat{U}_{\text{c}}(t)$ acts on the
total Hilbert space, while $\hat{\rho}_{\text{cav,I}}(t)$ acts only
on the cavity's Hilbert space. It is then understood that the part
of $\hat{U}_{\text{c}}(t)$ acting on the Hilbert space of the external
modes has no effect in these type of expressions.} $\hat{\rho}_{\text{cav,I}}(t)=\hat{U}_{\text{c}}^{\dagger}(t)\hat{\rho}_{\text{cav}}(t)\hat{U}_{\text{c}}(t)$
is the state of the cavity in the interaction picture. Finally, taking
into account that the initial condition term $\text{tr}_{\text{ext}}\{[\hat{H}_{\text{I}}(t),\hat{\rho}_{\text{I}}(0)]\}$
is proportional to $\text{tr}_{\text{ext}}\{\hat{b}(\omega)\bigotimes_{\omega}\hat{\rho}_{\mathrm{th}}(\bar{n})\}=0$,
and making the variable change $t^{\prime}=t-\tau$ in the time integral,
we get the following integro-differential equation for the reduced
density operator:
\begin{equation}
\partial_{t}\hat{\rho}_{\text{cav,I}}(t)=-\frac{1}{\hbar^{2}}\int_{0}^{t}d\tau\mathrm{tr}_{\mathrm{ext}}\{[\hat{H}_{\text{I}}(t),[\hat{H}_{\text{I}}(t-\tau),\hat{\rho}_{\text{I}}(t-\tau)]]\}.\label{ExactIntegroDiff}
\end{equation}
This equation is exact, but now we are going to introduce two important
approximations that will lead to huge simplifications, in particular
by keeping terms only up to quadratic order in the interaction Hamiltonian
$\hat{H}_{\text{I}}$. At a first sight, since (\ref{ExactIntegroDiff})
is explicitly quadratic in $\hat{H}_{\text{I}}$ already, it might
not seem obvious where the higher-order $\hat{H}_{\text{I}}$-dependence
comes from. In fact, it is not explicit, but implicit in $\hat{\rho}_{\text{I}}(t-\tau)$.
To see this, note that starting from the uncorrelated state $\hat{\rho}_{\text{I}}(0)=\hat{\rho}_{\mathrm{cav,I}}(0)\bigotimes_{\omega}\hat{\rho}_{\mathrm{th}}(\bar{n})$,
the interaction $\hat{H}_{\text{I}}$ creates correlations between
the intracavity and external fields. Hence, any deviations from an
uncorrelated state $\hat{\rho}_{\text{I}}(t)=\hat{\rho}_{\mathrm{cav,I}}(t)\otimes\hat{\rho}_{\mathrm{ext,I}}(t)$
will contribute as beyond-second order in the equation above. Neglecting
these correlations is known as the \emph{Born approximation}. In addition,
note that the external field is formed by an infinite number of modes
that form a huge system compared to the ``tiny'' intracavity mode.
Therefore, we expect the back-action of the cavity onto the external
field to occur only at a much larger order in the interaction $\hat{H}_{\text{I}}$,
so that we can make the approximation $\hat{\rho}_{\text{I}}(t)=\hat{\rho}_{\mathrm{cav,I}}(t)\bigotimes_{\omega}\hat{\rho}_{\text{th}}(\bar{n})$
in (\ref{ExactIntegroDiff}). A note on nomenclature: in the literature,
you will sometimes see that this absence of back-action is also included
in the definition of the Born approximation.

It is important to remark that we are not saying that $\hat{\rho}_{\text{I}}(t)=\hat{\rho}_{\mathrm{cav,I}}(t)\bigotimes_{\omega}\hat{\rho}_{\text{th}}(\bar{n})$
is the state of the system, but only that deviations from this state
appear as higher-order effects in the dynamics of $\hat{\rho}_{\mathrm{cav,I}}(t)$.
In fact, we will see that $\hat{\rho}_{\mathrm{cav,I}}(t)$ is mixed
in general, even when we start from a pure cavity state, and this
is only possible if it gets correlated with the environment during
the evolution. Also, we will see in the next chapter that the field
coming out of the cavity has properties such as anti-bunching and
squeezing (when an atom or a nonlinear dielectric are placed inside
the cavity), which means that the external modes are no longer in
the original displaced thermal state, but indeed receive corrections.

Using these approximations, Eq. (\ref{ExactIntegroDiff}) is turned
into
\begin{equation}
\partial_{t}\hat{\rho}_{\text{cav,I}}(t)=-\frac{1}{\hbar^{2}}\int_{0}^{t}d\tau\mathrm{tr}_{\mathrm{ext}}\{[\hat{H}_{\text{I}}(t),[\hat{H}_{\text{I}}(t-\tau),\hat{\rho}_{\mathrm{cav,I}}(t-\tau)\bigotimes_{\omega}\hat{\rho}_{\text{th}}(\bar{n})]]\}.\label{ExactIntegroDiff-AfterBorn}
\end{equation}
Note that the cavity state inside the integral is not evaluated at
time $t$, but all along its past. On the other hand, since the evolution
is produced by the interaction $\hat{H}_{\text{I}}$, this means that,
in principle, we are still keeping terms beyond second order on $\hat{H}_{\text{I}}$.
However, as we will see next, the Born and absence of backaction approximations,
together with the frequency-independent coupling $g(\omega)=\sqrt{\gamma/\pi}$,
are enough to remove all those higher-order terms. In contrast, we
will see later in this chapter that when the coupling is allowed to
depend on the frequency, an extra assumption known as \emph{Markov
approximation} is required to eliminate these higher-order terms.

Inserting the specific form of the Hamiltonian (\ref{Htilde}) into
(\ref{ExactIntegroDiff-AfterBorn}), we obtain an expression that
is proportional to second-order moments of the external operators,
which in the picture we are working on take the simple thermal form
\begin{align}
\mathrm{tr}_{\mathrm{ext}}\left\{ \hat{b}(\omega)\bigotimes_{\omega}\hat{\rho}_{\mathrm{th}}(\bar{n})\right\}  & =\mathrm{tr}_{\mathrm{ext}}\left\{ \hat{b}(\omega)\hat{b}(\omega^{\prime})\bigotimes_{\omega}\hat{\rho}_{\mathrm{th}}(\bar{n})\right\} =0,\label{VacExpExt}\\
\mathrm{tr}_{\mathrm{ext}}\left\{ \hat{b}^{\dagger}(\omega)\hat{b}(\omega^{\prime})\bigotimes_{\omega}\hat{\rho}_{\mathrm{th}}(\bar{n})\right\}  & =\bar{n}\delta(\omega-\omega'),\quad\mathrm{tr}_{\mathrm{ext}}\left\{ \hat{b}(\omega^{\prime})\hat{b}^{\dagger}(\omega)\bigotimes_{\omega}\hat{\rho}_{\mathrm{th}}(\bar{n})\right\} =(\bar{n}+1)\delta(\omega-\omega').\nonumber 
\end{align}
Now, after expanding the double commutator in Eq. (\ref{ExactIntegroDiff}),
we obtain 16 terms. 8 of these terms are proportional to $\text{tr}_{\text{ext}}\{\hat{b}(\omega)\hat{b}(\omega')\bigotimes_{\omega}\hat{\rho}_{\mathrm{th}}\}=0$
or $\text{tr}_{\text{ext}}\{\hat{b}^{\dagger}(\omega)\hat{b}^{\dagger}(\omega')\bigotimes_{\omega}\hat{\rho}_{\mathrm{th}}\}=0$,
and do not contribute. For example, one of such terms is
\begin{equation}
\frac{\gamma}{\pi}\int_{0}^{t}d\tau\int_{-\infty}^{+\infty}d\omega\int_{-\infty}^{+\infty}d\omega'\hat{a}_{\text{I}}(t)\hat{a}_{\text{I}}(t-\tau)\hat{\rho}_{\text{cav,I}}(t-\tau)e^{\mathrm{i}\omega t}e^{\mathrm{i}\omega'(t-\tau)}\text{tr}_{\text{ext}}\{\hat{b}^{\dagger}(\omega)\hat{b}^{\dagger}(\omega')\bigotimes_{\omega}\hat{\rho}_{\mathrm{th}}\}=0.
\end{equation}
The other 8 terms are either proportional to $\text{tr}_{\text{ext}}\{\hat{b}^{\dagger}(\omega)\hat{b}(\omega')\bigotimes_{\omega}\hat{\rho}_{\mathrm{th}}\}=\bar{n}\delta(\omega-\omega')$
or $\text{tr}_{\text{ext}}\{\hat{b}(\omega)\hat{b}^{\dagger}(\omega')\bigotimes_{\omega}\hat{\rho}_{\mathrm{th}}\}=(\bar{n}+1)\delta(\omega-\omega')$,
and hence provide a non-zero contribution. For example, one of such
terms is
\begin{align}
\frac{\gamma}{\pi}\int_{0}^{t}d\tau\int_{-\infty}^{+\infty}d\omega\int_{-\infty}^{+\infty}d\omega'\hat{a}_{\text{I}}(t)\hat{a}_{\text{I}}^{\dagger}(t-\tau)\hat{\rho}_{\text{cav,I}}(t-\tau)e^{\mathrm{i}\omega t}e^{-\mathrm{i}\omega'(t-\tau)}\text{tr}_{\text{ext}}\{\hat{b}(\omega)\hat{b}^{\dagger}(\omega')\bigotimes_{\forall\omega}\hat{\rho}_{\mathrm{th}}\}\\
=\frac{\gamma\bar{n}}{\pi}\int_{0}^{t}d\tau\hat{a}_{\text{I}}(t)\hat{a}_{\text{I}}^{\dagger}(t-\tau)\hat{\rho}_{\text{cav,I}}(t-\tau)\underset{2\pi\delta(\tau)}{\underbrace{\int_{-\infty}^{+\infty}d\omega e^{\mathrm{i}\omega\tau}}} & =\gamma\bar{n}\hat{a}_{\text{I}}(t)\hat{a}_{\text{I}}^{\dagger}(t)\hat{\rho}_{\text{cav,I}}(t),\nonumber 
\end{align}
where we have made use of the property (\ref{DiracDeltaExtremeInterval})
of the Dirac delta. After taking care of all this 16 terms, we end
up with the equation
\begin{equation}
\partial_{t}\hat{\rho}_{\text{cav,I}}=(\bar{n}+1)\gamma\left(2\hat{a}_{\text{I}}\hat{\rho}_{\text{cav,I}}\hat{a}_{\text{I}}^{\dagger}-\hat{a}_{\text{I}}^{\dagger}\hat{a}_{\text{I}}\hat{\rho}_{\text{cav,I}}-\hat{\rho}_{\text{cav,I}}\hat{a}_{\text{I}}^{\dagger}\hat{a}_{\text{I}}\right)+\bar{n}\gamma\left(2\hat{a}_{\text{I}}^{\dagger}\hat{\rho}_{\text{cav,I}}\hat{a}_{\text{I}}-\hat{a}_{\text{I}}\hat{a}_{\text{I}}^{\dagger}\hat{\rho}_{\text{cav,I}}-\hat{\rho}_{\text{cav,I}}\hat{a}_{\text{I}}\hat{a}_{\text{I}}^{\dagger}\right),\label{MasterInt}
\end{equation}
where all the operators are evaluated at the same time. Finally, coming
back to the Schrödinger picture, and including a Hamiltonian $\hat{H}_{\text{intra}}(t)$
that describes any additional intracavity processes (allowed within
the formalism as long as they are just a perturbation onto $\hat{H}_{\text{cav}}$),
we obtain
\begin{align}
\partial_{t}\hat{\rho}_{\mathrm{cav}} & =\left[\frac{\hat{H}_{\mathrm{cav}}+\hat{H}_{\mathrm{intra}}(t)+\hat{H}_{\mathrm{inj}}(t)}{\mathrm{i}\hbar},\hat{\rho}_{\mathrm{cav}}\right]\label{MasterEq}\\
 & \quad+(\bar{n}+1)\gamma\left(2\hat{a}\hat{\rho}_{\mathrm{cav}}\hat{a}^{\dagger}-\hat{a}^{\dagger}\hat{a}\hat{\rho}_{\mathrm{cav}}-\hat{\rho}_{\mathrm{cav}}\hat{a}^{\dagger}\hat{a}\right)+\bar{n}\gamma\left(2\hat{a}^{\dagger}\hat{\rho}_{\mathrm{cav}}\hat{a}-\hat{a}\hat{a}^{\dagger}\hat{\rho}_{\mathrm{cav}}-\hat{\rho}_{\mathrm{cav}}\hat{a}\hat{a}^{\dagger}\right)\nonumber 
\end{align}
This equation is known as the \textit{master equation} of the intracavity
mode. It contains two very different types of terms. First, a commutator
of the state with a Hamiltonian accounting for all the coherent processes
occurring within the cavity (which might get corrections from the
interaction with the external field, e.g., the one coming from the
injection of lasers). On the other hand, we have terms that cannot
be written as a commutator of the state with an operator, and account
for any incoherent processes occurring in the system mediated by the
external field (in this case, photons leaking out of the cavity irreversibly
or entering the cavity incoherently from the thermal external field).
In general, similarly to what we saw in the Heisenberg picture, the
coherent and incoherent processes reach a dynamical balance, leading
to an asymptotic state in the long term $t\rightarrow\infty$, which
does not depend on the initial state.

To finish this section, let us prove that the Hamiltonian after the
first change of picture
\begin{equation}
\hat{H}_{D}(t)=\hat{D}^{\dagger}(t)(\hat{H}_{\text{cav}}+\hat{H}_{\text{ext}}+\hat{H}_{\text{int}})\hat{D}(t)-\mathrm{i}\hbar\hat{D}^{\dagger}(t)\partial_{t}\hat{D}(t),\label{HDdef}
\end{equation}
takes the form (\ref{HD}). In order to find $\partial_{t}\hat{D}(t)$,
we first use the disentangling Baker-Cambell-Haussdorf lemma (\ref{DisentanglingBCH})
with $\hat{A}=\int_{-\infty}^{+\infty}d\omega\beta(\omega,t)\hat{b}^{\dagger}(\omega)$
and $\hat{B}=-\int_{-\infty}^{+\infty}d\omega\beta^{*}(\omega,t)\hat{b}(\omega)$,
so that using the commutation relations (\ref{ExternalCommutators})
of the external operators, we write
\begin{equation}
\hat{D}(t)=e^{-\int_{-\infty}^{+\infty}d\omega|\beta(\omega,t)|^{2}/2}e^{\int_{-\infty}^{+\infty}d\omega\beta(\omega,t)\hat{b}^{\dagger}(\omega)}e^{-\int_{-\infty}^{+\infty}d\omega\beta^{*}(\omega,t)\hat{b}(\omega)}.
\end{equation}
Applying the chain rule, we then obtain
\begin{align}
\partial_{t}\hat{D}(t) & =\int_{-\infty}^{+\infty}d\omega\left[\dot{\beta}(\omega,t)\partial_{\beta(\omega,t)}+\dot{\beta}^{*}(\omega,t)\partial_{\beta^{*}(\omega,t)}\right]\hat{D}(t)\label{dDdt}\\
 & =\int_{-\infty}^{+\infty}d\omega\left\{ \dot{\beta}(\omega,t)\left[\hat{b}^{\dagger}(\omega)-\frac{\beta^{*}(\omega,t)}{2}\right]\hat{D}(t)-\hat{D}(t)\dot{\beta}^{*}(\omega,t)\left[\hat{b}(\omega)+\frac{\beta(\omega,t)}{2}\right]\right\} \nonumber \\
 & =\hat{D}(t)\int_{-\infty}^{+\infty}d\omega\dot{\beta}(\omega,t)\left[\hat{b}^{\dagger}(\omega)+\frac{\beta^{*}(\omega,t)}{2}\right]-\text{H.c.},\nonumber 
\end{align}
here in the last step we have used $\hat{b}^{\dagger}(\omega)\hat{D}(t)=\hat{D}(t)[\hat{b}^{\dagger}(\omega)+\beta^{*}(\omega,t)]$,
which is just an alternative form of the displacement formula $\hat{D}^{\dagger}(t)\hat{b}(\omega)\hat{D}(t)=\hat{b}(\omega)+\beta(\omega,t)$,
trivially found by using the Baker-Cambell-Haussdorf lemma (\ref{BCHlemma-1}).
Applying this formula and (\ref{DisentanglingBCH}) to (\ref{HDdef}),
we then obtain
\begin{equation}
\hat{H}_{D}(t)=\hat{H}_{\text{cav}}+\hat{H}_{\text{ext}}+\hat{H}_{\text{int}}+\hat{H}_{\text{inj}}(t)+\int_{-\infty}^{+\infty}d\omega\left\{ \hbar[\omega\beta(\omega,t)-\mathrm{i}\dot{\beta}(\omega,t)]\left[\hat{b}^{\dagger}(\omega)+\frac{\beta^{*}(\omega,t)}{2}\right]+\text{H.c.}\right\} .
\end{equation}
The largest simplification of this Hamiltonian is then obtained by
asking the last term of this expression to vanish, obtaining the differential
equation $\dot{\beta}(\omega,t)=-\mathrm{i}\omega\beta(\omega,t)$.
Choosing the initial condition $\beta(\omega,0)=\alpha(\omega)$ as
the initial condition, in order to remove the initial displacement
of the environmental state (\ref{RhoExt0}), we then obtain $\beta(\omega,t)=\alpha(\omega)e^{-\mathrm{i}\omega t}$,
and the form (\ref{HD}) for $\hat{H}_{D}(t)$, as we wanted to prove.

\subsubsection{Master equations and their interpretation}

The form of the master equation above is not accidental, and indeed,
within the level of approximation that we are working with (Born-Markov),
it is possible (but beyond the scope of this introductory course,
although we'll get a glimpse of it at the end of the chapter) to show
that the most general evolution equation for the state $\hat{\rho}$
of the system, after tracing out the environment, takes the so-called
\emph{Lindblad} form
\begin{equation}
\partial_{t}\hat{\rho}(t)=\left[\frac{\hat{V}(t)}{\mathrm{i\hbar}},\hat{\rho}(t)\right]+\sum_{j}\kappa_{j}\left(2\hat{J}_{j}\hat{\rho}\hat{J}_{j}^{\dagger}-\hat{J}_{j}^{\dagger}\hat{J}_{j}\hat{\rho}-\hat{\rho}\hat{J}_{j}^{\dagger}\hat{J}_{j}\right).\label{GenMasterEq}
\end{equation}
Here, $\hat{V}(t)$ is a Hermitian operator that can be interpreted
as the Hamiltonian of the system (modified by interactions with the
environment). On the other hand, $\hat{J}_{j}$ are some operators
with related rates $\kappa_{j}>0$, all of which (including how many
we get) depend on the specific form of the interaction between the
system and its environments, as well as the properties of the latter.
This form is so common, that we typically define the so-called \emph{dissipative
superoperator} or \emph{dissipator} $\mathcal{D}_{J}$, which acts
on operators as
\begin{equation}
\mathcal{D}_{J}[\hat{\rho}]=2\hat{J}\hat{\rho}\hat{J}^{\dagger}-\hat{J}^{\dagger}\hat{J}\hat{\rho}-\hat{\rho}\hat{J}^{\dagger}\hat{J}.
\end{equation}
The word `superoperator' refers to the fact that $\mathcal{D}_{J}$
can be seen as a linear map from operators to operators, similarly
to how an operator is a map from kets to kets. In fact, the whole
master equation can be seen as the evolution equation generated by
the so-called \emph{Lindblad superoperator} or \emph{Lindbladian }$\mathcal{L}$,
so that $\partial_{t}\hat{\rho}=\mathcal{L}[\hat{\rho}]$. While all
this seems rather anecdotical at this point, we will see that it directly
inspires methods for the efficient numerical coding and simulation
of master equations, see \cite{CNB_Numerics}.

Note that the master equation (\ref{MasterEq}) for the open cavity
has exactly this Lindblad form with $\hat{V}(t)=\hat{H}_{\mathrm{cav}}+\hat{H}_{\mathrm{intra}}(t)+\hat{H}_{\mathrm{inj}}(t)$,
$\{\kappa_{1}=(\bar{n}+1)\gamma,\hat{J}_{1}=\hat{a}\}$ and $\{\kappa_{2}=\bar{n}\gamma,\hat{J}_{2}=\hat{a}^{\dagger}\}$.

Let us now provide a more explicit interpretation of the terms of
the master equation, and in particular of the so-called \emph{jump
operators} $\hat{J}_{j}$. For this, consider a simple master equation
with only one of such operators and a time-independent Hamiltonian
(the generalization is trivial). Defining an effective non-Hermitian
Hamiltonian
\begin{equation}
\hat{H}_{\text{eff}}=\hat{V}-\mathrm{i}\kappa\hat{J}^{\dagger}\hat{J},
\end{equation}
the master equation can be rewritten as
\begin{equation}
\partial_{t}\hat{\rho}=\frac{1}{\mathrm{i}\hbar}\left(\hat{H}_{\text{eff}}\hat{\rho}-\hat{\rho}\hat{H}_{\text{eff}}^{\dagger}\right)+2\kappa\hat{J}\hat{\rho}\hat{J}^{\dagger}.\label{Eff_Master}
\end{equation}
This form offers a very suggestive interpretation. The dynamics is
generated by two types of terms: the first one similar to a commutator,
$\hat{H}_{\text{eff}}\hat{\rho}-\hat{\rho}\hat{H}_{\text{eff}}^{\dagger}$;
the second one just the direct action of the jump operator on the
state, $\hat{J}\hat{\rho}\hat{J}^{\dagger}$. We now show that these
two terms are of extremely different nature. In particular, the first
one induces \emph{reversible} dynamics, while the second generates
\emph{irreversible} dynamics. In order to show this, consider the
following probabilistic protocol for the generation of $\hat{\rho}(t+dt)$
from $\hat{\rho}(t)$, which is shown at the end to be equivalent
to the master equation. With probability $p(t)=2\kappa dt\langle\hat{J}^{\dagger}\hat{J}\rangle$,
we apply a quantum jump to the state, so the state is transformed
as 
\begin{equation}
\hat{\rho}(t)\longrightarrow\frac{\hat{J}\hat{\rho}(t)\hat{J}^{\dagger}}{\text{tr}\{\hat{J}\hat{\rho}(t)\hat{J}^{\dagger}\}}\equiv\hat{\rho}_{\text{irrev}}(t).
\end{equation}
On the other hand, with the complementary probability $1-p(t)$, we
apply the effective time-evolution operator $\hat{U}_{\text{eff}}(t)=e^{\hat{H}_{\text{eff}}t/\mathrm{i}\hbar}$,
so that now the state is transformed as
\begin{equation}
\hat{\rho}(t)\longrightarrow\frac{\hat{U}_{\text{eff}}(dt)\hat{\rho}(t)\hat{U}_{\text{eff}}^{\dagger}(dt)}{\text{tr}\{\hat{U}_{\text{eff}}(dt)\hat{\rho}(t)\hat{U}_{\text{eff}}^{\dagger}(dt)\}}\equiv\hat{\rho}_{\text{rev}}(t).
\end{equation}
The point now is that while the inverse of the effective time-evolution
operator exists (we just need to change the sign of the Hamiltonian),
$\hat{U}_{\text{eff}}^{-1}(t)=\hat{U}_{\text{eff}}(-t)$, that's not
the general case for jump operators $\hat{J}$ (think of the annihilation
or creation operators, which are not invertible). Therefore, $\hat{\rho}_{\text{rev}}(t)$
can always be reversed into $\hat{\rho}(t)$, but once a jump happens,
and we have $\hat{\rho}_{\text{irrev}}(t)$, there is nothing trivial
we can do to come back to the original state\footnote{See however reference \cite{CatchJumps} for an impressive example
of the advances that modern experimental platforms, together with
a very mature theoretical understanding of quantum-jump processes,
are allowing in this field: even though the jumps are random, under
proper conditions they are preceded by a warning signal that allows
catching and reversing them.}. If we add to this that quantum jumps are random (they occur only
probabilistically), we conclude that the second term in the master
equation (\ref{Eff_Master}) is the one responsible for irreversible
dynamics. Note, in addition, that the evolution induced by $\hat{U}_{\text{eff}}(t)$
is continuous, while the one induced by quantum jumps is discontinuous
(hence the word `jump').

Let us now prove that, in the $dt\rightarrow0$ limit, this probabilistic
protocol is equivalent to the master equation (\ref{Eff_Master}).
According to the definition and interpretation of mixed state that
we saw in Section \ref{vonNeumannEntropy}, the quantum state right
after applying the protocol to $\hat{\rho}(t)$ can be written as
\begin{equation}
\hat{\rho}(t+dt)=p(t)\hat{\rho}_{\text{irrev}}+[1-p(t)]\hat{\rho}_{\text{rev}}.\label{MixedProtocolJumps}
\end{equation}
On the other hand, using the order-$dt$ expansions\begin{subequations}
\begin{align}
\hat{U}_{\text{eff}}(dt) & \approx1+\frac{dt}{\mathrm{i}\hbar}\hat{H}_{\text{eff}},\\
\hat{U}_{\text{eff}}^{\dagger}(dt)\hat{U}_{\text{eff}}(dt) & \approx1-2\kappa dt\hat{J}^{\dagger}\hat{J},
\end{align}
\end{subequations}and noting that the cyclic property of the trace
allows us to write\begin{subequations}
\begin{align}
\text{tr}\{\hat{J}\hat{\rho}(t)\hat{J}^{\dagger}\} & =\langle\hat{J}^{\dagger}\hat{J}\rangle(t)=p(t)/2\kappa dt,\\
\text{tr}\{\hat{U}_{\text{eff}}(dt)\hat{\rho}(t)\hat{U}_{\text{eff}}^{\dagger}(dt)\} & =\langle\hat{U}_{\text{eff}}^{\dagger}(dt)\hat{U}_{\text{eff}}(dt)\rangle(t)\approx1-2\kappa dt\langle\hat{J}^{\dagger}\hat{J}\rangle(t)=1-p(t),
\end{align}
 \end{subequations}we can rewrite (\ref{MixedProtocolJumps}) to
first order in $dt$ as
\begin{equation}
\hat{\rho}(t+dt)=\hat{\rho}(t)+\frac{dt}{\mathrm{i}\hbar}\left(\hat{H}_{\text{eff}}\hat{\rho}-\hat{\rho}\hat{H}_{\text{eff}}^{\dagger}\right)+2\kappa dt\hat{J}\hat{\rho}\hat{J}^{\dagger},
\end{equation}
which is the finite-differences version of the master equation (\ref{Eff_Master}),
and hence coincides with it in the $dt\rightarrow0$ limit, as we
wanted to prove.

\subsubsection{General dynamics of expectation values and example in the Schrödinger
picture: driven empty cavity\label{Sec:OpenCavityExample}}

In order to learn how to use the master equation, we study again the
simplest example possible: that of an empty cavity ($\hat{H}_{\text{intra}}=0$),
driven by a single monochromatic laser, so that $\mathcal{A}(t)=\mathcal{E}e^{-\mathrm{i}\omega_{\text{L}}t}$.
For this case, we already argued that the state of the cavity remains
Gaussian if it is Gaussian initially (otherwise, only the asymptotic
state will be Gaussian). With more generality, this happens whenever
the master equation is quadratic in annihilation and creation operators,
as we will prove next.

In order to find the evolution equations of the first and second order
moments required to determine the Gaussian state, we first consider
the time evolution of the expectation value of a generic operator
$\hat{A}$, that is, $\langle\hat{A}\rangle(t)=\text{\text{tr}}\{\hat{A}\hat{\rho}(t)\}$,
where we allow $\hat{\rho}$ to satisfy a generic master equation
(\ref{GenMasterEq}). We then have
\begin{equation}
\partial_{t}\langle\hat{A}\rangle=\text{tr}\left\{ \hat{A}\partial_{t}\hat{\rho}\right\} =\text{tr}\left\{ \hat{A}\left[\frac{\hat{V}}{\mathrm{i}\hbar},\hat{\rho}\right]\right\} +\sum_{j}\kappa_{j}\text{tr}\left\{ 2\hat{A}\hat{J}_{j}\hat{\rho}\hat{J}_{j}^{\dagger}-\hat{A}\hat{J}_{j}^{\dagger}\hat{J}_{j}\hat{\rho}-\hat{A}\hat{\rho}\hat{J}_{j}^{\dagger}\hat{J}_{j}\right\} .
\end{equation}
Using the cyclic property of the trace, we can rewrite this expression
in terms of commutators of $\hat{A}$ with the Hamiltonian and the
jump operators:\begin{subequations}
\begin{align}
\text{tr}\left\{ \hat{A}\left[\frac{\hat{V}}{\mathrm{i}\hbar},\hat{\rho}\right]\right\}  & =\text{tr}\left\{ \hat{A}\frac{\hat{V}}{\mathrm{i}\hbar}\hat{\rho}-\hat{A}\hat{\rho}\frac{\hat{V}}{\mathrm{i}\hbar}\right\} =\text{tr}\left\{ \left(\hat{A}\frac{\hat{V}}{\mathrm{i}\hbar}-\frac{\hat{V}}{\mathrm{i}\hbar}\hat{A}\right)\hat{\rho}\right\} =\left\langle \left[\hat{A},\frac{\hat{V}}{\mathrm{i}\hbar}\right]\right\rangle ,\\
\text{tr}\left\{ \hat{A}\hat{J}_{j}\hat{\rho}\hat{J}_{j}^{\dagger}-\hat{A}\hat{J}_{j}^{\dagger}\hat{J}_{j}\hat{\rho}\right\}  & =\text{tr}\left\{ \left(\hat{J}_{j}^{\dagger}\hat{A}-\hat{A}\hat{J}_{j}^{\dagger}\right)\hat{J}_{j}\hat{\rho}\right\} =\left\langle \left[\hat{J}_{j}^{\dagger},\hat{A}\right]\hat{J}_{j}\right\rangle ,\\
\text{tr}\left\{ \hat{A}\hat{J}_{j}\hat{\rho}\hat{J}_{j}^{\dagger}-\hat{A}\hat{\rho}\hat{J}_{j}^{\dagger}\hat{J}_{j}\right\}  & =\text{tr}\left\{ \hat{J}_{j}^{\dagger}\left(\hat{A}\hat{J}_{j}-\hat{J}_{j}\hat{A}\right)\hat{\rho}\right\} =\left\langle \hat{J}_{j}^{\dagger}\left[\hat{A},\hat{J}_{j}\right]\right\rangle ,
\end{align}
\end{subequations}leading to a final simple form of the evolution
equation
\begin{equation}
\partial_{t}\langle\hat{A}\rangle=\left\langle \left[\hat{A},\frac{\hat{V}}{\mathrm{i}\hbar}\right]\right\rangle +\sum_{j}\kappa_{j}\left(\left\langle \left[\hat{J}_{j}^{\dagger},\hat{A}\right]\hat{J}_{j}\right\rangle +\left\langle \hat{J}_{j}^{\dagger}\left[\hat{A},\hat{J}_{j}\right]\right\rangle \right).\label{GenOpenExpecEvo}
\end{equation}
Since for bosonic problems the commutator always reduces the order
in annihilation and creation operators by a factor 2, this expression
evidences that the equations of motion of first and second order moments
will form a closed set, only if the Hamiltonian is quadratic in annihilation
and creation operators and the jump operators are linear. In any other
case, information about higher-order moments will enter in the dynamical
equations of the mean vector and covariance matrix, and the state
will not remain Gaussian.

Considering the example at hand (driven empty cavity), we can particularize
the equation above to the first and second order moments of the cavity
mode. However, before diving into such calculations, it is always
advisable to move to a picture where the master equation becomes time
independent. Hence, we need to learn how a general master equation
(\ref{GenMasterEq}) is transformed under a change of picture with
associated unitary operator $\hat{U}_{\text{c}}(t)$. It is very easy
to see, just by taking the time derivative of the transformed state
$\hat{\rho}_{\text{I}}(t)=\hat{U}_{\text{c}}^{\dagger}(t)\hat{\rho}(t)\hat{U}_{\text{c}}(t)$,
that it still satisfies a Lindblad-form master equation
\begin{equation}
\partial_{t}\hat{\rho}_{\text{I}}(t)=\left[\frac{\hat{H}_{\text{I}}(t)}{\mathrm{i\hbar}},\hat{\rho}_{\text{I}}(t)\right]+\sum_{j}\kappa_{j}\left(2\hat{J}_{j,\text{I}}(t)\hat{\rho}_{\text{I}}\hat{J}_{j,\text{I}}^{\dagger}(t)-\hat{J}_{j,\text{I}}^{\dagger}(t)\hat{J}_{j,\text{I}}(t)\hat{\rho}_{\text{I}}-\hat{\rho}_{\text{I}}\hat{J}_{j,\text{I}}^{\dagger}(t)\hat{J}_{j,\text{I}}(t)\right),\label{GenMasterInteractionPicture}
\end{equation}
with the same intermediate-picture Hamiltonian $\hat{H}_{\text{I}}(t)=\hat{U}_{\text{c}}^{\dagger}(t)\hat{V}(t)\hat{U}_{\text{c}}(t)-\mathrm{i}\hbar\hat{U}_{\text{c}}^{\dagger}(t)\partial_{t}\hat{U}_{\text{c}}(t)$
as we found for closed systems (see Section \ref{Sec:ChangingPictures}),
and transformed jump operators $\hat{J}_{j,\text{I}}(t)=\hat{U}_{\text{c}}^{\dagger}(t)\hat{J}_{j}\hat{U}_{\text{c}}(t)$. 

In our master equation at hand, that of an empty open cavity (\ref{MasterEq}),
the only explicit time dependence comes from the injection Hamiltonian.
This can be easily removed by changing to a picture that rotates at
the laser frequency, with corresponding transformation operator $\hat{U}_{\text{c}}(t)=e^{-\mathrm{i}\omega_{\text{L}}t\hat{a}^{\dagger}\hat{a}}$.
This transforms the annihilation operator as $\hat{U}_{\text{c}}^{\dagger}(t)\hat{a}\hat{U}_{\text{c}}(t)=e^{-\mathrm{i}\omega_{\text{L}}t}\hat{a}$,
as we have seen several times. The transformed state $\hat{\rho}_{\text{I}}(t)=\hat{U}_{\text{c}}^{\dagger}(t)\hat{\rho}_{\text{cav}}(t)\hat{U}_{\text{c}}(t)$,
evolves then according to the master equation
\begin{equation}
\partial_{t}\hat{\rho}_{\text{I}}=\left[\mathrm{i}\Delta\hat{a}^{\dagger}\hat{a}+\mathcal{E}\hat{a}^{\dagger}-\mathcal{E}^{*}\hat{a},\hat{\rho}_{\text{I}}\right]+(\bar{n}+1)\gamma\mathcal{D}_{a}[\hat{\rho}_{\text{I}}]+\bar{n}\gamma\mathcal{D}_{a^{\dagger}}[\hat{\rho}_{\text{I}}],\label{MasterEqEx1}
\end{equation}
where we remind that $\Delta=\omega_{\text{L}}-\omega_{\text{c}}$
is the detuning between the cavity and the driving laser. Note that
the incoherent terms didn't change, because the jump operators, annihilation
and creation operators in this case, transform with simple $e^{\mp\text{i}\omega_{\text{L}}t}$
phases, which cancel in every term.

Let us further note one subtle issue. The relation between the expression
for the expectation of an operator $\hat{A}$ can be written either
either as $\text{tr}\{\hat{A}\hat{\rho}_{\text{cav}}(t)\}$ in the
Schrödinger picture or as $\text{tr}\{\hat{U}_{\text{c}}^{\dagger}(t)\hat{A}\hat{U}_{\text{c}}(t)\hat{\rho}_{\text{I}}(t)\}$
in the rotating picture. However, given the master equation (\ref{MasterEqEx1})
in the rotating picture, the natural object that can be evaluated
is what we will define as the \emph{intermediate-picture expectation
value} $\langle\hat{A}\rangle_{\text{I}}(t)=\text{tr}\{\hat{A}\hat{\rho}_{\text{I}}(t)\}=\text{tr}\{\hat{U}_{\text{c}}(t)\hat{A}\hat{U}_{\text{c}}^{\dagger}(t)\hat{\rho}_{\text{cav}}(t)\}$,
whose relation with the true expectation value $\text{tr}\{\hat{A}\hat{\rho}_{\text{cav}}(t)\}$
can only be found by working out the transformation $\hat{U}_{\text{c}}(t)\hat{A}\hat{U}_{\text{c}}^{\dagger}(t)$,
as we will see in the example.

We are now in conditions to proceed with the evaluation of the first
and second order moments. In particular, we now particularize (\ref{GenOpenExpecEvo})
to the master equation (\ref{MasterEqEx1}) and the operators needed
for these moments: $\hat{a}$, $\hat{a}^{2}$, and $\hat{a}^{\dagger}\hat{a}$.
Using the canonical commutation relations, we easily find\begin{subequations}\label{Example1Expectations}
\begin{align}
\partial_{t}\langle\hat{a}\rangle_{\text{I}} & =\mathcal{E}-(\gamma-\mathrm{i}\Delta)\langle\hat{a}\rangle_{\text{I}},\\
\partial_{t}\langle\hat{a}^{2}\rangle_{\text{I}} & =-2(\gamma-\mathrm{i}\Delta)\langle\hat{a}^{2}\rangle_{\text{I}}+2\mathcal{E}\langle\hat{a}\rangle_{\text{I}},\\
\partial_{t}\langle\hat{a}^{\dagger}\hat{a}\rangle_{\text{I}} & =2\text{Re}\left\{ \mathcal{E}^{*}\langle\hat{a}\rangle_{\text{I}}\right\} -2\gamma\langle\hat{a}^{\dagger}\hat{a}\rangle_{\text{I}}+2\bar{n}\gamma,
\end{align}
\end{subequations}Note that in terms of the fluctuation operator
$\delta\tilde{a}=\hat{a}-\langle\hat{a}\rangle_{\text{I}}$, the equations
for the second order moments are written as\begin{subequations}
\begin{align}
\partial_{t}\langle\delta\tilde{a}^{2}\rangle_{\text{I}} & =\partial_{t}\left(\langle\hat{a}^{2}\rangle_{\text{I}}-\langle\hat{a}\rangle_{\text{I}}^{2}\right)=\left(\partial_{t}\langle\hat{a}^{2}\rangle_{\text{I}}-2\langle\hat{a}\rangle_{\text{I}}\partial_{t}\langle\hat{a}\rangle_{\text{I}}\right)=-2(\gamma-\mathrm{i}\Delta)\langle\delta\tilde{a}^{2}\rangle_{\text{I}},\\
\partial_{t}\langle\delta\tilde{a}^{\dagger}\delta\tilde{a}\rangle_{\text{I}} & =\partial_{t}\left(\langle\hat{a}^{\dagger}\hat{a}\rangle_{\text{I}}-\langle\hat{a}\rangle_{\text{I}}^{*}\langle\hat{a}\rangle_{\text{I}}\right)=\left(\partial_{t}\langle\hat{a}^{\dagger}\hat{a}\rangle_{\text{I}}-2\text{Re}\left\{ \langle\hat{a}\rangle_{\text{I}}^{*}\partial_{t}\langle\hat{a}\rangle_{\text{I}}\right\} \right)=-2\gamma\langle\delta\tilde{a}^{\dagger}\delta\tilde{a}\rangle_{\text{I}}+2\bar{n}\gamma.
\end{align}
\end{subequations}Hence, we find a set of decoupled linear equations
for all the required moments, with solutions\begin{subequations}
\begin{align}
\langle\hat{a}\rangle_{\text{I}}(t) & =e^{-(\gamma-\mathrm{i}\Delta)t}\langle\hat{a}\rangle_{\text{I}}(0)+\mathcal{E}\int_{0}^{t}dt'e^{-(\gamma-\mathrm{i}\Delta)(t-t')}=e^{-(\gamma-\mathrm{i}\Delta)t}\langle\hat{a}\rangle_{\text{I}}(0)+\mathcal{E}\left(\frac{1-e^{-(\gamma-\mathrm{i}\Delta)t}}{\gamma-\mathrm{i}\Delta}\right),\\
\langle\delta\tilde{a}^{2}\rangle_{\text{I}}(t) & =e^{-2(\gamma-\mathrm{i}\Delta)t}\langle\delta\tilde{a}^{2}\rangle_{\text{I}}(0),\\
\langle\delta\tilde{a}^{\dagger}\delta\tilde{a}\rangle_{\text{I}}(t) & =e^{-2\gamma t}\langle\delta\tilde{a}^{\dagger}\delta\tilde{a}\rangle_{\text{I}}(0)+2\bar{n}\gamma\int_{0}^{t}dt'e^{-2\gamma(t-t')}=e^{-2\gamma t}\langle\delta\tilde{a}^{\dagger}\delta\tilde{a}\rangle_{\text{I}}(0)+\bar{n}\left(1-e^{-2\gamma t}\right),
\end{align}
\end{subequations}leading to the asymptotic solutions
\begin{equation}
\lim_{t\rightarrow\infty}\langle\hat{a}\rangle_{\text{I}}(t)=\frac{\mathcal{E}}{\gamma+\mathrm{i}\Delta},\hspace{1em}\lim_{t\rightarrow\infty}\langle\delta\tilde{a}^{2}\rangle_{\text{I}}(t)=0,\hspace{1em}\lim_{t\rightarrow\infty}\langle\delta\tilde{a}^{\dagger}\delta\tilde{a}\rangle_{\text{I}}(t)=\bar{n}.
\end{equation}
Finally, note that in our case $\hat{U}_{\text{c}}^{\dagger}(t)\hat{a}\hat{U}_{\text{c}}(t)=e^{-\mathrm{i}\omega_{\text{L}}t}\hat{a}$,
and hence, the relations between the true expectation values and the
interaction-picture expectation values are very simple in this case
\begin{equation}
\langle\hat{a}\rangle(t)=e^{-\mathrm{i}\omega_{\text{L}}t}\langle\hat{a}\rangle_{\text{I}}(t),\hspace{1em}\langle\delta\hat{a}^{2}\rangle(t)=e^{-2\mathrm{i}\omega_{\text{L}}t}\langle\delta\tilde{a}^{2}\rangle_{\text{I}}(t),\hspace{1em}\langle\delta\hat{a}^{\dagger}\delta\hat{a}\rangle(t)=\langle\delta\tilde{a}^{\dagger}\delta\tilde{a}\rangle_{\text{I}}(t),
\end{equation}
leading to the asymptotic moments
\begin{equation}
\lim_{t\rightarrow\infty}\left\langle \hat{a}(t)\right\rangle =\frac{\mathcal{E}}{\gamma+\mathrm{i}\Delta}e^{-\mathrm{i}\omega_{\text{L}}t},\hspace{1em}\lim_{t\rightarrow\infty}\left\langle \delta\hat{a}^{2}(t)\right\rangle =0,\hspace{1em}\lim_{t\rightarrow\infty}\left\langle \delta\hat{a}^{\dagger}(t)\delta\hat{a}(t)\right\rangle =\bar{n}.
\end{equation}
These are exactly the same expressions that we found with the quantum
Langevin equation in Section \ref{CavityExampleSS}.

\subsubsection{Extension to many modes and several environments\label{Sec:MultimodeCavity}}

\textbf{Soon...}

\subsubsection{Relation of the model parameters to physical parameters\label{Sec:PhysicalParametersDrivingDamping}}

Apart from the thermal number of excitations and intracavity parameters
such as the cavity frequency, the quantum Langevin and master equations
of an open cavity turned out to have only two more basic parameters,
$\mathcal{A}(t)$ and $\gamma$, describing the rates at which coherent
light is injected into the cavity and intracavity light is lost through
the partially transmitting mirror, respectively. In this section we
prove that these model parameters are connected to relevant physical
parameters like the transmissivity of the mirror $\mathcal{T}$ and
the power of the injected laser beam $P_{\text{inj}}$ through the
expressions (\ref{gamma_fromT}) and (\ref{El_fromPinj}) provided
above.

\textbf{Loss rate. }To determine the relation between $\gamma$ and
$\mathcal{T}$ it is enough to consider the undriven case ($\mathcal{A}=0$)
and follow classical arguments. We then neglect quantum fluctuations
and focus on the classical part of the intracavity field. Using (\ref{Acavity-1})
and (\ref{a(t)_HeisenbergExample}), the classical vector potential
of the cavity field of the mode of interest can then be written as
\begin{equation}
\mathbf{A}(z,t)=\bigl\langle\hat{\mathbf{A}}(z,t)\bigr\rangle=\mathbf{e}_{x}\sqrt{\frac{\hbar}{\varepsilon_{0}LS\omega_{\text{c}}}}e^{-(\gamma+\mathrm{i}\omega_{\text{c}})t}\alpha(0)\sin\left(k_{\text{c}}z\right)+\text{c.c.},\label{ClassicalIntracavityA}
\end{equation}
where $\alpha(0)=\langle\hat{a}(0)\rangle$ is the initial classical
amplitude of the cavity field. On the other hand, note that every
roundtrip a fraction $\mathcal{R}^{1/2}$ is lost through the cavity
mirror, with reflectivity $\mathcal{R}=1-\mathcal{T}$. Hence, after
$n\in\mathbb{N}$ roundtrips, we have the relation
\begin{equation}
\mathbf{A}\left(z,t_{n}\right)=\mathcal{R}^{n/2}\mathbf{A}\left(z,0\right)\hspace{1em}\Rightarrow\hspace{1em}e^{-(\gamma+\mathrm{i}\omega_{\text{c}})t_{n}}=\mathcal{R}^{n/2}\alpha(0),
\end{equation}
where $t_{n}=2nL/c$ is the time lapsed after $n$ roundtrips. Taking
into account that $\exp(\mathrm{i}\omega_{\text{c}}t_{n})=1$ by definition
of the resonance frequencies ($\omega_{\text{c}}=m\pi c/L$ for some
$m\in\mathbb{N}$), and assuming $\mathcal{T}\ll1$ so that $\ln\mathcal{R}=\log(1-\mathcal{T})\approx-\mathcal{T}$,
we obtain
\begin{equation}
\gamma=-\frac{c}{4L}\ln\mathcal{R}\simeq\frac{c\mathcal{T}}{4L}\text{,}\label{Gamma}
\end{equation}
as we wanted to prove.

\textbf{Injection parameter}. We follow again classical arguments,
but considering this time the injection of a resonant monochromatic
field (later we extend it to more spectral components), for which
$\langle\hat{b}_{0}(\omega)\rangle=\alpha(\omega_{\text{c}})\delta(\omega-\omega_{\text{c}})$,
so that $\mathcal{A}(t)=\mathcal{E}e^{-\mathrm{i}\omega_{\text{c}}t}$
with $\mathcal{E}=-\sqrt{\gamma/\pi}\alpha(\omega_{\text{c}})$, see
Eq. (\ref{A_polychromatic}). We remind that $\hat{b}_{0}(\omega)$
are the external annihilation operators the origin of time. We proceed
then by relating the coherent amplitude $\alpha(\omega_{\text{c}})$
to the power and phase of the injected beam. Since these are all quantities
that do not relate in any way to the cavity, it is enough to consider
the limit in which there is no interaction between the intracavity
and external fields, and hence the external field operators evolve
as $\hat{b}(\omega;t)=\hat{b}_{0}(\omega)e^{-\mathrm{i}\omega t}$.
Hence, according to (\ref{Aext}), the part of the classical vector
potential of the external field propagating towards the cavity, say
$\mathbf{A}_{\leftarrow}\left(z,t\right)$, reads
\begin{equation}
\mathbf{A}_{\leftarrow}\left(z,t\right)=\frac{\mathrm{i}}{2}\mathbf{e}_{x}\int_{0}^{\infty}d\omega\sqrt{\frac{\hbar}{\pi c\varepsilon_{0}S\omega}}\langle\hat{b}(\omega;t)\rangle e^{-\mathrm{i}k(z-L)}+\text{c.c.}=-\mathbf{e}_{x}\sqrt{\frac{\hbar}{\pi c\varepsilon_{0}S\omega_{\text{c}}}}|\alpha(\omega_{\text{c}})|\sin[\omega_{\text{c}}t+k_{\text{c}}(z-L)-\phi],
\end{equation}
where $\phi$ is the phase of $-\alpha(\omega_{\text{c}})$, which
we show next to be the phase of the laser's electric field. Indeed,
the electric and magnetic fields corresponding to this vector potential
read\begin{subequations}
\begin{align}
\mathbf{E}_{\leftarrow}\left(z,t\right) & =-\partial_{t}\mathbf{A}_{\leftarrow}\left(z,t\right)=\mathbf{e}_{x}\sqrt{\frac{\hbar\omega_{\text{c}}}{\pi c\varepsilon_{0}S}}|\alpha(\omega_{\text{c}})|\cos[\omega_{\text{c}}t+k_{\text{c}}(z-L)-\phi],\label{E_left_power}\\
\mathbf{B}_{\leftarrow}\left(z,t\right) & =\boldsymbol{\nabla}\times\mathbf{A}_{\leftarrow}\left(z,t\right)=-\mathbf{e}_{y}\sqrt{\frac{\hbar\omega_{\text{c}}}{\pi c^{3}\varepsilon_{0}S}}|\alpha(\omega_{\text{c}})|\cos[\omega_{\text{c}}t+k_{\text{c}}(z-L)-\phi].
\end{align}
\end{subequations}Consider now the instantaneous power impinging
the mirror, which we obtain by integrating the absolute value of the
Poynting vector $[\mathbf{E}_{\leftarrow}(z,t)\times\mathbf{B}_{\leftarrow}(z,t)]/\mu_{0}$
at $z=L$, obtaining 
\begin{equation}
P_{\leftarrow}(t)=\int_{\mathbb{R}^{2}}dxdy\frac{|\mathbf{E}_{\leftarrow}(z,t)\times\mathbf{B}_{\leftarrow}(z,t)|}{\mu_{0}}=\frac{\hbar\omega_{\text{c}}}{\pi}|\alpha(\omega_{\text{c}})|^{2}\cos^{2}[\omega_{\text{c}}t-\phi].
\end{equation}
Instead of the instantaneous power, it is common to use the power
averaged over an optical cycle, since resolving optical oscillations
is not possible in standard power measurements. Let us denote this
by $P_{\text{inj}}=\frac{\omega}{2\pi}\int_{-\omega/\pi}^{+\omega/\pi}dtP_{\leftarrow}(t)=\hbar\omega_{\text{c}}|\alpha(\omega_{\text{c}})|^{2}/2\pi$,
from which we obtain the relation
\begin{equation}
\mathcal{E}=\sqrt{\frac{\gamma}{\pi}}|\alpha(\omega_{\text{c}})|e^{\mathrm{i}\phi}=\sqrt{\frac{2\gamma}{\hbar\omega_{\text{c}}}P_{\text{inj}}}e^{\mathrm{i}\phi},
\end{equation}
which coincides with the relation that we wanted to prove, Eq. (\ref{El_fromPinj}),
particularized to a single frequency component.

Consider now the effect of having several frequency components in
the injected light. We are interested in the case in which each component
can be resolved by the cavity, that is, their spectral separation
is larger than $\gamma$ (but still below the free spectral range
$\pi c/L$, so that different components address the same cavity resonance).
It feels natural that, under such circumstances, the total power will
just be the direct sum of the power of each spectral component, which
can also have their own independent phase (for example, each spectral
component may come from a different laser). This brings us directly
to the expression provided in Eq. (\ref{El_fromPinj}).

\subsection{Incoherent atomic processes}

Having studied the model for an open cavity, let us now analyze another
equally paradigmatic and important open quantum optical system: an
atom interacting with the electromagnetic field in free space. As
we will see, this will allow us to introduce master equations following
a different route to the Markov approximation of even more general
use, and discuss some very relevant physics such as spontaneous emission
and the Lamb shift.

\subsubsection{Revisiting field quantization in free space}

In the previous section we made a continuous model for the field outside
the cavity. In principle, we could consider the interaction of an
atom with an electromagnetic continuum by following a similar route
in which we place the atom in a cavity and make the length go to infinity.
However, the way that we have quantized the cavity, with a mirror
place at $z=0$ would naturally lead to the interaction of an atom
with a field in half the real line, not the whole (1D) space. Moreover,
without reflective boundaries, modes with opposite wave vector should
be independent, while quantization in a cavity always assumes that
they cannot be excited independently. Hence, it is instructive and
timely to introduce here the quantization of the electromagnetic field
in free space starting from scratch. In addition, this will allow
us to learn a slightly different route towards quantization, based
on the normal variable of the harmonic oscillator instead of the position
and momentum.

Let us recall that the normal variable of an oscillator of frequency
$\omega$, mass $m$, position $q(t)$, and momentum $p(t)$, is defined
as $\nu(t)=q(t)+\mathrm{i}p(t)/m\omega$. Using the canonical equations
of motion (\ref{CanonicalEquationsHO}) of position and momentum,
it is easily proven to satisfy the evolution equation $\dot{\nu}=-\mathrm{i}\omega\nu$.
In addition, the Hamiltonian of the oscillator is written in terms
of this variable as $H_{\text{o}}=m\omega^{2}\nu^{*}\nu/2$. On the
other hand, in terms of true position and momentum operators instead
of quadratures, the annihilation operator reads $\hat{a}=\sqrt{m\omega/2\hbar}(\hat{q}+\mathrm{i}\hat{p}/m\omega)$,
which means that quantization can be carried away by symmetrizing
the Hamiltonian with respect $\nu$ and $\nu^{*}$ as $H_{\text{o}}=m\omega^{2}(\nu^{*}\nu+\nu\nu^{*})/4$,
making then the correspondence $\{\nu,\nu^{*}\}\rightarrow\sqrt{2\hbar/m\omega}\{\hat{a},\hat{a}^{\dagger}\}$.
Note that the symmetrization of the Hamiltonian can be overlooked
if we don't need to keep the constant $\hbar\omega/2$ contribution
of vacuum to the oscillator's energy. We will use these formulation
in what follows.

Within our usual quasi-1D approximation, let us start by expanding
the vector potential in plane waves with all the possible wave vectors,
that is,
\begin{equation}
\mathbf{A}(z,t)=\mathbf{e}_{x}\int_{\mathbb{R}}dk\mathcal{A}(k,t)e^{\mathrm{i}kz},\label{AfreespaceExpansion}
\end{equation}
where $\mathcal{A}(k,t)$ are complex expansion coefficients. Note
that we are not doubling the number of degrees of freedom of the electromagnetic
field by taking complex coefficients, since the amplitudes are related
by $\mathcal{A}(-k,t)=\mathcal{A}^{*}(k,t)$. This is easily proven
by using the reality of the vector potential, $\mathbf{A}=\mathbf{A}^{*}$,
which implies\footnote{In the following, like in the last equality of this equation, we will
be continuously inverting the sign of the wave vector as a variable
change in integrals, leading to $\int_{-\infty}^{+\infty}dkf(k)=\int_{-\infty}^{+\infty}dkf(-k)$.}
\begin{equation}
\int_{-\infty}^{+\infty}dk\mathcal{A}(k,t)e^{\mathrm{i}kz}=\int_{-\infty}^{+\infty}dk\mathcal{A}^{*}(k,t)e^{-\mathrm{i}kz}=\int_{-\infty}^{+\infty}dk\mathcal{A}^{*}(-k,t)e^{\mathrm{i}kz}.
\end{equation}
Operating with $\int_{-\infty}^{+\infty}dze^{\mathrm{i}k'z}$ and
using the definition of the Dirac delta function, $\int_{-\infty}^{+\infty}dze^{\mathrm{i}(k+k')z}=2\pi\delta(k+k')$,
we obtain the desired result. Plugging (\ref{AfreespaceExpansion})
into the wave equation for the vector potential, $(\partial_{t}^{2}-c^{2}\partial_{z}^{2})\mathbf{A}(z,t)=0$
, and applying $\int_{-\infty}^{+\infty}dze^{\mathrm{i}k'z}$ on the
expression, we obtain
\begin{equation}
\ddot{\mathcal{A}}(k,t)+c^{2}k^{2}\mathcal{A}(k,t)=0.
\end{equation}
While this is the equation of motion of a harmonic oscillator, the
amplitude $\mathcal{A}(k,t)$ cannot be interpreted as the position
of the oscillator because it can be complex. However, we next show
that a simple manipulation allows us to relate it to the normal variable
of a harmonic oscillator, which is complex. To this aim, we rewrite
the previous expression as 
\begin{equation}
(\partial_{t}+\mathrm{i}c|k|)(\partial_{t}-\mathrm{i}c|k|)\mathcal{A}(k,t)=0,
\end{equation}
so we can easily define a variable that evolves just like the normal
variable of a harmonic oscillator of frequency $c|k|$ as follows
\begin{equation}
\nu(k,t)=N_{k}^{-1}\left[\dot{\mathcal{A}}(k,t)-\mathrm{i}c|k|\mathcal{A}(k,t)\right]\hspace{1em}\Rightarrow\hspace{1em}\dot{\nu}(k,t)=-\mathrm{i}c|k|\nu(k,t).\label{FreeElectromagneticNormalVariable}
\end{equation}

Next we find the normalization factor $N_{k}$ by imposing that the
electromagnetic energy must be equal to the sum of the Hamiltonians
of each of the harmonic oscillators associated to the wave vectors
$k$. In particular, we use the expression 
\begin{equation}
E_{\text{em}}(t)=\frac{\varepsilon_{0}}{2}\int_{\mathbb{R}^{3}}d^{3}\mathbf{r}\left[\mathbf{E}^{2}(z,t)+c^{2}\mathbf{B}^{2}(z,t)\right],
\end{equation}
for the electromagnetic energy. Taking into account that
\begin{equation}
\mathbf{E}(z,t)=-\partial_{t}\mathbf{A}=-\mathbf{e}_{x}\int_{\mathbb{R}}dk\dot{\mathcal{A}}(k,t)e^{\mathrm{i}kz}\hspace{1em}\text{and}\hspace{1em}\mathbf{B}(z,t)=\boldsymbol{\nabla}\times\mathbf{A}=-\mathrm{i}\mathbf{e}_{x}\int_{\mathbb{R}}dkk\mathcal{A}(k,t)e^{\mathrm{i}kz},\label{EandBfreeIntermediate}
\end{equation}
the electric and magnetic contributions to the energy can be written
as\begin{subequations}
\begin{align}
\int_{-\infty}^{+\infty}dz\mathbf{E}^{2}(z,t) & =\int_{\mathbb{R}}dk\int_{\mathbb{R}}dk'\dot{\mathcal{A}}(k,t)\dot{\mathcal{A}}(k',t)\int_{-\infty}^{+\infty}dze^{\mathrm{i}(k+k')z}=2\pi\int_{\mathbb{R}}dk\dot{\mathcal{A}}(k,t)\dot{\mathcal{A}}(-k,t)=2\pi\int_{\mathbb{R}}dk|\dot{\mathcal{A}}(k,t)|^{2},\\
\int_{-\infty}^{+\infty}dz\mathbf{B}^{2}(z,t) & =-\int_{\mathbb{R}}\negthickspace dk\int_{\mathbb{R}}\negthickspace dk'\mathcal{A}(k,t)\mathcal{A}(k',t)kk'\int_{-\infty}^{+\infty}\negthickspace dze^{\mathrm{i}(k+k')z}=2\pi\int_{\mathbb{R}}\negthickspace dkk^{2}\mathcal{A}(k,t)\mathcal{A}(-k,t)=2\pi\int_{\mathbb{R}}\negthickspace dkk^{2}|\mathcal{A}(k,t)|^{2},
\end{align}
\end{subequations}leading to
\begin{equation}
E_{\text{em}}(t)=\varepsilon_{0}S\pi\int_{\mathbb{R}}dk\left[|\dot{\mathcal{A}}(k,t)|^{2}+c^{2}k^{2}|\mathcal{A}(k,t)|^{2}\right].
\end{equation}
In order to write this expression as a function of the normal variable
$\nu(k,t)$, we first invert (\ref{FreeElectromagneticNormalVariable})
as
\begin{equation}
\dot{\mathcal{A}}(k,t)=\frac{1}{2}\left[N_{k}\nu(k,t)+N_{-k}^{*}\nu^{*}(-k,t)\right],\hspace{1em}\mathcal{A}(k,t)=\frac{\mathrm{i}}{2c|k|}\left[N_{k}\nu(k,t)-N_{-k}^{*}\nu^{*}(-k,t)\right],\label{AtoNormal}
\end{equation}
leading to\begin{subequations}
\begin{align}
|\dot{\mathcal{A}}(k,t)|^{2} & =\frac{1}{4}\left[|N_{k}|^{2}|\nu(k,t)|^{2}+|N_{-k}|^{2}|\nu(-k,t)|^{2}+N_{k}N_{-k}^{*}\nu(k,t)\nu^{*}(-k,t)+N_{k}^{*}N_{-k}\nu^{*}(k,t)\nu(-k,t)\right],\\
c^{2}k^{2}|\mathcal{A}(k,t)|^{2} & =\frac{1}{4}\left[|N_{k}|^{2}|\nu(k,t)|^{2}+|N_{-k}|^{2}|\nu(-k,t)|^{2}-N_{k}N_{-k}^{*}\nu(k,t)\nu^{*}(-k,t)-N_{k}^{*}N_{-k}\nu^{*}(k,t)\nu(-k,t)\right],
\end{align}
\end{subequations}which implies
\begin{equation}
E_{\text{em}}(t)=\frac{\varepsilon_{0}S\pi}{2}\int_{\mathbb{R}}dk\left[|N_{k}|^{2}|\nu(k,t)|^{2}+|N_{-k}|^{2}|\nu(-k,t)|^{2}\right]=\varepsilon_{0}S\pi\int_{\mathbb{R}}dk|N_{k}|^{2}\nu^{*}(k,t)\nu(k,t).
\end{equation}
Choosing the normalization factor $N_{k}=-\mathrm{i}c|k|/\sqrt{2\pi\varepsilon_{0}S}$
(note that we can choose the phase at will, and this one will turn
out to be convenient for future purposes), we then find the electromagnetic
energy
\begin{equation}
E_{\text{em}}(t)=\int_{\mathbb{R}}dk\frac{c^{2}k^{2}}{2}\nu^{*}(k,t)\nu(k,t),
\end{equation}
which is the correct classical Hamiltonian for a collection of independent
harmonic oscillators labeled by a continuous index $k\in\mathbb{R}$
and with frequency $c|k|$.

Quantization is then performed through the correspondence $\{\nu(k,t),\nu^{*}(k,t)\}\rightarrow\sqrt{2\hbar/c|k|}\{\hat{a}(k,t),\hat{a}^{\dagger}(k,t)\}_{k\in\mathbb{R}}$
, with annihilation and creation operators satisfying canonical commutation
relations
\begin{equation}
[\hat{a}(k,t),\hat{a}(k',t)]=0,\quad[\hat{a}(k,t),\hat{a}^{\dagger}(k',t)]=\delta(k-k')\label{CCRwavevectors}
\end{equation}
, which in the previous section were proven to be the right commutation
relations for a set of independent oscillators labeled by a continuous
index (wave vector in our case). The Hamiltonian for light in free
(1D) space takes then the form
\begin{equation}
\hat{H}_{\text{L}}=\int_{-\infty}^{+\infty}dk\hbar c|k|\hat{a}^{\dagger}(k)\hat{a}(k)=\int_{0}^{+\infty}dk\hbar ck\left[\hat{a}^{\dagger}(k)\hat{a}(k)+\hat{a}^{\dagger}(-k)\hat{a}(-k)\right],\label{FreeSpaceHlight}
\end{equation}
with quantized electric field
\begin{equation}
\hat{\mathbf{E}}(z)=\mathrm{i}\mathbf{e}_{x}\int_{-\infty}^{+\infty}dk\sqrt{\frac{\hbar c|k|}{4\pi\varepsilon_{0}S}}\left[\hat{a}(k)e^{\mathrm{i}kz}-\hat{a}^{\dagger}(k)e^{-\mathrm{i}kz}\right]=\mathrm{i}\mathbf{e}_{x}\int_{0}^{+\infty}dk\sqrt{\frac{\hbar ck}{4\pi\varepsilon_{0}S}}\left[\hat{a}(k)e^{\mathrm{i}kz}+\hat{a}(-k)e^{-\mathrm{i}kz}-\text{H.c.}\right],\label{FreeSpaceE}
\end{equation}
where the right hand sides stress the fact that for a given frequency
$c|k|$ there are two contributions from modes with wave vector $\pm|k|$.
It is important to keep this in mind, so we don't confuse the integration
over negative wave vectors with the integration over fictitious negative
frequencies that we introduced in (\ref{HintExtendedNegativeFreqs})
as a convenient approximation.

\subsubsection{Atomic master equation in free space and spontaneous emission}

We can now consider the interaction of an atom with the electromagnetic
continuum in free space. As we explained in Section \ref{Sec:LigheAtomInteraction},
within the dipole approximation, we can write the interaction Hamiltonian
as $\hat{H}_{\text{LM}}=e\hat{\mathbf{E}}(z_{0})\cdot\hat{\mathbf{r}}_{\text{A}}$,
where $z_{0}$ is the position of the atom's center of mass (assumed
fixed) and $\hat{\mathbf{r}}_{\text{A}}$ is the atomic relative coordinate.
Using the expression for the electric field given above, focusing
on two atomic levels $\{|e\rangle,|g\rangle\}$ as usual for simplicity
(everything is easily generalized to more atomic levels under reasonable
approximations, similarly to the generalization of the single-mode
cavity to the multi-mode one), we then find
\begin{equation}
\hat{H}_{\text{LM}}=\mathrm{i}\int_{\mathbb{R}}dke\sqrt{\frac{\hbar c|k|}{4\pi\varepsilon_{0}S}}\left[\hat{a}(k)e^{\mathrm{i}kz_{0}}-\hat{a}^{\dagger}(k)e^{-\mathrm{i}kz_{0}}\right]\left[\langle g|\hat{x}_{A}|e\rangle\hat{\sigma}+\langle g|\hat{x}_{A}|e\rangle^{*}\hat{\sigma}^{\dagger}\right].
\end{equation}
Performing the rotating-wave approximation to neglect terms such as
$\hat{a}(k)\hat{\sigma}$, and defining the couplings
\begin{equation}
g(k)=-e\sqrt{\frac{c|k|}{4\pi\hbar\varepsilon_{0}S}}\langle g|\hat{x}_{A}|e\rangle e^{-\mathrm{i}kz_{0}},\label{UnphysCoupling}
\end{equation}
 which we assume to be real for simplicity, in particular setting
$z_{0}=0$ and taking $\langle g|\hat{x}_{A}|e\rangle\in\mathbb{R}$,
we obtain 
\begin{equation}
\hat{H}_{\text{LM}}=\mathrm{i}\hbar\int_{\mathbb{R}}dkg(k)\left[\hat{a}^{\dagger}(k)\hat{\sigma}-\hat{a}(k)\hat{\sigma}^{\dagger}\right].\label{HLMatomFreeSpace}
\end{equation}
The final Hamiltonian is then given by $\hat{H}=\hat{H}_{\text{A}}+\hat{H}_{\text{L}}+\hat{H}_{\text{LM}}$,
where $\hat{H}_{\text{L}}$ is given in Eq. (\ref{FreeSpaceHlight})
and $\hat{H}_{\text{A}}=\hbar\varepsilon\hat{\sigma}_{z}/2$ is the
atomic Hamiltonian within the two-level approximation.

Note that this simple model leads to an unphysical situation in which
the coupling $g(k)$ increases arbitrarily with the light frequency,
diverging for $|k|\rightarrow\infty$. Of course, in reality this
is not the case: the coupling remains close to $g(\varepsilon/c)$
in all the relevant frequency interval around the atomic transition
$\varepsilon$, and includes a natural cut off for large and small
frequencies that is not captured with the simple dipole Hamiltonian
$e\hat{\mathbf{E}}(z_{0})\cdot\hat{\mathbf{r}}_{\text{A}}$ (e.g.,
for really large frequencies such as X rays the wavelength enters
atomic scales, so the point-like approximation for the atom breaks
down). In any case, let us proceed with the expressions we have but
keeping this issue in mind.

In the following we will find a master equation for the reduced atomic
state $\hat{\rho}_{\text{A}}$. If we perform a frequency-independent
approximation $g(k)=g(\varepsilon/c)\equiv\sqrt{\gamma c/2\pi}$ for
the coupling (the factor 2 in the denominator comes from the fact
that now for each frequency $\hbar c|k|$ we have two modes with opposite
wave vectors $\pm|k|$ that contribute to the decay, while the factor
$c$ appears because we are labeling the modes by the wave vector
instead of the frequency), we can proceed just like we did with the
open cavity, finding exactly the same results at each step with the
obvious replacements ($\omega_{\text{c}}\rightarrow\varepsilon$,
$\hat{a}\rightarrow\hat{\sigma}$,...), obtaining the master equation
\begin{equation}
\partial_{t}\hat{\rho}_{\mathrm{A}}=\left[\frac{\hat{H}_{\mathrm{A}}+\hat{H}_{\mathrm{inj}}(t)}{\mathrm{i}\hbar},\hat{\rho}_{\mathrm{A}}\right]+(\bar{n}+1)\gamma\left(2\hat{\sigma}\hat{\rho}_{\mathrm{A}}\hat{\sigma}^{\dagger}-\hat{\sigma}^{\dagger}\hat{\sigma}\hat{\rho}_{\mathrm{A}}-\hat{\rho}_{\mathrm{A}}\hat{\sigma}^{\dagger}\hat{\sigma}\right)+\bar{n}\gamma\left(2\hat{\sigma}^{\dagger}\hat{\rho}_{\mathrm{A}}\hat{\sigma}-\hat{\sigma}\hat{\sigma}^{\dagger}\hat{\rho}_{\mathrm{A}}-\hat{\rho}_{\mathrm{A}}\hat{\sigma}\hat{\sigma}^{\dagger}\right),\label{AtomicMasterEq}
\end{equation}
where in this case
\begin{equation}
\hat{H}_{\mathrm{inj}}(t)=\mathrm{i}\hbar\left(\mathcal{A}(t)\hat{\sigma}^{\dagger}e^{-\mathrm{i}\omega_{\text{L}}t}-\mathcal{A}^{*}(t)\hat{\sigma}e^{\mathrm{i}\omega_{\text{L}}t}\right),\label{HinjAtom}
\end{equation}
is the Hamiltonian related to the coherent contribution (laser) of
the external fields. Just as in the case of the cavity, if the injection
is composed of multiple monochromatic components with wave vectors
$\{k_{\ell}\}_{\ell=1,2,...,L}$ and corresponding frequencies $\omega_{\ell}=c|k_{\ell}|$,
we have
\begin{equation}
\mathcal{A}(t)=\sum_{\ell=1}^{L}\mathcal{E}_{\ell}e^{-\mathrm{i}\omega_{\ell}t},
\end{equation}
where in this case the injection rates $\mathcal{E}_{\ell}$ can be
easily related the power $P_{\text{inj},\ell}$ (averaged over an
optical cycle) of each wave vector component by 
\begin{equation}
\mathcal{E}_{\ell}=-\mathrm{i}\frac{e}{\hbar}\langle e|\hat{x}_{A}|g\rangle\sqrt{\frac{P_{\text{inj},\ell}}{2\varepsilon_{0}cS}}e^{\mathrm{i}(k_{\ell}z_{0}+\phi_{\ell})},\label{EinjAtom}
\end{equation}
where $\phi_{\ell}$ is the electric field's phase of the corresponding
component, and we assume that the spectral separation between the
different components is larger than $\gamma$, but are still close
enough to drive the same atomic transition. This expression is easily
proven by noting that the injection Hamiltonian simply corresponds
to the interaction of the atomic dipole with the classical part of
the electric field, that is, the semiclassical Rabi Hamiltoinan $\hat{H}_{\mathrm{inj}}(t)=e\mathbf{E}(z_{0})\cdot\hat{\mathbf{r}}_{\text{A}}$.
On the other hand, for each wave vector component, the electric field
was shown in (\ref{E_left_power}) to be written as $\mathbf{E}\left(z_{0},t\right)=\sum_{\ell=1}^{L}\mathbf{e}_{x}\sqrt{P_{\text{inj},\ell}/c\varepsilon_{0}S}\cos(\omega_{\ell}t-k_{\ell}z_{0}-\phi_{\ell})$,
which substituted in the previous expression for $\hat{H}_{\mathrm{inj}}(t)$,
and after performing the rotating-wave approximation, leads to (\ref{HinjAtom})
and (\ref{EinjAtom}).

In order to interpret this master equation and its effect on the dynamics
of the atom, let us next focus on the specific situation of an atom
in the electromagnetic vacuum, that is, at zero temperature ($\bar{n}=0$)
and with no laser light shinned on it ($\mathcal{A}=0$). Following
what we saw in Section \ref{Sec:TwoLevelApprox}, we write the atomic
state in terms of its (complex) Bloch vector components $b_{z}(t)=\langle\hat{\sigma}_{z}\rangle(t)$
and $b(t)=\langle\hat{\sigma}\rangle(t)$ as
\[
\hat{\rho}_{\text{A}}(t)=\frac{1}{2}\left(\hat{I}+b_{z}(t)\hat{\sigma}_{z}+2b^{*}(t)\hat{\sigma}+2b(t)\hat{\sigma}^{\dagger}\right).
\]
The complex Bloch equations are easily found from (\ref{GenOpenExpecEvo})
and (\ref{AtomicMasterEq}) as\begin{subequations}
\begin{align}
\partial_{t}b=\partial_{t}\langle\hat{\sigma}\rangle & =-\mathrm{i}\frac{\varepsilon}{2}\left\langle \left[\hat{\sigma},\hat{\sigma}_{z}\right]\right\rangle +\gamma\left\langle \left[\hat{\sigma}^{\dagger},\hat{\sigma}\right]\hat{\sigma}\right\rangle =-\mathrm{i}\varepsilon\left\langle \hat{\sigma}\right\rangle +\gamma\left\langle \hat{\sigma}_{z}\hat{\sigma}\right\rangle =-(\gamma+\mathrm{i}\varepsilon)\left\langle \hat{\sigma}\right\rangle =-(\gamma+\mathrm{i}\varepsilon)b,\\
\partial_{t}b_{z}=\partial_{t}\langle\hat{\sigma}_{z}\rangle & =\gamma\left(\left\langle \left[\hat{\sigma}^{\dagger},\hat{\sigma}_{z}\right]\hat{\sigma}\right\rangle +\left\langle \hat{\sigma}^{\dagger}\left[\hat{\sigma}_{z},\hat{\sigma}\right]\right\rangle \right)=-4\gamma\left\langle \hat{\sigma}^{\dagger}\hat{\sigma}\right\rangle =-2\gamma\left(b_{z}+1\right),
\end{align}
\end{subequations}where we have used $\hat{\sigma}_{z}\hat{\sigma}=-\hat{\sigma}$
and $\hat{\sigma}_{z}=2\hat{\sigma}^{\dagger}\hat{\sigma}-1$, leading
to the solutions\begin{subequations}
\begin{align}
b(t) & =b(0)e^{-(\gamma+\mathrm{i}\varepsilon)t}\hspace{1em}\underset{\gamma t\rightarrow\infty}{\longrightarrow}\hspace{1em}0,\\
b_{z}(t) & =\left[b_{z}(0)+1\right]e^{-2\gamma t}-1\hspace{1em}\underset{\gamma t\rightarrow\infty}{\longrightarrow}\hspace{1em}-1,
\end{align}
\end{subequations}corresponding to the ground state. Hence, no matter
the initial state of the atom, it eventually emits a photon and ends
up in the ground state. We call this effect \emph{spontaneous emission}:
the electromagnetic vacuum fluctuations trigger emission of the atomic
excitation as a photon, which gets lost in the continuum of modes
and never comes back to the atom (shortly we will see this even more
explicitly). It's interesting to consider the case in which the atom
is initially in the excited state, $b_{z}(0)=1$ and $b(0)=0$. In
this case the state can be written as 
\begin{equation}
\hat{\rho}_{\text{A}}(t)=e^{-2\gamma t}|e\rangle\langle e|+\left(1-e^{-2\gamma t}\right)|g\rangle\langle g|,\label{AtomicStateEvoSpontaneousEmission}
\end{equation}
showing that starting in the excited state it decays to the ground
state by going through all possible mixed states of them (e.g., reaching
the maximally mixed state at time $t_{\text{MM}}=\ln(2)/2\gamma$).
This shows that except at the initial and final states of the evolution,
the atom becomes correlated with the electromagnetic field. This is
a direct evidence of something that we mentioned right before performing
the Born approximation in Eq. (\ref{ExactIntegroDiff-AfterBorn}):
even though we assume that the state can be approximately factorized
as $\hat{\rho}_{\text{A}}(t)\otimes_{k}|0\rangle\langle0|$ in the
dynamical equation of the reduced atomic state, the resulting equation
still keeps track (to some degree) of the effect that correlations
have on the dynamics of the system.

In order to get a deeper understanding of how the correlation comes
into this scenario, we proceed now to solve analytically the full
dynamical problem including both the atom and the field. This is possible
because, just as in the Jaynes-Cummings model, the interaction Hamiltonian
(\ref{HLMatomFreeSpace}) conserves the total number of excitations,
as first identified by Wigner and Weisskopf, from whom the upcoming
derivation takes the name. Starting with the field in vacuum and the
atom in the excited state, we can only reach states with one photon
and the atom in the ground state, that is, we are bound to the single-excitation
subspace spanned by $\{|0,e\rangle,\hat{a}^{\dagger}(k)|0,g\rangle\}_{k\in\mathbb{R}}$.
Let us then expand the state at any as
\begin{equation}
|\psi(t)\rangle=\alpha(t)e^{-\mathrm{i}\varepsilon t/2}|0,e\rangle+\int_{-\infty}^{+\infty}dk\beta(k;t)e^{-\mathrm{i}(c|k|-\varepsilon/2)t}\hat{a}^{\dagger}(k)|0,g\rangle,
\end{equation}
where $\alpha(t)$ and $\beta(k;t)$ are complex expansion coefficients.
Let us next consider the Schrödinger equation $\mathrm{i}\hbar\partial_{t}|\psi(t)\rangle=\hat{H}|\psi(t)\rangle$,
where we remind that the Hamiltonian is given by , where the first
term is the atomic Hamiltonian ... while the other two are provided
in Eqs. (\ref{FreeSpaceHlight}) and (\ref{HLMatomFreeSpace}). Projecting
the Schrödinger equation onto\footnote{In order to evaluate the action of the Hamiltonian onto the state,
simply use the commutation relations (\ref{CCRwavevectors}) to bring
annihilation operators to the right when needed, and use the property
$\hat{a}(k)|0\rangle=0$. The projections are then found from $\langle0,a|0,a'\rangle=\delta_{aa'}$,
$\langle0,a|\hat{a}^{\dagger}(k)|0,a'\rangle=0=\langle0,a|\hat{a}(k)|0,a'\rangle$,
and $\langle0,a|\hat{a}(k)\hat{a}^{\dagger}(k')|0,a'\rangle=\delta_{aa'}\delta(k-k')$. } $|e,0\rangle$ or $\hat{a}^{\dagger}(k)|g,0\rangle$, we obtain\begin{subequations}
\begin{align}
\dot{\alpha} & =-\int_{-\infty}^{+\infty}dkg(k)e^{-\mathrm{i}(c|k|-\varepsilon)t}\beta(k),\\
\dot{\beta}(k) & =g(k)e^{\mathrm{i}(c|k|-\varepsilon)t}\alpha.
\end{align}
\end{subequations}We proceed by formally integrating the second equation
as
\begin{equation}
\beta(k;t)=\int_{0}^{t}dt'g(k)e^{\mathrm{i}(c|k|-\varepsilon)t'}\alpha(t'),\label{FormalBeta}
\end{equation}
where we already took $\beta(k;0)=0$ as the initial condition. Inserted
in the first equation, and performing the integration-variable change
$t'=t-\tau$, leads to
\begin{equation}
\dot{\alpha}=-\int_{0}^{t}d\tau\alpha(t-\tau)\underbrace{\int_{-\infty}^{+\infty}dkg^{2}(k)e^{-\mathrm{i}(c|k|-\varepsilon)\tau}}_{C^{*}(\tau)e^{\mathrm{i}\varepsilon\tau}}.\label{AlphaDot}
\end{equation}
The function $C(\tau)$ will be shown in the next section to play
a fundamental role in the Markovian properties of the dynamics. For
now, however, let's simply evaluate it under the same approximations
that we used to derive the master equation above: the flat form of
the coupling, $g(k)=\sqrt{\gamma c/2\pi}$, and extending the lower
integration limit of the frequency integrals to $-\infty$. We then
obtain 
\begin{equation}
C^{*}(\tau)\approx\frac{\gamma c}{2\pi}\int_{-\infty}^{+\infty}dke^{-\mathrm{i}c|k|\tau}=\frac{\gamma c}{\pi}\int_{0}^{+\infty}dke^{-\mathrm{i}ck\tau}\approx\frac{\gamma}{\pi}\int_{-\infty}^{+\infty}d\omega e^{-\mathrm{i}ck\tau}=2\gamma\delta(\tau),
\end{equation}
where we have made the variable change $\omega=ck$ in the second
to last step, and extended the lower integration limit. Inserting
this expression into (\ref{AlphaDot}) and using the property (\ref{DiracDeltaExtremeInterval})
of the Dirac delta function, we then get
\begin{equation}
\dot{\alpha}=-\gamma\alpha\hspace{1em}\Rightarrow\hspace{1em}\alpha(t)=e^{-\gamma t},
\end{equation}
where we have used the initial condition $\alpha(0)=1$. We can now
come back to (\ref{FormalBeta}) and find
\begin{equation}
\beta(k;t)=\sqrt{\frac{\gamma c}{2\pi}}\int_{0}^{t}dt'e^{-[\gamma-\mathrm{i}(c|k|-\varepsilon)]t'}=\sqrt{\frac{\gamma c}{2\pi}}\frac{1-e^{-[\gamma-\mathrm{i}(c|k|-\varepsilon)]t}}{\gamma-\mathrm{i}(c|k|-\varepsilon)}.
\end{equation}
Hence, the probability of finding a photon with wave vector $k$ at
time $t$ is given by
\begin{equation}
|\beta(k;t)|^{2}=\frac{\gamma c}{2\pi}\frac{1+e^{-2\gamma t}-2e^{-\gamma t}\cos[(c|k|-\varepsilon)t]}{\gamma^{2}+(c|k|-\varepsilon)^{2}},
\end{equation}
which depends only on the frequency $c|k|$ and not the propagation
direction, as expected (there are no preferred directions in free
space). Moreover, it is interesting to remark that asymptotically,
this probability has a Lorentzian shape centered around the atomic
resonance as a function of the frequency,
\begin{equation}
\lim_{\gamma t\rightarrow\infty}|\beta(k;t)|^{2}=\frac{\gamma c/2\pi}{\gamma^{2}+(c|k|-\varepsilon)^{2}}.
\end{equation}
Hence, the atom can only emit photons with frequencies close to its
resonance, similarly to how we saw that the cavity can only be excited
by lasers close to one of its resonances. Finally, note that
\begin{equation}
\int_{-\infty}^{+\infty}dk|\beta(k;t)|^{2}=\frac{2}{c}\int_{0}^{+\infty}d\omega|\beta(\omega/c;t)|^{2}\approx\frac{\gamma}{\pi}\int_{-\infty}^{+\infty}d\omega\frac{1+e^{-2\gamma t}-2e^{-\gamma t}\cos[(\omega-\varepsilon)t]}{\gamma^{2}+(\omega-\varepsilon)^{2}}=1-e^{-2\gamma t}=1-|\alpha(t)|^{2},
\end{equation}
which is further prove of the consistency of the various approximations
we have used, since this relation must be satisfied at all times (unitary
evolution must keep the state normalized).

Hence, we see that starting from the excited state with no photons,
we end up with the atom in the ground state and a single photon distributed
among the electromagnetic frequencies following a Lorentzian distribution.
In order for the photon to not be able to be reabsorbed by the atom,
it is crucial to have a continuum of modes. If instead we would have
a discrete set, the dynamics would reveal revivals of the atomic population.
The extreme example is be that in which all modes can be ignored but
one, obtaining then a Jaynes-Cummings model with perfectly periodic
quantum Rabi oscillations. A (sufficiently) larger number of modes
would make the population collapse, and then revive in some fashion
that would depend on the specific relation between the frequencies
of the mode.

Interestingly, at intermediate times, we have a superposition between
the excitation being in the atom or the field, leading to an entangled
state of these. Moreover, using the relations
\begin{equation}
\langle0|\psi(t)\rangle=\alpha(t)e^{-\mathrm{i}\varepsilon t}|e\rangle,\hspace{1em}\langle0|\hat{a}(k)|\psi(t)\rangle=\beta(k;t)e^{-\mathrm{i}(c|k|-\varepsilon)t}|g\rangle,
\end{equation}
we obtain the following reduced atomic state
\begin{align}
\hat{\rho}_{\text{A}}(t) & =\text{tr}_{\text{L}}\left\{ |\psi(t)\rangle\langle\psi(t)|\right\} =\langle0|\psi(t)\rangle\langle\psi(t)|0\rangle+\int_{-\infty}^{\infty}dk\langle0|\hat{a}(k)|\psi(t)\rangle\langle\psi(t)|\hat{a}^{\dagger}(k)|0\rangle\\
 & =|\alpha(t)|^{2}|e\rangle\langle e|+\left[\int_{-\infty}^{+\infty}dk|\beta(k;t)|^{2}\right]|g\rangle\langle g|,\nonumber 
\end{align}
which leads precisely to the state (\ref{AtomicStateEvoSpontaneousEmission})
obtained with the master equation. Hence, we see that the origin of
the mixed atomic state is in quantum correlations (entanglement) of
the system (atom) with the environment (field).

\subsubsection{Frequency-dependent coupling: Markov approximation and the Lamb shift}

While the frequency-independent approximation for the coupling simplifies
things immensely, there are systems in which it is not possible to
do it. Hence, it is interesting to consider how to proceed when the
frequency dependence of the coupling cannot be ignored. Moreover,
as we are about to see, a very important effect is found when going
beyond such a raw approximation: the Lamb shift, whose experimental
observation in 1947 played a major role in the theoretical development
quantum electrodynamics (and renormalization in particular).

Let us, for simplicity, consider zero temperature for the electromagnetic
field (hence all the modes are in vacuum initially), and laser ignore
the injection, which we know we can add at the end as Hamiltonian
(\ref{HinjAtom}) in any case. All the way up to the Born and non-backaction
approximations, we proceed in the same way as before. In particular,
we obtain 
\begin{equation}
\partial_{t}\hat{\rho}_{\text{A,I}}(t)=-\frac{1}{\hbar^{2}}\int_{0}^{t}d\tau\mathrm{tr}_{\mathrm{L}}\{[\hat{H}_{\text{I}}(t),[\hat{H}_{\text{I}}(t-\tau),\hat{\rho}_{\text{A,I}}(t-\tau)\bigotimes_{\forall k}|0\rangle\langle0|]]\},
\end{equation}
where $\hat{\rho}_{\text{A,I}}(t)=\text{tr}_{\text{L}}\{\hat{\rho}_{\text{I}}(t)\}$
is the reduced atomic state in the interaction picture, $\hat{\rho}_{\text{I}}(t)=\hat{U}_{\text{c}}^{\dagger}(t)\hat{\rho}(t)\hat{U}_{\text{c}}(t)$
is the transformed state of the whole system, and we define the change
of picture through the unitary transformation is $\hat{U}_{\text{c}}(t)=e^{\hat{H}_{0}/\text{i}\hbar}$,
with $\hat{H}_{0}=\hat{H}_{\text{A}}+\hat{H}_{\text{L}}$, leading
to the interaction picture Hamiltonian
\begin{equation}
\hat{H}_{\text{I}}(t)=\hat{U}_{\text{c}}^{\dagger}(t)\left[\hat{H}_{0}+\hat{H}_{\text{LM}}\right]\hat{U}_{\text{c}}(t)-\hat{H}_{0}=\mathrm{i}\hbar\int_{-\infty}^{+\infty}dkg(|k|)\left[\hat{a}^{\dagger}(k)\hat{\sigma}e^{\mathrm{i}(c|k|-\varepsilon)t}-\hat{a}(k)\hat{\sigma}^{\dagger}e^{-\mathrm{i}(c|k|-\varepsilon)t}\right],\label{HintAtomFieldFreq}
\end{equation}
where we remark that the coupling depends solely on the magnitude
of the wave vector and not on its sign, since the atom doesn't have
a preferred orientation.

Out of the 16 terms appearing in this expression when expanding the
commutators and the products, only the terms proportional to $\text{tr}_{\text{L}}\{\hat{a}(k)\hat{a}^{\dagger}(k')\bigotimes_{\forall k}|0\rangle\langle0|\}=\delta(k-k')$
are nonzero. This leaves us with
\begin{align}
\partial_{t}\hat{\rho}_{\text{A,I}}(t) & =\int_{0}^{t}d\tau\int_{-\infty}^{+\infty}dkg^{2}(|k|)\left[\hat{\sigma}\hat{\rho}_{\text{A,I}}(t-\tau)\hat{\sigma}^{\dagger}e^{\mathrm{i}(c|k|-\varepsilon)\tau}-\hat{\rho}_{\text{A,I}}(t-\tau)\hat{\sigma}^{\dagger}\hat{\sigma}e^{\mathrm{i}(c|k|-\varepsilon)\tau}+\text{H.c.}\right]\label{AtomicMasterEq_NonMarkov}\\
 & =\int_{0}^{t}d\tau e^{-\mathrm{i}\varepsilon\tau}C(\tau)\left[\hat{\sigma}\hat{\rho}_{\text{A,I}}(t-\tau)\hat{\sigma}^{\dagger}-\hat{\rho}_{\text{A,I}}(t-\tau)\hat{\sigma}^{\dagger}\hat{\sigma}+\text{H.c.}\right],\nonumber 
\end{align}
where we have combined all the terms under the wave-vector integral
into the so-called \emph{environmental correlation function} (note
the usual change of variable in the integral, $\omega=c|k|$)\emph{
\begin{equation}
C(\tau)=\int_{0}^{+\infty}d\omega\underbrace{\frac{2}{c}g^{2}(\omega/c)}_{G(\omega)}e^{\mathrm{i}\omega\tau}.
\end{equation}
}This function contains all the information about the environment,
and its decay determines how far into the past the dynamics of the
atom can see. While we have found this expression for our particular
problem, it can be shown that even for completely general problems,
such a correlation function appears and can be written as half the
Fourier transform of the so-called \emph{environmental spectral density},
$G(\omega)$, which in our case takes the specific form written above.
When the spectral density is taken as constant $G(\omega)=\gamma/\pi$,
we are left with integrals of the type $C(\tau)\propto\int_{0}^{\infty}d\omega e^{\mathrm{i}\omega\tau}$,
which are proportional to $\delta(\tau)$ when we extend the lower
integration limit to $-\infty$. This would set $\tau=0$ in Eq. (\ref{AtomicMasterEq_NonMarkov}),
and lead to the master equation (\ref{AtomicMasterEq}) that we analyzed
in the previous section. However, when we cannot make this approximation,
we need to proceed in a different way.

In situations where $C(\tau)$ is known and tractable analytically,
one can try using Laplace transform techniques to solve the remaining
time non-local equation. However, in most situations this is not the
case, and we are forced to perform approximations. The most common
one is the so-called \emph{Markov approximation}, in which we assume
that the evolution of the state is independent of its history, that
is, we set $\hat{\rho}_{\text{A,I}}(t-\tau)$ to $\hat{\rho}_{\text{A,I}}(t)$
in the integral. Let us understand the conditions under which this
approximation holds. First, note that since we are working in the
interaction picture, changes in the state are fully due to the interaction,
that is, they are expected to occur at a rate of order $\pi G(\varepsilon)$.
On the other hand, while the correlation function $C(\tau)$ will
not be a delta function, we still expect it to decay with $\tau$,
quite fast if $G(\omega)$ varies slowly with $\omega$. Hence, if
the decay rate of $C(\tau)$ is larger than $\pi G(\varepsilon)$,
one can assume\footnote{Interestingly, note that when for long times we expect $\hat{\rho}_{\text{A,I}}(t)$
to reach a stationary state, this approximation becomes exact for
any finite decay rate of $C(\tau)$. Hence, even in situations where
the Markov approximation is not valid for the dynamics, one can still
use it to derive a reduced steady-state equation for stationary state
of the system.} that $\hat{\rho}_{\text{A,I}}(t-\tau)$ is approximately equal to
$\hat{\rho}_{\text{A,I}}(t)$ in (\ref{AtomicMasterEq_NonMarkov}).

Let us assume that the required conditions are met, and perform the
Markov approximation on (\ref{AtomicMasterNonLocal}), obtaining
\begin{equation}
\partial_{t}\hat{\rho}_{\text{A,I}}=\Gamma(t)\left(\hat{\sigma}\hat{\rho}_{\text{A,I}}\hat{\sigma}^{\dagger}-\hat{\rho}_{\text{A,I}}\hat{\sigma}^{\dagger}\hat{\sigma}\right)+\text{H.c.},\label{AtomicMasterAfterMarkov}
\end{equation}
with
\begin{equation}
\Gamma(t)=\int_{0}^{t}d\tau\int_{0}^{+\infty}d\omega G(\omega)e^{\mathrm{i}(\omega-\varepsilon)\tau}.
\end{equation}
If $\Gamma(t)$ was real, this would be exactly the same master equation
as (\ref{AtomicMasterEq}), just with a modified time-dependent rate
$\Gamma(t)$ instead of $\gamma$. However, as we will see shortly,
this parameter is complex in general. Actually, for short times, there
is no way of evaluating the integrals without the specific form of
the coupling. However, for long times we prove at the end of the section
that
\begin{equation}
\lim_{t\rightarrow\infty}\Gamma(t)=\underset{\gamma}{\underbrace{\pi G(\varepsilon)}}+\mathrm{i}\mathcal{P}\int_{0}^{+\infty}d\omega\frac{G(\omega)}{\omega-\varepsilon},\label{GammaAsymtptic}
\end{equation}
 where $\mathcal{P}\int$ refers to the integral's Cauchy principal
value\footnote{Given a function $f(x)$ that diverges at some point $x_{0}$ (trivially
extensible to multiple poles), the Cauchy principal value of the integral
$\int_{a}^{b}dxf(x)$ with $a<x_{0}<b$ is defined as $\mathcal{P}\int_{\mathbb{R}}dxf(x)=\lim_{\epsilon\rightarrow0}\left[\int_{a}^{x_{0}-\epsilon}dxf(x)+\int_{x_{0}+\epsilon}^{b}dxf(x)\right]$.
In other words, we perform the integral, but skipping the poles.}. Hence, in the long time limit the real part provides the same result
as we found with the frequency-independent approximation, but an imaginary
part also appears. Plugging this result (\ref{GammaAsymtptic}) into
(\ref{AtomicMasterAfterMarkov}), we get the asymptotic master equation
\begin{equation}
\partial_{t}\hat{\rho}_{\text{A,I}}(t)=-\mathrm{i}\frac{\Delta\varepsilon}{2}\left[\hat{\sigma}_{z},\hat{\rho}_{\text{A,I}}\right]+\gamma\left(2\hat{\sigma}\hat{\rho}_{\text{A,I}}\hat{\sigma}^{\dagger}-\hat{\rho}_{\text{A,I}}\hat{\sigma}^{\dagger}\hat{\sigma}-\hat{\sigma}^{\dagger}\hat{\sigma}\hat{\rho}_{\text{A,I}}\right),
\end{equation}
where we have used $\hat{\sigma}_{z}=2\hat{\sigma}^{\dagger}\hat{\sigma}-1$
in the commutator term, and we have defined
\begin{equation}
\Delta\varepsilon=-\mathcal{P}\int_{0}^{+\infty}d\omega\frac{G(\omega)}{\omega-\varepsilon},
\end{equation}
which provides a contribution to the atomic transition frequency known
as the \emph{Lamb shift}.

It's interesting to try to calculate this shift, or at least understand
its order of magnitude relative to the decay rate $\gamma$. The problem
is that the unphysical coupling (\ref{UnphysCoupling}) leads to a
spectral density $G(\omega)\propto\omega$, which makes the integral
diverge (and this is even worse in 2D and 3D, where $G(\omega)\sim\omega^{2}$
and $\omega^{3}$). A serious calculation of the Lamb shift would
require the use of relativistic quantum electrodynamics and renormalization.
However, in our weakly (logarithmically) divergent 1D case we can
make a rough estimation of $\Delta\varepsilon$ by perform a frequency-independent
approximation for the coupling, $G(\omega)=\gamma/\pi$, and introducing
a cutoff in the integral up to some frequency $\varepsilon+\Lambda$
that we will choose later. We then obtain
\begin{equation}
\Delta\varepsilon=-\frac{\gamma}{\pi}\lim_{\epsilon\rightarrow0}\left[\int_{0}^{\varepsilon-\epsilon}\frac{d\omega}{\omega-\varepsilon}+\int_{\varepsilon+\epsilon}^{\varepsilon+\Lambda}\frac{d\omega}{\omega-\varepsilon}\right]=-\frac{\gamma}{\pi}\lim_{\epsilon\rightarrow0}\underset{\log(\Lambda/\varepsilon)}{\underbrace{\left[\log(-\epsilon)-\log(-\varepsilon)+\log(\Lambda)-\log(\epsilon)\right]}}=-\frac{\gamma}{\pi}\log\frac{\Lambda}{\varepsilon}.
\end{equation}
Hence, we find a logarithmic divergence, which suggests that the Lamb
shift is of the same order as the decay rate $\gamma$. Indeed, let
us consider physical cutoffs. Of course, $\Lambda\gg\varepsilon$.
However, it feels that taking a cutoff beyond the energies where electron-positron
pair production makes little sense, which is on the order of 1MeV.
On the other hand, we can estimate the transition frequency of the
atom by considering, for example, the one between the two lowest lying
energy eigenstates of Hydrogen that we discussed in Chapter \ref{Sec:AtomicEnergySpectrum}.
We obtain a transition on the order of the Rydberg energy, which means
13eV. Hence, in this situation we obtain $\pi^{-1}\log(\Lambda/\varepsilon)\sim3.5$.
Even for transitions between hyperfine levels (usually on the order
of $10^{-6}$ times that of between states with different principal
atomic quantum number), the logarithmic dependence would still keep
this number on a similar order of magnitude. Hence, our rough estimation
tells us that the Lamb shift and the decay rate are on the same order,
which is indeed in agreement with experimental observations. Note,
however, that while the decay rate is easy to spot (just wait for
the atom to decay!), the Lamb shift is a tiny correction to the transition
frequency $\varepsilon$, and requires very well designed and sensitive
experiments to measure it.

Let us finally prove (\ref{GammaAsymtptic})... \textbf{soon}.

\newpage

\section{Analyzing the emission from open systems}

In the previous chapter we have introduced the model for two prototypical
open quantum optical systems: open cavities and radiating atoms. The
tools develop there allowed us to evaluate the state (mixed, in general)
of the cavity modes and the atoms. However, usually one is more interested
in the properties of the light radiated by these systems, since this
is the one that we can measure easily and use for applications. This
chapter is devoted to this issue. We will first learn how to easily
relate the field coming out of the system (cavity or atom) with system
operators that we know how to characterize via the master or quantum
Langevin equations. Next we will comment on the usual way in which
this output field is measured and characterized through specific observables
and correlation functions with strong physical meaning.

\subsection{Output fields}

\subsubsection{The output field from a cavity}

Let us start by discussing the output field associated with one mode
of an open cavity. It is convenient to work in the Heisenberg picture
for the upcoming calculations. Using the expression of the vector
potential outside the resonator (\ref{Aext}), we see that the part
of the field coming out from the cavity (hence propagating along the
positive $z$ direction) can be written as
\begin{equation}
\mathbf{\hat{A}}_{\text{out}}^{(+)}\left(z,t\right)=-\mathrm{i}\mathbf{e}_{x}\int_{0}^{\infty}d\omega\sqrt{\frac{\hbar}{4\pi c\varepsilon_{0}S\omega}}\hat{b}(\omega;t)e^{\mathrm{i}\omega(z-L)/c},
\end{equation}
where the operator $\hat{b}(\omega;t)$ is given by (\ref{bTOa})
in terms of the initial operators $\hat{b}_{0}(\omega)$ and the intracavity
mode $\hat{a}$. Now, given that only the frequencies around the cavity
resonance contribute to the dynamics as we argued all along the notes,
we can replace the slowly varying function of the frequency $1/\sqrt{\omega}$
by its value at $\omega_{\mathrm{c}}$, as well as extending the lower
integration limit to $-\infty$, arriving to
\begin{equation}
\mathbf{\hat{A}}_{\text{out}}^{(+)}\left(z,t\right)=-\mathrm{i}\mathbf{e}_{x}\sqrt{\frac{\hbar}{4\pi c\varepsilon_{0}S\omega_{\text{c}}}}\int_{-\infty}^{\infty}d\omega\hat{b}(\omega;t)e^{\mathrm{i}\omega(z-L)/c}\text{.}
\end{equation}
Introducing the solution (\ref{bTOa}) for $\hat{b}(\omega;t)$ in
this equation and defining the \textit{retarded time}\textbf{ }$t_{R}=t-(z-L)/c$,
we can write the output field as
\begin{align}
\mathbf{\hat{A}}_{\text{out}}^{(+)}\left(z,t\right)=-\mathrm{i}\mathbf{e}_{x}\sqrt{\frac{\hbar}{4\pi c\varepsilon_{0}S\omega_{\text{c}}}}\Biggl[\underset{-\sqrt{2\pi}\hat{b}_{\text{in}}(t_{R})}{\underbrace{\int_{-\infty}^{\infty}d\omega\hat{b}_{0}(\omega)e^{-\mathrm{i}\omega t_{R}}}}+\sqrt{\frac{\gamma}{\pi}}\int_{0}^{t}dt^{\prime}\hat{a}(t^{\prime})\underset{2\pi\delta(t'-t_{R})}{\underbrace{\int_{-\infty}^{\infty}d\omega e^{\mathrm{i}\omega(t^{\prime}-t_{R})}}}\Biggl]=-\mathrm{i}\mathbf{e}_{x}\sqrt{\frac{\hbar}{2c\varepsilon_{0}S\omega_{\mathrm{c}}}}\hat{b}_{\mathrm{out}}(t_{R}),\label{AoutCavity}
\end{align}
where we have defined the \textit{output operator}
\begin{equation}
\hat{b}_{\mathrm{out}}(t)=\sqrt{2\gamma}\hat{a}(t)-\hat{b}_{\mathrm{in}}(t).\label{Out-CavIn}
\end{equation}
The field coming out from the cavity is therefore a superposition
of the intracavity field leaking through the partially transmitting
mirror and the part of the input field which is reflected, just as
expected by the classical boundary conditions at the mirror (Fresnel
relations), see Fig. \ref{fOpenCavity}. In this context, we denote
(\ref{Out-CavIn}) by the \emph{input-output relation}.

A note on the interpretation of $\hat{n}_{\mathrm{out}}(t)=\hat{b}_{\text{out}}^{\dagger}(t)\hat{b}_{\text{out}}(t)$.
It is not difficult to show that the power coming out of the cavity
is essentially given by $\hbar\omega_{\text{c}}\langle\hat{n}_{\text{out}}(t)\rangle$.
Hence, we can interpret $\hat{n}_{\text{out}}(t)$ as the operator
associated to the number of photons per unit time that come out of
the cavity. Indeed, we will show in the next section that the output
annihilation and creation operators satisfy canonical commutation
relations in time, that is,
\begin{align}
[\hat{b}_{\mathrm{out}}(t),\hat{b}_{\mathrm{out}}^{\dagger}(t^{\prime})]=\delta(t-t^{\prime}),\qquad[\hat{b}_{\mathrm{out}}(t),\hat{b}_{\mathrm{out}}(t^{\prime})]=0,\label{OutputCCR}
\end{align}
reinforcing the interpretation of $\hat{n}_{\text{out}}(t)$ as a
number operator per unit time. Note that, similarly, we could interpret
$\hat{n}_{\mathrm{in}}(t)=\hat{b}_{\text{in}}^{\dagger}(t)\hat{b}_{\text{in}}(t)$
as the operator associated to the number of photons per unit time
impinging the cavity.

\subsubsection{Alternative form of the output operator and causality}

Let us now find an alternative form for the output operator, one that
will allow us to prove that the output operators satisfy the canonical
commutation relations in time of Eq. (\ref{d}), as well as proving
that this theory of inputs and outputs satisfies the expected causality\begin{subequations}\label{CausalityInOut}
\begin{align}
[\hat{c}(t),\hat{b}_{\text{in}}(t')] & =0\quad\text{if }t'>t,\\{}
[\hat{c}(t),\hat{b}_{\text{out}}(t')] & =0\quad\text{if }t'<t,
\end{align}
\end{subequations}where $\hat{c}(t)$ is any intracavity operator.
These simply state that the physics inside the cavity cannot depend
on a future input that hasn't yet arrived, while the output cannot
depend on intracavity processes yet to come. This will turn out to
be very relevant when discussing the detection and characterization
of the output field later.

We start by solving again formally the Heisenberg equation (\ref{EqsCavityAndExt2})
for the external operators $\hat{b}(\omega;t)$, but now in terms
of a final condition $\hat{b}(\omega;T)\equiv\hat{b}_{T}(\omega)$
at some long time $T>t$ (backwards integration). Using (\ref{GenLinEq})
we easily obtain
\begin{equation}
\hat{b}(\omega;t)=\hat{b}_{T}(\omega)e^{-\mathrm{i}\omega(t-T)}-\sqrt{\frac{\gamma}{\pi}}\int_{t}^{T}dt'\hat{a}(t')e^{-\mathrm{i}\omega(t-t')}.\label{bTOaFinalT}
\end{equation}
Performing a frequency integral, we obtain
\begin{equation}
\int_{-\infty}^{+\infty}d\omega\hat{b}(\omega;t)=\int_{-\infty}^{+\infty}d\omega\hat{b}_{T}(\omega)e^{-\mathrm{i}\omega(t-T)}-\sqrt{\frac{\gamma}{\pi}}\int_{t}^{T}dt'\hat{a}(t')\underset{2\pi\delta(t'-t)}{\underbrace{\int_{-\infty}^{+\infty}d\omega e^{\mathrm{i}\omega(t'-t)}}}=\int_{-\infty}^{+\infty}d\omega\hat{b}_{T}(\omega)e^{-\mathrm{i}\omega(t-T)}-\sqrt{\gamma\pi}\hat{a}(t).\label{HalfOutputInput}
\end{equation}
On the other hand, using the formal solution (\ref{bTOa}) in terms
of the initial condition $\hat{b}(\omega;0)=\hat{b}_{0}(\omega)$
at time $0<t$ (forward integration), we obtain
\begin{equation}
\int_{-\infty}^{+\infty}d\omega\hat{b}(\omega;t)=\underset{-\sqrt{2\pi}\hat{b}_{\text{in}}(t)}{\underbrace{\int_{-\infty}^{+\infty}d\omega\hat{b}_{0}(\omega)e^{-\mathrm{i}\omega t}}}+\sqrt{\frac{\gamma}{\pi}}\int_{0}^{t}dt'\hat{a}(t')\underset{2\pi\delta(t'-t)}{\underbrace{\int_{-\infty}^{+\infty}d\omega e^{\mathrm{i}\omega(t'-t)}}}=-\sqrt{2\pi}\hat{b}_{\text{in}}(t)+\sqrt{\gamma\pi}\hat{a}(t).\label{HalfInputOutput}
\end{equation}
Combining Eqs. (\ref{HalfOutputInput}) and (\ref{HalfInputOutput}),
we obtain
\begin{equation}
\frac{1}{\sqrt{2\pi}}\int_{-\infty}^{+\infty}d\omega\hat{b}_{T}(\omega)e^{-\mathrm{i}\omega(t-T)}=\sqrt{2\gamma}\hat{a}(t)-\hat{b}_{\text{in}}(t).
\end{equation}
Thus, comparing this equation with (\ref{Out-CavIn}), we obtain an
alternative form of the output operator
\begin{equation}
\hat{b}_{\text{out}}(t)=\frac{1}{\sqrt{2\pi}}\int_{-\infty}^{+\infty}d\omega\hat{b}_{T}(\omega)e^{-\mathrm{i}\omega(t-T)},
\end{equation}
in terms of the external field at a later time $T$. This allows us
to prove the canonical commutation relations in time of Eq. (\ref{d})
in a straightforward manner, simply using the equal-time canonical
commutation relations in frequency of the external bosonic operators,
particularized to the final time, that is, $[\hat{b}_{T}(\omega),\hat{b}_{T}^{\dagger}(\omega')]=\delta(\omega-\omega')$
and $[\hat{b}_{T}(\omega),\hat{b}_{T}(\omega')]=0$.

In order to discuss causality, it is interesting to write the quantum
Langevin equation for the intracavity operator in terms of the output
operator. To this aim, we simply insert $\hat{b}_{\text{in}}(t)=\sqrt{2\gamma}\hat{a}(t)-\hat{b}_{\text{out}}(t)$
in the quantum Langevin equation (\ref{QuantumLangevinBasic}), leading
to
\begin{equation}
\partial_{t}\hat{a}=\left(\gamma-\mathrm{i}\omega_{\mathrm{c}}\right)\hat{a}-\sqrt{2\gamma}\hat{b}_{\mathrm{out}}(t),\label{ReducedHeisenbergGeneralOut}
\end{equation}
where it is interesting to note that the damping term $-\gamma\hat{a}$
has turned into an amplification term $\gamma\hat{a}$. This is natural
once we realize that, for consistency with (\ref{bTOaFinalT}), this
equation must be solved backwards in time starting from a final condition
$\hat{a}(T)$. It is then common to call (\ref{ReducedHeisenbergGeneral})
and (\ref{ReducedHeisenbergGeneralOut}) the \emph{forward} and \emph{backwards}
\emph{quantum Langevin equations}, respectively.

With this at hand, we are in conditions to talk about causality in
this \emph{input-output theory}. In particular, note that since the
quantum Langevin equations are local in time, intracavity operators
can only depend on $\hat{b}_{\text{in}}(t')$ at times $t'\leq t$,
and on $\hat{b}_{\text{out}}(t')$ at times $t'\geq t$. Hence, since
both the input and output operators commute at different times, we
see that a generic intracavity operator $\hat{c}(t)$, which is therefore
a function of $\hat{a}(t)$ and $\hat{a}^{\dagger}(t)$, commutes
with $\hat{b}_{\text{in}}(t')$ for $t'>t$, and with $\hat{b}_{\text{out}}(t')$
when $t'<t$. This proves Eqs. (\ref{CausalityInOut}).

We can go even further, noting that when $t'>t$ we can use the \emph{input-output
relation} to write $[\hat{c}(t),\hat{b}_{\text{out}}(t')]=[\hat{c}(t),\sqrt{2\gamma}\hat{a}(t')-\hat{b}_{\text{in}}(t')]=\sqrt{2\gamma}[\hat{c}(t),\hat{a}(t')]$.
Similarly, when $t'<t$ we have $[\hat{c}(t),\hat{b}_{\text{in}}(t')]=[\hat{c}(t),\sqrt{2\gamma}\hat{a}(t')-\hat{b}_{\text{out}}(t')]=\sqrt{2\gamma}[\hat{c}(t),\hat{a}(t')]$.
In addition, when $t'=t$ we simply use (\ref{HalfInputOutput}),
together with the fact that external operators commute with intracavity
operators at equal times, obtaining $[\hat{c}(t),\hat{b}_{\text{in}}(t)]=\sqrt{\gamma/2}[\hat{c}(t),\hat{a}(t)]-\int_{-\infty}^{+\infty}d\omega[\hat{c}(t),\hat{b}(\omega;t)]/\sqrt{2\pi}=\sqrt{\gamma/2}[\hat{c}(t),\hat{a}(t)].$
And then, using the input-output relation (\ref{Out-CavIn}), $[\hat{c}(t),\hat{b}_{\text{out}}(t)]=\sqrt{2\gamma}[\hat{c}(t),\hat{a}(t)]-[\hat{c}(t),\hat{b}_{\text{in}}(t)]=\sqrt{\gamma/2}[\hat{c}(t),\hat{a}(t)]$.
Putting everything together, we can rewrite all commutators in terms
of intracavity operators only,
\begin{equation}
[\hat{c}(t),\hat{b}_{\text{in}}(t')]=\left\{ \begin{array}{cc}
\sqrt{2\gamma}[\hat{c}(t),\hat{a}(t')] & \text{for }t>t'\\
\sqrt{\gamma/2}[\hat{c}(t),\hat{a}(t)] & \text{for }t=t'\\
0 & \text{for }t<t'
\end{array}\right.,\label{CausalIn}
\end{equation}
and
\begin{equation}
[\hat{c}(t),\hat{b}_{\text{out}}(t')]=\left\{ \begin{array}{cc}
0 & \text{for }t>t'\\
\sqrt{\gamma/2}[\hat{c}(t),\hat{a}(t)] & \text{for }t=t'\\
\sqrt{2\gamma}[\hat{c}(t),\hat{a}(t')] & \text{for }t\leq t'
\end{array}\right..\label{CausalOut}
\end{equation}

\subsubsection{The output field radiated by an atom}

Let us now do a similar procedure for the field radiated by a two-level
atom. In this case, there is a subtlety related to the fact that the
atom radiates in all directions, whereas the cavity emits radiation
along a well defined optical axis defined by the partially transmitting
mirror. Usually, one doesn't have access to all this emitted light,
but only to part of it (this is evident in 3D, we cannot put detectors
filling the whole space!). In our 1D model, we then assume that we
have access only to light radiated towards the right (this will allow
us to understand the effect that missing part of the emitted has on
the detection), whose corresponding electric field we can write, using
(\ref{FreeSpaceE}), as
\begin{equation}
\hat{\mathbf{E}}_{\text{out}}^{(+)}(z,t)=\mathrm{i}\mathbf{e}_{x}\int_{0}^{+\infty}dk\sqrt{\frac{\hbar ck}{4\pi\varepsilon_{0}S}}\hat{a}(k,t)e^{\mathrm{i}kz}\approx\mathrm{i}\mathbf{e}_{x}\sqrt{\frac{\hbar\varepsilon}{4\pi\varepsilon_{0}S}}\int_{0}^{+\infty}dk\hat{a}(k,t)e^{\mathrm{i}kz},\label{Eout1}
\end{equation}
where we have again used the fact that $\sqrt{ck}$ is a slowly varying
function of $k$, so we can approximate it by its value $\sqrt{\varepsilon}$
at the atomic resonance $\varepsilon$. Following the same procedure
as with the field coming out of the cavity, we next use the Heisenberg
equations of motion of the field operators to relate $\hat{a}(k,t)$
with the fields at the initial time $\hat{a}(k,0)$, and with the
atomic operator $\hat{\sigma}(t)$. Reminding the Hamiltonians of
the field and the atom-field interaction, (\ref{FreeSpaceHlight})
and (\ref{HLMatomFreeSpace}), respectively, we obtain (for $k>0$)
\begin{equation}
\partial_{t}\hat{a}(k)=-\mathrm{i}cka(k)+\sqrt{\frac{\gamma c}{2\pi}}\hat{\sigma}\hspace{1em}\Rightarrow\hspace{1em}\hat{a}(k,t)=\hat{a}_{0}(k)e^{-\mathrm{i}ckt}+\sqrt{\frac{\gamma c}{2\pi}}\int_{0}^{t}dt'e^{\mathrm{i}ck(t'-t)}\hat{\sigma}(t'),
\end{equation}
where we have made the frequency-independent approximation for the
atom-field coupling and defined $\hat{a}_{0}(k)=\hat{a}(k,0)$, the
annihilation operators at the initial time. Introducing this in the
radiated field (\ref{Eout1}) we obtain
\begin{align}
\hat{\mathbf{E}}_{\text{out}}^{(+)}(z,t) & =\mathrm{i}\mathbf{e}_{x}\sqrt{\frac{\hbar\varepsilon}{4\pi\varepsilon_{0}S}}\int_{0}^{+\infty}dk\left[\hat{a}_{0}(k)e^{-\mathrm{i}ckt_{R}}+\sqrt{\frac{\gamma c}{2\pi}}\int_{0}^{t}dt'e^{\mathrm{i}ck(t'-t_{R})}\hat{\sigma}(t')\right].
\end{align}
Next we perform the same approximations as before. First, we extend
the lower integration limit to $-\infty$, understanding that the
negative values of $k$ do not refer to modes propagating to the left,
but to fictitious modes with negative frequency propagating to the
right. As usual, this is a convenient mathematical trick which does
not affect the physics because only modes with frequencies around
the atomic transition contribute. Defining the input operator
\begin{equation}
\hat{b}_{\text{in}}(t)=-\sqrt{\frac{c}{2\pi}}\int_{-\infty}^{+\infty}dk\hat{a}_{0}(k)e^{-\mathrm{i}ckt},
\end{equation}
which, just as the one defined for the open cavity, satisfies canonical
commutation relations in time (\ref{InCCR}) and the statistical properties
(\ref{ThermalAin}) for thermal fields, we then obtain
\begin{align}
\hat{\mathbf{E}}_{\text{out}}^{(+)}(z,t)=\mathrm{i}\mathbf{e}_{x}\sqrt{\frac{\hbar\varepsilon}{4\pi\varepsilon_{0}S}}\Biggl[\underset{-\sqrt{2\pi/c}\hat{a}_{\text{in}}(t_{R})}{\underbrace{\int_{-\infty}^{+\infty}dk\hat{a}_{0}(k)e^{-\mathrm{i}ckt_{R}}}}+\sqrt{\frac{\gamma c}{2\pi}}\int_{0}^{t}dt'\hat{\sigma}(t')\underset{2\pi\delta(t'-t_{R})/c}{\underbrace{\int_{-\infty}^{+\infty}dke^{\mathrm{i}ck(t'-t_{R})}}}\Biggl]=\mathrm{i}\mathbf{e}_{x}\sqrt{\frac{\hbar\varepsilon}{2c\varepsilon_{0}S}}\hat{b}_{\text{out}}(t_{R}),\label{EoutAtom}
\end{align}
with
\begin{equation}
\hat{b}_{\text{out}}(t)=\sqrt{\gamma}\hat{\sigma}(t)-\hat{b}_{\text{in}}(t).
\end{equation}
This result is similar to the one for the field leaking out of the
cavity except for one subtle issue: while in both cases $\gamma$
denotes the damping rate appearing in master or quantum Langevin equations,
in the case of the cavity we have a $\sqrt{2\gamma}\hat{a}$ contribution
to the output field, see (\ref{Out-CavIn}), while in the case of
the atom we have $\sqrt{\gamma}\hat{\sigma}$. This $\sqrt{2}$ mismatch
comes from the fact that we are considering only half of the radiation
emitted by the atom, so this output operator contains only half of
the field lost from the atom. This can have enormous consequences
for the detection and use of certain quantum properties of the output
field, as we shall see.

\subsection{Observing the output: photodetection, correlation functions, and
the quantum regression theorem}

\subsubsection{Photodetection and correlation functions}

Let us now move on to the characterization of the output field. The
traditional way of analyzing optical fields is by using photodetectors,
whose principle of action we pass to describe now. As we will see
with the example of homodyne detection later, any other measurement
technique aimed at characterizing optical fields makes use of photodetectors
in one way or another.

Photodetection is based on the photoelectric effect or variations
of it, see Fig. \textbf{ToDo}. The idea is that when the light beam
that we want to detect impinges the detector, photons are able to
release some of its bound electrons, which are accelerated towards
a series of metallic plates at increasing positive voltages, releasing
then more electrons which contribute to generate a measurable electric
pulse. Each photon is then capable of releasing one electron which
eventually produces a single pulse, named \emph{photopulse}. The combination
of all the generated photopulses leads to a macroscopic current $j(t)$,
which is a classical random\footnote{It is random, because the exact number of electrons ripped out during
the amplification stage, as well as the exact time when the photoelectron
and photopulse are generated, are all random, in the sense that fluctuate
from realization to realization.} signal that we can monitor and process. 

Though not trivially (see \cite{QO5,NavarretePhDthesis} and references
therein), when measuring the field radiated by an open system, it
is possible to show that the moments of this current are related to
moments of the number operator $\hat{n}_{\text{out}}(t)=\hat{b}_{\text{out}}^{\dagger}(t)\hat{b}_{\text{out}}(t)$
by
\begin{equation}
\overline{j(t_{1})j(t_{2})...j(t_{N})}\propto\langle:\hat{n}_{\text{out}}(t_{1})\hat{n}_{\text{out}}(t_{2})...\hat{n}_{\text{out}}(t_{N}):\rangle,\label{PhotoMoments}
\end{equation}
where the double dots denote normal ordering\footnote{Annihilation operators are moved to the right and creation operators
to the left as if they would commute. For example, $\langle:\hat{n}_{\text{out}}(t_{1})\hat{n}_{\text{out}}(t_{2}):\rangle=\langle\hat{b}_{\text{out}}^{\dagger}(t_{1})\hat{b}_{\text{out}}^{\dagger}(t_{2})\hat{b}_{\text{out}}(t_{2})\hat{b}_{\text{out}}(t_{1})\rangle$.} of the operators. Hence, the quantum statistics of the number of
photons coming out of the system per unit time are imprinted in the
stochastic fluctuations of the photocurrent.

Let us for the upcoming discussions introduce a nomenclature that
will gather the open cavity and radiating atom. We will refer to the
cavity mode or the atom as the \emph{system}. We will introduce an
operator $\hat{s}$ which will denote the annihilation operator $\hat{a}$
or the raising operator $\hat{\sigma}$ depending on the context.
On the other hand, we will call \emph{environment} to the field outside
the cavity and the field surrounding the atom, and we will introduce
a rate $\kappa$ which equals $2\gamma$ for the cavity case and $\gamma$
for the atomic case, so that the input-output relation takes the general
form
\begin{equation}
\hat{b}_{\text{out}}(t)=\sqrt{\kappa}\hat{s}(t)-\hat{b}_{\text{in}}.\label{d}
\end{equation}
Moreover, for reasons that will be obvious shortly, it is convenient
to work with an output operator from which we remove the coherent
part of the input operator, that is,
\begin{equation}
\hat{a}_{\text{out}}(t)=\sqrt{\kappa}\hat{s}(t)-\hat{a}_{\text{in}}=\hat{b}_{\text{out}}(t)+\langle\hat{b}_{\text{in}}(t)\rangle.\label{InputOutputGeneral}
\end{equation}
 Making use of lasers mixed with the output field and several detectors,
it is then clear that the most general quantity that we can observe
takes the form
\begin{equation}
G_{\text{out}}^{(N,M)}(t_{1},t_{2},...,t_{N+M})=\langle\hat{a}_{\text{out}}^{\dagger}(t_{1})\hat{a}_{\text{out}}^{\dagger}(t_{2})...\hat{a}_{\text{out}}^{\dagger}(t_{N})\hat{a}_{\text{out}}(t_{N+1})...\hat{a}_{\text{out}}(t_{N+M-1})\hat{a}_{\text{out}}(t_{N+M})\rangle.
\end{equation}
We call these the \emph{field correlation functions}, which were introduced
by Glauber in his seminal 1963 papers \cite{Glauber63a,Glauber63b,Glauber63c}
in order to produce a definitive quantum-mechanical explanation of
the coherence properties of lasers, masers, and thermal sources (the
last ones corresponding to the famous experiments of Hanbury-Brown
and Twiss \cite{HBT}). Without loss of generality, we can assume
time-ordering in this expression, that is, $t_{1}<t_{2}<...<t_{N-1}$
and $t_{N+M}<t_{N+M-1}<...<t_{N+1}$, because output operators at
different times commute. This will turn out to be very useful for
reasons that we explain next.

Note that, while we have defined correlation functions in terms of
the output operator because it is the one usually measured, the dynamics
is usually solved for operators of the system by using the master
or quantum Langevin equations. Hence, it is interesting to rewrite
the correlation functions in terms of system operators. In general,
the best we can do is using (\ref{InputOutputGeneral}) to write $G_{\text{out}}^{(N)}(t_{1},t_{2},...,t_{2N})$
in terms of correlation functions of system and input operators. However,
in the particular case in which the environment is at zero temperature,
this connection is straightforward. In particular, reminding that
$[\hat{s}(t),\hat{a}_{\text{in}}(t')]=0$ when $t<t'$ (causality)
and that operators are time ordered, we can bring all the input annihilation
operators to the right and input creation operators to the left, and
annihilate vacuum, so that the only remaining correlation function
is the one involving system operators,
\begin{equation}
G^{(N,M)}(t_{1},t_{2},...,t_{N+M})=\langle\hat{s}^{\dagger}(t_{1})\hat{s}^{\dagger}(t_{2})...\hat{s}^{\dagger}(t_{N})\hat{s}(t_{N+1})...\hat{s}(t_{N+M-1})\hat{s}(t_{N+M})\rangle,\label{CorrFuncG}
\end{equation}
and we obtain
\begin{equation}
G_{\text{out}}^{(N,M)}(t_{1},t_{2},...,t_{N+M})=\kappa^{(N+M)/2}G^{(N,M)}(t_{1},t_{2},...,t_{N+M}).\label{GoutG}
\end{equation}

Finally, it is interesting to consider normalized versions of the
correlation functions. In particular, given the interpretation of
$\hat{n}_{\text{out}}(t)$ as the number of output photons per unit
time, it is obvious that the absolute value of a given correlation
function will be larger the larger the emission of the system is,
as we will see through specific examples. Hence, in order to compare
the properties of sources that emit at different rates, we introduce
\begin{equation}
g_{\text{out}}^{(N)}(t_{1},t_{2},...,t_{2N})=\frac{G_{\text{out}}^{(N,N)}(t_{1},t_{2},...,t_{2N})}{\sqrt{G_{\text{out}}^{(1,1)}(t_{1},t_{1})G_{\text{out}}^{(1,1)}(t_{2},t_{2})...G_{\text{out}}^{(1,1)}(t_{2N},t_{2N})}}.
\end{equation}
Of course, we can also define the same type of normalized correlation
function for the system
\begin{equation}
g^{(N)}(t_{1},t_{2},...,t_{2N})=\frac{G^{(N,N)}(t_{1},t_{2},...,t_{2N})}{\sqrt{G^{(1,1)}(t_{1},t_{1})G^{(1,1)}(t_{2},t_{2})...G^{(1,1)}(t_{2N},t_{2N})}}.
\end{equation}
Note that whenever (\ref{GoutG}) holds, these two normalized correlation
functions coincide. Note also that $G_{\text{out}}^{(1)}(t,t)=\langle\hat{n}_{\text{out}}(t)\rangle$
is nothing but the average photon number radiated per unit time, and
hence the normalization factors simply help bringing to the same value
correlation functions of sources with different emission rates but
same qualitative properties otherwise. For example, we shall see shortly
that systems emitting coherent light will all have $g_{\text{out}}^{(N)}=1$,
irrespective of how much light they emit.

\subsubsection{Quantum regression theorem}

Let us now focus on the correlation functions of the system (\ref{CorrFuncG}).
In particular, we consider two-time correlators of various types,
which are the most important for our purposes and in typical experiments,
as we will see. Given three system operators $\{\hat{A},\hat{B},\hat{C}\}$,
we will focus on the correlation function $\langle\hat{A}(t)\hat{B}(t')\hat{C}(t)\rangle$.
The aim of this section is to relate this Heisenberg picture expression
to a Schrödinger picture one. This is known as the \emph{quantum regression
theorem}, and, while it might sound like a trivial exercise, we will
see that it provides a route towards efficient and simple ways towards
the evaluation of such correlation functions, as well as towards their
interpretation.

In order to make the notation lighter, we will not make an explicit
distinction between Heisenberg or Schrödinger operators, instead the
picture will be evident from the time dependence of the operators:
time-dependent states must be understood as Schrödinger picture, while
time-independent ones as Heisenberg picture; similarly, operators
(other than the state) must be understood as Heisenberg picture or
Schrödinger picture depending on whether they are explicitly time
dependent or not, respectively.

With this in mind, consider an open system evolving according to a
master equation with a general time-dependent Lindbladian $\mathcal{L}^{(t)}$,
where we denote the time-dependence as a superscript for later convenience.
The quantum regression theorem reads then
\begin{equation}
\langle\hat{A}(t)\hat{B}(t+\tau)\hat{C}(t)\rangle=\text{tr}\left\{ \hat{B}\mathcal{U}^{(\tau)}[\hat{C}\hat{\rho}(t)\hat{A}]\right\} ,\label{QRT}
\end{equation}
where
\begin{equation}
\mathcal{U}^{(\tau)}=1+\sum_{n=1}^{\infty}\frac{1}{(\mathrm{i}\hbar)^{n}}\int_{0}^{\tau}dt_{1}\int_{0}^{t_{1}}dt_{2}...\int_{0}^{t_{n-1}}dt_{n}\mathcal{L}^{(t_{1})}\mathcal{L}^{(t_{2})}...\mathcal{L}^{(t_{n})}=\mathcal{T}\left\{ e^{\int_{0}^{\tau}dt\mathcal{L}^{(t)}}\right\} ,
\end{equation}
is the time-evolution superoperator associated to the master evolution
induced by the master equation, and $\mathcal{T}$ is the time-ordering
symbol, which in this case orders superoperators in chronological
order
\begin{equation}
\mathcal{T}\left\{ \mathcal{L}^{(t)}\mathcal{L}^{(t')}\right\} =\left\{ \begin{array}{cc}
\mathcal{L}^{(t)}\mathcal{L}^{(t')} & \text{for }t>t'\\
\mathcal{L}^{(t')}\mathcal{L}^{(t)} & \text{for }t<t'
\end{array}\right..
\end{equation}
Note that in the case of a time-independent Lindbladian, the time-evolution
superoperator simply reads
\begin{equation}
\mathcal{U}^{(\tau)}=e^{\mathcal{L}\tau}.
\end{equation}
The quantum regression theorem is very suggestive, and allows for
an interesting interpretation of the two-time correlation function:
at time $t$, operators $\hat{C}$ and $\hat{A}$ effect a change
on the state, which then evolves for a time $\tau$; this leads to
an unnormalized state, which we then use to evaluate the expectation
value of $\hat{B}$.

Before we prove the theorem, it is important to remark that the quantum
regression theorem can be adapted to any intermediate picture we choose
to work on. That is, if we work in a picture defined by some unitary
transformation $\hat{U}_{\text{c}}(t)$, where the transformed state
$\hat{\rho}_{\text{I}}(t)=\hat{U}_{\text{c}}^{\dagger}(t)\hat{\rho}(t)\hat{U}_{\text{c}}(t)$
evolves according to a master equation $\partial_{t}\hat{\rho}_{\text{I}}=\mathcal{L}_{\text{I}}^{(t)}[\hat{\rho}_{\text{I}}]$,
with $\mathcal{L}_{\text{I}}^{(t)}$ obtained from $\mathcal{L}^{(t)}$
as explained in Section \ref{Sec:OpenCavityExample}, we can turn
(\ref{QRT}) into
\begin{equation}
\langle\hat{A}(t)\hat{B}(t+\tau)\hat{C}(t)\rangle=\text{tr}\left\{ \hat{B}_{\text{I}}\mathcal{U}_{\text{I}}^{(\tau)}[\hat{C}_{\text{I}}\hat{\rho}_{\text{I}}(t)\hat{A}_{\text{I}}]\right\} ,\label{QRTintermediate}
\end{equation}
where $\hat{A}_{\text{I}}(t)=\hat{U}_{\text{c}}^{\dagger}(t)\hat{A}\hat{U}_{\text{c}}(t)$
is an intermediate-picture operator (similarly for $\hat{B}$ and
$\hat{C}$), and $\mathcal{U}_{\text{I}}^{(\tau)}=\mathcal{T}\left\{ e^{\int_{0}^{\tau}dt\mathcal{L}_{\text{I}}^{(t)}}\right\} $.
This is a useful expression, as it allows us to evaluate the correlation
function in the most convenient picture, for example, one in which
the Lindbladian is time-independent, and the system reaches a time-independent
(stationary) asymptotic state $\lim_{t\rightarrow\infty}\hat{\rho}_{\text{I}}(t)=\bar{\rho}_{\text{I}}$.
Under such circumstances, we obtained the steady-state correlation
functions
\begin{equation}
\lim_{t\rightarrow\infty}\langle\hat{A}(t)\hat{B}(t+\tau)\hat{C}(t)\rangle=\text{tr}\left\{ \hat{B}_{\text{I}}e^{\mathcal{L}_{\text{I}}\tau}[\hat{C}_{\text{I}}\bar{\rho}_{\text{I}}\hat{A}_{\text{I}}]\right\} ,\label{QRTstationary}
\end{equation}
which will be the expression that we will use the most, since this
is the typical object that experiments are interested in.

Let us now pass to prove the quantum regression theorem. In order
to do this, let us introduce some notation. First, we denote the time-evolution
operator of the whole universe (system and environment) by $\hat{U}_{\text{SE}}(t)$.
Similarly, we denote by $\hat{\rho}_{\text{SE}}$ the state of the
whole universe, while $\hat{\rho}_{\text{E}}$ is the reduced state
of the environment (taken as a thermal state within the Born approximation
in our examples), leaving the name $\hat{\rho}$ for the state of
the system. Consequently, we will denote the trace over the whole
universe by $\text{tr}_{\text{SE}}$, over the environment by $\text{tr}_{\text{E}}$,
and over the system simply by $\text{tr}$. Consider then an operator
$\hat{W}$ acting on the whole universe. Within the framework introduced
in the previous chapter (Born-Markov, non-backaction, and Markov approximations),
it is clear that we can make the following approximation for the reduced
dynamics 
\begin{equation}
\text{tr}_{\text{E}}\left\{ \hat{U}_{\text{SE}}(\tau)\hat{W}\hat{U}_{\text{SE}}^{\dagger}(\tau)\right\} =\mathcal{U}^{(\tau)}[\text{tr}_{\text{E}}\{\hat{W}\}],\label{IntegratedMaster}
\end{equation}
which is just the integral form of the master equation. Keeping this
in mind, it is then straightforward to prove the quantum regression
theorem as

\begin{align}
\langle\hat{A}(t)\hat{B}(t+\tau)\hat{C}(t)\rangle & =\text{tr}_{\text{SE}}\left\{ \hat{\rho}_{\text{SE}}\hat{U}_{\text{SE}}^{\dagger}(t)\hat{A}\hat{U}_{\text{SE}}(t)\hat{U}_{\text{SE}}^{\dagger}(t+\tau)\hat{B}\hat{U}_{\text{SE}}(t+\tau)\hat{U}_{\text{SE}}^{\dagger}(t)\hat{C}\hat{U}_{\text{SE}}(t)\right\} \nonumber \\
 & =\text{tr}_{\text{SE}}\left\{ \hat{B}\hat{U}_{\text{SE}}(\tau)\hat{C}\hat{\rho}_{\text{SE}}(t)\hat{A}\hat{U}_{\text{SE}}^{\dagger}(\tau)\right\} \approx\text{tr}\left\{ \hat{B}\mathcal{U}^{(\tau)}[\text{tr}_{\text{E}}\{\hat{C}\hat{\rho}_{\text{SE}}(t)\hat{A}\}]\right\} ,\label{QRTproof}
\end{align}
where in the first step we just written the Heisenberg evolution of
the operators explicitly, in the second step we have used the cyclic
property of the trace and the composition property of the time-evolution
operator, $\hat{U}_{\text{SE}}^{\dagger}(t_{2})\hat{U}_{\text{SE}}(t_{1})=\hat{U}_{\text{SE}}^{\dagger}(t_{2}-t_{1})$,
and in the final step we have applied (\ref{IntegratedMaster}). The
proof is completed by noting that $\text{tr}_{\text{E}}\{\hat{C}\hat{\rho}_{\text{SE}}(t)\hat{A}\}=\hat{C}\text{tr}_{\text{E}}\{\hat{\rho}_{\text{SE}}(t)\}\hat{A}=\hat{C}\hat{\rho}(t)\hat{A}$,
since $\hat{A}$ and $\hat{C}$ act as the identity on the Hilbert
space of the environment (hence, $\hat{A}\otimes\hat{I}$ would be
a more rigorous notation, which we avoid when there is no room for
confusion). This is easily proven by representing the state in a basis
$\{|e_{m}^{\text{S}}\rangle\otimes|e_{n}^{\text{E}}\rangle\}_{m=1,...,d_{\text{S}}}^{n=1,...,d_{\text{E}}}$
of the full universe 
\begin{align}
\text{tr}_{\text{E}}\{\hat{C}\hat{\rho}_{\text{SE}}(t)\hat{A}\} & =\text{tr}_{\text{E}}\left\{ \hat{C}\otimes\hat{I}\left(\sum_{mm'=1}^{d_{\text{S}}}\sum_{nn'=1}^{d_{\text{E}}}\rho_{mm';nn'}(t)|e_{m}^{\text{S}}\rangle\langle e_{m'}^{\text{S}}|\otimes|e_{n}^{\text{E}}\rangle\langle e_{n'}^{\text{E}}|\right)\hat{A}\otimes\hat{I}\right\} \\
 & \text{tr}_{\text{E}}\left\{ \left(\sum_{mm'=1}^{d_{\text{S}}}\sum_{nn'=1}^{d_{\text{E}}}\rho_{mm';nn'}(t)\hat{C}|e_{m}^{\text{S}}\rangle\langle e_{m'}^{\text{S}}|\hat{A}\otimes|e_{n}^{\text{E}}\rangle\langle e_{n'}^{\text{E}}|\right)\right\} \nonumber \\
 & =\sum_{mm'=1}^{d_{\text{S}}}\sum_{lnn'=1}^{d_{\text{E}}}\rho_{mm';nn'}(t)\hat{C}|e_{m}^{\text{S}}\rangle\langle e_{m'}^{\text{S}}|\hat{A}\underbrace{\langle e_{l}^{\text{E}}|e_{n}^{\text{E}}\rangle}_{\delta_{ln}}\underbrace{\langle e_{n'}^{\text{E}}|e_{l}^{\text{E}}\rangle}_{\delta_{ln'}}\nonumber \\
 & =\hat{C}\underbrace{\left[\sum_{mm'=1}^{d_{\text{S}}}\left(\sum_{l=1}^{d_{\text{E}}}\rho_{mm';ll'}(t)\right)|e_{m}^{\text{S}}\rangle\langle e_{m'}^{\text{S}}|\right]}_{\text{tr}_{\text{E}}\{\hat{\rho}_{\text{SE}}(t)\}}\hat{A}.\nonumber 
\end{align}

As a final remark, note that this proof and the quantum regression
theorem can be generalized to higher-order correlators with not too
much effort. In particular, it's easy to prove, along the lines of
the proof in Eqs. (\ref{QRTproof}), the following connection between
multi-time correlation functions in the Heisenberg and Shcrödinger
pictures
\begin{align}
\langle\hat{A}_{0}(t_{0})\hat{A}_{1}(t_{1})...\hat{A}_{n}(t_{n})\hat{C}_{n}(t_{n})...\hat{C}_{1}(t_{1})\hat{C}_{0}(t_{0})\rangle\label{QRTmulti-time}\\
=\text{tr}\left\{ \hat{A}_{n}\hat{C}_{n}\mathcal{U}^{(t_{n}-t_{n-1})}\right. & \left.\left[\hat{C}_{n-1}\mathcal{U}^{(t_{n-1}-t_{n-2})}\left[...\hat{C}_{1}\mathcal{U}^{(t_{1}-t_{0})}\left[\hat{C}_{0}\hat{\rho}(t_{0})\hat{A}_{0}\right]\hat{A}_{1}...\right]\hat{A}_{n-1}\right]\right\} ,\nonumber 
\end{align}
where $t_{0}<t_{1}<...<t_{n}$. This is the most general type of multi-time
correlation function that one can consider. Of course, this expression
can also be written with all the operators, the state, and the time-evolution
superoperator written in any intermediate picture, as we did in Eq.
(\ref{QRTintermediate}).

\subsubsection{Quantum regression formula}

We can use the quantum regression theorem to develop a way of finding
the correlation functions from the evolution equations of expectation
values of system operators. In particular, consider a \emph{closed
set }of system operators $\{\hat{B}_{j}\}_{j=1,2,...,L}$, `closed'
in the sense that their expectation values evolve according to a closed
linear system
\begin{equation}
\partial_{\tau}\langle\hat{\mathbf{B}}(\tau)\rangle=M(\tau)\langle\hat{\mathbf{B}}(\tau)\rangle,\label{CompleteSetEqs}
\end{equation}
where $\hat{\mathbf{B}}=(\hat{B}_{1},\hat{B}_{2},...,\hat{B}_{L})^{T}$,
$M(\tau)$ is some $L\times L$ matrix (generically time dependent),
and we denote the time by $\tau$ for later convenience. Note that,
for a Hilbert space of dimension $d$, this is always possible with
$L\sim d^{2}$ in the worst case (number of elements of the density
operator). But, of course, the method will only be practical (either
analytically or numerically) when $L$ is small enough, either naturally,
or by doing some approximate truncation to the most relevant operators.
We will see examples of this shortly, but as a generic one, we already
saw in the previous chapter that the first and second order moments
of bosonic modes form a closed set of linear equations when the master
equation is quadratic in annihilation and creation operators (Gaussian
dynamics).

Next, we prove that the two-time correlators $\langle\hat{A}(t)\hat{B}_{j}(t+\tau)\hat{C}(t)\rangle$
at a given time $t$ satisfy the same equations of motion as the expectation
values, that is,
\begin{equation}
\partial_{\tau}\langle\hat{A}(t)\hat{\mathbf{B}}(t+\tau)\hat{C}(t)\rangle=M(\tau)\langle\hat{A}(t)\hat{\mathbf{B}}(t+\tau)\hat{C}(t)\rangle,\label{QRF}
\end{equation}
with the difference that now the initial condition is given by $\langle\hat{A}(t)\hat{\mathbf{B}}(t)\hat{C}(t)\rangle=\text{tr}\left\{ \hat{A}\hat{\mathbf{B}}\hat{C}\hat{\rho}(t)\right\} $.
This expression, known as the \emph{quantum regression formula}, allows
us to find any desired two-time correlator by solving a linear system,
which we will find to be highly practical.

Let us now pass to prove (\ref{QRF}). It follows easily by rewriting
the left-hand-side and right-hand-side of (\ref{CompleteSetEqs})
in the Schrödinger picture as\begin{subequations}
\begin{align}
\partial_{\tau}\text{tr}\left\{ \hat{\mathbf{B}}\hat{\rho}(\tau)\right\}  & =\text{tr}\left\{ \hat{\mathbf{B}}\partial_{\tau}\mathcal{U}^{(\tau)}[\hat{\rho}]\right\} =\text{tr}\left\{ \hat{\mathbf{B}}\mathcal{L}^{(\tau)}\mathcal{U}^{(\tau)}\left[\hat{\rho}\right]\right\} ,\label{QRFintermediate}\\
M(\tau)\langle\hat{\mathbf{B}}(\tau)\rangle & =M(\tau)\text{tr}\left\{ \hat{\mathbf{B}}\hat{\rho}(\tau)\right\} =M(\tau)\text{tr}\left\{ \hat{\mathbf{B}}\mathcal{U}^{(\tau)}[\hat{\rho}]\right\} .
\end{align}
\end{subequations}. Note that this expression holds for any choice
of initial-condition operator $\hat{\rho}$, and therefore the relation
between the upper and lower lines in the previous equations must be
a property of the set $\hat{\mathbf{B}}$ and the Lindblad superoperator
alone. It is not even required that $\hat{\rho}$ is Hermitian, positive,
or even normalized. In other words, the relation
\begin{equation}
\partial_{\tau}\text{tr}\left\{ \hat{\mathbf{B}}\mathcal{U}^{(\tau)}\hat{S}\right\} =M(\tau)\text{tr}\left\{ \hat{\mathbf{B}}\mathcal{U}^{(\tau)}[\hat{S}]\right\} ,
\end{equation}
must hold for any system operator $\hat{S}$ (as long as the operator
inside the trace remains in the trace class, of course). Choosing
$\hat{S}=\hat{A}\hat{\rho}(t)\hat{C}$, we then obtain
\begin{equation}
\partial_{\tau}\text{tr}\{\hat{\mathbf{B}}\mathcal{U}^{(\tau)}[\hat{A}\hat{\rho}(t)\hat{C}]\}=M(\tau)\text{tr}\{\hat{\mathbf{B}}\mathcal{U}^{(\tau)}[\hat{A}\hat{\rho}(t)\hat{C}]\},
\end{equation}
which using the quantum regression theorem (\ref{QRT}) leads to the
desired expression (\ref{QRF}).

\subsubsection{Interpretation of the photodetection two-time correlation function}

Before we move on to examples, it is interesting to provide an interpretation
for the two-time correlators in a physical situation. In particular,
we consider the two-time correlator associated with the photocurrent
obtained by direct photodetection of the output field in the stationary
limit $t\rightarrow\infty$. This is the simplest dynamical object
that one can study for a source after it reaches its asymptotic state,
and we will see shortly that it has a clear and interesting physical
interpretation. According to (\ref{PhotoMoments}), the corresponding
correlation function is
\begin{equation}
\lim_{t\rightarrow\infty}\overline{j(t)j(t+\tau)}\propto\lim_{t\rightarrow\infty}\langle\hat{s}^{\dagger}(t)\hat{s}^{\dagger}(t+\tau)\hat{s}(t+\tau)\hat{s}(t)\rangle=\lim_{t\rightarrow\infty}G^{(2,2)}(t,t+\tau,t+\tau,t)\equiv\bar{G}^{(2)}(\tau),\label{G2tau}
\end{equation}
where we have assumed $\tau\geq0$ for definiteness, a vacuum state
for the environment so we can use (\ref{GoutG}), and we introduce
the notation $\bar{G}^{(2)}(\tau)$ for this correlation function
because we will use it a lot in what follows. Note that this has exactly
the form given in (\ref{QRTproof}) and (\ref{QRF}), with $\hat{A}=\hat{s}^{\dagger}$,
$\hat{B}=\hat{s}^{\dagger}\hat{s}$, and $\hat{C}=\hat{s}$.

The interpretation of this correlation function follows straightforwardly
from the atomic case, for which $\hat{s}=\hat{\sigma}=|g\rangle\langle e|$,
and considering a time-independent Lindbladian, so the system reaches
a stationary state asymptotically. Using the quantum regression theorem
(\ref{QRTstationary}), we can write the correlation function as
\begin{equation}
\bar{G}^{(2)}(\tau)=\text{tr}\biggl\{\underbrace{\hat{\sigma}^{\dagger}\hat{\sigma}}_{|e\rangle\langle e|}e^{\mathcal{L}\tau}\underbrace{[\hat{\sigma}\bar{\rho}\hat{\sigma}^{\dagger}]}_{|g\rangle\langle e|\bar{\rho}|e\rangle\langle g|}\biggr\}=\bar{p}_{e}\times\underset{\left.p_{e}(\tau)\right|_{|\psi(0)\rangle=|g\rangle}}{\underbrace{\langle e|\left(e^{\mathcal{L}\tau}[|g\rangle\langle g|]\right)|e\rangle}}=\bar{p}_{e}\times\left.p_{e}(\tau)\right|_{|\psi(0)\rangle=|g\rangle}.\label{G2atom}
\end{equation}
Where $\bar{p}_{e}=\langle e|\bar{\rho}|e\rangle$ is the probability
of finding the atom in the excited state once its state has reached
its asymptotic limit, while $\left.p_{e}(\tau)\right|_{|\psi(0)\rangle=|g\rangle}$
is the probability of finding the atom excited at time $\tau$, starting
from the ground state at time $0$, since $e^{\mathcal{L}\tau}[|g\rangle\langle g|]$
is the state evolved during a time $\tau$ from the ground state.
Now, when the atom is in the ground state, the probability of recording
a photopulse in the detector is clearly zero. Hence, assuming a perfect
photodetector that turns all photons into photopulses without errors,
the probability of being excited is equivalent to the probability
of recording a \emph{click!} in the photodetector (after which the
atom must be in the ground state according to collapse axiom of quantum
mechanics). Therefore, the two-time correlation function (\ref{G2tau})
arising naturally from direct photodetection must be interpreted as
the probability of getting two consecutive photodetection events separated
by a time interval $\tau$. In other words, this correlation function
provides the waiting statistics for the arrival of photons to the
photodetector. Hence, we will call \emph{coincidence correlation function}
to (\ref{G2tau}).

This interpretation is easily generalized to higher-order correlation
functions. In particular, consider the multi-time correlation function
\begin{align}
\lim_{t\rightarrow\infty}\overline{j(t)j(t+\tau_{1})...j(t+\tau_{N-1})} & \propto\lim_{t\rightarrow\infty}\langle\hat{s}^{\dagger}(t)\hat{s}^{\dagger}(t+\tau_{1})...\hat{s}^{\dagger}(t+\tau_{N-1})\hat{s}(t+\tau_{N-1})...\hat{s}(t+\tau_{1})\hat{s}(t)\rangle\label{GNtau}\\
 & =\lim_{t\rightarrow\infty}G^{(N,N)}(t,t+\tau_{1},...,,t+\tau_{N-1},t+\tau_{N-1},...,t+\tau_{1},t)\equiv\bar{G}^{(N)}(\tau_{1},...,\tau_{N-1}),\nonumber 
\end{align}
where we have assumed $\tau_{1}\leq\tau_{2}\leq...\leq\tau_{N-1}$
for definiteness. Particularizing this expression again to the atomic
case with a time-independent Lindbladian, and using the multi-time
form of the quantum regression theorem of Eq. (\ref{QRTmulti-time}),
it's easy to find
\begin{equation}
\bar{G}^{(N)}(\tau_{1},\tau_{2},...,\tau_{N-1})=\bar{p}_{e}\times\left.p_{e}(\tau_{1})\right|_{|\psi(0)\rangle=|g\rangle}\times\left.p_{e}(\tau_{2}-\tau_{1})\right|_{|\psi(0)\rangle=|g\rangle}\times...\times\left.p_{e}(\tau_{N-1}-\tau_{N-2})\right|_{|\psi(0)\rangle=|g\rangle},
\end{equation}
which is then interpreted as the probability of getting $N$ consecutive
photodetection events separated by time intervals $\{\tau_{j}-\tau_{j-1}\}_{j=1,2,...,N-1}$
(with $\tau_{0}=0$), once the atom has reached its asymptotic state.

\subsection{Open cavity: coherent sources and photon bunching}

As a first example we consider the light emitted by an empty, monochromatically-driven
cavity. We saw in Section \ref{Sec:OpenCavityExample} that the corresponding
master equation, in a picture rotating at the laser frequency, is
given by
\begin{equation}
\partial_{t}\hat{\rho}_{\text{I}}=\left[\mathrm{i}\Delta\hat{a}^{\dagger}\hat{a}+\mathcal{E}\hat{a}^{\dagger}-\mathcal{E}^{*}\hat{a},\hat{\rho}_{\text{I}}\right]+(\bar{n}+1)\gamma\left(2\hat{a}\hat{\rho}_{\text{I}}\hat{a}^{\dagger}-\hat{a}^{\dagger}\hat{a}\hat{\rho}_{\text{I}}-\hat{\rho}_{\text{I}}\hat{a}^{\dagger}\hat{a}\right)+\bar{n}\gamma\left(2\hat{a}^{\dagger}\hat{\rho}_{\text{I}}\hat{a}-\hat{a}\hat{a}^{\dagger}\hat{\rho}_{\text{I}}-\hat{\rho}_{\text{I}}\hat{a}\hat{a}^{\dagger}\right)\equiv\mathcal{L}_{\text{I}}[\hat{\rho}_{\text{I}}],\label{MasterEqEx1corr}
\end{equation}
which has a time-independent Lindbladian $\mathcal{L}_{\text{I}}$.
On the other hand, note that the interaction-picture annihilation
operators read as $\hat{a}_{\text{I}}(t)=\hat{U}_{\text{c}}^{\dagger}(t)\hat{a}\hat{U}_{\text{c}}(t)=e^{-\mathrm{i}\omega_{\text{L}}t}\hat{a}$,
so that using Eq. (\ref{QRTstationary}), the coincidence correlation
function (\ref{G2tau}) can be written as
\begin{equation}
\bar{G}^{(2)}(\tau)=\text{tr}\left\{ \hat{a}_{\text{I}}^{\dagger}\hat{a}_{\text{I}}e^{\mathcal{L}_{\text{I}}\tau}[\hat{a}_{\text{I}}\bar{\rho}_{\text{I}}\hat{a}_{\text{I}}^{\dagger}]\right\} =\text{tr}\left\{ \hat{N}e^{\mathcal{L}_{\text{I}}\tau}[\hat{a}\bar{\rho}_{\text{I}}\hat{a}^{\dagger}]\right\} ,
\end{equation}
where we remind that $\bar{\rho}_{\text{I}}=\lim_{t\rightarrow\infty}\tilde{\rho}(t)$
is the stationary state in the rotating picture, that is, that satisfying
$\mathcal{L}_{\text{I}}[\bar{\rho}_{\text{I}}]=0$.

As we saw in Section \ref{Sec:OpenCavityExample}, the stationary
state is the displaced thermal state, $\bar{\rho}_{\text{I}}=\hat{D}(\bar{\alpha})\hat{\rho}_{\text{th}}(\bar{n})\hat{D}^{\dagger}(\bar{\alpha})$
with $\bar{\alpha}=\mathcal{E}/(\gamma-\mathrm{i}\Delta)$. It is
interesting though, to consider separately the case in which the environment
is at zero temperature ($\bar{n}=0$), and the case in which the driving
is zero ($\mathcal{E}=0$).

\subsubsection{Zero temperature: Poissonian light}

In the $\bar{n}=0$ case, the steady state is coherent $\bar{\rho}=|\bar{\alpha}\rangle\langle\bar{\alpha}|$
and we thus easily find
\begin{equation}
\bar{G}^{(2)}(\tau)=\text{tr}\biggl\{\hat{N}e^{\mathcal{L}_{\text{I}}\tau}\underset{|\bar{\alpha}|^{2}|\bar{\alpha}\rangle\langle\bar{\alpha}|}{[\underbrace{\hat{a}|\bar{\alpha}\rangle\langle\bar{\alpha}|\hat{a}^{\dagger}}]}\biggr\}=|\bar{\alpha}|^{2}\text{tr}\biggl\{\underset{\underbrace{|\bar{\alpha}\rangle\langle\bar{\alpha}|}_{\begin{array}{c}
\text{by definition}\\
\text{of steady state}
\end{array}}}{\hat{a}^{\dagger}\hat{a}\underbrace{e^{\mathcal{L}_{\text{I}}\tau}[|\bar{\alpha}\rangle\langle\bar{\alpha}|]}}\biggr\}=|\bar{\alpha}|^{2}\langle\bar{\alpha}|\hat{N}|\bar{\alpha}\rangle=|\bar{\alpha}|^{4}.
\end{equation}
We observe that the probability of getting two consecutive clicks
in the photodetector is independent of their time delay. This means
that the photodetection events are statistically independent: photons
arrive to the detector at random times, with no correlation whatsoever.
Indeed, this is consistent with the fact that the photon number probability
distribution is Poissonian for a coherent state, since this distribution
is the one corresponding to statistically independent events. On the
other hand, we have $\lim_{t\rightarrow\infty}G^{(1,1)}(t,t)=\lim_{t\rightarrow\infty}\langle\hat{N}(t)\rangle=\langle\bar{\alpha}|\hat{N}|\bar{\alpha}\rangle=|\bar{\alpha}|^{2},$
leading to a normalized correlation function
\begin{equation}
\bar{g}^{(2)}(\tau)\equiv\lim_{t\rightarrow\infty}g^{(2)}(t,t+\tau,t+\tau,t)=\frac{\bar{G}^{(2)}(\tau)}{\left[\lim_{t\rightarrow\infty}\langle\hat{a}^{\dagger}(t)\hat{a}(t)\rangle\right]^{2}}=1.
\end{equation}
Moreover, the generalization of this calculation to higher-order correlators
is straightforward, and leads to the result $\bar{G}^{(N)}(\tau_{1},...,\tau_{N-1})=|\bar{\alpha}|^{2N}$,
so that
\begin{equation}
\bar{g}^{(N)}(\tau_{1},...,\tau_{N-1})\equiv\lim_{t\rightarrow\infty}g^{(N)}(t,t+\tau_{1},...,t+\tau_{N-1},t+\tau_{N-1},...,t+\tau_{1},t)=1.\label{gNcoherent}
\end{equation}
This value of 1 for the normalized photodetection correlation functions
is not unique to coherent states, but, more generally, it is characteristic
of sources whose steady-state number distribution is Poissonian. This
is easy to prove as follows. First, note that the most general state
whose photon-number distribution is Poissonian must be necessarily
written as a mixture of coherent states with the same amplitude $\alpha$
but arbitrary phases, say
\begin{equation}
\hat{\rho}_{\text{Poisson}}=\int_{0}^{2\pi}d\phi P(\phi)|\alpha e^{\mathrm{i}\phi}\rangle\langle\alpha e^{\mathrm{i}\phi}|,\qquad\text{with }P(\phi)>0\,\forall\phi\text{ and }\int_{0}^{2\pi}d\phi P(\phi)=1.\label{rhoPoisson}
\end{equation}
Note that the a mixture of $K$ coherent states with phases $\{\phi_{k}\}_{k=1,2,...,K}$
is obtained setting $P(\phi)=\sum_{k=1}^{K}w_{k}\delta(\phi-\phi_{k})$,
with $w_{k}>0\:\forall k$ and $\sum_{k=1}^{K}w_{k}=1$. Indeed, note
that the photon number distribution $\langle n|\hat{\rho}_{\text{Poisson}}|n\rangle=\exp(-\alpha^{2})\alpha^{2n}/n!$
is Poissonian for any choice of $P(\phi)$. This is obviously not
the case for any other type of state. Using the Schrödinger-picture
expression (\ref{QRTmulti-time}) for the $\bar{G}^{(N)}(\tau_{1},...,\tau_{N-1})$
correlation function (\ref{GNtau}), assuming that the stationary
state of the system has the form (\ref{rhoPoisson}), so that $\lim_{t\rightarrow\infty}\hat{\rho}(t)=\hat{\rho}_{\text{Poisson}}$
with $\mathcal{L}[\hat{\rho}_{\text{Poisson}}]=0$, it is then immediate
to obtain $\bar{G}^{(N)}(\tau_{1},...,\tau_{N-1})=\alpha^{2N}$ for
any choice of $P(\phi)$, which together with $\lim_{t\rightarrow\infty}G^{(1,1)}(t,t)=\text{tr}\{\hat{N}\hat{\rho}_{\text{Poisson}}\}=\alpha^{2}$,
leads to (\ref{gNcoherent}).

\subsubsection{Non-driven cavity: thermal light and bunching}

Let us now move on to the $\mathcal{E}=0$ case, such that the steady
state of the cavity is thermal, $\bar{\rho}=\hat{\rho}_{\text{th}}(\bar{n})$
with $\lim_{t\rightarrow\infty}G^{(1)}(t,t)=\text{tr}\left\{ \hat{a}^{\dagger}\hat{a}\hat{\rho}_{\text{th}}(\bar{n})\right\} =\bar{n}$.
Note that without driving, the master equation is time-independent
in the original picture, so there is no need to refer everything to
a rotating picture.

This case provides a perfect example for the use the quantum regression
formula. In particular, note that the expectation value of the photon
number satisfies the evolution equation
\begin{equation}
\partial_{t}\left\langle \hat{a}^{\dagger}\hat{a}\right\rangle =-2\gamma\left\langle \hat{a}^{\dagger}\hat{a}\right\rangle +2\bar{n}\gamma,
\end{equation}
as we proved in (\ref{Example1Expectations}). We can rewrite this
equation in the form (\ref{CompleteSetEqs}) by choosing $\hat{B}_{1}=\hat{a}^{\dagger}\hat{a}$
and $\hat{B}_{2}=\bar{n}$ (constant), which form then a closed set
with
\begin{equation}
\mathcal{M}=\left(\begin{array}{cc}
-2\gamma & 2\gamma\\
0 & 0
\end{array}\right).
\end{equation}
Therefore, according to the quantum regression formula (\ref{QRF})
with $\hat{A}=\hat{a}^{\dagger}$ and $\hat{C}=\hat{a}$, we get
\begin{equation}
\partial_{\tau}\left(\begin{array}{c}
\langle\hat{a}^{\dagger}(t)\hat{a}^{\dagger}(t+\tau)\hat{a}(t+\tau)\hat{a}(t)\rangle\\
\bar{n}\langle\hat{a}^{\dagger}(t)\hat{a}(t)\rangle
\end{array}\right)=\mathcal{M}\left(\begin{array}{c}
\langle\hat{a}^{\dagger}(t)\hat{a}^{\dagger}(t+\tau)\hat{a}(t+\tau)\hat{a}(t)\rangle\\
\bar{n}\langle\hat{a}^{\dagger}(t)\hat{a}(t)\rangle
\end{array}\right),
\end{equation}
which taking the $t\rightarrow\infty$ limit, leads to 
\begin{equation}
\partial_{\tau}\left(\begin{array}{c}
\bar{G}^{(2)}(\tau)\\
\bar{n}^{2}
\end{array}\right)=\mathcal{M}\left(\begin{array}{c}
\bar{G}^{(2)}(\tau)\\
\bar{n}^{2}
\end{array}\right)\Rightarrow\frac{d\bar{G}^{(2)}}{d\tau}=-2\gamma\bar{G}^{(2)}+2\bar{n}^{2}\gamma.
\end{equation}
Using (\ref{GenLinEq}), we obtain the solution
\begin{equation}
\bar{G}^{(2)}(\tau)=e^{-2\gamma\tau}\bar{G}^{(2)}(0)+\bar{n}^{2}\left(1-e^{-2\gamma\tau}\right).\label{G2t-thermal}
\end{equation}
The initial condition of this expression is given by
\begin{equation}
\bar{G}^{(2)}(0)=\text{tr}\left\{ \hat{a}^{\dagger2}\hat{a}^{2}\hat{\rho}_{\text{th}}(\bar{n})\right\} =\underbrace{\left|\text{tr}\left\{ \hat{a}^{2}\bar{\rho}_{\text{th}}(\bar{n})\right\} \right|^{2}}_{0}+2\underbrace{\text{tr}\left\{ \hat{a}^{\dagger}\hat{a}\bar{\rho}_{\text{th}}(\bar{n})\right\} ^{2}}_{\bar{n}^{2}}=2\bar{n}^{2},\label{G20}
\end{equation}
where we have used the fact that the thermal state is Gaussian with
$\langle\hat{a}\rangle=0$, and hence it satisfies the Gaussian moment
theorem (\ref{GaussianMomentTheoremQuantum}) with $\hat{L}_{1}=\hat{L}_{2}=\hat{a}^{\dagger}$
and $\hat{L}_{3}=\hat{L}_{4}=\hat{a}$. Inserting then (\ref{G20})
in (\ref{G2t-thermal}), we obtain the normalized coincidence correlation
function
\begin{equation}
\bar{g}^{(2)}(\tau)=\frac{\bar{G}^{(2)}(\tau)}{\bar{n}^{2}}=1+e^{-2\gamma\tau}.
\end{equation}
Interestingly, in this case $\bar{g}^{(2)}(\tau)$ starts at 2 for
$\tau=0$ and decays monotonically to 1 as time goes by. In other
words, it is more likely to detect two photons at the same time, than
spaced in time: photons arrive in bunches. We then call this phenomenon
\emph{bunching}.

\subsection{Resonance fluorescence and antibunching}

For the next example, let us discuss now the properties of the light
radiated by a driven two-level atom, a problem known as \emph{resonance
fluorescence}. The general master equation of this problem is given
by Eq. (\ref{AtomicMasterEq}), but here we will focus on the zero-temperature
case ($\bar{n}=0$) for simplicity, and assume the driving to be monochromatic,
that is $\mathcal{A}(t)=\mathcal{E}e^{-\mathrm{i}\omega_{\text{L}}t}$.

Let us start by moving to a picture rotating at the laser frequency,
where the Hamiltonian becomes time independent. This Hamiltonian is
defined the transformation operator $\hat{U}_{\text{c}}(t)=e^{-\mathrm{i}\omega_{\text{L}}t\hat{\sigma}_{z}/2}$,
leading to an interaction-picture lowering operator $\hat{\sigma}_{\text{I}}(t)=\hat{U}_{\text{c}}^{\dagger}(t)\hat{\sigma}\hat{U}_{\text{c}}(t)=e^{-\mathrm{i}\omega_{\text{L}}t}\hat{\sigma}$,
and keeping $\hat{\sigma}_{z}$ invariant. The transformed state $\hat{\rho}_{\text{I}}(t)=\hat{U}_{\text{c}}^{\dagger}(t)\hat{\rho}(t)\hat{U}_{\text{c}}(t)$
evolves then according to the master equation
\begin{equation}
\partial_{t}\hat{\rho}_{\text{I}}=\left[\mathrm{i}\frac{\Delta}{2}\hat{\sigma}_{z}+\mathcal{E}\hat{\sigma}^{\dagger}-\mathcal{E}^{*}\hat{\sigma},\hat{\rho}_{\text{I}}\right]+\gamma\left(2\hat{\sigma}\hat{\rho}_{\text{I}}\hat{\sigma}^{\dagger}-\hat{\sigma}^{\dagger}\hat{\sigma}\hat{\rho}_{\text{I}}-\hat{\rho}_{\text{I}}\hat{\sigma}^{\dagger}\hat{\sigma}\right)=\mathcal{L}_{\text{I}}[\hat{\rho}_{\text{I}}],\label{AtomicMasterEq-1}
\end{equation}
where $\Delta=\omega_{\text{L}}-\varepsilon$.

In order to find the coincidence correlation function, we could proceed
by showing that the operators $\{\hat{\sigma}_{x},\hat{\sigma}_{y},\hat{\sigma}_{z},\hat{I}\}$
form a closed set (Bloch equations), applying then the quantum regression
formula similarly to how we did in the previous example. However,
in the atomic case we have an easier route based on (\ref{G2atom}):
we only need to evaluate the evolution of the excited-state probability.
We will then follow this route.

We start by writing down the Bloch equations. Using the same notation
as in Sec. \ref{CavityExampleSS} for expectation values in the interaction
picture, that is, $\tilde{b}(t)=\langle\hat{\sigma}\rangle_{\text{I}}=\text{tr}\left\{ \hat{\sigma}\hat{\rho}_{\text{I}}(t)\right\} $,
and noting that $b_{z}(t)=\text{tr}\left\{ \hat{\sigma}_{z}\hat{\rho}(t)\right\} =\text{tr}\left\{ \hat{\sigma}_{z}\hat{\rho}_{\text{I}}(t)\right\} $,
we can use the general expression (\ref{GenOpenExpecEvo}) to find\begin{subequations}
\begin{align}
\partial_{t}\tilde{b} & =\mathrm{i}\frac{\Delta}{2}\left\langle \left[\hat{\sigma},\hat{\sigma}_{z}\right]\right\rangle _{\text{I}}+\mathcal{E}\left\langle \left[\hat{\sigma},\hat{\sigma}^{\dagger}\right]\right\rangle _{\text{I}}+\gamma\left\langle \left[\hat{\sigma}^{\dagger},\hat{\sigma}\right]\hat{\sigma}\right\rangle _{\text{I}}=-\mathcal{E}b_{z}-(\gamma-\mathrm{i}\Delta)\tilde{b},\\
\partial_{t}b_{z} & =\mathcal{E}\left\langle \left[\hat{\sigma}_{z},\hat{\sigma}^{\dagger}\right]\right\rangle _{\text{I}}-\mathcal{E}^{*}\left\langle \left[\hat{\sigma}_{z},\hat{\sigma}\right]\right\rangle _{\text{I}}+\gamma\left(\left\langle \left[\hat{\sigma}^{\dagger},\hat{\sigma}_{z}\right]\hat{\sigma}\right\rangle _{\text{I}}+\left\langle \hat{\sigma}^{\dagger}\left[\hat{\sigma}_{z},\hat{\sigma}\right]\right\rangle _{\text{I}}\right)=4\text{Re}\{\mathcal{E}^{*}\tilde{b}\}-4\gamma b_{z}-4\gamma.
\end{align}
\end{subequations}Since we are ultimately interested in the excited-state
population, it is best to write this equations in terms of it already.
Using $p_{e}(t)=[1+b_{z}(t)]/2$, we obtain\begin{subequations}
\begin{align}
\partial_{t}\tilde{b} & =-(\gamma-\mathrm{i}\Delta)\tilde{b}-2\mathcal{E}p_{e}+\mathcal{E},\label{dbIdt}\\
\partial_{t}p_{e} & =2\text{Re}\{\mathcal{E}^{*}\tilde{b}\}-4\gamma p_{e}.\label{dpedt}
\end{align}
\end{subequations}

Before solving the full dynamics, it is interesting to evaluate the
asymptotic state predicted by these equations. Since the Lindbladian
is time independent, we expect a such state to be stationary. Then,
we simply set the derivatives to zero, and solve the remaining linear
system in a straightforward manner (just write $\tilde{b}$ in terms
of $p_{e}$ from the first equation, and substitute it in the second
equation). Introducing the parameter $P=|\mathcal{E}|^{2}/(\gamma^{2}+\Delta^{2})$,
we obtain
\begin{equation}
\bar{p}_{e}\equiv\lim_{t\rightarrow\infty}p_{e}(t)=\frac{P}{2(1+P)}\hspace{1em}\text{and}\hspace{1em}\lim_{t\rightarrow\infty}\tilde{b}(t)=\frac{\sqrt{P}}{1+P}e^{\mathrm{i}\left(\arg\{\mathcal{E}\}+\arg\{\gamma+\mathrm{i}\Delta\}\right)}.
\end{equation}
This shows that the atom remains in the ground state when the driving
rate is negligible with respect to the damping rate or the detuning
($P\ll1$), while in the limit where it dominates ($P\gg1$), it can
only bring half the population to the excited state\footnote{Why only half? Since $|\mathcal{E}|$ dominates, the atom starts performing
perfect Rabi oscillations between the ground and excited states. However,
on a much longer time scale, dissipation kicks in, and the corresponding
noise starts shifting randomly the phase of the oscillations, until
eventually they average out to half the population on each state.}. This is interesting because it tells us that just through driving,
it is impossible to fully invert the steady-state population of an
atom, so more ingenious ways are required.

Let us now move on to the coincidence correlation function (\ref{G2atom}),
focusing on the normalized form
\begin{equation}
\bar{g}^{(2)}(\tau)=\frac{\left.p_{e}(\tau)\right|_{|\psi(0)\rangle_{\text{I}}=|g\rangle}}{\bar{p}_{e}},
\end{equation}
where we have taken into account that $\lim_{t\rightarrow\infty}\langle\hat{\sigma}^{\dagger}(t)\hat{\sigma}(t)\rangle=\text{tr}\{\hat{\sigma}^{\dagger}\hat{\sigma}\bar{\rho}\}=\bar{p}_{e}$
for the normalization. There are very interesting things that we can
see already in this expression without the need to evaluate it at
all times. First, note that $\bar{g}^{(2)}(\tau\rightarrow\infty)=1$,
while $\bar{g}^{(2)}(\tau=0)=0$, since the probability of being in
the excited state at time zero when starting from the ground state
vanishes. Hence, we see that the probability of obtaining two simultaneous
photons is identically zero, while the photodetection events become
independent for long times. This is a phenomenon known as \emph{antibunching},
and it is rooted in the quantization of the atom: once it decays,
it requires some time to get re-excited (we call it \emph{reloading
time}), which means that it can emit only one photon at a time.

Let us now find the full time evolution of the coincidence correlation
function. This analytically difficult in the general case, but quite
easy in the resonant case, $\Delta=0$, which we consider next. Taking
the derivative of Eq. (\ref{dpedt}), and using also Eq. (\ref{dbIdt})
to write $\tilde{b}$ and $\partial_{t}\tilde{b}$ as a function of
$p_{e}$ and $\partial_{t}p_{e}$, it is straightforward to find
\begin{equation}
\ddot{p}_{e}+5\gamma\dot{p}_{e}+4\gamma^{2}(1+P)p_{e}=2\gamma^{2}P.
\end{equation}
Together with the initial conditions $p_{e}(0)=0=\dot{p}_{e}(0)$,
this is a second order differential equation with constant coefficients
that is easily solved, leading to
\begin{equation}
\bar{g}^{(2)}(\tau)=1-e^{-5\gamma\tau/2}\left[\cosh(r\gamma\tau/2)-\frac{5\sinh(r\gamma\tau/2)}{r}\right],
\end{equation}
where we have defined $r=\sqrt{9-16P}$. We show this function in
Fig. \textbf{soon}, for different values of $P$. As expected, the
function starts at 0 and goes to 1 for long times. For intermediate
times, the behavior depends on $P$. The most interesting case occurs
under strong driving conditions, that is, $P\gg1$ or, equivalently,
$|\mathcal{E}|\gg\gamma$. In this case, $r\approx4\mathrm{i}|\mathcal{E}|$
becomes imaginary, and leads to
\begin{equation}
\bar{g}^{(2)}(\tau)\approx1-e^{-5\gamma\tau/2}\cos(2|\mathcal{E}|\tau).
\end{equation}
This result is very much connected to Rabi oscillations, which are
clearly visible for several cycles, until dissipation washes them
out. Hence, in order to measure Rabi oscillations, it is not required
to initialize the atom in the ground state and monitoring the populations.
Instead, we can simply let the strongly-driven atom relax to its asymptotic
state, and then check the Rabi oscillations in the delay between photodetection
events.

\subsection{Homodyne detection and output squeezing}

We have seen that direct photodetection provides normally-ordered
moments of the output number operator. Hence, in loose terms, direct
photodetection allows to measure only the number operator and related
photon counting statistics. However, we can design clever measurement
schemes involving different sources and photodetectors such that the
moments of the combined photocurrents give us access to the moments
of diffferent observables of the output light. As a most relevant
example, in this section we study homodyne detection. As we will see,
this type of measurement will allow us to determine the statistics
of the quadratures of the field. We will also apply it to one specific
source, the so-called below-threshold optical parametric oscillator,
which can be modeled as an open cavity whose intracavity mode is fed
via down-conversion in a nonlinear medium.

\subsubsection{Homodyne detection and the squeezing spectrum}

The scheme for homodyne detection is shown in Fig. \textbf{ToDo}.
The output field of the source that we want to characterize is mixed
in a balanced beam splitter with a strong laser field (dubbed \emph{local
oscillator}); the two fields emerging from the beam splitter are photodetected
and their photocurrents are subtracted, leading to the so-called \emph{homodyne
photocurrent }$j_{H}(t)$. In the following we show that the stochastic
correlation functions of this photocurrent are proportional to normally-ordered
correlation functions of output quadratures.

We start by modeling the local oscillator field. We assume that it
is the output of a monochromatic coherent source (that is, a source
whose asymptotic state is coherent). This is indeed the commonly used
model of a laser\footnote{However, as we shall see in a later chapter, lasers are a bit more
subtle, and this model only holds under some specific conditions.}. We write the vector potential of the corresponding output field
as
\begin{equation}
\mathbf{\hat{A}}_{\mathrm{LO}}^{(+)}\left(z,t\right)=\mathbf{e}_{x}\sqrt{\frac{\hbar}{2c\varepsilon_{0}S\omega_{\mathrm{LO}}}}\hat{a}_{\mathrm{LO}}(t),
\end{equation}
characterized by the correlation functions
\begin{equation}
\langle\hat{a}_{\text{LO}}^{\dagger}(t_{1})...\hat{a}_{\text{LO}}^{\dagger}(t_{N})\hat{a}_{\text{LO}}(t_{N+1})...\hat{a}_{\text{LO}}(t_{N+M})\rangle=e^{\mathrm{i}\omega_{\text{LO}}(t_{1}+...+t_{N}-t_{N+1}-...-t_{N+M})}|\alpha_{\text{LO}}|^{N+M}e^{\mathrm{i}\phi_{\text{LO}}(M-N)},\label{LOstatistics}
\end{equation}
where $\omega_{\text{LO}}$ is the frequency of the coherent source
(laser frequency) and $\alpha_{\text{LO}}=|\alpha_{\text{LO}}|e^{\mathrm{i}\phi_{\text{LO}}}$
is a complex amplitude (specifically, if we denote by $|\bar{\beta}e^{-\mathrm{i}\omega_{\text{LO}}t}\rangle$
the amplitude of the asymptotic coherent state reached by the cavity
associated to this source, with decay rate $\Gamma$, then $\alpha_{\text{LO}}=\sqrt{2\Gamma}\bar{\beta}$,
as we have seen in previous results). We assume that both of this
quantities are controllable in experiments.

Let us also note that, obviously, these local oscillator operators
commute with those of the source at all times, since they act on different
Hilbert spaces and do not interact. Moreover, since they are generated
from independent sources, the states of the output field we want to
study and the local oscillator fields are not correlated, meaning
that the expectation value of the product of any output operator $\hat{B}_{\text{out}}(t)$
of the relevant source with any local oscillator operator $\hat{B}_{\text{LO}}(t')$
factorizes as $\langle\hat{B}_{\text{out}}(t)\hat{B}_{\text{LO}}(t')\rangle=\langle\hat{B}_{\text{out}}(t)\rangle\langle\hat{B}_{\text{LO}}(t')\rangle$. 

Let us now consider the fields emerging from the beam splitter. Considering
a balanced beam splitter (50\% transmission and reflection), these
fields are just the sum and difference of the fields entering its
input ports. We denote them by
\begin{equation}
\mathbf{\hat{A}}_{\pm}\left(z,t\right)=\frac{1}{\sqrt{2}}\left[\mathbf{\hat{A}}_{\mathrm{out}}\left(z,t\right)\pm\mathbf{\hat{A}}_{\mathrm{LO}}\left(z,t\right)\right].
\end{equation}
Obviously, in order for the local oscillator to have any effect on
the source that we want to characterize, $\omega_{\text{LO}}$ must
be close to the central frequency of the light emitted by that source
(in particular, usually in experiments one makes them match). Under
such circumstances, we can write the these fields in the usual form
\begin{equation}
\mathbf{\hat{A}}_{\pm}^{(+)}\left(z,t\right)\approx\mathbf{e}_{x}\sqrt{\frac{\hbar}{2c\varepsilon_{0}S\omega_{\mathrm{LO}}}}\hat{a}_{\pm}(t_{R}),
\end{equation}
with
\begin{equation}
\hat{a}_{\pm}(t)=\frac{1}{\sqrt{2}}\left[\hat{a}_{\mathrm{out}}\left(t\right)\pm\hat{a}_{\mathrm{LO}}\left(t\right)\right].
\end{equation}
Let us write the number operators associated to any of these fields
as
\begin{equation}
\hat{n}_{\pm}(t)=\hat{a}_{\pm}^{\dagger}(t)\hat{a}_{\pm}(t)=\frac{1}{2}\left[\hat{n}_{\text{out}}(t)+\hat{n}_{\text{LO}}(t)\pm\hat{h}(t)\right],\label{PMnumber}
\end{equation}
where we have defined the operator
\begin{equation}
\hat{h}(t)=\hat{a}_{\text{LO}}^{\dagger}(t)\hat{a}_{\text{out}}(t)+\hat{a}_{\text{LO}}(t)\hat{a}_{\text{out}}^{\dagger}(t).
\end{equation}
Note that, using (\ref{LOstatistics}), it is simple to show that
\begin{equation}
\langle:\hat{h}(t_{1})\hat{h}(t_{2})...\hat{h}(t_{N}):\rangle=|\alpha_{\text{LO}}|^{N}\langle:\tilde{X}_{\text{out}}^{\phi_{\text{LO}}}(t_{1})\tilde{X}_{\text{out}}^{\phi_{\text{LO}}}(t_{2})...\tilde{X}_{\text{out}}^{\phi_{\text{LO}}}(t_{N}):\rangle,
\end{equation}
where we have defined slowly-varying quadratures of the output field
\begin{equation}
\tilde{X}_{\text{out}}^{\phi}(t)=e^{\mathrm{i}\omega_{\text{LO}}t-\mathrm{i}\phi}\hat{a}_{\text{out}}(t)+e^{-\mathrm{i}\omega_{\text{LO}}t+\mathrm{i}\phi}\hat{a}_{\text{out}}^{\dagger}(t).
\end{equation}
We call them 'slowly varying' because the factors $e^{\pm\mathrm{i}\omega_{\text{LO}}t}$
cancel the fast oscillations of the output operators. Hence, we see
two things. First, the detection scheme is only sensitive to the slowly-varying
envelopes of the field (this is quite common to all measurement schemes:
resolving optical oscillations is hard and only reserved to very specific
state-of-the-art techniques). Second, we see that the operator $\hat{h}$
encodes the information about the statistics of the output quadratures.
Essentially, the rest of the steps in the detection scheme are designed
to remove all information but the one corresponding to this operator,
as we see next by analyzing the homodyne photocurrent, for what we
need one more result from photodetection theory.

Similarly to the case of direct photodetection, it is possible to
prove (though not trivially) that the cross correlation between two
photocurrents is related to quantum cross correlation functions between
their corresponding detected photon number operators,
\begin{equation}
\overline{j_{m}(t)j_{m'}(t')}\propto\langle:\hat{n}_{m}(t)\hat{n}_{m'}(t'):\rangle.
\end{equation}
This is indeed natural, because we expect two photodetection signals
to be statistically independent only when the underlaying measured
fields are.

We can now proceed to evaluate the first and second order stochastic
correlation functions of the homodyne photocurrent $j_{H}(t)=j_{+}(t)-j_{-}(t)$.
We obtain\begin{subequations}
\begin{align}
\overline{j_{H}(t)} & =\overline{j_{+}(t)}+\overline{j_{-}(t)}\propto\langle\hat{n}_{+}(t)\rangle-\langle\hat{n}_{-}(t)\rangle,\\
\overline{j_{H}(t)j_{H}(t')} & =\overline{j_{+}(t)j_{+}(t')}+\overline{j_{-}(t)j_{-}(t')}-\overline{j_{+}(t)j_{-}(t')}-\overline{j_{-}(t)j_{+}(t')}\\
 & \propto\langle:\hat{n}_{+}(t)\hat{n}_{+}(t'):\rangle+\langle:\hat{n}_{-}(t)\hat{n}_{-}(t'):\rangle-\langle:\hat{n}_{+}(t)\hat{n}_{-}(t'):\rangle-\langle:\hat{n}_{-}(t)\hat{n}_{+}(t'):\rangle.\nonumber 
\end{align}
\end{subequations}Expanding each term using (\ref{PMnumber}) and
(\ref{LOstatistics}), it is then straightforward to obtain\begin{subequations}
\begin{align}
\overline{j_{H}(t)} & \propto|\alpha_{\text{LO}}|\langle\tilde{X}_{\text{out}}^{\phi_{\text{LO}}}(t)\rangle,\\
\overline{j_{H}(t)j_{H}(t')} & \propto|\alpha_{\text{LO}}|^{2}\langle:\tilde{X}_{\text{out}}^{\phi_{\text{LO}}}(t)\tilde{X}_{\text{out}}^{\phi_{\text{LO}}}(t'):\rangle,
\end{align}
\end{subequations}which is precisely the result we advanced above.

It is usual to work with photocurrent fluctuations $\delta j_{H}(t)=j_{H}(t)-\overline{j_{H}(t)}$,
since in quantum mechanics we are usually interested in how noise
is distributed around the average value of an observable, as we have
discussed before. The expressions above allow us to write then
\begin{equation}
\overline{\delta j_{H}(t)\delta j_{H}(t')}\propto|\alpha_{\text{LO}}|^{2}\langle:\delta\tilde{X}_{\text{out}}^{\phi_{\text{LO}}}(t)\delta\tilde{X}_{\text{out}}^{\phi_{\text{LO}}}(t'):\rangle.
\end{equation}

Before moving on to an example where homodyne detection plays an important
role, it is important to point out that most photodetectors do not
allow to measure instantaneous photocurrents (some state-of-the-art
ones do, though). In particular, a more detailed model of the photodetection
process for common photodetectors \cite{NavarretePhDthesis} would
show that the variance of the photocurrent, $\overline{\delta j_{H}^{2}(t)}$
in this case, is infinite at all times, meaning that the instantaneous
signal $j_{H}(t)$ is completely hidden by a massive noise wall. Therefore,
most experiments deal with a different quantity, namely the power
spectrum, given by
\begin{equation}
\frac{1}{T}\int_{0}^{T}dt\int_{0}^{T}dt'\cos[\Omega(t-t')]\overline{\delta j_{H}(t)\delta j_{H}(t')},\label{PowerSpectrum}
\end{equation}
where $T$ is the \emph{detection time} and $\Omega$ is commonly
known as the \emph{noise} or \emph{detection frequency.} The massive
instantaneous noise is now confined to the spectral region $\Omega\in[0,\sim T^{-1}]$,
and hence one gets a sensible signal for noise frequencies outside
that region.

Consider a source that reaches a stationary state in which $\lim_{t\rightarrow\infty}\overline{\delta j_{H}(t)\delta j_{H}(t+\tau)}$
is a function only of $|\tau|$, which is the most common case. If
we also consider sufficiently long detection times, it is then simple
to prove that the previous expression can be written as the simple
Fourier transform
\begin{equation}
\int_{-\infty}^{+\infty}d\tau e^{-\mathrm{i}\Omega\tau}\lim_{t\rightarrow\infty}\overline{\delta j_{H}(t)\delta j_{H}(t+\tau)}.
\end{equation}
Related to the power spectrum of the homodyne photocurrent, we then
define in quantum optics the so-called \emph{quadrature noise spectrum},
given by
\begin{equation}
V^{\phi}(\Omega)=\int_{-\infty}^{+\infty}d\tau e^{-\mathrm{i}\Omega\tau}\lim_{t\rightarrow\infty}\langle\delta\tilde{X}_{\text{out}}^{\phi}(t)\delta\tilde{X}_{\text{out}}^{\phi}(t+\tau)\rangle,\label{NoiseSpectrum}
\end{equation}
which is the usual object measured in an experiment involving homodyne
detection. Note that we have assumed that the two-time correlation
function of the quadrature depends only on $|\tau|$, as it is usually
the case in stationary quantum optical problems; if that's not the
case, then the simple Fourier integral must be replaced by the more
general integral (\ref{PowerSpectrum}). Using the commutation relations
of the output operators (\ref{d}), we can write the quadrature product
in normal order as $\delta\tilde{X}_{\text{out}}^{\phi}(t)\delta\tilde{X}_{\text{out}}^{\phi}(t')=\delta(t-t')+:\delta\tilde{X}_{\text{out}}^{\phi}(t)\delta\tilde{X}_{\text{out}}^{\phi}(t'):$.
Using then the input-output relations under the assumption of a zero-temperature
environment (so that the terms with input operators vanish), we then
obtain the quadrature noise spectrum in terms of system correlation
functions,
\begin{equation}
V^{\phi}(\Omega)=1+\underset{S^{\phi}(\Omega)}{\underbrace{\kappa\int_{-\infty}^{+\infty}d\tau e^{-\mathrm{i}\Omega\tau}\lim_{t\rightarrow\infty}\langle:\delta\tilde{X}^{\phi}(t)\delta\tilde{X}^{\phi}(t+\tau):\rangle}},\label{NoiseSpectrumSystem}
\end{equation}
where we have defined the slowly-varying quadratures of the system
\begin{equation}
\tilde{X}^{\phi}(t)=e^{\mathrm{i}\omega_{\text{LO}}t-\mathrm{i}\phi}\hat{s}(t)+e^{-\mathrm{i}\omega_{\text{LO}}t+\mathrm{i}\phi}\hat{s}^{\dagger}(t).\label{Xslow}
\end{equation}
The normally-ordered part of the noise spectrum, $S^{\phi}(\Omega)$,
is known as \emph{squeezing spectrum}.

Finally, let us discuss the concept of squeezing in the output light
of an open system. As explained in Section \ref{Sec:SqueezedStates},
some of the most common applications of light are sensing or communication.
In both cases, a signal is encoded as a low-frequency (compared to
optical frequencies) modulation of some quadrature. This signal would
then appear as a peak at the corresponding frequency in the power
spectrum of a homodyne signal, or, equivalently, in the quadrature
noise spectrum. On the other hand, note that for a vacuum or coherent
stationary state of the system we get $V^{\phi_{\text{LO}}}(\Omega)=1$
for all \emph{noise frequencies} $\Omega$ and all local oscillator's
phases $\phi_{\text{LO}}$. This then sets a base value that limits
the signal-to-noise ratio with which we will be able to observe the
peak (see Fig. \textbf{ToDo}a). We call this the \emph{shot noise.
}Therefore, in order to increase the signal-to-noise ratio it is desirable
to work with states that have $V^{\phi_{\text{LO}}}(\Omega)<1$ for
some values of the noise frequency and the local oscillator's phase
(see Fig. \textbf{ToDo}b). This is what we call \emph{squeezed states}
in the context of the light radiated by a open systems.

It's interesting to point out that, similarly to the position and
momentum of a single harmonic oscillator, we can derive a sort of
uncertainty relation between the noise spectra of orthogonal quadratures,
namely
\begin{equation}
V^{\phi}(\Omega)V^{\phi+\pi/2}(\Omega)\geq1.
\end{equation}
Hence, the squeezing of an output quadrature must be accompanied by
the antisqueezing of its canonical conjugate. Let us prove this relation...
\textbf{soon}.

Next we proceed to put all this in action with a paradigmatic example:
the optical parametric oscillator.

\subsubsection{Squeezing from the optical parametric oscillator}

Consider an open cavity at zero temperature containing a dielectric
medium with second order nonlinearity such as the one we studied in
Section \ref{Sec:NonlinearDielectric}. Such system is known as optical
parametric oscillator (we will fully understand the reason for the
name in a chapter to come), and we now show now that it allows for
the generation of highly-squeezed output light.

We worked out the previous examples through different approaches in
the Schrödinger picture. Hence, in order to show a different method,
we work in the Heisenberg picture in this case. We then consider the
quantum Langevin equations (\ref{ReducedHeisenbergGeneral}) for the
cavity mode, using
\[
\hat{H}_{\text{intra}}(t)=\mathrm{i}\frac{\hbar g}{2}(e^{\mathrm{i}\omega_{2}t}\hat{a}^{\dagger2}-e^{\mathrm{i}\omega_{2}t}\hat{a}^{\dagger2})
\]
as the intracavity Hamiltonian, corresponding to down-conversion under
the parametric approximation. Note that, for convenience, we have
chosen a different phase for the pump compared to the Hamiltonian
of Eq. (\ref{HPDC}). The quantum Langevin equation then becomes
\begin{equation}
\partial_{t}\hat{a}=-(\gamma+\mathrm{i}\omega_{0})\hat{a}+ge^{\mathrm{i}\omega_{2}t}\hat{a}^{\dagger}+\sqrt{2\gamma}\hat{a}_{\text{in}}(t).
\end{equation}
In order to turn the equation into a time-independent one, we define
the slowly-varying operator\footnote{Equivalent to moving to a picture rotating at half the pump frequency.}
$\tilde{a}(t)=e^{\mathrm{i}\omega_{2}t/2}\hat{a}(t)$, which evolves
according to the equation
\begin{equation}
\partial_{t}\tilde{a}=-(\gamma+\mathrm{i}\Delta)\tilde{a}+g\tilde{a}^{\dagger}+\sqrt{2\gamma}\tilde{a}_{\text{in}}(t),\label{DOPOlangevinBT}
\end{equation}
where $\Delta=\omega_{0}-\omega_{2}/2$ and $\tilde{a}_{\text{in}}(t)=e^{\mathrm{i}\omega_{2}t/2}\hat{a}_{\text{in}}(t)$
is a slowly-varying input operator, which satisfies the same statistical
properties (\ref{ThermalAin}) and commutation relations (\ref{d})
as the original one. It is also interesting to note that if we use
a local oscillator of frequency $\omega_{\text{LO}}=\omega_{2}/2$,
then the slowly-varying quadratures of the system (\ref{Xslow}) can
be written directly in terms of the slowly-varying operators as
\begin{equation}
\tilde{X}^{\phi}(t)=e^{-\mathrm{i}\phi}\tilde{a}(t)+e^{\mathrm{i}\phi}\tilde{a}^{\dagger}(t).
\end{equation}

In order to simplify the problem and avoid spurious technical details,
we consider the resonant case, $\Delta=0$. In such case, the position
$\tilde{X}^{0}\equiv\tilde{X}$ and momentum $\tilde{X}^{\pi/2}\equiv\tilde{P}$
quadratures obey independent equations that can be easily treated.
In particular, simply adding and subtracting (\ref{DOPOlangevinBT})
and it's Hermitian conjugate, we find\begin{subequations}
\begin{align}
\partial_{t}\tilde{X} & =-(\gamma-g)\tilde{X}+\sqrt{2\gamma}\tilde{X}_{\text{in}}(t),\\
\partial_{t}\tilde{P} & =-(\gamma+g)\tilde{P}+\sqrt{2\gamma}\tilde{P}_{\text{in}}(t),
\end{align}
\end{subequations}where we have defined the input quadratures $\tilde{X}_{\text{in}}^{\phi}(t)=e^{-\mathrm{i}\phi}\tilde{a}_{\text{in}}(t)+e^{\mathrm{i}\phi}\tilde{a}_{\text{in}}^{\dagger}(t)$,
which satisfy the statistical properties\begin{subequations}\label{XinStatistics}
\begin{align}
\langle\tilde{X}_{\text{in}}^{\phi}(t)\rangle & =0,\\
\langle\tilde{X}_{\text{in}}^{\phi}(t)\tilde{X}_{\text{in}}^{\phi'}(t')\rangle & =e^{\mathrm{i}(\phi'-\phi)}\delta(t-t').
\end{align}
\end{subequations}These equations are readily solved using (\ref{GenLinEq})
as usual, obtaining\begin{subequations}
\begin{align}
\tilde{X}(t) & =\tilde{X}(0)e^{-(\gamma-g)t}+\sqrt{2\gamma}\int_{0}^{t}dt'e^{-(\gamma-g)(t-t')}\tilde{X}_{\text{in}}(t'),\\
\tilde{P}(t) & =\tilde{P}(0)e^{-(\gamma+g)t}+\sqrt{2\gamma}\int_{0}^{t}dt'e^{-(\gamma+g)(t-t')}\tilde{P}_{\text{in}}(t').
\end{align}
\end{subequations}Interestingly, we see that when $g<\gamma$ (damping
dominates over down conversion) the initial condition is washed out
for long times, and the cavity mode reaches a stationary state as
usual. However, when $g>\gamma$ the position quadrature increases
exponentially with time, showing that we enter a special regime where,
apparently, no stationary state is reached. We will see in one of
the next chapters that what happens is that the parametric approximation
breaks down for $g\geq\gamma$, and we have to consider pump depletion.
Hence, in what follows we focus on the $g<\gamma$ regime, known as
\emph{below-threshold regime} for reasons that will become obvious
along the next chapters. In this regime, we can write the asymptotic
solution as\begin{subequations}\label{XsolsStationary}
\begin{align}
\lim_{t\rightarrow\infty}\tilde{X}(t) & =\sqrt{2\gamma}\lim_{t\rightarrow\infty}\int_{0}^{t}d\tau e^{-(\gamma-g)\tau}\tilde{X}_{\text{in}}(t-\tau),\\
\lim_{t\rightarrow\infty}\tilde{P}(t) & =\sqrt{2\gamma}\lim_{t\rightarrow\infty}\int_{0}^{t}d\tau e^{-(\gamma+g)\tau}\tilde{P}_{\text{in}}(t-\tau).
\end{align}
\end{subequations}

Before commenting on the squeezing properties of the output field,
it is interesting to understand the type of asymptotic state reached
by the intracavity mode. Since the quantum Langevin equations are
linear, we know that the state will be Gaussian, and hence we only
need to find the asymptotic mean vector and covariance matrix as usual.
Using the statistical properties of the input quadratures (\ref{XinStatistics}),
we get\begin{subequations}
\begin{align}
\lim_{t\rightarrow\infty}\langle\tilde{X}(t)\rangle & =0=\lim_{t\rightarrow\infty}\langle\tilde{P}(t)\rangle,\\
\lim_{t\rightarrow\infty}\langle\tilde{X}^{2}(t)\rangle & =2\gamma\lim_{t\rightarrow\infty}\int_{0}^{t}d\tau\int_{0}^{t}d\tau'e^{-(\gamma-g)(\tau+\tau')}\underset{\delta(\tau-\tau')}{\underbrace{\langle\tilde{X}_{\text{in}}(t-\tau)\tilde{X}_{\text{in}}(t-\tau')\rangle}}=2\gamma\lim_{t\rightarrow\infty}\int_{0}^{t}d\tau e^{-2(\gamma-g)\tau}=\frac{\gamma}{\gamma-g},\\
\lim_{t\rightarrow\infty}\langle\tilde{P}^{2}(t)\rangle & =2\gamma\lim_{t\rightarrow\infty}\int_{0}^{t}d\tau\int_{0}^{t}d\tau'e^{-(\gamma+g)(\tau+\tau')}\underset{\delta(\tau-\tau')}{\underbrace{\langle\tilde{P}_{\text{in}}(t-\tau)\tilde{P}_{\text{in}}(t-\tau')\rangle}}=2\gamma\lim_{t\rightarrow\infty}\int_{0}^{t}d\tau e^{-2(\gamma+g)\tau}=\frac{\gamma}{\gamma+g},\\
\lim_{t\rightarrow\infty}\left\langle \left(\tilde{X}(t)\tilde{P}(t)\right)^{(s)}\right\rangle  & =2\gamma\lim_{t\rightarrow\infty}\int_{0}^{t}d\tau\int_{0}^{t}d\tau'e^{-(\gamma-g)\tau-(\gamma+g)\tau'}\Bigl(\underset{\mathrm{i}\delta(\tau-\tau')}{\underbrace{\langle\tilde{X}_{\text{in}}(t-\tau)\tilde{P}_{\text{in}}(t-\tau')\rangle}}+\underset{-\mathrm{i}\delta(\tau-\tau')}{\underbrace{\langle\tilde{P}_{\text{in}}(t-\tau')\tilde{X}_{\text{in}}(t-\tau)\rangle}}\Bigl)=0,
\end{align}
\end{subequations}, where we remind that $(\tilde{X}\tilde{P})^{(s)}=(\tilde{X}\tilde{P}+\tilde{X}\tilde{P})/2$
denotes symmetrization. This expressions lead to a (rotating-picture)
Gaussian state with zero mean and covariance matrix
\begin{equation}
\lim_{t\rightarrow\infty}V(t)=\left(\begin{array}{cc}
(1-\sigma)^{-1} & 0\\
0 & (1+\sigma)^{-1}
\end{array}\right).
\end{equation}
Here we have defined the parameter $\sigma=g/\gamma$, which determines
the distance to threshold. For $\sigma=0$ (no pump) we then obtain
a vacuum state as expected. On the other hand, as the threshold condition
$\sigma=1$ is approached, the momentum quadrature gets more and more
squeezed until its variance reaches half the value found in vacuum
(50\% of quantum noise reduction), that is, $\Delta\tilde{P}^{2}\rightarrow0.5$.
At the same time, the position quadrature gets antisqueezed, but far
more dramatically as threshold is approached, where $\Delta\tilde{X}\rightarrow\infty$.
Hence, the down-converted mode is in a squeezed state for $\sigma>0$,
but not a minimum-uncertainty one, since $\Delta\tilde{X}\Delta\tilde{P}>1$.

We are now in conditions of evaluating the noise spectrum of the output
field. In this case, instead of using the normally ordered expression
(\ref{NoiseSpectrumSystem}), the calculation is easier if we use
the general definition (\ref{NoiseSpectrum}) together with the input-output
relations (\ref{Out-CavIn}), which allow us to write
\begin{equation}
\tilde{X}_{\text{out}}^{\phi}(t)=\sqrt{2\gamma}\tilde{X}^{\phi}(t)-\tilde{X}_{\text{in}}^{\phi}(t).
\end{equation}
In addition, we will combine the position and momentum solutions (\ref{XsolsStationary})
in the single expression
\begin{equation}
\lim_{t\rightarrow\infty}\tilde{X}^{\varphi}(t)=\sqrt{2\gamma}\lim_{t\rightarrow\infty}\int_{0}^{t}d\tau'e^{-\lambda_{\varphi}\tau'}\tilde{X}_{\text{in}}^{\varphi}(t-\tau'),\hspace{1em}\hspace{1em}\varphi=0,\frac{\pi}{2}.\label{XsolGen}
\end{equation}
where $\lambda_{0}=\gamma-g$ and $\lambda_{\pi/2}=\gamma+g$. Let
us start by evaluating the asymptotic two-time correlation functions
of the output quadratures, which can be written as the $t\rightarrow\infty$
limit of
\begin{equation}
\langle\tilde{X}_{\text{out}}^{\varphi}(t)\tilde{X}_{\text{out}}^{\varphi}(t+\tau)\rangle=\left[2\gamma\langle\tilde{X}^{\varphi}(t)\tilde{X}^{\varphi}(t+\tau)\rangle-\sqrt{2\gamma}\langle\tilde{X}^{\varphi}(t)\tilde{X}_{\text{in}}^{\varphi}(t+\tau)+\tilde{X}_{\text{in}}^{\varphi}(t)\tilde{X}^{\varphi}(t+\tau)\rangle+\langle\tilde{X}_{\text{in}}^{\varphi}(t)\tilde{X}_{\text{in}}^{\varphi}(t+\tau)\rangle\right].\label{TwoTimeOutX}
\end{equation}
We evaluate the different terms for the position and momentum quadratures
by using (\ref{XsolGen}) and (\ref{XinStatistics}). For the terms
involving the input quadrature we get\begin{subequations}\label{XXin}
\begin{align}
\langle\tilde{X}_{\text{in}}^{\varphi}(t)\tilde{X}_{\text{in}}^{\varphi}(t+\tau)\rangle & =\delta(\tau)\hspace{1em}\forall(\varphi,t),\\
\lim_{t\rightarrow\infty}\langle\tilde{X}^{\varphi}(t)\tilde{X}_{\text{in}}^{\varphi}(t+\tau)\rangle & =\sqrt{2\gamma}\lim_{t\rightarrow\infty}\int_{0}^{t}d\tau'e^{-\lambda_{\varphi}\tau'}\underset{\delta(\tau+\tau')}{\underbrace{\langle\tilde{X}_{\text{in}}^{\varphi}(t-\tau')\tilde{X}_{\text{in}}^{\varphi}(t+\tau)\rangle}}=\left\{ \begin{array}{cc}
\sqrt{2\gamma}e^{\lambda_{\varphi}\tau} & \tau<0\\
\sqrt{\gamma/2} & \tau=0\\
0 & \tau>0
\end{array}\right.,\\
\lim_{t\rightarrow\infty}\langle\tilde{X}_{\text{in}}^{\varphi}(t)\tilde{X}^{\varphi}(t+\tau)\rangle & =\sqrt{2\gamma}\lim_{t\rightarrow\infty}\int_{0}^{t}d\tau'e^{-\lambda_{\varphi}\tau'}\underset{\delta(\tau-\tau')}{\underbrace{\langle\tilde{X}_{\text{in}}^{\varphi}(t)\tilde{X}_{\text{in}}^{\varphi}(t+\tau-\tau')\rangle}}=\left\{ \begin{array}{cc}
0 & \tau<0\\
\sqrt{\gamma/2} & \tau=0\\
\sqrt{2\gamma}e^{-\lambda_{\varphi}\tau} & \tau>0
\end{array}\right..
\end{align}
\end{subequations}On the other hand, the term involving only intracavity
quadratures requires a bit more attention
\begin{align}
\langle\tilde{X}^{\varphi}(t)\tilde{X}^{\varphi}(t+\tau)\rangle & =2\gamma\lim_{t\rightarrow\infty}\int_{0}^{t}d\tau_{1}\int_{0}^{t+\tau}d\tau_{2}e^{-\lambda_{\varphi}(\tau_{1}+\tau_{2})}\underset{\delta(\tau_{1}+\tau-\tau_{2})}{\underbrace{\langle\tilde{X}_{\text{in}}^{\varphi}(t-\tau_{1})\tilde{X}_{\text{in}}^{\varphi}(t+\tau-\tau_{2})\rangle}}.
\end{align}
Since the domain of integration of $\tau_{2}$ is different than that
of $\tau_{1}$, we have to be careful with the application of the
delta function, which provides the condition $\tau_{1}=\tau_{2}-\tau$
only as long as the domains of both sides of the expression overlap
(otherwise gives a zero). When $\tau>0$, the domain of $\tau_{2}$
is larger than the domain of $\tau_{1}$ (see Fig. \textbf{ToDo}a).
In particular, the condition $\tau_{1}=\tau_{2}-\tau$ cannot be satisfied
when $\tau_{2}\in[0,\tau[$ since $\tau_{1}\in[0,t]$. Hence we integrate
only over the latter domain
\begin{equation}
\langle\tilde{X}^{\varphi}(t)\tilde{X}^{\varphi}(t+\tau)\rangle=2\gamma\lim_{t\rightarrow\infty}\int_{0}^{t}d\tau_{1}e^{-2\lambda_{\varphi}\tau_{1}-\lambda_{\varphi}\tau}=2\gamma\lim_{t\rightarrow\infty}\frac{1-e^{-2\lambda_{\varphi}t}}{2\lambda_{\varphi}}e^{-\lambda_{\varphi}\tau}=\frac{\gamma}{\lambda_{\varphi}}e^{-\lambda_{\varphi}\tau}.
\end{equation}
On the other hand, when $\tau<0$, the domain of $\tau_{1}$ is larger
than that of $\tau_{2}$ (see Fig. \textbf{ToDo}a). Now the condition
$\tau_{1}=\tau_{2}-\tau=\tau_{2}+|\tau|$ cannot be satisfied when
$\tau_{1}\in[0,|\tau|[$ since $\tau_{2}\in[0,t-|\tau|]$. Hence,
in this case we integrate over the domain of $\tau_{2}$
\begin{equation}
\langle\tilde{X}^{\varphi}(t)\tilde{X}^{\varphi}(t+\tau)\rangle=2\gamma\lim_{t\rightarrow\infty}\int_{0}^{t+\tau}d\tau_{2}e^{-2\lambda_{\varphi}\tau_{2}+\lambda_{\varphi}\tau}=2\gamma\lim_{t\rightarrow\infty}\frac{1-e^{-2\lambda_{\varphi}(t+\tau)}}{2\lambda_{\varphi}}e^{\lambda_{\varphi}\tau}=\frac{\gamma}{\lambda_{\varphi}}e^{\lambda_{\varphi}\tau}.
\end{equation}
Therefore, putting both results together, we obtain
\begin{equation}
\langle\tilde{X}^{\varphi}(t)\tilde{X}^{\varphi}(t+\tau)\rangle=\frac{\gamma}{\lambda_{\varphi}}e^{-\lambda_{\varphi}|\tau|}.\label{XX}
\end{equation}
Combining (\ref{XXin}) and (\ref{XX}), we write the asymptotic two-time
correlation function of the output quadratures (\ref{TwoTimeOutX})
as
\begin{equation}
\lim_{t\rightarrow\infty}\langle\tilde{X}_{\text{out}}^{\varphi}(t)\tilde{X}_{\text{out}}^{\varphi}(t+\tau)\rangle=2\gamma\left(\frac{\gamma}{\lambda_{\varphi}}-1\right)e^{-\lambda_{\varphi}|\tau|}+\delta(\tau).
\end{equation}
Note that this function is independent of $t$ and it is symmetric
in $\tau$. Therefore, we can use the Fourier-transform expression
(\ref{NoiseSpectrum}) of the quadrature noise spectrum, obtaining
\begin{align}
V^{\varphi}(\Omega) & =\int_{-\infty}^{+\infty}d\tau e^{-\mathrm{i}\Omega\tau}\left[2\gamma\left(\frac{\gamma}{\lambda_{\varphi}}-1\right)e^{-\lambda_{\varphi}|\tau|}+\delta(\tau)\right]\\
 & =1+2\gamma\left(\frac{\gamma}{\lambda_{\varphi}}-1\right)\left[\int_{-\infty}^{0}d\tau e^{(\lambda_{\varphi}-\mathrm{i}\Omega)\tau}+\int_{0}^{+\infty}d\tau e^{-(\lambda_{\varphi}+\mathrm{i}\Omega)\tau}\right]\nonumber \\
 & =1+2\gamma\left(\frac{\gamma}{\lambda_{\varphi}}-1\right)\left(\frac{1}{\lambda_{\varphi}-\mathrm{i}\Omega}+\frac{1}{\lambda_{\varphi}+\mathrm{i}\Omega}\right)=1+\frac{4\gamma\left(\gamma-\lambda_{\varphi}\right)}{\lambda_{\varphi}^{2}+\Omega^{2}}.\nonumber 
\end{align}
Hence, introducing again the parameter $\sigma=g/\gamma$, we finally
get the quadrature noise spectrum of the position and momentum quadratures\begin{subequations}
\begin{align}
V^{0}(\Omega) & =1+\frac{4\sigma}{(1-\sigma)^{2}+(\Omega/\gamma)^{2}},\\
V^{\pi/2}(\Omega) & =1-\frac{4\sigma}{(1+\sigma)^{2}+(\Omega/\gamma)^{2}}.
\end{align}
\end{subequations}For $\sigma=0$ both spectra are equal to 1 for
all noise frequencies $\Omega$, which is consistent with the fact
that the intracavity mode is in vacuum. On the other hand, for any
other value of $\sigma$ we get $V^{\pi/2}(\Omega)<1$ and $V^{0}(\Omega)>1$,
showing that the output field is in a momentum-squeezed state. However,
contrary to the intracavity field, as the threshold $\sigma=1$ is
approached, the squeezing of the output field becomes perfect at zero
noise frequency, $V^{\pi/2}(\Omega=0)\rightarrow0$. Moreover, again
in contrast to the intracavity field, the output field is in a minimum
uncertainty state, as we see that $V^{0}(\Omega)V^{\pi/2}(\Omega)=1\hspace{1em}\forall(\Omega,\sigma)$.
Finally, note that the shape of the spectra is Lorentzian, with a
width given by the cavity decay rate, which sets the spectral bandwidth
around which we can expect quantum noise reduction.

Let us conclude by pointing out that these devices, optical parametric
oscillators, are the ones currently used in quantum optics and quantum
information experiments for the production of the highest quality
squeezed light, reaching impressive squeezing levels (the current
record standing at 97\% of quantum noise reduction \cite{97squeezing}).

\newpage

\section{Elimination of spurious degrees of freedom: effective models}

Strictly speaking, all the models that we have used so far (and essentially
all the models used in quantum optics, condensed matter, atomic physics,
etc...) are effective, in the sense that they do not describe all
their fundamental constituents and elementary interactions, but rather
consider a few relevant degrees of freedom that emerge from them.
While these relevant degrees of freedom appear very natural in many
situations, it is interesting to understand more deeply the reasons
under which they emerge, and design systematic ways of deriving the
corresponding effective models.

We have already seen a few situations leading to effective models.
First, when describing an atom, we implicitly assumed that the strong
force confining quarks inside protons and neutrons, and even the strong
nuclear force binding the nucleons into atomic nuclei, are so large
compared with the frequencies and the powers of common lasers, that
we can simply take the whole nuclei as a single particle. This is
an example of strongly correlated systems in which many degrees of
freedom merge into just a few effective ones. We went through another
example when discussing open systems and the master equation in particular:
the environment (external field) was so large as compared with the
system (atom or cavity mode), that the latter had little-to-none effect
on the former. Moreover, the dynamics of the environment occur on
a much faster time scale than that of the system, as shown by the
fact that environmental correlation functions decayed instantaneously
within the Markov approximation. Hence, in such situations, one can
find an effective description for the dynamics of the system, for
which the environment is essentially frozen in some fixed state.

Hence, we see that the concept of effective theories is rooted at
the heart of quantum optics, or more generally, quantum mechanics
at low energies. Moreover, in recent decades we have stopped looking
at effective theories as some kind of natural phenomena, but rather
as something that we can exploit to engineer desired (effective) models
in our experimental systems.

In this section we are going to analyze this scenario, showing systematic
methods for the derivation of effective models both for closed and
open systems. We apply these methods to two very relevant examples:
the engineering of effective motional potentials and zero-temperature
effective environments for an atom via its coupling to light.

\subsection{Elimination in closed systems}

There are many methods designed to find effective Hamiltonians in
closed systems. Some of them are quite general, while others are more
applied to particular scenarios. Which one of them is better, actually
depends on the specific problem. However, there are two methods which
have found a wide range of applicability. The first one, which we
introduce next, is based on projector operators, and works very well
when we can identify the sector of the Hilbert space we are interested
in and captures most (ideally all) the system excitations. The second
one, known as Schrieffer-Wolff or Frölich transformation, is essentially
equivalent to Hamiltonian-eigensystem perturbation theory, but leading
to explicit and simple expressions for the different perturbation
terms directly in the Hamiltonian (\textbf{not shown yet in the notes}).

\subsubsection{Projection operators method}

With full generality, consider a closed system evolving according
to a Hamiltonian $\hat{H}$, that is, its state $|\psi(t)\rangle$
satisfies the Schrödinger equation $\mathrm{i}\hbar\partial_{t}|\psi(t)\rangle=\hat{H}|\psi(t)\rangle$.
The idea of the method relies on the fact that we can divide the Hilbert
space into a relevant sector (whose effective dynamics we want to
describe) and an irrelevant one (whose dynamics is trivial, typically
because it stays unpopulated). We then define the projector operator
$\hat{P}=\hat{P}^{2}$, which projects onto the relevant subspace,
and its complement $\hat{Q}=1-\hat{P}$. Applying this projectors
onto the Schrödinger equation, we get the equations\begin{subequations}
\begin{align}
\mathrm{i}\hbar\partial_{t}\hat{P}|\psi(t)\rangle & =\hat{P}\hat{H}(\underset{1}{\underbrace{\hat{P}+\hat{Q}}})|\psi(t)\rangle=\hat{P}\hat{H}\hat{P}|\psi(t)\rangle+\hat{P}\hat{H}\hat{Q}|\psi(t)\rangle,\\
\mathrm{i}\hbar\partial_{t}\hat{Q}|\psi(t)\rangle & =\hat{Q}\hat{H}(\underset{1}{\underbrace{\hat{P}+\hat{Q}}})|\psi(t)\rangle=\hat{Q}\hat{H}\hat{Q}|\psi(t)\rangle+\hat{Q}\hat{H}\hat{P}|\psi(t)\rangle.
\end{align}
\end{subequations}which are two coupled equations for the projections
$\hat{P}|\psi(t)\rangle$ and $\hat{Q}|\psi(t)\rangle$. We can formally
solve the second equation\footnote{Note that it's formally like (\ref{GenLinEq}), but with operators.
With full generality, consider the equation $\partial_{t}|\psi\rangle=\hat{V}|\psi\rangle+|\chi(t)\rangle$.
The solution is found following similar steps as with (\ref{GenLinEq}).
First we make the change $|\Psi(t)\rangle=e^{-\hat{V}t}|\psi(t)\rangle$,
so that $\partial_{t}|\Psi\rangle=e^{-\hat{V}t}(\partial_{t}|\psi\rangle-\hat{V}|\psi\rangle)=e^{-\hat{V}t}|\chi(t)\rangle\Rightarrow|\Psi(t)\rangle=|\Psi(t_{0})\rangle+\int_{t_{0}}^{t}dt'e^{-\hat{V}t'}|\chi(t')\rangle$.
Then, undoing the variable change as $|\psi(t)\rangle=e^{\hat{V}t}|\Psi(t)\rangle$,
we find
\begin{equation}
|\psi(t)\rangle=e^{\hat{V}(t-t_{0})}|\psi(t_{0})\rangle+\int_{t_{0}}^{t}dt'e^{\hat{V}(t-t')}|\chi(t')\rangle.\label{GenLinVecEq}
\end{equation}
}, obtaining
\begin{equation}
\hat{Q}|\psi(t)\rangle=\frac{1}{\mathrm{i}\hbar}e^{\hat{Q}\hat{H}t/\mathrm{i}\hbar}\hat{Q}|\psi(0)\rangle+\int_{0}^{t}\frac{dt'}{\mathrm{i}\hbar}e^{\hat{Q}\hat{H}(t-t')/\mathrm{i}\hbar}\hat{Q}\hat{H}\hat{P}|\psi(t')\rangle.\label{FormalQpsi}
\end{equation}
Naturally, we assume that the system is in the relevant subspace initially,
so that $\hat{Q}|\psi(0)\rangle=0$. Hence, substituting this formal
solution into the first equation, we obtain
\begin{equation}
\mathrm{i}\hbar\partial_{t}\hat{P}|\psi(t)\rangle=\hat{P}\hat{H}\hat{P}|\psi(t)\rangle+\int_{0}^{t}\frac{d\tau}{\mathrm{i}\hbar}\hat{P}\hat{H}e^{\hat{Q}\hat{H}\tau/\mathrm{i}\hbar}\hat{Q}\hat{H}\hat{P}|\psi(t-\tau)\rangle,
\end{equation}
where we have made the integration variable change $t'=t-\tau$. We
can rewrite this expression as an effective Schrödinger equation on
the relevant part of the state, $\hat{P}|\psi(t)\rangle$. In particular,
we know that $|\psi(t-\tau)\rangle=e^{-\hat{H}\tau/\mathrm{i}\hbar}|\psi(t)\rangle$,
which allows us to write
\begin{align}
\mathrm{i}\hbar\partial_{t}\hat{P}|\psi(t)\rangle & =\hat{P}\hat{H}\hat{P}|\psi(t)\rangle+\int_{0}^{t}\frac{d\tau}{\mathrm{i}\hbar}\hat{P}\hat{H}e^{\hat{Q}\hat{H}\tau/\mathrm{i}\hbar}\hat{Q}\hat{H}\hat{P}e^{-\hat{H}\tau/\mathrm{i}\hbar}|\psi(t)\rangle,\\
 & =\hat{P}\hat{H}\hat{P}|\psi(t)\rangle+\int_{0}^{t}\frac{d\tau}{\mathrm{i}\hbar}\hat{P}\hat{H}e^{\hat{Q}\hat{H}\tau/\mathrm{i}\hbar}\hat{Q}\hat{H}\hat{P}e^{-\hat{H}\tau/\mathrm{i}\hbar}(\underset{1}{\underbrace{\hat{P}+\hat{Q}}})|\psi(t)\rangle.\nonumber 
\end{align}
Reintroducing (\ref{FormalQpsi}) and iterating this process, we then
obtain $\mathrm{i}\hbar\partial_{t}\hat{P}|\psi(t)\rangle=\hat{H}_{\text{eff}}(t)\hat{P}|\psi(t)\rangle$,
with an effective Hamiltonian given by
\begin{equation}
\hat{H}_{\text{eff}}(t)=\sum_{n=0}^{\infty}\hat{P}\hat{H}\left[\int_{0}^{t}\frac{d\tau}{\mathrm{i}\hbar}e^{\hat{Q}\hat{H}\tau/\mathrm{i}\hbar}\hat{Q}\hat{H}\hat{P}e^{-\hat{H}\tau/\mathrm{i}\hbar}\right]^{n}\hat{P}.\label{Heff}
\end{equation}
There are three interesting things to note about this expression.
First, it is not Hermitian. This seems to be at odds with the fact
that we take it as an effective Hamiltonian. In fact, this is telling
us that, in general, one cannot simply write the effective dynamics
of a closed system as another closed system, but rather as an open
system, that is, as a mixed state evolving according to a master equation.
In fact, we already saw examples of this in Chapter \ref{Sec:OpenSystems},
where we started from a system+environment model described by a Hamiltonian,
but ended up with a master equation for the system, not with an effective
Hamiltonian. However, in many situations it indeed occurs that (\ref{Heff})
becomes approximately Hermitian under the physical conditions we work
on, as we will see in the example we treat in the next section. In
general terms, we can expect this to happen when the dynamics and
physical conditions do not allow excitations of the system to leak
out from the relevant subspace, which is clearly not the case for
the open systems considered in Chapter \ref{Sec:OpenSystems}.

The second interesting thing is that (\ref{Heff}) seems to be time-dependent.
Hence, it looks like the effective Hamiltonian is time dependent,
even if the original Hamiltonian wasn't. However, in most situations
one can show that the time-dependent terms are negligible within rotating-wave
like approximations. This will again become clear with the example
of the next section.

The final and most interesting thing to note is that, in general we
can decompose the Hamiltonian as $\hat{H}=\hat{H}_{0}+\hat{H}_{1}$.
$\hat{H}_{0}$ is denoted by \emph{free Hamiltonian} and contains
all the terms that do not connect the relevant and irrelevant subspaces,
that is, $\hat{P}\hat{H}_{0}\hat{Q}=0=\hat{Q}\hat{H}_{0}\hat{P}$.
On the other hand, $\hat{H}_{1}$ is denoted by \emph{interaction
Hamiltonian }and gathers the rest of the terms. Note that we can even
assume without loss of generality that $\hat{P}\hat{H}_{1}\hat{P}=0$,
that is, the interaction Hamiltonian does not connect states within
the relevant subspace. It is always possible to ensure such a property,
for if that's not the case, we just need to redefine the interaction
and free Hamiltonians as $\hat{H}_{1}-\hat{P}\hat{H}_{1}\hat{P}$
and $\hat{H}_{0}+\hat{P}\hat{H}_{1}\hat{P}$, which we can do because
$\hat{Q}(\hat{P}\hat{H}_{1}\hat{P})\hat{P}=0=\hat{P}(\hat{P}\hat{H}_{1}\hat{P})\hat{Q}$.
Effective theories are meaningful whenever one can treat $\hat{H}_{1}$
as a perturbation with respect to $\hat{H}_{0}$. Hence, we don't
need to consider the full effective Hamiltonian (\ref{Heff}), but
rather terms up to a desired order in the interaction.

Terms of order larger than two require a Dyson expansion of the exponentials,
which is not difficult, but rather cumbersome. Let us then consider
here terms up to second order, which we show now to appear by taking
up to $n=1$ in the sum, leading to
\begin{equation}
\hat{H}_{\text{eff}}(t)=\hat{P}\hat{H}\hat{P}+\int_{0}^{t}\frac{d\tau}{\mathrm{i}\hbar}\hat{P}\hat{H}e^{\hat{Q}\hat{H}\tau/\mathrm{i}\hbar}\hat{Q}\hat{H}\hat{P}e^{-\hat{H}\tau/\mathrm{i}\hbar}\hat{P}.
\end{equation}
Expanding the first exponential in Taylor series, and recalling $\hat{P}\hat{H}_{0}\hat{Q}=0=\hat{Q}\hat{H}_{0}\hat{P}$,
we see that the $\hat{H}$'s appearing in the second term can be replaced
by $\hat{H}_{1}$. Hence, this second term is order two in the interaction
already without counting the exponentials; hence, we can set the Hamiltonians
in the exponentials to $\hat{H}_{0}$, since the $\hat{H}_{1}$ contribution
will lead to terms of order larger than two overall. We then obtain
\begin{equation}
\hat{H}_{\text{eff}}(t)=\hat{P}\hat{H}_{0}\hat{P}+\int_{0}^{t}\frac{d\tau}{\mathrm{i}\hbar}\hat{P}\hat{H}_{1}e^{\hat{Q}\hat{H}_{0}\tau/\mathrm{i}\hbar}\hat{Q}\hat{H}_{1}\hat{P}e^{-\hat{H}_{0}\tau/\mathrm{i}\hbar}\hat{P},
\end{equation}
where we have further used $\hat{P}\hat{H}_{1}\hat{P}=0$. This expression
can be simplified even further by using $[\hat{P},\hat{H}_{0}]=0=[\hat{Q},\hat{H}_{0}]$,
which follows directly from $0=\hat{P}\hat{H}_{0}\hat{Q}-\hat{Q}\hat{H}_{0}\hat{P}=\hat{P}\hat{H}_{0}(\hat{1}-\hat{P})-(\hat{1}-\hat{P})\hat{H}_{0}\hat{P}=[\hat{P},\hat{H}_{0}]$.
This ensures that the relevant subspace can be chosen as a subset
of the eigenspace of the free Hamiltonian. Using this property, we
can then move $\hat{Q}$ down from the first exponential as

\begin{align}
e^{\hat{Q}\hat{H}_{0}\tau/\mathrm{i}\hbar} & =\sum_{k=0}^{\infty}\frac{1}{k!}\left(\frac{\tau}{\mathrm{i}\hbar}\right)^{k}\underset{k\text{ times}}{\underbrace{\hat{Q}\hat{H}_{0}\hat{Q}\hat{H}_{0}...\hat{Q}\hat{H}_{0}}}=1+\sum_{k=1}^{\infty}\frac{1}{k!}\left(\frac{\tau}{\mathrm{i}\hbar}\right)^{k}\underset{\hat{Q}}{\underbrace{\hat{Q}^{k}}}\hat{H}_{0}^{k}=1+\hat{Q}\underbrace{\sum_{k=1}^{\infty}\frac{1}{k!}\left(\frac{\tau}{\mathrm{i}\hbar}\right)^{k}\hat{H}_{0}^{k}}_{e^{\hat{H}_{0}\tau/\mathrm{i}\hbar}-1}=\hat{P}+\hat{Q}e^{\hat{H}_{0}\tau/\mathrm{i}\hbar},
\end{align}
where we have used $\hat{P}\hat{Q}=0$, leading to
\begin{equation}
\hat{H}_{\text{eff}}(t)=\hat{P}\hat{H}_{0}\hat{P}+\int_{0}^{t}\frac{d\tau}{\mathrm{i}\hbar}\hat{P}\hat{H}_{1}\hat{Q}e^{\hat{H}_{0}\tau/\mathrm{i}\hbar}\hat{Q}\hat{H}_{1}\hat{P}e^{-\hat{H}_{0}\tau/\mathrm{i}\hbar}\hat{P}.
\end{equation}
Using next $\hat{P}\hat{H}_{1}\hat{Q}=\hat{P}\hat{H}_{1}(1-\hat{P})=\hat{P}\hat{H}_{1}$,
and similarly $\hat{Q}\hat{H}_{1}\hat{P}=\hat{H}_{1}\hat{P}$, we
obtain the final form of the effective Hamiltonian up to second order
in the interaction
\begin{equation}
\hat{H}_{\text{eff}}(t)=\hat{P}\hat{H}_{0}\hat{P}+\int_{0}^{t}\frac{d\tau}{\mathrm{i}\hbar}\hat{P}\hat{H}_{1}\tilde{H}_{1}(t)\hat{P},\label{Heff2}
\end{equation}
where
\begin{equation}
\tilde{H}_{1}(\tau)=e^{\hat{H}_{0}\tau/\mathrm{i}\hbar}\hat{H}_{1}e^{-\hat{H}_{0}\tau/\mathrm{i}\hbar}.
\end{equation}
This is a very simple expression whose use we put in practice next
with an important example.

\subsubsection{Optical motional potential on an atom}

Let us now consider a particular example: an atom whose center of
mass is free to move, interacts with a classical optical field very
far detuned from the atomic transition. At a first sight, it might
seem like nothing will happen, since the field is far detuned. However,
we will see that a perturbative effective motional potential with
the shape of the intensity of the optical beam is generated on the
atomic motion.

The problem is defined by two degrees of freedom. First, the center-of-mass
motion, whose position and momentum operators we denote, respectively,
by $\hat{z}$ and $\hat{p}$ (as usual, we stick to the quasi-1D model
we have been using), and evolves according to the Hamiltonian $\hat{p}^{2}/2m$,
where $m$ is the mass of the atom. The other degree of freedom refers
to the relative atomic coordinate (sometimes denoted by \emph{internal}
or \emph{electronic} degree of freedom), which we treat in the usual
way defined in Chapter \ref{Sec:Atoms} within the two-level approximation,
evolving according to the the Hamiltonian $\hbar\varepsilon\hat{\sigma_{z}}/2$.
We connect these two degrees of freedom by using a laser whose power
we allow to vary spatially, leading to an interaction Hamiltonian
$\hbar\left[\Omega^{*}(\hat{z})e^{\mathrm{i}\omega_{\text{L}}t}\hat{\sigma}+\Omega(\hat{z})e^{-\mathrm{i}\omega_{\text{L}}t}\hat{\sigma}\right]/2$
within the rotating-wave approximation. Moving to a picture rotating
at the laser frequency as usual, we then obtain the time-independent
Hamiltonian
\begin{equation}
\hat{H}=\underset{\hat{H}_{0}}{\underbrace{\frac{\hat{p}^{2}}{2m}+\frac{\hbar\Delta}{2}\hat{\sigma}_{z}}}+\underset{\hat{H}_{1}}{\underbrace{\frac{\hbar}{2}\left[\Omega^{*}(\hat{z})\hat{\sigma}+\Omega(\hat{z})\hat{\sigma}^{\dagger}\right]}},
\end{equation}
where we have defined the detuning as $\Delta=\varepsilon-\omega_{\text{L}}$.
We are interested in the effective dynamics of the system when the
laser is far off resonance, that is, $|\Delta|\gg|\Omega|$. As we
learned in previous chapters, we don't expect anything to happen to
the internal dynamics of the atom. In particular, initializing the
atom in the ground state, we expect it to remain in that state, and
hence we define the projector $\hat{P}=|g\rangle\langle g|$, which
acts as the identity in the motional space. We will show, however,
that some nontrivial effective dynamics will be induced in the center
of mass motion of the atom.

Let us first note that $\hat{P}\hat{H}_{1}\hat{P}\propto\langle g|\hat{H}_{1}|g\rangle=0$
in this case. On the other hand, defining the kinetic evolution operator
$\hat{K}(\tau)=\exp\left(-\mathrm{i}\hat{p}^{2}\tau/2m\hbar\right)$,
we can write
\begin{equation}
\tilde{H}_{1}(\tau)=\frac{\hbar}{2}\left[\hat{K}(\tau)\Omega^{*}(\hat{z})\hat{K}^{\dagger}(\tau)\hat{\sigma}e^{\mathrm{i}\Delta\tau}+\hat{K}(\tau)\Omega(\hat{z})\hat{K}^{\dagger}(\tau)\hat{\sigma}^{\dagger}e^{-\mathrm{i}\Delta\tau}\right],
\end{equation}
and hence
\begin{equation}
\langle g|\hat{H}_{1}\tilde{H}_{1}(\tau)|g\rangle=\frac{\hbar^{2}}{4}e^{-\mathrm{i}\Delta\tau}\Omega^{*}(\hat{z})\hat{K}(\tau)\Omega(\hat{z})\hat{K}^{\dagger}(\tau),
\end{equation}
where we have used $\hat{\sigma}^{2}=0=\langle g|\hat{\sigma}^{\dagger}\hat{\sigma}|g\rangle$
and $\langle g|\hat{\sigma}\hat{\sigma}^{\dagger}|g\rangle=1$. We
then write the effective Hamiltonian (\ref{Heff2}) as
\begin{equation}
\hat{H}_{\text{eff}}(t)=\Biggl[\underbrace{-\frac{\hbar\Delta}{2}}_{\frac{\hbar\Delta}{2}\langle g|\hat{\sigma}_{z}|g\rangle}+\frac{\hat{p}^{2}}{2m}-\frac{\mathrm{i}\hbar}{4}\int_{0}^{t}d\tau e^{-\mathrm{i}\Delta\tau}\Omega^{*}(\hat{z})\hat{K}(\tau)\Omega(\hat{z})\hat{K}^{\dagger}(\tau)\Biggr]|g\rangle\langle g|.
\end{equation}
We finally assume that the atoms have been cooled down, so their kinetic
energy is small in the sense $\langle\hat{p}^{2}\rangle/2m\ll\hbar|\Delta|$.
We can then neglect the kinetic evolution inside the integral under
a rotating-wave approximation, finally obtaining
\begin{equation}
\hat{H}_{\text{eff}}(t)\approx\left[-\frac{\hbar\Delta}{2}+\frac{\hat{p}^{2}}{2m}-\frac{\hbar|\Omega(\hat{z})|^{2}}{4\Delta}\left(1-e^{-\mathrm{i}\Delta t}\right)\right]|g\rangle\langle g|\approx\left[-\frac{\hbar\Delta}{2}+\frac{\hat{p}^{2}}{2m}-\frac{\hbar|\Omega(\hat{z})|^{2}}{4\Delta}\right]|g\rangle\langle g|,
\end{equation}
where in the last step we have have dropped out the $e^{-\mathrm{i}\Delta t}$
term, again within a rotating-wave approximation, since $|\Omega(\hat{z})|\ll|\Delta|$.
Hence, we see that the laser induces an effective potential $V_{\text{opt}}(\hat{z})=-\hbar|\Omega(\hat{z})|^{2}/4\Delta$
on the atom's motion, which is known as \emph{optical potential}.
Alternatively, this effect can be understood as an energy shift $V_{\text{opt}}(\hat{z})$
of the ground-state energy, a phenomenon known as \emph{AC Stark shift},
which occurs even when the atom's center of mass is fixed. Hence,
even though the laser is not capable of transferring population from
the ground to the excited state, it has a perturbative effect on the
atomic energy levels in the form of a shift (it can be easily proven
that the excited state feels an opposite AC Stark shift $-V_{\text{opt}}$).

\subsection{Elimination in open systems}

\subsubsection{Projection superoperators method}

Let us now generalize the previous methods for obtaining effective
models to open systems. The same kind of general principles can be
applied to the von Neumann or master equations. Here, we focus on
the generalization of the projector-operator method. Consider hence
a system whose state $\hat{\rho}$ evolves according to a master equation
$\partial_{t}\hat{\rho}=\mathcal{L}[\hat{\rho}]$, with a time-independent
Liouvillian for simplicity as usual. Let us define a projector superoperator
$\mathcal{P}=\mathcal{P}^{2}$, as well as its complement $\mathcal{Q}=1-\mathcal{P}$.
In this case, it is more difficult to interpret such objects, since
they do not directly project in the Hilbert space, but in the space
of operators. Sometimes, though, their action is similar to the one
of the projector operators that we encounter in the previous section:
they simply divide the Hilbert space in relevant states and irrelevant
states. However, since now we are dealing with mixed states, in general
we need to do that by defining a set of relevant and irrelevant (density)
operators. In any case, the meaning of these projection superoperators
will become clearer through examples later. Acting on the master equation
with this projectors, we then find\begin{subequations}
\begin{align}
\partial_{t}\mathcal{P}[\hat{\rho}] & =\mathcal{P}\mathcal{L}\underbrace{(\mathcal{P}+\mathcal{Q})}_{1}[\hat{\rho}]=\mathcal{P}\mathcal{L}\mathcal{P}[\hat{\rho}]+\mathcal{P}\mathcal{L}\mathcal{Q}[\hat{\rho}],\\
\partial_{t}\mathcal{Q}[\hat{\rho}] & =\mathcal{Q}\mathcal{L}\underbrace{(\mathcal{P}+\mathcal{Q})}_{1}[\hat{\rho}]=\mathcal{Q}\mathcal{L}\mathcal{Q}[\hat{\rho}]+\mathcal{Q}\mathcal{L}\mathcal{P}[\hat{\rho}].
\end{align}
\end{subequations}Our goal is finding an effective evolution equation
for the relevant part of the state $\mathcal{P}[\hat{\rho}]$. For
that, we simply integrate formally the second equation\footnote{We simply need to generalize (\ref{GenLinVecEq}) by interpreting
operators as vectors in the space of operators, and superoperators
as operators acting on such space. Hence, the solution to $\partial_{t}\hat{\rho}=\mathcal{M}[\hat{\rho}]+\hat{\chi}(t)$
is easily found as
\begin{equation}
\hat{\rho}(t)=e^{\mathcal{M}(t-t_{0})}[\hat{\rho}(t_{0})]+\int_{t_{0}}^{t}dt'e^{\mathcal{M}(t-t')}[\hat{\chi}(t')].\label{GenLinOpEq}
\end{equation}
}, obtaining
\begin{equation}
\mathcal{Q}[\hat{\rho}(t)]=e^{\mathcal{QL}t}\mathcal{Q}[\hat{\rho}(0)]+\int_{0}^{t}dt'e^{\mathcal{QL}(t-t')}\mathcal{Q}\mathcal{L}\mathcal{P}[\hat{\rho}(t')].
\end{equation}
The first term depends on the initial condition and can be eliminated
under either one of the following physical assumptions: either the
initial state belongs already to the relevant subspace, or we are
interested in the asymptotic dynamics only, which should not depend
on the initial condition for well-behaved open systems, as we have
seen in previous chapters. Hence, substituting this expression into
the first equation we obtain
\begin{align}
\partial_{t}\mathcal{P}[\hat{\rho}] & =\mathcal{P}\mathcal{L}\mathcal{P}[\hat{\rho}]+\int_{0}^{t}d\tau\mathcal{P}\mathcal{L}e^{\mathcal{QL}\tau}\mathcal{Q}\mathcal{L}\mathcal{P}[\hat{\rho}(t-\tau)],
\end{align}
where we made the integration variable change $\tau=t-t'$. Up to
here, the derivation has followed exactly the same steps as with the
closed system, just with the obvious generalizations. To continue
the equivalence, we would now need to write the state $\hat{\rho}(t-\tau)$
as $\hat{\rho}(t)$ evolved backwards during a time $\tau$. While
for closed systems this is useful because the time-evolution is very
simple, for open systems it doesn't really lead us to a practical
form of the effective Liouvillian. Hence, it is better to proceed
for now with this time-nonlocal equation.

Similarly to the closed system, we now expand the Liouvillian as $\mathcal{L}=\mathcal{L}_{0}+\mathcal{L}_{1}$.
The first term, denoted by \emph{free Liouvillian}, contains only
terms that do not connect the relevant and irrelevant subspaces, that
is, $\mathcal{P}\mathcal{L}_{0}\mathcal{Q}=0=\mathcal{Q}\mathcal{L}_{0}\mathcal{P}$.
The second term, denoted by \emph{interaction Liouvillian}, contains
the rest of terms, and we can assume that $\mathcal{P}\mathcal{L}_{1}\mathcal{P}=0$
with full generality, following the arguments given for closed systems.
Just as with the latter, also for open systems effective theories
only make sense as long as the interaction Liouvillian can be treated
as a perturbation. We then truncate to second order in $\mathcal{L}_{1}$
and follow similar steps as we did for the closed case, obtaining
the effective (time-nonlocal) master equation
\begin{equation}
\partial_{t}\mathcal{P}[\hat{\rho}]\approx\mathcal{P}\mathcal{L}_{0}\mathcal{P}[\hat{\rho}]+\int_{0}^{t}d\tau\mathcal{P}\mathcal{L}_{1}e^{\mathcal{L}_{0}\tau}\mathcal{L}_{1}\mathcal{P}[\hat{\rho}(t-\tau)].\label{EffMEgen}
\end{equation}

In order to proceed forward, we consider now one specific situation
that it's usually the most interesting one in open quantum optical
settings. We consider a bipartite scenario consisting on a \emph{system}
and an \emph{environment} with Hilbert spaces $\mathcal{H}_{\text{S}}$
and $\mathcal{H}_{\text{E}}$ respectively, so that the total Hilbert
space is $\mathcal{H}=\mathcal{H}_{\text{S}}\otimes\mathcal{H}_{\text{E}}$.
The notation already suggests that we will treat the case in which
the environment remains essentially frozen from the point of view
of the system, so that we can find an effective master equation for
the latter in a similar way how we did it in Chapter \ref{Sec:OpenSystems}.
However, in contrast to that chapter, we will not assume here that
the environment is a huge system, but rather study the conditions
under which such effective theory can be developed. Indeed, we will
see that even when the environment is `small', one can still use it
to induce effective dynamics on the system as long as it satisfies
certain conditions (e.g., couples weakly to the system and decays
fast enough).

Let us then write the free Liouvillian as $\mathcal{L}_{0}=\mathcal{L}_{\text{S}}+\mathcal{L}_{\text{E}}$,
where $\mathcal{L}_{\text{S}}$ acts nontrivially only onto system
operators and $\mathcal{L}_{\text{E}}$ onto environment operators,
so that $[\mathcal{L}_{\text{S}},\mathcal{L}_{\text{E}}]=0$. As the
reference state for the environment we take its asymptotic state $\bar{\rho}_{\text{E}}$,
defined by $\mathcal{L}_{\text{E}}[\bar{\rho}_{\text{E}}]=0$. Consequently,
we take as the relevant subspace the one defined by density operators
that match this reference one for the environment, leading to a projector
superoperator $\mathcal{P}$ defined by the following action on any
operator $\hat{Y}$:
\begin{equation}
\mathcal{P}[\hat{Y}]=\text{tr}_{\text{E}}\{\hat{Y}\}\otimes\bar{\rho}_{\text{E}}.
\end{equation}
Denoting the system's state by $\hat{\rho}_{\text{S}}=\text{tr}_{\text{E}}\{\hat{\rho}\}$,
we then rewrite the effective master equation (\ref{EffMEgen}) as
\begin{equation}
\partial_{t}\hat{\rho}_{\text{S}}\approx\mathcal{L}_{\text{S}}[\hat{\rho}_{\text{S}}]+\int_{0}^{t}d\tau\text{tr}_{\text{E}}\left\{ \mathcal{L}_{1}e^{\mathcal{L}_{0}\tau}\mathcal{L}_{1}[\hat{\rho}_{\text{S}}(t-\tau)\otimes\bar{\rho}_{\text{E}}]\right\} ,\label{EffMEintermediate}
\end{equation}
where we have performed a trace over the environment. In order to
simplify further, we need to give a form to the interaction. Usually,
we are interested in cases in which the system and the environment
are connected by a Hamiltonian, say $\hat{H}_{1}$. We will assume
this in the following, although the next steps are easily generalizable
to any other situation. Moreover, we assume that the interaction admits
some kind of polynomial expansion in terms of products of system and
environment operators, that is,
\begin{equation}
\hat{H}_{1}/\hbar=\sum_{m=1}^{M}g_{m}\hat{S}_{m}\otimes\hat{E}_{m},\label{H1}
\end{equation}
where we note that $\hat{H}_{1}=\hat{H}_{1}^{\dagger}$ imposes a
restriction on the set of operators, that is, each product must either
be Hermitian or have its Hermitian-conjugate counterpart in the set.
The interaction Liouvillian acts then as 
\begin{equation}
\mathcal{L}_{1}[\hat{Y}]=\left[\frac{\hat{H}_{1}}{\mathrm{i}\hbar},\hat{Y}\right]=-\mathrm{i}\sum_{m=1}^{M}g_{m}[\hat{S}_{m}\otimes\hat{E}_{m},\hat{Y}],
\end{equation}
which introduced in (\ref{EffMEintermediate}) leads to
\begin{align}
\text{tr}_{E}\left\{ \mathcal{L}_{1}e^{\mathcal{L}_{0}\tau}\mathcal{L}_{1}[\hat{\rho}_{\text{S}}(t-\tau)\otimes\bar{\rho}_{\text{E}}]\right\}  & =\sum_{m,n=1}^{M}g_{m}g_{n}\biggl(\text{tr}_{\text{E}}\left\{ \hat{E}_{n}e^{\mathcal{L}_{\text{E}}\tau}\left[\bar{\rho}_{\text{E}}\hat{E}_{m}\right]\right\} \hat{S}_{n}e^{\mathcal{L}_{\text{S}}\tau}\left[\hat{\rho}_{\text{S}}(t-\tau)\hat{S}_{m}\right]\\
 & \hspace{1em}\hspace{1em}\hspace{1em}\hspace{1em}\hspace{1em}\hspace{1em}-\text{tr}_{\text{E}}\left\{ \hat{E}_{n}e^{\mathcal{L}_{\text{E}}\tau}\left[\hat{E}_{m}\bar{\rho}_{\text{E}}\right]\right\} \hat{S}_{n}e^{\mathcal{L}_{\text{S}}\tau}\left[\hat{S}_{m}\hat{\rho}_{\text{S}}(t-\tau)\right]\biggr)+\text{H.c.}.\nonumber 
\end{align}
Invoking now the quantum regression theorem (\ref{QRTstationary}),
we see that the terms involving traces over the environment are nothing
but asymptotic two-time correlators that we denote by\begin{subequations}\label{CKcorrs}
\begin{align}
\text{tr}_{\text{E}}\left\{ \hat{E}_{n}e^{\mathcal{L}_{\text{E}}\tau}\left[\bar{\rho}_{\text{E}}\hat{E}_{m}\right]\right\}  & =\lim_{t\rightarrow\infty}\langle\hat{E}_{m}(t)\hat{E}_{n}(t+\tau)\rangle_{\text{E}}=C_{mn}(\tau),\label{Ccorrs}\\
\text{tr}_{\text{E}}\left\{ \hat{E}_{n}e^{\mathcal{L}_{\text{E}}\tau}\left[\hat{E}_{m}\bar{\rho}_{\text{E}}\right]\right\}  & =\lim_{t\rightarrow\infty}\langle\hat{E}_{n}(t+\tau)\hat{E}_{m}(t)\rangle_{\text{E}}=K_{nm}(\tau),\label{Kcorrs}
\end{align}
\end{subequations}where the subscript `E' in the expectation value
denotes that it is related to the free-environment Liouvillian, not
to the whole Liouvillian. Note that $K_{nm}(\tau)=\lim_{t\rightarrow\infty}\langle\hat{E}_{n}(t)\hat{E}_{m}(t-\tau)\rangle_{\text{E}}=C_{nm}(-\tau)$,
which is sometimes useful. Inserting these correlators in the previous
expression, we get to
\begin{equation}
\text{tr}_{\text{E}}\left\{ \mathcal{L}_{1}e^{\mathcal{L}_{0}\tau}\mathcal{L}_{1}[\hat{\rho}_{\text{S}}(t-\tau)\otimes\bar{\rho}_{\text{E}}]\right\} \approx\sum_{m,n=1}^{M}g_{m}g_{n}\left(C_{mn}(\tau)\hat{S}_{n}e^{\mathcal{L}_{\text{S}}\tau}\left[\hat{\rho}_{\text{S}}(t-\tau)\hat{S}_{m}\right]-K_{nm}(\tau)\hat{S}_{n}e^{\mathcal{L}_{\text{S}}\tau}\left[\hat{S}_{m}\hat{\rho}_{\text{S}}(t-\tau)\right]\right)+\text{H.c.}.
\end{equation}

We arrive then to the crucial assumption that will allow us to find
a time-local effective master equation for the system: we assume that
the correlation functions $C_{mn}(\tau)$ and $K_{nm}(\tau)$ are
either zero or decay as a function of $\tau$, at a much faster rate
than any process affecting the system, except for some Hamiltonian
term that we denote by $\hat{H}_{\text{S}}$. Typically $\hat{H}_{\text{S}}$
corresponds to some Hamiltonian terms of $\mathcal{L}_{\text{S}}$.
On the other hand, we must check this approximation carefully, estimating
the rate of the processes taken into account both by $\mathcal{L}_{\text{S}}$
and the final effective Liouvillian, in a self-consistent manner.
This will become clear in the example of the next section. Applied
to the previous expression, this approximation amounts to\begin{subequations}
\begin{align}
\hat{\rho}_{\text{S}}(t-\tau) & \approx e^{-\hat{H}_{\text{S}}\tau/\mathrm{i}\hbar}\hat{\rho}_{\text{S}}(t)e^{\hat{H}_{\text{S}}\tau/\mathrm{i}\hbar},\\
e^{\mathcal{L}_{\text{S}}\tau}[\hat{S}] & \approx e^{\hat{H}_{\text{S}}\tau/\mathrm{i}\hbar}\hat{S}e^{-\hat{H}_{\text{S}}\tau/\mathrm{i}\hbar},
\end{align}
\end{subequations}where $\hat{S}$ can be any system operators, finally
obtaining the effective time-local master equation
\begin{equation}
\partial_{t}\hat{\rho}_{\text{S}}\approx\mathcal{L}_{\text{S}}[\hat{\rho}_{\text{S}}]+\left(\sum_{m,n=1}^{M}g_{m}g_{n}\int_{0}^{t}d\tau\left[C_{mn}(\tau)\hat{S}_{n}\hat{\rho}_{\text{S}}(t)\tilde{S}_{m}(\tau)-K_{nm}(\tau)\hat{S}_{n}\tilde{S}_{m}(\tau)\hat{\rho}_{\text{S}}(t)\right]+\text{H.c.}\right),\label{EffMasterEQ}
\end{equation}
with $\tilde{S}(\tau)=e^{\hat{H}_{S}\tau/\mathrm{i}\hbar}\hat{S}e^{-\hat{H}_{S}\tau/\mathrm{i}\hbar}$.

Note that, in general, this effective master equation is time dependent,
even if we started with a time-independent problem. However, in most
situations the time-dependent terms can be neglected by using either
one of the following two arguments: either a rotating-wave approximation
naturally arising from the fact that the interaction Liouvillian is
small (similarly to the effective Hamiltonian of the previous section);
or the fact that we are interested in the asymptotic limit, where
the time dependent terms have already died off. This, and all the
other steps of the procedure, will become clear with the example that
we analyze next.

\subsubsection{Cooling and Purcell enhancement of the atomic decay}

Consider the following scenario. We have a two-level atom at a much
higher temperature than we'd like, meaning that it is described by
a master equation (\ref{AtomicMasterEq}) with a relatively large
$\bar{n}$, and radiating equally in all spatial directions (further,
we assume that it is not driven, $\mathcal{E}=0$). We now show how
to cool down the atom effectively just by (strongly\footnote{This refers to the \emph{strong coupling regime}, which in this context
is defined as that in which the coupling between light and matter
exceeds the atomic spontaneous emission rate. The term \emph{ultrastrong
coupling regime} is reserved for when the coupling reaches the atomic
transition energies, which has become available in recent implementations
based on superconducting circuits.}) coupling it to an open cavity at zero temperature, which will also
have the effect of directing most of the atomic spontaneous emission
towards the direction defined by the cavity.

Consider then the Jaynes-Cummings Hamiltonian (\ref{JCH}). We will
work in a picture rotating at the atomic frequency both for the atom
and the cavity mode, defined by the transformation operator $\hat{U}_{\text{c}}(t)=e^{\hat{H}_{\text{c}}t/\mathrm{i}\hbar}$
with $\hat{H}_{\text{c}}=\hbar\varepsilon(\hat{a}^{\dagger}\hat{a}+\hat{\sigma}_{z}/2)$.
Defining the detuning $\Delta=\omega-\varepsilon$, and including
the interaction of the cavity and the atom with their respective environments,
the transformed state $\tilde{\rho}(t)=\hat{U}_{\text{c}}^{\dagger}(t)\hat{\rho}(t)\hat{U}_{\text{c}}(t)$
evolves according to the master equation
\begin{equation}
\frac{d\tilde{\rho}}{dt}=\underbrace{-\mathrm{i}\left[\Delta\hat{a}^{\dagger}\hat{a},\tilde{\rho}\right]+\kappa\mathcal{D}_{a}[\tilde{\rho}]}_{\mathcal{L}_{\text{E}}[\tilde{\rho}]}+\underbrace{\gamma(\bar{n}+1)\mathcal{D}_{\sigma}[\tilde{\rho}]+\gamma\bar{n}\mathcal{D}_{\sigma^{\dagger}}[\tilde{\rho}]}_{\mathcal{L}_{\text{S}}[\tilde{\rho}]}\underbrace{-\mathrm{i}\biggl[\underbrace{g(\hat{a}\sigma^{\dagger}+\hat{a}^{\dagger}\hat{\sigma})}_{\hat{H}_{1}/\hbar},\tilde{\rho}\biggr]}_{\mathcal{L}_{1}}.
\end{equation}
Note that we have chosen a different phase for the coupling constant
$g$ with respect to (\ref{JCH}), which means that in this case we
take the dipole matrix element $\langle g|\hat{x}_{A}|e\rangle$ as
purely imaginary. This will simplify the upcoming derivations but
makes no fundamental difference for the results we will find. Note
also that we have denoted the cavity decay rate by $\kappa$.

Let us first argue qualitatively how these equations predict that
the atom will get cooled down. Set $\Delta=0$ for simplicity, and
imagine that before the interaction is switched on, we have the atom
at the equilibrium thermal state, and the cavity empty. Starting in
such state, once we switch on the interaction, what will happen is
that the $\hat{a}^{\dagger}\hat{\sigma}$ Jaynes-Cummings Hamiltonian
will destroy an atomic excitation and create a photon in the cavity.
Now, if $\kappa\gg g$, the photon will leave the cavity before the
atom can reabsorb it. On the other hand, the environment will try
to bring the atom to the thermal state, but if $g\gg\gamma$, then
we expect the Jaynes-Cummings interaction to transfer any atomic excitation
to the cavity before the atom can equilibrate with its environment.
Hence, provided the hierarchy $\kappa\gg g\gg\gamma$ is satisfied,
the atom will be effectively decoupled from its thermal environment,
and loosing all excitations through the cavity.

Let us prove rigorously that the physical picture explained above
is correct. Note that vacuum is the stationary state associated to
the free optical problem, $\bar{\rho}_{\text{E}}=|0\rangle\langle0|$.
This implies that $\mathcal{PL}_{1}\mathcal{P}=0$, since, acting
on any general operator $\hat{Y}$, we obtain
\begin{equation}
\mathcal{PL}_{1}\mathcal{P}[\hat{Y}]=\mathcal{PL}_{1}\left[\text{tr}_{\text{E}}\{\hat{Y}\}\otimes|0\rangle\langle0|\right]=-\mathrm{i}g\biggl[\underbrace{\langle0|\hat{a}|0\rangle}_{0}\sigma^{\dagger}+\underbrace{\langle0|\hat{a}^{\dagger}|0\rangle}_{0}\hat{\sigma},\text{tr}_{E}\{\hat{Y}\}\biggr]\otimes|0\rangle\langle0|.
\end{equation}
The interaction Hamiltonian can be written as (\ref{H1}) with the
choices $g_{1}=g_{2}=g$, $\hat{E}_{1}=\hat{a}=\hat{E}_{2}^{\dagger}$,
and $\hat{S}_{2}=\hat{\sigma}=\hat{S}_{1}^{\dagger}$. In order to
evaluate the correlation functions (\ref{CKcorrs}), the easiest in
this case is to resort to the quantum regression formula (\ref{QRF}).
In order to lighten up the notation, let us not write tildes on top
of the expectation values taken with respect the rotating-picture
state, that is, we simply denote $\text{tr}\{\hat{B}\tilde{\rho}\}$
by $\langle\hat{B}\rangle$. The annihilation operator forms a closed
set by itself,
\begin{align}
\frac{d\langle\hat{a}(\tau)\rangle_{\text{E}}}{d\tau} & =\text{tr}\{\hat{a}\mathcal{L}_{\text{E}}[\tilde{\rho}]\}=-(\kappa+\mathrm{i}\Delta)\langle\hat{a}(\tau)\rangle_{\text{E}}\hspace{1em}\Rightarrow\hspace{1em}\left\{ \begin{array}{c}
\langle\hat{a}(\tau)\rangle_{\text{E}}=\langle\hat{a}(0)\rangle_{\text{E}}e^{-(\kappa+\mathrm{i}\Delta)\tau}\\
\langle\hat{a}^{\dagger}(\tau)\rangle_{\text{E}}=\langle\hat{a}^{\dagger}(0)\rangle_{\text{E}}e^{-(\kappa-\mathrm{i}\Delta)\tau}
\end{array}\right.,
\end{align}
so that for any two environment operators $\hat{C}$ and $\hat{D}$,
the quantum regression formula (\ref{QRF}) tells us that ($\tau>0$)\begin{subequations}
\begin{align}
\lim_{t\rightarrow\infty}\langle\hat{C}(t)\hat{a}(t+\tau)\hat{D}(t)\rangle_{\text{E}} & =\langle0|\hat{C}\hat{a}\hat{D}|0\rangle e^{-(\kappa+\mathrm{i}\Delta)\tau},\\
\lim_{t\rightarrow\infty}\langle\hat{C}(t)\hat{a}^{\dagger}(t+\tau)\hat{D}(t)\rangle_{\text{E}} & =\langle0|\hat{C}\hat{a}^{\dagger}\hat{D}|0\rangle e^{-(\kappa-\mathrm{i}\Delta)\tau}.
\end{align}
\end{subequations}Applied to the environmental correlation functions
(\ref{Ccorrs}), we obtain\begin{subequations}
\begin{align}
C_{11}(\tau) & =\lim_{t\rightarrow\infty}\langle\hat{a}(t)\hat{a}(t+\tau)\rangle_{\text{E}}=\langle0|\hat{a}\hat{a}\hat{I}|0\rangle e^{-(\kappa+\mathrm{i}\Delta)\tau}=0,\\
C_{22}(\tau) & =\lim_{t\rightarrow\infty}\langle\hat{a}^{\dagger}(t)\hat{a}^{\dagger}(t+\tau)\rangle_{\text{E}}=\langle0|\hat{a}^{\dagger}\hat{a}\hat{I}|0\rangle e^{-(\kappa-\mathrm{i}\Delta)\tau}=0,\\
C_{21}(\tau) & =\lim_{t\rightarrow\infty}\langle\hat{a}^{\dagger}(t)\hat{a}(t+\tau)\rangle_{\text{E}}=\langle0|\hat{a}^{\dagger}\hat{a}\hat{I}|0\rangle e^{-(\kappa+\mathrm{i}\Delta)\tau}=0,\\
C_{12}(\tau) & =\lim_{t\rightarrow\infty}\langle\hat{a}(t)\hat{a}^{\dagger}(t+\tau)\rangle_{\text{E}}=\langle0|\hat{a}\hat{a}^{\dagger}\hat{I}|0\rangle e^{-(\kappa-\mathrm{i}\Delta)\tau}=e^{-(\kappa-\mathrm{i}\Delta)\tau}.
\end{align}
\end{subequations}In the case of the correlators (\ref{Kcorrs}),
we get\begin{subequations}
\begin{align}
K_{11}(\tau) & =\lim_{t\rightarrow\infty}\langle\hat{a}(t+\tau)\hat{a}(t)\rangle_{\text{E}}=\langle0|\hat{I}\hat{a}\hat{a}|0\rangle e^{-(\kappa+\mathrm{i}\Delta)\tau}=0,\\
K_{22}(\tau) & =\lim_{t\rightarrow\infty}\langle\hat{a}^{\dagger}(t+\tau)\hat{a}^{\dagger}(t)\rangle_{\text{E}}=\langle0|\hat{I}\hat{a}^{\dagger}\hat{a}^{\dagger}|0\rangle e^{-(\kappa-\mathrm{i}\Delta)\tau}=0,\\
K_{21}(\tau) & =\lim_{t\rightarrow\infty}\langle\hat{a}^{\dagger}(t+\tau)\hat{a}(t)\rangle_{\text{E}}=\langle0|\hat{I}\hat{a}^{\dagger}\hat{a}|0\rangle e^{-(\kappa-\mathrm{i}\Delta)\tau}=0,\\
K_{12}(\tau) & =\lim_{t\rightarrow\infty}\langle\hat{a}(t+\tau)\hat{a}^{\dagger}(t)\rangle_{\text{E}}=\langle0|\hat{I}\hat{a}\hat{a}^{\dagger}|0\rangle e^{-(\kappa+\mathrm{i}\Delta)\tau}=e^{-(\kappa+\mathrm{i}\Delta)\tau}.
\end{align}
\end{subequations}We can combine all these results as\begin{subequations}\label{CKcorrsExample}
\begin{align}
C_{mn}(\tau) & =e^{-(\kappa-\mathrm{i}\Delta)\tau}\delta_{m1}\delta_{n2},\\
K_{nm}(\tau) & =e^{-(\kappa+\mathrm{i}\Delta)\tau}\delta_{n1}\delta_{m2},
\end{align}
\end{subequations}both of which decay at rate $\kappa$. Before substituting
these expressions onto the effective master equation (\ref{EffMasterEQ}),
note that in this case $\tilde{S}_{m}(\tau)=\hat{S}_{m}$, since there
is no free atomic Hamiltonian in the picture we are working in. Remember
that it is important to come back and check this approximation at
the end. Taking also into account that 
\begin{equation}
\int_{0}^{t}e^{-(\kappa\pm\mathrm{i}\Delta)\tau}=\frac{1-e^{-(\kappa\pm\mathrm{i}\Delta)t}}{\kappa\pm\mathrm{i}\Delta},
\end{equation}
and assuming that we are only interested in times $t\gg\kappa^{-1}$,
we obtain the effective master equation
\begin{align}
\partial_{t}\tilde{\rho}_{\text{S}} & \approx\mathcal{L}_{\text{S}}[\tilde{\rho}_{\text{S}}]+\frac{g^{2}}{\kappa-\mathrm{i}\Delta}\hat{\sigma}\tilde{\rho}_{\text{S}}\hat{\sigma}^{\dagger}-\frac{g^{2}}{\kappa+\mathrm{i}\Delta}\hat{\sigma}^{\dagger}\hat{\sigma}\tilde{\rho}_{\text{S}}+\text{H.c.}\\
 & =\mathrm{i}\underbrace{\frac{g^{2}\Delta}{\kappa^{2}+\Delta^{2}}}_{-\Delta\varepsilon}[\hat{\sigma}^{\dagger}\hat{\sigma},\tilde{\rho}_{\text{S}}]+\underbrace{\left(\frac{g^{2}\kappa}{\kappa^{2}+\Delta^{2}}+\gamma(\bar{n}+1)\right)}_{\Gamma_{-}}\mathcal{D}_{\sigma}[\tilde{\rho}_{\text{S}}]+\underbrace{\gamma\bar{n}}_{\Gamma_{+}}\mathcal{D}_{\sigma^{\dagger}}[\tilde{\rho}_{\text{S}}].\nonumber 
\end{align}
The notation stresses that $\Delta\varepsilon$ is an atomic energy
shift, $\Gamma_{-}$ the rate of loss of atomic excitations, and $\Gamma_{+}$
the rate of incoherent gain of atomic excitations. We can come back
to the Schrödinger picture and write it in the even more suggestive
form
\begin{equation}
\partial_{t}\hat{\rho}_{\text{S}}=-\mathrm{i}\left[\frac{\varepsilon+\Delta\varepsilon}{2}\hat{\sigma}_{z},\hat{\rho}_{\text{S}}\right]+\Gamma_{\text{eff}}(\bar{n}_{\text{eff}}+1)\mathcal{D}_{\sigma}[\hat{\rho}_{\text{S}}]+\Gamma_{\text{eff}}\bar{n}_{\text{eff}}\mathcal{D}_{\sigma^{\dagger}}[\hat{\rho}_{\text{S}}],\label{EffMasterExample}
\end{equation}
with\begin{subequations}
\begin{align}
\Gamma_{\text{eff}} & =\Gamma_{-}-\Gamma_{+}=\gamma\left[1+\frac{g^{2}/\kappa\gamma}{1+(\Delta/\kappa)^{2}}\right],\\
\bar{n}_{\text{eff}} & =\frac{\Gamma_{+}}{\Gamma_{-}-\Gamma_{+}}=\frac{\bar{n}}{1+\frac{g^{2}/\kappa\gamma}{1+(\Delta/\kappa)^{2}}}.
\end{align}
\end{subequations}This form shows that the atom feels now a different
effective thermal environment, with effective thermal excitations
$\bar{n}_{\text{eff}}$ and effective rate $\Gamma_{\text{eff}}$.
It is common to define the cooperativity $C=g^{2}/\kappa\gamma$.
Whenever this is large enough, in particular $C\gg\{\bar{n},1+(\Delta/\kappa)^{2}\}$,
we obtain $\bar{n}_{\text{eff}}\ll1$, which is exactly what we wanted:
an effective temperature close to zero. Moreover, note that $\Gamma_{\text{eff}}$
is the sum of two terms. The first one, $\gamma$, is related to losses
of the atom to its original environment. On the other hand, the second
term $\gamma C/[1+(\Delta/\kappa)^{2}]$ comes from losses through
the optical mode into the environment of the cavity, that is, emission
radiated through the partially transmitting mirror. Hence, together
with cooling, we find that the emission gets redirected mostly in
the direction defined by the cavity. This phenomenon is known as \emph{Purcell
effect}.

Finally, we need to be careful and check self-consistently that all
the required approximations are satisfied in the limit of large cooperativity
that we want to work with. In particular, in this case we essentially
care about two approximations. First, we need the interaction term
to be much smaller than the free term so that our perturbation theory
makes sense. Since the dominant scale of the free terms is $\kappa$,
we then require $\kappa\gg g$. Next, we need to make sure that the
incoherent rate of evolution of the atom is much smaller than the
correlation time of the environment, which is $\kappa$ as well according
to (\ref{CKcorrsExample}). This means that we require $\kappa\gg\Gamma_{\text{eff}}<\gamma C=g^{2}/\kappa$,
a condition that is always satisfied since $\kappa\gg g$ is already
assumed. Related to this, there is also one subtle condition: since
we have neglected the evolution provided by the effective energy shift
in (\ref{EffMasterExample}), we also require $\kappa\gg\Delta\varepsilon$.
This condition is always satisfied, as can be easily argued by inspecting
the limits $|\Delta|\ll\kappa$ and $|\Delta|\gg\kappa$. However,
note that when $|\Delta|\gg|\kappa|$, the large cooperativity condition
$(\Delta/\kappa)^{2}\ll C=g^{2}/\kappa\gamma$ can only be achieved
by increasing the coupling $g$ with respect to the atomic spontaneous
emission rate $\gamma$, which becomes unfeasible experimentally beyond
some limit. In any case, while there is no fundamental problem with
this cooling method, it is indeed very impractical, because the large
cooperativity limit requires $g/\gamma\gg\bar{n}(\kappa/g)$, which
is beyond what can be accomplished experimentally, except in some
special platforms (certainly not atoms and light).

\newpage{}
\section{Key questions and exercises}

This final chapter contains the exercise sheets that I provide to the students. The sheet for each chapter is divided in two parts. First, I provide the key questions that I think students should know how to answer, which form the backbone of the exam. The goal of these questions is to help students know what they should study and get comfortable with. Second, I have designed exercises that can be worked out with the tools developed in the chapter, but complement the topics we saw there and should help students to develop independence when applying quantum mechanics to problems.

\subsection{Brief review of quantum mechanics}

\subsubsection{Key questions of the chapter}
\begin{enumerate}
\item The theory of Hilbert spaces is the basic language of quantum mechanics.
It's important that you feel comfortable with it, for which you should
study carefully Appendix A.2 (which hopefully will be just a reminder
of concepts you have already studied in detail before). After that,
you should be able to answer confidently and correctly the following
questions:
\begin{enumerate}
\item What is a complex vector space?
\item What's an inner product? and an outer product?
\item What's a \emph{ket}? and a \emph{bra}?
\item What's a Euclidean space? is it the same a Hilbert space?
\item What's a linear operator? Enumerate the most important types of operators
and their properties? How about the most important operations between
or on them?
\item What's a basis of a Hilbert space? Is it unique?
\item Given an infinite-dimensional Euclidean space, what's the defining
property of the vectors contained in the corresponding Hilbert space?
Are the remaining vectors not contained in the Hilbert space still
useful? Can they form a basis of the Hilbert space? What's a continuous
representations?
\item Do you understand the most characteristic examples of infinite-dimensional
spaces, denoted by $l^{2}(\infty)$ and $\text{L}^{2}(x)$ spaces?
What do we mean when we say that they are isomorphic to each other?
And when we say that \emph{all} infinite-dimensional Hilbert spaces
are isomorphic to them?
\item What's a tensor product?
\end{enumerate}
\item Quantum mechanics is the theory with which we will approach the electromagnetic
field and matter in this course. Hence, you should also be very much
acquainted with its basic postulates and how to apply them (in fact,
this course will hopefully serve you to master them even further).
In particular, after studying carefully Chapter I and Appendix A.3
(or even before, if you remember well your quantum mechanics lectures),
you should be able to answer correctly and confidently the following
questions:
\begin{enumerate}
\item How do we describe the state of the degrees of freedom of a system
in quantum mechanics?
\item How about measurable quantities (\emph{observables})? What are the
possible outcomes of a measurement of such a quantity? Can you predict
the measurement outcomes with certainty? How about its statistics?
\item Can we prepare systems in such a state that we will know with full
certainty the outcomes of measurements of all observables?
\item What are the \emph{canonical commutation relations}? What are they
useful for?
\item Why do we care about the so-called \emph{expectation value} of an
observable? And about its \emph{variance}?
\item What's the Schrödinger equation? and the Heisenberg equation? What
are they used for in quantum mechanics? In which sense are they equivalent?
\item What are composite systems? How do we build their Hilbert space from
the Hilbert spaces of the subsystems that compose them? What are the
main properties of the tensor product map? Can you build a basis of
the total Hilbert space form the bases of the subspaces? How about
building operators acting on the total Hilbert space, from operators
acting on the subspaces?
\item What are entangled states? Why can't the correlations present on these
states be present in systems described by classical physics?
\item What are the properties of density operators?And their ensemble decomposition?
Is the latter unique?
\item Show that density operators naturally represent the mixed state of
noisy systems, for which state preparation is imperfect.
\item Given a composite system, show that the reduced state, obtained by
tracing out one of the subsystems, represents the state of the remaining
parts of the system.
\item Show that, even when the state of the full system is pure, the reduced
state can be mixed if there is entanglement between the subsystems.
Use this approach to show that ensemble decompositions are not unique,
even though the reduced state is.
\item Provide an informational interpretation of mixedness by arguing that
it represents our ignorance about information that has `leaked out'
of the system. What are maximally-mixed states and how do they reinforce
this interpretation in terms of ignorance? Can mixedness be properly
quantified?
\item When the system is in a given mixed state $\hat{\rho}$, how can you
evaluate expectation values of operators? And the probability of obtaining
a certain outcome when measuring an observable? How about the evolution
of the state of the system? What is the von Neumann equation?
\end{enumerate}
\end{enumerate}
\newpage{}

\subsubsection{Graded exercise 1a: Practical quantum mechanics with pure states}

Consider a 3-dimensional Hilbert space a basis $\{|b_{j}\rangle\}_{j=1,2,3}$
defined on it. Consider also a vector $|\phi\rangle=N(2|b_{1}\rangle+2|b_{2}\rangle+|b_{3}\rangle)$,
where $N$ is an arbitrary positive number.
\begin{enumerate}
\item \textbf{{[}2.5{]}}\emph{ Show that $N=1/3$ if we want $|\phi\rangle$
to be a valid quantum state.}\\
Hint: remember that the probabilistic interpretation of quantum mechanics
requires states to be normalized to a specific value.
\item \textbf{{[}2.5{]}}\emph{ Show that the representation of $|\phi\rangle$
in the basis }$\{|b_{j}\rangle\}_{j=1,2,3}$\emph{ is given by
\begin{equation}
\vec{\phi}=\frac{1}{3}\left(\begin{array}{c}
2\\
2\\
1
\end{array}\right).
\end{equation}
}Hint: just check the definition of representations in the notes.
\item \textbf{{[}7.5{]}}\emph{ Consider the Hamiltonian $\hat{H}=\hbar(|b_{1}\rangle\langle b_{2}|+|b_{2}\rangle\langle b_{1}|)$.
Show that it possess the following eigenvectors and eigenvalues}
\begin{equation}
\begin{array}{ll}
|e_{1}\rangle=\frac{1}{\sqrt{2}}(|b_{1}\rangle+|b_{2}\rangle), & E_{1}=\hbar,\\
|e_{2}\rangle=\frac{1}{\sqrt{2}}(|b_{1}\rangle-|b_{2}\rangle),\text{ } & E_{2}=-\hbar,\\
|e_{3}\rangle=|b_{3}\rangle, & E_{3}=0.
\end{array}
\end{equation}
Hint: Use the matrix representation of $\hat{H}$, denoted by $H$,
and your knowledge of linear algebra.
\item \textbf{{[}7.5{]}}\emph{ Use the Schrödinger equation to show that,
starting from $|\psi(0)\rangle=|\phi\rangle$, the state of the system
at any other time is given by
\begin{equation}
|\psi(t)\rangle=\frac{1}{3}\left(2e^{-it}|b_{1}\rangle+2e^{-it}|b_{2}\rangle+|b_{3}\rangle\right)\quad\Leftrightarrow\quad\vec{\psi}(t)=\frac{1}{3}\left(\begin{array}{c}
2e^{-it}\\
2e^{-it}\\
1
\end{array}\right),\label{psi(t)}
\end{equation}
where the representation is in the }$\{|b_{j}\rangle\}_{j=1,2,3}$\emph{
basis.}\\
Hint: Represent the Schrödinger equation in any of the bases that
you know, and solve the linear differential equations with any of
the tools that you have learned in your math courses.
\item \textbf{{[}10{]}}\emph{ Show that the unitary evolution operator $\hat{U}(t)=\exp(\hat{H}t/\mathrm{i\hbar})$
has the following representation in the} $\{|b_{j}\rangle\}_{j=1,2,3}$\emph{
basis
\begin{equation}
U(t)=\left(\begin{array}{ccc}
\cos t & -\mathrm{i}\sin t & 0\\
-\mathrm{i}\sin t & \cos t & 0\\
0 & 0 & 1
\end{array}\right).\label{U(t)QM}
\end{equation}
How about its representation in the} $\{|e_{j}\rangle\}_{j=1,2,3}$\emph{
basis?}\\
Hint: Use the spectral theorem given your knowledge of the eigensystem
of $\hat{H}$ (check the spectral theorem in Appendix A.2.b of the
notes if you are not familiar with it).
\item \textbf{{[}5{]}}\emph{ Find Eq. }(\ref{psi(t)})\emph{ again, this
time using $|\psi(t)\rangle=\hat{U}(t)|\psi(0)\rangle$.}\\
Hint: As usual, it comes handy to use representations.
\item \textbf{{[}20{]}}\emph{ Show that the probabilities of obtaining the
outcomes $E_{j}$ when measuring the energy at some time $t$, denoted
by $p_{j}(t)$, are given by
\begin{equation}
p_{1}(t)=\frac{8}{9},\quad p_{2}(t)=0,\quad p_{3}(t)=\frac{1}{9}.
\end{equation}
Show also that, at any time, the mean energy is equal to $8\hbar/9$,
with a variance equal to $8\hbar^{2}/81$.}\\
\emph{Is there any special reason why all these statistical objects
are time independent, even though the state evolves nontrivially in
time?}\\
Hint: Just use the various definitions given in the notes. Again,
representations might come handy.
\item \textbf{{[}25{]}}\emph{ Consider the observable $\hat{A}=\mathrm{i}(|b_{3}\rangle\langle b_{2}|-|b_{2}\rangle\langle b_{3}|)+|b_{1}\rangle\langle b_{1}|$.
Show that its eigensystem is given by}
\begin{equation}
\begin{array}{ll}
|a_{1}\rangle=|b_{1}\rangle, & a_{1}=1,\\
|a_{2}\rangle=\frac{1}{\sqrt{2}}(|b_{2}\rangle+\mathrm{i}|b_{3}\rangle), & a_{2}=1,\\
|a_{3}\rangle=\frac{1}{\sqrt{2}}(|b_{2}\rangle-\mathrm{i}|b_{3}\rangle),\text{ } & a_{3}=-1.
\end{array}
\end{equation}
\emph{Given the state $|\psi(t)\rangle$, show that the probabilities
$P_{\pm1}(t)$ of obtaining the outcomes $\pm1$ in a measurement
of $\hat{A}$ at time $t$, are given by
\begin{equation}
P_{1}(t)=\frac{13}{18}+\frac{2}{9}\sin t,\quad P_{-1}(t)=\frac{5}{18}-\frac{2}{9}\sin t.
\end{equation}
Finally, show that the expectation value of $\hat{A}$ evolves as
$4(1+\sin t)/9$, with a variance given by $1-[4(1+\sin t)/9]^{2}$.}\\
Hint: Keep in mind that the outcome can be 1 through two possible
eigenvalues, such that you'll have to add up their individual probabilities.
\item \textbf{{[}10{]}}\emph{ Probably you used the Schrödinger picture
to find the expectation value of $\hat{A}$ and its variance }(\emph{otherwise,
do it here}). \emph{Use the time-evolution operator that you found
in }(\ref{U(t)QM})\emph{ to show that the representation of the time-evolved
operator $\hat{A}(t)$ is given by
\begin{equation}
A(t)=\left(\begin{array}{ccc}
\cos^{2}t & -\frac{\mathrm{i}}{2}\sin2t & \sin t\\
\frac{\mathrm{i}}{2}\sin2t & \sin^{2}t & -\mathrm{i}\cos t\\
\sin t & \mathrm{i}\cos t & 0
\end{array}\right),
\end{equation}
in the} $\{|b_{j}\rangle\}_{j=1,2,3}$\emph{ basis. Evaluate then
again the expectation value and variance of the previous section in
this picture.}\\
Hint: Don't use the Heisenberg equations of motion, unless you really
like to work unnecessarily hard. Just stick to the definitions in
terms of the evolution operator and use your preferred representation.
\item \textbf{{[}10{]}}\emph{ Evaluate the commutator $[\hat{A},\hat{H}]$
and check that the uncertainty relations are not violated at any time
$t$.}\\
Hint: Just stick to your preferred representation and you'll be fine.
\end{enumerate}
\newpage{}

\subsubsection{Graded exercise 1b: Practical quantum mechanics with mixed states}

Consider a composite system described by the Hilbert space $\mathcal{H}=\mathcal{A}\otimes\mathcal{B}$,
and the orthonormal bases $\{|a_{j}\rangle\}_{j=1,2}$ and $\{|b_{l}\rangle\}_{l=1,2,3}$
of $\mathcal{A}$ and $\mathcal{B}$, respectively.
\begin{enumerate}
\item \textbf{{[}5{]}}\emph{ What are the dimensions of the Hilbert spaces
$\mathcal{A}$, $\mathcal{B}$, and $\mathcal{H}$?}
\item \textbf{{[}10{]}}\emph{ Suppose that the composite system is in the
pure state
\begin{equation}
|\psi\rangle=\sqrt{2p}|\varphi_{1}\rangle\otimes|\phi_{1}\rangle+|\varphi_{2}\rangle\otimes|\phi_{2}\rangle+|\varphi_{3}\rangle\otimes|\phi_{3}\rangle,\label{PsiAB}
\end{equation}
with
\begin{equation}
|\phi_{1}\rangle=|b_{1}\rangle,\quad|\phi_{2}\rangle=\frac{1}{\sqrt{2}}(|b_{2}\rangle+|b_{3}\rangle),\quad|\phi_{3}\rangle=\frac{1}{\sqrt{2}}(|b_{1}\rangle+|b_{2}\rangle),
\end{equation}
and
\begin{equation}
|\varphi_{1}\rangle=\frac{1}{\sqrt{2}}(|a_{1}\rangle+|a_{2}\rangle),\quad|\varphi_{2}\rangle=\left(\frac{\sqrt{p}-\sqrt{1-p}}{\sqrt{2}}\right)|a_{1}\rangle+\left(\frac{\sqrt{p}+\sqrt{1-p}}{\sqrt{2}}\right)|a_{2}\rangle=-|\varphi_{3}\rangle,
\end{equation}
and where $p\in[0,1]$. Show that $|\psi\rangle$ is normalized to
}1\emph{.}
\item \textbf{{[}15{]}} \emph{Show that the state $|\psi\rangle$ of Eq.
}(\ref{PsiAB})\emph{ can be written in the form
\begin{equation}
|\psi\rangle=\sum_{m=1}^{3}\sqrt{w_{m}}|\psi_{m}\rangle\otimes|b_{m}\rangle,\label{psiABb}
\end{equation}
with $w_{1}=w_{3}=1/2$ and $w_{2}=0$, and find the corresponding
states $\{|\psi_{m}\rangle\}_{m=1,2,3}$.}
\item \textbf{{[}15{]}} \emph{Show that the representation of the reduced
density operator }$\hat{\rho}_{\mathcal{A}}=\text{tr}_{\mathcal{B}}\{|\psi\rangle\langle\psi|\}$\emph{
in the basis $\{|a_{j}\rangle\}_{j=1,2}$ is given by
\begin{equation}
\rho_{\mathcal{A}}=\frac{1}{2}\left(\begin{array}{cc}
1 & 2p-1\\
2p-1 & 1
\end{array}\right).
\end{equation}
}
\item \textbf{{[}15{]}}\emph{ Find the the von Neumann entropy }$S[\hat{\rho}_{\mathcal{A}}]=-\text{tr}_{\mathcal{A}}\{\hat{\rho}_{\mathcal{A}}\log\hat{\rho}_{\mathcal{A}}\}$\emph{,
and discuss the result as a function of $p$.}
\item \textbf{{[}10{]}}\emph{ Show that the state $|\psi\rangle$ of Eqs.
}(\ref{PsiAB})\emph{ or }(\ref{psiABb})\emph{ can also be written
in the alternative form
\begin{equation}
|\psi\rangle=\sum_{m=1}^{2}\sqrt{v_{m}}|a_{m}\rangle\otimes|\chi_{m}\rangle,\label{psiABa}
\end{equation}
with $v_{1}=v_{2}=1/2$, and states\begin{subequations}
\begin{align}
|\chi_{1}\rangle & =\left(\frac{\sqrt{p}+\sqrt{1-p}}{\sqrt{2}}\right)|b_{1}\rangle+\left(\frac{\sqrt{p}-\sqrt{1-p}}{\sqrt{2}}\right)|b_{3}\rangle,\nonumber \\
|\chi_{2}\rangle & =\left(\frac{\sqrt{p}-\sqrt{1-p}}{\sqrt{2}}\right)|b_{1}\rangle+\left(\frac{\sqrt{p}+\sqrt{1-p}}{\sqrt{2}}\right)|b_{3}\rangle.
\end{align}
\end{subequations}}
\item \textbf{{[}15{]}}\emph{ Use expression }(\ref{psiABa})\emph{ of the
state to show that the representation of the reduced density operator
}$\hat{\rho}_{\mathcal{B}}=\text{tr}_{\mathcal{A}}\{|\psi\rangle\langle\psi|\}$\emph{
in the basis $\{|b_{j}\rangle\}_{j=1,2,3}$ is given by
\begin{equation}
\rho_{\mathcal{B}}=\frac{1}{2}\left(\begin{array}{ccc}
1 & 0 & 2p-1\\
0 & 0 & 0\\
2p-1 & 0 & 1
\end{array}\right),
\end{equation}
and show then that $S[\hat{\rho}_{\mathcal{B}}]=S[\hat{\rho}_{\mathcal{A}}]$.
Do you think that this equality is just a coincidence particular to
our choice of state $|\psi\rangle$, or is there something deeper
to it? Provide physical arguments to support your answer.}
\item \textbf{{[}15{]}}\emph{ Can you provide an ensemble decomposition
for $\hat{\rho}_{\mathcal{A}}$ different than $\{w_{m},|\psi_{m}\rangle\}_{m=1,2,3}$}?
\end{enumerate}
\newpage{}

\subsection{Quantization of the electromagnetic field as a collection of harmonic oscillators}

\subsubsection{Key questions of the chapter}
\begin{enumerate}
\item Starting from the Maxwell equations, derive the wave equation for
the vector potential. Explain the quasi-1D approximation and, assuming
perfectly conducting boundary conditions, write the corresponding
vector potential in terms of appropriate mode functions. Interpret
physically the conditions that such boundaries impose on the expansion
coefficients and the allowed wave vectors.
\item Using the wave equation, prove that the expansion coefficients satisfy
the dynamical equations of independent harmonic oscillators. Show
that choosing a proper normalization factor in the vector potential
expansion, the electromagnetic energy turns into the Hamiltonian for
the independent oscillators. Quantize the electromagnetic field, write
down the fields in terms of annihilation and creation operators, as
well as the Hamiltonian governing the evolution of the empty cavity,
and introduce the concept of photon.
\item Take the Hamiltonian of a harmonic oscillator in terms of its quadratures,
$\hat{H}_{\text{o}}=\hbar\omega(\hat{X}^{2}+\hat{P}^{2})$ with $[\hat{X},\hat{P}]=2\mathrm{i}$.
Use the expansion of the quadratures in terms of annihilation and
creation operators to diagonalize the Hamiltonian, proving that the
oscillator's Hilbert space is infinite dimensional. Discuss the main
physical consequences that quantum physics imposes on the oscillator:
energy quantization, zero-point fluctuations, and absence of well-defined
phase-space trajectories.
\item Can we describe quantum mechanics through a probability density function
in phase space? Motivated by this question, introduce physically and
mathematically the \emph{Wigner function} and its main properties.
Show in particular that it has the right marginals for quadrature
measurements. Explain why it cannot be interpreted as a true probability
density function and argue that the statistics of quantum mechanics
cannot be simulated with standard noise, and hence quantum physics
goes beyond classical physics.
\item Define \emph{Gaussian states}, discuss their Wigner function, and
give meaning to the mean vector and covariance matrix required to
parametrize them. Introduce a general criterion based on coherent
states that provides a necessary and sufficient condition for a state
$\hat{\rho}$ to be Gaussian. Introduce also the weaker criterion
based on Gaussian states being connected to other Gaussian states
through Hamiltonians quadratic in the quadratures, and explain what
this criterion means for the possibility of generating non-Gaussian
states experimentally.
\item Introduce \emph{coherent states} as displaced vacuum states, and show
that they are eigenstates of the annihilation operator. Argue that
they are Gaussian and find the corresponding Wigner function, interpreting
its shape in phase space. Use it to discuss the concepts of \emph{amplitude}
and \emph{phase} of the oscillator, and how their fluctuations can
be related to a specific pair of quadratures under certain conditions.
Find the representation of coherent states in the Fock-state basis,
and discuss the photon-number distribution they lead to. Show also
that coherent states form an overcomplete basis of the Hilbert space.
Discuss how do coherent states provide the connection between quantum
and classical physics.
\item Introduce the \emph{quantum shot-noise} limiting sensing and information
encoding, and introduce \emph{squeezed states}, explaining how they
allow us to go beyond such limit. Discuss the concept of \emph{minimum
uncertainty states}, and show that coherent states are of this type,
but not Fock states with $n>0$. Introduce the \emph{squeezing operator}
and find the transformation that it induces on the mean vector and
covariance matrix of general states. Apply it to the vacuum state
to generate the \emph{squeezed vacuum state}, and argue that the state
is Gaussian, finding the corresponding Wigner function and interpreting
it in phase space. How can you transform this state into \emph{amplitude-squeezed}
or \emph{phase-squeezed} states? Represent the squeezed vacuum state
into the Fock basis, and discuss the properties of the corresponding
distribution of quanta.
\item Explain why the concept of maximally-mixed state must be extended
to infinite dimension by adding an energy constrain, and show that
this leads to \emph{thermal states}. Discuss why we call them thermal
states, through their connection to statistical physics. Apply the
definitions to the harmonic oscillator Hamiltonian $\hat{H}=\hbar\omega\hat{a}^{\dagger}\hat{a}$,
writing the corresponding thermal state in the Fock basis. Prove that
the state is Gaussian and compare its Wigner function with that of
vacuum.
\end{enumerate}
\newpage{}

\subsubsection{Graded exercise 2: Cat states}

Consider the state $|\alpha_{\text{cat}}\rangle=N_{\alpha}(|\alpha\rangle+|-\alpha\rangle)$,
where $|\pm\alpha\rangle$ are coherent states, we take $\alpha\in\mathbb{R}$,
and $N$ is a suitable normalization constant. This state is known
as a \emph{cat state} because it is a quantum superposition of two
coherent states, which are the most classical quantum states in the
sense discussed in the lectures.
\begin{enumerate}
\item \textbf{{[}10{]} }\emph{Find the factor $N_{\alpha}$ that normalizes
the state.}\\
Hint: just operate on $\langle\alpha_{\text{cat}}|\alpha_{\text{cat}}\rangle=1$
\item \textbf{{[}20{]} }\emph{Find the Fock-state representation of the
state and discuss the corresponding photon-number probability distribution.}\\
Hint: just write the Fock representations of each coherent state and
simplify.
\item \textbf{{[}20{]} }\emph{Find the quantum characteristic function}
$\chi_{|\alpha_{\text{cat}}\rangle}(\mathbf{r})=\langle\hat{D}(x,p)\rangle$
\emph{and show that it is not Gaussian.}\\
Hint: Use the complex representation of the displacement operator
defining $\beta=(x+\mathrm{i}p)/2$, together with the normal form
$\hat{D}(\beta)=e^{-|\beta|^{2}/2}e^{\beta\hat{a}^{\dagger}}e^{-\beta^{*}\hat{a}}$.
Use then the fact that coherent states are eigenstates of the annihilation
and creation operators, and you should be able to write the characteristic
function as the sum of four exponentials quadratic in $x$ and $p$. 
\item \textbf{{[}20{]} }\emph{Find the Wigner function $W_{|\alpha_{\text{cat}}\rangle}(\mathbf{r})$,
and try to interpret each of the terms that add up to it. Set $x=0$
and discuss the shape and negativities that you can observe along
the $p$ axis.}\\
Hint: Just Fourier-transform the characteristic function with the
help of the Gaussian integral $\int_{\mathbb{R}}dze^{Bz-z^{2}/2A}=\sqrt{2\pi A}e^{AB^{2}/2}$
valid for any $A>0$ and $B\in\mathbb{C}$.
\item \textbf{{[}30{]} }\emph{Find the probability density functions for
measurements of the $\hat{X}$ and $\hat{P}$ quadratures. Compare
them with the ones corresponding to the mixed state $\hat{\rho}_{\alpha}=N_{\alpha}^{2}\left[|\alpha\rangle\langle\alpha|+|-\alpha\rangle\langle-\alpha|+2\exp(-2\alpha^{2})|0\rangle\langle0|\right]$.}\\
Hint: Find the marginals of the Wigner function with the help of the
Gaussian integral of the previous question. Note that the definition
of the Wigner function is linear in the density operator, and hence,
the Wigner function of the mixed state $\hat{\rho}_{\alpha}$ is just
a mixture of the Wigner function of the states that form $\hat{\rho}_{\alpha}$.
\end{enumerate}
\newpage{}

\subsection{Quantum theory of atoms and the two-level approximation}

\subsubsection{Key questions of the chapter}
\begin{enumerate}
\item Explain what we understand by ``matter'' in quantum optics. Using
the example of the hydrogen atom and atoms with a single-valence electron,
show that the spectrum of matter systems is highly non-uniform.
\item Define the parity operator as the unitary transformation that inverts
the sign of the relative atomic coordinate. Show that it also inverts
the sign of the relative momentum. Assuming that the atomic Hamiltonian
is symmetric under parity transformations, show that energy eigenstates
with the same parity cannot be connected by the relative-coordinate
operator.
\item Justify the two-level approximation using intuitive arguments based
on how the atom gets excited when irradiated with a monochromatic
field. Introduce the Pauli pseudo-spin operators associated to the
two atomic states, and write the atomic Hamiltonian in terms of $\hat{\sigma}_{z}$
within this approximation.
\item Using the defining properties of density operators, show that a general
atomic state can be written as $\hat{\rho}=(\hat{I}+\mathbf{b}^{T}\hat{\boldsymbol{\sigma}})/2$,
where $b_{j}=\langle\hat{\sigma}_{j}\rangle$ and $|\mathbf{b}|\leq1$.
Argue that $\mathbf{b}=0$ corresponds to the maximally-mixed state,
while states are pure if and only if $|\mathbf{b}|=1$. Introduce
the Bloch space, and identify the points where the eigenstates of
the Pauli operators are located.
\item Write down a general time-dependent Hamiltonian and the corresponding
Bloch equations (both in ordinary and complex form).
\item Consider an atom evolving freely. Solve the corresponding Bloch equations
and interpret the solution as pseudo-spin precession, explaining how
the dynamics looks in Bloch space.
\item Consider now the semiclassical Rabi Hamiltonian. Using the description
of light-matter interaction within the dipole approximation introduced
in the next chapter, argue that this corresponds to the interaction
of the two-level atomic system with a classical monochromatic field.
\item Write down the complex Bloch equations associated to the semiclassical
Rabi Hamiltonian and introduce the rotating-wave approximation, justifying
it by averaging the equations over an optical cycle.
\item Particularize the solution provided in the notes to the case in which
the atom starts in the ground state, explaining how the dynamics looks
in the Bloch sphere. Discuss the frequency and amplitude of the oscillations
performed by the excited-state population. What happens when the detuning
between the atomic transition and the light beam is very large? How
does this justify the two-level approximation?
\end{enumerate}
\newpage{}

\subsubsection{Graded exercise 3: Beyond the rotating-wave approximation.}

Consider the semiclassical Rabi Hamiltonian $\hat{H}(t)=\hbar\varepsilon\hat{\sigma}_{z}/2+\hbar\Omega\cos(\omega t)\hat{\sigma}_{x}$,
acting on a two-level atom initially in the ground state.
\begin{enumerate}
\item \textbf{{[}15{]}}\emph{ Writing the state as $|\psi(t)\rangle=a(t)e^{\mathrm{i}\varepsilon t/2}|g\rangle+b(t)e^{-\mathrm{i}\varepsilon t/2}|e\rangle$
with $a(0)=1$ and $b(0)=0$, use the Schrödinger equation to show
that the amplitudes obey the evolution equations}\begin{subequations}\emph{\label{BeyondRWAeqs-1}}
\begin{align}
\dot{a} & =-\mathrm{i}\Omega e^{-\mathrm{i}\varepsilon t}\cos(\omega t)b,\\
\dot{b} & =-\mathrm{i}\Omega e^{\mathrm{i}\varepsilon t}\cos(\omega t)a.
\end{align}
\end{subequations}
\item \textbf{{[}20{]}}\emph{ Within the rotating-wave approximation, neglect
fast-oscillating terms in these equations (or use the rotating-wave
Hamiltonian $\hat{H}(t)=\hbar\varepsilon\hat{\sigma}_{z}/2+\hbar\Omega\left(e^{\mathrm{i}\omega t}\hat{\sigma}+e^{-\mathrm{i}\omega t}\hat{\sigma}^{\dagger}\right)/2$
to find the equations of motion), and solve them to find the solution}\begin{subequations}\emph{\label{RWAsol-1}
\begin{align}
b(t) & =-\mathrm{i}\frac{\Omega}{\sqrt{\Delta^{2}+\Omega^{2}}}e^{-\mathrm{i}\Delta t/2}\sin\left(\frac{\sqrt{\Delta^{2}+\Omega^{2}}}{2}t\right),\label{fail-1}\\
a(t) & =e^{\mathrm{i}\Delta t/2}\left[\cos\left(\frac{\sqrt{\Delta^{2}+\Omega^{2}}}{2}t\right)-\mathrm{i}\frac{\Delta}{\sqrt{\Delta^{2}+\Omega^{2}}}\sin\left(\frac{\sqrt{\Delta^{2}+\Omega^{2}}}{2}t\right)\right],
\end{align}
}\end{subequations}\emph{with $\Delta=\omega-\varepsilon$.}\\
Hint: Starting from the equations for $a$ and $b$, you can find
a second order differential equation for $b$ with constant coefficients,
which you can easily solve by, e.g., an exponential ansatz.
\item \textbf{{[}20{]}}\emph{ Next you are going to apply a perturbative
approach on the original equations }(\ref{BeyondRWAeqs-1})\emph{,
under the assumption that $\Omega$ is small. Expanding the amplitudes
as the power series
\begin{equation}
a(t)=\sum_{n=0}^{\infty}\Omega^{n}a^{(n)}(t),\hspace{1em}\text{and}\hspace{1em}b(t)=\sum_{n=0}^{\infty}\Omega^{n}b^{(n)}(t),
\end{equation}
plug these into the evolution equations and match the terms of the
same power in $\Omega$ to show that they lead to the recurrent equations}\begin{subequations}\emph{
\begin{align}
\dot{a}^{(n)} & =-\mathrm{i}\cos(\omega t)e^{-\mathrm{i}\varepsilon t}b^{(n-1)}(t),\hspace{1em}\hspace{1em}\hspace{1em}\hspace{1em}[n=1,2,3,...]\\
\dot{b}^{(n)} & =-\mathrm{i}\cos(\omega t)e^{\mathrm{i}\varepsilon t}a^{(n-1)}(t),
\end{align}
}\end{subequations}\emph{starting at $a^{(0)}(t)=1$ and $b^{(0)}(t)=0$.
Argue that the initial conditions of these equations are $a^{(n>0)}(t)=0=b^{(n>0)}(t)$.}
\item \textbf{{[}25{]}}\emph{ Working on resonance, $\omega=\varepsilon$,
prove that to second order in $\Omega$ we get}\begin{subequations}\emph{
\begin{align}
b(t) & =-\frac{\mathrm{i}}{2}\Omega t+\frac{\Omega}{4\varepsilon}\left(1-e^{2\mathrm{i}\varepsilon t}\right),\\
a(t) & =1-\frac{\Omega^{2}t^{2}}{8}-\left(\frac{\Omega}{4\varepsilon}\right)^{2}\left(1-e^{2\mathrm{i}\varepsilon t}+2\mathrm{i}\varepsilon te^{-2\mathrm{i}\varepsilon t}\right).
\end{align}
}\end{subequations}
\item \textbf{{[}20{]}}\emph{ Compare this solution with the rotating-wave
solutions }(\ref{RWAsol-1})\emph{ for times $t\ll\Omega^{-1}$. Use
this comparison to argue that effects beyond the rotating-wave approximation
are suppressed by powers of $\Omega/\varepsilon$.}
\end{enumerate}
\newpage{}

\subsection{Light-matter interaction}

\subsubsection{Key questions of the chapter}
\begin{enumerate}
\item Introduce the quantum description of light-matter interactions based
on the dipole approximation for matter.
\item Apply this description to the interaction between the electromagnetic
field of a cavity and a single atom, and derive the quantum-Rabi and
Jaynes-Cummings Hamiltonians, explaining the conditions under which
these are expected to hold.
\item Introduce the dressed states as the eigenstates of the Jaynes-Cummings
Hamiltonian. Discuss the corresponding energy spectrum, and compare
it with that of the non-interacting case. Introduce also the concept
of avoided crossing from the analysis of the energy transitions between
the ground state and the single-excitation manifold.
\item Find the evolution of an initial state consisting of an atom in the
ground state and the field in an arbitrary pure state. Evaluate the
excited-state population, and introduce the concept of quantum Rabi
oscillations by assuming that the field starts in a Fock state.
\item Discuss the evolution of the excited-state population for a field
starting in a coherent state. Make a qualitative plot showing the
relevant physical phenomena: semiclassical Rabi oscillations, collapses,
and revivals. Use approximations and reasonable arguments to find
the time scale for each of these phenomena. How do these results reconcile
the quantum and semiclassical pictures?
\item Introduce the dipole model of a dielectric. Use it to show how Maxwell
equations inside of them are modified with respect to the ones in
vacuum.
\item Discuss the effect that the linear term of the polarization density
has on the field entering the dielectric: wavelength and amplitude
reduction.
\item Use the macroscopic Maxwell equations to derive a wave equation for
the electric field where the nonlinear terms of the polarization density
field act as a forcing or source term. Discuss then the frequency
conversion phenomena linked to second-order nonlinearity: second-harmonic
generation, sum-frequency generation, difference-frequency generation,
and down-conversion.
\item Write down the general interaction Hamiltonian of the electromagnetic
field with the nonlinear dielectric. Assume that only two cavity modes
with frequencies $\omega_{0}$ and $\omega_{2}\approx2\omega_{0}$
and orthogonal polarization are relevant. Use energy and momentum
conservation arguments to write down the down-conversion Hamiltonian.
\item Perform a classical (parametric) approximation for the pump mode,
and find the Hamiltonian of the down-converted mode within this approximation.
Introduce the Bogoliubov mode, and use it to diagonalize the Hamiltonian
in the $|\Delta|>g$ region (explaining the meaning of the eigenstates),
and to show that the Hamiltonian is non-diagonalizable in the $|\Delta|<g$
limit. What happens when $|\Delta|=g$? Starting from the vacuum state
of the electromagnetic field, find the evolution of the number of
photons in the stable ($|\Delta|>g$) and unstable ($|\Delta|<g$)
regions.
\end{enumerate}
\newpage{}

\subsubsection{Graded exercise 4: Third-order (Kerr) nonlinearity}

Consider an optical cavity containing a dielectric medium with inversion
symmetry, for which $\mathbf{P}^{(2)}(\mathbf{r},t)$ vanishes identically.
In these media $\mathbf{P}^{(3)}(\mathbf{r},t)$ becomes then the
dominant nonlinear term of the polarization density.
\begin{enumerate}
\item \textbf{{[}10{]} }\emph{Neglecting all cavity modes but one with frequency
$\omega$ polarized along the $x$ direction }(\emph{you can further
assume that the phase $\varphi$ of its spatial profile inside the
dielectric is zero})\emph{, show that
\begin{equation}
\hat{P}_{j}^{(3)}(\mathbf{r})=-\mathrm{i}\frac{\chi_{jxxx}^{(3)}}{\varepsilon_{0}^{1/2}}\left(\frac{\hbar\omega}{n_{x}LS}\right)^{3/2}\sin^{3}(n_{x}kz)(\hat{a}-\hat{a}^{\dagger})^{3}.
\end{equation}
}\\
Hint: Just take the form of the single-mode field that we introduced
in the lectures inside a dielectric medium, and apply the definition
of the third-order nonlinear polarization density.
\item \textbf{{[}25{]} }\emph{Assume that the dielectric medium has length
$l$ and is placed right at the surface of the left mirror, hence
covering the space $z\in[0,l]$ inside the cavity (as usual, we assume
that the transverse dimensions of the dielectric medium cover the
whole cavity). Use the light-matter interaction Hamiltonian within
the dipole approximation, together with the rotating-wave approximation
and the physical condition $n_{x}kl\gg1$, to show that the dynamics
of the optical mode is ruled by the Hamiltonian
\begin{equation}
\hat{H}=\hbar\left(\omega+g\right)\hat{N}+\hbar g\hat{N}^{2},
\end{equation}
with
\begin{equation}
g=-\frac{9\chi_{xxxx}^{(3)}l\hbar\omega^{2}}{4\varepsilon_{0}Sn_{x}^{2}L^{2}}.
\end{equation}
}\\
Hint: Just apply the definition of the light-matter interaction Hamiltonian
within the dipole approximation. Write the result of the spatial integral
along $z$ in terms of sinc functions that fall to zero when their
argument is larger than one. Then, after keeping the energy-conserving
terms within the rotating-wave approximation, you'll need to use the
canonical commutation relations to write the Hamiltonian in terms
of the number operator.
\item \textbf{{[}20{]} }\emph{What are the eigenstates and eigenenergies
of this Hamiltonian? Make a graphic representation of the energy levels
for a given $|g|\ll\omega$, considering both the positive and negative
cases, comparing it with the noninteracting situation $g=0$.}
\item \textbf{{[}45{]} }\emph{Move to a picture rotating at frequency $\omega+g$,
so that the Hamiltonian in the new picture reads as $\hat{H}_{\mathrm{I}}=\hbar g\hat{N}^{2}$.
Consider a coherent state $|\alpha\rangle$ as the initial state $|\psi(0)\rangle$.
Find the state $|\psi(t)\rangle_{\mathrm{I}}$ in the rotating picture
at subsequent times and use it to show that (take $g>0$ from now
on): }(a) \emph{the state is periodic, with period $T=2\pi/g$; }(b)\emph{
the photon-number distribution doesn't change in time; }(c)\emph{
at times $t_{k}=\pi(2k+1)/2g$ with $k=0,2,4,...$, the state is equivalent
to the cat superposition
\begin{equation}
|\psi(t_{k})\rangle_{\mathrm{I}}=\frac{e^{-\mathrm{i}\pi/4}}{\sqrt{2}}\left(|\alpha\rangle+\mathrm{i}|-\alpha\rangle\right).
\end{equation}
}\\
Hint: Work in the Fock state basis.
\end{enumerate}
\newpage{}

\subsection{Quantum optics in open systems}

\subsubsection{Key questions of the chapter}
\begin{enumerate}
\item Introduce a model for the field outside a cavity with a partially-transmitting
mirror (open cavity). For this, consider an external cavity that shares
the partially-transmitting mirror with the main cavity. Write down
the vector potential associated to the external cavity, and show that
the infinite-length limit leads to a field composed of a continuous
set of harmonic oscillators.
\item Focusing on a single mode of the main cavity, introduce the Hamiltonian
that describes the whole ``cavity + external field'' system. In particular,
argue that the interaction through the partially-transmitting mirror
can be modeled with a simple photon tunneling Hamiltonian (sometimes
referred to as \emph{beam splitter}), with a coupling parameter that
can be taken independent of the frequency provided the mirror's reflectivity
is large and slowly-varying with the frequency. Explain as well why
we can introduce fictitious negative-frequency external modes that
simplify enormously the math.
\item Working in the Heisenberg picture, introduce the equations that model
the dynamics of the cavity mode (quantum Langevin equation), showing
that a damping term appears, together with an input operator that
contains information about the state of the external field. Show that
the presence of a coherent component in the external field (laser
injection) can be accounted for with a time-dependent Hamiltonian
linear in the intracavity annihilation and creation operators. Evaluate
also the type of statistics (expectation values and two-time correlators)
that the input operator has when the external field is in a thermal
state (argue why you can assume the same thermal state for all external
modes, that is, a frequency-independent thermal excitation number).
\item Working from the Schrödinger picture, derive an approximate evolution
equation for the cavity mode. Start with the von Neumann equation
for the density operator of the whole system, and make two picture
changes: first to a displaced picture where the state of the environment
is a thermal state with no displacement, and then to an interaction
picture that discounts the evolution owed to all terms but the weak
interaction. Introduce the Born and non-backaction approximations,
and show that they lead to an evolution equation that describes the
dynamics of the cavity mode up to second order in the interaction.
Trace out the external field using a thermal state as its initial
state, finding the final Lindblad form for the master equation that
rules the dynamics of the cavity mode. Interpret the meaning of the
jump operators, and how they bring irreversibility about.
\item Argue that the state of an empty cavity mode is Gaussian at all times
(provided it starts in a Gaussian state). Find the first and second
moments both through the quantum Langevin equations and through the
master equation. Show that they reach an asymptotic value (independent
of the initial condition), and interpret the corresponding state as
a displaced thermal state.
\item Argue that a similar kind of procedure allows finding a master equation
for an atom in free space. Write down such equation, derive the corresponding
Bloch equations of the atom, and use them to introduce the concept
of spontaneous emission. Show that starting in the excited state,
the atomic density operator evolves through all the ground-excited
mixed states until it reaches the ground state. 
\item Starting from the Hamiltonian describing the interaction between the
atom and the field in free space, and assuming that initially the
atom is excited and the field is in vacuum, find the state of the
whole system at any later time. For this, you can assume that the
coupling is independent of the wave vector of the external modes,
and extend the lower limit of all frequency integrals to $-\infty$.
How does the probability of generating a photon with wave vector $k$
look like in the asymptotic $t\rightarrow\infty$ limit? What about
the probability that a photon is emitted to towards the right or left
at any time? How are the entanglement properties of the state throughout
the evolution?
\item Consider now the situation in which the $k$-dependence of the coupling
cannot be neglected, and show that the non-Markovianity of the master
equation (that is, how far into the past the evolution of the atomic
state sees) is determined by a so-called environmental correlation
function. Introduce then the Markov approximation, explaining its
range of applicability. Under such an approximation, show that, in
the long-time limit, the master equation takes the standard Lindblad
form, but including a correction to the atomic transition frequency
(Lamb shift), that can be of the same order as the damping rate.
\end{enumerate}
\newpage{}

\subsubsection{Graded exercise 5: Atomic dephasing}

While spontaneous emission has traditionally been the most relevant
incoherent process occurring in atoms from a theoretical point of
view, there is another incoherent process which is even more important
in practical terms: the so-called \emph{dephasing}. Many different
physical processes contribute to it, and in this exercise we will
study it with a generic model: the atomic transition is affected by
random fluctuations (coming, for example, from stray random or thermal
magnetic fields in the laboratory).
\begin{enumerate}
\item \textbf{{[}20{]} }\emph{Consider the Hamiltonian $\hat{H}=\hbar[\varepsilon+\varphi(t)]\hat{\sigma}_{z}/2$,
where $\varphi(t)$ is a random variable that accounts for random
fluctuations in the atomic transition frequency. Use the complex Bloch
equations to show that the state, averaged over stochastic realizations,
can be written as 
\begin{equation}
\hat{\rho}(t)=\frac{1}{2}\left[\hat{I}+b_{z}(0)\hat{\sigma}_{z}+2e^{\mathrm{i}\varepsilon t-\Gamma^{*}(t)}b^{*}(0)\hat{\sigma}+2e^{-\mathrm{i}\varepsilon t-\Gamma(t)}b(0)\hat{\sigma}^{\dagger}\right],\label{DephasedState}
\end{equation}
where we have defined
\begin{equation}
e^{-\Gamma(t)}=\overline{e^{-\mathrm{i}\int_{0}^{t}dt'\varphi(t')}},
\end{equation}
with $\Gamma(t)\in\mathbb{C}$ in general and where the overbar denotes
stochastic average.}
\item \textbf{{[}20{]} }\emph{Define $\Phi(t)=\int_{0}^{t}dt'\varphi(t')$.
Provide arguments in favor of the identity $\overline{\cos\Phi(t)}^{2}+\overline{\sin\Phi(t)}^{2}\leq1$,
and use it to prove that $\mathrm{Re}\{\Gamma(t)\}\geq0$. Assuming
$e^{-\text{\ensuremath{\mathrm{Re}}}\{\Gamma(t)\}}$ decays to zero
monotonically as a function of time, explain the trajectory followed
in Bloch space by the initial superposition state $|\psi(0)\rangle=\sqrt{p}|g\rangle+\sqrt{1-p}|e\rangle$
with $p\in[0,1]$. Show that the final state is the incoherent mixture
\begin{equation}
\lim_{t\rightarrow\infty}\hat{\rho}(t)=p|g\rangle\langle g|+(1-p)|e\rangle\langle e|,
\end{equation}
so that the populations are left untouched, but the coherence has
been destroyed. }\\
Hint: Just evaluate the Bloch vector components at $t=0$ using $|\psi(0)\rangle$,
plug them in (\ref{DephasedState}), and infer the Bloch vector components
at any other time from it. In order to visualize the trajectory, it
is useful to first forget about the fast $z$-precession at frequency
$\varepsilon$; the remaining trajectory should just be a straight
line (which becomes a spiral after including the precession).
\item \textbf{{[}20{]}} \emph{Show that the state }(\ref{DephasedState})\emph{
obeys the master equation
\begin{equation}
\partial_{t}\hat{\rho}=\left[-\mathrm{\frac{i}{2}}\left(\varepsilon+\mathrm{Im}\left\{ \dot{\Gamma}(t)\right\} \right)\hat{\sigma}_{z},\hat{\rho}\right]+\frac{\mathrm{Re}\left\{ \dot{\Gamma}(t)\right\} }{2}\left(\hat{\sigma}_{z}\hat{\rho}\hat{\sigma}_{z}-\hat{\rho}\right),
\end{equation}
where the overdot denotes time derivative.}\\
Hint: Just evaluate the left- and right-hand sides independently,
and show that they match.
\item \textbf{{[}20{]}} \emph{Consider the case in which $\varphi(t)$ comes
from a Gaussian stochastic process and has zero mean, $\overline{\varphi(t)}=0$.
Defining $\Phi(t)=\int_{0}^{t}dt'\varphi(t')$, use the Gaussian-moment
formula
\begin{equation}
\overline{\Phi^{n}(t)}=\left\{ \begin{array}{cc}
0 & n\in\mathrm{odd}\\
(n-1)!!\overline{\Phi^{2}(t)}^{n/2} & n\in\mathrm{even}
\end{array}\right.,
\end{equation}
to show that
\begin{equation}
\overline{e^{-\mathrm{i}\Phi(t)}}=e^{-\overline{\Phi^{2}(t)}/2}.\label{GaussianExponentialMoment}
\end{equation}
}\\
Hint: Simply expand the exponential on the left-hand side in Taylor
series, use the formula, and simplify the expression to turn the elements
of the sum into the ones required to obtain the right-hand-side's
exponential.
\item \textbf{{[}20{]}} \emph{Consider the white-noise limit, that is, $\overline{\varphi(t)\varphi(t')}=\gamma_{\varphi}\delta(t-t')$,
where $\gamma_{\varphi}$ is some constant rate. Show that $\Gamma(t)=\gamma_{\varphi}t/2$
in this case, leading to a master equation
\begin{equation}
\frac{d\hat{\rho}}{dt}=\left[-\mathrm{\frac{i}{2}}\varepsilon\hat{\sigma}_{z},\hat{\rho}\right]+\frac{\gamma_{\varphi}}{4}\left(\hat{\sigma}_{z}\hat{\rho}\hat{\sigma}_{z}-\hat{\rho}\right),
\end{equation}
which is the usual master equation commonly used in quantum optics
and quantum information to account for dephasing effects.}\\
Hint: Just use the definition of $\Gamma(t)$ and $\Phi(t)$, together
with (\ref{GaussianExponentialMoment}).
\end{enumerate}
\newpage{}

\subsection{Analyzing the emission of open systems}

\subsubsection{Key questions of the chapter}
\begin{enumerate}
\item Working in the Heisenberg picture, and using reasonable approximations
such as setting slowly-varying functions of the frequency to their
value at the cavity resonance or ignoring retardation effects, write
down the field coming out of an open cavity as a function of the intracavity
and input annihilation operators (input-output relations). 
\item Using the model of the previous chapter for the interaction between
the cavity mode and the external modes, derive a formal solution for
the external annihilation operators in terms of their final conditions
at some future late time and the cavity annihilation operator. Use
this solution to find an alternative form of the output field's annihilation
operator as a Fourier transform of the external field's late-time
condition. Use it to show that the output annihilation and creation
operators satisfy canonical commutation relations in time. Prove that
the structure of the forward and backward quantum Langevin equations
make the open-cavity model satisfy causality.
\item Introduce the qualitative model of a photodetector, explaining how
a single photon is capable of creating a measurable electronic pulse.
Explain that the combination of all the pulses generated by different
photons creates a macroscopic stochastic current, and write down the
relation between its moments and the quantum moments of the number
operator associated to the output field coming from a source. Inspired
by this relation, define the correlation functions of the output field,
and use the input-output relations to show that, at zero temperature,
they are directly proportional to correlation functions of the source.
\item Introduce, without proving but explaining them carefully, the quantum
regression theorem and the quantum regression formula for two-time
correlators. Using the example of a radiating atom, provide a physical
interpretation for the two-time correlation function associated to
direct photodetection (coincidence correlation function) as the probability
of getting two consecutive photodetection events separated by a given
time interval.
\item Taking the master equation of the monochromatically-driven cavity
as the starting point (in a picture rotating at the laser frequency),
evaluate the (normalized) coincidence correlation function for two
situations: zero temperature ($\bar{n}=0$) and zero drive ($\mathcal{A}=0$).
You may use the properties of coherent states for the first one, and
the quantum regression formula for the second. Interpret the results
as photons arriving to the detector at random in the first case, and
photons arriving in bunches in the second case.
\item Consider an atom subject to resonant driving and spontaneous emission
(at zero temperature). Using the master equation in a picture rotating
at the laser frequency, write down the Bloch equations. Turn them
into a second-order differential equation for the excited-state population
and find its value in the asymptotic, stationary limit. Show that
it is never possible to obtain population inversion (that is, more
population in the excited state than in the ground state), reaching
a limit of 1/2 for infinite driving. Find a expression of the (normalized)
coincidence correlation function as a function of the excited-state
population and solve its equation of motion to find the full time-dependence
of the correlation function. Use the result to introduce the concept
of antibunching, and to show that (damped) Rabi oscillations appear
in the correlation function.
\item Introduce homodyne detection for the field coming out of a source,
and show that the first and second moments of the corresponding photocurrent
are proportional to normally-ordered moments of the quadratures. Introduce
also the quadrature noise spectrum and the definition of squeezing
for the fields coming out of an open source.
\item Write down the quantum Langevin equations of an open cavity containing
a second-order nonlinear medium that provides down-conversion for
the cavity mode (use the parametric approximation). Assuming zero
detuning for the down-conversion process and working below threshold
(damping rate larger than the down-conversion rate), write down decoupled
equations for two orthogonal quadratures and solve them. Argue that
the state is Gaussian, and study the asymptotic state by evaluating
the mean vector and covariance matrix. How much squeezing can you
get inside the cavity? Is the state a minimum-uncertainty one? How
about the state of the output field? Evaluate the noise spectrum of
the independent quadratures and answer this question.
\end{enumerate}
\newpage{}

\subsubsection{Graded exercise 6: Coincidence correlation function of the below-threshold
OPO}

We have studied in class the squeezing properties of the light coming
out of an optical parametric oscillator (OPO) below threshold and
at resonance. Here we will now evaluate the coincidence correlation
function, which is essential to understand the photon statistics of
the field emitted by any system.
\begin{enumerate}
\item \emph{Can you guess, based on physical arguments, whether he photons
emitted by the OPO will arrive at the photodetector bunched, antibunched,
or randomly?}
\item \emph{Let us remind the solution for the slowly-varying position and
momentum quadratures,}
\begin{equation}
\lim_{t\rightarrow\infty}\tilde{X}^{\varphi}(t)=\sqrt{2\gamma}\lim_{t\rightarrow\infty}\int_{0}^{t}d\tau'e^{-\lambda_{\varphi}\tau'}\tilde{X}_{\text{in}}^{\varphi}(t-\tau'),\hspace{1em}\hspace{1em}\left[\varphi=0,\frac{\pi}{2}\right],\label{XsolGen}
\end{equation}
\emph{with $\lambda_{0}=\gamma-g$, $\lambda_{\pi/2}=\gamma+g$, $g\in[0,\gamma[$,
and where the input operators satisfy the statistical properties $\langle\tilde{X}_{\text{in}}^{\phi}(t)\rangle=0$
and $\langle\tilde{X}_{\text{in}}^{\phi}(t)\tilde{X}_{\text{in}}^{\phi'}(t')\rangle=e^{\mathrm{i}(\phi'-\phi)}\delta(t-t').$
Use it to show that
\begin{equation}
\lim_{t\rightarrow\infty}\langle\tilde{X}^{\varphi}(t)\tilde{X}^{\varphi'}(t+\tau)\rangle=\frac{2\gamma}{\lambda_{\varphi}+\lambda_{\varphi'}}e^{\mathrm{i}(\varphi'-\varphi)}\left\{ \begin{array}{cc}
e^{-\lambda_{\varphi'}\tau} & \text{for }\tau>0\\
e^{\lambda_{\varphi}\tau} & \text{for }\tau<0
\end{array}\right..\label{XXcorrOPO}
\end{equation}
}\\
Hint: In class we evaluated $\lim_{t\rightarrow\infty}\langle\tilde{X}^{\varphi}(t)\tilde{X}^{\varphi}(t+\tau)\rangle$.
Just follow similar steps.
\item \emph{In class we saw the formula}
\begin{equation}
\langle\delta\hat{L}_{1}\delta\hat{L}_{2}\delta\hat{L}_{3}\delta\hat{L}_{4}\rangle=\langle\delta\hat{L}_{1}\delta\hat{L}_{2}\rangle\langle\delta\hat{L}_{3}\delta\hat{L}_{4}\rangle+\langle\delta\hat{L}_{1}\delta\hat{L}_{3}\rangle\langle\delta\hat{L}_{2}\delta\hat{L}_{4}\rangle+\langle\delta\hat{L}_{1}\delta\hat{L}_{4}\rangle\langle\delta\hat{L}_{2}\delta\hat{L}_{3}\rangle,
\end{equation}
\emph{valid for Gaussian states when the operators $\hat{L}_{j}$
are all linear in position and momentum, and $\delta\hat{L}_{j}=\hat{L}_{j}-\langle\hat{L}_{j}\rangle$.
Argue that for the OPO below threshold (and any other problem described
by linear quantum Langevin equations), this formula applies even when
considering Heisenberg-picture operators at different times, that
is, for the set $\{\hat{L}_{1}(t_{1}),\hat{L}_{2}(t_{2}),\hat{L}_{3}(t_{3}),\hat{L}_{4}(t_{4})\}$.}
\item \emph{Use the previous formula to show that the coincidence correlation
function of the OPO below threshold can be written as
\begin{equation}
\bar{G}^{(2)}(\tau)=\lim_{t\rightarrow\infty}\left[\left|\left\langle \hat{a}^{\dagger}(t)\hat{a}^{\dagger}(t+\tau)\right\rangle \right|^{2}+\left|\left\langle \hat{a}^{\dagger}(t)\hat{a}(t+\tau)\right\rangle \right|^{2}+\left\langle \hat{a}^{\dagger}(t)\hat{a}(t)\right\rangle ^{2}\right].
\end{equation}
Use then }(\ref{XXcorrOPO})\emph{ to obtain}\begin{subequations}\emph{
\begin{align}
\lim_{t\rightarrow\infty}\left\langle \hat{a}^{\dagger}(t)\hat{a}^{\dagger}(t+\tau)\right\rangle  & =\frac{1}{4}\left[\frac{\sigma}{1-\sigma}e^{-(1-\sigma)\gamma\tau}+\frac{\sigma}{1+\sigma}e^{-(1+\sigma)\gamma\tau}\right],\\
\lim_{t\rightarrow\infty}\left\langle \hat{a}^{\dagger}(t)\hat{a}(t+\tau)\right\rangle  & =\frac{1}{4}\left[\frac{\sigma}{1-\sigma}e^{-(1-\sigma)\gamma\tau}-\frac{\sigma}{1+\sigma}e^{-(1+\sigma)\gamma\tau}\right],
\end{align}
}\end{subequations}\emph{where $\sigma=g/\gamma\in[0,1[$. Show from
these expressions that $\bar{G}^{(2)}(\tau)$ is a monotonically decreasing
function of time. Does this match your answer in the first question
of the exercise?}\\
Hint: For the first step, just apply the identity $\langle\hat{A}\hat{B}\rangle=\langle\hat{B}^{\dagger}\hat{A}^{\dagger}\rangle^{*}$.
For the second part, just write the annihilation and creation operators
as a function of the position and momentum quadratures, and use (\ref{XXcorrOPO}).
\end{enumerate}
\newpage{}

\subsection{Effective models: elimination of spurious degrees of freedom}

\subsubsection{Key questions of the chapter}
\begin{enumerate}
\item What is an effective model? Can you explain in which situations we
might expect them to appear?
\item Consider a closed system described by a time-independent Hamiltonian.
Define projector operators that divide the Hilbert space into relevant
and irrelevant sectors, and use them to find an equation of motion
for the state projected on the relevant sector. Define formally an
effective Hamiltonian from these equations, showing that in general
we obtain a non-Hermitian and time-dependent expression for it.
\item While for some problems the effective Hamiltonian becomes Hermitian
under suitable conditions, for other it does not. For the latter,
how do you interpret the fact that we cannot find a Hermitian effective
Hamiltonian?
\item Show that the original Hamiltonian of the system can be split into
a term that doesn't connect the relevant and irrelevant subspaces
(\emph{free Hamiltonian}), and another that does (\emph{interaction
Hamiltonian}). Find a simple expression for the effective Hamiltonian
up to second order in the interaction. Use it to show that a far-detuned
monochromatic optical beam can be used to generate a motional potential
on the center of mass motion of an atom.
\item Consider now an open system described by a master equation with time-independent
Liouvillian. Define projector superoperators that divide the space
of operators into relevant and irrelevant sectors, and use them to
find an effective (time-nonlocal) master equation for the relevant
part of the state. Show that the original Liouvillian can be split
into a term that does not connect the relevant and irrelevant subspaces
(free Liouvillian), and another one that does (interaction Liouvillian).
Approximate the effective master equation to second order in the latter.
\item Consider a bipartite Hilbert space structure, consisting in the tensor
product of two degrees of freedom that we call system and environment.
Consider physical situations in which, from the point of view of the
system, the environment remains approximately frozen in its free-Liouvillian
equilibrium state. Define a projector superoperator adapted to such
situation. Particularize the previous effective master equation to
this choice of projector, and show how it can be approximated by a
time-local effective master equation provided that the correlation
functions of the environment decay fast enough compared to all incoherent
rates of the system's evolution.
\item Consider an atom at high temperature emitting in all directions. Consider
as well one mode of an open cavity at zero temperature. Argue qualitatively
how the atom can be effectively cool down and its emission directed
by coupling it to the cavity via a Jaynes-Cummings interaction, provided
that the coupling is strong (i.e., larger than the atomic spontaneous
emission rate) but still much smaller than the cavity decay rate.
Use the effective master equation derived above (taking the atom as
the system and the cavity as the environment) to make this statement
mathematically rigorous.
\end{enumerate}
\newpage{}

\subsubsection{Graded exercise 7a: Some important effective Hamiltonians}

\begin{figure}[b]
\includegraphics[width=0.7\textwidth]{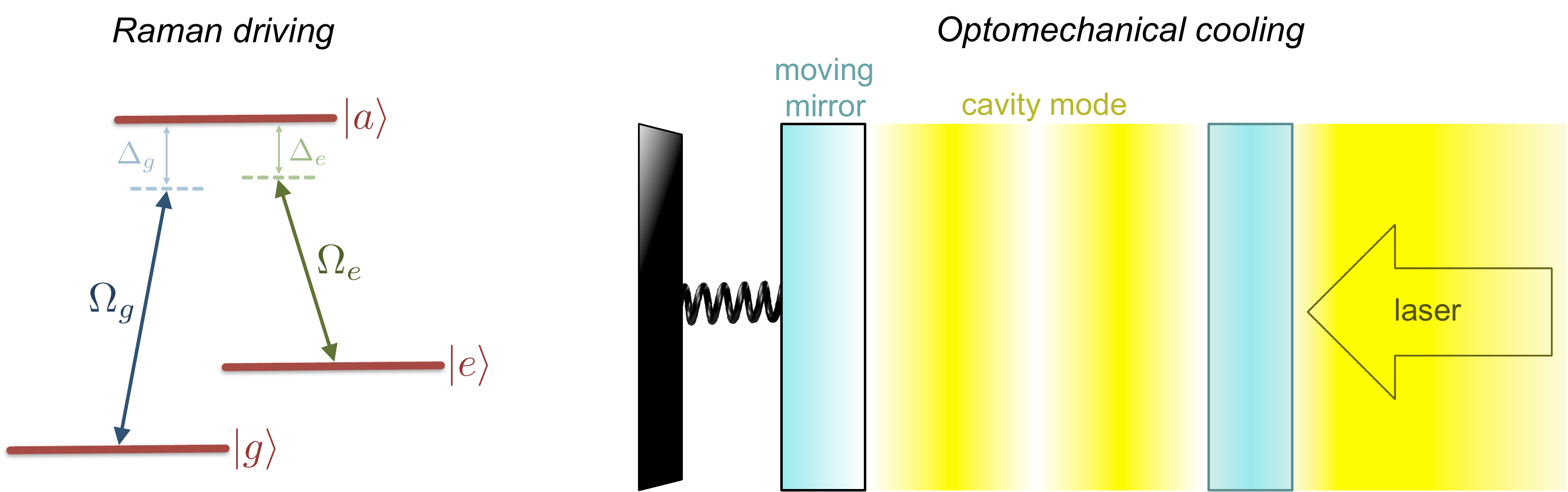}

\caption{Sketches of the Raman driving and optomechanical cooling schemes.}
\label{FigEffective}
\end{figure}

\begin{enumerate}
\item \textbf{\emph{Kerr from down-conversion. }}\emph{Consider the down-conversion
Hamiltonian
\begin{equation}
\hat{H}=\hbar\omega_{0}\hat{a}^{\dagger}\hat{a}+\hbar\omega_{2}\hat{b}^{\dagger}\hat{b}+\mathrm{i\hbar}\frac{g_{0}}{2}(\hat{b}\hat{a}^{\dagger2}-\hat{b}^{\dagger}\hat{a}^{2}),
\end{equation}
in the large-detuning limit, $|\Delta|\gg g_{0}$, with $\Delta=\omega_{0}-\omega_{2}/2$.
Find an effective Hamiltonian for the down-converted mode, under the
assumption that the pump mode is not populated initially. In particular,
argue that $\hat{P}=|0\rangle_{b}\langle0|$ is the appropriate projector
in this case, with $\hat{b}|0\rangle_{b}=0$, and show that the effective
Hamiltonian has the Kerr form
\begin{equation}
\hat{H}_{\mathrm{eff}}=\hbar\left(\omega-g\right)\hat{n}+\hbar g\hat{n}^{2},
\end{equation}
with $\hat{n}=\hat{a}^{\dagger}\hat{a}$ and $g=-g_{0}^{2}/8\Delta$.}
\item \textbf{\emph{Raman driving. }}\emph{Consider the following problem:
we would like to couple two atomic energy eigenstates whose transition
does not couple to a laser (e.g., because they have the same parity
or their transition frequency is not in the optical domain, but in
the microwave domain, as usually happens with hyperfine states). You
are going to show that there is an effective way of connecting them
by using an additional atomic state and two far-detuned laser beams
that couple it to the original states. Consider specifically the situation
depicted in Fig. \ref{FigEffective}, described by the Hamiltonian
\begin{equation}
\hat{H}(t)=-\hbar\varepsilon_{g}|g\rangle\langle g|-\hbar\varepsilon_{e}|e\rangle\langle e|+\hbar\left(\Omega_{g}e^{-\mathrm{i}\omega_{g}t}|a\rangle\langle g|+\Omega_{g}^{*}e^{\mathrm{i}\omega_{g}t}|g\rangle\langle a|+\Omega_{e}e^{-\mathrm{i}\omega_{e}t}|a\rangle\langle e|+\Omega_{e}^{*}e^{\mathrm{i}\omega_{e}t}|e\rangle\langle a|\right),
\end{equation}
where we have taken the energy origin in the auxiliary level $|a\rangle$,
and chosen the signs so that $\varepsilon_{j}>0$.}\\
\emph{Show that in a new picture defined by the transformation $\hat{U}_{\mathrm{c}}(t)=e^{\hat{H}_{\mathrm{c}}t/\mathrm{i}\hbar}$
with $\hat{H}_{\mathrm{c}}=-\hbar\omega_{g}|g\rangle\langle g|-\hbar\omega_{e}|e\rangle\langle e|$,
the Hamiltonian reads
\begin{equation}
\tilde{H}=\hbar\Delta_{g}|g\rangle\langle g|+\hbar\Delta_{e}|e\rangle\langle e|+\hbar\left(\Omega_{g}|a\rangle\langle g|+\Omega_{g}^{*}|g\rangle\langle a|+\Omega_{e}|a\rangle\langle e|+\Omega_{e}^{*}|e\rangle\langle a|\right),
\end{equation}
with $\Delta_{j}=\omega_{j}-\varepsilon_{j}$.}\\
\emph{In the far-detuned limit $|\Delta_{j}|\gg|\Omega_{j}|$, argue
that starting from an unpopulated auxiliary level, the relevant sector
of the Hilbert space is defined by the projector $\hat{P}=|g\rangle\langle g|+|e\rangle\langle e|$,
leading to the effective Hamiltonian
\begin{equation}
\tilde{H}_{\mathrm{eff}}\approx\hbar\Delta_{g}|g\rangle\langle g|+\hbar\Delta_{e}|e\rangle\langle e|+\hbar\Omega_{\mathrm{eff}}|e\rangle\langle g|+\hbar\Omega_{\mathrm{eff}}^{*}|g\rangle\langle e|,
\end{equation}
with $\Omega_{\mathrm{eff}}=\Omega_{g}\Omega_{e}^{*}/\Delta_{g}$.}\\
\emph{Come back to the original picture and show that the Hamiltonian
can be written in the common form (you'll need to make an energy shift)
\begin{equation}
\hat{H}_{\mathrm{eff}}=\frac{\hbar\varepsilon}{2}\hat{\sigma}_{z}+\hbar\left(\Omega_{\mathrm{eff}}e^{-\mathrm{i}\omega_{\mathrm{eff}}t}\hat{\sigma}^{\dagger}+\Omega_{\mathrm{eff}}^{*}e^{\mathrm{i}\omega_{\mathrm{eff}}t}\hat{\sigma}\right),
\end{equation}
where $\omega_{\mathrm{eff}}=\omega_{e}-\omega_{g}$, $\varepsilon=\varepsilon_{g}-\varepsilon_{e}$,
and the Pauli operators are defined in the usual way.}\\
Hint: Just follow the steps we have developed in class, and be careful
that at the end you will need to justify the approximation $\Omega_{g}\Omega_{e}^{*}/\Delta_{g}\approx\Omega_{g}\Omega_{e}^{*}/\Delta_{e}$
to get a Hermitian Hamiltonian.
\end{enumerate}
\newpage

\subsubsection{Graded exercise 7b: Optomechanical sideband cooling}

In the last exercise of the course we will study one research-level
problem: how to cool down a mechanical resonator by using laser light.
It is research level in the sense that the theory behind it, which
is what you we will do in this exercise, was developed only about
a decade ago, and the related experiments started only around the
same time. Altogether, these developments gave birth to the now well-established
field of quantum optomechanics, which is currently a subject of active
investigation by many groups all around the world.
\begin{enumerate}
\item \emph{Consider an optical cavity with an the end mirror coupled to
a fixed wall through a spring (Fig. \ref{FigEffective}), so that
it can oscillate at some frequency $\Omega$ (much smaller than optical
frequencies). Take the position of the mirror around its equilibrium
location as a dynamical variable $\hat{z}$, whose evolution (in the
absence of coupling to other systems) is provided by a harmonic oscillator
Hamiltonian $\hat{H}_{\mathrm{m}}=\frac{1}{2m}\hat{p}_{z}^{2}+\frac{m\Omega^{2}}{2}\hat{z}^{2}$,
where $m$ is the mirror's mass and $\hat{p}_{z}$ its momentum, so
that $[\hat{z},\hat{p}_{z}]=\mathrm{i}\hbar$. Consider now one mode
of the cavity with annihilation operator $\hat{a}$ and resonance
frequency $\omega_{\mathrm{c}}$ when the cavity has length $L$.
Using the fact that the cavity's length is $L+\hat{z}$, and assuming
that the mirror's deviations from equilibrium are small compared to
$L$, show that the Hamiltonian of the system can be approximated
by
\begin{equation}
\hat{H}=\hbar\Omega\hat{b}^{\dagger}\hat{b}+\hbar\omega_{\mathrm{c}}\hat{a}^{\dagger}\hat{a}+\hbar g_{0}\hat{a}^{\dagger}\hat{a}(\hat{b}^{\dagger}+\hat{b}),
\end{equation}
where we have introduced the annihilation and creation operators for
the mechanical oscillator, $\hat{b}$ and $\hat{b}^{\dagger}$, respectively,
and the so-called optomechanical coupling rate is given
\begin{equation}
g_{0}=-\omega_{\mathrm{c}}\frac{z_{_{\mathrm{zpf}}}}{L},
\end{equation}
where $z_{\mathrm{zpf}}=\sqrt{\hbar/2\Omega m}$ refers to the zero-point
fluctuations of the mirror's position.}\\
Hint: You will need to use the expression that we saw at the beginning
of the lectures for the resonance frequencies inside a cavity as a
function of its length.
\item \emph{Assume now that the temperature is low enough for the number
of thermal photons at optical frequencies to be negligible, but large
enough for the mechanical oscillator to have a lot of thermal excitations
or phonons $\bar{n}$ (this is what naturally happens at room temperature,
since typical optical frequencies are hundreds of THz while mechanical
oscillations are never larger than GHz). Moreover, assume as well
that the cavity is driven by a laser, so that the master equation
of the whole system reads
\begin{equation}
\frac{d\hat{\rho}}{dt}=\left[\frac{\hat{H}}{\mathrm{i}\hbar}+\left(\mathcal{E}e^{-\mathrm{i}\omega_{\mathrm{L}}t}\hat{a}^{\dagger}-\mathcal{E}^{*}e^{\mathrm{i}\omega_{\mathrm{L}}t}\hat{a}\right),\hat{\rho}\right]+\kappa\mathcal{D}_{a}[\hat{\rho}]+\gamma(\bar{n}+1)\mathcal{D}_{b}[\hat{\rho}]+\gamma\bar{n}\mathcal{D}_{b^{\dagger}}[\hat{\rho}].
\end{equation}
Show that there exists a transformation operator $\hat{U}_{\mathrm{c}}(t)$
such that the state $\hat{\rho}'(t)=\hat{U}_{\mathrm{c}}^{\dagger}(t)\hat{\rho}(t)\hat{U}_{\mathrm{c}}(t)$
in the new picture satisfies the time-independent master equation
\begin{equation}
\frac{d\hat{\rho}'}{dt}=\left[-\mathrm{i}\Omega\hat{b}^{\dagger}\hat{b}+\mathrm{i}\Delta_{\mathrm{c}}\hat{a}^{\dagger}\hat{a}+\mathcal{E}\hat{a}^{\dagger}-\mathcal{E}^{*}\hat{a}-\mathrm{i}g_{0}\hat{a}^{\dagger}\hat{a}(\hat{b}^{\dagger}+\hat{b}),\hat{\rho}'\right]+\kappa\mathcal{D}_{a}[\hat{\rho}']+\gamma(\bar{n}+1)\mathcal{D}_{b}[\hat{\rho}']+\gamma\bar{n}\mathcal{D}_{b^{\dagger}}[\hat{\rho}'],\label{OMmasterEq}
\end{equation}
where $\Delta_{\mathrm{c}}=\omega_{\mathrm{L}}-\omega_{\mathrm{c}}$.}
\item \emph{Assume that the state is coherent, that is, $\hat{\rho}'(t)=|\beta(t)\rangle\langle\beta(t)|\otimes|\alpha(t)\rangle\langle\alpha(t)|$,
and show that the coherent amplitudes satisfy the equations of motion}\begin{subequations}\label{OMclassEqs}\emph{
\begin{align}
\dot{\alpha} & =\mathcal{E}-\left[\kappa-\mathrm{i}\Delta_{\mathrm{c}}+\mathrm{i}g_{0}(\beta+\beta^{*})\right]\alpha,\\
\dot{\beta} & =-(\gamma+\mathrm{i}\Omega)\beta-\mathrm{i}g_{0}|\alpha|^{2}.
\end{align}
}\end{subequations}Hint: Find the equations of motion of $\langle\hat{a}\rangle$
and $\langle\hat{b}\rangle$ like we did in class for other problems
starting from the master equation, and use the fact that coherent
states are right and left eigenstates of the annihilation and creation
operators, respectively.
\item \emph{Consider now a displacement operator with time-dependent amplitude
\begin{equation}
\hat{D}_{c}[\chi(t)]=e^{\chi(t)\hat{c}^{\dagger}-\chi^{*}(t)\hat{c}},
\end{equation}
where $\hat{c}$ and $\hat{c}^{\dagger}$ are annihilation and creation
operators, respectively, that is, $[\hat{c},\hat{c}^{\dagger}]=1$.
Show that its time-derivative can be written as
\begin{equation}
\frac{d\hat{D}_{c}[\chi(t)]}{dt}=\hat{D}_{c}[\chi(t)]\left(\dot{\chi}\hat{c}^{\dagger}-\dot{\chi}^{*}\hat{c}+\frac{\dot{\chi}\chi^{*}-\chi\dot{\chi}^{*}}{2}\right).\label{DtimeDerivative}
\end{equation}
}\\
Hint: Take $\chi$ and $\chi^{*}$ as independent variables, so that
you can write the time derivative as $\frac{d}{dt}=\dot{\chi}\frac{\partial}{\partial\chi}+\dot{\chi}^{*}\frac{\partial}{\partial\chi^{*}}$,
and apply it to the normally-ordered form of the displacement operator
$\hat{D}_{c}[\chi(t)]=e^{-|\chi(t)|^{2}/2}e^{\chi(t)\hat{c}^{\dagger}}e^{-\chi^{*}(t)\hat{c}}$.
\item \emph{Move to another picture defined by the transformation operator
$\hat{U}(t)=\hat{D}_{b}[\beta(t)]\hat{D}_{a}[\alpha(t)]$, where the
amplitudes $\alpha$ and $\beta$ satisfy the classical equations
of motion }(\ref{OMclassEqs})\emph{. Show that the transformed state
$\tilde{\rho}(t)=\hat{U}^{\dagger}(t)\hat{\rho}'\hat{U}(t)$ satisfies
the master equation
\begin{equation}
\frac{d\tilde{\rho}}{dt}=\left[-\mathrm{i}\Omega\hat{b}^{\dagger}\hat{b}+\mathrm{i}\Delta(t)\hat{a}^{\dagger}\hat{a}-\mathrm{i}[g(t)\hat{a}^{\dagger}+g^{*}(t)\hat{a}+g_{0}\hat{a}^{\dagger}\hat{a}](\hat{b}^{\dagger}+\hat{b}),\tilde{\rho}\right]+\kappa\mathcal{D}_{a}[\tilde{\rho}]+\gamma(\bar{n}+1)\mathcal{D}_{b}[\tilde{\rho}]+\gamma\bar{n}\mathcal{D}_{b^{\dagger}}[\tilde{\rho}],\label{OmmasterEq2}
\end{equation}
where we have defined the dressed detuning $\Delta(t)=\Delta_{\mathrm{c}}-g_{0}\left[\beta(t)+\beta^{*}(t)\right]$
and the dressed optomechanical coupling $g(t)=g_{0}\alpha(t)$.}\\
Hint: Make the time derivative of $\tilde{\rho}(t)$ and use (\ref{DtimeDerivative}),
(\ref{OMmasterEq}), and the displacement formulas $\hat{U}^{\dagger}(t)\hat{a}\hat{U}(t)=\hat{a}+\alpha(t)$
and $\hat{U}^{\dagger}(t)\hat{b}\hat{U}(t)=\hat{b}+\beta(t)$, to
rewrite it as (\ref{OmmasterEq2}). You should obtain extra terms
in the Hamiltonian linear in the annihilation and creation operators,
but this terms vanish identically thanks to the classical equations
(\ref{OMclassEqs}). 
\item \emph{Next we are going to find an effective master equation for the
mechanical oscillator. Hence, we make the choices (we assume from
now on that the classical solution is time-independent)}\begin{subequations}\emph{
\begin{align}
\mathcal{L}_{\mathrm{S}}[\tilde{\rho}] & =\left[-\mathrm{i}\Omega\hat{b}^{\dagger}\hat{b},\tilde{\rho}\right]+\gamma(\bar{n}+1)\mathcal{D}_{b}[\tilde{\rho}]+\gamma\bar{n}\mathcal{D}_{b^{\dagger}}[\tilde{\rho}],\\
\mathcal{L}_{\mathrm{E}}[\tilde{\rho}] & =\left[\mathrm{i}\Delta\hat{a}^{\dagger}\hat{a},\tilde{\rho}\right]+\kappa\mathcal{D}_{a}[\tilde{\rho}],\\
\mathcal{L}_{1}[\tilde{\rho}] & =\biggl[-\mathrm{i}\underbrace{(g\hat{a}^{\dagger}+g^{*}\hat{a}+g_{0}\hat{a}^{\dagger}\hat{a})\hat{X}}_{\hat{H}_{1}/\hbar},\tilde{\rho}\biggl],
\end{align}
}\end{subequations}\emph{where we have defined the mechanical position
quadrature $\hat{X}=\hat{b}^{\dagger}+\hat{b}$. Show that the stationary
state of the free cavity mode is vacuum, $\bar{\rho}_{\mathrm{E}}=|0\rangle\langle0|$,
and then prove that
\begin{equation}
\mathcal{PL}_{1}\mathcal{P}[\hat{Y}]=0,\label{L1cond}
\end{equation}
for any operator $\hat{Y}$, where the projector superoperator acts
as $\mathcal{P}[\hat{Y}]=\mathrm{tr}_{\mathrm{E}}\{\hat{Y}\}\otimes\bar{\rho}_{\mathrm{E}}$.}\\
Hint: Just apply the definition of the projector to (\ref{L1cond}),
and you should be able to write it as something proportional to $\langle0|\hat{a}|0\rangle$
and $\langle0|\hat{a}^{\dagger}\hat{a}|0\rangle$.
\item \emph{Note that the Hamiltonian can be written as $\hat{H}_{1}/\hbar=\sum_{m=1}^{3}g_{m}\hat{S}_{m}\otimes\hat{E}_{m}$,
with the choices $g_{2}=g=g_{1}^{*}$, $g_{3}=g_{0}$, $\hat{S}_{1}=\hat{S}_{2}=\hat{S}_{3}=\hat{X}$,
$\hat{E}_{1}=\hat{a}=\hat{E}_{2}^{\dagger}$, and $\hat{E}_{3}=\hat{a}^{\dagger}\hat{a}$.
Show that the environmental operators are all closed on their own,
and use the quantum regression formula to show that ($\tau>0$)}\begin{subequations}\emph{
\begin{align}
\lim_{t\rightarrow\infty}\langle\hat{A}(t)\hat{a}(t+\tau)\hat{C}(t)\rangle_{\mathrm{E}} & =\langle0|\hat{A}\hat{a}\hat{C}|0\rangle_{\mathrm{E}}e^{-(\kappa-\mathrm{i}\Delta)\tau},\\
\lim_{t\rightarrow\infty}\langle\hat{A}(t)\hat{a}(t+\tau)\hat{C}(t)\rangle_{\mathrm{E}} & =\langle0|\hat{A}\hat{a}^{\dagger}\hat{C}|0\rangle_{\mathrm{E}}e^{-(\kappa+\mathrm{i}\Delta)\tau}\\
\lim_{t\rightarrow\infty}\langle\hat{A}(t)\hat{a}^{\dagger}(t+\tau)\hat{a}(t+\tau)\hat{C}(t)\rangle_{\mathrm{E}} & =\langle0|\hat{A}\hat{a}^{\dagger}\hat{a}\hat{C}|0\rangle_{\mathrm{E}}e^{-2\kappa\tau},
\end{align}
}\end{subequations}\emph{for any environmental operators $\hat{A}$
and $\hat{C}$. Then, show that the cavity correlators required for
the mechanical effective master equation read as
\begin{equation}
C_{mn}(\tau)=e^{-(\kappa+\mathrm{i}\Delta)\tau}\delta_{m1}\delta_{n2},\hspace{1em}\hspace{1em}K_{nm}(\tau)=e^{-(\kappa-\mathrm{i}\Delta)\tau}\delta_{n1}\delta_{m2}.
\end{equation}
}\\
Hint: Follow the steps followed in the example of the lectures.
\item \emph{Take $\hat{H}_{\mathrm{S}}=\hbar\Omega\hat{b}^{\dagger}\hat{b}$
as the only part of the mechanical evolution which is relevant within
the decay of the cavity correlators. Show that
\begin{equation}
\tilde{X}(\tau)=e^{\hat{H}_{\mathrm{S}}\tau/\mathrm{i}\hbar}\hat{X}e^{-\hat{H}_{\mathrm{S}}\tau/\mathrm{i}\hbar}=e^{\mathrm{i}\Omega\tau}\hat{b}+e^{-\mathrm{i}\Omega\tau}\hat{b}^{\dagger},
\end{equation}
and then prove that the effective mechanical master equation
\begin{equation}
\frac{d\tilde{\rho}_{\mathrm{S}}}{dt}=\mathcal{L}_{\mathrm{S}}[\tilde{\rho}_{\mathrm{S}}]+|g|^{2}\int_{0}^{t}d\tau\left[C_{12}(\tau)\hat{X}\tilde{\rho}_{\mathrm{S}}\hat{X}(\tau)-K_{12}(\tau)\hat{X}\hat{X}(\tau)\tilde{\rho}_{\mathrm{S}}+\mathrm{H.c.}\right],
\end{equation}
can be written in the $t\gg\kappa^{-1}$ limit as
\begin{equation}
\frac{d\tilde{\rho}_{\mathrm{S}}}{dt}=\mathcal{L}_{\mathrm{S}}[\tilde{\rho}_{\mathrm{S}}]+\mathcal{L}_{-}[\tilde{\rho}_{\mathrm{S}}]+\mathcal{L}_{+}[\tilde{\rho}_{\mathrm{S}}]+\mathcal{L}_{\mathrm{NRW}}[\tilde{\rho}_{\mathrm{S}}],
\end{equation}
with}\begin{subequations}\emph{
\begin{align}
\mathcal{L}_{-}[\tilde{\rho}_{\mathrm{S}}] & =\Gamma_{-}^{\mathrm{opt}}\mathcal{D}_{b}[\tilde{\rho}_{\mathrm{S}}]-\mathrm{i}[\delta\Omega_{-}\hat{b}^{\dagger}\hat{b},\tilde{\rho}_{\mathrm{S}}],\hspace{1em}\mathcal{L}_{+}[\tilde{\rho}_{\mathrm{S}}]=\Gamma_{+}^{\mathrm{opt}}\mathcal{D}_{b^{\dagger}}[\tilde{\rho}_{\mathrm{S}}]-\mathrm{i}[\delta\Omega_{+}\hat{b}^{\dagger}\hat{b},\tilde{\rho}_{\mathrm{S}}],\\
\mathcal{L}_{\mathrm{NRW}}[\tilde{\rho}_{\mathrm{S}}] & =|g|^{2}\left[\frac{\hat{b}\tilde{\rho}_{\mathrm{S}}\hat{b}}{\kappa+\mathrm{i}(\Delta-\Omega)}+\frac{\hat{b}^{\dagger}\tilde{\rho}_{\mathrm{S}}\hat{b}^{\dagger}}{\kappa+\mathrm{i}(\Delta+\Omega)}-\frac{\hat{b}^{2}\tilde{\rho}_{\mathrm{S}}}{\kappa-\mathrm{i}(\Delta+\Omega)}-\frac{\tilde{\rho}_{\mathrm{S}}\hat{b}^{\dagger2}}{\kappa-\mathrm{i}(\Delta-\Omega)}+\mathrm{H.c.}\right],
\end{align}
}\end{subequations}\emph{and
\[
\Gamma_{\mp}^{\mathrm{opt}}=\frac{|g|^{2}/\kappa}{1+\left(\frac{\Delta\pm\Omega}{\kappa}\right)^{2}},\hspace{1em}\delta\Omega_{\mp}^{\mathrm{opt}}=\frac{|g|^{2}\left(\Delta\pm\Omega\right)}{\kappa^{2}+\left(\Delta\pm\Omega\right)^{2}}.
\]
}\\
Hint: Just perform the time integrals (taking the limit $t/\kappa\gg1$),
and reorder the terms in the proper way.
\item \emph{Argue that $\mathcal{L}_{\mathrm{NRW}}$ can be neglected within
a rotating-wave approximation provided that
\begin{equation}
\frac{|g|^{2}}{\sqrt{\kappa^{2}+(\Delta\pm\Omega)^{2}}}\ll2\Omega.
\end{equation}
Within this approximation, show that the effective mechanical master
equation can be written as
\begin{equation}
\frac{d\tilde{\rho}_{\mathrm{S}}}{dt}=\left[-\mathrm{i}(\Omega+\delta\Omega_{-}+\delta\Omega_{+})\hat{b}^{\dagger}\hat{b},\tilde{\rho}_{\mathrm{S}}\right]+\Gamma_{-}\mathcal{D}_{b}[\tilde{\rho}_{\mathrm{S}}]+\Gamma_{+}\mathcal{D}_{b^{\dagger}}[\tilde{\rho}_{\mathrm{S}}],
\end{equation}
with
\begin{equation}
\Gamma_{-}=\gamma(\bar{n}+1)+\Gamma_{-}^{\mathrm{opt}},\hspace{1em}\Gamma_{+}=\gamma\bar{n}+\Gamma_{+}^{\mathrm{opt}},
\end{equation}
which can be rewritten in the standard form
\begin{equation}
\frac{d\tilde{\rho}_{\mathrm{S}}}{dt}=\left[-\mathrm{i}\Omega_{\mathrm{eff}}\hat{b}^{\dagger}\hat{b},\tilde{\rho}_{\mathrm{S}}\right]+\Gamma_{\mathrm{eff}}(\bar{n}_{\mathrm{eff}}+1)\mathcal{D}_{b}[\tilde{\rho}_{\mathrm{S}}]+\Gamma_{\mathrm{eff}}\bar{n}_{\mathrm{eff}}\mathcal{D}_{b^{\dagger}}[\tilde{\rho}_{\mathrm{S}}],
\end{equation}
with}\begin{subequations}\emph{
\begin{align}
\Omega_{\mathrm{eff}} & =\Omega+\delta\Omega_{-}+\delta\Omega_{+},\\
\Gamma_{\mathrm{eff}} & =\Gamma_{-}-\Gamma_{+}=\gamma\left[1+\frac{C}{1+\left(\frac{\Delta+\Omega}{\kappa}\right)^{2}}-\frac{C}{1+\left(\frac{\Delta-\Omega}{\kappa}\right)^{2}}\right],\\
\bar{n}_{\mathrm{eff}} & =\frac{\Gamma_{+}}{\Gamma_{-}-\Gamma_{+}}=\frac{\bar{n}+\frac{C}{1+\left(\frac{\Delta-\Omega}{\kappa}\right)^{2}}}{1+\frac{C}{1+\left(\frac{\Delta+\Omega}{\kappa}\right)^{2}}-\frac{C}{1+\left(\frac{\Delta-\Omega}{\kappa}\right)^{2}}},
\end{align}
}\end{subequations}\emph{where we have defined the cooperativity
$C=|g|^{2}/\gamma\kappa$.}
\item \emph{Inspect $\bar{n}_{\mathrm{eff}}$, and argue that the best cooling
is obtained when $\Delta=-\Omega$ (red sideband driving) and $\Omega\gg\kappa$
(resolved sideband regime). Under these conditions, show that when
$C\gg1$, then we have
\begin{equation}
\Gamma_{\mathrm{eff}}\approx\gamma C,\hspace{1em}\hspace{1em}\bar{n}_{\mathrm{eff}}\approx\frac{\bar{n}}{C}+\frac{\kappa^{2}}{4\Omega^{2}},
\end{equation}
so that when $C\gg\bar{n}$, we obtain the minimum effective thermal
occupation $\bar{n}_{\mathrm{eff}}\approx\frac{\kappa^{2}}{4\Omega^{2}}$.}
\item \emph{Self-consistency check: show that all the approximations required
to derive the effective master equation are compatible with the cooling
conditions.}
\item \emph{Finally, argue that the asymptotic state of this equation in
the original Schrödinger picture corresponds to the displaced thermal
state}
\begin{equation}
\lim_{t\rightarrow\infty}\hat{\rho}_{\mathrm{S}}(t)=\hat{D}_{b}(\beta)\bar{\rho}_{\mathrm{th}}(\bar{n}_{\mathrm{eff}})\hat{D}_{b}^{\dagger}(\beta).
\end{equation}
\end{enumerate}

\newpage{}

\appendix

\section{Review of classical mechanics, Hilbert spaces, and quantum mechanics\label{QuantumMechanics}}

The purpose of this appendix is introducing the fundamental principles
of quantum mechanics as are used throughout the lectures. The quantum
framework is far from being intuitive, but somehow feels reasonable
(and even inevitable) once one understands the context in which it
was created, specifically: (\textit{i}) the theories that were used
to describe physical systems prior to its development, (\textit{ii})
the experiments which did not fit in this context, and (\textit{iii})
the mathematical language that accommodates the new quantum formulation
of physical phenomena. We will then briefly review these context prior
to introducing and discussing the principles conforming the quantum-mechanical
framework.

\subsection{Classical mechanics\label{ClassicalMechanics}}

In this section we go through a brief introduction to classical mechanics\footnote{For a deeper reading I recommend Goldstein's book \cite{GoldsteinBook},
as well as Greiner's books \cite{GreinerMechanicsBook,GreinerMechanics2Book},
or the one of Hand and Finch \cite{HandFinchBook}.}, with emphasis in analyzing the Hamiltonian formalism and how it
treats observable magnitudes. We will see that a proper understanding
of this formalism will make the transition to quantum mechanics more
natural.

\subsubsection{The Lagrangian formalism}

In classical mechanics the state of a system is specified by the position
of its constituent particles at all times, $\mathbf{r}_{j}(t)=[x_{j}(t),y_{j}(t),z_{j}(t)]$
with $j=1,2,...,N$, being $N$ the number of particles. Defining
the kinetic momentum of the particles as $\mathbf{P}_{j}=m_{j}\mathbf{\dot{r}}_{j}$
($m_{j}$ is the mass of particle $j$), the evolution of the system
is found from a set of initial positions and velocities by solving
the Newton equations of motion $\mathbf{\dot{P}}_{j}=\mathbf{F}_{j}$,
being $\mathbf{F}_{j}$ the forces acting onto particle $j$.

Most physical systems have further constraints that have to fulfill
(for example, the distance between the particles of a rigid body cannot
change with time, that is, $\vert\mathbf{r}_{j}-\mathbf{r}_{l}\vert=const$),
and therefore the positions $\{\mathbf{r}_{j}\}_{j=1,...,N}$ are
no longer independent, which makes Newton equations hard to solve.
This calls for a new simpler theoretical framework: the so-called
\textit{analytical mechanics}. In the following we review this framework,
but assuming, for simplicity, that the constrains are \textit{holonomic}\footnote{A constrain is \textit{holonomic} when it can be written as $f(\mathbf{r}_{1},...,\mathbf{r}_{N},t)=0$.
Nonholonomic constrains correspond, for example, to the boundary imposed
by a wall that particles cannot cross, which is usually expressed
in terms of inequalities \cite{GoldsteinBook}, and require a much
more careful treatment.} and \textit{scleronomous}\footnote{A constrain is called \textit{scleronomous} when the it does not depend
explicitly on time. Time-dependent constrains are called \textit{rheonomous}
and correspond, for example, to a situation in which the motion of
the particles is restricted to a moving surface or curve \cite{GoldsteinBook}.}, meaning that they can always be written in the form $f(\mathbf{r}_{1},...,\mathbf{r}_{N})=0$.

In analytical mechanics the state of the system at any time is specified
by a vector $\mathbf{q}(t)=[q_{1}(t),q_{2}(t),...,q_{n}(t)]$. $n$
is the number of degrees of freedom of the system (the total number
of coordinates, $3N$, minus the number of constraints), and the $q_{j}$'s
are called the \textit{generalized coordinates} of the system, which
are compatible with the constraints and related to the usual coordinates
of the particles through some smooth functions $\mathbf{q}(\{\mathbf{r}_{j}\}_{j=1,...,N})\Leftrightarrow\{\mathbf{r}_{j}(\mathbf{q})\}_{j=1,...,N}$.
The space formed by the generalized coordinates is called \textit{coordinate
space}, and $\mathbf{q}(t)$ describes a \textit{trajectory} on it.

The basic object in analytical mechanics is the \textit{Lagrangian},
$L[\mathbf{q}(t),\mathbf{\dot{q}}(t),t]$, which is a function of
the generalized coordinates and velocities, and can even have some
explicit time dependence. In general, the Lagrangian must be built
based on general principles like symmetries. However, if the forces
acting on the particles of the system are conservative, that is, $\mathbf{F}_{j}=\boldsymbol{\nabla}_{j}V[\{\mathbf{r}_{l}\}_{l=1,...,N}]=(\partial_{x_{j}}V,\partial_{y_{j}}V,\partial_{z_{j}}V)$
for some \textit{potential} $V[\{\mathbf{r}_{l}\}_{l=1,...,N}]$,
one can choose a Lagrangian with the simple form $L=T(\mathbf{\dot{q}},\mathbf{q})-V(\mathbf{q})$,
being $T(\mathbf{\dot{q}},\mathbf{q})=\sum_{j=1}^{N}m_{j}\mathbf{\dot{r}}_{j}^{2}(\mathbf{q})/2$
the kinetic energy of the system and $V(\mathbf{q})=V[\{\mathbf{r}_{l}(\mathbf{q})\}_{l=1,...,n}]$.
The dynamical equations of the system are then formulated as a \textit{variational
principle} on the \textit{action} 
\begin{equation}
S=\int_{t_{1}}^{t_{2}}dtL[\mathbf{q}(t),\mathbf{\dot{q}}(t),t],
\end{equation}
by asking the trajectory of the system $\mathbf{q}(t)$ between two
fixed points $\mathbf{q}(t_{1})$ and $\mathbf{q}(t_{2})$ to be such
that the action is an extremal, $\delta S=0$. From this principle,
it is straightforward to arrive to the well known Euler-Lagrange equations
\begin{equation}
\frac{\partial L}{\partial q_{j}}-\frac{d}{dt}\frac{\partial L}{\partial\dot{q}_{j}}=0,\label{Euler-Lagrange}
\end{equation}
which are a set of second order differential equations for the generalized
coordinates $\mathbf{q}$, and together with the conditions $\mathbf{q}(t_{1})$
and $\mathbf{q}(t_{2})$ provide the trajectory $\mathbf{q}(t)$.

\subsubsection{The Hamiltonian formalism}

As we have seen, the Euler-Lagrange equations are a set of second
order differential equations which allows us to find the trajectory
$\mathbf{q}(t)$ in coordinate space. We could reduce the order of
the differential equations by taking the velocities $\mathbf{\dot{q}}$
as dynamical variables, arriving then to a set of $2n$ first-order
differential equations. This is, however, a very naïve way of reducing
the order, which leads to a non-symmetric system of equations for
$\mathbf{q}$ and $\mathbf{\dot{q}}$. In this section we review Hamilton's
approach to analytical mechanics, which leads to a symmetric-like
first-order system of equations, and will play a major role in understanding
the transition from classical to quantum mechanics.

Instead of using the velocities, the Hamiltonian formalism considers
the \textit{generalized momenta}
\begin{equation}
p_{j}=\frac{\partial L}{\partial\dot{q}_{j}},
\end{equation}
as the dynamical variables. Note that this definition establishes
a relation between these generalized momenta and the velocities $\mathbf{\dot{q}}(\mathbf{q},\mathbf{p})\Leftrightarrow\mathbf{p}(\mathbf{q},\mathbf{\dot{q}})$.
Note also that when the usual Cartesian coordinates of the system's
particles are taken as the generalized coordinates, these momenta
coincide with those of Newton's approach.

The theory is then built in terms of a new object called the \textit{Hamiltonian},
which is defined as a Legendre transform of the Lagrangian,
\begin{equation}
H(\mathbf{q},\mathbf{p})=\mathbf{p\dot{q}}(\mathbf{q},\mathbf{p})-L[\mathbf{q},\mathbf{\dot{q}}(\mathbf{q},\mathbf{p}),t],
\end{equation}
and coincides with the total energy\footnote{The general conditions under which the Hamiltonian coincides with
the system's energy can be found in \cite{GoldsteinBook}.} for conservative systems with scleronomous constrains, that is, $H(\mathbf{q},\mathbf{p})=T(\mathbf{q},\mathbf{p})+V(\mathbf{q})$,
with $T(\mathbf{q},\mathbf{p})=T[\mathbf{q},\mathbf{\dot{q}}(\mathbf{q},\mathbf{p})]$.
Differentiating this expression and using the Euler-Lagrange equations
(or using again the variational principle on the action), it is then
straightforward to obtain the equations of motion for the generalized
coordinates and momenta (the \textit{canonical equations}),
\begin{equation}
\dot{q}_{j}=\frac{\partial H}{\partial p_{j}}\text{ \ \ \ \ and \ \ \ \ }\dot{p}_{j}=-\frac{\partial H}{\partial q_{j}},\label{CanEqs}
\end{equation}
which together with some initial conditions $\{\mathbf{q}(t_{0}),\mathbf{p}(t_{0})\}$
allow us to find the trajectory $\{\mathbf{q}(t),\mathbf{p}(t)\}$
in the space formed by the generalized coordinates and momenta, which
is known as \textit{phase space}.

Another important object in the Hamiltonian formalism is the \textit{Poisson
bracket}. Given two functions of the coordinates and momenta $F(\mathbf{q},\mathbf{p})$
and $G(\mathbf{q},\mathbf{p})$, their Poisson bracket is defined
as
\begin{equation}
\{F,G\}=\sum_{j=1}^{n}\frac{\partial F}{\partial q_{j}}\frac{\partial G}{\partial p_{j}}-\frac{\partial F}{\partial p_{j}}\frac{\partial G}{\partial q_{j}}.\label{PB}
\end{equation}
The importance of this object is reflected in the fact that the evolution
equation of any quantity $g(\mathbf{q},\mathbf{p},t)$ can be written
as
\begin{equation}
\frac{dg}{dt}=\{g,H\}+\frac{\partial g}{\partial t},\label{HamObsEvo}
\end{equation}
and hence, if the quantity does not depend explicitly on time and
its Poisson bracket with the Hamiltonian is zero, it is a \textit{constant
of motion}.

Of particular importance for the transition to quantum mechanics are
the \textit{canonical Poisson brackets}, that is, the Poisson brackets
of the coordinates and momenta,
\begin{equation}
\{q_{j},p_{l}\}=\delta_{jl}\text{, }\{q_{j},q_{l}\}=\{p_{j},p_{l}\}=0,\label{funPB}
\end{equation}
which define the mathematical structure of phase space.

\subsubsection{Observables and their mathematical structure}

In this last section concerning classical mechanics we will discuss
about the mathematical structure in which observables are embedded
within the Hamiltonian formalism. We will see that the mathematical
objects corresponding to physical observables form a well defined
mathematical structure, a real Lie algebra. Moreover, the position
and momentum will be shown to be the generators of a particular Lie
group, the Heisenberg group. Understanding this internal structure
of \textit{classical observables} will give us the chance to introduce
the quantum description of observables in a reasonable way. Let us
start by defining the concept of Lie algebra.

A \textit{real} \textit{Lie algebra} is a real vector space\footnote{The concept of complex vector space is defined in the next section.
The definition of a \textit{real} vector space is the same, but replacing
by real numbers the complex numbers that appear in the definitions
there.} $\mathcal{L}$ equipped with an additional operation, the \textit{Lie
product}, which takes two vectors $f$ and $g$ from $\mathcal{L}$,
to generate another vector also in $\mathcal{L}$ denoted by\footnote{Note that we are using the same notation for the general definition
of the Lie product and for the Poisson bracket, which are a particular
case of Lie product as we will learn shortly.} $\{f,g\}$. This operation must satisfy the following properties:
\begin{enumerate}
\item $\{f,g+h\}=\{f,g\}+\{f,h\}$ (linearity)
\item $\{f,f\}=0\underset{\text{with 1}}{\overset{\text{together}}{\Longrightarrow}}\{f,g\}=-\{g,f\}$
(anticommutativity)
\item $\{f,\{g,h\}\}+\{g,\{h,f\}\}+\{h,\{f,g\}\}=0$ (Jacobi identity) 
\end{enumerate}
Hence, in essence a real Lie algebra is a vector space equipped with
a linear, non-commutative, non-associative product. They have been
a subject of study for many years, and now we know a lot about the
properties of these mathematical structures. They appear in many branches
of physics and geometry, specially connected to continuous symmetry
transformations, whose associated mathematical structures are actually
called \textit{Lie groups}. In particular, it is possible to show
that given any Lie group with $p$ parameters (like, e.g., the three-parameter
groups of translations or rotations in real space), any transformation
onto the system in which it is acting can be generated from a set
of $p$ elements of a Lie algebra $\{g_{1},g_{2},...,g_{p}\}$, called
the \textit{generators }of the Lie group, which satisfy some particular
relations 
\begin{equation}
\{g_{j},g_{k}\}=\sum_{l=1}^{p}c_{jkl}g_{l}.
\end{equation}
These relations are called the \textit{algebra-group relations}, and
the \textit{structure constants} $c_{jkl}$ are characteristic of
the particular Lie group (for example, the generators of translations
and rotations in real space are the momenta and angular momenta, respectively,
and the corresponding structure constants are $c_{jkl}=0$ for the
translation group and $c_{jkl}=\epsilon_{jkl}$ for the rotation group\footnote{$\epsilon_{j_{1}j_{2}...j_{M}}$ with all the subindices going from
1 to $M$ is the Levi-Civita symbol in dimension $M$, which has $\epsilon_{12...M}=1$
and is completely antisymmetric, that is, changes its sign after permutation
of any pair of indices.}).

Coming back to the Hamiltonian formalism, we start by noting that
\textit{observables}, being \textit{measurable} quantities, must be
given by continuous, real functions in phase space. Hence they form
a real vector space with respect to the usual addition of functions
and multiplication of a function by a real number. It also appeared
naturally in the formalism a linear, non-commutative, non-associative
operation between phase-space functions, the Poisson bracket, which
applied to real functions gives another real function. It is easy
to see that the Poisson bracket satisfies all the requirements of
a Lie product, and hence, observables form a Lie algebra within the
Hamiltonian formalism.

Moreover, the fundamental Poisson brackets (\ref{funPB}) show that
the generalized coordinates $\mathbf{q}$ and momenta $\mathbf{p}$,
together with the identity in phase space, satisfy particular algebra-group
relations, namely\footnote{Ordering the generators as $\{\mathbf{q},\mathbf{p},1\}$, the structure
constants associated to this algebra-group relations are explicitly
\begin{equation}
c_{jkl}=\left\{ \begin{array}{cc}
\Omega_{jk}\delta_{l,2n+1} & j,k=1,2,...,2n\\
0 & j=2n+1\text{ or }k=2n+1
\end{array}\right.\text{, }
\end{equation}
being $\Omega=\left(\begin{array}{cc}
0_{n\times n} & I_{n\times n}\\
-I_{n\times n} & 0_{n\times n}
\end{array}\right)$, with $I_{n\times n}$ and $0_{n\times n}$ the $n\times n$ identity
and null matrices, respectively.} $\{q_{j},p_{k}\}=\delta_{jk}1$ and $\{q_{j},1\}=\{p_{j},1\}=\{1,1\}=0$,
and hence can be seen as the generators of a Lie group. This group
is known as the \textit{Heisenberg group}, and was introduced by Weyl
when trying to prove the equivalence between the Schrödinger and Heisenberg
pictures of quantum mechanics (about which we will learn later). It
was later shown to have connections with the symplectic group, which
is the basis of many physical theories. Note that we could have taken
the Poisson brackets between the angular momenta associated to the
possible rotations in the system of particles (which are certainly
far more intuitive transformations than the one related to the Heisenberg
group) as the fundamental ones. However, we have chosen the Lie algebra
associated to the Heisenberg group just because it deals directly
with position and momenta, allowing for a simpler connection to quantum
mechanics.

Therefore, we arrive to the main conclusion of this review in classical
mechanics: $\bigskip$

$\lceil$The mathematical framework of Hamiltonian mechanics associates
physical observables with elements of a Lie algebra, being the phase-space
coordinates themselves the generators of the Heisenberg group.$\rfloor$

\bigskip{}

Maintaining this structure for observables will help us introducing
the laws of quantum mechanics in a coherent way.

\subsection{The mathematical language of quantum mechanics\label{Section:HilbertSpaces}}

Just as classical mechanics is formulated in terms of the mathematical
language of differential calculus and its extensions, quantum mechanics
takes linear algebra (and Hilbert spaces in particular) as its fundamental
grammar. In this section we will review the concept of Hilbert space,
and discuss the properties of some operators which play important
roles in the formalism of quantum optics.

\subsubsection{Finite-dimensional Hilbert spaces}

In essence, a Hilbert space is a \textit{complex vector space} in
which an \textit{inner product} is defined. Let us define first these
terms as we use them in these lectures.

A \textit{complex vector space} is a set $\mathcal{V}$, whose elements
will be called \textit{vectors} or \textit{kets} and will be denoted
by $\left\{ \vert a\rangle,\vert b\rangle,\vert c\rangle,...\right\} $
($a$, $b$, and $c$ may correspond to any suitable label), in which
the following two operations are defined: the \textit{vector addition},
which takes two vectors $\vert a\rangle$ and $\vert b\rangle$ and
creates a new vector inside $\mathcal{V}$ denoted by $\vert a\rangle+\vert b\rangle$;
and the \textit{multiplication by a scalar}, which takes a complex
number $\alpha\in\mathbb{C}$ (in this section Greek letters will
represent complex numbers) and a vector $\vert a\rangle$ to generate
a new vector in $\mathcal{V}$ denoted by $\alpha\vert a\rangle$.

The following additional properties must be satisfied: 
\begin{enumerate}
\item The vector addition is commutative and associative, that is, $\vert a\rangle+\vert b\rangle=\vert b\rangle+\vert a\rangle$
and $(\vert a\rangle+\vert b\rangle)+\vert c\rangle=\vert a\rangle+(\vert b\rangle+\vert c\rangle)$ 
\item There exists a null vector $\vert null\rangle$ such that $\vert a\rangle+\vert null\rangle=\vert a\rangle$ 
\item $\alpha(\vert a\rangle+\vert b\rangle)=\alpha\vert a\rangle+\alpha\vert b\rangle$ 
\item $(\alpha+\beta)\vert a\rangle=\alpha\vert a\rangle+\beta\vert a\rangle$ 
\item $(\alpha\beta)\vert a\rangle=\alpha(\beta\vert a\rangle)$ 
\item $1\vert a\rangle=\vert a\rangle$ 
\end{enumerate}
From these properties it can be proved that the null vector is unique,
and can be built from any vector $\vert a\rangle$ as $0\vert a\rangle$;
hence, in the following we denote it simply by $\vert null\rangle\equiv0$.
It is also readily proved that any vector $\vert a\rangle$ has a
unique \textit{antivector} $\vert\hspace{-1mm}-\hspace{-0.7mm}a\rangle$
such that $\vert a\rangle+\vert\hspace{-1mm}-\hspace{-0.7mm}a\rangle=0$,
which is given by $(-1)\vert a\rangle$ or simply $-\vert a\rangle$.

An \textit{inner product} is an additional operation defined in the
complex vector space $\mathcal{V}$, which takes two vectors $\vert a\rangle$
and $\vert b\rangle$ and associates them with a complex number. It
will be denoted by $\langle a|b\rangle$ or sometimes also by $(\vert a\rangle,\vert b\rangle)$,
and must satisfy the following properties:
\begin{enumerate}
\item $\langle a|a\rangle>0$ \hspace{2mm}if\hspace{2mm} $\vert a\rangle\neq0$
\item $\langle a|b\rangle=\langle b|a\rangle^{\ast}$
\item $(\vert a\rangle,\alpha\vert b\rangle)=\alpha\langle a|b\rangle$
\item $(\vert a\rangle,\vert b\rangle+\vert c\rangle)=\langle a|b\rangle+\langle a|c\rangle$ 
\end{enumerate}
The following additional properties can be proved from these ones:
\begin{itemize}
\item $\langle null\vert null\rangle=0$
\item $(\alpha\vert a\rangle,\vert b\rangle)=\alpha^{\ast}\langle a|b\rangle$
\item $(\vert a\rangle+\vert b\rangle,\vert c\rangle)=\langle a|c\rangle+\langle b|c\rangle$
\item $\vert\langle a|b\rangle\vert^{2}\leq\langle a|a\rangle\langle b|b\rangle$
\hspace{3mm}(Cauchy-Schwarz) 
\end{itemize}
Note that for any vector $\vert a\rangle$, one can define the object
$\langle a\vert\equiv(\vert a\rangle,\cdot)$, which will be called
a \textit{dual vector} or a \textit{bra}, and which takes a vector
$\vert b\rangle$ to generate the complex number $(\vert a\rangle,\vert b\rangle)\in\mathbb{C}$.
It can be proved that the set formed by all the dual vectors corresponding
to the elements in $\mathcal{V}$ is also a vector space, which will
be called the \textit{dual space} and will be denoted by $\mathcal{V}^{+}$.
Within this picture, the inner product can be seen as an operation
which takes a bra $\langle a\vert$ and a ket $\vert b\rangle$ to
generate the complex number $\langle a|b\rangle$, a \textit{bracket}.
This whole \textit{bra-c-ket} notation is due to Dirac \cite{DiracBook30}.

In the following we assume that any time a bra $\langle a\vert$ is
applied to a ket $\vert b\rangle$, the complex number $\langle a|b\rangle$
is formed, so that objects like $|b\rangle\hspace{-0.4mm}\langle a|$
generate kets when applied to kets from the left, $(|b\rangle\hspace{-0.4mm}\langle a|)|c\rangle=(\langle a|c\rangle)|b\rangle$,
and bras when applied to bras from the right, $\langle c|(|b\rangle\hspace{-0.4mm}\langle a|)=(\langle c|b\rangle)\langle a|$.
Technically, $|b\rangle\langle a|$ is called an \textit{outer product}.

A vector space equipped with an inner product is called a \textit{Euclidean
space} \cite{PrugoveckyBook71}. In the following we give some important
definitions and properties which are needed in order to understand
the concept of Hilbert space:
\begin{itemize}
\item The vectors $\left\{ \vert a_{1}\rangle,\vert a_{2}\rangle,...,\vert a_{m}\rangle\right\} $
are said to be \textit{linearly independent} if the relation $\alpha_{1}\vert a_{1}\rangle+\alpha_{2}\vert a_{2}\rangle+...+\alpha_{m}\vert a_{m}\rangle=0$
is satisfied only for $\alpha_{1}=\alpha_{2}=...=\alpha_{m}=0$, as
otherwise one of them can be written as a linear combination of the
rest.
\item The \textit{dimension} of the vector space is defined as the maximum
number of linearly independent vectors that can be found in the space,
and can be finite or infinite.
\item If the dimension of a Euclidean space is $d<\infty$, it is always
possible to build a set of $d$ orthonormal vectors $E=\left\{ \vert e_{j}\rangle\right\} _{j=1,2,..,d}$
satisfying $\langle e_{j}|e_{l}\rangle=\delta_{jl}$, such that any
other vector $\vert a\rangle$ can be written as a linear superposition
of them, that is, $\vert a\rangle=\sum_{j=1}^{d}a_{j}\vert e_{j}\rangle$,
being the $a_{j}$'s some complex numbers. This set is called an \textit{orthonormal
basis} of the Euclidean space $\mathcal{V}$, and the coefficients
$a_{j}$ of the expansion can be found as $a_{j}=\langle e_{j}|a\rangle$.
The column formed with the expansion coefficients, which is denoted
by $\text{col}(a_{1},a_{2},...,a_{d})$, is called a \textit{representation}
of the vector $\vert a\rangle$ in the basis $E$.

Note that the set $E^{+}=\left\{ \langle e_{j}\vert\right\} _{j=1,2,..,d}$
is an orthonormal basis in the dual space $\mathcal{V}^{+}$, so that
any bra $\langle a\vert$ can be expanded then as $\langle a\vert=\sum_{j=1}^{d}a_{j}^{\ast}\langle e_{j}\vert$.
The representation of the bra $\langle a\vert$ in the basis $E$
corresponds to the row formed by its expansion coefficients, and is
denoted by $(a_{1}^{\ast},a_{2}^{\ast},...,a_{n}^{\ast})$. Note that
if the representation of $\vert a\rangle$ is seen as a $d\times1$
matrix, the representation of $\langle a\vert$ can be obtained as
its $1\times d$ conjugate-transpose matrix.

Note finally that the inner product of two vectors $\vert a\rangle$
and $\vert b\rangle$ reads $\langle a|b\rangle=\sum_{j=1}^{d}a_{j}^{\ast}b_{j}$
when represented in the same basis, which is the matrix product of
the representations of $\langle a\vert$ and $\vert b\rangle$. 
\end{itemize}
For finite dimension, a Euclidean space is a \textit{Hilbert space}.
However, in most applications of quantum mechanics one has to deal
with infinite-dimensional vector spaces. We will treat them after
the following section.

\subsubsection{Linear operators in finite dimensions\label{LinearOperators}}

We now discuss the concept of linear operator, as well as analyze
the properties of some important classes of operators. Only finite-dimensional
Hilbert spaces are considered in this section, and we will generalize
the discussion to infinite-dimensional Hilbert spaces in the next
section.

We are interested in maps $\hat{L}$ (operators are denoted by a `hat'
throughout the monograph) which associate to any vector $\vert a\rangle$
of a Hilbert space $\mathcal{H}$ another vector denoted by $\hat{L}\vert a\rangle$
in the same Hilbert space. If the map satisfies
\begin{equation}
\hat{L}(\alpha\vert a\rangle+\beta\vert b\rangle)=\alpha\hat{L}\vert a\rangle+\beta\hat{L}\vert b\rangle,
\end{equation}
then it is called a \textit{linear operator}. For our purposes this
is the only class of interesting operators, and hence we will simply
call them \textit{operators} in the following.

Before discussing the properties of some important classes of operators,
we need some definitions:
\begin{itemize}
\item Given an orthonormal basis $E=\left\{ \vert e_{j}\rangle\right\} _{j=1,2,..,d}$
in a Hilbert space $\mathcal{H}$ with dimension $d<\infty$, any
operator $\hat{L}$ has a matrix representation. While bras and kets
are represented by $d\times1$ and $1\times d$ matrices (rows and
columns), respectively, an operator $\hat{L}$ is represented by a
$d\times d$ matrix with \textit{elements} $L_{jl}=(\vert e_{j}\rangle,\hat{L}\vert e_{l}\rangle)\equiv\langle e_{j}\vert\hat{L}\vert e_{l}\rangle$.
An operator $\hat{L}$ can then be expanded in terms of the basis
$E$ as $\hat{L}=\sum_{j,l=1}^{d}L_{jl}\vert e_{j}\rangle\hspace{-0.4mm}\langle e_{l}\vert$.
It follows that the representation of the vector $\vert b\rangle=\hat{L}\vert a\rangle$
is just the matrix multiplication of the representation of $\hat{L}$
by the representation of $\vert a\rangle$, that is, $b_{j}=\sum_{l=1}^{d}L_{jl}a_{l}$.
\item The \textit{addition} and \textit{multiplication} of two operators
$\hat{L}$ and $\hat{K}$, denoted by $\hat{L}+\hat{K}$ and $\hat{L}\hat{K}$,
respectively, are defined by their action onto any vector $\vert a\rangle$:
$(\hat{L}+\hat{K})\vert a\rangle=\hat{L}\vert a\rangle+\hat{K}\vert a\rangle$
and $\hat{L}\hat{K}\vert a\rangle=\hat{L}(\hat{K}\vert a\rangle)$.
It follows that the representation of the addition and the product
are, respectively, the sum and the multiplication of the corresponding
matrices, that is, $(\hat{L}+\hat{K})_{jl}=L_{jl}+K_{jl}$ and $(\hat{L}\hat{K})_{jl}=\sum_{k=1}^{d}L_{jk}K_{kl}$.
\item Note that while the addition is commutative, the product is not in
general. This leads us to the notion of \textit{commutator}, defined
for two operators $\hat{L}$ and $\hat{K}$ as $[\hat{L},\hat{K}]=\hat{L}\hat{K}-\hat{K}\hat{L}$.
When $[\hat{L},\hat{K}]=0$, we say that the operators \textit{commute}.
\item Given an operator $\hat{L}$, its \textit{trace} is defined as the
sum of the diagonal elements of its matrix representation, that is,
$\mathrm{tr}\{\hat{L}\}=\sum_{j=1}^{d}L_{jj}$. It may seem that this
definition is basis-dependent, as in general the elements $L_{jj}$
are different in different bases. However, we will see later that
the trace is invariant under any change of basis.

The trace has two important properties. It is \textit{linear} and
\textit{cyclic}, that is, given two operators $\hat{L}$ and $\hat{K}$,
$\mathrm{tr}\{\hat{L}+\hat{K}\}=\mathrm{tr}\{\hat{L}\}+\mathrm{tr}\{\hat{K}\}$
\hspace{2mm}and\hspace{2mm} $\mathrm{tr}\{\hat{L}\hat{K}\}=\mathrm{tr}\{\hat{K}\hat{L}\}$,
as is trivially proved.
\item Given an operator $\hat{L}$, we define its \textit{determinant} as
the determinant of its matrix representation, that is, $\text{det}\{\hat{L}\}=\sum_{j_{1},j_{2},...,j_{d}=1}^{d}\epsilon_{j_{1}j_{2}...j_{d}}L_{1j_{1}}L_{2j_{2}}...L_{dj_{d}}$.
Just as the trace, we will see that it does not depend on the basis
used to represent the operator.

The determinant is a multiplicative map, that is, given two operators
$\hat{L}$ and $\hat{K}$, the determinant of the product is the product
of the determinants, $\text{det}\{\hat{L}\hat{K}\}=\text{det}\{\hat{L}\}\text{det}\{\hat{K}\}$.
\item We say that a vector $\vert l\rangle$ is an \textit{eigenvector}
of an operator $\hat{L}$ if there exists a $\lambda\in\mathbb{C}$
(called its associated \textit{eigenvalue}) such that $\hat{L}\vert l\rangle=\lambda\vert l\rangle$.
The set of all the eigenvalues of an operator is called its \textit{spectrum}. 
\end{itemize}
\bigskip{}

We can pass now to describe some classes of operators which play important
roles in quantum mechanics.

\bigskip{}

\textbf{The identity operator.} The \textit{identity operator}, denoted
by $\hat{I}$, is defined as the operator which maps any vector onto
itself. Its representation in any basis is then $I_{jl}=\delta_{jl}$,
so that it can be expanded as
\begin{equation}
\hat{I}=\sum_{j=1}^{d}\vert e_{j}\rangle\hspace{-0.4mm}\langle e_{j}\vert\text{.}
\end{equation}
This expression is known as the \textit{completeness relation} of
the basis $E$; alternatively, it is said that the set $E$ forms
a \textit{resolution of the identity}.

Note that the expansion of a vector $\vert a\rangle$ and its dual
$\langle a\vert$ in the basis $E$ is obtained just by application
of the completeness relation from the left and the right, respectively.
Similarly, the expansion of an operator $\hat{L}$ is obtained by
application of the completeness relation both from the right and the
left at the same time.

\bigskip{}

\textbf{The inverse of an operator.} The \textit{inverse} of an operator
$\hat{L}$, denoted by $\hat{L}^{-1}$, is defined as that satisfying
$\hat{L}^{-1}\hat{L}=\hat{L}\hat{L}^{-1}=\hat{I}$. Not every operator
has an inverse. An inverse exists if and only if the operator does
not have a zero eigenvalue, or, equivalently, when $\det\{\hat{L}\}\neq0$.

\bigskip{}

\textbf{An operator function}. Consider a real, analytic function
$f(x)$ which can be expanded in powers of $x$ as $f(x)=\sum_{m=0}^{\infty}f_{m}x^{m}$.
Given an operator $\hat{L}$, we define the \textit{operator} \textit{function}
$\hat{f}(\hat{L})=\sum_{m=0}^{\infty}f_{m}\hat{L}^{m}$, where $\hat{L}^{m}$
means the product of $\hat{L}$ with itself $m$ times.

\bigskip{}

\textbf{The adjoint of an operator.} Given an operator $\hat{L}$,
we define its \textit{adjoint}, and denote it by $\hat{L}^{\dagger}$,
as that satisfying $(\vert a\rangle,\hat{L}\vert b\rangle)=(\hat{L}^{\dagger}\vert a\rangle,\vert b\rangle)$
for any two vectors $\vert a\rangle$ and $\vert b\rangle$. Note
that the representation of $\hat{L}^{\dagger}$ corresponds to the
conjugate transpose of the matrix representing $\hat{L}$, that is,
$(\hat{L}^{\dagger})_{jl}=L_{lj}^{\ast}$. Note also that the adjoint
of a product of two operators $\hat{K}$ and $\hat{L}$ is given by
$(\hat{K}\hat{L})^{\dagger}=\hat{L}^{\dagger}\hat{K}^{\dagger}$.

\bigskip{}

\textbf{Self-adjoint operators.} We say that $\hat{H}$ is a \textit{self-adjoint}
when it coincides with its adjoint, that is, $\hat{H}=\hat{H}^{\dagger}$.
A property of major importance for the construction of the laws of
quantum mechanics is that the spectrum $\left\{ h_{j}\right\} _{j=1,2,...,d}$
of a self-adjoint operator is real. Moreover, its associated eigenvectors\footnote{For simplicity, we will assume that the spectrum of any operator is
non-degenerate, that is, all the eigenvectors possess a distinctive
eigenvalue.} $\left\{ \vert h_{j}\rangle\right\} _{j=1,2,...,d}$ form an orthonormal
basis of the Hilbert space.

The representation of any operator function $\hat{f}(\hat{H})$ in
the \textit{eigenbasis} of $\hat{H}$ is then $[\hat{f}(\hat{H})]_{jl}=f(h_{j})\delta_{jl}$,
from which it follows
\begin{equation}
\hat{f}(\hat{H})=\sum_{j=1}^{d}f(h_{j})\vert h_{j}\rangle\hspace{-0.4mm}\langle h_{j}\vert.
\end{equation}
This result is known as the \textit{spectral theorem}.

\bigskip{}

\textbf{Unitary operators.} We say that $\hat{U}$ is a \textit{unitary
operator} when $\hat{U}^{\dagger}=\hat{U}^{-1}$. The interest of
this class of operators is that they preserve inner products, that
is, for any two vectors $\vert a\rangle$ and $\vert b\rangle$ the
inner product $(\hat{U}\vert a\rangle,\hat{U}\vert b\rangle)$ coincides
with $\langle a|b\rangle$. Moreover, it is possible to show that
given two orthonormal bases $E=\left\{ \vert e_{j}\rangle\right\} _{j=1,2,..,d}$
and $E^{\prime}=\{|e_{j}^{\prime}\rangle\}_{j=1,2,..,d}$, there exists
a unique unitary matrix $\hat{U}$ which connects them as $\{|e_{j}^{\prime}\rangle=\hat{U}\vert e_{j}\rangle\}_{j=1,2,..,d}$,
and then any basis of the Hilbert space is unique up to a unitary
transformation.

We can now prove that both the trace and the determinant of an operator
are basis-independent. Let us denote by $\mathrm{tr}\{\hat{L}\}_{E}$
the trace of an operator $\hat{L}$ in the basis $E$. The trace of
this operator in the transformed basis can be written then as $\mathrm{tr}\{\hat{L}\}_{E^{\prime}}=\mathrm{tr}\{\hat{U}^{\dagger}\hat{L}\hat{U}\}_{E}$,
which, using the cyclic property of the trace and the unitarity of
$\hat{U}$, is rewritten as $\mathrm{tr}\{\hat{U}\hat{U}^{\dagger}\hat{L}\}_{E}=\mathrm{tr}\{\hat{L}\}_{E}$,
proving that the trace is equal in both bases. Similarly, in the case
of the determinant we have $\text{det}\{\hat{L}\}_{E'}=\text{det}\{\hat{U}^{\dagger}\hat{L}\hat{U}\}_{E}$,
which using the multiplicative property of the determinant is rewritten
as $\text{det}\{\hat{U}^{\dagger}\}_{E}\text{det}\{\hat{L}\}_{E}\text{det}\{\hat{U}\}_{E}=\text{det}\{\hat{L}\}_{E}$,
where we have used $\text{det}\{\hat{U}^{\dagger}\}_{E}\text{det}\{\hat{U}\}_{E}=1$
as follows from $\hat{U}^{\dagger}\hat{U}=\hat{I}$.

Note finally that a unitary operator $\hat{U}$ can always be written
as the complex exponential a self-adjoint operator $\hat{H}$, that
is, $\hat{U}=\exp(\mathrm{i}\hat{H})$.

\bigskip{}

\textbf{Projection operators.} In general, any self-adjoint operator
$\hat{P}$ satisfying $\hat{P}^{2}=\hat{P}$ is called a \textit{projector}.
We are interested only on those projectors which can be written as
the outer product of a vector $\vert a\rangle$ with itself, that
is, rank-1 projectors\footnote{The term ``rank'' refers to the number of non-zero eigenvalues.}
$\hat{P}_{a}=\vert a\rangle\hspace{-0.4mm}\langle a\vert$. When applied
to a vector $\vert b\rangle$, this gets \textit{projected} along
the `direction' of $\vert a\rangle$ as $\hat{P}_{a}\vert b\rangle=\langle a|b\rangle\hspace{1mm}\vert a\rangle$.

Note that given an orthonormal basis $E$, we can use the projectors
$\hat{P}_{j}=\vert e_{j}\rangle\hspace{-0.4mm}\langle e_{j}\vert$
to extract the components of a vector $\vert a\rangle$ as $\hat{P}_{j}\vert a\rangle=a_{j}\vert e_{j}\rangle$.
Note also that the completeness and orthonormality of the basis $E$
implies that $\sum_{j=1}^{d}\hat{P}_{j}=\hat{I}$ and $\hat{P}_{j}\hat{P}_{l}=\delta_{jl}\hat{P}_{j}$,
respectively.

\bigskip{}

\textbf{Density operators.} A self-adjoint operator $\hat{\rho}$
is called a \textit{density operator} when it has unit trace and it
is \textit{positive semidefinite}, that is, $\langle a\vert\hat{\rho}\vert a\rangle\geq0$
for any vector $\vert a\rangle$.

The interesting property of density operators is that they `contain'
probability distributions in the diagonal of its representation. To
see this just note that given an orthonormal basis $E$, the self-adjointness
and positivity of $\hat{\rho}$ ensure that all its diagonal elements
$\left\{ \rho_{jj}\right\} _{j=1,2,...,d}\ $are either positive or
zero, that is, $\rho_{jj}\geq0$ $\forall j$, while the unit trace
makes them satisfy $\sum_{j=1}^{d}\rho_{jj}=1$. Hence, the diagonal
elements of a density operator have all the properties required by
a \textit{probability distribution}.

\subsubsection{Generalization to infinite dimensions\label{InfiniteHilbert}}

Unfortunately, not all the previous concepts and objects that we have
introduced for the finite-dimensional case are trivially generalized
to infinite dimensions. In this section we discuss this generalization.

The first problem that we meet when dealing with infinite-dimensional
Euclidean spaces is that the existence of a basis $\left\{ \vert e_{j}\rangle\right\} _{j=1,2,...}$
in which any other vector can be represented as $\vert a\rangle=\sum_{j=1}^{\infty}a_{j}\vert e_{j}\rangle$
is not granted, because not all infinite sequences converge. The class
of infinite-dimensional Euclidean spaces in which these infinite but
countable bases exist are called \textit{separable Hilbert spaces},
and are the ones relevant for the quantum description of physical
systems.

The conditions that ensure that an infinite-dimensional Euclidean
space is indeed a Hilbert space\footnote{From now on we will assume that all the Hilbert spaces we refer to
are ``separable'', even if we don't write it explicitly.} can be found in, for example, reference \cite{PrugoveckyBook71}.
Here we just want to stress that, quite intuitively, any infinite-dimensional
Hilbert space\footnote{An example of infinite-dimensional complex Hilbert space consists
in the vector space formed by the complex functions of real variable,
say $|f\rangle=f(x)$ with $x\in\mathbb{R}$, with integrable square,
that is 
\begin{equation}
\int_{\mathbb{R}}dx|f(x)|^{2}<\infty,
\end{equation}
together with the inner product 
\begin{equation}
\langle g|f\rangle=\int_{\mathbb{R}}dxg^{\ast}(x)f(x).
\end{equation}
This Hilbert space is usually denoted by $\mathrm{L}^{2}(x)$.} is \textit{isomorphic} to the space called $l^{2}(\infty)$, which
is formed by the column vectors $\vert a\rangle=\text{col}(a_{1},a_{2},...)$
where the set $\{a_{j}\in\mathbb{C}\}_{j=1,2,...}$ satisfies the
restriction $\sum_{j=1}^{\infty}|a_{j}|^{2}<\infty$, and has the
operations $\vert a\rangle+\vert b\rangle=\text{col}(a_{1}+b_{1},a_{2}+b_{2},...)$,
$\alpha\vert a\rangle=\text{col}(\alpha a_{1},\alpha a_{2},...)$,
and $\langle a|b\rangle=\sum_{j=1}^{\infty}a_{j}^{\ast}b_{j}$.

Most of the previous definitions are directly generalized to Hilbert
spaces by taking $d\rightarrow\infty$ (dual space, representations,
operators,...). However, there is one crucial property of self-adjoint
operators which does not hold in this case: its eigenvectors may not
form an orthonormal basis in the Hilbert space. The remainder of this
section is devoted to deal with this problem.

Just as in finite dimension, given an infinite-dimensional Hilbert
space $\mathcal{H}$, we say that one of its vectors $|d\rangle$
is an eigenvector of the self-adjoint operator $\hat{H}$ if $\hat{H}|d\rangle=\delta|d\rangle$,
where $\delta\in\mathbb{R}$ is called its associated eigenvalue.
Nevertheless, it can happen in infinite-dimensional spaces that some
vector $|c\rangle$ not contained in $\mathcal{H}$ also satisfies
the condition $\hat{H}|c\rangle=\chi|c\rangle$, in which case we
call it a \textit{generalized eigenvector}, being $\chi$ its \textit{generalized
eigenvalue}\footnote{In $\mathrm{L}^{2}(x)$ we have two simple examples of self-adjoint
operators with eigenvectors not contained in $\mathrm{L}^{2}(x)$:
the so-called $\hat{X}$ (\textit{position}) and $\hat{P}$ (\textit{momentum}),
which, given an arbitrary vector $|f\rangle=f(x)$, act as $\hat{X}|f\rangle=xf(x)$
and $\hat{P}|f\rangle=-\mathrm{i}\partial_{x}f$, respectively. This
is simple to see, as the equations
\begin{equation}
xf_{X}(x)=Xf_{X}(x)\hspace{0.6cm}\text{and}\hspace{0.6cm}-\mathrm{i}\partial_{x}f_{P}(x)=Pf_{P}(x),
\end{equation}
have 
\begin{equation}
f_{X}(x)=\delta(x-X)\hspace{0.6cm}\text{and}\hspace{0.6cm}f_{P}(x)=\exp(\mathrm{i}Px),
\end{equation}
as solutions, which are not square-integrable, and hence do not belong
to $\mathrm{L}^{2}(x)$.}. The set of all the eigenvalues of the self-adjoint operator is called
its \textit{discrete }(or \textit{point})\textit{ spectrum} and it
is a countable set, while the set of all its generalized eigenvalues
is called its \textit{continuous spectrum} and it is uncountable,
that is, forms a continuous set \cite{PrugoveckyBook71} (see also
\cite{GalindoBook90}).

In this lectures we only deal with two extreme cases: either the observable,
say $\hat{H}$, has a pure discrete spectrum $\{h_{j}\}_{j=1,2,...}$;
or the observable, say $\hat{X}$, has a pure continuous spectrum
$\{x\}_{x\in\mathbb{R}}$. It can be shown that in the first case
the eigenvectors of the observable form an orthonormal basis of the
Hilbert space, so that we can build a resolution of the identity as
$\hat{I}=\sum_{j=1}^{\infty}\vert h_{j}\rangle\hspace{-0.4mm}\langle h_{j}\vert$,
and proceed along the lines of the previous sections.

In the second case, the set of generalized eigenvectors cannot form
a basis of the Hilbert space in the strict sense, as they do not form
a countable set and do not even belong to the Hilbert space. Fortunately,
there are still ways to treat the generalized eigenvectors of $\hat{X}$
`as if' they were a basis of the Hilbert space. The idea was introduced
by Dirac \cite{DiracBook30}, who realized that normalizing the generalized
eigenvectors as\footnote{$\delta(x)$ is the so-called \textit{Dirac delta distribution} which
is defined by the conditions 
\begin{equation}
\int_{x_{1}}^{x_{2}}dx\delta(x-y)=\left\{ \begin{array}{cc}
1 & \text{if }y\in\left[x_{1},x_{2}\right]\\
0 & \text{if }y\notin\left[x_{1},x_{2}\right]
\end{array}\right..
\end{equation}
} $\langle x|y\rangle=\delta(x-y)$, one can define the following integral
operator 
\begin{equation}
\hat{I}_{\mathrm{c}}=\int_{\mathbb{R}}dx|x\rangle\hspace{-0.4mm}\langle x\vert,
\end{equation}
which acts as the identity onto the generalized eigenvectors, that
is, $\hat{I}_{\mathrm{c}}|y\rangle=|y\rangle$. It is then conjectured
that $\hat{I}_{\mathrm{c}}$ coincides with the identity in $\mathcal{H}$,
so that any other vector $|a\rangle$ or operator $\hat{L}$ defined
in the Hilbert space can be expanded as 
\begin{equation}
|a\rangle=\int_{\mathbb{R}}dxa(x)|x\rangle\hspace{8mm}\text{and}\hspace{8mm}\hat{L}=\int_{\mathbb{R}^{2}}dxdyL(x,y)|x\rangle\hspace{-0.4mm}\langle y\vert,
\end{equation}
where the elements $a(x)=\langle x|a\rangle$ and $L(x,y)=\langle x|\hat{L}|y\rangle$
of these \textit{continuous representations} form complex functions
defined in $\mathbb{R}$ and $\mathbb{R}^{2}$, respectively. From
now on, we will call \textit{continuous basis} to the set $\left\{ |x\rangle\right\} _{x\in\mathbb{R}}$.

Dirac introduced this continuous representation as a `limit to the
continuum' of the countable case. Even though this approach was very
intuitive, it lacked mathematical rigor. Some decades after Dirac's
proposal, Gel'fand showed how to generalize the concept of Hilbert
space to include these generalized representations in full mathematical
rigor \cite{GelfandBook64}. The generalized spaces are called \textit{rigged
Hilbert spaces} (in which the algebra of Hilbert spaces joins forces
with the theory of continuous probability distributions), and working
on them it is possible to show that given any self-adjoint operator,
one can use its eigenvectors and generalized eigenvectors to expand
any vector of the Hilbert space, just as we did above. In other words,
within the framework of rigged Hilbert spaces, one can prove the identity
$\hat{I}_{c}=\hat{I}$ rigorously.

Note finally that given two vectors $|a\rangle$ and $|b\rangle$
of the Hilbert space, and a continuous basis $\left\{ |x\rangle\right\} _{x\in\mathbb{R}}$,
we can use their generalized representations to write their inner
product as
\begin{equation}
\langle a|b\rangle=\int_{\mathbb{R}}dxa^{\ast}(x)b(x).
\end{equation}
It is also easily proved that the trace of any operator $\hat{L}$
can be evaluated from its continuous representation on $\left\{ |x\rangle\right\} _{x\in\mathbb{R}}$
as
\begin{equation}
\mathrm{tr}\{\hat{L}\}=\int_{\mathbb{R}}dxL(x,x).
\end{equation}
This has important consequences for the properties of density operators,
say $\hat{\rho}$ for the discussion which follows. We explained at
the end of the last section that when represented on an orthonormal
basis of the Hilbert space, its diagonal elements (which are real
owed to its self-adjointness) can be seen as a probability distribution,
because they satisfy $\sum_{j=1}^{\infty}\rho_{jj}=1$ and $\rho_{jj}\geq0\hspace{2mm}\forall j$.
Similarly, because of its unit trace and positivity, the diagonal
elements of its continuous representation satisfy $\int_{\mathbb{R}}dx\rho(x,x)=1$
and $\rho(x,x)\geq0\hspace{2mm}\forall x$, and hence, the real function
$\rho(x,x)$ can be seen as a \textit{probability density function}.

\subsubsection{Composite Hilbert spaces\label{CompositeHilbertSpaces}}

In many moments of this lectures, we will find the need to associate
a Hilbert space to a composite system, the Hilbert spaces of whose
parts we know. In this section we show how to build a Hilbert space
$\mathcal{H}$ starting from a set of Hilbert spaces $\left\{ \mathcal{H}_{A},\mathcal{H}_{B},\mathcal{H}_{C},...\right\} $.

Let us start with only two Hilbert spaces $\mathcal{H}_{A}$ and $\mathcal{H}_{B}$
with dimensions $d_{A}$ and $d_{B}$, respectively (which might be
infinite); the generalization to an arbitrary number of Hilbert spaces
is straightforward. Consider a vector space $\mathcal{V}$ with dimension
$\mathrm{dim}(\mathcal{V})=d_{A}\times d_{B}$. We define a map called
the \textit{tensor product} which associates to any pair of vectors
$|a\rangle\in\mathcal{H}_{A}$ and $|b\rangle\in\mathcal{H}_{B}$
a vector in $\mathcal{V}$ which we denote by $|a\rangle\otimes|b\rangle\in\mathcal{V}$.
This tensor product must satisfy the following properties:
\begin{enumerate}
\item $(|a\rangle+|b\rangle)\otimes|c\rangle=|a\rangle\otimes|c\rangle+|b\rangle\otimes|c\rangle$
\item $|a\rangle\otimes(|b\rangle+|c\rangle)=|a\rangle\otimes|b\rangle+|a\rangle\otimes|c\rangle$
\item $(\alpha|a\rangle)\otimes|b\rangle=|a\rangle\otimes(\alpha|b\rangle)$ 
\end{enumerate}
If we endorse the vector space $\mathcal{V}$ with the inner product
$(|a\rangle\otimes|b\rangle,|c\rangle\otimes|d\rangle)=\langle a|c\rangle\hspace{-0.4mm}\langle b|d\rangle$,
it is easy to show it becomes a Hilbert space, which in the following
will be denoted by $\mathcal{H}=\mathcal{H}_{A}\otimes\mathcal{H}_{B}$.
Given the bases $E_{A}=\{|e_{j}^{A}\rangle\}_{j=1,2,...,d_{A}}$ and
$E_{B}=\{|e_{j}^{B}\rangle\}_{j=1,2,...,d_{B}}$ of the Hilbert spaces
$\mathcal{H}_{A}$ and $\mathcal{H}_{B}$, respectively, a basis of
the \textit{tensor product Hilbert space }$\mathcal{H}_{A}\otimes\mathcal{H}_{B}$
can be built as $E=E_{A}\otimes E_{B}=\{|e_{j}^{A}\rangle\otimes|e_{l}^{B}\rangle\}_{l=1,2,...,d_{B}}^{j=1,2,...,d_{A}}$
(note that the notation after the first equality is symbolic).

We may use a more economic notation for the tensor product, namely
$|a\rangle\otimes|b\rangle=|a,b\rangle$, except when the explicit
tensor product symbol is needed for some special reason. With this
notation the basis of the tensor product Hilbert space is written
as $E=\{|e_{j}^{A},e_{l}^{B}\rangle\}_{l=1,2,...,d_{B}}^{j=1,2,...,d_{A}}$.

The tensor product also maps operators acting on $\mathcal{H}_{A}$
and $\mathcal{H}_{B}$ to operators acting on $\mathcal{H}$. Given
two operators $\hat{L}_{A}$ and $\hat{L}_{B}$ acting on $\mathcal{H}_{A}$
and $\mathcal{H}_{B}$, the \textit{tensor product operator} $\hat{L}=\hat{L}_{A}\otimes\hat{L}_{B}$
is defined in $\mathcal{H}$ as that satisfying $\hat{L}|a,b\rangle=(\hat{L}_{A}|a\rangle)\otimes(\hat{L}_{B}|b\rangle)$
for any pair of vectors $|a\rangle\in\mathcal{H}_{A}$ and $|b\rangle\in\mathcal{H}_{B}$.
When explicit subindices making reference to the Hilbert space on
which operators act on are used, so that there is no room for confusion,
we will use the shorter notations $\hat{L}_{A}\otimes\hat{L}_{B}=\hat{L}_{A}\hat{L}_{B}$,
$\hat{L}_{A}\otimes\hat{I}=\hat{L}_{A}$, and $\hat{I}\otimes\hat{L}_{B}=\hat{L}_{B}$.

Note that the tensor product preserves the properties of the operators;
for example, given two self-adjoint operators $\hat{H}_{A}$ and $\hat{H}_{B}$,
unitary operators $\hat{U}_{A}$ and $\hat{U}_{B}$, or density operators
$\hat{\rho}_{A}$ and $\hat{\rho}_{B}$, the operators $\hat{H}_{A}\otimes\hat{H}_{B}$,
$\hat{U}_{A}\otimes\hat{U}_{B}$, and $\hat{\rho}_{A}\otimes\hat{\rho}_{B}$
are self-adjoint, unitary, and a density operator acting on $\mathcal{H}$,
respectively. But keep in mind that this does not mean that all self-adjoint,
unitary, or density operators acting on $\mathcal{H}$ can be written
in a simple tensor product form $\hat{L}_{A}\otimes\hat{L}_{B}$.

\subsection{The quantum-mechanical framework\label{Axioms}}

In this section we review the basic postulates that describe how quantum
mechanics treats physical systems. As the building blocks of the theory,
these principles cannot be \textit{proved}. They can only be formulated
following \textit{plausibility arguments} based on the \textit{observation}
of physical phenomena and the \textit{connection} of the theory with
previous theories which are known to work in some limit. We will try
to motivate (and justify to a point) these principles as much as possible,
starting with a brief historical introduction to the context in which
they were created\footnote{For a thorough historical overview of the birth of quantum physics
see \cite{WhitakerBook}.}.

\subsubsection{A brief historical introduction}

By the end of the XIX century there was a great feeling of safety
and confidence among the physics community: analytical mechanics (together
with statistical mechanics) and Maxwell's electromagnetism (in the
following \textit{classical physics} altogether) seem to explain the
whole range of physical phenomena that one could observe, and hence,
in a sense, the foundations of physics were complete. There were,
however, a couple of experimental observations which lacked explanation
within this `definitive' framework, which actually led to the construction
of a whole new way of understanding physical phenomena: quantum mechanics.

Among these experimental evidences, the shape of the high-energy spectrum
of the radiation emitted by a black body, the photoelectric effect
which showed that only light exceeding some frequency can release
electrons from a metal irrespective of its intensity, and the discrete
set of spectral lines of hydrogen, were the principal triggers of
the revolution to come in the first quarter of the XX century. The
first two led Planck and Einstein, respectively, to suggest that electromagnetic
energy is not continuous but divided in small packets of energy $\hbar\omega$
($\omega$ being the angular frequency of the radiation), while Bohr
succeeded in explaining the Hydrogen spectrum by assuming that the
electron orbiting the nucleus can occupy only a discrete set of orbits
with angular momenta proportional to $\hbar$. The constant $\hbar=h/2\pi\sim10^{-34}\mathrm{J\cdot s}$,
where $h$ is now known as the Planck constant, appeared in both cases
as somehow the `quantization unit', the value separating the quantized
values that energy or angular momentum are able to take.

Even though the physicists of the time tried to understand this quantization
of the physical magnitudes within the framework of classical physics,
it was soon realized that a completely new theory was required. The
first attempts to build such a theory (which actually worked for some
particular scenarios) were based on applying ad-hoc quantization rules
to various mechanical variables of systems, but with a complete lack
of physical interpretation for such rules \cite{Waerden68}. However,
between 1925 and 1927 the first real formulations of the needed theory
were developed: the \textit{wave mechanics} of Schrödinger \cite{Schrodinger26}
and the \textit{matrix mechanics} of Heisenberg, Born and Jordan \cite{Heisenberg25,Born25,Born26}
(see \cite{Waerden68} for English translations), which received also
independent contributions by Dirac \cite{Dirac26}. Even though in
both theories the quantization of various observable quantities appeared
naturally and in correspondence with experiments, they seemed completely
different, at least until Schrödinger showed the equivalence between
them both.

The new theory was later formalized mathematically using vector spaces
by Dirac \cite{DiracBook30} (though not entirely rigorously), and
a little later by von Neumann with more mathematical rigor using Hilbert
spaces \cite{vonNeumann32} (\cite{vonNeumann55} for an English version).
They developed the laws of \textit{quantum mechanics} basically as
we know them today \cite{CohenTannoudjiBookI,CohenTannoudjiBookII,GreinerQuantumBook1,GreinerQuantumBook2,Basdevant02,Ballentine98}.
In the next sections we will introduce these rules in the form of
six principles that will set the structure of the theory of quantum
mechanics as we will use it along the lectures.

\subsubsection{Principle I: Observables and measurement outcomes}

The experimental evidence for the quantization of some observable
physical quantities motivates the first principle:

\bigskip{}

\textbf{Principle I. }$\lceil$Any physical observable quantity \textsf{A}
corresponds to a self-adjoint operator $\hat{A}$ acting on an abstract
Hilbert space. After a measurement of \textsf{A}, the only possible
outcomes are the eigenvalues of $\hat{A}$.$\rfloor$

\bigskip{}

The quantization of physical observables is therefore directly introduced
within the theory by this postulate. Note that it does not say anything
about the dimension $d$ of the Hilbert space corresponding to a given
observable, and it even leaves open the possibility of observables
having a continuous spectrum, rather than a discrete one. The problem
of how to make the proper correspondence between observables and self-adjoint
operators will be addressed in an principle to come.

In this lectures we use the name `observable' both for the physical
quantity \textsf{A} and its associated self-adjoint operator $\hat{A}$
indistinctly. Observables having purely-discrete or purely-continuous
spectra will be referred to as \textit{countable} and \textit{continuous
observables}, respectively.

\subsubsection{Principle II: State of the system and statistics of measurements}

\label{AxiomII}

The next principle follows from the following question: according
to the previous principle the eigenvalues of an observable are the
only values that can appear when measuring it, but what about the
statistics of such a measurement? We know a class of operators in
Hilbert spaces which act as probability distributions for the eigenvalues
of any self-adjoint operator, density operators. This motivates the
second principle:

\bigskip{}

\textbf{Principle II. }$\lceil$The state of the system is completely
specified by a density operator $\hat{\rho}$. When measuring a countable
observable \textsf{A} with eigenvectors $\{|a_{j}\rangle\}_{j=1,2,...,d}$
($d$ might be infinite), it is associated to the possible outcomes
$\{a_{j}\}_{j=1,2,...,d}$ a probability distribution $\{p_{j}=\rho_{jj}\}_{j=1,2,...,d}$
which determines the statistics of the experiment (\textit{Born rule}).
Similarly, when measuring a continuous observable \textsf{X} with
eigenvectors $\{|x\rangle\}_{x\in\mathbb{R}}$, the probability density
function $P(x)=\rho(x,x)$ is associated to the possible outcomes
$\{x\}_{x\in\mathbb{R}}$ in the experiment.$\rfloor$

\bigskip{}

This principle has deep consequences that we analyze now. Contrary
to classical mechanics (and intuition), even if the system is prepared
in a given state, the value of an observable is in general not well
defined. We can only specify with what probability a given value of
the observable will come out in a measurement. Hence, this principle
proposes a change of paradigm; determinism must be abandoned: the
theory is no longer able to predict with certainty the outcome of
a single run of an experiment in which an observable is measured,
but rather gives the statistics that will be extracted after a large
number of runs.

To be fair, there is a case in which the theory allows us to predict
the outcome of the measurement of an observable with certainty: When
the system is prepared such that its state is an eigenvector of the
observable. This seems much like when in classical mechanics the system
is prepared with a given value of its observables. However, we will
show that it is impossible to find a common eigenvector to \textit{all}
the available observables of a system, and hence the difference between
classical and quantum mechanics is that in the later it is impossible
to prepare the system in a state which would allow us to predict with
certainty the outcome of a measurement of each of its observables.
Let us try to elaborate on this in a more rigorous fashion.

Let us define the \textit{expectation value} of a given operator $\hat{B}$
as
\begin{equation}
\langle\hat{B}\rangle=\mathrm{tr}\{\hat{\rho}\hat{B}\}\text{.}
\end{equation}
In the case of a countable observable $\hat{A}$ or a continuous observable
$\hat{X}$, this expectation value can be written in their own eigenbases
as
\begin{equation}
\langle\hat{A}\rangle=\sum_{j=1}^{d}\rho_{jj}a_{j}\hspace{6mm}\text{and}\hspace{6mm}\langle\hat{X}\rangle=\int_{-\infty}^{+\infty}dx\rho(x,x)x\text{,}
\end{equation}
which correspond to the mean value of the outcomes registered in large
number of measurements of the observables. We define also the \textit{variance}
$V(A)$ of the observable as the expectation value of the square of
its \textit{fluctuation operator }$\delta\hat{A}=\hat{A}-\langle\hat{A}\rangle$,
that is,
\begin{equation}
V(A)=\mathrm{tr}\{\hat{\rho}(\delta\hat{A})^{2}\}=\langle\hat{A}^{2}\rangle-\langle\hat{A}\rangle^{2}\text{,}
\end{equation}
from which we obtain the \textit{standard deviation} or \textit{uncertainty}
as $\Delta A=\sqrt{V(A)}$, which measures how much the outcomes of
the experiment deviate from the mean, and hence, somehow specifies
how `well defined' is the value of the observable \textsf{A}.

Note that the probability of obtaining the outcome $a_{j}$ when measuring
\textsf{A} can be written as the expectation value of the projection
operator $\hat{P}_{j}=|a_{j}\rangle\hspace{-0.4mm}\langle a_{j}|$,
that is $p_{j}=\langle\hat{P}_{j}\rangle$. Similarly, the probability
density function associated to the possible outcomes $\{x\}_{x\in\mathbb{R}}$
when measuring \textsf{X} can be written as $P(x)=\langle\hat{P}(x)\rangle$,
where $\hat{P}(x)=|x\rangle\hspace{-0.4mm}\langle x|$.

Having written all these objects (probabilities, expectation values,
and variances) in terms of traces is really useful, since the trace
is invariant under basis changes, and hence can be evaluated in any
basis we want to work with, see Section \ref{LinearOperators}.

These principles have one further counterintuitive consequence. It
is possible to prove that irrespectively of the state of the system,
the following relation between the variances of two non-commuting
observables \textsf{A} and \textsf{B} is satisfied:
\begin{equation}
\Delta A\Delta B\geq\frac{1}{2}|\langle[\hat{A},\hat{B}]\rangle|\text{.}
\end{equation}
According to this inequality, known as the \textit{uncertainty principle}
(which was first derived by Heisenberg), in general, the only way
in which the observable \textsf{A} can be perfectly defined ($\Delta A\rightarrow0$)
is by making completely undefined observable \textsf{B} ($\Delta B\rightarrow\infty$),
or vice-versa. Hence, in the quantum formalism one cannot, in general,
prepare the system in a state in which all its observables are well
defined, what is completely opposite to our everyday experience.

Before moving to the third principle, let us comment on a couple more
things related to the state of the system. It is possible to show
that a density operator can always be expressed as a \textit{statistical}
or \textit{convex mixture} of projection operators, that is, $\hat{\rho}=\sum_{m=1}^{M}w_{m}\vert\varphi_{m}\rangle\hspace{-0.4mm}\langle\varphi_{m}\vert$,
where $\{w_{m}\}_{m=1,2,...,M}$ is a probability distribution and
the vectors $\left\{ \vert\varphi_{m}\rangle\right\} _{m=1}^{M}$
are normalized to one, but do not need to be orthogonal (note that
in fact $M$ does not need to be equal to $d$). Hence, another way
of specifying the state of the system is by a set of normalized vectors
together with some statistical rule for mixing them, that is, the
set $\{w_{m},|\varphi_{m}\rangle\}_{m=1,2,...,M}$, known as an \textit{ensemble
decomposition} of the state $\hat{\rho}$. Such decompositions are
not unique, in the sense that different ensembles can lead to the
same $\hat{\rho}$. It can be proved though \cite{NielsenChuangBook}
that two ensembles $\{w_{m},|\varphi_{m}\rangle\}_{m=1,2,...,M}$
and $\{v_{n},|\psi_{n}\rangle\}_{n=1,2,...,N}$ (we take $M\leq N$
for definiteness) give rise to the same density operator $\hat{\rho}$
if and only if there exists a left-unitary matrix\footnote{$U$ is left-unitary if $U^{\dagger}U=I$ but $UU^{\dagger}$ might
not be $I$, where $I$ is the identity matrix of the corresponding
dimension. It is easy to prove that finite-dimensional left-unitary
matrices are unitary.} $U$ with elements $\{U_{mn}\}_{m,n=1,2,...,N}$ such that \cite{NielsenChuangBook}
\begin{equation}
\sqrt{w_{m}}|\varphi_{m}\rangle=\sum_{n=1}^{N}U_{mn}\sqrt{v_{n}}|\psi_{n}\rangle,\hspace{8mm}m=1,2,...,N,
\end{equation}
where if $M\neq N$, $N-M$ zeros must be included in the ensemble
with less states, so that $\mathcal{U}$ is a square matrix.

When only one vector $\vert a\rangle$ contributes to the mixture,
$\hat{\rho}=|\varphi\rangle\hspace{-0.4mm}\langle\varphi|$ is completely
specified by just this single vector, and we say that the density
operator is \textit{pure}; otherwise, we say that it is \textit{mixed}.
A necessary and sufficient condition for $\hat{\rho}$ to be pure
is $\hat{\rho}^{2}=\hat{\rho}$. Along the lectures we will learn
that the mixedness of a state always comes from the fact that some
of the information of the system has been lost to some other inaccessible
system with which it has interacted for a while before becoming isolated
or is interacting continuously. In other words, the state of a system
is pure only when it has no correlations at all with other systems.

Note, finally, that when the state of the system is in a \textit{pure
state} $|\psi\rangle$, the expectation value of an operator $\hat{B}$
takes the simple form $\langle\psi|\hat{B}|\psi\rangle$. Moreover,
the pure state can be expanded in the countable and continuous bases
of two observables $\hat{A}$ and $\hat{X}$ as
\begin{equation}
\vert\psi\rangle=\sum_{j=1}^{d}\psi_{j}\vert a_{j}\rangle\hspace{6mm}\text{and}\hspace{6mm}\vert\psi\rangle=\int_{-\infty}^{+\infty}dx\psi(x)|x\rangle,
\end{equation}
respectively, being $\psi_{j}=\langle a_{j}|\psi\rangle$ and $\psi(x)=\langle x|\psi\rangle$.
In this case, the probability distribution for the discrete outcomes
$\{a_{j}\}_{j=1,2,...,d}$ is given by $\{p_{j}=|\psi_{j}|^{2}\}_{j=1,2,...,d}$,
while the probability density function for the continuous outcomes
$\{x\}_{x\in\mathbb{R}}$ is given by $P(x)=|\psi(x)|^{2}$.

\subsubsection{Principle III: Composite systems}

The next principle specifies how the theory accommodates dealing with
composite systems within its mathematical framework. Of course, a
composition of two systems is itself another system subject to the
laws of quantum mechanics; the question is how can we construct it.

\bigskip{}

\textbf{Principle III. }$\lceil$Consider two systems A and B with
associated Hilbert spaces $\mathcal{H}_{A}$ and $\mathcal{H}_{B}$.
Then, the state of the composite system $\hat{\rho}_{AB}$ as well
as its observables act onto the tensor product Hilbert space $\mathcal{H}_{AB}=\mathcal{H}_{A}\otimes\mathcal{H}_{B}$.$\rfloor$

\bigskip{}

This principle has the following consequence. Imagine that the systems
\textit{A} and \textit{B} interact during some time in such a way
that they cannot be described anymore by independent states $\hat{\rho}_{A}$
and $\hat{\rho}_{B}$ acting on $\mathcal{H}_{A}$ and $\mathcal{H}_{B}$,
respectively, but by a state $\hat{\rho}_{AB}$ acting on the joint
space $\mathcal{H}_{AB}$. After the interaction, system \textit{B}
is kept isolated from any other system, but system \textit{A} is given
to an observer, who is therefore able to measure observables defined
in $\mathcal{H}_{A}$ only, and might not even know that system \textit{A}
is part of a larger system. The question is, is it possible to reproduce
the statistics of the measurements performed on system \textit{A}
with some state $\hat{\rho}_{A}$ acting on $\mathcal{H}_{A}$ only?
This question has a positive and \textit{unique} answer: this state
is given by the \textit{reduced density operator} $\hat{\rho}_{A}=\mathrm{tr}_{B}\{\hat{\rho}_{AB}\}$,
that is, by performing the partial trace\footnote{Given an orthonormal basis $\{|b_{j}\rangle\}_{j}$ of $\mathcal{H}_{B}$,
this is defined by
\begin{equation}
\mathrm{tr}_{B}\{\hat{\rho}_{AB}\}=\sum_{j}\langle b_{j}|\hat{\rho}_{AB}|b_{j}\rangle,
\end{equation}
which is indeed an operator acting on $\mathcal{H}_{A}$.} with respect system's \textit{B} subspace onto the joint state.

\subsubsection{Principle IV: Quantization rules}

\label{Axiom4}

The introduction of the fourth principle is motivated by the following
fact. The class of self-adjoint operators forms a real vector space
with respect to the addition of operators and the multiplication of
an operator by a real number. Using the commutator we can also build
an operation that takes two self-adjoint operators $\hat{A}$ and
$\hat{B}$ to generate another self-adjoint operator $\hat{C}=\mathrm{i}[\hat{A},\hat{B}]$,
which, in addition, satisfies all the properties required by a Lie
product. Hence, even if classical and quantum theories seem fundamentally
different, it seems that observables are treated similarly within
their corresponding mathematical frameworks: they are elements of
a Lie algebra.

On the other hand, we saw that the generalized coordinates and momenta
have a particular mathematical structure in the Hamiltonian formalism,
they are the generators of the Heisenberg group. It seems then quite
reasonable to ask for the same in the quantum theory, so that at least
in what concerns to observables both theories are equivalent. This
motivates the fourth principle:

\bigskip{}

\textbf{Principle IV. }$\lceil$Consider a physical system which is
described classically within a Hamiltonian formalism by a set of generalized
coordinates $\mathbf{q}=\{q_{j}\}_{j=1}^{n}$ and momenta $\mathbf{p}=\{p_{j}\}_{j=1}^{n}$
at a given time. Within the quantum formalism, the corresponding observables
$\mathbf{\hat{q}}=\{\hat{q}_{j}\}_{j=1}^{n}$ and $\mathbf{\hat{p}}=\{\hat{p}_{j}\}_{j=1}^{n}$
satisfy the \textit{canonical commutation relations}\vspace{-0.3mm}
 
\begin{equation}
[\hat{q}_{j},\hat{p}_{l}]=\mathrm{i}\hbar\delta_{jl}\hspace{6mm}\text{and}\hspace{6mm}[\hat{q}_{j},\hat{q}_{l}]=[\hat{p}_{j},\hat{p}_{l}]=0.\hspace{2mm}\rfloor\label{CanComPosMom}
\end{equation}

\bigskip{}

The constant $\hbar$ is included because, while the Poisson bracket
$\{q_{j},p_{l}\}$ has no units, the commutator $[\hat{q}_{j},\hat{p}_{l}]$
has units of action. That it is exactly $\hbar$ the proper constant
can be seen only once the theory is compared with experiments.

We can now discuss how to build the self-adjoint operator corresponding
to a given observable. In general, meaningful observables are built
from symmetry principles \cite{Ballentine98}, e.g., the kinetic and
angular momenta as the generators of space translations and rotations,
respectively. An alternative route might be taken when the observable
is well-defined classically. Suppose that in the Hamiltonian formalism
the observable \textsf{A} is represented by the real phase-space function
$A(\mathbf{q},\mathbf{p})$. It seems quite natural to use then $A(\mathbf{\hat{q}},\mathbf{\hat{p}})$
as the corresponding quantum operator. However, this correspondence
faces a lot of troubles resulting from the fact that, while coordinates
and momenta commute in classical mechanics, they do not in quantum
mechanics. For example, given the classical observable $A=qp=pq$,
we could be tempted to assign to it any of the quantum operators $\hat{A}_{1}=\hat{q}\hat{p}$
or $\hat{A}_{2}=\hat{p}\hat{q}$. These two operators are not equivalent
(they do not commute) and they are not even self-adjoint, and hence
cannot represent observables. One possible solution to this problem,
at least for observables with a series expansion, is to always symmetrize
the classical expressions with respect to coordinates and momenta,
so that the resulting operator is self-adjoint. Applied to our previous
example, we should take $\hat{A}=(\hat{p}\hat{q}+\hat{q}\hat{p})/2$
according to this rule. This simple procedure leads to the correct
results most of the times, and when it fails (for example, if the
classical observable does not have a series expansion) it was proved
by Groenewold \cite{Groenewold46} that it is possible to make a faithful
systematic correspondence between classical observables and self-adjoint
operators by using more sophisticated correspondence rules.

Of course, when the observable corresponds to a degree of freedom
which is not defined in a classical context (like \textit{spin}),
it must be built from scratch based on experimental observations and/or
first principles.

Note that the commutation relations between coordinates and momenta
makes them satisfy the uncertainty relation $\Delta q\Delta p\geq\hbar/2$,
and hence, if one of them is well defined in the system, the other
must have statistics very spread around the mean.

\subsubsection{Principle V: Free evolution of the system}

The previous principles have served to define the mathematical structure
of the theory and its relation to physical systems. We have not said
anything yet about how quantum mechanics treats the evolution of the
system. As we are about to see, the formalism treats very differently
the evolution due to a measurement performed by an observer, and the
\textit{free} evolution of the system when it is not subject to observation.
The following principle specifies how to deal with the latter case.
Just as with the previous principle, it feels pretty reasonable to
keep the analogy with the Hamiltonian formalism, a motivation which
comes also from the fact that, as stated, quantum mechanics must converge
to classical mechanics in some limit. In the Hamiltonian formalism,
observables evolve according to (\ref{HamObsEvo}), so that making
the correspondence between the classical and quantum Lie products
as in the previous principle, we enunciate the fifth principle:

\bigskip{}

\textbf{Principle V. }$\lceil$The evolution of an observable $\hat{A}(\mathbf{\hat{q}},\mathbf{\hat{p}},...;t)$
is given by 
\begin{equation}
\mathrm{i}\hbar\frac{d\hat{A}}{dt}=[\hat{A},\hat{H}]+\frac{\partial\hat{A}}{\partial t},
\end{equation}
which is known as the Heisenberg equation, and where $\hat{H}(\mathbf{\hat{q}},\mathbf{\hat{p}},...;t)$
is the self-adjoint operator corresponding to the Hamiltonian of the
system. Note that the notation ``$\mathbf{\hat{q}},\mathbf{\hat{p}},...$''
emphasizes the fact that the observable may depend on fundamental
operators other than the generalized coordinates, e.g., purely-quantum
degrees of freedom such as spin.$\rfloor$

\bigskip{}

For the case of an observable and a Hamiltonian with no explicit time-dependence
(as will be assumed from now on), this evolution equation admits the
explicit solution 
\begin{equation}
\hat{A}(t)=\hat{U}^{\dagger}(t)\hat{A}(0)\hat{U}(t),\hspace{3mm}\text{being}\hspace{1.5mm}\hat{U}(t)=\exp[\hat{H}t/\mathrm{i}\hbar],
\end{equation}
a unitary operator called the \textit{evolution operator}. For explicitly
time-dependent Hamiltonians it is still possible to solve formally
the Heisenberg equation as a \textit{Dyson series}, but we will not
worry about this case, as it does not appear throughout the lectures.
Let us remark that this type of evolution ensures that if the canonical
commutation relations (\ref{CanComPosMom}) are satisfied at some
time, they will be satisfied at all times.

Note that within this formalism the state $\hat{\rho}$ of the system
is fixed in time, the observables are the ones which evolve. On the
other hand, we have seen that, on what concerns to observations (experiments),
only expectation values of operators are relevant; and for an observable
$\hat{A}$ at time $t$, these can be written as 
\begin{equation}
\langle\hat{A}(t)\rangle=\mathrm{tr}\{\hat{\rho}\hat{A}(t)\}=\mathrm{tr}\{\hat{U}(t)\hat{\rho}\hat{U}^{\dagger}(t)\hat{A}(0)\},
\end{equation}
where in the last equality we have used the cyclic property of the
trace. This expression shows that, instead of treating the observable
as the evolving operator, we can define a new state at time $t$ given
by 
\begin{equation}
\rho(t)=\hat{U}(t)\hat{\rho}(0)\hat{U}^{\dagger}(t),
\end{equation}
while keeping fixed the operator. In differential form, this expression
reads 
\begin{equation}
\mathrm{i}\hbar\frac{d\hat{\rho}}{dt}=[\hat{H},\hat{\rho}],
\end{equation}
which is known as the \textit{von Neumann equation}. When the system
is in a pure state $|\psi\rangle$, the following evolution equation
is derived for the state vector itself 
\begin{equation}
\mathrm{i}\hbar\frac{d}{dt}|\psi\rangle=\hat{H}|\psi\rangle\text{,}
\end{equation}
which is known as the \textit{Schrödinger equation}, from which the
state at time $t$ is found as $|\psi(t)\rangle=\hat{U}(t)|\psi(0)\rangle$.

Therefore, we have two different but equivalent evolution formalisms
or \textit{pictures}. In one, which we shall call \textit{Heisenberg
picture}, the state of the system is fixed, while observables evolve
according to the Heisenberg equation. In the other, which we will
denote by \textit{Schrödinger picture}, observables are fixed, while
states evolve according to the von Neumann equation.

We can even define intermediate pictures in which both the state and
the observables evolve, the so-called \textit{interaction pictures}.
To show how this is done, let us denote by $\hat{A}$ and $\hat{\rho}\left(t\right)$
an observable and the state of the system in the Schrödinger picture
(the same kind of transformation can be performed from the Heisenberg
picture). The expectation value of an observable \textsf{A} is written
as $\mathrm{tr}\{\hat{\rho}(t)\hat{A}\}$. We can define a unitary
operator $\hat{U}_{\mathrm{c}}=\exp[\hat{H}_{\mathrm{c}}t/\mathrm{i}\hbar]$,
with $\hat{H}_{\mathrm{c}}$ some self-adjoint operator, and then
a transformed state $\tilde{\rho}=\hat{U}_{\mathrm{c}}^{\dagger}\hat{\rho}\hat{U}_{\mathrm{c}}$
and a transformed observable $\tilde{A}=\hat{U}_{\mathrm{c}}^{\dagger}\hat{A}\hat{U}_{\mathrm{c}}$.
This transformation leaves invariant the expectation value, which
can be evaluated as $\mathrm{tr}\{\tilde{\rho}\left(t\right)\tilde{A}\left(t\right)\}$,
but now the evolution equations of the state and the observable read
\begin{equation}
\mathrm{i}\hbar\frac{d\tilde{\rho}}{dt}=[\hat{H}_{\text{I}},\tilde{\rho}]\text{ \ \ \ \ and \ \ \ \ }\mathrm{i}\hbar\frac{d\tilde{A}}{dt}=[\tilde{A},\hat{H}_{\mathrm{c}}],
\end{equation}
so that within this new picture states evolve according to the \textit{interaction-picture
Hamiltonian} $\hat{H}_{\mathrm{I}}=\hat{U}_{\mathrm{c}}^{\dagger}\hat{H}\hat{U}_{\mathrm{c}}-\hat{H}_{\mathrm{c}}$,
while observables evolve according to the \textit{transformation Hamiltonian}
$\hat{H}_{\mathrm{c}}$.

The same kind of game can be played from the Heisenberg picture. Let
us denote in this case by $\hat{A}(t)$ and $\hat{\rho}$ an observable
and the state in the Heisenberg picture. The transformed state $\tilde{\rho}=\hat{U}_{\mathrm{c}}^{\dagger}\hat{\rho}\hat{U}_{\mathrm{c}}$
and observable $\tilde{A}=\hat{U}_{\mathrm{c}}^{\dagger}\hat{A}_{\mathrm{}}\hat{U}_{\mathrm{c}}$
satisfy then the evolution equations
\begin{equation}
\mathrm{i}\hbar\frac{d\tilde{\rho}}{dt}=[\hat{H}_{\text{c}},\tilde{\rho}]\text{ \ \ \ \ and \ \ \ \ }\mathrm{i}\hbar\frac{d\tilde{A}}{dt}=[\tilde{A},\hat{H}_{\mathrm{I}}].
\end{equation}

\subsubsection{Principle VI: Post-measurement state}

\label{AxiomVI}

The previous postulate specifies how the free evolution of the system
is taken into account in the quantum mechanical formalism. It is then
left to specify how the state of the system evolves after a measurement
is performed on it. For reasons that we will briefly review after
enunciating the principle, this is probably the most controversial
point in the quantum formalism. Indeed, while in classical physics
we assume that it is possible to perform measurements onto the system
without disturbing its state, this final quantum-mechanical principle
states:

\bigskip{}

\textbf{Principle VI. }$\lceil$If upon a measurement of a countable
observable \textsf{A} the outcome $a_{m}$ is obtained, then immediately
after the measurement the state of the system \textit{collapses} to
$|a_{m}\rangle$.$\rfloor$

\bigskip{}

The principle assumes that, after the measurement, the observer gains
knowledge about the measurement outcome, what we will denote as a
\textit{selective} measurement. However, suppose that for some reason
the user interface of the measurement device does not allow us to
distinguish between a set of outcomes $\{a_{m_{k}}\}_{k=1,2,...,K}$,
what we will denote by a \textit{partially-selective} measurement.
Then, after the corresponding experimental outcome is obtained, the
best estimate that the observer can assign to the post-measurement
state is the ensemble decomposition $\{\bar{p}_{m_{k}},\vert a_{m_{k}}\rangle\}_{k=1,2,...,K}$
with relative probabilities $\bar{p}_{m_{k}}=p_{m_{k}}/\sum_{k=1}^{K}p_{m_{k}}$,
since the real outcome is unknown, but the \textit{a priori} probabilities
$p_{m}$ of the possible outcomes are known. Hence, in such case we
would assign the post-measurement state $\hat{\rho}_{\{m_{1},m_{2},...,m_{K}\}}=\sum_{k=1}^{K}\bar{p}_{m_{k}}\vert a_{m_{k}}\rangle\langle a_{m_{k}}|$
to the system. The extreme case in which the outcome of the measurement
is simply not recorded, so that we cannot know which outcome occurred
and the best estimate for the post-measurement state is $\hat{\rho}'=\sum_{m=1}^{d}p_{m}\vert a_{m}\rangle\langle a_{m}|$,
is known as a \textit{non-selective} measurement.

So far we have considered the post-measurement state in the case of
measuring a countable observable. The continuous case is tricky, since,
as mentioned in Sec. \ref{InfiniteHilbert}, the eigenvectors of a
continuous observable cannot correspond to physical states (they cannot
be normalized). On the other hand, one can always argue that the detection
of a single definite value out of the spectrum $\{x\}_{x\in\mathbb{R}}$
of a continuous observable $\hat{X}$ would require an infinite precision,
whereas detectors always have some finite precision. Consequently,
there are two natural ways of dealing with such problem:
\begin{itemize}
\item Accepting that measuring continuous observables is simply not possible,
and what is measured in real experiments is always some countable
version of them, which only in some unphysical limit reproduce the
continuous measurement precisely. An example of this consists in the
process of \textit{bining} the continuous observable, which assumes
that the detector can only distinguish between pixels with width $\Delta_{x}$
centered at certain points $\{x_{k}=k\Delta_{x}\}_{k\in\mathbb{Z}}$
in the spectrum of the continuous observable, so what is measured
is instead the countable observable 
\begin{equation}
\hat{X}_{\text{count}}=\sum_{k=-\infty}^{\infty}x_{k}\vert x_{k}\rangle\hspace{-0.4mm}\langle x_{k}\vert\text{,}\hspace{6mm}\text{with}\hspace{2mm}\vert x_{k}\rangle=\frac{1}{\sqrt{\Delta_{x}}}\int_{x_{k}-\Delta_{x}/2}^{x_{k}+\Delta_{x}/2}\hspace{-0.2cm}dx\vert x\rangle.
\end{equation}
\item Allowing for the possible outcomes of the measurement to still be
continuous, but with an uncertainty given by the precision of the
measurement device. In the case of starting with a pure state, this
would simply mean that the post-measurement state is not an eigenvector
of the continuous operator, but a normalizable superposition of several
of them, spanning around the measured value with a width given by
the measurement's precision. This intuitive approach can be formalized
with the theory of generalized quantum measurements \cite{WisemanBook,Jacobs06}. 
\end{itemize}
In any case, it is sometimes useful for theoretical calculations to
proceed as if perfectly precise measurements were possible, with the
system collapsing to one eigenvector of the continuous observable.
However, it is important to keep in mind that this is just an unphysical
idealization, whose corrections have to be taken into account when
applying it to a real situation.

We can pass now to discuss the controversial aspects of this principle,
of which a pedagogical introduction can be found in \cite{Basdevant02}
(see also \cite{Bassi03} and Appendix E of \cite{GalindoBook90}).
In short, the problem is that, even though it leads to predictions
which fully agree with the observations, the principle somehow creates
an \textit{inconsistency} in the theory because of the following argument.
According to Principle V, the unitary evolution of a system not subject
to observation is \textit{reversible}, that is, one can always change
the sign of the relevant terms of the Hamiltonian which contributed
to the evolution (at least conceptually), and come back to the original
state. On the other hand, the \textit{collapse} principle claims that
when the system is put in contact with a measurement device and an
observable is measured, the state of the system collapses to some
other state in an \textit{irreversible} way\footnote{Note that in the literature the terms \textit{reversible} and \textit{irreversible}
are sometimes replaced by \textit{linear} and \textit{nonlinear},
referring to the fact that unitary evolution comes from a linear equation
(Schrödinger or von Neumann equations), while measurement-induced
dynamics becomes nonlinear through its dependence on the probability
of the possible outcomes, which in turn depend on the state.}. However, coming back to Principle V, the whole measurement process
could be described reversibly by considering, in addition to the system's
particles, the evolution of all the particles forming the measurement
device (or even the human who is observing the measurement outcome
if needed!), that is, the Hamiltonian for the whole `observed system
+ measurement device' scenario. Hence, it seems that, when including
the collapse principle, quantum mechanics allows for two completely
different descriptions of the measurement process, one reversible
and one irreversible, without giving a clear rule for when to apply
each. It is in this sense that the theory contains an inconsistency.

There are three\footnote{But each with many sub-interpretations differing in subtle, or even
not so subtle points. In essence, one can find a lot of truth to the
say ``Give me a room with $N$ physicists and I'll find you $N+1$
different interpretations of quantum mechanics''.} main positions that physicists have taken regarding how this inconsistency
might be solved, which we will refer too as \textit{objective}, \textit{subjective},
and \textit{apparent} collapse interpretations, and whose (highly
simplified) main ideas we hereby discuss\footnote{One interpretation of quantum mechanics that is formulated with a
completely different set of principles, and hence does not fit this
list is \textit{Bohmian mechanics} \cite{HollandBook93,MompartBook12},
in which particles follow deterministic trajectories, but determined
from a \textit{guiding wave} that obeys the Schrödinger equation.}:
\begin{itemize}
\item \textbf{Objective collapse.} There is a clear boundary (yet to be
found) between the quantum and the classical worlds. In the classical
world, to which measurement devices and observers belong, there exists
some \textit{decoherence mechanism} that prevents systems from being
in a superposition of states corresponding to mutually exclusive values
of their observables. When the measurement device enters in contact
with the quantum system, the latter becomes a part of the classical
world, and the aforementioned decoherence mechanism forces its collapse.
Hence, within this interpretation the collapse is pretty much a real
physical process that we still need to understand along with the quantum/classical
boundary. There are several \textit{collapse theories} available at
the moment \cite{Bassi03}, part of which we expect to be able to
falsify or confirm in the near future with modern quantum technologies
based on, for example, opto-, electro-, or magneto-mechanics \cite{RomeroIsart11}. 
\item \textbf{Subjective collapse.} The state is simply a mathematical object
which conveniently describes the statistics of experiments, but that
otherwise has no physical significance. As such, what the quantum
formalism provides is simply a set of rules for how to update our
best estimate to the state according to the information that we have
about the system. In this sense, the collapse is just the way that
an observer subjectively updates the state of the system after gathering
the information concerning the outcome. Quantum Bayesianism or \textit{QBism}
\cite{Fuchs14,Fuchs12} is possibly the most refined of such interpretations,
which has gathered a lot of momentum in the last years. 
\item \textbf{Apparent collapse.} The measurement can be described without
the need of abandoning the framework of Principle V as a joint unitary
transformation onto the system and the measurement device (even including
the observer), leading to a final entangled state\footnote{We will discuss in detail about the concept of entanglement towards
the end of this chapter.} between these in which the eigenstates of the system's observable
are in one-to-one correspondence with a set of macroscopically distinct
\textit{pointer} states of the measurement device \cite{Basdevant02}.
Hence, after a measurement, reality splits into many \textit{branches}
where observers experience different outcomes and which stay in a
quantum superposition, and collapse appears just an illusion coming
from the fact that we only see the effective dynamics projected into
the corresponding branch that we are experiencing. This approach finds
its best-developed expression in the so-called \textit{many-worlds
interpretation} \cite{Vaidman15}, which describes the quantum-mechanical
framework as unitary evolution of a pure state (\textit{wave function})
of the whole universe, which includes all the branches or \textit{worlds}\footnote{Believers of the many-worlds interpretation, or even people who are
not sure of their quantum-mechanical beliefs (better safe than sorry),
are strongly encouraged to use the \textit{quantum world splitter}
when in need of choosing between equally reasonable options in life:
www.qol.tau.ac.il.}. 
\end{itemize}
In any case whether of objective, subjective, or apparent value, it
is clear that the collapse principle is of great \textit{operational}
value, that is, it is currently the easiest successful way of analyzing
schemes involving measurements, and hence we will apply it when needed.

\subsubsection{Classical versus quantum correlations: entanglement\label{Sec:Entanglement}}

Consider two systems $A$ and $B$ (named after Alice and Bob, two
observers which are able to interact locally with their respective
system) with associated Hilbert spaces $\mathcal{H}_{A}$ and $\mathcal{H}_{B}$
of dimension $d_{A}$ and $d_{B}$, respectively. The systems are
prepared in some state $\hat{\rho}_{AB}$ acting on the joint space
$\mathcal{H}_{A}\otimes\mathcal{H}_{B}$. Recall from the discussion
after Principle III that Alice and Bob can reproduce the statistics
of measurements performed on their subsystems via the reduced states
$\hat{\rho}_{A}=\mathrm{tr}_{B}\{\hat{\rho}_{AB}\}$ and $\hat{\rho}_{B}=\mathrm{tr}_{A}\{\hat{\rho}_{AB}\}$,
respectively.

When the state of the joint system is of the type $\hat{\rho}_{AB}^{(\mathrm{prod})}=\hat{\rho}_{A}\otimes\hat{\rho}_{B}$,
that is, a tensor product of two arbitrary density operators, the
actions performed by Alice on system $A$ won't affect Bob's system,
the statistics of which are given by $\hat{\rho}_{B}$, no matter
the actual state $\hat{\rho}_{A}$. In this case $A$ and $B$ are
completely \textit{uncorrelated}. For any other type of joint state,
$A$ and $B$ will share some kind of correlation.

Correlations are not strange in classical mechanics. Hence, a problem
of paramount relevance in quantum mechanics is understanding which
type of correlations can appear at a classical level, and which are
purely quantum. This is because only if the latter are present, one
can expect the correlated systems to be useful for quantum-mechanical
applications which go beyond what is classically possible, the paradigmatic
example being the exponential speed up of computational algorithms.

Intuitively, the state of the system will induce only classical correlations
between $A$ and $B$ when it can be written in the form 
\begin{equation}
\hat{\rho}_{AB}^{(\mathrm{sep})}=\sum_{k=1}^{M}w_{k}\hat{\rho}_{A}^{(k)}\otimes\hat{\rho}_{B}^{(k)},\label{RhoSep}
\end{equation}
where the $\hat{\rho}^{(k)}$'s are density operators, $\{w_{k}\}_{k=1,2,...,M}$
is a probability distribution, $M$ can be infinite, and the index
$k$ can even be continuous in some range, in which case the sum turns
into an integral in that range and the probability distribution into
a probability density function. Indeed, a state of the type (\ref{RhoSep})
can be prepared by a protocol involving only local actions and classical
correlations: Alice and Bob can share a classical machine which randomly
picks a value of $k$ according to the distribution $\{w_{k}\}_{k=1,2...,M}$
to trigger the preparation of the states $\hat{\rho}_{A}^{(k)}$ and
$\hat{\rho}_{B}^{(k)}$, what can be done locally, and hence cannot
induce further correlations. If the process is automatized so that
Alice and Bob do not learn the outcome $k$ of the random number generator,
the best estimate that they can assign to the state in order to reproduce
the statistics of any forthcoming experiments is the mixture $\hat{\rho}_{AB}^{(\mathrm{sep})}$.
In other words, the state does not contain quantum correlations if
it can be prepared using only \textit{local operations and classical
communication}.

There is yet another intuitive way of justifying that states which
cannot be written in the separable form (\ref{RhoSep}) will make
$A$ and $B$ share quantum correlations. The idea is that arguably
the most striking difference between classical and quantum mechanics
is the \textit{superposition principle}, that is, the possibility
of states corresponding to \textit{mutually exclusive} properties
of the system to \textit{interfere} (e.g., two different colors, $\vert\text{blue}\rangle+\vert\text{red}\rangle$).
Hence, it is intuitive that correlations should have a quantum nature
only when they come from some kind of superposition of joint states
corresponding to mutually exclusive properties of the correlated systems
(e.g., color in one system and flavor in the other, $\vert\text{blue}\rangle\otimes\vert\text{sweet}\rangle+\vert\text{red}\rangle\otimes\vert\text{sour}\rangle$),
in which case the state cannot be written as a tensor product of two
independent states ($\vert\text{blue}\rangle\otimes\vert\text{sour}\rangle$)
or as a purely classical statistical mixture of these ($\vert\text{blue}\rangle\hspace{-0.4mm}\langle\text{blue}\vert\otimes\vert\text{sweet}\rangle\hspace{-0.4mm}\langle\text{sweet}\vert+\vert\text{red}\rangle\hspace{-0.4mm}\langle\text{red}\vert\otimes\vert\text{sour}\rangle\hspace{-0.4mm}\langle\text{sour}\vert$).

States of the type $\hat{\rho}_{AB}^{(\mathrm{sep})}$ are called
\textit{separable}. Any state which cannot be written in this form
will induce quantum correlations between $A$ and $B$. These correlations
which cannot be generated by classical means are known as \textit{entanglement},
and states which are not separable are called \textit{entangled states}.

\newpage{}


\begin{thebibliography}{10}
\bibitem{QO1} W. H. Louisell, \emph{Quantum statistical properties
of radiation} (John Wiley \& Sons, 1973).

\bibitem{QO2}C. Cohen-Tannoudji, J. Dupont-Roc, and J. Grynberg,
\emph{Photons and atoms: Introduction to quantum electrodynamics}
(John Wiley \& Sons, 1989).

\bibitem{QO3}H. J. Carmichael, \emph{An open systems approach to
quantum optics} (Springer Verlag, 1993).

\bibitem{QO4}D. F. Walls and G. J. Milburn, \emph{Quantum optics}
(Springer, 1994). 

\bibitem{QO5}L. Mandel and E. Wolf, \emph{Optical coherence and quantum
optics} (Cambridge University Press, 1995).

\bibitem{QO6}M. O. Scully and M. S. Zubairy, \emph{Quantum optics}
(Cambridge University Press, 1997). 

\bibitem{QO7}H. J. Carmichael, \emph{Statistical methods in quantum
optics 1: Master equations and Fokker\textendash Planck equations}
(Springer Verlag, 1999).

\bibitem{QO8}C. W. Gardiner and P. Zoller, \emph{Quantum Noise} (Springer-Verlag,
2000).

\bibitem{SchleichBook} W. P. Schleich, \textit{Quantum optics in
phase space} (Wiley-VCH, 2001).

\bibitem{QO9}P. D. Drummond and Z. Ficek (Eds.), \emph{Quantum squeezing}
(Springer Verlag, 2004).

\bibitem{QO10}C. C. Gerry and P. L. Knight, \emph{Introductory quantum
optics} (Cambridge University Press, 2005). 

\bibitem{QO11}H. J. Carmichael, \emph{Statistical methods in quantum
optics 2: Non-classical fields} (Springer Verlag, 2008).

\bibitem{QO12}G. Grynberg, A. Aspect, and C. Fabre, \emph{Introduction
to quantum optics} (Cambridge University Press, 2010). 

\bibitem{QO13}C. W. Gardiner, P. Zoller, \emph{The Quantum World
of Ultra-Cold Atoms and Light, Book I} (Imperial College press, 2014).

\bibitem{QO14}P. D. Drummond and M. Hillery, \emph{The quantum theory
of nonlinear optics} (Cambridge University press, 2014).

\bibitem{BraunsteinVanLoockReview} S. L. Braunstein and P. van Loock,
\textit{Quantum information with continuous variables}, Rev. Mod.
Phys. \textbf{77}, 513 (2005). 

\bibitem{Weedbrook12} C. Weedbrook, S. Pirandola, R. García-Patrón,
N. J. Cerf, T. C. Ralph, J. H. Shapiro, and S. Lloyd, \textit{Gaussian
quantum information}, Rev. Mod. Phys. \textbf{84}, 621 (2012). 

\bibitem{ParisBook} A. Ferraro S. Olivares, and M. G A Paris, \textit{Gaussian
States in Quantum Information} (Napoli Series on Physics and Astrophysics,
2005). 

\bibitem{CerfBook07}N. J. Cerf, G. Leuchs, and E. S. Polzik (Eds.),
\emph{Quantum Information with Continuous Variables of Atoms and Light}
(Imperial College Press, 2007).

\bibitem{ShapiroNotes}J. H. Shapiro, \emph{Quantum optical communication
Lecture notes,} https://ocw.mit.edu/courses/electrical-engineering-and-computer-science/6-453-quantum-optical-communication-fall-2016/readings-and-lecture-slides/index.htm.

\bibitem{KokBook10}P. Kok and B. W. Lovett, \emph{Introduction to
Optical Quantum Information Processing} (Cambridge University Press,
2010).

\bibitem{GoldsteinBook} H. Goldstein, C. Poole, and J. Safko, \textit{Classical
mechanics} (Addison Wesley, 2001). 

\bibitem{GreinerMechanicsBook} W. Greiner, \textit{Classical mechanics:
Point particles and relativity} (Springer Verlag, 1989). 

\bibitem{GreinerMechanics2Book} W. Greiner and J. Reinhardt, \textit{Classical
mechanics: Systems of particles and Hamiltonian dynamics} (Springer
Verlag, 1989).

\bibitem{HandFinchBook}L. N. Hand and J. D. Finch, \emph{Analytical
Mechanics} (Cambridge University Press, 1998).

\bibitem{DiracBook30} P. A. M. Dirac, \textit{The principles of quantum
mechanics} (Oxford university press, 1930). 

\bibitem{PrugoveckyBook71} E. Prugove\v{c}ky, \textit{Quantum mechanics
in Hilbert space} (Academic Press, 1971). 

\bibitem{GalindoBook90} A. Galindo and P. Pascual, \textit{Quantum
Mechanics I} (Springer Verlag, 1990). 

\bibitem{GelfandBook64} I. M. Gelfand and N. Y. Vilenkin, \textit{Generalized
Functions, Vol. IV} (Academic Press, 1964). 

\bibitem{WhitakerBook} A. Whitaker, \textit{Einstein, Bohr, and the
Quantum Dilemma} (Cambridge University Press, 1996). 

\bibitem{Waerden68} B. L. van der Waerden, \textit{Sources of Quantum
Mechanics} (Dover Publications, 1968). 

\bibitem{Schrodinger26} E. Schrödinger, \textit{An undulatory theory
of the mechanics of atoms and molecules}, Phys. Rev. \textbf{28},
1049 (1926). 

\bibitem{Heisenberg25} W. Heisenberg, \textit{Über quantentheoretische
umdeutung kinematischer und mechanischer beziehungen}, Zs. f. Phys.
\textbf{33}, 879 (1925). 

\bibitem{Born25} M. Born and P. Jordan, \textit{Zur quantenmechanik},
Zs. f. Phys., \textbf{34}, 858 (1925). 

\bibitem{Born26} M. Born, W. Heisenberg, and P. Jordan, \textit{Zur
quantenmechanik ii}, Zs. f. Phys. \textbf{35}, 557 (1926). 

\bibitem{Dirac26} P. A. M. Dirac, \textit{The fundamental equations
of quantum mechanics}, Proc. Roy. Soc. A \textbf{109}, 642 (1926). 

\bibitem{vonNeumann32} J. von Neumann, \textit{Mathematische Grundlagen
der Quantenmechanik} (Springer Verlag, 1932). 

\bibitem{vonNeumann55} J. von Neumann, \textit{Mathematical foundations
of quantum mechanics} (Princeton university press, 1955). 

\bibitem{CohenTannoudjiBookI} C. Cohen-Tannoudji, B. Diu, and F.
Laloë, \textit{Quantum mechanics, Vol. I} (Hermann and John Wiley
\& Sons, 1977). 

\bibitem{CohenTannoudjiBookII} C. Cohen-Tannoudji, B. Diu, and F.
Laloë \textit{Quantum mechanics, Vol. II} (Hermann and John Wiley
\& Sons, 1977) 

\bibitem{GreinerQuantumBook1} W. Greiner and B. Müller, \textit{Quantum
mechanics: An introduction} (Springer Verlag, 1989). 

\bibitem{GreinerQuantumBook2} W. Greiner, \textit{Quantum mechanics:
Symmetries} (Springer Verlag, 1989). 

\bibitem{Basdevant02} J.-L. Basdevant and J. Dalibard, \textit{Quantum
mechanics} (Springer Verlag, 2002). 

\bibitem{Ballentine98}L. E. Ballentine, \emph{Quantum Mechanics:
A Modern Development} (World Scientific, 1998)

\bibitem{NielsenChuangBook} M. A. Nielsen and I. L. Chuang, \textit{Quantum
information and quantum computation} (Cambridge University Press,
2000). 

\bibitem{Groenewold46} H. J. Groenewold, \textit{On the principles
of elementary quantum mechanics}, Physica \textbf{12}, 405 (1946). 

\bibitem{WisemanBook} H. M. Wiseman and G. J. Milburn, \textit{Quantum
measurement and control} (Cambridge University Press, 2009). 

\bibitem{Jacobs06} K. Jacobs and D. A. Steck, \textit{A Straightforward
Introduction to Continuous Quantum Measurement}, Contemporary Phys.
\textbf{47}, 279 (2006).

\bibitem{Bassi03} A. Bassi and G. C. Ghirardi, \textit{Dynamical
reduction models}, Phys. Rep. \textbf{379}, 257 (2003). 

\bibitem{HollandBook93}P. R. Holland, \emph{The Quantum Theory of
Motion} (CambridgeUniversityPress, 1993).

\bibitem{MompartBook12} X. Oriols and J. Mompart, \emph{Applied Bohmian
Mechanics: From Nanoscale Systems to Cosmology} (Pan Stanford Publishing,
2012).

\bibitem{RomeroIsart11} O. Romero-Isart, \textit{Quantum superposition
of massive objects and collapse models}, Phys. Rev. A \textbf{84},
052121 (2011). 

\bibitem{Fuchs14} C. A. Fuchs,N. D. Mermin, and R. Schack, \textit{An
Introduction to QBism with an Application to the Locality of Quantum
Mechanics}, Am. J. Phys. \textbf{82}, 749 (2014).

\bibitem{Fuchs12}C. A. Fuchs, \emph{Interview with a quantum Bayesian,}
arXiv:1207.2141 (2012).

\bibitem{Vaidman15} L. Vaidman, \textit{Many-worlds interpretation
of quantum mechanics} (The Stanford Encyclopedia of Philosophy, Spring
2015 Edition); http://plato.stanford.edu/entries/qm-manyworlds.

\bibitem{Maxwell1865}J. C. Maxwell, \emph{A dynamical theory of the
electromagnetic field}, Philosophical Transactions of the Royal Society
of London \textbf{155}, 459512 (1865).

\bibitem{Hertz1893}H. R. Hertz, \emph{Electric waves: being researches
on the propagation of electric action with finite velocity through
space} (Macmillan and Co., 1893).

\bibitem{Heaviside893Book}O. Heaviside, \emph{Electrical papers,
Volume II} (Macmillan and Co., 1894).

\bibitem{Jackson62book}J. Jackson, \emph{Classical electrodynamics}
(John Wiley \& Sons, 1962).

\bibitem{NavarretePhDthesis} C. Navarrete-Benlloch, \textit{Contributions
to the quantum optics of multi-mode optical parametric oscillators}
(PhD dissertation, 2011), arXiv:1504.05917.

\bibitem{Griffiths99book}D. J. Griffiths, \emph{Introduction to electrodynamics}
(Prentice Hall, 1999).

\bibitem{NavarreteBook15}C. Navarrete-Benlloch, \emph{An introduction
to the formalism of quantum information with continuous variables
}(Morgan \& Claypool and IOP publishing, 2015).

\bibitem{Glauber63a}R. J. Glauber, \emph{Photon correlations}, Phys.
Rev. Lett. 10, 84 (1963).

\bibitem{Glauber63b}R. J. Glauber, \emph{The quantum theory of optical
coherence}, Phys. Rev. \textbf{130}, 2529 (1963).

\bibitem{Glauber63c}R. J. Glauber, \emph{Coherent and incoherent
states of the electromagnetic field}, Phys. Rev. \textbf{131}, 2766
(1963).

\bibitem{GlauberNobelLecture}R. J. Glauber, Nobel Lecture (2006),
www.nobelprize.org.

\bibitem{PuriBook01}R. R. Puri, \emph{Mathematical methods of quantum
optics }(Springer, 2001).

\bibitem{Lo90} C.F. Lo, \emph{Eigenvalues and eigenfunctions of squeeze
operators}, Phys. Rev. A \textbf{42}, 6752 (1990).

\bibitem{HBT}R. Hanbury-Brown and R. Q. Twiss, Nature \textbf{177},
27 (1956); Proc. Roy. Soc. (London) A \textbf{243}, 291 (1957).

\bibitem{97squeezing}H. Vahlbruch, M. Mehmet, K. Danzmann, and R.
Schnabel, \emph{Detection of 15 dB Squeezed States of Light and their
Application for the Absolute Calibration of Photoelectric Quantum
Efficiency}, Phys. Rev. Lett. \textbf{117}, 110801 (2016).

\bibitem{PeskinSchroeder}M. E. Peskin and D. V. Schroeder, \emph{An
introduction to quantum field theory} (Perseus Books, 1995)

\bibitem{GreinerQFT}W. Greiner and J. Reinhardt, \emph{Field quantization}
(Springer Verlag, 1996)

\bibitem{GreinerQED}W. Greiner and J. Reinhardt, \emph{Quantum electrodynamics}
(Springer Verlag, 1992)

\bibitem{BornWolf}M. Born and E. Wolf, \emph{Principles of optics}
(Pergamon, 1980)

\bibitem{Boyd}R. W. Boyd, \emph{Nonlinear optics} (Elsevier, 2003)

\bibitem{BraakRabi}D. Braak, \emph{On the integrability of the Rabi
model}, Phys. Rev. Lett. \textbf{107}, 100401 (2011)

\bibitem{QRM-review}Q. Xie, H. Zhong, M. T. Batchelor, C. Lee, \emph{The
quantum Rabi model: solution and dynamics}, J. Phys. A: Math. Theor.
\textbf{50}, 113001 (2017)

\bibitem{CNB_Numerics}C. Navarrete-Benlloch, \emph{Open systems dynamics:
Simulating master equations in the computer}, arXiv:1504.05266

\bibitem{CatchJumps}Z.K. Minev, S.O. Mundhada, S. Shankar, P. Reinhold,
R. Gutierrez-Jauregui, R.J. Schoelkopf, M. Mirrahimi, H.J. Carmichael,
and M.H. Devoret, \emph{To catch and reverse a quantum jump mid-flight},
Nature \textbf{570}, 200 (2019)
\end{thebibliography}
\end{document}